\documentclass[11pt]{article}

\usepackage{epsfig,a4wide,amsmath,amssymb,amsfonts,mathrsfs,chicago}
\RequirePackage{amssymb}
\RequirePackage{doublespace}
\RequirePackage{amsfonts}
\RequirePackage{sectsty}
\RequirePackage{bbox}
\RequirePackage{latexsym}
\numberwithin{equation}{section}

\font\vec=cmbrbx10
\font\smvec=cmbrbx10 scaled 700
\font\grvec=omsegrb
\font\smgrvec=omsegrb scaled 700
\font\oneone=bbm12

\sectionfont{\nohang\centering\huge}
\subsectionfont{\Large}

\newcommand{\n}{\hbox{\vec n}}

\renewcommand{\d}{\hbox{\vec d}}
\newcommand{\e}{\hbox{\vec e}}
\renewcommand{\v}{\hbox{\vec v}}

\renewcommand{\l}{\hbox{\vec l}}

\newcommand{\la}{\lambda}
\renewcommand{\k}{\hbox{\vec k}}

\newcommand{\m}{\hbox{\vec m}}

\newcommand{\mbar}{\bar{\hbox{\vec m}}}

\newcommand{\scite}{\shortciteN}
\newcommand{\lcite}{\citeN}
\renewcommand{\r}{\tilde{r}}
\renewcommand{\t}{\tilde{t}}

\newcommand{\hR}{\hat{\rho}}

\def\artanh{\mathop{\rm tanh}^{-1}\nolimits}

\begin{document}

\newcounter{count}

\thispagestyle{empty}
\begin{center}
\huge 
University of Southampton\\[10pt]

\vspace{1cm}
\Huge
Non-linear numerical Schemes in\\ General Relativity\\[10pt]

\vspace{1cm}
\huge
by \\[10pt]

\vspace{1cm}
Ulrich Sperhake \\[10pt]

\vspace{1.5cm}
\huge 
Submitted for the degree of Doctor of Philosophy\\[10pt]

\vspace{1.5cm}
\huge
Faculty of Mathematical Studies\\[10pt]

\vspace{0cm}
September, 2001

%\vspace{2.0cm}
%\large
%This thesis was submitted for examination in September, 2001. It does not
%necessarily represent the final form of the thesis as deposited in the
%University after examination.
 
\end{center}

\newpage
\thispagestyle{empty}
\begin{center}
\setstretch{1.0}
\Large
UNIVERSITY OF SOUTHAMPTON \\[6pt]
\underline{ABSTRACT} \\[6pt]
FACULTY OF MATHEMATICAL STUDIES \\[6pt]
MATHEMATICS \\[6pt]
\underline{Doctor of Philosophy}\\[6pt]
NON-LINEAR NUMERICAL SCHEMES IN GENERAL RELATIVITY\\[6pt]
by Ulrich Sperhake\\[6pt]
\end{center}

\setstretch{1.0}
\normalsize
This thesis describes the application of numerical techniques to solve
Einstein's field equations in three distinct cases.\\
First we present the first long-term stable second order convergent
Cauchy characteristic matching code in
cylindrical symmetry including both gravitational degrees of freedom.
Compared with previous work we achieve a substantial simplification
of the evolution equations as well as the relations at the interface
by applying the method of Geroch decomposition to both the inner and the
outer region. We use analytic vacuum solutions with one and two
gravitational degrees of freedom to demonstrate the accuracy
and convergence properties of the code.

\setlength{\parindent}{0.5cm}
In the second part we numerically solve the equations for static and dynamic
cosmic strings of infinite length coupled to gravity and provide
the first fully non-linear evolutions of cosmic strings in curved spacetimes.
The inclusion
of null infinity as part of the numerical grid allows us to apply
suitable boundary conditions on the metric and the matter fields to
suppress unphysical divergent solutions. The resulting code is checked for
internal consistency by a convergence analysis and also by verifying that
static cosmic string initial data remain constant when evolved. The
dynamic code is also shown to reproduce analytic vacuum solutions with
high accuracy. We then study the interaction between a Weber-Wheeler
pulse of gravitational radiation with an initially static string. The
interaction causes the string to oscillate with frequencies proportional
to the masses of its scalar and vector field. After the pulse has
largely radiated away, the string continues to ring but the oscillations
slowly decay and eventually the variables return to their equilibrium
values.
 
In the final part of the thesis we probe a new numerical
approach for highly accurate
evolutions of neutron star oscillations in the case of radial oscillations
of spherically symmetric stars. For this purpose we decompose
the problem into a
static background governed by the Tolman-Oppenheimer-Volkoff equations
and time dependent perturbations. In
contrast to conventional treatments, the fully non-linear form
of the resulting perturbative equations is used. In an Eulerian formulation 
of the problem the movement of the surface of the star relative to the 
fixed numerical grid leads to difficulties in the numerical as well as 
the algebraic analysis. In order to alleviate the surface problem
we use a simplified neutron star model
to study the non-linear coupling of eigenmodes. By virtue of the high
accuracy of our numerical method we are able to analyse the excitation
of eigenmodes
% due to non-linear effects
over a wide range of initial
amplitudes. We find two distinct regimes, a weakly non-linear
regime where the coefficients of higher order eigenmodes increase
quadratically with the initial amplitude and a moderately non-linear
regime where this increase steepens and an initially present mode
of order $j$ couples more efficiently to modes of order $2j$, $3j$
and so on.

We conclude this work with 
the development of a fully non-linear perturbative Lagrangian code. We 
demonstrate how the difficulties at the surface of the star that arise 
in an Eulerian framework are naturally resolved in the Lagrangian 
formulation. This code is used to study the formation of discontinuities
near the surface for initial data of low amplitude.
%\setstretch{1.5}\normalsize
\setlength{\parindent}{0cm}

\setstretch{1.0}
\normalsize

%\listoffigures
\pagenumbering{roman}

\newpage
\thispagestyle{plain}
\tableofcontents
%\thispagestyle{empty} 
%\addtocontents{toc}{\protect\thispagestyle{empty}}\tableofcontents

\newpage
\begin{center}
  \Large {\bf Acknowledgements}\\[10pt]
\end{center}

In the vain hope of achieving completeness I would like to
express my gratitude to the following people. \\[10pt]
My supervisor Ray d'Inverno for making this whole project possible and for
all support and suggestions I have received in the course of
the last three years.\\
Nils Andersson, Carsten Gundlach and James Vickers for many suggestions,
discussions, encouragement and constructive criticism of my work.\\
All other members of the Southampton relativity group who provided
a phantastic working environment and were barely ever short of encouraging
comments, support and ideas. \\
My colleague Robert Sj{\"o}din with whom I worked jointly on the cosmic
string project. The derivation of the equations for a cosmic string
in section 4 has been done in collaboration with him and he reformulated
the analytic vacuum solutions used in section 3 and 4 in the way
they are implemented in the codes. \\
Last but not least Philippos Papadopoulos of the University of Portsmouth
for his many suggestions, ideas and support concerning the work in section 5.

\newpage
\begin{center}
  \Large {\bf Notation}\\[10pt]
\end{center}

Unless stated otherwise, the following conventions apply. Greek indices run
from 0 to 3, whereas Latin indices are used for 3-dimensional
quantities. We will generally represent vectors and tensors of higher rank
with boldfaced letters (e.g. $\hbox{\vec{T}}$).
Sometimes we will denote vectors, i.e. tensors of rank (1,0), by
partial differential operators (e.g.
$\bbox{\partial}_t$). If we need to distinguish between a one-form and
a vector, the one-form will be marked with a tilde
(e.g. $\tilde{\hbox{\vec{u}}}$).
If a one-form is the exterior derivative of a scalar
function $f$, it will be denoted by $\d f$ and the tilde will be
omitted. If ${\v}$ is a vector, then $\tilde{\v}$ is the associated
one-form, i.e. $\tilde{\v} = \hbox{\vec g}({\hbox{\vec v}},.)$.
In coordinate free language
the contraction of a one-form $\tilde{\hbox{\vec{u}}}$ with a
vector $\hbox{\vec{v}}$ will be written as
$\langle \tilde{\hbox{\vec u}},{\hbox{\vec v}}\rangle$.
The 4-dimensional Riemann tensor
and its contractions will be denoted by the standard $\hbox{\vec R}$. For the
3-dimensional Riemann tensor we always use $\mathcal{R}$.
We will use square brackets to denote the commutator as is done in
quantum mechanics, so for example $[\nabla_{\alpha}, \nabla_{\beta}]
= \nabla_{\alpha} \nabla_{\beta} - \nabla_{\beta} \nabla_{\alpha}$.
 Throughout
this work we will use natural units with $c=1=G$ and
the sign convention ``$-++\,+$'' for the metric.\\

\newpage
\pagestyle{headings}
\pagenumbering{arabic}
\setcounter{page}{1}
%===========================================================================
\section{Introduction}
In 1915 Albert Einstein published a geometrical theory of gravitation:
{\it The General Theory of Relativity}. He presented a fundamentally new
description of gravity in the sense that the relative acceleration of
particles is not viewed as a
consequence of gravitational forces but results from the curvature of the
spacetime in which the particles are moving. As long as no non-gravitational
forces act on a particle, it is always moving on a ``straight line''.
If we consider curved manifolds there is still a concept of
straight lines which are called {\em geodesics}, but these will
not necessarily have the properties we intuitively associate with straight
lines from our experience in flat Euclidean geometry. It is, for example,
a well known fact that two distinct straight lines in 2-dimensional
flat geometry will
intersect each other exactly once unless they are parallel in which case
they do not intersect each other at all. These
ideas result from the fifth Euclidean
postulate of geometry which
plays a special role in the formulation of geometry.
It is a well known fact that one needs to impose it
separately from the first four Euclidean postulates in order to
obtain flat Euclidean geometry. It was not realised until
the work of Gauss, Lobachevsky, Bolyai and Riemann in the 19th century
that the omission of the fifth postulate leads to an entirely new
class of non-Euclidean geometries in curved manifolds. A fundamental
feature of non-Euclidean geometry is that straight lines in curved
manifolds can intersect each other more than once and correspondingly
diverge from and converge towards each other several times. In order
to illustrate how these properties give rise to effects we commonly
associate with forces such as gravitation, we consider two
observers on the earth's surface, say one in Southampton and one in
Hamburg. We assume that these two observers start moving due south
in ``straight lines'' as for example guided by an idealised compass
exactly pointing towards the south pole. If we follow their separate
paths we will discover exactly the ideas outlined above. As long as both
observers are in the northern hemisphere the proper distance between
them will increase and reach a maximum when they reach the equator. From
then on they will gradually approach each other and their paths
will inevitably cross at the south pole. In the framework of Newtonian
physics the observers will attribute the relative acceleration of
their positions to the action of a force. It is clear, however, that no
force is acting in the east-west direction on either observer
at any stage of their journey. In a geometric description the
relative movement of the observers finds a qualitatively new interpretation
in terms of the curvature of the manifold they are moving in, the curvature
of the earth's surface. With the development of general relativity
Einstein provided the exact mathematical foundation for applying these
ideas to the forces of gravitation in 4-dimensional spacetime. One may
ask why such a geometrical interpretation has only been
developed for gravitation. Or in other words which feature distinguishes
gravitation from the other three fundamental interactions? The answer
lies in the ``gravitational charge'', the mass. It is a common observation
that the gravitational mass $m_{\rm G}$ which determines the coupling of a
particle to the gravitational field is virtually identical to the inertial
mass $m_{\rm I}$ which describes the particle's
kinematic reaction to an external force. High precision experiments have
been undertaken to measure the difference between these two types of masses.
All these results are compatible with the assumption that the masses are
indeed equal. The mass will therefore drop out of the Newtonian equations
governing the dynamics of a particle subject exclusively to gravitational
forces $m a = GmM/r^2 $, where $a$ is the acceleration of the particle,
$G$ the gravitational constant, $M$ the mass of an external source and
$r$ the distance from this source. The particle mass $m$ can be factored
out so that the movement of the particle is described in purely
kinematic terms. The redundancy of the concept of a gravitational
force is naturally incorporated into a geometric theory of
gravity such as general relativity.
It is important to note that this behaviour distinguishes gravity from the
other fundamental interactions which are associated with different types
of charges, such as electric charge in the case of electromagnetic interaction.
It is not obvious how and whether it is possible to obtain similar
geometric formulations for the electromagnetic, weak and strong interaction.
The unification of these three fundamental forces with gravity in the
framework of quantum theory is one of the important areas of ongoing
research. \\
In order to formalize the ideas mentioned above, general relativity views
spacetime as a 4-dimensional manifold equipped with a metric
$\hbox{\vec g}_{\alpha \beta}$ of Lorentzian signature
where the Greek indices range from 0 to 3.
At any given point in the manifold the
signature enables one to distinguish between time-like, space-like
and null directions. The metric further induces a whole range of
higher level geometric concepts on the manifold. It defines a
scalar product between vectors which leads to the measurement
of length and the idea of orthogonality. From the metric and
its derivatives one can derive a {\em connection} on the manifold which
facilitates the definition of a {\em covariant derivative}. The notion
of a derivative is more complicated in a curved manifold than
in the common case of flat geometry and Cartesian coordinates because
the basis vectors will in general vary
from point to point in the manifold. It is therefore no longer possible
to identify the derivative of a tensor with the derivative of its
components. Instead one obtains extra terms involving the derivatives
of the basis vectors. In terms of a covariant derivative these
terms are represented by the connection. In general relativity one
uses a metric-compatible connection defined by
\begin{align*}
  \Gamma^{\gamma}_{\alpha \beta} &= \frac{1}{2} \hbox{\vec g}^{\gamma \delta}
      (\partial_{\alpha} \hbox{\vec g}_{\beta \delta} + \partial_{\beta}
      \hbox{\vec g}_{\alpha \delta} - \partial_{\delta}
      \hbox{\vec g}_{\alpha \beta}),
\end{align*}
where the {\em Einstein summation convention}, according to which
one sums over repeated upper and lower indices, has been used.
These connection coefficients
are also known as the {\em Christoffel symbols} and define a covariant
derivative of tensors of arbitrary rank by
\begin{align*}
  \nabla_{\delta} \hbox{\vec T}^{\alpha \beta}{}_{\gamma}
      &= \partial_{\delta} \hbox{\vec T}^{\alpha \beta}{}_{\gamma}
         + \Gamma^{\alpha}_{\rho \delta} \hbox{\vec T}^{\rho \beta}{}_{\gamma}
         + \Gamma^{\beta}_{\rho \delta} \hbox{\vec T}^{\alpha \rho}{}_{\gamma}
         - \Gamma^{\rho}_{\gamma \delta} \hbox{\vec T}^{\alpha \beta}{}_{\rho},
\end{align*}
where $\partial_{\delta}$ represents the standard partial derivative with
respect to the coordinate $x^{\delta}$.
So for each upper index one adds a term containing the connection
coefficients and for each lower index a corresponding term is subtracted.
With the definition of a covariant derivative we can finally write
down the exact definition of a ``straight line'' in a curved manifold.
A geodesic is defined as the integral curve of a vector field
$\hbox{\vec v}$ which is parallel transported along itself
\begin{align*}
  \hbox{\vec v}^{\alpha} \nabla_{\alpha} \hbox{\vec v}^{\beta} &= 0.
\end{align*}
Based on the covariant derivative we can also give a precise definition
of curvature. For this purpose the {\em Riemann tensor} is defined
by
\begin{align*}
  \hbox{\vec R}^{\alpha}{}_{\beta \gamma \delta} &= \partial_{\gamma}
      \Gamma^{\alpha}_{\delta \beta} - \partial_{\delta}
      \Gamma^{\alpha}_{\gamma \beta} + \Gamma^{\alpha}_{\gamma \rho}
      \Gamma^{\rho}_{\delta \beta} - \Gamma^{\alpha}_{\delta \rho}
      \Gamma^{\rho}_{\gamma \beta}.
\end{align*}
If we use a coordinate basis, i.e. $\hbox{\vec e}_{\alpha}
= \partial/\partial x^{\alpha}$,
this definition can be shown to imply that for any vector field
$\hbox{\vec v}^{\alpha}$
\begin{align*}
  \hbox{\vec R}^{\alpha}{}_{\beta \gamma \delta} \hbox{\vec v}^{\beta}
     &= \nabla_{\gamma} \nabla_{\delta} \hbox{\vec v}^{\alpha}
       -\nabla_{\delta} \nabla_{\gamma} \hbox{\vec v}^{\alpha},
\end{align*}
which is commonly interpreted by saying that a vector
$\hbox{\vec v}$ is changed by being
parallel transported around a closed loop unless the curvature vanishes
(see for example \citeNP{Misner1973}). In order to describe the effect
of the matter distribution on the geometry of spacetime one defines the
{\em Ricci tensor} as the {\em contraction} of the Riemann tensor
$\hbox{\vec R}_{\beta \delta} = \hbox{\vec R}^{\alpha}{}_{\beta \alpha
\delta}$, where again the Einstein summation convention for repeated
indices has been used. The geometry and the matter are then related by
the Einstein field equations
\begin{align*}
  \hbox{\vec G}_{\alpha \beta}
      := \hbox{\vec R}_{\alpha \beta} -1/2\,R\,\hbox{\vec g}_{\alpha \beta}\,
      = 8\pi \hbox{\vec T}_{\alpha \beta},
\end{align*}
where $R=\hbox{\vec R}^{\alpha}{}_{\alpha}$ is the {\em Ricci scalar}
and $\hbox{\vec T}_{\alpha \beta}$ the {\em energy momentum tensor}.
The interaction between the matter distribution and the geometry of
spacetime can be summed up in the words of \citeANP{Misner1973}:
{\em ``Space acts on matter, telling it how to move. In turn, matter 
reacts back on space, telling it how to curve''}. \\
Although the field equations look rather neat in the
compact notation we have given above, this should not hide the fact that
the {\em Einstein tensor}
$\hbox{\vec G}_{\alpha \beta}$ is in fact a complicated function of the
metric $\hbox{\vec g}_{\alpha \beta}$ and its first and second derivatives.
Due to the symmetry of the Einstein tensor and the energy momentum tensor
the field equations represent 10 coupled, non-linear partial differential
equations, which written explicitly
may contain of the order of 100,000 terms in the general case.
It therefore came as quite a surprise
when Karl Schwarzschild found a non-trivial, analytic
solution to these equations just some months after their
publication. Since then many analytic solutions have been found and
a whole branch of the studies of general relativity is concerned with
their classification. Enormous insight into the structure of general
relativity has been gained from these analytic solutions, but due to the
complexity of the field equations these solutions are normally
idealized and restricted by symmetry assumptions. In order
to obtain accurate descriptions of astrophysically relevant scenarios
one may therefore have to go beyond purely analytic studies. A
particularly important area of research connected with general relativity
that has emerged in recent years concerns the detection of
{\em gravitational waves}. In analogy to the prediction of electromagnetic
waves by the Maxwell equations of electrodynamics, the Einstein field
equations admit radiative solutions with a characteristic propagation
speed given by the speed of light. Due to the weak coupling constant
of the gravitational interaction, which is a factor of $10^{40}$ smaller
than the electromagnetic coupling constant, gravitational waves will
have an extremely small effect on the movement of matter and are
correspondingly difficult to detect. If one considers for example a metal
bar of a length of several kilometres, estimates have shown that the detection
of gravitational waves requires one to measure
changes in length orders of magnitude smaller than the diameter of an atomic
nucleus. Even though attempts to detect gravitational radiation go
back to the work of Joe Weber in the early sixties, it is only the
recent advance of computer and laser technology that provides
scientists with a realistic chance of success. The current generation of
gravitational wave detectors GEO-600, LIGO, TAMA and VIRGO that have
been constructed for this purpose are complex multi-national collaborations
and have recently gone online or are expected to go online in the near
future. Due to the extreme smallness of the signals, the accumulation of data
over several years is expected to improve the chances of a positive
identification of signals from extra-galactic sources. \\
Confidence in the
existence of gravitational waves has been significantly boosted by the
Nobel prize winning discovery of the
binary neutron star system PSR1913$+$16 (\citeNP{Hulse1975},
\citeNP{Taylor1989}). The spin-down of this system has
been found to agree remarkably well with the energy-loss predicted
by general relativity due to the emission of gravitational waves and
is generally accepted as indirect proof of the existence of gravitational
radiation. \\
In order to simplify the enormous task of detecting gravitational waves, it is
vital to obtain information about the structure of the signals one
is looking for. It is necessary for this purpose to accurately
model the astrophysical scenarios that are considered likely
sources of gravitational waves
and extract the corresponding signals from these models.
According to {\em Birkhoff's} \citeyear{Birkhoff1923} {\em theorem}
the Schwarzschild solution, which describes a static, spherically symmetric
vacuum spacetime, is the only spherically symmetric,
asymptotically flat solution to the Einstein vacuum field equations.
As a consequence a spherically symmetric spacetime,
even if it contains a radially
pulsating object, will necessarily have an exterior static region and
be non-radiating.
%According
%to a well known result of general relativity a spherically symmetric
%spacetime will necessarily be non-radiating, so that
It is necessary, therefore,
to use less restrictive symmetry assumptions in the modelling of
astrophysical sources of gravitational waves.
In fact the most promising
sources of gravitational waves currently under consideration are the
in-spiralling and merger of two compact bodies (neutron stars or black holes)
and complicated oscillation modes of neutron stars that increase in amplitude
due to the emission of gravitational waves
by extracting energy from the rotation of the star. Even though
a great deal of information about these scenarios has been gained
from approximative studies, such as the
{\em post-Newtonian} formalism or the use of
{\em perturbative techniques}, a detailed simulation will require the
solution of the Einstein equations in three dimensions.
The complicated
structure of the corresponding models in combination with the enormous
advance in computer technology has given rise to
{\em numerical relativity}, the computer based
generation of solutions to Einstein's field equations. \\
In order to numerically solve Einstein's field equations
it is necessary to cast the equations in a form suitable for a
computer based treatment. Among the formulations proposed for this
purpose by far the most frequently applied
is the canonical ``3+1'' decomposition of \lcite{Arnowitt1962},
commonly referred to as the ADM formalism. In this approach spacetime
is decomposed into a 1-parameter family of 3-dimensional space-like
hypersurfaces and the Einstein equations are put into the form
of an initial value problem. Initial data is provided on one
hypersurface in the form of the spatial 3-metric and its time derivative and
this data is evolved subject to certain constraints and the specification
of gauge choices.
It is a known problem, however, that the ADM formalism does not result
in a strictly hyperbolic formulation of the Einstein equations and in
combination with its complicated structure the stability properties
of the ensuing finite differencing schemes remain unclear. These difficulties
have given rise to the development of modified versions of the
ADM formulation in which the Einstein equations are written as a
hyperbolic system. These and similar modifications of the canonical ADM
scheme have been successfully tested, but
%hyperbolic ``3+1'' formulations (see for example \shortciteNP{Bona1995},
%\shortciteNP{Friedrich1996}, \shortciteNP{Anderson1997}) which
%in turn facilitate
%the application of powerful mathematical theorems.
%The conformal decomposition of \lcite{Shibata1995} and \lcite{Baumgarte1999}
%(``BSSN'' for short)
%aims in the same direction and has been successfully implemented
%in several cases (see for example \shortciteNP{Baumgarte1999},
%\shortciteNP{Alcubierre2001}).
%The final answer as to the optimal ``3+1''
%formalism has not yet been found, but attention has definitely shifted
%in recent years from the standard ADM-formulation to alternative schemes
%such as the ``BSSN''-scheme. The majority of these schemes, however,
%are based
%on the standard ADM decomposition of spacetime, so that a description of
%the canonical ADM-formalism is essential for their understanding.
an optimal ``3+1'' formulation has yet to be found
and it may well be possible
%that there is not {\em one} optimal formulation but instead
that an optimal ``3+1''-strategy depends sensitively on the problem that needs
to be solved. \\
An entirely different approach to the field equations
is based on the decomposition of spacetime into families of null-surfaces,
the characteristic surfaces of the propagation of gravitational
radiation. The Einstein field equations are again formulated
as an initial value problem and by virtue of a suitable choice of
characteristic coordinates one obtains
a natural classification of the equations into
evolution and hypersurface equations.
The characteristic initial value problem was first formulated 
by \scite{Bondi1962} and \lcite{Sachs1962} in order to facilitate
a rigorous analysis of gravitational radiation which is properly described
at null infinity only. It is a generic drawback of ``3+1'' formulations
that null infinity cannot be included in the numerical grid by means
of compactifying spacetime and instead outgoing radiation boundary conditions
need to be used at finite radius. Aside from the non-rigorous analysis
of gravitational radiation at finite distances these artificial boundary
conditions give rise to spurious numerical reflections. A characteristic
formulation resolves these problems in a natural way but is itself
vulnerable to the formation of caustics in regions of strong curvature.
It is these properties of ``3+1'' formulations and the characteristic method
that resulted in the idea of {\em Cauchy characteristic matching} (CCM), 
i.e. the combination of a ``3+1'' scheme applied in the interior and a
characteristic formalism in the outer vacuum region. This allows one to
make use of the advantages of both methods as we will illustrate in more
detail below.\\
This thesis consists of four parts. First we will investigate the Einstein
field equations from the numerical point of view. This includes a detailed
description of the ADM and the characteristic Bondi-Sachs formalism as
well as a general discussion of finite difference methods and numerical
concepts such as stability and convergence. Section \ref{ccm} is concerned with
Cauchy characteristic matching as a numerical tool to solve the field
equations. In particular we present
a long term stable CCM code for cylindrically symmetric
vacuum spacetimes containing both gravitational degrees of freedom.
In section \ref{cstr} we investigate the behaviour of static and
dynamic cosmic strings in cylindrical symmetry. 
The numerical codes developed for the analysis are described together
with a detailed study of the oscillations of a cosmic string
excited by gravitational radiation.
Finally in section \ref{pert} we present a fully non-linear perturbative
approach to study non-linear radial oscillations of neutron stars.
The perturbative formulation enables us to study non-linear oscillations over
a large amplitude range with high precision. In an Eulerian formulation,
however, the surface of the star gives rise to numerical difficulties
which leads us to investigate a simplified neutron star model instead.
The section is concluded with the development of a Lagrangian formulation
of dynamic spherically symmetric stars
in which the surface problems are resolved in a natural way.
% The resulting
%code is tested for second order convergence and shown to accurately
%reproduce analytic solutions.
We use the exact treatment of the
surface for the analysis of shock formation near the surface for initial
data of low amplitude.

\newpage
%==============================================================
\section{The field equations from a numerical point of view}
We have already mentioned that the Einstein field equations have to be put
into an appropriate initial value form before they can be integrated
numerically. In this section we will describe in detail the ``3+1''
decomposition of \lcite{Arnowitt1962} and the characteristic formalism
introduced by \scite{Bondi1962} and \lcite{Sachs1962}. The section is completed
by a discussion of general numerical aspects and the description of some
finite differencing schemes used later in this work.
% We start, however, 
%with some general comments on the notation used throughout this work.

%=========================================================================
%\subsection{Notation}
%
%
%Unless stated otherwise, the following conventions apply. Greek indices run
%from 0 to 3, whereas Latin indices are used for 3-dimensional
%quantities. We will generally represent vectors and tensors of higher rank
%with boldfaced letters (e.g. $\hbox{\vec{T}}$).
%Sometimes we will denote vectors, i.e. tensors of rank (1,0), by
%partial differential operators (e.g.
%$\bbox{\partial}_t$). If we need to distinguish between a one-form and
%a vector, the one-form will be marked with a tilde
%(e.g. $\tilde{\hbox{\vec{u}}}$).
%If a one-form is the exterior derivative of a scalar
%function $f$, it will be denoted by $\d f$ and the tilde will be
%omitted. If ${\v}$ is a vector, then $\tilde{\v}$ is the associated
%one-form, i.e. $\tilde{\v} = \hbox{\vec g}({\hbox{\vec v}},.)$.
%In coordinate free language
%the contraction of a one-form $\tilde{\hbox{\vec{u}}}$ with a 
%vector $\hbox{\vec{v}}$ will be written as
%$\langle \tilde{\hbox{\vec u}},{\hbox{\vec v}}\rangle$.
%The 4-dimensional Riemann tensor
%and its contractions will be denoted by the standard $\hbox{\vec R}$. For the
%3-dimensional Riemann tensor we always use $\mathcal{R}$. Throughout
%this work we will use natural units with $c=1=G$ and
%the sign convention ``$-+++$'' for the metric.\\

%=========================================================================
\subsection{The ``3+1'' decomposition of spacetime}
\label{threep1}
%
%
%Our description of the ``3+1'' formalism, is largely inspired by
%the work of \lcite{York1979}. We will start by decomposing a 4-dimensional
%manifold into a 1-parameter family of 3-dimensional hypersurfaces. 

%=========================================================================
\subsubsection{The foliation}
Following \lcite{York1979} we start the discussion of the ``3+1'' formalism
with a 4-dimensional manifold $M$
with coordinates $x^{\alpha}$. Then a suitable function $t(x^{\alpha})$
defines a 1-parameter family of 3-dimensional hypersurfaces by
\begin{align}
  t(x^{\alpha}) &= {\rm const} .
\end{align}
We will refer to these hypersurfaces as $\Sigma_t$. Geometrically
they are represented by the one-form ${\hbox{\vec d}}t$. Next we consider a
3-parameter family of curves threading the family of hypersurfaces.
By threading we mean
\begin{list}{\rm{(\arabic{count})}}{\usecounter{count}
             \labelwidth1cm \leftmargin1.5cm \labelsep0.4cm \rightmargin1cm
             \parsep0.5ex plus0.2ex minus0.1ex \itemsep0ex plus0.2ex}
  \item the curves do not intersect each other,
  \item the tangent vectors $\v$ of the curves are nowhere tangent
        to $\Sigma_t$, i.e. $\langle {\d}t,{\v}\rangle \neq 0$ everywhere.
\end{list}

\begin{figure}[t]
  \centering
  %%%%%%%%%%%%%%%%%%%%%%%%%%%%%%%%%%%%%%%%%%%%%%%%%%%%%%%%%%%%%%%%%%%%%
\begin{picture}(0,0)%
\epsfig{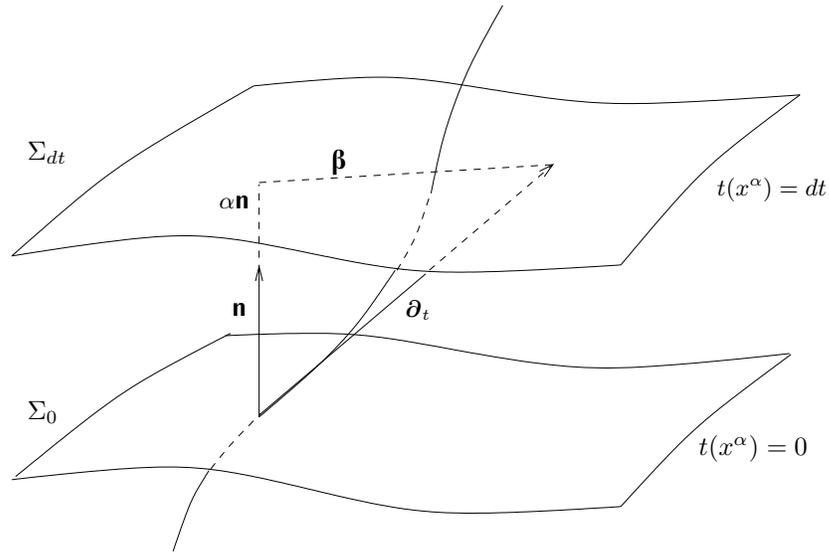}%
\end{picture}%
\setlength{\unitlength}{2960sp}%
\begingroup\makeatletter\ifx\SetFigFont\undefined%
\gdef\SetFigFont#1#2#3#4#5{%
  \reset@font\fontsize{#1}{#2pt}%
  \fontfamily{#3}\fontseries{#4}\fontshape{#5}%
  \selectfont}%
\fi\endgroup%
\begin{picture}(6624,4599)(1489,-3898)
\put(7276,-3061){\makebox(0,0)[lb]{\smash{\SetFigFont{10}{12.0}{\rmdefault}{\mddefault}{\updefault}$t(x^ {\alpha})=0$}}}
\put(1651,-586){\makebox(0,0)[lb]{\smash{\SetFigFont{10}{12.0}{\rmdefault}{\mddefault}{\updefault}$\Sigma_{dt}$}}}
\put(1651,-2761){\makebox(0,0)[lb]{\smash{\SetFigFont{10}{12.0}{\rmdefault}{\mddefault}{\updefault}$\Sigma_0$}}}
\put(7426,-886){\makebox(0,0)[lb]{\smash{\SetFigFont{9}{10.8}{\rmdefault}{\mddefault}{\updefault}$t(x^ {\alpha})=dt$}}}
\put(3256,-991){\makebox(0,0)[lb]{\smash{\SetFigFont{9}{10.8}{\rmdefault}{\mddefault}{\updefault}$\alpha \n$}}}
\put(4201,-661){\makebox(0,0)[lb]{\smash{\SetFigFont{9}{10.8}{\rmdefault}{\mddefault}{\updefault}$\hbox{\grvec b}$}}}
\put(3361,-1891){\makebox(0,0)[lb]{\smash{\SetFigFont{9}{10.8}{\rmdefault}{\mddefault}{\updefault}$\n$}}}
\put(4816,-1906){\makebox(0,0)[lb]{\smash{\SetFigFont{9}{10.8}{\rmdefault}{\mddefault}{\updefault}$\bbox{\partial}_t$}}}
\end{picture}
%%%%%%%%%%%%%%%%%%%%%%%%%%%%%%%%%%%%%%%%%%%%%%%%%%%%%%%%%%%%%%%%%%%
  \caption{Two hypersurfaces of the foliation of spacetime in the
           ``3+1'' formalism. $\bbox{\partial}_t$ is the
           tangent vector field to the
           curves threading the foliation and $\hbox{\vec n}$ the hypersurface
           orthogonal vector field. The relation between these
           vectors is defined by the lapse function $\alpha$ and
           the shift vector $\hbox{\grvec b}$.}
  \label{Foliation}
\end{figure}

In this case the curves are parameterized by $t$ and the tangent
vector with respect to this parameterization is $\bbox{\partial}_t$
which satisfies $\langle {\d}t,\bbox{\partial}_t\rangle = 1$. 
This foliation is illustrated
graphically in Fig.\,\ref{Foliation}. 
We are now in the position to construct basis
vector fields in the manifold $M$. For each slice $\Sigma$ we choose three
vector fields $\e_a$, so that they are linearly independent at each
point of $\Sigma$ and satisfy the condition $\langle \hbox{\vec d}t,
\hbox{\vec e}_i \rangle=0$. Then at each point $P$ of $M$, the set of vectors
$\{\bbox{\partial}_t, \hbox{\vec e}_i\}$ is a
basis of the tangent space $T_P$ at this
particular point. We note that no use of a ``metric'' has been made so far.
All we have done is to foliate $M$ into a 1-parameter family of
3-dimensional slices and to choose suitable basis vectors at each point.\\
%

%======================================================================
\subsubsection{Gauge freedom}
\label{ADM_GAUGE}
Without a metric, the concepts of length and orthogonality are not defined.
It will, therefore, be an essential step in the construction of a metric to
give meaning to these notions. We let $\hbox{\vec g}$ be a symmetric rank two
tensor field, choose a vector field ${\hbox{\vec n}}$ with
$\langle \hbox{\vec d}t, \hbox{\vec n}\rangle \ne 0$ and demand
\begin{align}
   \hbox{\vec{g}}(\hbox{\vec{n}}, \hbox{\vec{\n}})
   &= -1 \hspace{1cm} (\hbox{\vec{n}}\,\,{\rm  is\,\, a\,\, unit
   \,\, vector}),
   \label{AMD_NORMN} \\[10pt]
   \forall_i \, \hbox{\vec{g}}(\hbox{\vec{e}}_i, \hbox{\vec{n}}) &= 0
   \hspace{1.31cm} (\hbox{\vec{n}}\,\,{\rm is\,\, orthogonal\,\, to}\,\,
   \Sigma), \label{ADM_ORTHOGONALITY} \\[10pt]
   \hbox{\vec{g}}(\hbox{\vec{e}}_i,
   \hbox{\vec{e}}_j) &= \hbox{\grvec{g}}_{ij}, \label{gammaij}
\end{align}
where $\hbox{\grvec g}_{ij}$ is a positive definite metric inside the
hypersurfaces $\Sigma$. At this stage the 3-metric $\hbox{\grvec g}$ is
unknown and below we shall see that its components
are the dynamic variables of the ADM ``3+1'' scheme and thus need to
be specified on the initial slice (subject to certain constraints).
It is important to note the minus sign in Eq.\,(\ref{AMD_NORMN}). It is
this choice in combination with the positive definiteness of the 3-metric
$\hbox{\grvec g}$ which determines the spatial nature of the 3-dimensional
hypersurfaces and the time-like character of the normal vector
$\hbox{\vec n}$.
To what extent we have now specified the metric will become clearer if
we use the basis $\{\bbox{\partial}_t,\hbox{\vec e}_i\}$.
Furthermore we will introduce the lapse function
$\alpha$ and the shift vector $\hbox{\grvec b}^i$ %(Figure \ref{Foliation})
defined by
\begin{align}
  \bbox{\partial}_t &= \alpha \hbox{\vec{n}}
      + \hbox{\grvec b}^i\hbox{\vec{e}}_i,
  \label{alphabeta1} \\
  \hbox{\vec{n}} &= \frac{1}{\alpha}(\bbox{\partial}_t
      - \hbox{\grvec b}^i\hbox{\vec{e}}_i).
  \label{alphabeta2}
\end{align}
Then the components of the metric become
\begin{align}
  \begin{split}
  \hbox{\vec{g}}_{00} &= \hbox{\vec{g}}(\bbox{\partial}_t,\bbox{\partial}_t)
      = \hbox{\vec{g}}(\alpha \hbox{\vec{n}}+\hbox{\grvec b}^i\hbox{\vec{e}}_i,
      \alpha \hbox{\vec{n}}+\hbox{\grvec b}^i\hbox{\vec{e}}_i) \\
         &= -\alpha^2 + \hbox{\grvec b}^i\hbox{\grvec b}_i,
  \end{split} \\
  \begin{split}
  \hbox{\vec{g}}_{0i} &= \hbox{\vec{g}}(\bbox{\partial}_t,\hbox{\vec{e}}_i)
       = \hbox{\vec{g}}(\alpha \hbox{\vec{n}}+\hbox{\grvec b}^j
         \hbox{\vec{e}}_j,
         \hbox{\vec{e}}_i) \\
      &= \hbox{\grvec b}_i,
  \end{split} \\
  \hbox{\vec g}_{ij} &= \hbox{\grvec g}_{ij},
\end{align}
which corresponds to the canonical ``3+1'' line element
\begin{align}
  ds^2 &= (-\alpha^2+\hbox{\grvec b}_i\hbox{\grvec b}^i)dt^2
          +2\hbox{\grvec b}_i dt dx^i+\hbox{\grvec g}_{ij}
          dx^i dx^j. \label{ADMlineel}
\end{align}
From this equation we can see
that the metric component $\hbox{\vec g}_{tt}$ will be negative
unless a large shift vector is chosen. In the remainder of this discussion
we will assume a sufficiently small shift vector and therefore
consider $t$ the time-like coordinate. In contrast the positive definite
nature of the 3-metric $\hbox{\grvec g}$ implies that the
$x^i$ are space-like coordinates. \\
In order to investigate the remaining gauge freedom we will now consider the
implications of a different choice of lapse $\alpha$ and
shift $\hbox{\grvec b}$.
According to Eq.\,(\ref{alphabeta1}) such a different choice
would result in a modified
relation between $\hbox{\vec{n}}$ and $\bbox{\partial}_t$, i.e. a different
family of curves
threading the foliation. This, however, merely corresponds to a
coordinate transformation (relabelling of the points in the
manifold) and we see that lapse and shift represent
the coordinate or gauge freedom of general relativity. They can
in principle be chosen arbitrarily without affecting the resulting 
spacetime.\\
The lapse can be interpreted as the proper time measured by an Eulerian
observer, that is an observer moving with 4-velocity \hbox{\vec{n}}.
If we consider two hypersurfaces $\Sigma_t$, 
$\Sigma_{t+\delta t}$, the difference in coordinate time is by definition
$\langle \d t, \delta t\cdot\bbox{\partial}_t\rangle
 = \delta t$. An illustrative
way of describing this result is to say that
$\delta t\cdot\bbox{\partial}_t$ points from $\Sigma_t$ to
$\Sigma_{t+\delta t}$.
On the other hand we know from Eq.\,(\ref{alphabeta2}) that
$\langle \d t, \n\rangle  = 1/\alpha$.
So the vector connecting the two hypersurfaces in the normal direction is
$\alpha\cdot \delta t\cdot \n$. The proper length of this vector is given by
$ds^2=-\alpha^2 \delta t^2$ and the proper time experienced by
travelling along the integral
curve of $\n$ from $\Sigma_t$ to $\Sigma_{t+\delta t}$
is $\alpha\cdot \delta t$.
In this sense, the lapse allows us to measure the length of vectors
pointing outside the hypersurfaces.
In numerical relativity the lapse can be used to control the advance of
proper time in different regions of spacetime as the numerical code is evolved
into the future. Suitable choices for $\alpha$ and $\hbox{\grvec b}$
will be discussed
in section \ref{lapseshift}. \\
The shift vector on the other hand introduces the concept of orthogonality
relative to the spatial hypersurfaces $\Sigma$. For this purpose it is
necessary to define the scalar product between the spatial
basis vectors $\hbox{\vec e}_i$ and vectors pointing out of the hypersurface.
The shift vector which is given by $\hbox{\grvec b}_i=
\hbox{\vec g}(\bbox{\partial}_t, \hbox{\vec e}_i)$ introduces this
scalar product. As a result $\bbox{\partial}_t
- \hbox{\grvec b}^i\hbox{\vec{e}}_i$ is orthogonal to $\Sigma$
in the sense that its scalar product with any vector tangent to 
$\Sigma$ vanishes. We can then use the lapse function to rescale this
vector to unit length and thus recover Eq.\,(\ref{ADM_ORTHOGONALITY}).

%========================================================================
\subsubsection{Extrinsic curvature $\hbox{\vec{K}}$ and the 
               3-metric $\hbox{\grvec{g}}$}
\begin{figure}[t]
  \centering
  \epsfig{file=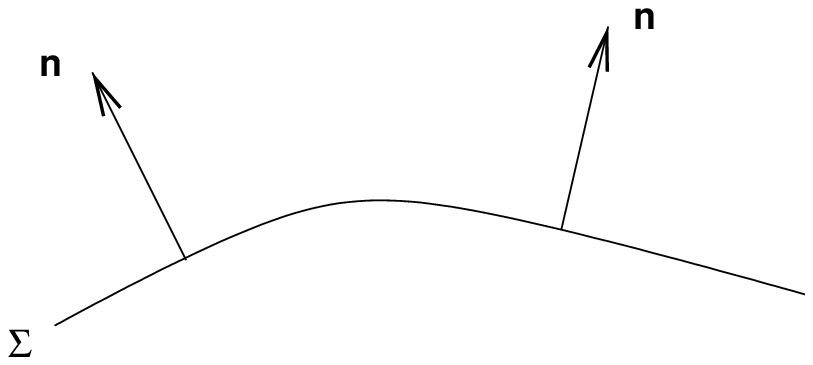, height=150pt, width=75pt, angle=-90}
  \epsfig{file=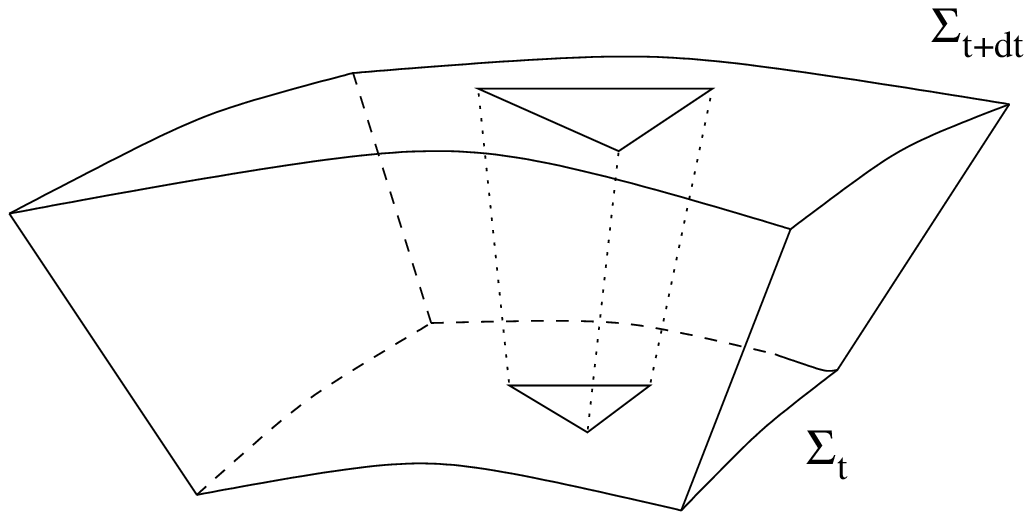, height=200pt, width=100pt, angle=-90}
  \caption{Illustration of the effect of a non-zero extrinsic
           curvature on the embedding of the hypersurface $\Sigma$.
           In the left plot we see that \n\ points in different
           directions at different points of $\Sigma$. In the right plot
           distances increase or decrease as an observer moves from
           one hypersurface to another.}
  \label{ExtrCurv}
\end{figure}

Even though we have determined a basis adapted to our foliation of spacetime,
it is convenient to describe the Cauchy initial value problem in a general
basis. Following \lcite{York1979},
we introduce the projection operator $\bbox{\bot}$
and a shorthand notation for the  projection of a tensor of arbitrary rank
${\bbox \bot} \hbox{\vec{T}}$ by
\begin{align}
  {\bbox \bot}^{\mu}{}_{\nu} &= \delta^{\mu}{}_{\nu} +
       \hbox{\vec{n}}^{\mu}\hbox{\vec{n}}_{\nu}, \\
  {\bbox \bot} \hbox{\vec{T}}^{\lambda}{}_{\mu \nu} &= 
       {\bbox \bot}^{\lambda}{}_{\alpha}
       {\bbox \bot}^{\beta}{}_{\mu}{\bbox \bot}^{\gamma}{}_{\nu}
       \hbox{\vec{T}}^{\alpha}{}_{\beta \gamma}.
\end{align}
We can use this definition to write the 3-metric $\hbox{\grvec g}$ as the
projection of the 4-metric $\hbox{\vec g}$ onto $\Sigma$
\begin{align}
  \hbox{\grvec{g}}_{\mu \nu} &= {\bbox \bot} \hbox{\vec{g}}_{\mu \nu}
       = \hbox{\vec g}_{\mu \nu} + \hbox{\vec n}_{\mu}
       \hbox{\vec n}_{\nu},
\end{align}
which in the ``3+1'' basis reduces to
\begin{align}
  \hbox{\grvec{g}}_{ij} &= {\bbox \bot}^{\mu}{}_i {\bbox \bot}^{\nu}{}_j
                 \hbox{\vec{g}}_{\mu \nu} = \hbox{\vec{g}}_{ij}, \\
  \hbox{\grvec{g}}_{0\mu} &= 0.
\end{align}
The 3-metric $\hbox{\grvec{g}}$ completely describes 
the intrinsic properties
of the 3-dimensional manifold $\Sigma$. 
In particular, the connection on $\Sigma$ which
for a vector $\hbox{\vec v}$ tangent to the slice is defined by
\begin{align}
  D_{\mu} \hbox{\vec{v}} ^{\nu} &= {\bbox \bot}^{\alpha}{}_{\mu}
                          {\bbox \bot}^{\nu}{}_{\beta}
                          \nabla_{\alpha} \hbox{\vec{v}}^{\beta},
\end{align}
with obvious extension to general tensors,
turns out to be the Christoffel connection of $\hbox{\grvec{g}}_{ij}$ 
if we restrict ourselves to spatial quantities and use the ``3+1'' basis
$\{\bbox{\partial}_t,\hbox{\vec e}_i\}$.
Furthermore we define the 3-dimensional Riemann tensor $\mathcal{R}$ by
\begin{align}
  [D_{\mu},D_{\nu}]\hbox{\vec{v}} - D_{[\hbox{\vec{e}}_{\mu},
  \hbox{\vec{e}}_{\nu}]}\hbox{\vec{v}} &=
  \mathcal{R}(\hbox{\vec{e}}_{\mu},\hbox{\vec{e}}_{\nu})\hbox{\vec{v}}, \\
  \mathcal{R}(\hbox{\vec{e}}_{\mu},\hbox{\vec{e}}_{\nu})\hbox{\vec{n}}
      &= 0.
\end{align}
Again, this amounts to the usual definition in terms of $\hbox{\grvec{g}}_{ij}$
if the ``3+1'' basis is used. \\
In order to describe the embedding of $\Sigma$ into $M$,
we define the extrinsic curvature
\begin{align}
  \hbox{\vec{K}}_{\mu \nu} &= -{\bbox \bot} \nabla_{\mu}
     \hbox{\vec{n}}_{\nu}. \label{BACKGR_EXTRCURV}
\end{align}
This can be shown to be equivalent to
\begin{align}
  \hbox{\vec{K}}_{\mu \nu} &= -\frac{1}{2} \bbox{\bot}
     {\mathcal{L}_{\hbox{\smvec{n}}}} \,\hbox{\vec{g}}_{\mu \nu}
                    = -\frac{1}{2}{\mathcal{L}_{\hbox{\smvec{n}}}}
                      \,\hbox{\grvec{g}}_{\mu \nu},
\end{align}
where $\mathcal{L}_{\hbox{\smvec n}}$ is the Lie-derivative along the unit
normal vector field $\hbox{\vec n}$.
In particular this equation implies that $\hbox{\vec{K}}$ is a symmetric
tensor.
The effect of a non-vanishing extrinsic curvature is
schematically illustrated in
Fig.\,\ref{ExtrCurv} by the following two examples.
\begin{list}{\rm{(\arabic{count})}}{\usecounter{count}
             \labelwidth1cm \leftmargin1.5cm \labelsep0.4cm \rightmargin1cm
             \parsep0.5ex plus0.2ex minus0.1ex \itemsep0ex plus0.2ex}
\item At different points of $\Sigma$,
      the unit normal vector \hbox{\vec{n}} points
      in different directions because of the embedding:
      $\bot \nabla \hbox{\vec{n}} \neq 0$.
\item Due to the extrinsic curvature an observer moving along $\hbox{\vec n}$
      from one hypersurface to another observes an increase or
      decrease in distance between points with fixed spatial coordinates.
      This corresponds to a change of the 3-metric $\hbox{\grvec{g}}$:
      ${\mathcal{L}_{\hbox{\smvec{n}}}} \,\hbox{\grvec{g}} \neq 0$.
\end{list}
In section \ref{FieldEqs} we will see 
that the extrinsic curvature $\hbox{\vec{K}}$ and the 3-metric
$\hbox{\grvec{g}}$ are the dynamic variables of the ADM scheme
and need to be specified on an initial hypersurface
$\Sigma_0$. With an appropriate choice of lapse function and shift vector
we will then be able to evolve the 4-metric over some region of the manifold.

%========================================================================
\subsubsection{The projections of the Riemann tensor}
In order to derive the equations that will finally determine the evolution
of the metric, we follow \lcite{Stachel1962} and look at the projections of the
Riemann tensor. Given the 3-dimensional hypersurfaces and the unit normal
vector field $\hbox{\vec n}$ there are three non-trivial projections of
$\hbox{\vec R}_{\mu \nu \kappa \lambda}$:
\begin{list}{\rm{(\arabic{count})}}{\usecounter{count}
             \labelwidth1cm \leftmargin1.5cm \labelsep0.4cm \rightmargin1cm
             \parsep0.5ex plus0.2ex minus0.1ex \itemsep0ex plus0.2ex}
\item all four components are projected onto $\Sigma$:
      $\bbox{\bot} \hbox{\vec{R}}_{\mu \nu \kappa \lambda}$,
\item three times onto $\Sigma$, once onto \hbox{\vec{n}}:
      $\bbox{\bot} \hbox{\vec R}_{\mu \nu \kappa \sigma} \n^{\sigma}$,
\item twice onto $\Sigma$, twice onto \n:
      $\bbox{\bot} \hbox{\vec R}_{\mu \rho \kappa \sigma}
      \hbox{\vec n}^{\rho} \hbox{\vec n}^{\sigma}$.
\end{list}

These are all non-trivial projections we can construct
since projecting three or more components
onto \hbox{\vec n} yields zero because of the symmetry properties
of $\hbox{\vec R}$. 
It is a remarkable fact that the first two projections
are entirely determined by the
initial data according to the {\em Gauss-Codacci equations}
\begin{align}
  \bbox{\bot} \hbox{\vec R}_{\mu \nu \kappa \lambda}
         &= \mathcal{R}_{\mu \nu \kappa \lambda}
       + \hbox{\vec K}_{\mu \kappa} \hbox{\vec K}_{\nu \lambda}
       - \hbox{\vec K}_{\mu \lambda} \hbox{\vec K}_{\nu \kappa},
       \label{GAUSSCODACCI1} \\[10pt]
  \bbox{\bot} \hbox{\vec R}_{\mu \nu \kappa \sigma} \hbox{\vec n}^{\sigma} &=
       D_{\nu} \hbox{\vec K}_{\mu \kappa} - D_{\mu}
       \hbox{\vec K}_{\nu \kappa}. \label{GAUSSCODACCI2}
\end{align}
These equations determine 14 of the 20 independent
components of the 4-dimensional Riemann tensor. The remaining 6
components are contained in the third projection of $\hbox{\vec R}$
according to the {\em Mainardi equation}
\begin{align}
  \bbox{\bot} \hbox{\vec R}_{\mu \rho \kappa \sigma} \hbox{\vec n}^{\rho}
       \hbox{\vec n}^{\sigma} &=
       {\mathcal{L}}_{\hbox{\smvec n}} \hbox{\vec K}_{\mu \kappa}
       + \hbox{\vec K}_{\mu \sigma} \hbox{\vec K}^{\sigma}{}_{\kappa}
       + \frac{1}{\alpha} D_{\mu} D_{\kappa} \alpha. \label{MAINARDI}
\end{align}
If we assume that the 3-metric $\hbox{\grvec g}$ and the extrinsic curvature
$\hbox{\vec K}$ are given on some initial slice we are able to
derive 14 of the 20 components of the 4-dimensional Riemann tensor from these
initial data. The Lie derivative of the extrinsic curvature
${\mathcal{L}}_{\hbox{\smvec n}} \hbox{\vec K}_{\mu \kappa}$,
however, is not known at this
stage and as a consequence we cannot determine the remaining 6 components
of $\hbox{\vec R}_{\mu \rho \kappa \sigma}$ nor can we evolve the
extrinsic curvature and the 3-metric forward in time. We therefore need
an additional source of information that relates the Lie-derivative
${\mathcal{L}}_{\hbox{\smvec n}} \hbox{\vec K}_{\mu \kappa}$,
i.e. the time derivative
of the extrinsic curvature, to the initial data. In general relativity this
extra information is given in the form of the field equations
\begin{align}
  \hbox{\vec R}_{\mu \nu} -\frac{1}{2} R\, \hbox{\vec g}_{\mu \nu}
      &= 8\pi\hbox{\vec T}_{\mu \nu}, \label{FIELDEQ}
\end{align}
where the {\em Ricci tensor} $\hbox{\vec R}_{\mu \nu}
= \hbox{\vec R}^{\rho}{}_{\mu \rho \nu}$ and the {\em Ricci scalar}
$R = \hbox{\vec R}^{\mu}{}_{\mu}$ describe the geometry
and the energy-momentum tensor $\hbox{\vec T}_{\mu \nu}$
is determined by the distribution of matter in spacetime. The terms on the
left hand side of this equation are often combined into the
{\em Einstein tensor} $\hbox{\vec G}_{\mu \nu}$.

%========================================================================
\subsubsection{The role of the field equations}
\label{FieldEqs}
It is important to note that the field equations have not been used
so far. We have seen that the initial data $\hbox{\vec K}$
and $\hbox{\grvec g}$ determine a
substantial part of the 4-dimensional Riemann tensor, but 6 components,
or put another way, the second time derivatives of the
3-metric $\hbox{\grvec g}$
remain unknown. It is Einstein's field equations that allow us
to express the undetermined projections of the Riemann tensor
$\bbox{\bot} \hbox{\vec R}_{\mu \rho \kappa \sigma} \hbox{\vec n}^{\rho}
\hbox{\vec n}^{\sigma}$ in terms of the
other projections $\bbox{\bot} \hbox{\vec R}_{\mu \nu \kappa \lambda}$ and
$\bbox{\bot} \hbox{\vec R}_{\mu \nu \kappa \sigma}
\hbox{\vec n}^{\sigma}$ and the matter distribution
on $\Sigma$. That allows us to calculate the 4-dimensional Riemann tensor
$\hbox{\vec R}_{\mu \nu \kappa \lambda}$
on the initial slice $\Sigma_0$. Furthermore we can calculate the time
derivatives of $\hbox{\grvec g}$ and $\hbox{\vec K}$
and evolve the variables onto the next slice
$\Sigma_{dt}$. Then the process is repeated on each new slice and eventually
we have (in principle) determined the geometry of the whole
spacetime. Lapse and shift provide the remaining information for the components
of the 4-metric $\hbox{\vec g}$. Before we look at the field
equations in more detail, however, we
have to turn our attention to the matter distribution.
% which is described by the energy momentum tensor.

\vspace{0.5cm}
{\em a) The energy-momentum tensor} \\[5pt]
We have already mentioned that the energy-momentum tensor
represents the matter distribution in spacetime.
We illustrate this by considering the components of
$\hbox{\vec T}$ in a coordinate system $x^{\alpha}$. One can then
interprete the component $\hbox{\vec T}^{\mu \nu}$ as the $\nu$-component
of flux of $\mu$-momentum as measured by an observer at rest in the
coordinate system. In the case of spatial components this
is commonly referred to as the $(\mu, \nu)$-component of the ``stress''. The
concept extends to the time component, so that
$\hbox{\vec T}^{\mu 0}$ describes the flux of $\mu$-momentum across
surfaces $t=\mathrm{const}$ which is just the density of
$\mu$-momentum.
As a special case $\hbox{\vec T}^{00}$ represents the energy density.
Similarly $\hbox{\vec T}^{0 \mu}$ is the energy flux across surfaces
$x^{\mu}=\mathrm{const}$.
It can be shown that the energy flux
$\hbox{\vec T}^{0\mu}$ is equal to the momentum density
$\hbox{\vec T}^{\mu 0}$ and that the stress components $\hbox{\vec T}^{ij}$
are symmetric (see for example \shortciteNP{Misner1973}). As a consequence
the energy momentum tensor is symmetric: $\hbox{\vec T}^{\mu \nu}
=\hbox{\vec T}^{\nu \mu}$. \\
Below we will see that projecting the Einstein equations in the same way
as the Riemann tensor will naturally divide the equations into two
different groups, the constraints and the evolution equations.
In the previous section we have studied the projections of the Riemann tensor,
which determines the left hand side of the field equations (\ref{FIELDEQ}),
onto $\hbox{\vec n}$ and the hypersurfaces $\Sigma$.
It remains therefore to calculate the corresponding projections
of the right hand side of the equations given by the energy-momentum tensor.
For this purpose we define the energy and momentum density and 
the stress tensor by
\begin{align}
  \rho &= \hbox{\vec T}_{\mu \nu} \hbox{\vec n}^{\mu}
      \hbox{\vec n}^{\nu}, \label{ADM_RHO} \\
  \hbox{\vec j}_{\mu} &= \bot \hbox{\vec T}_{\mu \nu}
       \hbox{\vec n}^{\nu}, \label{ADM_J} \\
  \hbox{\vec S}_{\mu \nu} &= \bot \hbox{\vec T}_{\mu \nu}. \label{ADM_S}
\end{align}
%
%In the same way that the time evolution of the extrinsic curvature is
%determined by the Einstein field equations,
The evolution of the matter variables
follows from the conservation
of energy and momentum $\nabla_{\nu} \hbox{\vec T}^{\mu \nu}=0$
\begin{align}
  {\mathcal{L}}_{\bbox{\partial}_t} \rho &= -\alpha D_{\nu} \hbox{\vec j}^{\nu}
    +\alpha(\hbox{\vec S}^{\mu \nu} \hbox{\vec K}_{\mu \nu}
    +\rho\, {\rm tr}\, \hbox{\vec K})
    -2 \hbox{\vec j}^{\nu} D_{\nu} \alpha
    + {\mathcal{L}}_{\hbox{\smgrvec b}} \rho, \\
  {\mathcal{L}}_{\bbox{\partial}_t} \hbox{\vec j}^{\mu} &= -\alpha D_{\nu}
    \hbox{\vec S}^{\mu \nu} +\alpha(2 \hbox{\vec K}^{\mu \nu} 
    \hbox{\vec j}_{\nu} + \hbox{\vec j}^{\mu}\, {\rm tr}\, \hbox{\vec K})
    -\hbox{\vec S}^{\mu \nu} D_{\nu} \alpha - \rho D^{\mu} \alpha
    + {\mathcal{L}}_{\hbox{\smgrvec b}} \hbox{\vec j}^{\mu}.
\end{align}
In order to determine the time
derivatives of \hbox{\vec S} extra information is required
which usually comes in the form of an equation of state.

\vspace{0.5cm}
{\em b) The evolution equations} \\[5pt]
With the projections of the Riemann tensor given by
Eqs.\,(\ref{GAUSSCODACCI1})-(\ref{MAINARDI}) and those of the energy-momentum
tensor given by Eqs.\,(\ref{ADM_RHO})-(\ref{ADM_S})
we are now in a position to project the field
equations onto $\Sigma$ and $\hbox{\vec n}$. First we consider the projection
of both components onto $\Sigma$
\begin{align}
  \bot \hbox{\vec G}_{\mu \nu} &= 8\pi \bbox{\bot} \hbox{\vec T}_{\mu \nu}.
\end{align}
Inserting the projections of $\hbox{\vec T}$ and $\hbox{\vec G}$
and solving for 
the time derivative of $\hbox{\vec K}$, we obtain
\begin{align}
  {\mathcal{L}}_{\bbox{\partial_t}} \hbox{\vec K}_{\mu \nu} &=
      -D_{\mu} D_{\nu} \alpha + \alpha[ \mathcal{R}_{\mu \nu}
      -2\hbox{\vec K}_{\mu \rho} \hbox{\vec K}^{\rho}_{\,\,\,\nu}
      + \hbox{\vec K}_{\mu \nu}\, {\rm tr}\, \hbox{\vec K}
      -4\pi\left( 2\hbox{\vec S}_{\mu \nu}- \hbox{\grvec g}_{\mu \nu}\,
      {\rm tr}\,\hbox{\vec S}+\rho \hbox{\grvec g}_{\mu \nu}\right)]
      + {\mathcal{L}}_{\hbox{\smgrvec b}}
       \hbox{\vec K}_{\mu \nu}, \label{ADM_evolK} \\
  {\mathcal{L}}_{\bbox{\partial_t}} \hbox{\grvec g}_{\mu \nu} &=
      -2\alpha \hbox{\vec K}_{\mu \nu}
      + {\mathcal{L}}_{\hbox{\smgrvec b}} \hbox{\grvec g}_{\mu \nu},
      \label{ADM_evolGAMMA}
\end{align}
where the evolution equations for the 3-metric are simply the definition of
the extrinsic curvature. It is this set of equations which forms the
core of the ADM-evolution of the metric. Given appropriate initial data
on some initial slice $\Sigma_0$ for the extrinsic curvature
$\hbox{\vec K}_{\mu \nu}$ and the 3-metric $\hbox{\grvec g}_{\mu \nu}$
we can evolve these functions into the future. The 4-dimensional
Riemann tensor and thus the geometry of the spacetime is determined
at any time according to Eqs.\,(\ref{GAUSSCODACCI1})-(\ref{MAINARDI}).
The appearance of Greek indices in the evolution equations should not hide
the fact that there are only six components each for the extrinsic curvature
and the 3-metric $\hbox{\grvec g}$. This becomes clear when we use the
adapted basis $\{\bbox{\partial}_t, \hbox{\vec e}_i\}$ in which case all
Greek indices can be replaced by Latin indices
in Eqs.\,(\ref{ADM_evolK}), (\ref{ADM_evolGAMMA}). We can also see then that
there are no evolution equations for $\hbox{\vec g}_{0\mu}$
or, put another way,
in this basis the field equations do not contain second time derivatives
of the $\hbox{\vec g}_{0\mu}$. In this sense the problem is
under-determined. \\

\newpage
{\em c) The constraint equations} \\[5pt]
If we consider the remaining projections of the field equations,
we find that they can be expressed in terms of the initial data only
\begin{align}
  \hbox{\vec G}_{\mu \nu} \hbox{\vec n}^{\mu} \hbox{\vec n}^{\nu} &=
      \frac{1}{2}\left[\mathcal{R} + ({\rm tr}\,\hbox{\vec K})^2 -
        \hbox{\vec K}_{\mu \nu} \hbox{\vec K}^{\mu \nu} \right]
      = 8\pi \rho,  \label{econstr}\\
  \bot \hbox{\vec G}^{\mu \nu} \hbox{\vec n}_{\nu} &= -D_{\nu}
      \hbox{\vec K}^{\mu \nu}
      + D^{\mu}\, {\rm tr}\,\hbox{\vec K} = 8\pi \hbox{\vec j}^{\mu}.
      \label{mconstr}
\end{align}
These equations impose conditions that need to be satisfied by the
hypersurface data for all values of $t$. They are called the energy or
Hamiltonian constraint (\ref{econstr})
and the momentum constraints (\ref{mconstr}).
In this sense, the problem is over-determined. However, it
can be shown that by virtue of the contracted Bianchi identities
$\nabla_{\mu} \hbox{\vec G}^{\mu \nu}=0$ the
constraints are satisfied for all values of $t$ if they are satisfied by the
initial data.

\vspace{0.5cm}
{\em d) The initial data problem} \\[5pt]
The problem we are facing now is to find initial data for $\hbox{\grvec g}$ and
$\hbox{\vec K}$ that satisfy the constraint
equations. A systematic approach to solving this problem is given in
\lcite{OMurchadha1974}. We will illustrate their method in the vacuum case
with ``maximal slicing'' (cf. section \ref{lapseshift}), where the vanishing of
${\rm tr}\,\hbox{\vec K}$ leads to a decoupling of the constraint equations.
\citeANP{OMurchadha1974} start by introducing a conformal
3-metric and extrinsic curvature according to
\begin{align}
  \hbox{\grvec g}_{ij} &= \phi^4 \hat{\hbox{\grvec g}}_{ij},
      \label{INIDATA_GAMMAIJ}  \\
  \hbox{\vec K}_{ij} &= \phi^{-2} \hat{\hbox{\vec K}}_{ij}.
\end{align}
In the case of maximal slicing the constraint equations can then be written
in the form
\begin{align}
  \hat{\Delta}\phi - \frac{1}{8} \phi \hat{\mathcal{R}} + \frac{1}{8}\phi^{-7}
    \hat{\hbox{\vec K}}_{ij} \hat{\hbox{\vec K}}^{ij} &= 0,
    \label{Ini_econstr} \\
  \hat{D}_j \hat{\hbox{\vec K}}^{ij} &= 0, \label{INIDATA2}
\end{align}
where $\hat{D}$ is the covariant derivative with respect to
$\hat{\hbox{\grvec g}}$
and $\hat{\Delta} = \hat{\hbox{\grvec g}}^{ij}\hat{D}_i \hat{D}_j$ is the
conformal Laplace operator. The conformal transformation of the
3-dimensional curvature scalar is given by
\begin{align}
  \mathcal{R} &= \phi^{-4}\, \hat{\mathcal{R}} - 8 \phi^{-5}\hat{\Delta} \phi.
\end{align}
One can further split the traceless
$\hat{\hbox{\vec K}}_{ij}$ according to
\begin{align}
  \hat{\hbox{\vec K}}^{ij} &= \hat{\hbox{\vec A}}_{*}^{ij}
     + \hat{D}^i W^j + \hat{D}^j W^i - \frac{2}{3} \hat{\hbox{\grvec g}}^{ij}
     \hat{D}_k W^k. \label{INIDATA2b}
\end{align}
Here $\hat{\hbox{\vec A}}_{*}^{ij}$ is the transverse traceless part of
the conformal extrinsic curvature $\hat{\hbox{\vec K}}^{ij}$ satisfying
\begin{align}
  \hat{D}_j \hat{\hbox{\vec A}}_{*}^{ij} &=
     {\rm tr}\,\hat{\hbox{\vec A}}_{*}^{ij} = 0,
\end{align}
and the vector $W$ is to be determined by Eq.\,(\ref{INIDATA2}) which
in the case of maximal slicing can be written as
\begin{align}
  \hat{\Delta} W^i + \frac{1}{3} \hat{D}^i \hat{D}_j W^j
     + \hat{\mathcal{R}}^i{}_jW^j &= 0.
\end{align}
In this formulation of the initial data problem the conformal 3-metric
$\hat{\hbox{\grvec g}}$ and the transverse traceless part
$\hat{\hbox{\vec A}}_{*}$ are regarded
as given. Then the momentum constraint (\ref{INIDATA2b}) has
to be solved to obtain $W$ and the conformal factor $\phi$ results from
the energy constraint (\ref{Ini_econstr}).
By means of the conformal decomposition
we have thus isolated $\phi$ and $W^j$
as the four variables determined
by the constraint equations on the initial hypersurface. \\
Much of the work that has gone into the calculation of initial
data has been based on the
conformally flat approach of \lcite{Bowen1980}. In this approach one
assumes the spatial 3-metric to be conformally flat, so that
$\hbox{\grvec g}_{ij} = \phi^4 \delta_{ij}$. However,
recent work has cast doubt on the suitability of this approach in the case
of black hole initial data. The difficulties arise from the fact that
there exist no conformally flat space-like slices of the Kerr spacetime
(\citeNP{Garat2000}). The initial data resulting from the conformally
flat approach will therefore represent distorted Kerr black holes which
generally radiate off a burst of gravitational waves which contaminates
the evolution of binary black holes or perturbed Kerr spacetimes
(``close limit'' calculations). Recent efforts have therefore gone into
the calculation of more realistic initial data which is not based on the
conformally flat approach (see for example \citeNP{Marronetti2000}). \\
A comprehensive description of the general initial
value problem and more details on solving the
constraint equations can be found in \lcite{York1983}.

%==========================================================================
\subsubsection{The kinematic degrees of freedom: lapse and shift}
\label{lapseshift}
In the previous section we have seen that there are no evolution
equations for the components $\hbox{\vec g}_{0\mu}$ of the metric
if we use the adapted basis $\{\bbox{\partial}_t, \hbox{\vec e}_i\}$. 
The line element (\ref{ADMlineel}), however, shows that the
$\hbox{\vec g}_{0\mu}$ are completely determined by the lapse $\alpha$ and the
shift vector $\hbox{\grvec b}$ and these can be chosen arbitrarily
without affecting the metric. Nevertheless the choice has
a substantial impact on the performance of a numerical scheme. For example
a poor choice of coordinates can result in a code which runs into a
singularity before interesting results are computed.
% Several choices for
%the lapse function and shift vector have been suggested in the past. Following
%\lcite{Piran1983} we discuss some of the more popular ones.
A large number of gauge choices have been suggested in the past, some
of which we will describe below. A more comprehensive discussion can
be found in \lcite{Piran1983}.

\vspace{0.5cm}
{\bf The lapse function}

\begin{figure}[t]
\centering
%%%%%%%%%%%%%%%%%%%%%%%%%%%%%%%%%%%%%%%%%%%%%%%%%%%%%%%%%%%%%%%%%%%%%%
\begin{picture}(0,0)%
\epsfig{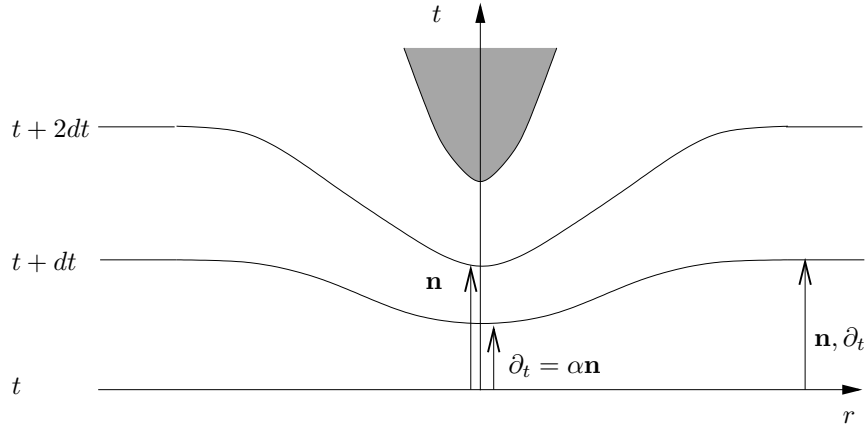}%
\end{picture}%
\setlength{\unitlength}{3158sp}%
\begingroup\makeatletter\ifx\SetFigFont\undefined%
\gdef\SetFigFont#1#2#3#4#5{%
  \reset@font\fontsize{#1}{#2pt}%
  \fontfamily{#3}\fontseries{#4}\fontshape{#5}%
  \selectfont}%
\fi\endgroup%
\begin{picture}(6702,3330)(526,-2536)
\put(6826,-1936){\makebox(0,0)[lb]{\smash{\SetFigFont{10}{12.0}{\rmdefault}{\mddefault}{\updefault}${\bf n}, \partial_t$}}}
\put(526,-2311){\makebox(0,0)[lb]{\smash{\SetFigFont{10}{12.0}{\rmdefault}{\mddefault}{\updefault}$t$}}}
\put(526,-1336){\makebox(0,0)[lb]{\smash{\SetFigFont{10}{12.0}{\rmdefault}{\mddefault}{\updefault}$t+dt$}}}
\put(7051,-2536){\makebox(0,0)[lb]{\smash{\SetFigFont{10}{12.0}{\rmdefault}{\mddefault}{\updefault}$r$}}}
\put(526,-286){\makebox(0,0)[lb]{\smash{\SetFigFont{10}{12.0}{\rmdefault}{\mddefault}{\updefault}$t+2dt$}}}
\put(3826,614){\makebox(0,0)[lb]{\smash{\SetFigFont{10}{12.0}{\rmdefault}{\mddefault}{\updefault}$t$}}}
\put(4426,-2146){\makebox(0,0)[lb]{\smash{\SetFigFont{10}{12.0}{\rmdefault}{\mddefault}{\updefault}$\partial_t=\alpha {\bf n}$}}}
\put(3776,-1486){\makebox(0,0)[lb]{\smash{\SetFigFont{10}{12.0}{\rmdefault}{\mddefault}{\updefault}${\bf n}$}}}
\end{picture}
%%%%%%%%%%%%%%%%%%%%%%%%%%%%%%%%%%%%%%%%%%%%%%%%%%%%%%%%%%%%%%%%%%%%%%
  \caption{In order to avoid the code entering a singularity that forms after
           a finite amount of time (the shaded region indicates an associated
           horizon) the advance of proper
           time is delayed in the central region by
           use of an appropriate lapse function. For convenience we have
           set the shift vector $\hbox{\grvec b} = 0$.}
  \label{slicing}
\end{figure}

\vspace{0.2cm}
{\em (a) Geodesic slicing} \\[5pt]
In geodesic slicing $\alpha$ is set to 1 everywhere. 
This means that the coordinate time is identical
to the proper time of Eulerian observers. Although this slicing
condition appears to be quite natural it does not lead to any significant
simplifications of the equations and, worse, it is singularity seeking.
We illustrate this behaviour in the case of the Schwarzschild spacetime
in Kruskal coordinates (\citeNP{Smarr1978}),
%In their notation we take $v=0$ as the initial hypersurface and 
by considering an Eulerian observer
close to the black hole. An Eulerian observer does not initially move in
the spatial hypersurface and will fall into the
singularity on a time scale $\pi M$, where $M$ is the mass of the
black hole. Choosing the orthogonal time of an Eulerian observer as
coordinate time will therefore cause the code to crash on a coordinate
time scale of $\pi M$. Far away from the black hole, however, where
Eulerian proper time is close to the proper time of an astronomical
observer we would basically like the code to advance up to $t\to
\infty$. One way to accomplish this is to slow down the advance of
proper time near the formation of a singularity as illustrated
in Fig.\,\ref{slicing}. This, however, implies a different choice for the
lapse function $\alpha$. \\
An alternative way of avoiding the code to encounter singularities
consists in cutting off the singularity from the calculation
assuming that it is hidden inside an apparent
horizon and thus no information is lost in the excision (\citeNP{Thornburg1987},
\citeNP{Seidel1992}). This approach has attracted a lot of attention in
recent years and has been successfully implemented in the evolution of black
holes (see \shortciteNP{Alcubierre2001} for example). In this work, however,
we will not make use of these methods and therefore restrict this
discussion to conventional techniques for avoiding singularities.

\vspace{0.6cm}
{\em (b) Maximal slicing} \\[5pt]
The restrictions arising from geodesic slicing were recognised long ago
by \lcite{Lichnerowicz1944} who showed that a much more suitable choice for
$\alpha$ is obtained if one requires that the trace of the extrinsic curvature
vanishes: ${\rm tr}\, \hbox{\vec K} = 0$. This choice has been termed
maximal slicing since the volume of an arbitrary region $\Omega$
of a hypersurface $\Sigma$ will be maximal with respect to all other
hypersurfaces that are identical with $\Sigma$ outside $\Omega$ if
${\rm tr}\,\hbox{\vec K} = 0$ (see for example \citeNP{York1979}).
If we insert the energy constraint (\ref{econstr}) into the
evolution equation for ${\rm tr}\, \hbox{\vec K}$
[obtained from Eq.\,(\ref{ADM_evolK})]
we obtain the following condition for $\alpha$
\begin{align}
  \Delta \alpha &= \alpha \left[ \hbox{\vec K}_{ij} \hbox{\vec K}^{ij}
                   +4\pi ({\rm tr}\,\hbox{\vec S} + \rho) \right].
  \label{MaxSlicing}
\end{align}
A number of useful properties have made maximal slicing one of the most
popular choices in numerical relativity.
\begin{list}{(\arabic{count})}{\usecounter{count}
             \labelwidth1cm \leftmargin1.5cm \labelsep0.4cm \rightmargin1cm
             \parsep0.5ex plus0.2ex minus0.1ex \itemsep0ex plus0.2ex}
\item It avoids singularities.
\item The constraint equations in the initial data problem are decoupled
      (cf.\ section \ref{FieldEqs}).
\item It leads to some simplification of the evolution equations.
\end{list}
The major drawback is that we have to solve the elliptic partial differential
equation (PDE) (\ref{MaxSlicing}) on each time slice.

\vspace{0.6cm}
{\em (c) Hyperbolic slicing} \\[5pt]
Hyperbolic slicing is a generalised version of maximal slicing.
The trace of the extrinsic curvature is required to be constant
but not necessarily to vanish:
${\rm tr}\, \hbox{\vec K} = {\rm const}$.
The major difference is that the hypersurfaces
asymptotically extend to future or past null infinity, depending on the sign of
${\rm tr}\, \hbox{\vec K}$, instead of spatial infinity as
in the case of maximal \mbox{slicing}. This property makes it an interesting
choice for the analysis of gravitational radiation.

\vspace{0.6cm}
{\em (d) Polar slicing} \\[5pt]
Another slicing condition where the lapse function is determined by enforcing
a condition on the extrinsic curvature is polar slicing (see
\citeNP{Bardeen1983} for a detailed discussion).
Using polar coordinates $(r,\theta,\phi)$, one demands that
\begin{align}
  {\rm tr}\,\hbox{\vec K} = \hbox{\vec K}^r{}_{r}
              &\Leftrightarrow \hbox{\vec K}^{\theta}{}_{\theta}
              + \hbox{\vec K}^{\phi}{}_{\phi} = 0.
\end{align}
This condition leads to a parabolic PDE for the lapse function $\alpha$ which,
in general, is
easier to solve than the elliptic PDE that appears for example in
maximal slicing. Furthermore
polar slicing is strongly singularity avoiding as we will
illustrate in the evolution of a spherically symmetric dust sphere
in Lagrangian gauge and polar slicing in section \ref{LAGR}.
The main drawback of polar slicing is the irregular behaviour of the
lapse function in the non-spherically symmetric case (\citeNP{Bardeen1983}).
This problem can be overcome by using an alternative condition, for example
maximal slicing, near the origin and implementing a gradual transition
to polar slicing outside a finite radius $r$.

\vspace{0.6cm}
{\em (e) Harmonic slicing} \\[5pt]
In harmonic slicing one requires that $t$ is a harmonic time coordinate
\begin{align}
  \Box t &= 0.
\end{align}
In terms of the lapse function $\alpha$ this condition results in
equations similar to those of maximal slicing
\begin{center}
\begin{tabular}{cc}
  harmonic sl.& maximal sl. \\[6.5pt]
  $\nabla_{\mu} \frac{\n^{\mu}}{\alpha} = 0$, &
  \qquad \qquad  $-{\rm tr}\,\hbox{\vec K} = \nabla_{\mu} \n^{\mu} = 0$,\\[6.5pt]
  $\partial_t \sqrt{\hbox{\grvec g}} = 0$, &
  $\partial_t \frac{\sqrt{\hbox{\grvec g}}}{\alpha} = 0$.\\
\end{tabular}
\end{center}

Harmonic slicing is another singularity avoiding condition and was used by 
\lcite{Bona1992} to write the Einstein equations as
a hyperbolic system of balance laws. The same authors and coworkers
have shown that many other slicing conditions suit this purpose as well
(\shortciteNP{Bona1997}).

\vspace{0.6cm}
{\em (f) approximate coordinate conditions, driver conditions} \\[5pt]
The suggestion of so-called driver conditions by \scite{Balakrishna1996}
arises from the fact that one is normally interested in the ensuing properties
of the numerical evolution rather than the exact shape of the lapse (or shift)
function. In this respect one has to note that the field equations are
intrinsically coordinate independent and thus there is no need to
implement a specific coordinate condition exactly if an approximate
implementation leads to a stable evolution.
\shortciteANP{Balakrishna1996}
illustrate this effect in the case of maximal slicing
${\rm tr}\,\hbox{\vec K}=0$, where the important property is the vanishing
of the trace of the extrinsic curvature. They
demonstrate how this condition is actually satisfied with higher numerical
accuracy if one imposes the ``$\hbox{\vec K}$-driver'' slicing condition
$\partial_t ({\rm tr}\,\hbox{\vec K}) + c\cdot{\rm tr}\,\hbox{\vec K}=0$
where $c$ is a
positive constant. This condition will result in an exponential decay in
any deviation from ${\rm tr}\,\hbox{\vec K}=0$, whereas the original
implementation of maximal slicing has no such built-in correction mechanism.
The lapse function $\alpha$ is determined in this case by an elliptic equation
similar to Eq.\,(\ref{MaxSlicing}) in maximal slicing. The only difference
is the appearance of the term $c\cdot {\rm tr}\,\hbox{\vec K}$ on the
right hand side of the equation.
\shortciteANP{Balakrishna1996} demonstrate the superior performance
of the ``$K$-driver'' condition in the cases of flat space and a
self-gravitating scalar field.\\
A related proposal by \shortciteANP{Balakrishna1996} concerning elliptic
coordinate conditions in general is also based on the suitability of
approximate implementations of coordinate conditions. Instead of
solving the elliptic equation directly, which in general is computationally
expensive, they suggest ``evolving the elliptic equations'' by
rewriting them in parabolic form which is similar to the relaxation method
of solving elliptic PDEs (see for example \shortciteNP{Press1989}). \\
We have listed these methods under the heading of slicing
conditions, but the same principles apply to the shift vector.

\vspace{0.6cm}
{\em (g) New slicing conditions used in black hole evolutions} \\[5pt]
In recent work on 3-dimensional black hole excision
\scite{Alcubierre2001} have achieved substantial progress
in terms of stability and accuracy by using a new type of evolution equation
for the lapse function
in combination with ``Gamma freezing'' conditions for the shift vector
(see below).
\shortciteANP{Alcubierre2001} propose to evolve the lapse $\alpha$
according to
\begin{align}
  \partial_t^2 \alpha &= - \alpha^2f(\alpha)
      \partial_t ({\rm tr}\,\hbox{\vec K}),
\end{align}
where $f(\alpha)$ is a positive function of $\alpha$ which they normally
set to $2/\alpha$. The key feature of this choice is that the
trace of the extrinsic curvature becomes time independent for the
final state of a stationary black hole (see their paper for details).

\vspace{0.7cm}
{\bf The shift vector}

\vspace{0.2cm}
{\em (a) Normal coordinates} \\[5pt]
In normal coordinates the shift vector is set to zero
\begin{align}
  \hbox{\grvec b}^i &= 0,
\end{align}
which implies that the coordinate vector $\bbox{\partial}_t$ is normal
to the hypersurfaces $\Sigma$.
Normal coordinates have the advantage that they do not become singular
as long as the hypersurfaces have a regular intrinsic and extrinsic geometry
(\citeNP{Bardeen1983b}). They do not, however, facilitate a substantial
simplification of the field equations.

\vspace{0.5cm}
{\em (b) Minimal shear gauge}\\[5pt]
The minimal shear condition suggested by \lcite{Smarr1978}
leads to elliptic equations for the components of $\hbox{\grvec b}^i$.
\citeANP{Smarr1978} find this gauge choice particularly
useful for the description of
gravity in the wave zone.
The major drawbacks are the complexity of the elliptic
equations for $\hbox{\grvec b}^i$ and the fact that it barely
simplifies the field equations.

\vspace{1.5cm}
{\em (c) Simplifying gauge choices} \\[5pt]
This is actually a whole class of gauge choices. The idea is to impose
algebraic relations on the metric components on the initial slice
\begin{align}
  f(\hbox{\grvec g}_{ij}, x^i) &= q(x^i), \label{SimplyfGauge}
\end{align}
and to choose the shift vector so that these algebraic relations hold on all
future hypersurfaces. The three components of the shift vector allow us
to impose three relations of this kind. 
In particular, we can choose
up to three metric components to vanish identically. Solving the
resulting equations for $\hbox{\grvec b}^i$, however, is non-trivial
and it cannot even be guaranteed that such a solution does
exist. Popular examples of this gauge choice are
\begin{list}{\rm{(\arabic{count})}}{\usecounter{count}
             \labelwidth1cm \leftmargin1.5cm \labelsep0.4cm \rightmargin1cm
             \parsep0.5ex plus0.2ex minus0.1ex \itemsep0ex plus0.2ex}
\item {\em Diagonal gauge}, where the 3-metric $\hbox{\grvec g}$
      is diagonalized.
\item {\em Radial gauge}, which employs polar coordinates $(r,\theta,\phi)$
      and imposes the conditions $\hbox{\grvec g}_{r\theta}
      =\hbox{\grvec g}_{r\phi}=0$
      and $\hbox{\grvec g}_{\theta \theta}\hbox{\grvec g}_{\phi \phi}
          - \hbox{\grvec g}_{\theta \phi}^{\,\,\,2} = r^4\sin ^2 \theta$.
      Radial gauge simplifies the field equations significantly and
      results in parabolic equations for the $\hbox{\grvec b}^i$.
\item {\em Isothermal gauge} is similar to radial gauge, except
      that the third condition on the metric components is now
      $\hbox{\grvec g}^{rr} = \hbox{\grvec g}^{\theta \theta}$.
      The simplifications are
      not as substantial as in radial gauge, but isothermal gauge
      can be used for a more general class of physical scenarios.
\end{list}

{\em (d) ``Gamma freezing conditions''} \\[5pt]
We have already mentioned the substantial improvements that
\scite{Alcubierre2001} have achieved in their 3-dimensional black hole
evolutions using new gauge conditions. In combination with the slicing
condition mentioned above under {\em (g)} they relate the shift vector
to the evolution of the conformal connection functions
$\hat{\Gamma}^i$ introduced by \lcite{Baumgarte1999} and
\lcite{Shibata1995}. In their simulations they use a condition of the
form
\begin{align}
  \partial_t^2 \hbox{\grvec b}^i &= \frac{k}{\phi^4}\partial_t
      \hat{\Gamma}^i - \eta \partial_t \hbox{\grvec b}^i,
\end{align}
where $k=0.75$, $\eta=3/M$, $M$ is the initial ADM mass of the system
and $\phi$ is the conformal factor introduced in the discussion of the initial
value problem in section \ref{FieldEqs}.
\shortciteANP{Alcubierre2001} call these conditions ``Gamma freezing''
because they are related to the elliptic operator for $\hbox{\grvec b}^i$
in the ``Gamma freezing condition'' $\partial_t \hat{\Gamma}^i=0$.

\vspace{0.7cm}
A more detailed description of different gauge choices can be found in
\lcite{Piran1983}.

%=======================================================================
\subsubsection{The current state of ``3+1'' formulations: recent progress
               and limitations}
The standard ``3+1'' decomposition we have described above was first
formulated by \lcite{Arnowitt1962}. In the course of time numerous codes
have been developed on the basis of this formulation. The structure
of the ADM evolution equations (\ref{ADM_evolK}),
(\ref{ADM_evolGAMMA}), however, has been a constant cause of concern.
It is well known that these equations do not satisfy any known hyperbolicity
condition and the stability properties of the corresponding numerical
implementations remain obscure. In the course of the
1990s attention
shifted towards modifying the canonical ADM-formalism in order to
obtain strictly hyperbolic formulations of the Einstein equations
(see for example \shortciteNP{Bona1995}, \citeNP{Friedrich1996},
\shortciteNP{Anderson1997}).
%Large efforts have been undertaken to convert the field
%equations into so-called flux-conservative form
%$\partial_t\hbox{\vec u} + \nabla \cdot \hbox{\vec F}(\hbox{\vec u})
%= \hbox{\vec S}$ (see for example \shortciteNP{Bona1995}).
%This type of equations appears in many branches of
%physics, such as fluid dynamics, and correspondingly powerful
%techniques have been developed for their numerical solution.
%These transformations of the Einstein equations
%into hyperbolic form do often come at the price of introducing
%a large number of auxiliary
%variables and equations which makes the numerical treatment
%computationally expensive.
The question to what extent these formulations
result in a superior numerical performance and thus whether the difficulties
encountered in the ADM formalism are entirely due to a possible
non-hyperbolicity has not yet been answered. \\
An alternative modification of the ADM-formulation which has attracted
a great deal of attention recently is based on a conformal decomposition
of the original ADM-equations (\citeNP{Shibata1995}, \citeNP{Baumgarte1999}).
In this ``BSSN''-formulation one starts with a conformal transformation
analogous to that used in the initial-value problem in
section \ref{FieldEqs} {\em(d)}. The 3-metric $\hbox{\grvec g}_{ij}$
is decomposed into the conformal metric $\hat{\hbox{\grvec g}}_{ij}$
and the conformal factor $\phi$ according to Eq.\,(\ref{INIDATA_GAMMAIJ}).
Similarly
the extrinsic curvature is split up into the trace ${\rm tr}\,\hbox{\vec K}$
and the conformal traceless extrinsic curvature $\hat{\hbox{\vec A}}_{ij}$.
The set
of fundamental variables is completed by the conformal connection
coefficients $\hat{\Gamma}^i=\hat{\hbox{\grvec g}}^{jk}
\hat{\Gamma}^i_{jk}$. In terms of these variables \citeANP{Baumgarte1999}
have obtained significantly improved stability properties as compared
with the standard ADM-equations. The ``BSSN''-formalism
has also been successfully implemented by \shortciteNP{Alcubierre2001}. \\
Significant progress in ``3+1'' numerical relativity has been achieved
by the implementation of new slicing conditions
and shift vectors in 3-dimensional evolutions of black holes
(\shortciteNP{Alcubierre2001}). We have included these new gauge conditions
in the list in the previous section. \\
In spite of the progress achieved in recent years, there
remain some difficulties intrinsic to any ``3+1'' formulation. These
are generally concerned with the restriction to a finite grid in numerical
computations. A lot of interest in the modelling of complicated astrophysical
scenarios in the framework of general relativity is motivated by the
advent of highly sensitive gravitational wave detectors. One of the
fundamental requirements of a numerical simulation is therefore the
extraction of gravitational waves and the generation of predicted
gravitational wave templates. It is a well known fact, however, that
gravitational waves are unambiguously defined at null infinity only.
\lcite{Penrose1963} has shown how it is possible to describe infinity
in terms of finite
coordinate values which enables one to incorporate null infinity in a
finite coordinate grid. In numerical relativity, however,
this ``compactification'' is only practical if the coordinates are adapted
to the characteristics of the underlying equations and it is not entirely
clear how to implement this technique in ``3+1'' formulations.
Consequently approximating techniques are used
to interprete gravitational waves at finite
radii. Furthermore outgoing radiation boundary conditions need to be
specified at the outer grid boundaries. These will normally give
rise to spurious
reflections which contaminate the numerical evolution. \\
The difficulties concerning the interpretation of gravitational waves
in ``3+1'' formulations have been known for a long time and
motivated the development of alternative decompositions of spacetime
as early as the early sixties (\shortciteNP{Bondi1962},
\shortciteNP{Sachs1962}).
In the next section we will discuss this {\em characteristic formulation}
in more detail. A generic problem of this approach, however, arises
from the fact that light rays are deflected by matter. In regions of strong
curvature the focusing of light rays may give rise to so-called caustics.
If that is the case the characteristic foliation
of spacetime which is based on the
null-geodesics will break down. Regions of strong curvature
are generally restricted to small regions around the astrophysical sources.
In this sense the ``3+1'' and the characteristic formalisms complement
each other which has given rise to the idea of {\em Cauchy-characteristic
matching}, i.e. the use of a ``3+1'' scheme for an interior region containing
the astrophysical source and a characteristic method in the outer
vacuum region including null infinity. In section \ref{ccm} we will discuss
these ideas in more detail and develop a Cauchy-characteristic matching code
in cylindrical symmetry.

%=======================================================================
\subsection{The characteristic initial value problem}
In section \ref{threep1} we have seen how one can decompose
spacetime into a 1-parameter family of 3-dimensional space-like hypersurfaces.
An alternative way to foliate spacetime is
based on the characteristic surfaces of the vacuum field equations
which can be shown to be the null surfaces of the underlying
spacetime (\citeNP{Pirani1965}).
Gravitational waves will as a matter of course travel along null geodesics
and the characteristic approach is thus particularly suitable for the 
analysis of gravitational waves. It is this property which provided the
main motivation for the ground breaking work by
\scite{Bondi1962} and \lcite{Sachs1962} which we will follow in our
description of the characteristic formalism. In this discussion
we will consider the vacuum case of the field equations
$\hbox{\vec R}_{\mu \nu}=0$. In the case of Cauchy-characteristic matching
this is normally no restriction since matter is assumed to be present in the
inner Cauchy region only.

%=======================================================================
\subsubsection{Characteristic coordinates}
We start our discussion with a 4-dimensional manifold $M$ and
assume that $M$ is equipped with a metric $\hbox{\vec g}$ of signature
+2. In the Bondi-Sachs formalism the gauge freedom of general relativity
is used to impose the following conditions on the coordinates.
\begin{list}{\rm{(\arabic{count})}}{\usecounter{count}
             \labelwidth1cm \leftmargin1.0cm \labelsep0.4cm \rightmargin0.0cm
             \parsep1.5ex plus0.2ex minus0.1ex \itemsep0.5ex plus0.2ex}
\item It is assumed that there exists a scalar function $u$ with the
      property $ \hbox{\vec g}(\hbox{\vec d}u, \hbox{\vec d}u)  = 0$,
      which means that the
      surfaces $u={\rm const}$ are null surfaces. Such null surfaces
      will always exist if the field equations admit wave-like solutions
      since the corresponding characteristic surfaces can be shown to
      be null (\citeNP{Pirani1965}).
\item A normal direction to these surfaces is defined by
      $\tilde{\k}:=\d u$. It follows that $\langle \tilde{\k},\k\rangle=
      0$ and $\nabla_{\hbox{\smvec k}}\hbox{\vec k} = 0$,
      i.e. the tangent curves of $\k$ are
      null-geodesics. They are normal to the surfaces $u={\rm const}$
      [any vector $\v$ in that surface satisfies $\hbox{\vec g}(\k, \v)=0$]
      and lie in these surfaces ($\langle \d u, \k\rangle  =0$).
\item In order to eliminate coordinate irregularities, the normal vector
      $\hbox{\vec k}_{\mu}$ is assumed to satisfy the conditions
      \begin{align}
        \rho&:=\nabla_{\mu}\hbox{\vec k}^{\mu} \neq 0,\\[10pt]
        |\sigma|^2&:=\frac{1}{2}(\nabla_{\nu}\hbox{\vec k}_{\mu})
               (\nabla^{\nu}\hbox{\vec k}^{\mu})-\rho^2\neq \rho^2,
      \end{align}
      where $\rho$ can be interpreted as the expansion and $\sigma$ as the
      shear of the congruences of null geodesics.
\item The next step consists of labelling the geodesics. For this
      purpose we will use standard angular coordinates
      $\theta$ and $\phi$. These can always be chosen so that
      \begin{align}
        \langle \hbox{\vec d} \theta,\hbox{\vec k} \rangle
             &= \langle\hbox{\vec d} \phi, \hbox{\vec k} \rangle=0, \\[7pt]
        D&:= \hbox{\vec g}_{\theta \theta} \cdot
            \hbox{\vec g}_{\phi \phi}
            - \hbox{\vec g}_{\theta\phi}^2  \neq 0.
      \end{align}
      The first condition implies that the coordinates $\theta$ and $\phi$
      are constant along a geodesic and the second condition ensures
      a non-degenerate 2-dimensional volume element
      \hbox{$\det{(\hbox{\vec g}_{AB})}\ne 0$}, where upper case Latin
      indices run from 2 to 3 corresponding to the coordinates $\theta$
      and $\phi$.
      % on surfaces $u=\mathrm{const}$,
      %$r=\mathrm{const}$.
\item Finally the null geodesics labelled by $u, \theta, \phi$
      are parametrized by a function
      $r(u,\theta,\phi)$. In order to obtain
      a regular parametrization it is necessary that the Jacobian matrix
      of $r$ vanish nowhere. The conditions imposed in (3) on the
      expansion and shear ensure that this will be the case. Bondi and Sachs
      further require the coordinate $r$ to satisfy the relation
      \begin{align}
        r^4 := D(\sin^2\theta )^{-1}.
      \end{align}
      As a consequence the area of the 2-spheres defined by
      $u,r={\rm const}$ is given by $4\pi r^2$ and $r$ is the so-called
      areal radius. This condition corresponds to the radial gauge condition
      discussed in section \ref{lapseshift}.
\end{list}
\begin{figure}[t]
  \centering
  %%%%%%%%%%%%%%%%%%%%%%%%%%%%%%%%%%%%%%%%%%%%%%%%%%%%%%%%%%%%%%%%%%%
\begin{picture}(0,0)%
\epsfig{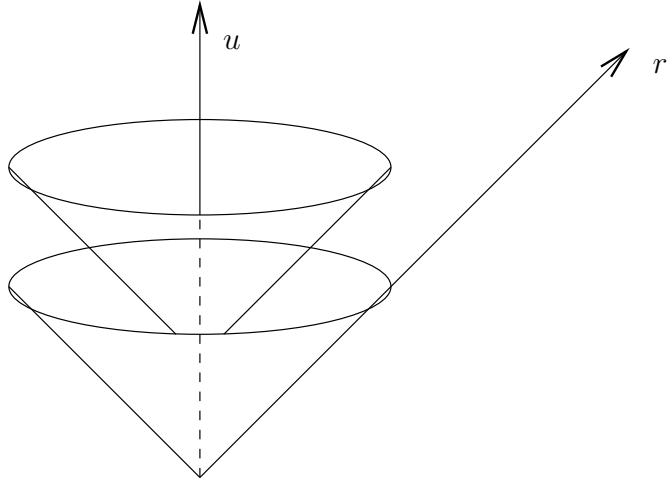}%
\end{picture}%
\setlength{\unitlength}{3947sp}%
\begingroup\makeatletter\ifx\SetFigFont\undefined%
\gdef\SetFigFont#1#2#3#4#5{%
  \reset@font\fontsize{#1}{#2pt}%
  \fontfamily{#3}\fontseries{#4}\fontshape{#5}%
  \selectfont}%
\fi\endgroup%
\begin{picture}(4062,3024)(1189,-2323)
\put(2551,389){\makebox(0,0)[lb]{\smash{\SetFigFont{12}{14.4}{\rmdefault}{\mddefault}{\updefault}$u$}}}
\put(5251,239){\makebox(0,0)[lb]{\smash{\SetFigFont{12}{14.4}{\rmdefault}{\mddefault}{\updefault}$r$}}}
\end{picture}
  %%%%%%%%%%%%%%%%%%%%%%%%%%%%%%%%%%%%%%%%%%%%%%%%%%%%%%%%%%%%%%%%%%%
  \caption{Characteristic coordinates in the case of a null time-like
           foliation.}
  \label{CharCoords}
\end{figure}
The coordinate lines $u={\rm const}$ and $r={\rm const}$ are schematically
illustrated in Fig.\,\ref{CharCoords}
in the case of a time-like $\bbox{\partial}_u$ and a null vector 
$\bbox{\partial}_r$.% ($\hbox{\vec g}_{uu} < 0,\, \hbox{\vec g}_{rr} = 0$).

%=======================================================================
\subsubsection{The Bondi-Sachs line element}
With the coordinate conditions of the previous paragraph the gauge
freedom of general relativity has been used to constrain
the form of the metric. This process is analogous to specifying lapse
and shift in the ``3+1'' formalism. The result can be shown
to be the Bondi-Sachs line element
\begin{align}
  ds^2 &= V\frac{e^{2\beta}}{r}du^2 - 2e^{2\beta} du dr
           +r^2h_{AB}(dx^A -U^Adu)(dx^B-U^Bdu),
\end{align}
where upper case Latin indices again run from 2 to 3 and $h_{AB}$ is defined by
\begin{align}
  & 2h_{AB}dx^A dx^B = (e^{2\gamma}+e^{2\delta})d\theta^2
    +4 \sin\theta \sinh(\gamma-\delta ) d\theta d\phi
    +\sin^2\theta (e^{-2\gamma}+e^{-2\delta})d\phi^2. 
\end{align}
We note that the metric $\hbox{\vec g}$ as a geometric object is still
completely undetermined. This is represented by the six unknowns
$V,U^A,\beta,\gamma,\delta$ which correspond to the six unknown functions
$\hbox{\grvec g}_{ij}$ in the ``3+1'' decomposition. We shall see below
that the characteristic formulation leads to a natural classification
of the field equations and the two gravitational degrees of freedom
are contained in the functions $\gamma$ and $\delta$. The remaining
quantities are determined on each hypersurface irrespective of their
history.

%======================================================================
\subsubsection{Introduction of a tetrad}
In order to classify the field equations, it is convenient to introduce 
basis vectors $\k ,\l ,\m ,\mbar$, where $\hbox{\vec l}$ is a real and
$\m ,\mbar$ are
complex null-vectors and $\k$ is the null-vector field introduced above.
These vectors are required to satisfy the relations
\begin{align}
  \k \cdot \l &= 1, \\
  \m \cdot \mbar &= 1, \\
  \l \cdot \l &= \k \cdot \k = \m \cdot \m = \l \cdot \m = \k \cdot \m = 0.
\end{align}
If we use the complex conjugate of the last equation we further obtain
\begin{align}
  \mbar \cdot \mbar &= \l \cdot \mbar = \k \cdot \mbar = 0.
\end{align}
With the corresponding one forms the metric can now be written as
\begin{align}
  \hbox{\vec g} &= \tilde{\k} \otimes \tilde{\l}+\tilde{\l} \otimes \tilde{\k}
     + \tilde{\mbar} \otimes \tilde{\m} + \tilde{\m} \otimes \tilde{\mbar}.
\end{align}
We note that in spite of the use of complex vectors
eventually all results will be real. In fact if we write the
complex vector as $\hbox{\vec m}=\hbox{\grvec m} + i\hbox{\grvec n}$,
it follows directly from the conditions imposed on $\hbox{\vec m}$,
that $\hbox{\grvec m}$ and $\hbox{\grvec n}$ are space-like vectors
orthogonal to the null-vectors $\hbox{\vec k}$ and $\hbox{\vec l}$.
We conclude that $\hbox{\vec k}$ represents the
null-surfaces $u=\mathrm{const}$, $\hbox{\vec l}$ determines a unique
null-direction out of these hypersurfaces and the complex vector
$\hbox{\vec m}$ defines two spatial directions orthogonal to both
$\hbox{\vec k}$ and $\hbox{\vec l}$. The only remaining freedom is the
phase of $\hbox{\vec m}$ which is normally fixed by relating
$\bar{\hbox{\vec m}}$ to the shear $\sigma$ (see \shortciteANP{Sachs1962}
for details). The benefit of this particular basis
is that it provides a convenient way to create linear
combinations of the vacuum field equations that can be
classified in a natural way.

%========================================================================
\subsubsection{The field equations}
We have already mentioned that the two gravitational degrees of freedom are
contained in the metric functions $\gamma$ and $\delta$. It is a remarkable
property of the characteristic formalism that it naturally leads
to a classification of the field equations which reflects the isolation
of the gravitational degrees of freedom. As originally shown by Bondi the
field equations can be grouped into
\begin{list}{\rm{(\roman{count})}}{\usecounter{count}
             \labelwidth1cm \leftmargin1.5cm \labelsep0.4cm \rightmargin1cm
             \parsep0.5ex plus0.2ex minus0.1ex \itemsep0.5ex plus0.2ex}
\item 6 main equations:\\[7pt]
      {\rm (a)} 4 hypersurface equations:
            $\hbox{\vec R}_{\mu \mu} \hbox{\vec k}^{\mu}
            \hbox{\vec k}^{\nu}
            = \hbox{\vec R}_{\mu \nu} \hbox{\vec k}^{\mu}
            \hbox{\vec m}^{\nu}
            = \hbox{\vec R}_{\mu \nu} \hbox{\vec m}^{\mu}
            {\bar{\hbox{\vec m}}}^{\nu}=0$, \\[7pt]
      {\rm (b)} 2 evolution equations:
            $\hbox{\vec R}_{\mu \nu} \hbox{\vec m}^{\mu}
            \hbox{\vec m}^{\nu}=0$,
\item 1 trivial equation: $\hbox{\vec R}_{\mu \nu}
      \hbox{\vec k}^{\mu} \hbox{\vec l}^{\nu} = 0$,
\item 3 supplementary equations: $\hbox{\vec R}_{\mu \nu}
      \hbox{\vec l}^{\mu} \hbox{\vec m}^{\nu}
      = \hbox{\vec R}_{\mu \nu} \hbox{\vec l}^{\mu}
      \hbox{\vec l}^{\nu} = 0$.
\end{list}
The reasoning for this classification is as follows. If we suppose that
the main equations are satisfied, it can be shown %(\citeANP{Sachs1962})
that
\begin{list}{\rm{(\arabic{count})}}{\usecounter{count}
             \labelwidth1cm \leftmargin1.5cm \labelsep0.4cm \rightmargin1cm
             \parsep0.5ex plus0.2ex minus0.1ex \itemsep0.5ex plus0.2ex}
\item The trivial equation is satisfied: $\hbox{\vec R}_{\mu \nu}
      \hbox{\vec l}^{\mu} \hbox{\vec k}^{\nu} = 0$.
\item $\hbox{\vec R}_{\mu \nu}\hbox{\vec l}^{\mu}\hbox{\vec m}^{\nu}$ 
      vanishes along a null-geodesic (integral curve of $\hbox{\vec k}$)
      either everywhere or nowhere.
\item If all equations except $\hbox{\vec R}_{\mu \nu}
      \hbox{\vec l}^{\mu}\hbox{\vec l}^{\nu}=0$ 
      are satisfied, it follows from the Bianchi identities that
      $\partial_r (r^2 \hbox{\vec R}_{\mu \nu}
      \hbox{\vec l}^{\mu} \hbox{\vec l}^{\nu}) = 0$.
\end{list}
We conclude that the trivial equation is an algebraic consequence of the
main equations. The supplementary equations are satisfied everywhere
if they are satisfied at some value $r={\rm const}$ and
the main equations are satisfied.
As far as the main equations are concerned, we note that
\begin{list}{\rm{(\arabic{count})}}{\usecounter{count}
             \labelwidth1cm \leftmargin1.5cm \labelsep0.4cm \rightmargin1cm
             \parsep0.5ex plus0.2ex minus0.1ex \itemsep0ex plus0.2ex}
\item the hypersurface equations do not contain any derivatives of the metric
functions with respect to $u$,
\item the evolution equations contain the derivatives $\gamma_{,u}$ and
      $\delta_{,u}$ (although in several forms, e.g.
      $\gamma_{,ur}$).
\end{list}
%

%========================================================================
\subsubsection{Boundary conditions}
The boundary conditions are determined by the requirements that
\begin{list}{\rm{(\arabic{count})}}{\usecounter{count}
             \labelwidth1cm \leftmargin1.5cm \labelsep0.4cm \rightmargin1cm
             \parsep0.5ex plus0.2ex minus0.1ex \itemsep0ex plus0.2ex}
\item the spacetime has Euclidean topology at large distance from the source,
\item the spacetime is asymptotically flat,
\item gravitational radiation obeys an outgoing radiation boundary condition.
\end{list}
As shown by \lcite{Sachs1962} these requirements are necessarily satisfied
if the following boundary conditions are imposed.
\begin{list}{\rm{(\arabic{count})}}{\usecounter{count}
             \labelwidth1cm \leftmargin1.5cm \labelsep0.4cm \rightmargin1cm
             \parsep0.5ex plus0.2ex minus0.1ex \itemsep0ex plus0.2ex}
\item For any choice of $u$ one can go to the limit $r\rightarrow \infty$
      along each ray.
\item For this $u$ and any choice of $\theta, \phi$ we have\\[6.0pt]
      $\lim_{r\rightarrow \infty} V/r = -1$\\[6.0pt]
      $\lim_{r\rightarrow \infty} (rU^A) = \lim_{r\rightarrow \infty} \beta =
      \lim_{r\rightarrow \infty} \gamma = \lim_{r\rightarrow \infty} \delta =0$.
\item For $u_0\leq u\leq u_1$, $r_0 \leq r\leq \infty$, $0\leq \theta \leq \pi$,
      $0\leq \phi \leq 2\pi$ all metric components and quantities of interest
      can be expressed as a series in $r^{-1}$ with at most a finite pole at
      $r=\infty$.
\end{list}
\begin{figure}[t]
\centering
%%%%%%%%%%%%%%%%%%%%%%%%%%%%%%%%%%%%%%%%%%%%%%%%%%%%%%%%%%%%%%%%%%%%
\begin{picture}(0,0)%
\epsfig{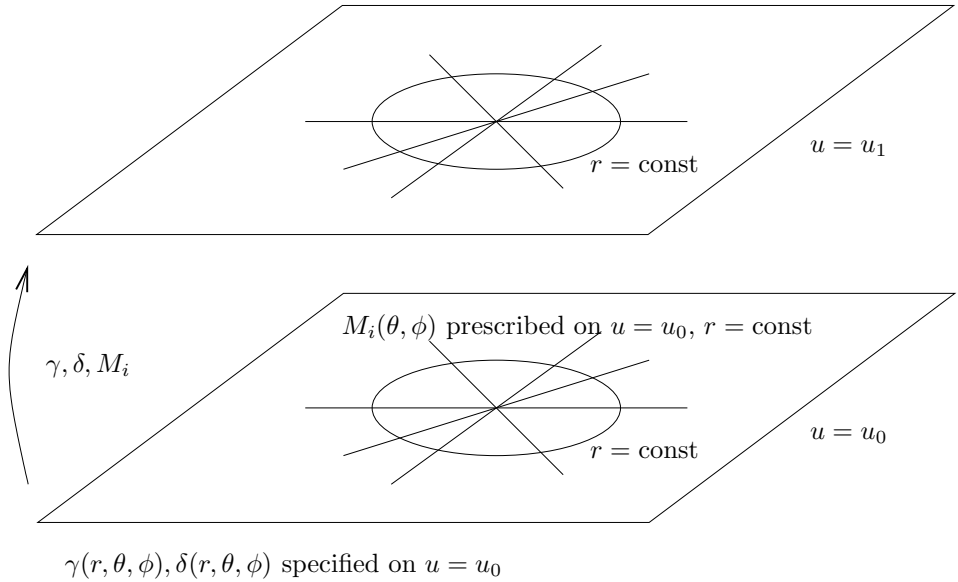}%
\end{picture}%
\setlength{\unitlength}{3158sp}%
\begingroup\makeatletter\ifx\SetFigFont\undefined%
\gdef\SetFigFont#1#2#3#4#5{%
  \reset@font\fontsize{#1}{#2pt}%
  \fontfamily{#3}\fontseries{#4}\fontshape{#5}%
  \selectfont}%
\fi\endgroup%
\begin{picture}(7449,4507)(1264,-4394)
\put(3901,-2461){\makebox(0,0)[lb]{\smash{\SetFigFont{10}{12.0}{\rmdefault}{\mddefault}{\updefault}$M_i(\theta,\phi)$ prescribed on $u=u_0$, $r={\rm const}$}}}
\put(1726,-4336){\makebox(0,0)[lb]{\smash{\SetFigFont{10}{12.0}{\rmdefault}{\mddefault}{\updefault}$\gamma(r,\theta,\phi), \delta(r,\theta,\phi)$ specified on $u=u_0$}}}
\put(5851,-3436){\makebox(0,0)[lb]{\smash{\SetFigFont{10}{12.0}{\rmdefault}{\mddefault}{\updefault}$r={\rm const}$}}}
\put(5851,-1186){\makebox(0,0)[lb]{\smash{\SetFigFont{10}{12.0}{\rmdefault}{\mddefault}{\updefault}$r={\rm const}$}}}
\put(7576,-1036){\makebox(0,0)[lb]{\smash{\SetFigFont{10}{12.0}{\rmdefault}{\mddefault}{\updefault}$u=u_1$}}}
\put(7576,-3286){\makebox(0,0)[lb]{\smash{\SetFigFont{10}{12.0}{\rmdefault}{\mddefault}{\updefault}$u=u_0$}}}
\put(1576,-2761){\makebox(0,0)[lb]{\smash{\SetFigFont{10}{12.0}{\rmdefault}{\mddefault}{\updefault}$\gamma, \delta, M_i$}}}
\end{picture}
%%%%%%%%%%%%%%%%%%%%%%%%%%%%%%%%%%%%%%%%%%%%%%%%%%%%%%%%%%%%%%%%%%%%
\caption{Evolution of the initial data in the characteristic formalism.}
\label{EvolChar}
\end{figure}
%

%========================================================================
\subsubsection{Initial data and the integration of the field equations}
The evolution of the metric variables $V$, $U^A$, $\beta$, $\gamma$
and $\delta$ can be split up into four steps.
In the discussion of these steps it will become obvious what
type of initial data we need to specify in order to start the
evolution of the metric. We have graphically illustrated the integration
of the field equations from time slice $u_0$ to $u_1$ in Fig.\,\ref{EvolChar}.\\

{\bf 1.)} We start by providing initial data for $\gamma$ and $\delta$
on a hypersurface $u=u_0$.
This means that we need to specify two functions of
$(r,\theta,\phi)$.\\

{\bf 2.)} Next the hypersurface equations are integrated along $r$ to
obtain $\beta$, $V$, $U^A$ on
the initial hypersurface. For this purpose we need to specify
three functions of integration $M_i(\theta, \phi)$.
A potential fourth function of integration for $\beta$ is fixed by the
boundary condition $\lim_{r\rightarrow \infty} \beta=0$.\\

{\bf 3.)} We use the evolution equations in order to calculate
$\gamma$ and $\delta$ on the future hypersurface $u = u_1$.
The evolution equations contain the $u$-derivatives of $\gamma$ and $\delta$
in the form $\gamma_{,ur}$, $\delta_{,ur}$. Consequently the solution
requires in principle the integration over $r$ to obtain the corresponding
$u$-derivatives. For this purpose we need to specify two functions
of $(u,\theta,\phi)$ as functions of integration.
These functions are commonly introduced as the complex {\em news function}
$\displaystyle{\frac{\partial c}{\partial u}}(u,\theta,\phi)$.
Below we will illustrate the meaning of news function in more detail.\\

{\bf 4.)} Finally, the supplementary equations are used to evolve
the $M_i(\theta,\phi)$ onto the hypersurface
$u=u_1$.\\

\begin{figure}[t]
\centering
%%%%%%%%%%%%%%%%%%%%%%%%%%%%%%%%%%%%%%%%%%%%%%%%%%%%%%%%%%%%%%%%%%%%%%%%
\begin{picture}(0,0)%
\epsfig{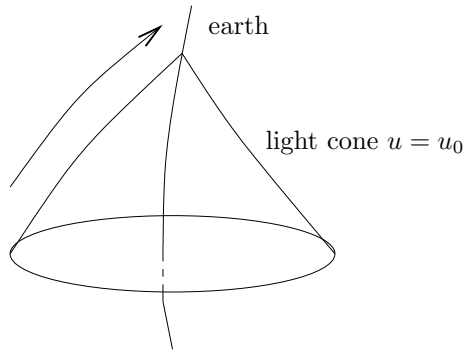}%
\end{picture}%
\setlength{\unitlength}{3158sp}%
\begingroup\makeatletter\ifx\SetFigFont\undefined%
\gdef\SetFigFont#1#2#3#4#5{%
  \reset@font\fontsize{#1}{#2pt}%
  \fontfamily{#3}\fontseries{#4}\fontshape{#5}%
  \selectfont}%
\fi\endgroup%
\begin{picture}(2574,2724)(664,-3223)
\put(2251,-736){\makebox(0,0)[lb]{\smash{\SetFigFont{10}{12.0}{\rmdefault}{\mddefault}{\updefault}earth}}}
\put(2701,-1636){\makebox(0,0)[lb]{\smash{\SetFigFont{10}{12.0}{\rmdefault}{\mddefault}{\updefault}light cone $u=u_0$}}}
\end{picture}
%%%%%%%%%%%%%%%%%%%%%%%%%%%%%%%%%%%%%%%%%%%%%%%%%%%%%%%%%%%%%%%%%%%%%%
\caption{Information on a past light cone is insufficient to determine the
         future of the earth.}
\label{NewsFunc}
\end{figure}

We complete the description of the characteristic formalism
with an explanation why the news function needs to be specified
for all values of $u$.
%Let us discuss further the necessity of prescribing
%$\displaystyle{\frac{\partial c}{\partial u}}(u,\theta,\phi)$
%for all $u$. We will look at past light cones only, but
%the same idea applies to future light cones.
For this purpose we consider the path of an object, e.g. the
earth, in spacetime as illustrated in Fig.\,\ref{NewsFunc}. 
Even if we have complete data on the past light cone $t+r=u_0$, we can still
not determine the future of the earth. There may be waves outside $u=u_0$,
that have not yet reached the planet.
$\displaystyle{\frac{\partial c}{\partial u}}(u,\theta,\phi)$ provides this
extra information and is, therefore, called the news function. This
is to be contrasted with the ``3+1'' decomposition discussed above,
where the initial data on a slice $t=\mathrm{const}$
completely determines the evolution up to the specification of boundary
conditions.\\
In sections \ref{ccm} and \ref{cstr} we will use a similar characteristic
formulation with a different gauge choice to evolve cylindrically
symmetric vacuum spacetimes and dynamic cosmic strings. The presence of
matter in the latter case does not result in any significant complications
compared with the vacuum case described in this section.

%=========================================================================
\subsection{Numerical methods}
In order to numerically solve a set of differential equations, the equations
have to be cast into a form suitable for a computer based treatment.
The most common method used for this purpose
is {\em finite differencing} which replaces
derivatives with finite difference expressions and thus converts
differential equations into large sets of algebraic equations. Alternative
methods, as for example {\em spectral} or {\em finite element methods}
have been used successfully in various cases.
In this thesis, however, we will use finite difference
methods throughout and therefore restrict our description to this approach. In
particular, we will concentrate on finite differencing in the case of
two dimensions, time and one spatial dimension, which we will label
by the coordinates $t$ and $x$.

%======================================
\subsubsection{The numerical grid}
\begin{figure}[t]
  \centering
  \epsfig{file=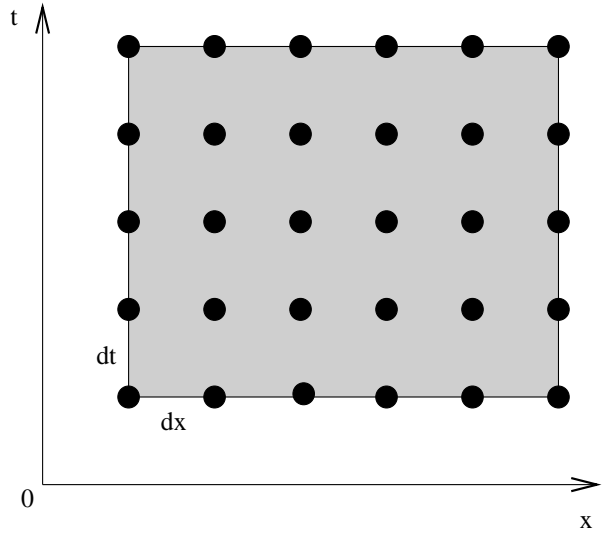, height=225pt, width=200pt, angle=-90}
  \caption{A 2-dimensional grid with constant spacing. We note that the
           domain does not have to be rectangular and different values
           for $dt$ and $dx$ may be used.}
  \label{FDE_grid}
\end{figure}

Given a system of differential equations, our aim is to determine the solution
$f$ in a subset $\Omega \subset \hbox{\oneone R}^2$. In finite differencing the
domain of $f$ is replaced by a set of discrete grid points as
illustrated in Fig.\,\ref{FDE_grid} and
the numerical scheme will provide values for $f$ at these grid points
only. If information of the function $f$ is required between the
grid points we will derive the corresponding values from interpolation. \\
Throughout this work, we will only use uniform grids which means that
the distance $dx$ between neighbouring grid points is independent of
position $x$ and time $t$.
At any given value of $t$ the interval $[x_0,x_K]$ will therefore
be replaced by the set of points $(x_0$, $x_0+dx$, $x_0+2dx$,$\ldots$,
$x_K)$ with
\begin{align}
  dx &= \frac{x_K-x_0}{K}.
\end{align}
In section \ref{pert} we will demonstrate how
a coordinate transformation to a new spatial coordinate $y$ can be used
to simulate an inhomogeneous grid in terms of the original coordinate $x$
without abandoning the concept of a uniform grid. \\
For the presentation of finite difference expressions
it is convenient to introduce a short hand notation for the
function values at the grid points. For this purpose we define
$f_k^n :=f(x_k, t_n)$. If the meaning is obvious we may omit either index.

%=======================================
\subsubsection{Derivatives and finite differences}
We describe the approximation of derivatives with finite differences in the
case of spatial derivatives. The same ideas apply to time derivatives.
Suppose a function $f$ is given at positions $x_0,\ldots,x_K$
for fixed time and
we want to calculate $\displaystyle{\frac{\partial^m f}{\partial x^m}}$
at $x_k$.
For this purpose we expand $f$ in a Taylor series about $x_k$ which allows us
to express $f_{k}$, $f_{k-1}$, $f_{k+1}$,$\ldots$ in terms of $f$ and
its derivatives at $x_k$. Next the derivative that needs
to be calculated is
expressed as a linear combination of the function values at neighbouring
grid points. The required finite difference expression is then
obtained from inserting the Taylor expansions for the $f_k$, $f_{k-1}$,
$f_{k+1},\ldots$
and comparing the
coefficients on both sides of the equations. The number of grid points
that needs to be included in this calculation depends
on the degree of the derivative and the order of accuracy to be achieved.\\
We illustrate these ideas by calculating the second derivative $f_k''$
with second order accuracy. We assume that the function $f$ is known at
the grid points $x_k$, $x_{k-1}$, $x_{k-2}$ and $x_{k-3}$.
By Taylor expanding $f$
around $x_k$ we can relate the function values to $f$ and its derivatives
at $x_k$
\begin{align}
  f_k &= f_k, \label{FDE_FN} \\
  f_{k-1} &= f_k-f_k'dx +\frac{1}{2}f_k''dx^2-\frac{1}{6}f_k'''dx^3
             +\mathscr{O}(dx^4), \label{FDE_FNM1} \\
  f_{k-2} &= f_k-f_k'2dx+\frac{1}{2}f_k''4dx^2-\frac{1}{6}f_k'''8dx^3
             +\mathscr{O}(dx^4), \label{FDE_FNM2} \\
  f_{k-3} &= f_k-f_k'3dx+\frac{1}{2}f_k''9dx^2-\frac{1}{6}f_k'''27dx^2
             +\mathscr{O}(dx^4). \label{FDE_FNM3}
\end{align}
Next we write $f_k''$ as a linear combination of the function values
\begin{align}
  dx^2 \cdot f_k'' &= A f_k + B f_{k-1} + C f_{k-2} + D f_{k-3}.
\end{align}
If we insert Eqs.\,(\ref{FDE_FN})-(\ref{FDE_FNM3}) for the function values
$f_{k-3},\ldots ,f_k$ and compare
the coefficients of both sides of the equation, we obtain the system of
linear equations
\begin{align}
  \begin{split}
  A+B+C+D &= 0, \\
  B+2C+3D &= 0, \\
  B+4C+9D &= 2, \\
  B+8C+27D &= 0. 
  \end{split}
\end{align}
The solution is $A=2$, $B=-5$, $C=4$, $D=-1$ and we can approximate the
derivative $f_k''$ with second order accuracy by
\begin{align}
  f_k'' &= \frac{2f_k - 5f_{k-1} +4f_{k-2} - f_{k-3}}{dx^2}
      + \mathscr{O}(dx^2).
  \label{ddfone}
\end{align}
In general, a one sided calculation as used in this example yields less
accurate estimates of the derivative and two sided approximations are to
be preferred. In our case the centred finite difference expression is
given by
\begin{align}
  f_k'' &= \frac{f_{k+1} - 2f_k + f_{k-1}}{dx^2} + \mathscr{O}(dx^2).
  \label{ddfcent}
\end{align}
If we substitute expressions corresponding to (\ref{ddfone}) or
(\ref{ddfcent}) for all derivatives, the differential equation is
replaced by a large set of algebraic equations.

%=======================================================================
\subsubsection{The leapfrog scheme}   \label{FDE_LEAP}
The {\em leapfrog} scheme is a second order in space and time 
finite differencing scheme in which three successive time-levels are used at each integration step.
If we assume that the differential equation can be written in the form
\begin{align}
  f_{,t} &= H(f, f_{,x}, f_{,xx},...,x,t),
\end{align}
the right hand side
can be evaluated on the $n^{th}$ time slice.
The time derivative, on the other hand, is approximated by
\begin{align}
  \frac{\partial f}{\partial t} &= \frac{f^{n+1}-f^{n-1}}{2dt},
\end{align}
and the difference equation can be explicitly solved for $f^{n+1}$.
Because of the centred finite difference approximation for
$f_{,t}$, three
time slices are involved in the calculation. As an example
\begin{figure}
  \centering
  \epsfig{file=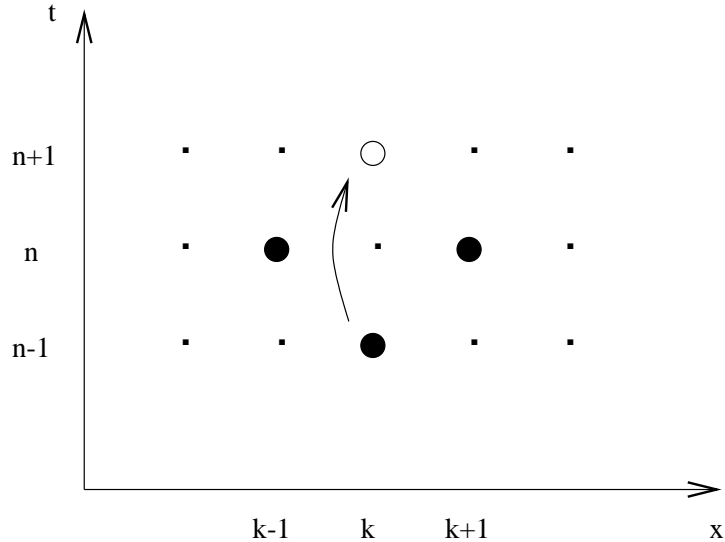, height=270pt, width=202pt, angle=-90}
  \caption{The leapfrog scheme: In the evolution one slice is leapt over.}
  \label{leapfrog}
\end{figure}
we consider the special case where $H=f_{,x}$. At the spatial position
$x_k$ the finite difference equation is then given by
\begin{align}
  f^{n+1}_k &= f^{n-1}_k + \frac{dt}{dx}(f^n_{k+1} - f^n_{k-1}).
  \label{FDE_leapfrog}
\end{align}
The value of $f$ is taken on slice $n-1$ and we ``leap'' across
slice $n$ to calculate $f^{n+1}$. This property is schematically
illustrated in Fig.\,\ref{leapfrog} and has given the scheme
its characteristic name. The need to store the function values of
two time slices makes this scheme more memory intensive than 2-level
schemes such as the McCormack scheme discussed in the next section.
Second order accurate two-level schemes, on the other hand, involve more
complicated finite difference expressions and are therefore more
CPU-intensive. \\
A potential problem of the leap-frog scheme is its vulnerability to the
so-called {\em mesh-drifting} effect, an instability that results from
the decoupling of odd and even mesh points. This instability can often
be cured by evolving some of the variables on a separate grid translated
with respect to the original one
by half a grid step ({\em staggered} leap-frog) or introducing artificial
dissipation which couples odd and even grid points. In our application of 
this scheme in section \ref{ccm}, however, we do not encounter this
problem and have no need to use either of the remedies. \\
We finally note that in Eq.\,(\ref{FDE_leapfrog}) the function
value on the new slice $f^{n+1}_k$ is expressed explicitly in terms
of known function values on previous slices. Finite differencing schemes with
this property are called {\em explicit} schemes. In section \ref{FDE_CN} we
will, by contrast, introduce an {\em implicit} scheme where this is
in general not possible for non-linear partial differential equations
and iterative methods or linear solvers are used to determine
the $f^{n+1}_k$.

%=========================================================================
\subsubsection{The McCormack scheme}
\label{McCormack}
The {\em McCormack} scheme is another second order 
accurate explicit finite differencing
method. In contrast to the leapfrog scheme it is a two-level method, i.e.
requires storage of one previous slice only. However, this comes at the expense
of two computation steps in the calculation 
of the new values, a {\em predictor}
and a {\em corrector} step. We illustrate this method by considering
the partial differential equation
\begin{align}
  f_{,t} &= H(f, f_{,x}, x, t).
\end{align}
In the first step preliminary values on the new time slice are calculated
according to
\begin{align}
  \tilde{f}^{n+1}_k &= f^n_k + \Delta t \cdot H^n_{k,k-1},
\end{align}
where $H^n_{k,k-1}$ is the source term evaluated to second order accuracy at
$x_{k-1/2}$ by using $f^n_{k-1}$
and $f^n_k$. This predictor step itself is a first order accurate scheme,
but the terms of first order truncation error are eliminated in the
corrector step
\begin{align}
  f^{n+1}_k &= f^n_k + \Delta t \cdot \frac{1}{2} \left( H^n_{k,k-1}
               + \tilde{H}^{n+1}_{k+1,k} \right),
\end{align}
where $\tilde{H}^{n+1}_{k+1,k}$ is the source term evaluated from the
preliminary values $\tilde{f}^{n+1}_k$ and $\tilde{f}^{n+1}_{k+1}$.
The extension to systems with more functions is obvious.\\

%=========================================================================
\subsubsection{Relaxation}
\label{relaxation}
{\em Relaxation} is a method for solving so-called {\em two point boundary
value problems}, that is ordinary differential equations (ODEs) where boundary
conditions are given at different locations on the grid. A straightforward
integration to obtain the solution is not possible in these cases and one
needs to resort to more sophisticated techniques. One such technique, of which
we will make extensive use in this work is numerical relaxation. In the
case of ordinary differential equations 
only one independent coordinate is present which can be visualised
in Fig.\,\ref{FDE_grid} by suppressing the time dimension so that we have
only one row of grid points.
It is straightforward to see that any ordinary differential equation
can be written as a first order system. Without loss of generality we
will therefore restrict our discussion to this case.
Suppose for example that we have a system of 4 ODEs for 4 functions
$A(x)$, $B(x)$, $C(x)$ and $D(x)$ given by
\begin{align}
  G_i(A, A_{,x}, B, B_{,x}, C, C_{,x}, D, D_{,x}) &= 0,  \qquad i=1\ldots 4.
  \label{FDE_ode}
\end{align}
A numerical solution consists of $4K$ function values $A_1$, $B_1$, $C_1$,
$D_1$, $A_2$, $B_2$ and so on. It is convenient to introduce a vector $f_j$ to
label these values, i.e. $f_1:=A_1$, $f_2:=B_1$ and so on.
For each pair of grid points $k$, $k-1$ we
apply centred finite differencing according to
\begin{align}
  A &= \frac{1}{2}(A_k + A_{k-1}),  \label{relax_FDE1} \\[10pt]
  A_{,x} &= \frac{A_k-A_{k-1}}{\Delta x}, \label{relax_FDE2}
\end{align}
and likewise for the other functions. In combination with Eq.\,(\ref{FDE_ode})
this
amounts up to $4(K-1)$ algebraic equations for the 4K variables $f_j$. This
set is completed by 4 boundary conditions for $A$, $B$, $C$ and $D$ and
we arrive at 4K algebraic equations which we write in the form
\begin{align}
  F_i(f_j) &= 0.
  \label{FDE_Ff}
\end{align}
In general these equations are non-linear and we have to resort to iterative
techniques to obtain a solution $f_j$. For this purpose we assume that $f_j$
is a solution of (\ref{FDE_Ff}) and $f^0_j$ is a sufficiently close guess.
Then $4K$-dimensional Taylor expansion yields
\begin{align}
  0 &= F_i(f_j) \approx F_i(f^0_j) + \sum_j{\frac{\partial F_i}{\partial f_j}
       \Delta f_j},
\end{align}
where $\Delta f_j = f_j - f^0_j$. This is simply a system of linear
equations which we can write as
\begin{align}
  {\bf A}\Delta{\bf f} = {\bf b},
\end{align}
where
\begin{align}
  {\bf A}_{ij} = \frac{\partial F_i}{\partial f_j}, \\[15pt]
  {\bf b}_i = -F_i(f^0_j).
\end{align}
Even though the Jacobi matrix {\bf A} is a $4K$ by $4K$ matrix,
it is a sparse matrix which greatly simplifies its inversion. If the equations
$F_i = 0$ are ordered appropriately, {\bf A} has block diagonal structure
and can be inverted by standard methods (see for example
\shortciteNP{Press1989}).
Starting with an initial guess $f^0_j$, we can calculate the correction
$\Delta f_j$ which leads to an improved approximation $f^1_j$ and the
process is repeated until the norm $||\Delta f_j||$ satisfies some
convergence criterion. This iteration scheme is the Newton-Raphson
method generalized to $4K$ dimensions and usually converges fast.

%========================================================================
\subsubsection{The Crank-Nicholson scheme}
\label{FDE_CN}
\begin{figure}[t]
\centering
%%%%%%%%%%%%%%%%%%%%%%%%%%%%%%%%%%%%%%%%%%%%%%%%%%%%%%%%%%%%%%%%%%%%%%
\begin{picture}(0,0)%
\epsfig{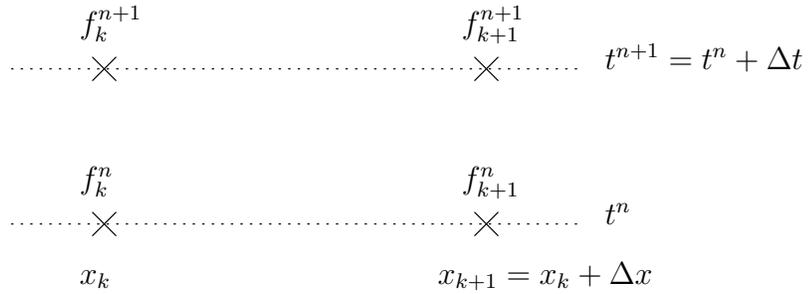}%
\end{picture}%
\setlength{\unitlength}{3947sp}%
\begingroup\makeatletter\ifx\SetFigFont\undefined%
\gdef\SetFigFont#1#2#3#4#5{%
  \reset@font\fontsize{#1}{#2pt}%
  \fontfamily{#3}\fontseries{#4}\fontshape{#5}%
  \selectfont}%
\fi\endgroup%
\begin{picture}(3762,1770)(589,-1111)
\put(1051,464){\makebox(0,0)[lb]{\smash{\SetFigFont{12}{14.4}{\rmdefault}{\mddefault}{\updefault}$f^{n+1}_k$}}}
\put(3451,464){\makebox(0,0)[lb]{\smash{\SetFigFont{12}{14.4}{\rmdefault}{\mddefault}{\updefault}$f^{n+1}_{k+1}$}}}
\put(1051,-511){\makebox(0,0)[lb]{\smash{\SetFigFont{12}{14.4}{\rmdefault}{\mddefault}{\updefault}$f^n_k$}}}
\put(3451,-511){\makebox(0,0)[lb]{\smash{\SetFigFont{12}{14.4}{\rmdefault}{\mddefault}{\updefault}$f^n_{k+1}$}}}
\put(4351,-736){\makebox(0,0)[lb]{\smash{\SetFigFont{12}{14.4}{\rmdefault}{\mddefault}{\updefault}$t^n$}}}
\put(4351,239){\makebox(0,0)[lb]{\smash{\SetFigFont{12}{14.4}{\rmdefault}{\mddefault}{\updefault}$t^{n+1}=t^n+\Delta t$}}}
\put(3301,-1111){\makebox(0,0)[lb]{\smash{\SetFigFont{12}{14.4}{\rmdefault}{\mddefault}{\updefault}$x_{k+1}=x_k + \Delta x$}}}
\put(1051,-1111){\makebox(0,0)[lb]{\smash{\SetFigFont{12}{14.4}{\rmdefault}{\mddefault}{\updefault}$x_k$}}}
\end{picture}
%%%%%%%%%%%%%%%%%%%%%%%%%%%%%%%%%%%%%%%%%%%%%%%%%%%%%%%%%%%%%%%%%%%%
\caption{The numerical stencil used in the Crank Nicholson scheme to obtain
         centred second order accurate expressions for $f$, $f_{,x}$
         and $f_{,t}$ at position $(x_k+\Delta x/2, t^n+\Delta t/2)$.}
\label{FDE_cngrid}
\end{figure}

The {\em Crank-Nicholson} scheme is a two-level evolution scheme for partial
differential equations and can be considered a generalization
of the relaxation scheme. Again the system of equations
is rewritten as a first order system by introducing auxiliary variables.
For convenience we will illustrate the scheme for one equation and one
function $f$ only. The extension to more functions is obvious.
Consider the PDE
\begin{align}
  \label{FDE_cnpde}
  G(f, f_{,t}, f_{,x}, x, t) &= 0
\end{align}
on a grid of the type shown in Fig.\,\ref{FDE_grid} with $K$ points on each
slice $t={\rm const}$. We can use a stencil of the type shown in
Fig.\,\ref{FDE_cngrid} to obtain second order centred finite
difference expressions for the functions and their derivatives
according to
\begin{align}
  \label{FDE_cnfde}
  f   &= \frac{1}{4}(f^{n+1}_k + f^{n+1}_{k-1} + f^n_k + f^n_{k-1}),\\[15pt]
  f_{,x} &= \frac{f^{n+1}_k -f^{n+1}_{k-1} + f^n_k-f^n_{k-1}}{2\Delta x},
  \\[15pt] \label{cn_fu}
  f_{,t} &= \frac{f^{n+1}_k + f^{n+1}_{k-1} - f^n_k - f^n_{k-1}}{2\Delta t}.
\end{align}
Inserting these relations into Eq.\,(\ref{FDE_cnpde}) we obtain
$K-1$ algebraic equations for the $K$ unknown values $f^{n+1}_k$ in terms
of the known $f^n_k$. The set is completed by the boundary condition for $f$
and we are in exactly the same situation as in Eq.\,(\ref{FDE_Ff}) in
the relaxation scheme. Note that each algebraic equation involves two
unknown values $f^{n+1}_k$, $f^{n+1}_{k-1}$, so it is in general not possible
to obtain explicit expressions similar to Eq.\,(\ref{FDE_leapfrog}) in the
leapfrog-scheme. Therefore methods like the Crank-Nicholson scheme are
called {\em implicit} and a solution is obtained by using
iterative methods. The initial guess for the values on the new
slice is usually taken from the previous slice. \\
An explicit variation of the Crank-Nicholson scheme which has attracted
a good deal of attention recently is the so-called
{\em iterative Crank-Nicholson} method. There one calculates intermediate
values $\tilde{u}^{n+1}$ on the new time slice according to the unconditionally
unstable {\em forward time centred space} method and averages these values
with the data on the old slice $n$ to obtain
$\bar{u}^{n+1/2}=(\tilde{u}^{n+1} + u^n)/2$. These averaged values are then
used to calculate the source terms in the partial differential equation
at time $t^{n+1/2}$ and can be used to evolve the data with centred finite
differencing of the time derivatives in a second order scheme.
In fact this iteration process can be repeated arbitrarily often and
the number of iterations significantly affects the stability properties of
the scheme. In particular \lcite{Teukolsky2000} has shown that
the smallest number of iterations required for a stable method is two
and that any further iterations do not lead to any superior performance in
terms of stability and accuracy. \\
Before we apply these numerical schemes to general relativistic scenarios,
we discuss some general properties of numerical evolution schemes.

%=======================================
\subsubsection{Consistency}
\label{consistency}
If we take the difference equations and calculate the $f_k^n$ as a
Taylor series about some fixed grid-point, we will again arrive at a
differential equation for $f$.
The difference between this differential equation and the original one is
the {\em truncation error}. The numerical scheme is said to be
{\em consistent} if the truncation error vanishes in the limit
$dx,dt\rightarrow0$ (see for example \citeNP{LeVeque1992}).
Assuming that $dx$ and $dt$ differ
by a constant factor in the limit $dx \to 0$, 
the scheme is of {\em $n^{th}$ order accuracy} if
the leading term of the truncation error vanishes as $dx^n$.

%=======================================
\subsubsection{Stability}
\label{stability}
The concept of stability is concerned with an exponentially
increasing deviation of the numerical solution from the solution
of the underlying differential equation. If such a deviation is present
either due to the initial data or round off errors, it will quickly
swamp the entire numerical solution and make the code practically
useless. The stability of a code can depend on many properties. Often
changing the grid parameters $dx$, $dt$ has a substantial effect on the
stability.\\
In the case of linear partial differential equations one can use the
{\em von Neumann stability analysis} in order to test
finite differencing schemes for stability. For this purpose
we assume that the numerical grid is uniform, i.e. $dx$ and $dt$ are
constant. The solution of the
difference equation can then be expanded as a Fourier series
\begin{align}
  f^n(x) &= \sum_{{\kappa}} \hat{f}^n(\kappa) e^{i\kappa x},
\end{align}
where $\kappa$ is a spatial wave vector (1-dimensional in our case).
It is sufficient to consider one mode $\hat{f}^n(\kappa)e^{i{\kappa} x}$
which can be written as
\begin{align}
  \hat{f}^n_{{\kappa}} &=\xi({\kappa})^n e^{i{\kappa} x}, \label{vonNeumann}
\end{align}
if the coefficients of the difference equations show sufficiently weak
variation in space and time and can be considered nearly constant.
The important aspect is that the amplitude at some time is obtained from
that of the preceding time step by multiplication with a {\em
time independent} factor $\xi({\kappa})$. If $|\xi({\kappa})| > 1$
the scheme is unstable. In practice, Eq.\,(\ref{vonNeumann}) 
is inserted into the difference equations
which then is solved for $\xi$. For many applications, the result is the
well known {\em Courant-Friedrichs-Lewy condition} (CFL-condition)
\begin{align}
  \left| \frac{\lambda_i\cdot dt}{dx} \right| \leq 1,
\end{align}
where the $\lambda_i$ are the slopes of the characteristics
of the underlying system of PDEs. An intuitive
interpretation of this result is that the numerical domain of
dependency of the point where $f$ is to be calculated must contain
the physical one. Indeed this condition was recognised as a
necessary stability condition for {\em any} numerical scheme by
\lcite{Courant1928} (See \shortciteNP{Courant1967} for an
English translation). The CFL condition is therefore commonly used in
non-linear codes to determine the permissible Courant factor $dt/dx$.
We will illustrate the use of this criterion in the evolution
of non-linear radial oscillations of neutron stars in
section \ref{PERT_NUMERICS}.

%=====================================
\subsubsection{Convergence}
It is necessary to carefully distinguish between {\em consistency},
{\em stability} and {\em convergence} of a code. The convergence of
a numerical method is a stronger requirement than consistency or
stability. It is quite obvious, for example that a consistent method will
not be convergent if it is unstable. In order to define convergence,
we consider a solution $f$ of the system of differential equations and
a solution $F$ of the corresponding
difference equations. We note that $F$ is never obtained in practice due to
round off errors. A scheme is said to {\em converge} if
\begin{align}
  |f_k^n-F_k^n| \rightarrow 0 \qquad {\rm as}
  \qquad dx,dt \rightarrow 0. \label{FDE_CONVERGENCE}
\end{align}
In the case of linear equations convergence can be ensured by the
{\em Lax Equivalence Theorem} which states:
{\em Given a properly posed linear initial value problem and a
finite difference approximation to it that satisfies the consistency
condition, the stability is a necessary and sufficient condition for
convergence} (see for example \citeNP{Richtmyer1967}).\\
In the case of non-linear equations there is no corresponding theorem but in
some cases we will be able to check our codes for convergence by
comparing the results with known analytic solutions. If such analytic
solutions are not available, we need to use reference solutions
obtained for high resolutions instead. We will thus be able to ensure the
{\em Cauchy convergence} of the numerical scheme. This is, however,
a weaker statement than Eq.\,(\ref{FDE_CONVERGENCE}) and does not
strictly guarantee convergence to the solution of the differential equations.

\newpage
%=====================================================================
\section{Cauchy characteristic matching in cylindrical symmetry}
\label{ccm}
%
%

%=========================================================================
\subsection{The idea of Cauchy characteristic matching}
\label{CCM}
\begin{figure}[b]
  \centering
  \epsfig{file=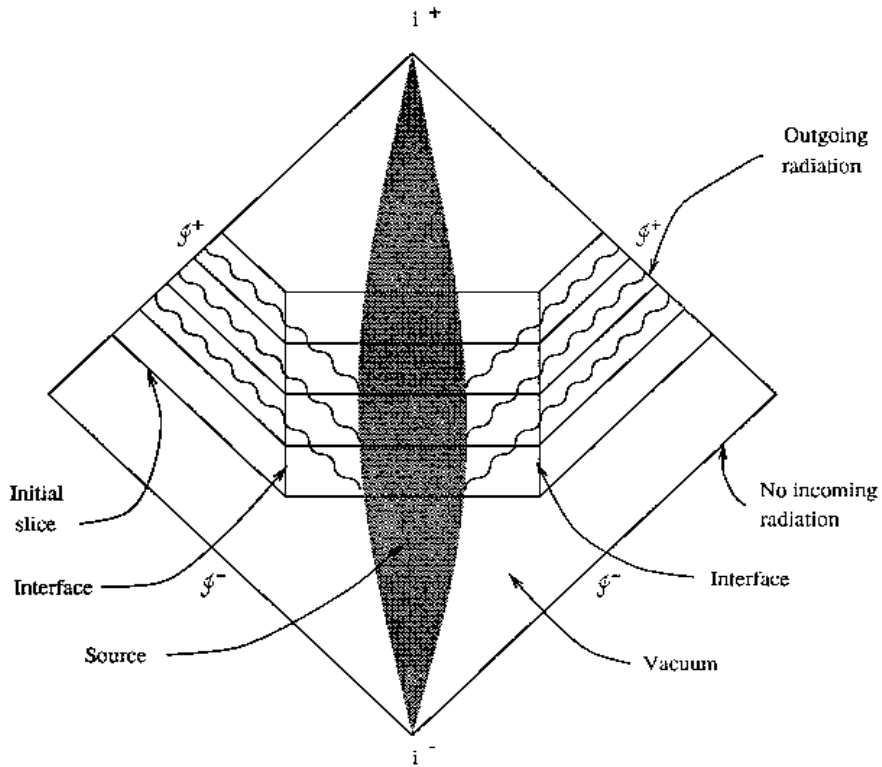, height=300pt, width=350pt, angle=0}
  \caption{In this conformal diagram Cauchy characteristic
           matching is schematically illustrated.
           In the inner region matter is evolved with
           a ``3+1'' scheme whereas characteristic coordinates
           defined by the null geodesics are used in the
           outer vacuum region. The two formalisms are matched
           at the interface at finite radius.}
  \label{CCM_illu}
\end{figure}
Cauchy characteristic matching (CCM) is a method that simultaneously
makes use of the beneficial properties of the ``3+1'' and the characteristic
formalism. In section \ref{threep1} we have seen that in the ``3+1''
case spacetime is decomposed into 3-dimensional
space-like hypersurfaces threaded by a one parameter family of curves. The
dynamic variables are the
components $\hbox{\grvec g}_{ij}$ of the 3-metric of the hypersurfaces.
A complete set of initial data
consists of values for $\hbox{\grvec g}_{ij}$ and their time derivatives
on some initial hypersurface. The second order evolution equations then
determine the 4-metric of the spacetime up to gauge transformations.
This type of initial value problem is known as
a {\em Cauchy problem} and has been extensively used for the numerical
solution of Einstein's field equations.
It is however not suitable for the
analysis of gravitational radiation since it is not clear how to
incorporate null infinity into a finite numerical grid via conformal
compactification. Instead one uses approximating techniques to
extract information about the gravitational radiation at finite
radii and imposes {\em outgoing radiation
boundary conditions} in order to prevent incoming
gravitational waves. Unfortunately attempts to implement
these boundary conditions give rise
to spurious reflected numerical waves. Characteristic formalisms solve
this problem in an elegant way. Spacetime is decomposed into a
2-parameter family of 2-dimensional space-like surfaces threaded by two
1-parameter families of curves. At least one of these families
consists of null geodesics, the {\em characteristics} of the propagation
of radiation. The spacetime can be compactified by standard methods,
exact boundary conditions
can be applied at future or past null infinity and gravitational radiation
can be properly analysed. In regions of strong curvature, however, caustics
can form and the foliation along null geodesics breaks down.\\
A possible remedy for this problem
consists in using both a ``3+1'' and a characteristic formulation,
each in its preferred region.
Normally an astrophysical scenario is approximated as a finite inner region
containing all the matter (a neutron star, for example) and the outer
vacuum region with an observer located at future null infinity. In CCM
a ``3+1'' scheme is used for the evolution of the interior and
a characteristic formulation for the evolution of the exterior region.
At a finite radius an interface facilitates the transfer of information
between these two regions. The method is illustrated
in Fig.\,\ref{CCM_illu} where the dark shaded area represents the
astrophysical source. Gravitational waves emitted from this source
travel along null geodesics which are given by straight lines at an angle
of 45 degrees in this figure. In the outer region the null geodesics
are used to define the characteristic coordinate axis. \\
The feasibility of combining Cauchy algorithms with characteristic
methods in order to evolve the gravitational field was first studied
by \lcite{Bishop1992}. The first attempts at obtaining numerical evolutions
have been carried out in one spatial dimension. The work of
the Southampton CCM-group in cylindrical symmetry will be discussed in
detail in the next section. The Pittsburgh relativity
group studied CCM in spherical symmetry by evolving the
Einstein-Klein-Gordon system
(\shortciteNP{Gomez1996}). They have demonstrated second order
convergence and found no indications of back reflection or instabilities
at the interface. After the demonstration of the viability
of CCM in one dimension attention shifted towards higher-dimensional problems.
The Southampton relativity group focused their studies on the
axisymmetric case. After laying the theoretical foundations
(\citeNP{dInverno1996}, \citeNP{dInverno1997}) a great deal
of work has gone into the development of an
axisymmetric CCM code (see \citeNP{Pollney2000} for details). This
code has now been completed and is currently being evaluated and tested.
In contrast the Pittsburgh group has immediately turned their
attention towards the general 3-dimensional case.
\scite{Bishop1996} and \scite{Bishop1997} have probed the use
of Cauchy-characteristic matching in three dimensions by evolving
non-linear scalar waves in a flat space-time.
The application of these ideas to 3-dimensional problems
in general relativity has resulted in a module for the combination
of Cauchy and characteristic codes for the evolution of a binary black
hole (\shortciteNP{Bishop1998}). A more comprehensive overview of the
ongoing research using Cauchy-characteristic matching can be found in
\lcite{Winicour2001}.

%======================================================================
\subsection{The Southampton CCM-project}
The Southampton CCM-project is a long term project devoted to the
study of Cauchy-charac-teristic matching in scenarios of
decreasing symmetry assumptions
(\citeNP{dInverno2000}). The first step was to demonstrate
the viability of the approach. That was done by \citeN{Clarke1994} by
evolving the wave equation in flat spacetime. Attention then turned
towards gravitational waves in cylindrical symmetry. The theoretical 
foundations were laid by \shortciteN{Clarke1995} and the resulting
code of \shortciteN{Dubal1995}
showed good agreement with analytic solutions containing one gravitational
degree of freedom. Furthermore \shortciteANP{Dubal1995} demonstrated the
superior performance
of the CCM-method as compared with the use of artificial outer boundary
conditions in ``3+1'' schemes. \scite{dInverno2000b} presented
a generalisation of this
code to also include the rotational degree of freedom. They find,
however, that the convergence of the code drops to first
order level in later stages of the evolutions. In this work we will present
a new code that allows us
to include the rotational degree of freedom in terms of natural geometrical
variables with regular behaviour at null infinity. This reformulation
resulted in improved accuracy, long term stability and
ensures second order convergence over long evolution times.
We will demonstrate the improved quality by comparing the numerical results
with analytic solutions possessing both gravitational degrees of freedom. \\
The Southampton CCM-project has continued meanwhile with the
development of the axisymmetric code mentioned in the previous section.
%The latest stage in the Southampton CCM-project is the axisymmetric
%code mentioned in the . This code has been the subject
%of intensive work in recent years (see \citeNP{Pollney2000} for a
%detailed description) and is currently being evaluated and tested.

%======================================================================
\subsection{The original code} \label{CCM_org}
In this section we will describe the cylindrically symmetric Cauchy
characteristic matching code
developed by the Southampton Relativity Group (\shortciteNP{Clarke1995},
\shortciteNP{Dubal1995}). This code was used to reproduce the analytic
solution by \lcite{Weber1957}, which possesses one gravitational
degree of freedom,
with high accuracy and second order convergence. \scite{dInverno2000b}
presented an extension of this code based on the formulation
of \shortciteANP{Clarke1995} to also include the rotational degree of freedom.
Their difficulties in obtaining a long term stable second order convergent
code motivated the reinvestigation of the problem described in this thesis. \\
%The reformulation we will present below has resulted in a significant
%simplification of the
%relations at the interface. A new code has been written to implement
%these ideas and is compared with analytic solutions with one or two
%gravitational degrees of freedom. \\
In their derivation of the equations \shortciteANP{Clarke1995} find
it necessary to decompose spacetime according to the methods
of \lcite{Geroch1970} in order to eliminate irregularities of the
equations in the characteristic region.
The Geroch decomposition plays a crucial role in our reformulation
and will also be used in section \ref{cstr}
when we numerically simulate cosmic strings. 
Before we turn our attention to the cylindrically symmetric CCM code,
we will therefore describe the Geroch decomposition in more detail.

%========================================
\subsubsection{The Geroch decomposition}
\label{Gerochdec}
A problem generally faced in cylindrical symmetry is that the spacetime
is not asymptotically flat due to the infinite extension in the $z$-direction.
The decomposition of \lcite{Geroch1970} solves this problem by factoring out the
Killing direction and reformulating the 4-dimensional problem in terms
of two scalar fields on an asymptotically flat 3-dimensional spacetime.
Suppose, the spacetime admits a Killing 
field $\hbox{\grvec x}^{\mu}$ which in
the case of cylindrical symmetry simply is $\bbox{\partial}_z$.
Then we define the norm of the Killing vector
\begin{align}
  \nu &= \hbox{\grvec x}^{\alpha} \hbox{\grvec x}_{\alpha},
  \label{Geroch_norm}
\end{align}
and the Geroch twist
\begin{align}
  \hbox{\grvec t}_{\alpha} &= -\hbox{\grvec e}_{\alpha \beta \rho \sigma}
                   \hbox{\grvec x}^{\beta} 
                   \nabla^{\rho} \hbox{\grvec x}^{\sigma},
  \label{GEROCH_TWIST}
\end{align}
where $\hbox{\grvec e}_{\alpha \beta \rho \sigma}$ is the completely
antisymmetric Levi-Cevita tensor.
These fields are well defined on the 3-dimensional space $\mathcal{S}$
given by $z=\mathrm{const}$
with the resulting metric
\begin{align}
  \hbox{\vec h}_{\alpha \beta} &= \hbox{\vec g}_{\alpha \beta}
  - \frac{1}{\nu} \hbox{\grvec x}_{\alpha} \hbox{\grvec x}_{\beta}.
  \label{Geroch_3m}
\end{align}
If $D_{\mu}$ denotes the covariant derivative associated with the
3-metric $\hbox{\vec h}$, one can show that
\begin{align}
  D_{[\rho}\hbox{\grvec t}_{\sigma ]} &= \hbox{\grvec e}_{\rho \sigma
      \alpha \beta} \hbox{\grvec x}^{\alpha}
      \hbox{\vec R}^{\beta}_{\,\,\, \lambda} \hbox{\grvec x}^{\lambda}.
      \label{Geroch_cond}
\end{align}
In vacuum the right hand side vanishes so that $\hbox{\grvec t}_{\sigma}$
is curl free and can be expressed in terms of a potential
\begin{align}
  \hbox{\grvec t}_{\sigma} &= D_{\sigma} \tau.
  \label{Geroch_pot}
\end{align}
It is a remarkable fact that the right hand side of Eq.\,(\ref{Geroch_cond})
will also vanish in some non-vacuum cases. In the discussion of cosmic strings
in section \ref{cstr} we will encounter such an example.\\
Geroch has then shown that the Einstein equations for the metric
$\hbox{\vec g}$ of the
4-dimensional \mbox{spacetime} can be written in terms of the two scalar fields
$\tau$ and $\nu$ and the 3-metric $\hbox{\vec h}$
\begin{eqnarray}
 {\cal R}_{ab} &=& \frac{1}{2}\nu^{-2}[(D_a\tau)
                   (D_b\tau)-\hbox{\vec h}_{ab}(D_m\tau)
                   (D^m\tau)] + \frac{1}{2}\nu^{-1}
                   D_aD_b\nu - \frac{1}{4}\nu^{-2}(D_a\nu)
                   (D_b\nu)\nonumber\\
                && +8\pi \hbox{\vec h}_a{}^\alpha \hbox{\vec h}_b{}^{\beta}
                   (\hbox{\vec T}_{\alpha\beta}-\frac{1}{2}
                   \hbox{\vec g}_{\alpha\beta}\hbox{\vec T}),
                   \label{Geroch3d} \\[10pt]
 D^2 \nu &=& \frac{1}{2}\nu^{-1}(D_m\nu) (D^m\nu) - \nu^{-1}(D_m\tau)
                      (D^m\tau) +16\pi (\hbox{\vec T}_{\alpha\beta}-
                      \frac{1}{2}\hbox{\vec g}_{\alpha\beta}\hbox{\vec T})
                      \hbox{\grvec x}^\alpha\hbox{\grvec x}^\beta, 
             \label{D91}\\[10pt]
 D^2 \tau &=& \frac{3}{2}\nu^{-1}(D_m\tau)
             (D^m\nu), \label{Geroch_tau}
\end{eqnarray}
where Latin indices run from 0 to 2 and $\mathcal{R}_{ab}$
is the Ricci tensor associated with the 3-metric $\hbox{\vec h}$.
Note that even in the
case of a vanishing energy-momentum tensor $\hbox{\vec T}$,
the scalar fields $\nu$ and $\tau$
present source terms in the field equations
(\ref{Geroch3d}) for the 3-metric $\hbox{\vec h}$. \\
In the vacuum case $\hbox{\vec T}_{\alpha \beta}=0$, \scite{Sjodin2000}
have shown how
it is possible to reformulate the Einstein-Hilbert
Lagrangian in terms of $\nu$, $\tau$ and the conformal 3-metric
$\tilde{\hbox{\vec h}}_{ab} = \nu \hbox{\vec h}_{ab}$.
This leads directly to the 3-dimensional energy-momentum tensor
\begin{equation}
\mathcal{T}_{ab} = \frac{1}{2}\nu^{-2}[{\tilde D}_a \tau{\tilde
                   D_b}\tau-\frac{1}{2}\tilde{\hbox{\vec h}}_{ab}
                   \tilde{ \hbox{\vec h}}^{cd}(\tilde
                   D_c\tau)(\tilde D_d\tau)
                   +{\tilde D}_a\nu{\tilde
                   D_b}\nu-\frac{1}{2}\tilde{ \hbox{\vec h}}_{ab}
                   \tilde{ \hbox{\vec h}}^{cd}(\tilde
                   D_c\nu)(\tilde D_d\nu)]
                   \label{Geroch_confT},
\end{equation}
where $\tilde{D}_a$ is the covariant derivative associated with the
conformal 3-metric $\tilde{\hbox{\vec h}}$.
Since the Weyl-curvature vanishes identically in three dimensions, the
curvature is completely determined by the Ricci tensor $\mathcal{R}_{ab}$,
i.e. the energy-momentum tensor $\mathcal{T}_{ab}$ which in turn is
determined by $\nu$ and $\tau$. Thus the gravitational
degrees of freedom of the original 4-dimensional spacetime are represented
by the scalar fields $\nu$ and $\tau$.
If matter is present in the
4-dimensional spacetime, there are extra terms on the right hand side of
(\ref{Geroch_confT}).

%=======================================================================
\subsubsection{The equations of the original code}
\label{Dubal_equations}
We will now turn our attention to the original cylindrically symmetric
CCM code of the Southampton relativity group. An extensive description
of this code and the derivation of the equations can be found in
\scite{Clarke1995} and \scite{Dubal1995}. In order to illustrate the
effects of our reformulation, we will include here a rather
detailed description of their equations and choice of variables.
They start with the metric in Jordan, Ehlers, Kundt and
Kompaneets (JEKK) form (\shortciteNP{Jordan1960}, \citeNP{Kompaneets1958})
\begin{align}
  ds^2 &= e^{2(\gamma - \psi)}(-dt^2+dr^2) + r^2e^{-2\psi}d\phi^2
          +e^{2\psi} (\omega d\phi + dz)^2,
  \label{JEKK}
\end{align}
which describes a general cylindrically symmetric vacuum spacetime. The
metric functions $\psi$, $\omega$ and $\gamma$ are functions of $(r,t)$
only. In terms of the gauge freedom discussed in section \ref{ADM_GAUGE}
this choice implies a vanishing shift vector and the lapse is determined
by the requirement that $\hbox{\vec g}_{tt}=-\hbox{\vec g}_{rr}$. As a
consequence the null geodesics are given by the simple relations
$t\pm r = \mathrm{const}$.
In the outer characteristic region, the line element is rewritten by
transforming to the coordinates
\begin{align}
  u &= t-r, \\[10pt]
  y &= \frac{1}{\sqrt{r}},
  \label{y}
\end{align}
and the regions are matched at $r=1=y$. \shortciteANP{Clarke1995} 
find, however, that the
compactified field equations cannot be made regular in this way.
Therefore they factor out the $z$-direction in the outer region according to
the Geroch decomposition described above. This leads to
a reformulation of the problem in terms of the variables
\begin{align}
  m &= \frac{\nu - 1}{y}, \label{DUBAL_M} \\[10pt]
  w &= \frac{\tau}{y},
\end{align}
where $\tau$ is the Geroch potential and $\nu$ the
norm of the $z$-Killing vector. These are related to the metric functions
$\psi$ and $\omega$ by Eqs.\,(\ref{Geroch_norm}) and (\ref{GEROCH_TWIST})
which in this particular case become
\begin{align}
  \nu &= e^{2\psi}, \label{DUBAL_PSI} \\[10pt]
  \tau_{,y} &= y^2 e^{4\psi} \omega_{,u}.
\end{align}
With this choice of variables one obtains two
evolution equations for $\psi$ and $\omega$ in the interior
Cauchy region and a constraint equation for $\gamma$. \shortciteANP{Dubal1995}
write this set of equations as a first order system
\begin{align}
  \psi_{,t} &= \frac{1}{r} \tilde{L}, \label{DUBAL_PSIT} \\[10pt]
  \omega_{,t} &= -2 e^{-4\psi} L^{\phi}_z, \label{DUBAL_OMEGAT} \\[10pt]
  \tilde{L}_{,t} &= \frac{1}{r}[ r^2\psi_{,rr} + r\psi_{,r} -
     \frac{1}{2} e^{4\psi} \omega^2_{,r} + 2 e^{-4\psi}(L^{\phi}_z)^2],
     \label{DUBAL_LT} \\[10pt]
  L^{\phi}_{z,t} &= \frac{1}{r}e^{4\psi} (\frac{1}{2} \omega_{,r} -
     \frac{1}{2} r \omega_{,rr} - 2r \psi_{,r} \omega_{,r}),
     \label{DUBAL_LPZT} \\[10pt]
  \chi_{,r} &= \frac{1}{4r} e^{4\psi} \omega^2_{,r} - \psi_{,r} +
     r\psi^2_{,r} + \frac{1}{r}[\tilde{L}^2 + e^{-4\psi}(L^{\phi}_z)^2],
    \label{DUBAL_CHIR}
\end{align}
where $\chi=\gamma - \psi$. The corresponding set of equations in the
characteristic region is given by two evolution equations for $m$ and $w$
and a hypersurface equation for $\gamma$ which is again written as a first
order system
\begin{align}
  m_{,u} = \,\,&\nu M,     \label{DUBAL_MU} \\[10pt]
  w_{,u} = \,\,&\nu W,     \label{DUBAL_WU} \\[10pt]
  \begin{split}
  M_{,y} = &-\frac{1}{\nu}(yw)_,{y}W + \frac{1}{4\nu}\biggl[
            -y(m+y^2m_{,yy} + 3ym_{,y})\\[10pt]
           & +\frac{1}{\nu}y^2(m^2
            +2ymm_{,y} - w^2 - 2yww_{,y} + y^2m^2_{,y} -
            y^2w^2_{,y})\biggr], \label{DUBAL_MY}
  \end{split} \\[10pt]
  \begin{split}
  W_{,y} = \,\,&\frac{1}{\nu}(yw)_{,y} M + \frac{1}{4\nu} \biggl[
            -y(w+y^2w_{,yy}+3yw_{,y})\\[10pt]
            & +\frac{1}{\nu}2y^2
            (mw+ymw_{,y}+ywm_{,y}+y^2m_{,y}+y^2m_{,y}w_{,y}) \biggr],
            \label{DUBAL_WY}
  \end{split} \\[10pt]
  \gamma_{,y} = & -\frac{1}{8\nu^2}y[m^2+w^2+2y(mm_{,y}+ww_{,y})
            +y^2(m^2_{,y} + w^2_{,y})]. \label{DUBAL_GAMMAY}
\end{align}
The transformation between the two pairs
of variables $(\psi, \omega)$ and $(m,w)$ and their derivatives is
implemented at the interface at $r=1=y$ according to the relations
\begin{eqnarray}
  \tilde{L} &=& \frac{M}{2y}, \label{DUBAL_TILDEL} \\[10pt]
  \omega_{,r} &=& \frac{W}{y\nu}, \label{DUBAL_OMEGAR}\\[10pt]
  m &=& \frac{1}{y} (e^{2\psi}-1),\label{Dubal_m}  \\[10pt]
  (yw)_{,y} &=& -\frac{2}{r} L_z^{\phi}. \label{DUBAL_YWY}
\end{eqnarray}
The problematic relations are (\ref{DUBAL_OMEGAR}) and
(\ref{DUBAL_YWY}) which
involve the spatial derivative of $\omega$ and $w$.
The presence of spatial derivatives in combination with the
interpolation techniques applied at the interface make the implementation
of these relations a rather subtle issue. \\
The code of \shortciteANP{Dubal1995} formed the starting point for our
investigation of the problem. This code has been
well checked in the non-rotating case but did not include the implementation
of Eqs.\,(\ref{DUBAL_OMEGAR}) and (\ref{DUBAL_YWY}) for
the rotational variables
$\omega$ and $w$ at the interface. In this work we therefore started with
the addition of these missing modules to the original code. In order to
describe our implementation it is necessary to first discuss the numerical
techniques, in particular those underlying the transmission of
information from the Cauchy to the characteristic region and vice versa.

%========================================================================
\subsubsection{The numerical implementation}
\label{start_int}
We will now discuss the numerical implementation of
Eqs.\,(\ref{DUBAL_PSIT})-(\ref{DUBAL_YWY}).
%has been described in detail in \shortciteNP{Dubal1995}. Because of the
%importance of the numerical methods in particular at the interface, we
%summarise of the essential steps.
The numerical grid used for the evolution consists of an inner
Cauchy region which covers the range $0 \leq r \leq 1$ and the 
outer characteristic region extending from $r=1$ to infinity which 
corresponds to the range $1 \geq y \geq 0$.
The evolution equations in these regions
are discretized in a straightforward way using the leapfrog scheme
described in section \ref{FDE_LEAP} while second order centered finite
differencing is used for the constraints. 
If we assume that all functions are known on the time slices $n$, $n-1$ and
$n-2$, a full evolution cycle consists of the following steps.
\begin{list}{\rm{(\arabic{count})}}{\usecounter{count}
             \labelwidth1cm \leftmargin1.5cm \labelsep0.4cm \rightmargin1cm
             \parsep0.5ex plus0.2ex minus0.1ex \itemsep0ex plus0.2ex}
  \item Evolution of $\psi$, $\omega$, $\tilde{L}$ and $L_z^{\phi}$
        at the interior grid points of the Cauchy region according to
        Eqs.\,(\ref{DUBAL_PSIT})-(\ref{DUBAL_LPZT}).
  \item Update of these variables at the origin according to the inner
        boundary conditions $\psi_{,r}=\omega_{,r}=\tilde{L}_{,r}
        =L^{\phi}_{z,r}=0$.
  \item Evolution of $\psi$ and $\omega$ at the outer boundary of the Cauchy
        grid ($r=1$) according to Eqs.\,(\ref{DUBAL_PSIT}),
        (\ref{DUBAL_OMEGAT}).
  \item Extraction of $\psi$ and $\omega$ from the interface at $1+dr$ on
        the Cauchy grid on time slice $n$.
  \item Evolution of $\tilde{L}$, $L_z^{\phi}$ at the outer boundary
        of the Cauchy grid ($r=1$) according to
        Eqs.\,(\ref{DUBAL_LT}), (\ref{DUBAL_LPZT}).
  \item Calculation of $\chi$ on the Cauchy grid via quadrature according to
        Eq.\,(\ref{DUBAL_CHIR}).
  \item Evolution of $m$ and $w$ in the characteristic region according to
        Eqs.\,(\ref{DUBAL_MU}), (\ref{DUBAL_WU}).
  \item Extraction of $m$ and $w$ from the interface at $1+dy$ on the
        characteristic grid on time slice $n+1$.
  \item Calculation of $M$, $W$ and $\gamma$ on the characteristic grid
        via quadrature according to Eqs.\,(\ref{DUBAL_MY})-(\ref{DUBAL_GAMMAY}).
\end{list}
The crucial steps which provide the flow of information through the
interface are (4) and (8). These steps together with the start up procedure
required to get the leap-frog scheme running will now be discussed in more
detail. We start with the interface. \\
We first note that the interface is fixed at the radial position $r=1=y$.
Since we always have the freedom to rescale the radial coordinate $r$ by
a constant factor, this implies no loss of generality. From a numerical point
of view the need of an interface arises from the calculation of spatial
derivatives at $r=1$ on the Cauchy grid and $y=1$ on the characteristic grid.
The centred finite differencing used for the leapfrog scheme as illustrated
in Eq.\,(\ref{FDE_leapfrog}) requires knowledge of the Cauchy variables
at $r=1+dr$ and the characteristic variables at $y=1+dy$ for this purpose.
In order to obtain these values, they need to be calculated with interpolation
techniques using Eqs.\,(\ref{DUBAL_TILDEL})-(\ref{DUBAL_YWY}). We will
describe this process in the case of the direction ``char$\rightarrow$Cauchy''
corresponding to step (4). The reverse direction in step (8) works in
complete analogy. The situation is graphically illustrated in
Fig.\,\ref{Interface}. 
\begin{figure}
  \centering
  %%%%%%%%%%%%%%%%%%%%%%%%%%%%%%%%%%%%%%%%%%%%%%%%%%%%%%%%%%%%%%%%
\begin{picture}(0,0)%
\epsfig{file=Interface3.pstex}%
\end{picture}%
\setlength{\unitlength}{3947sp}%
\begingroup\makeatletter\ifx\SetFigFont\undefined%
\gdef\SetFigFont#1#2#3#4#5{%
  \reset@font\fontsize{#1}{#2pt}%
  \fontfamily{#3}\fontseries{#4}\fontshape{#5}%
  \selectfont}%
\fi\endgroup%
\begin{picture}(6087,4737)(301,-4036)
\put(1201,-2761){\makebox(0,0)[lb]{\smash{\SetFigFont{12}{14.4}{\familydefault}{\mddefault}{\updefault}$dt$}}}
\put(6301,-961){\makebox(0,0)[lb]{\smash{\SetFigFont{12}{14.4}{\rmdefault}{\mddefault}{\updefault}$-y$}}}
\put(2176,-2161){\makebox(0,0)[lb]{\smash{\SetFigFont{12}{14.4}{\rmdefault}{\mddefault}{\updefault}$r_{K-1}$}}}
\put(3076,-2161){\makebox(0,0)[lb]{\smash{\SetFigFont{12}{14.4}{\rmdefault}{\mddefault}{\updefault}$r_K$}}}
\put(2851,-286){\makebox(0,0)[lb]{\smash{\SetFigFont{12}{14.4}{\rmdefault}{\mddefault}{\updefault}$r=1=y$}}}
\put(3376,-736){\makebox(0,0)[lb]{\smash{\SetFigFont{12}{14.4}{\rmdefault}{\mddefault}{\updefault}$t,u$}}}
\put(5176,-886){\makebox(0,0)[lb]{\smash{\SetFigFont{12}{14.4}{\rmdefault}{\mddefault}{\updefault}$du$}}}
\put(2701,-3736){\makebox(0,0)[lb]{\smash{\SetFigFont{12}{14.4}{\rmdefault}{\mddefault}{\updefault}$dr$}}}
\put(3676,-3661){\makebox(0,0)[lb]{\smash{\SetFigFont{12}{14.4}{\rmdefault}{\mddefault}{\updefault}$dy$}}}
\put(901,-4036){\makebox(0,0)[lb]{\smash{\SetFigFont{12}{14.4}{\rmdefault}{\mddefault}{\updefault}$-r$}}}
\put(4276,-2236){\makebox(0,0)[lb]{\smash{\SetFigFont{12}{14.4}{\rmdefault}{\mddefault}{\updefault}$r_{K+1}$}}}
\put(301,-3886){\makebox(0,0)[lb]{\smash{\SetFigFont{12}{14.4}{\rmdefault}{\mddefault}{\updefault}$n-2$}}}
\put(301,-3136){\makebox(0,0)[lb]{\smash{\SetFigFont{12}{14.4}{\rmdefault}{\mddefault}{\updefault}$n-1$}}}
\put(301,-2386){\makebox(0,0)[lb]{\smash{\SetFigFont{12}{14.4}{\rmdefault}{\mddefault}{\updefault}$n$}}}
\end{picture}
  %%%%%%%%%%%%%%%%%%%%%%%%%%%%%%%%%%%%%%%%%%%%%%%%%%%%%%%%%%%%%%%%
  \caption{The interface in the direction from the characteristic to the
           Cauchy region. See the text for details.}
  \label{Interface}
\end{figure}
The derivatives of a function $f$ at $r=1$ can be calculated
to second order accuracy by centred finite differencing
\begin{align}
  f_{,r}|_K &= \frac{f_{K+1} - f_{K-1}}{2 dr}, \label{Interface_fr} \\[10pt]
  f_{,rr}|_K &= \frac{f_{K+1} -2f_K + f_{K-1}}{dr^2}, \label{Interface_frr}
\end{align}
if $f_{K+1}$ is obtained from interpolation to fourth order accuracy
in the characteristic region. For this purpose $\psi$, $\omega_{,r}$,
$\tilde{L}$ and $L^{\phi}_z$
are calculated in terms of the characteristic variables according to
Eqs.\,(\ref{DUBAL_TILDEL})-(\ref{DUBAL_YWY}) at the 12 points
of the characteristic 
region (including 3 points at the interface) indicated by filled
circles in Fig.\,\ref{Interface}. These values can then
be used to obtain the function
values $\psi_{K+1}$ and $\omega_{K+1}$ at location $r_{K+1}$ with
the required accuracy. \\
An alternative to this method consists in using the
same interpolation technique to calculate
the $r$-derivatives $\psi_{,r}$ and
$\omega_{,r}$ at grid point $K+1$ instead of the function values $\psi$
and $\omega$. We can then 
calculate the $r$-derivatives at the interface from
\begin{align}
  f_{,r}|_K &= \frac{f_{,r}|_{K+1} + f_{,r}|_{K-1}}{2}, 
  \label{Interface_fr2} \\[10pt]
  f_{,rr}|_K &= \frac{f_{,r}|_{K+1} - f_{,r}|_{K-1}}{2 dr}.
  \label{Interface_frr2}
\end{align}
Even though this alternative looks natural for the transformation
between $\omega$ and $w$ because these variables are related via
their derivatives according to Eq.\,(\ref{DUBAL_OMEGAR}),
it does not lead to any improvement of the performance of the code. \\
The second point we need to discuss is the so-called start-up problem.
It is an intrinsic difficulty of 3-level schemes such as the leap-frog
algorithm that the specification of initial data on one time slice
will not be sufficient to start the numerical engine. Instead
different techniques need to be used to obtain data on auxiliary
time slices. Due to the requirements of the fourth-order interpolation
at the interface we need information on two additional slices. The data
on these auxiliary slices are calculated in three steps.
\begin{list}{\rm{(\arabic{count})}}{\usecounter{count}
             \labelwidth1cm \leftmargin1.5cm \labelsep0.4cm \rightmargin1cm
             \parsep0.5ex plus0.2ex minus0.1ex \itemsep0ex plus0.2ex}
  \item The first order Euler scheme (see for example \shortciteNP{Press1989})
        is used to calculate data at $t=t_0-dt/2$.
  \item This auxiliary time slice is then used to determine the variables
        at $t=t_0-dt$ according to the leapfrog scheme.
  \item In another leapfrog step, this time using the full time step $dt$,
        data is calculated at $t_0-2dt$.
\end{list}
An alternative treatment at the interface is required for this
start-up procedure, because
the necessary three time-slices are not available at this stage. For
this purpose the
Cauchy grid is extended into the characteristic region by 10 grid
points. The derivatives of the Cauchy variables
can thus be calculated at $r=1$ using
centred finite differencing and the derivatives of the characteristic
variables follow from chain-rule. The treatment of the outer boundary of the
Cauchy grid is irrelevant for the numerical evolution, since the spurious
signal cannot travel across the additional 10 grid points during the
three evolution steps at the start-up procedure and these points
are not used in the remaining evolution.

%========================================================================
\subsubsection{Including the rotational degree of freedom $\omega$}
In our first attempt to include the rotational degree of freedom we have
made use of the set of variables of section \ref{Dubal_equations}, 
namely $\psi$, $\omega$ and $\chi$ in the inner 
and $m$, $w$ and $\gamma$ in the outer region. For this purpose
we have extended the interface of the original code
to also include the transformations
between $\omega$ and $w$ as described in the previous section.
In order to test the code we use the analytic solution from
\lcite{Xanthopoulos1986} which we will discuss in more detail
in section \ref{as}. In Eqs.\,(\ref{as_Q})-(\ref{as_ygamma}) we give
analytic expressions for this solution in terms of the
Killing vector $\nu$, the Geroch potential $\tau$
and the metric function $\gamma$. The corresponding results for the
variables $\psi$, $m$ and $w$ are obtained straightforwardly from their
definitions (\ref{DUBAL_M})-(\ref{DUBAL_PSI}). The transformation
into values for the function $\omega$ is more complicated. The result
is given by \scite{Sjodin2000}
\begin{align}
  \omega(t,r) &= \sqrt{a^2+1}(X+Q-2)\frac{Z-Y}{2aZ},
\end{align}
where the auxiliary functions $Q$, $X$, $Y$ and $Z$ are defined
in Eqs.\,(\ref{as_Q})-(\ref{as_Z}).
We have not been able, however, to obtain a long term stable evolution
in this formulation of the problem. 
For 300 grid points in each region
instability set in after less than 1000 time steps and from the
pattern of the noise it is clear that the problems originate at the
interface. In our attempts to overcome the instability we have
varied the obvious parameters such as the Courant factor and
the number of grid points over a large range, but no improvement
has been achieved. We have also used the alternative implementation
of the interface according to Eqs.\,(\ref{Interface_fr2}),
(\ref{Interface_frr2}). Even though this alternative looks quite natural
at least for the transformation between $\omega$ and $w$ which are
related via their derivatives according to (\ref{DUBAL_OMEGAR}), we
did not achieve a significantly better performance with this method.
Finally we have
changed the start time of the numerical evolution and, thus, the
initial data. The obvious choice $t=0$ is not possible because some
derivatives of Xanthopoulos' solution are discontinuous at $t=0$,
but any positive value large enough to ensure that the start-up procedure does
not extend to negative times can be chosen. Again the code
became unstable after less than 1000 time steps. We have therefore
decided to restart the investigation of this problem by looking
for alternative sets of variables.

%========================================================================
\subsection{A reformulation of the problem}
A striking peculiarity of the formulation described above is the
drastically different treatment of the Cauchy and the characteristic region.
In view of the numerical subtleties associated with the interface one may
question the wisdom of factoring out the $z$-direction in one region
and work in the framework of the 4-dimensional spacetime in the other.
It rather seems natural to look for as homogeneous a description of the whole
spacetime as possible. In this context it is worth noting that the
restriction of the Geroch decomposition to the characteristic region
was a voluntary choice and not enforced at any stage of the derivation
of the equations. We have therefore decided to factor out the $z$-direction
in the Cauchy region as well and thus Geroch decomposed the whole spacetime.
This enables us to use the same set of fundamental variables throughout
spacetime and thus obtain almost trivial interface relations. A closer
investigation of the equations suggests that aside from the metric function
$\gamma$ the geometric variables $\nu$ and $\tau$ are
the natural variables to describe the cylindrically symmetric spacetime.
With this choice the equations in the Cauchy region can be written as
\begin{align}
  \nu_{,tt} &= \frac{1}{\nu}(\nu_{,t}^2 -\nu_{,r}^2+\tau_{,r}^2-\tau_{,t}^2)
      +\nu_{,rr} +\frac{\nu_{,r}}{r}, \label{CCM_NUTT} \\[10pt]
  \tau_{,tt} &= \frac{2}{\nu}(\tau_{,t} \nu_{,t} - \tau_{,r} \nu_{,r})
      + \tau_{,rr}+\frac{1}{r}\tau_{,r}, \label{CCM_TAUTT} \\[10pt]
  \gamma_{,r} &= \frac{r}{4\nu^2}(\nu_{,r}^2 +\nu_{,t}^2+\tau_{,r}^2
      +\tau_{,t}^2). \label{CCMgammar}
\end{align}
In practice we use $\nu_{,t}$ and $\tau_{,t}$ as
auxiliary variables in order to write Eqs.\,(\ref{CCM_NUTT}),
(\ref{CCM_TAUTT}) as a first order system.
If we transform to the new set of variables 
the equations in the characteristic region become
\begin{align}
  \nu_{,u} &= y\nu M, \\[10pt]
  \tau_{,u} &= y\nu W, \\[10pt]
  M_{,y} &= -\frac{y}{4\nu} \left[y\nu_{,yy} + \nu_{,y} + \frac{y}{\nu}
            (\tau_{,y}^2 - \nu_{,y}^2)\right]-W\frac{\tau_{,y}}{\nu}, \\[10pt]
  W_{,y} &= -\frac{y}{4\nu} \left(y\tau_{,yy} + \tau_{,y}-2\frac{y}{\nu}\nu_{,y}
            \right) + M\frac{\tau_{,y}}{\nu}, \\[10pt]
  \gamma_{,y} &= -\frac{y}{8\nu^2}(\tau_{,y}^2 +\nu_{,y}^2). \label{CCMgammay}
\end{align}
Finally the non-trivial relations at the interface are now given by
\begin{align}
  \nu_{,t} &= y\nu M, \\[10pt]
  \tau_{,t} &= y\nu W.
\end{align}
We have developed a code using the numerical techniques of section
\ref{start_int} based on these evolution equations and interface 
relations.
% In the next section we will discuss the performance of this
% new scheme.

%======================================================================
\subsection{Testing the code}
In order to test the performance of the new code, we will check it against 
analytic solutions with one and two gravitational degrees of freedom.
Furthermore we will demonstrate its internal consistency with a
time dependent convergence analysis. \\
We have already mentioned the vacuum solution by \lcite{Weber1957} that
was successfully used by \shortciteANP{Dubal1995} 
to test their CCM code. A solution with
both gravitational degrees of freedom was derived by \lcite{Xanthopoulos1986}.
Both these solutions can be rewritten in terms of our
variables $\nu$, $\tau$ and $\gamma$ and thus compared with the numerical
results.

%====================================
\subsubsection{The Weber-Wheeler wave}
\label{CCM_ww}
\begin{figure}[t]
  \centering
  \epsfig{file=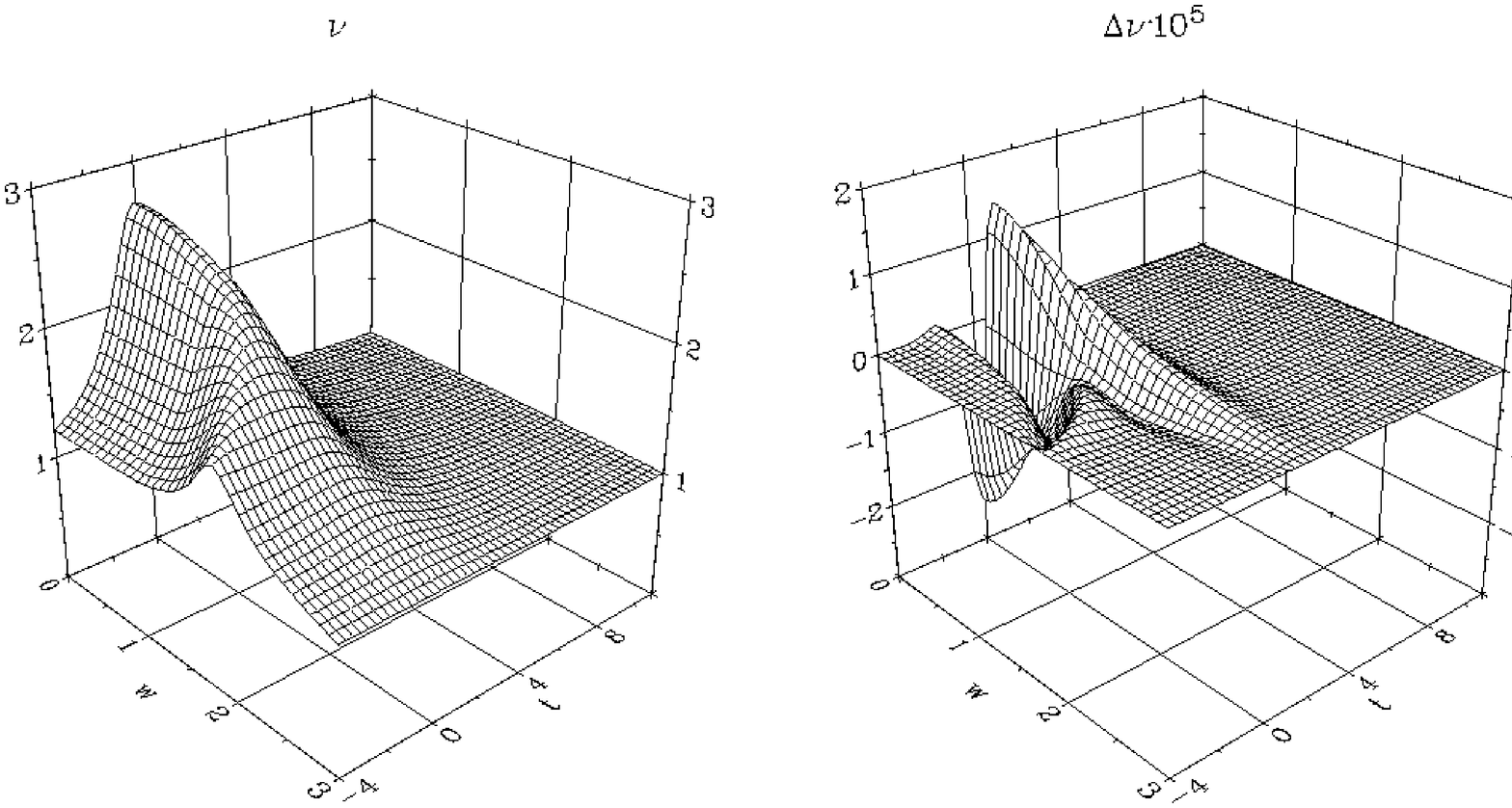, height=400pt, width=175pt, angle=-90}
  \epsfig{file=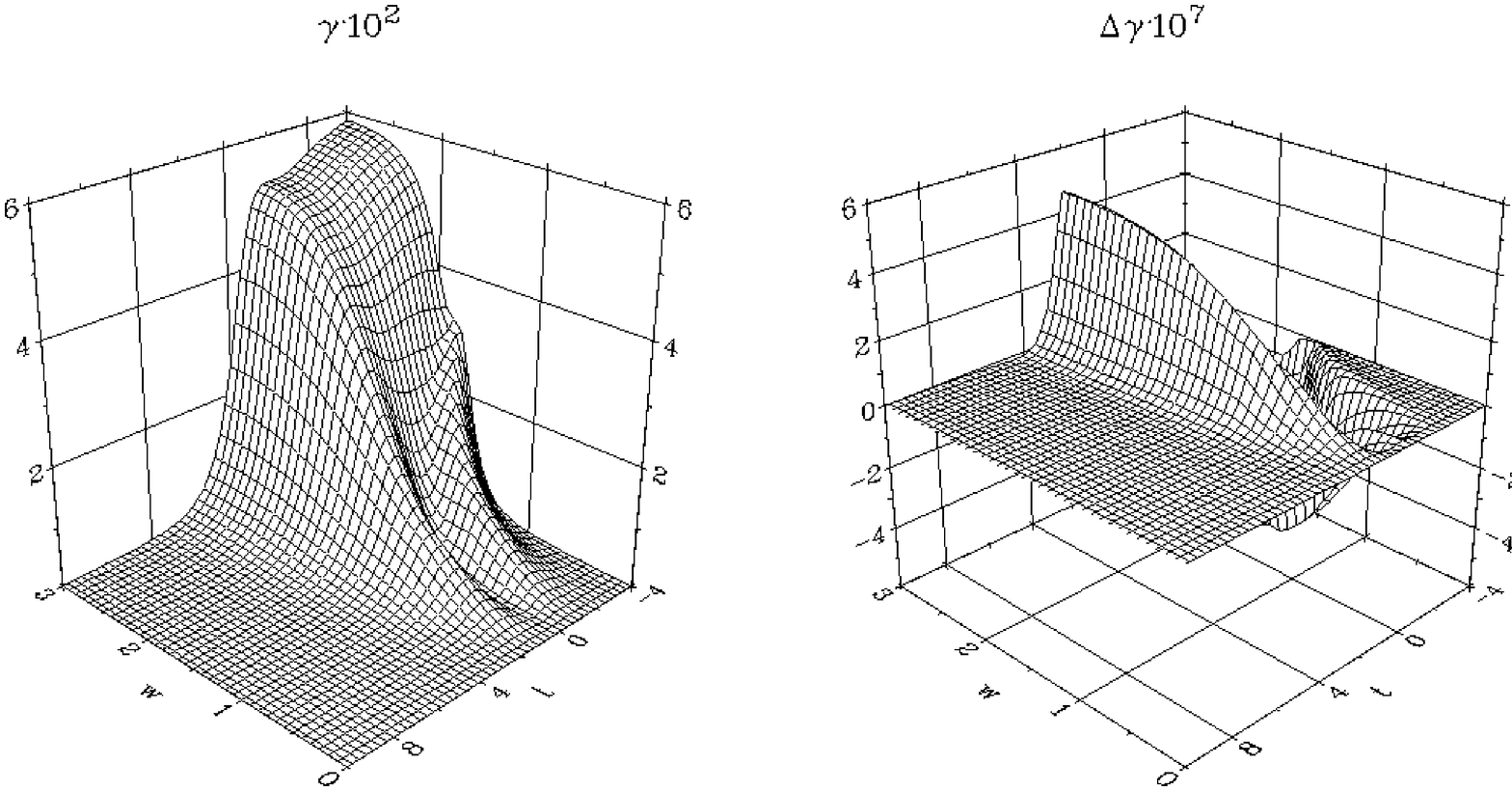, height=400pt, width=175pt, angle=-90}
  \caption{The numerical solutions for $\nu$ and $\gamma$ of the
           Weber-Wheeler wave for $a=0.5$, $b=2$ obtained with
           1200 grid points in each region (left panels). In the right
           panels the corresponding deviation from the analytic result
           is amplified by $10^5$ and $10^6$, respectively. For presentation
           purposes $\nu$ and $\gamma$ are viewed from
           different angles.}
  \label{CCMng}
\end{figure}

The analytic solution by \citeANP{Weber1957} describes a gravitational
pulse of the ``$+$'' polarization mode that moves in from past
null infinity, implodes on the axis and emanates away to future
null infinity. The analytic expressions in terms of $\nu$
and $\gamma$
have been derived in \scite{Sjodin2000}. In the Cauchy region it
is convenient to introduce the auxiliary quantities
\begin{align}
  X &= a^2 + r^2 - t^2, \label{ww_X} \\[10pt]
  Y &= X^2+4a^2t^2,
\end{align}
and the Weber-Wheeler wave can be written as
\begin{align}
  \nu &= \exp \left[ 2b \sqrt{\frac{2(X+\sqrt{Y})} {Y}}
            \right], \\[10pt]
  \gamma &= \frac{b^2}{2a^2} \left[ 1 - 2a^2r^2 \frac{X^2-4a^2t^2}
            {Y^2} - \frac{a^2+t^2-r^2}{\sqrt{Y}} \right],
  \label{ww_gamma}
\end{align}
where $a$ and $b$ are constants representing the width and amplitude 
of the pulse. The corresponding result in terms of the characteristic 
coordinates $u$, $y$ is
\begin{align}
  \tilde{X} &= a^2 y^2 - u^2 y^2 - 2u, \label{ww_yX} \\[10pt]
  \tilde{Y} &= \tilde{X}^2 + 4a^2(uy^2+1), \\[10pt]
  \nu &= \exp \left[ 2by \sqrt{\frac{2(\tilde{X}
              +\sqrt{\tilde{Y}})} {\tilde{Y}}} \right], \\[10pt]
  \gamma &= \frac{b^2}{2a^2} \left[ 1 - 2a^2 \frac{\tilde{X}^2-4a^2(uy^2+1)^2}
            {\tilde{Y}^2} - \frac{a^2y^2+u^2y^2+2u}{\sqrt{\tilde{Y}}} \right].
  \label{ww_ygamma}
\end{align}
The initial values for $\nu$ and its time
derivative are prescribed according
to these equations whereas $\gamma$ on the initial slice is calculated 
via quadrature from the constraint equations (\ref{CCMgammar}) and
(\ref{CCMgammay}). In order
to plot the solution for $0\le r < \infty$ we introduce the radial variable
\begin{align}
  w &= \left\{ \parbox{4cm}
               {
               $r \hspace{1.21cm} {\rm for}\,\, 0\le r \le 1 \\[5pt]
               3-\frac{2}{\sqrt{r}} \hspace{0.27cm} {\rm for}\,\, r > 1. $
               } \right. \label{w}
\end{align}
In Fig.\,\ref{CCMng} we show the numerical results for $\nu$
and $\gamma$
and their deviation from the analytic values obtained for $a=2$ and $b=0.5$
using 1200 grid points in each region and a Courant factor of 0.45. 
As in the case of the original code from \shortciteANP{Dubal1995} we find that
a Courant factor $<0.5$ is required for a stable evolution.
The plots show the incoming pulse in $\nu$ which is reflected at
the origin and then moves outwards to null infinity. The relatively
large number of grid points is required to achieve a high accuracy 
at early times in modelling the steep gradients of the incoming pulse.
If the calculation starts at a later time or a smaller parameter $a$ for 
the width of the pulse is used, the same accuracy is obtained with
significantly fewer grid points. 
We also see that longer runs do not reveal any 
new features as the metric variables approach their Minkowskian values
after $t\approx 5$. This solution, however,
does not provide a test for the rotational degree of freedom. For that purpose
we need an analytic solution with both gravitational degrees of freedom.

%=======================================================================
\subsubsection{Xanthopoulos' rotating solution}
\label{as}
\begin{figure}[t]
  \centering
  \epsfig{file=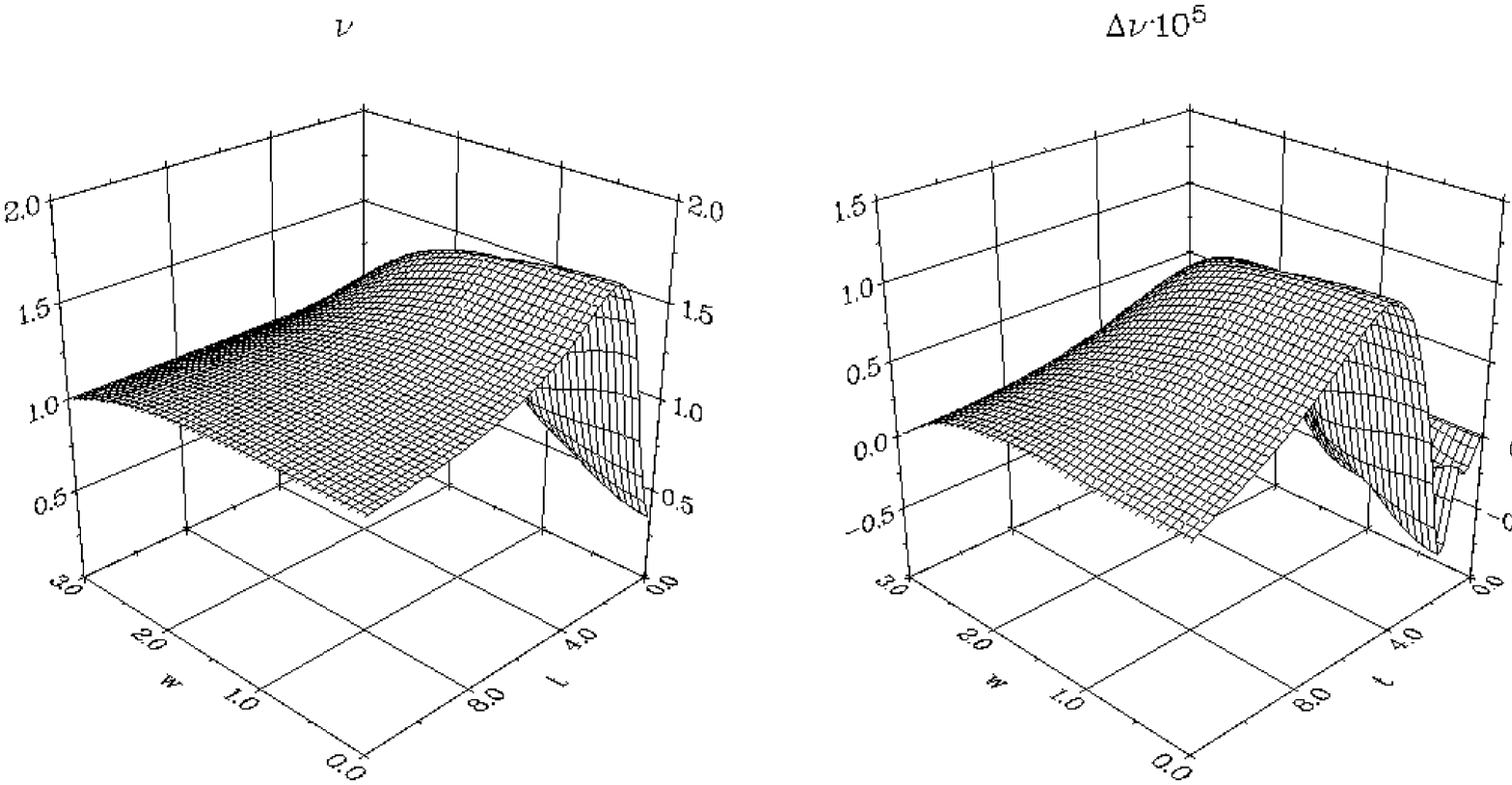, height=400pt, width=175pt, angle=-90}
  \epsfig{file=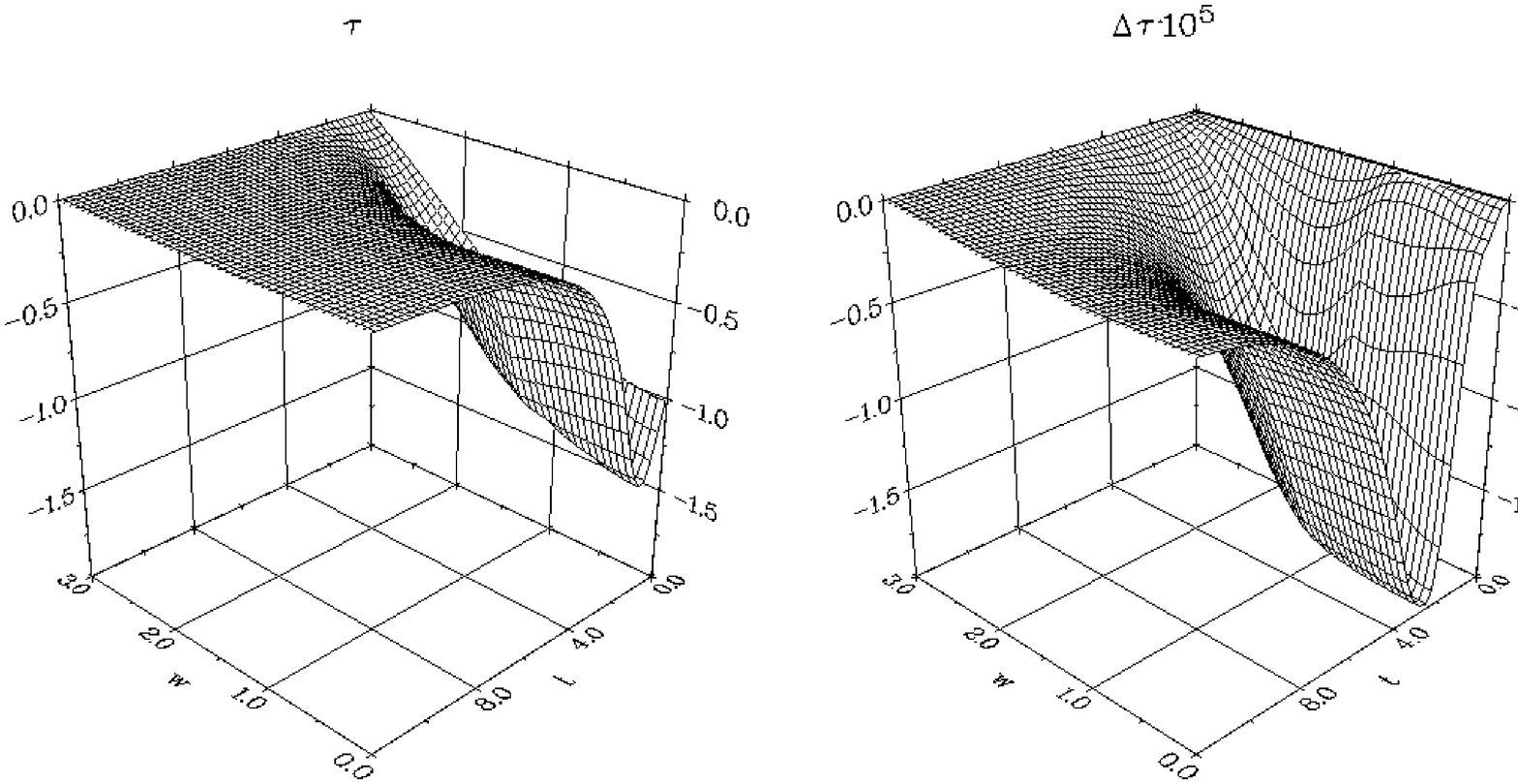, height=400pt, width=175pt, angle=-90}
  \epsfig{file=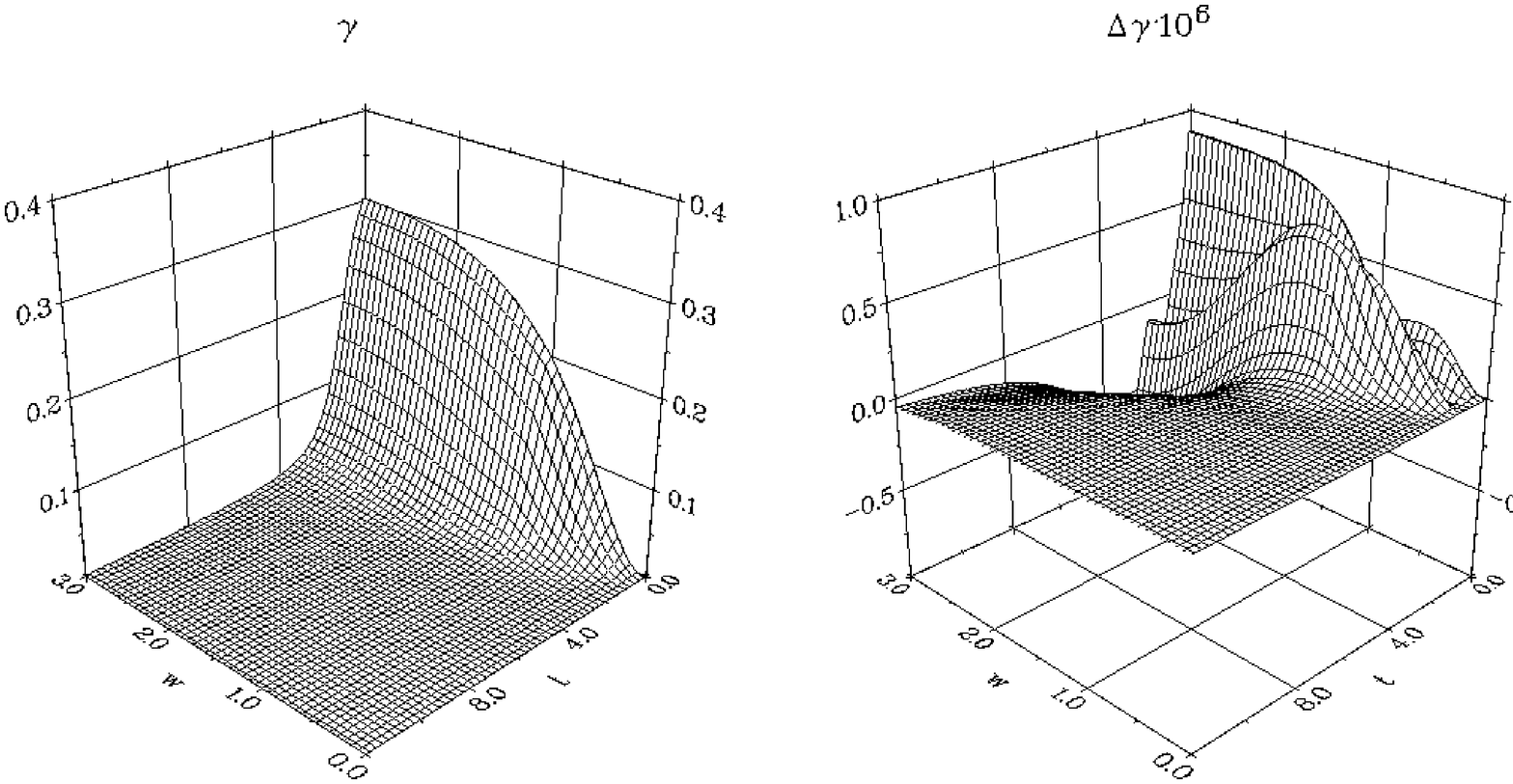, height=400pt, width=175pt, angle=-90}
  \caption{The numerical solutions for $\nu$,
           $\tau$ and $\gamma$ of
           Xanthopoulos' spacetime for $a=1$ obtained with
           300 grid points in each region (left panels). In the right
           panels the corresponding deviation from the analytic result
           is amplified by $10^5$ and $10^6$, respectively.
           The spatial coordinate $w$ is defined in Eq.\,(\ref{w}).}
  \label{CCMntg}
\end{figure}

The next solution we consider is one due to \scite{Xanthopoulos1986}
which has a conical singularity on the $z$-axis and therefore
describes a rotating vacuum solution with a cosmic string type
singularity. The solution has been rewritten in terms of our variables
by \scite{Sjodin2000}. Again it is convenient to introduce auxiliary
quantities
\begin{align}
    Q&=r^2-t^2+1, \label{as_Q} \\[10pt]
    X&=\sqrt{Q^2+4t^2}, \label{as_X} \\[10pt]
    Y&=\frac{1}{2}[(2a^2+1)X+Q]+1-a\sqrt{2(X-Q)},\\[10pt]
    Z&=\frac{1}{2}[(2a^2+1)X+Q]-1, \label{as_Z}
\end{align}
where $a$ is a free parameter which can take on any non-zero value.
The solution derived by Xanthopoulos then becomes
\begin{align}
    \nu(t,\rho)&=\frac{Z}{Y}, \\[10pt]
    \tau(t,\rho)&=-\frac{\sqrt{2(a^2+1)}\sqrt{X+Q}}{Y}, \\[10pt]
    \gamma(t,\rho)&=\frac{1}{2}\ln \frac{Z}{a^2X}.
\end{align}
In the outer region where we use the coordinates $(u,y)$ the result is
\begin{align}
    \tilde{Q} &= y^2-u^2y^2-2u, \\[10pt]
    \tilde{X} &= \sqrt{\tilde{Q}^2+4(uy^2+1)^2}, \\[10pt]
    \tilde{Y} &= \frac{1}{2}[(2a^2+1)\tilde{X}+\tilde{Q}]+y^2
                 -ay\sqrt{2(\tilde{X}-\tilde{Q})}, \\[10pt]
    \tilde{Z} &= \frac{1}{2}[(2a^2+1)\tilde{X}+\tilde{Q}]-y^2, \\[10pt]
    \nu(u,y)  &= \frac{\tilde{Z}}{\tilde{Y}}, \\[10pt]
    \tau(u,y) &= -\frac{\sqrt{2(a^2+1)}\sqrt{\tilde{X}
                 +\tilde{Q}}}{\tilde{Y}}, \\[10pt]
    \gamma(u,y) &= \frac{1}{2}\ln \frac{\tilde{Z}}{a^2\tilde{X}}.
    \label{as_ygamma}
\end{align}
In Fig.\,\ref{CCMntg} we show the numerical results and the deviation from the
analytic values obtained for $a=1$ and a Courant factor of 0.45. 
In this solution no steep gradients are present
and 300 grid points in each region are sufficient
to reproduce the analytic values to within a relative error of about
$10^{-5}$. Again longer runs do not reveal any further features as
the metric settles down into Minkowskian values. We conclude that the
code reproduces analytic solutions with one or two
gravitational degrees of freedom with high accuracy over the dynamically
relevant time intervals.

%========================================================================
\subsubsection{Time dependent convergence analysis}
\label{CCM_CONVERGENCE}
Even though the accuracy and long term stability of the code has been 
demonstrated in the previous sections, we still have to make sure that
it is also second order convergent. In particular the
start-up procedure described in section \ref{start_int} and the use
therein of the Euler scheme to calculate the auxiliary time slice
at $-dt/2$ might raise questions in this respect. \\
% We shall see that
%such doubts are unjustified. \\
For the convergence analysis we define the
$\ell_2$-norm of the deviation of a numerical solution $\Psi^K$ as a function 
of time
\begin{align}
  \Delta \Psi^K_k &= \Psi^K_k - \Psi(x_k), \\[10pt]
  \ell_2[\Delta\Psi^K](t) &= \sqrt{\frac{\sum_k{\left[ \Delta \Psi^K_k(t)
      \right]^2}}{K}}.
  \label{l2norm}
\end{align}
Here $\Psi_k$ is the exact and $\Psi^K_k$ the numerical value at
\begin{figure}[t]
  \centering
  \epsfig{file=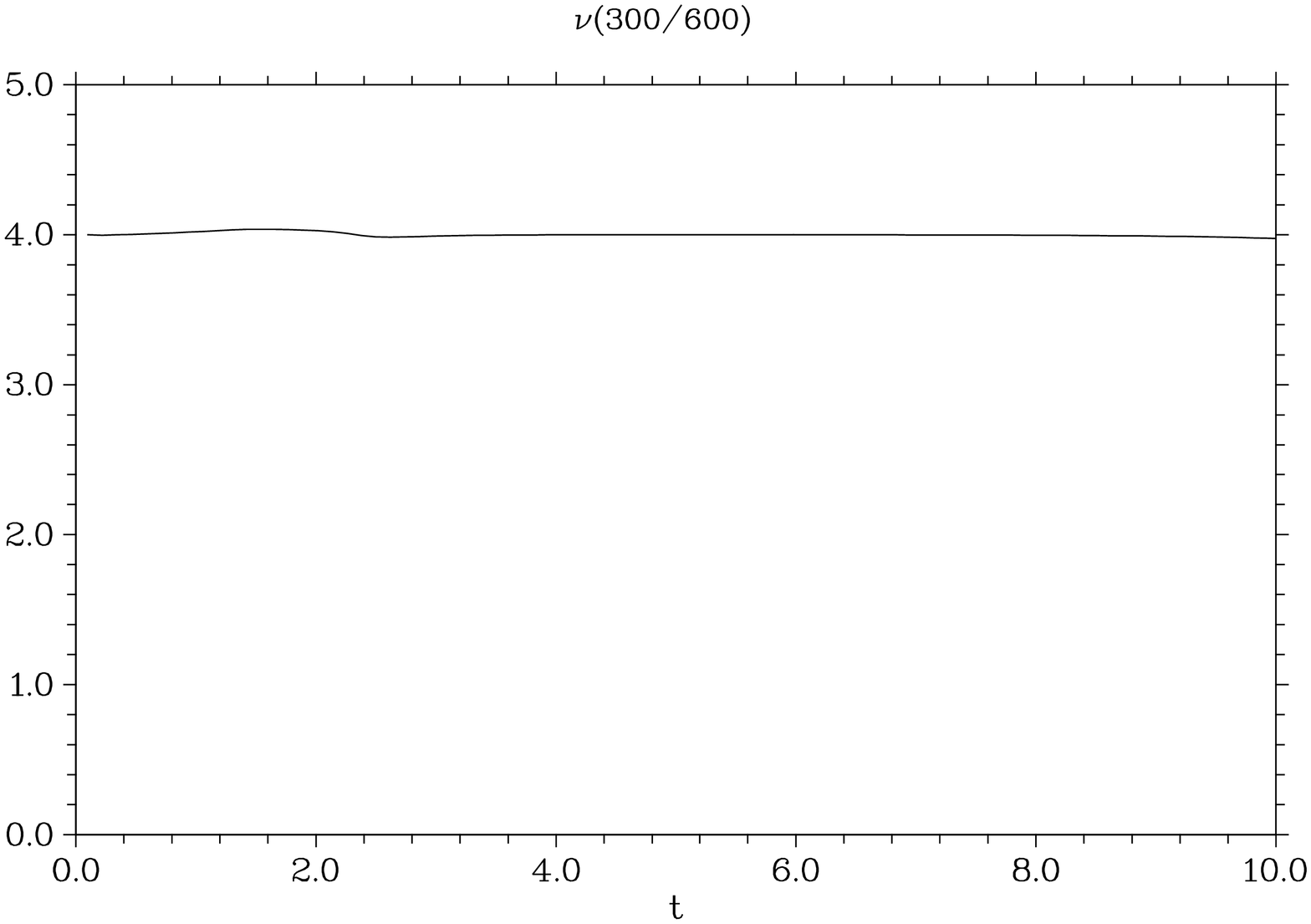, height=200pt, width=150pt, angle=-90}
  \epsfig{file=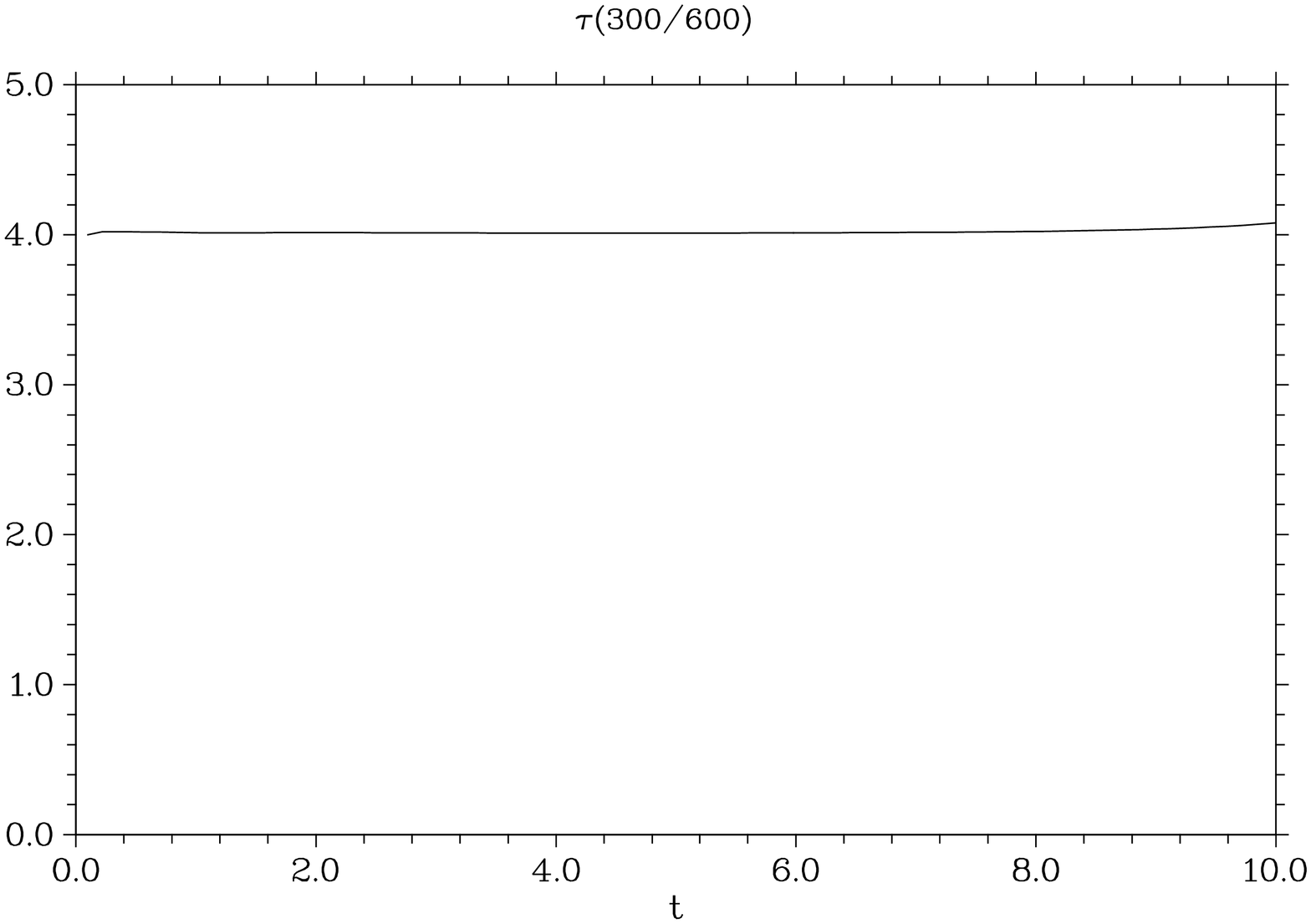, height=200pt, width=150pt, angle=-90}
  \epsfig{file=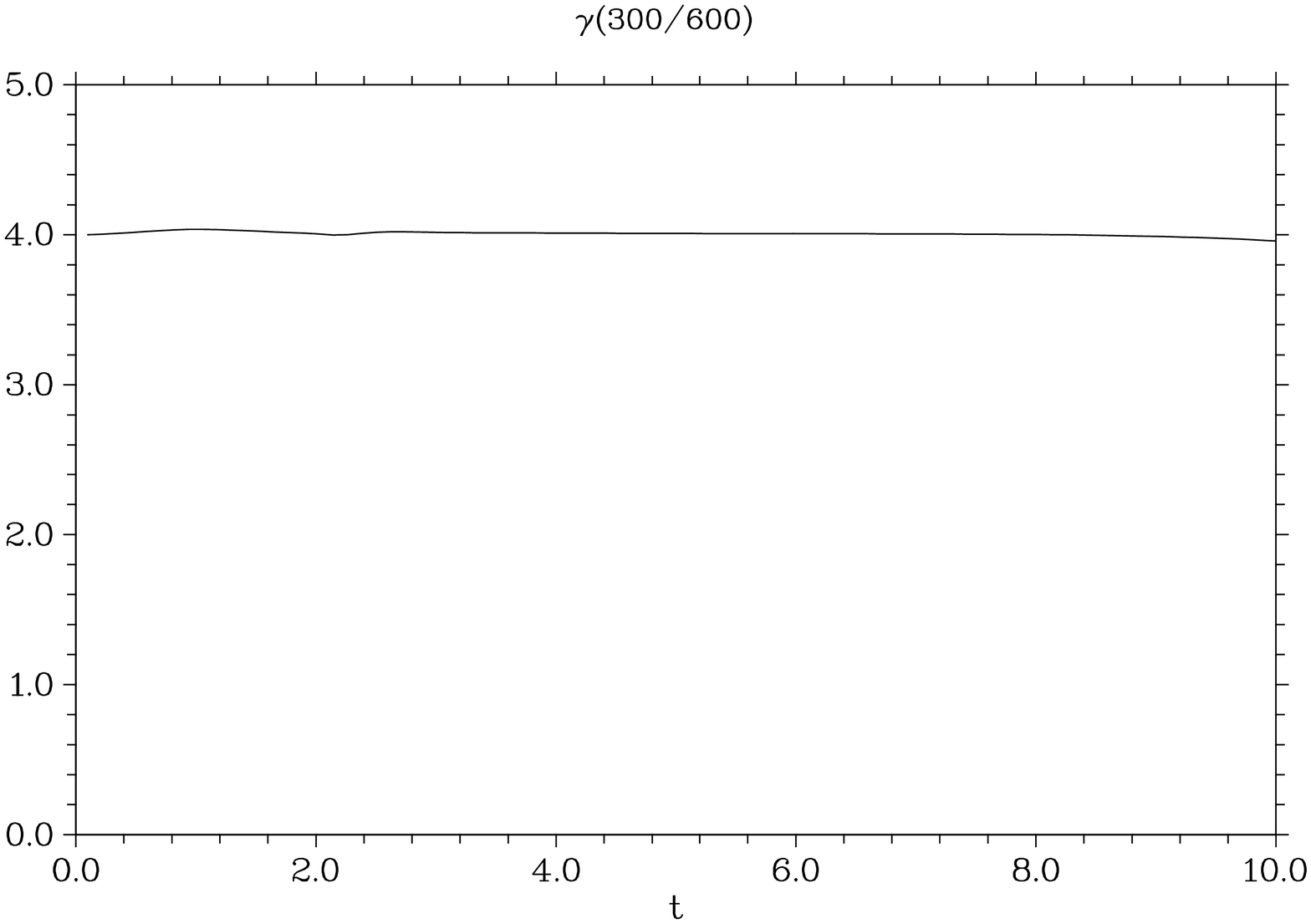, height=200pt, width=150pt, angle=-90}
  \caption{The convergence factor $\ell_2[\Psi^{300}]/\ell_2[\Psi^{600}]$ is
           plotted as a function of time for the variables $\nu$,
           $\tau$ and $\gamma$. For our second order scheme
           we obtain a constant convergence factor of 4 expected for
           doubling the grid resolution.}
  \label{as_ccm_conv}
\end{figure}
grid point $k$ obtained for a total of $K$ grid points. 
We have calculated the $\ell_2$ norm for the Xanthopoulos solution
of the previous section using 300 and 600 grid points 
in each region. In Fig.\,\ref{as_ccm_conv} we plot the
quotient as a function of time. Corresponding to the increase of the grid
resolution by a factor of 2 we expect a convergence factor of 4 for
the second order scheme. In spite of the use of the first order
Euler method for the start-up, second order convergence is clearly maintained
throughout the dynamically relevant evolution.

\newpage
%=========================================================================
\section{Numerical evolution of excited cosmic strings}
\label{cstr}
%
%
%==================================
\subsection{Introduction}
According to the standard ``big bang'' model of cosmology, the universe
is continuously expanding and cooling and was extremely hot and dense in its 
early stages. The grand unified theories (GUT) of elementary particle
physics predict phase transitions to occur as a result of this cooling
process in the early universe. These result in topological defects, regions
with the ``old symmetry'' surrounded by ``new symmetry''. The topology
of the defects depends on the symmetry groups characterising the
involved fields before and after the symmetry breaking. Cosmic strings
are a 1-dimensional, ``string-like'' version of these topological defects.
The type of strings usually considered from the astrophysical point of view
has a mass per unit length $\mu \approx 10^{-6}$ in natural units
($\hbar = G = c = 1$). The corresponding phase transitions are predicted
to have occurred at the GUT energy scale $10^{15}$ GeV. Strings with
significantly higher mass created at higher energy scales cannot be ruled
out, however, and their treatment can no longer be achieved in the
weak-field limit. \\
Numerical simulations by \scite{Vachaspati1984} show that 
cosmic strings are created in the form of a network of infinitely
long or loop like strings. In this work we will focus on infinitely
long strings which are modelled in the framework of cylindrical symmetry. \\
Cosmic strings have caught the interest of astrophysicists and
relativists for several reasons. Most importantly the suggestion
that cosmic strings be seeds for galaxy formation by \scite{Zeldovich1980}
has given rise to intense efforts to understand the evolution of the
resulting density perturbations (see e.g. \citeNP{Turok1986}).
Cosmic strings are also thought to be sources of gravitational radiation
(\citeNP{Vilenkin1994}). Below we will study the interaction of an
infinitely long cosmic string with a wave pulse with one gravitational
degree of freedom. 
Cosmic strings have also been considered of astrophysical relevance
because of the bending of light rays that arises from the conical structure
of the resulting spacetime.
It has been shown by \scite{Vilenkin1981} that the
geometry around an isolated cosmic string is Minkowskian minus
a wedge, the ``deficit angle'', and consequently cosmic strings
may act as gravitational lenses. \\
Even though static cosmic strings in cylindrical symmetry have been
studied extensively in the past either in Minkowskian or curved
spacetime (see e.g. \shortciteNP{Laguna1987}, 
\citeNP{Garfinkle1985}), no solution
has been obtained, to our knowledge, for a dynamic cosmic string
coupled to gravity via the fully non-linear Einstein equations. Below we
will present a numerical solution of this scenario and investigate
the behaviour of a cosmic string excited by gravitational radiation.
After presenting the mathematical description of a cosmic string
in the next section we will derive the equations of a dynamic
cosmic string coupled to gravity. In section \ref{CSnum} we will
describe the numerical treatment of these equations.
The simple scenario of a static cosmic string in Minkowski spacetime
presents already most of the subtleties involved
in solving the general problem and is therefore suitable for illustrating our
numerical methods. Subsequently we address a static string in curved
spacetime and finally present the dynamic code. This code is
extensively tested in section \ref{CStest} before we investigate
the time evolution in section \ref{CSevol}. \\
The results and techniques presented in this section can also be found in
\scite{Sperhake2000}. \\
We conclude this introduction with some
comments on the numerical formulations used in this section. We have seen
above how the combination of an interior Cauchy evolution with a
characteristic evolution in the exterior region leads to a stable
accurate simulation of cylindrically symmetry vacuum spacetimes. In a
natural extension of this project we studied the inclusion of matter in
the form of a cylindrically symmetric cosmic string. Such an
extension of the CCM-code of the previous section has been
developed, but no long term stable evolutions have been achieved
with that code. Consequently we have restarted the investigation.
For convenience this has been done in a purely characteristic
framework and finally
resulted in the long-term stable, accurate code described below. In the
course of this work we have isolated the existence of exponentially
diverging solutions and the corresponding difficulties at the outer boundary
as the source of the problems. We will describe how these difficulties
can be naturally controlled with the use of implicit numerical techniques.
The use of such techniques, however, is by no means restricted to
characteristic methods and we have no reason to believe that an implicit
Cauchy-characteristic matching code would perform less satisfactorily.
Such an implicit CCM code has been tested in the simple case
of a cylindrically symmetric vacuum spacetime with vanishing rotation
and has lead to an accurate long-term stable evolution of the
Weber-Wheeler wave. From this point of view the choice of a characteristic
formulation for the work described in this section is merely a
consequence of the chronology in which progress has been achieved. \\

%=========================================================================
\subsection{Mathematical description of a cosmic string}
In the following work we will use cylindrical coordinates $r$, $\phi$,
$z$. Here $z$ is the Killing direction corresponding to cylindrical
symmetry and $r$, $\phi$ are standard polar coordinates. In 4-dimensional
spacetime the time coordinate is $t$, but we will apply a characteristic
formalism for the numerical solution and therefore also use the retarded
time $u=t-r$.
The simplest model of a cosmic string consists of a scalar field
$\Phi$ coupled to a $U(1)$-gauge field $\hbox{\vec A}_{\mu}$.
The Lagrangian for these coupled fields is given by
\begin{align}
  L_M &= -|(\nabla_{\mu}+ie\hbox{\vec A}_{\mu})\Phi|^2 - V(\Phi)
         - \frac{1}{4} \hbox{\vec F}_{\mu \nu}\hbox{\vec F}^{\mu \nu}
         \label{LM}.
\end{align}
Here $e$ is a constant, which describes the coupling between the
scalar and the vector field.
The self-coupling potential $V(\Phi)$ has the ``Mexican-hat'' shape
predicted by the standard model of elementary particle physics and
$\hbox{\vec F}_{\mu \nu}$ is the field tensor
\begin{align}
  \hbox{\vec F}_{\mu \nu} &= \nabla_{\mu}\hbox{\vec A}_{\nu}
      - \nabla_{\nu}\hbox{\vec A}_{\mu}, \\[10pt]
  V(\Phi) &= 2 \lambda (\Phi^2 - \langle \Phi \rangle^2)^2,
\end{align}
where $\lambda$ is the self-coupling constant of the scalar field.
It turns out to be useful to introduce the Higgs vacuum expectation
value of the scalar field as a parameter $\eta = 2 \langle 
\Phi \rangle^2$. 
Generalizing the notation of \scite{Garfinkle1985} we write the fields as
\begin{align}
  \Phi    &= \frac{S}{\sqrt{2}} e^{i\psi}, \label{Phi} \\[10pt]
  \hbox{\vec A}_{\mu} &= \frac{1}{e} (P-1) \nabla_{\mu} \phi , \label{A}
\end{align}
where $P$, $S$ and $\psi$ are functions of $u$, $r$, $\phi$.
From now on, however, we will make the simplifying assumption of
cylindrical symmetry. Then $P$ and $S$ are functions of
$u$, $r$ only and $\psi = n\phi$, where $n$ is the winding number. In this
work we will only consider the case $n=1$, so $\psi = \phi$.
We can calculate the energy momentum tensor $\hbox{\vec T}^{\mu \nu}$
from the Lagrangian according to
\begin{align}
  \hbox{\vec T}^{\mu \nu} &= \frac{2}{\sqrt{-\hbox{\vec g}}}
      \frac{\delta \mathcal{L}_M} {\delta \hbox{\vec g}_{\mu \nu}},
\end{align}
where $\mathcal{L}_M = \sqrt{-\hbox{\vec g}} L_M$ is the Lagrange density.
Summarising the variables and parameters, we have
\begin{list}{\rm{(\arabic{count})}}{\usecounter{count}}{
             \labelwidth1cm \leftmargin1.5cm \labelsep0.4cm \rightmargin1cm
             \parsep0.5ex plus0.2ex minus0.1ex \itemsep0ex plus0.2ex}
  \item the amplitude of the scalar field $S(u,r)$,
  \item the amplitude of the $U(1)$ gauge field $P(u,r)$,
  \item the constant $e$ which describes the coupling between
        the scalar and vector field,
  \item the self-coupling constant $\lambda$ of the scalar field,
  \item the vacuum expectation value of the scalar field $\eta$.
\end{list}
If we substitute Eqs.\,(\ref{Phi}), (\ref{A}) in (\ref{LM}) we obtain the
Lagrangian and the energy momentum tensor in terms of these quantities
\begin{align}
  L_M &= -\frac{1}{2} \hbox{\vec g}^{\mu \nu} (\nabla_{\mu} S) (\nabla_{\nu} S)
         -\frac{1}{2} S^2 \hbox{\vec g}^{\mu \nu} (\nabla_{\mu} \phi
         + e\hbox{\vec A}_{\mu})
         (\nabla_{\nu} \phi +e\hbox{\vec A}_{\nu}) - \lambda (S^2-\eta^2)^2
         -\frac{1}{4} \hbox{\vec F}_{\mu \nu} \hbox{\vec F}^{\mu \nu}, \\[10pt]
  \hbox{\vec T}_{\mu \nu} &= (\nabla_{\mu}S) (\nabla_{\nu}S)
       + S^2(\nabla_{\mu} \phi + e\hbox{\vec A}_{\mu})(\nabla_{\nu} \phi
       +e\hbox{\vec A}_{\nu})
       +\hbox{\vec g}_{\mu \nu} L_M \label{CSemtensor} .
\end{align}
%

%=========================================================================
\subsection{The field equations}
We start again with the line element in Jordan, Ehlers, Kundt
and Kompaneets (JEKK) form (\ref{JEKK}) for a cylindrically
symmetric spacetime. This form of the metric, however, is not compatible 
with the cosmic string energy momentum tensor so we follow 
\scite{Marder1958} by introducing an extra variable $\mu$ into the metric
\begin{equation}
  ds^2 = e^{2(\gamma-\psi)}(-d\tilde{t}^2 + d\tilde{r}^2) 
         + \tilde{r}^2e^{-2\psi}d\phi^2
         + e^{2(\psi+\mu)}(\omega d\phi+dz)^2,
\end{equation}
where the tilde is used to reserve the names
$t$ and $r$ for rescaled coordinates below.
This choice enables us to compare our numerical solutions
with the results of the Cauchy-characteristic matching code
described in section \ref{ccm}. We have already
noted that this metric has a zero shift vector and the lapse is determined
by the requirement $\hbox{\vec g}_{\tilde{t}\tilde{t}}=
\hbox{\vec g}_{\tilde{r}\tilde{r}}$. 
The function $\mu$, however, introduces
the extra gauge freedom of relabelling the radial null surfaces:
$\tilde{u} \rightarrow f(\tilde{u})$ and $\tilde{v} 
\rightarrow g(\tilde{v})$. We may fix this
by specifying the initial values for $\mu$ and either its time derivative
in a ``3+1'' formalism or its boundary conditions in a characteristic
formalism. We will follow the second approach and below we will see that
the function $\mu$ is uniquely determined in the static case and 
the boundary conditions follow from regularity assumptions
of the metric. The further requirement that the dynamic results reduce
to the static ones in the case of vanishing time dependence therefore
fixes the gauge. \\
It turns out that we can eliminate one of the free parameters and simplify
the equations if we introduce rescaled quantities according to
\begin{align}
  t &= \sqrt{\lambda}\eta \t,  \label{rescale_t} \\[10pt]
  r &= \sqrt{\lambda}\eta \r, \\[10pt]
  X  &= \frac{S}{\eta},  \label{rescale_X}\\[10pt]
  \alpha &= \frac{e^2}{\lambda}.
\end{align}
Thus $\alpha$ represents the relative strength of the coupling between
scalar and vector field compared to the self-coupling. Furthermore
we use the retarded time $u = t - r$ so that the line element becomes
\begin{align}
  ds^2 &= \frac{e^{2(\gamma-\psi)}}{\lambda \eta^2}
          (-du^2 - 2du dr) + r^2\frac{e^{-2\psi}}
          {\lambda \eta^2}d\phi^2 + e^{2(\psi+\mu)}(\omega d\phi+dz)^2.
\end{align}
In section \ref{Gerochdec} we have described the Geroch decomposition which
can be used to factor out the Killing direction $\bbox{\partial}_z$ even
if the Killing field is not hypersurface-orthogonal. It is a remarkable
fact that the right hand side of equation (\ref{Geroch_cond}) still
vanishes for spacetimes with a cosmic string energy-momentum tensor
(\ref{CSemtensor}) (\shortciteNP{Sjodin2000}),
so that the Geroch twist can be described by a
potential according to Eq.\,(\ref{Geroch_pot}). The other geometrical variable, 
the norm of the $z$-Killing vector (\ref{Geroch_norm}) becomes
\begin{align}
  \nu = e^{2(\psi + \mu)},
\end{align}
and the 3-dimensional line element (\ref{Geroch_3m}) is
\begin{align}
  ds^2 &= \frac{1}{\lambda \eta^2 \nu} \left[e^{2(\gamma-\psi)}
          (-du^2 - 2du dr) + r^2e^{2\mu}
          d\phi^2 \right] \label{CS_3dline}.
\end{align}
With the energy momentum tensor given by (\ref{CSemtensor}) and the
3-dimensional line element (\ref{CS_3dline}) we are now in a position to
calculate the field equations according to equations 
(\ref{Geroch3d})-(\ref{Geroch_tau}). We obtain
\begin{align}
  \begin{split}
  \Box \nu =& \,\, \nu_{,r} \mu_{,r} + \frac{\tau_{,r}^2-\nu_{,r}^2}{\nu}
        -\nu_{,u} \mu_{,r} - \nu_{,r} \mu_{,u}
        +2\frac{\nu_{,u} \nu_{,r} - \tau_{,u} \tau_{,r}}{\nu} \\[10pt]
      & +8\pi \eta^2\left[ 2e^{2(\gamma+\mu)} (X^2-1)^2
        +e^{-2\mu} \nu^2
        \frac{2P_{,u} P_{,r} - P_{,r}^2}{\alpha r^2} \right], \label{nuur} 
  \end{split} \\[10pt]
  \Box \tau =&\,\, \tau_{,r}\mu_{,r} - 2\frac{\tau_{,r} \nu_{,r}}{\nu}
        - \tau_{,u} \mu_{,r}-\tau_{,r} \mu_{,u}
        + 2\frac{\tau_{,r} \nu_{,u} +\tau_{,u} \nu_{,r}}{\nu}, \\[10pt]
  \Box \mu =&\,\, \mu_{,r}^2 +\frac{\mu_{,r}}{r}
        - \frac{\mu_{,u}}{r} -2\mu_{,u} \mu_{,r} +8\pi\eta^2 \left[
        2\frac{e^{2(\gamma + \mu)}}{\nu} (X^2-1)^2 + e^{2\gamma}
        \frac{X^2P^2}{r^2} \right] \\[10pt]
  0 =&\,\, 2\gamma_{,r} +2r\gamma_{,r} \mu_{,r} -r\mu_{,rr} + r\mu_{,r}^2
        -\frac{r}{2\nu^2} (\tau_{,r}^2 + \nu_{,r}^2) - 8\pi \eta^2 \left[
        r X_{,r}^2 + \frac{1}{\alpha} e^{-2\mu} \nu \frac{P_{,r}^2}{r} \right]
        \label{gammar},
\end{align}
where we have introduced the flat-space d'Alembert operator
\begin{equation}
  \Box = 2\frac{\partial^2}{\partial u \partial r}
         - \frac{\partial^2}{\partial r^2} - \frac{1}{r} \left(
         \frac{\partial}{\partial r} - \frac{\partial}{\partial u} \right).
\end{equation}
This set of equations is supplemented by the matter evolution equations
obtained either from conservation of energy-momentum 
$\nabla_{\mu} \hbox{\vec T}^{\mu \nu} = 0$ or variation of the Lagrange density
$\mathcal{L}_M$ with respect to the matter fields $P$ and $X$. The result is
\begin{align}
  \Box P &= 2\frac{P_{,u}-P_{,r}}{r} - P_{,r} \mu_{,r}
        + P_{,r}\frac{\nu_{,r}}{\nu} + P_{,r} \mu_{,u} + P_{,u} \mu_{,r}
        - \frac{P_{,r} \nu_{,u} + P_{,u} \nu_{,r}}{\nu}
        - \alpha \frac{e^{2(\gamma + \mu)}}{\nu} PX^2,
        \label{pur} \\[10pt]
  \Box X &= X_{,r} \mu_{,r} - X_{,u} \mu_{,r} - X_{,r} \mu_{,u}
        - 4 \nu^{-1} e^{2(\gamma + \mu)} X(X^2-1)
        - e^{2\gamma} \frac{XP^2}{r^2}. \label{xur}
\end{align}
Note that in equations (\ref{nuur})-(\ref{gammar}) the matter terms
exclusively appear with a factor $\eta^2$. Consequently $\eta$ describes
the effect of the string on the spacetime geometry and, thus, represents the
string's mass.
There are two further Einstein equations which can be shown to be
a direct consequence of (\ref{nuur})-(\ref{xur}) and their derivatives.
These equations have only been used to provide a check on the accuracy
of the code.
Finally we have to supplement the equations by
boundary conditions on the axis. For the 4-dimensional metric
variables the simplest condition is to require the metric to be
$C^2$ on the axis so that we have a well defined curvature tensor.
The resulting boundary conditions are (\citeNP{Sjodin2001b})
\begin{align}
  \nu(t,r) &= a_1(t) + \mathscr{O}(r^2), \label{bound_nutr} \\[10pt]
  \tau(t,r) &= \mathscr{O}(r^2), \\[10pt]
  \mu(t,r) &= a_2(t) + \mathscr{O}(r^2), \label{bound_mutr} \\[10pt]
  \gamma(t,r) &= \mathscr{O}(r). \label{gammatr}
\end{align}
The boundary conditions for $S$ and $P$ are (\citeNP{Garfinkle1985})
\begin{align}
  P(t,r) &= 1+\mathscr{O}(r^2), \\[10pt]
  X(t,r) &= \mathscr{O}(r). \label{bound_xtr}
\end{align}
The numerical implementation of these boundary conditions as well as regularity
requirements at null infinity will be discussed in section \ref{dynstring}.

%=======================================================================
\subsection{Numerical methods}
\label{CSnum}
In order to solve the above field equations we have developed two
independent codes. The first is based on
the Cauchy characteristic matching code described in section \ref{ccm}.
This code performs well in the absence of matter and has been used 
to study several cylindrically symmetric vacuum solutions
(see also \shortciteNP{Sjodin2000}). However, this CCM code performed less 
satisfactorily in the evolution of the cosmic string. This
is due to the existence of
unphysical solutions to the evolution equations (\ref{nuur})-(\ref{xur}) which
diverge exponentially as $r\rightarrow \infty$. Controlling the time evolution
near null infinity by means of a sponge function enabled us to select the
physical solutions with regular behaviour at $I^+$, but the sponge function
itself introduced noise which eventually gave rise to instabilities.
We therefore implemented a second implicit,
purely characteristic, code which
allows us to directly control the behaviour of the solutions at the 
boundaries and thus suppress diverging solutions. The main problem with
the system of differential equations is the irregularity of the equations
at both the origin and null infinity. It is the implicit nature of the
scheme that provides a
simple way of implementing the boundary conditions and thus circumventing
all problems with these irregularities. A purely characteristic formulation
has been used for the second code for convenience rather than
numerical necessity and we believe that an implicit CCM scheme would
produce similar accuracy, convergence and long term stability.
It is interesting that the irregularity problems are 
already present in the calculation of the static
cosmic string in Minkowski spacetime. We will,
therefore, first describe the numerical scheme used in the static Minkowskian
case where the equations are fairly simple. We then present the modifications
necessary for the static and dynamic case coupled to the gravitational
field.

%========================================================================
\subsubsection{The static cosmic string in Minkowski spacetime} 
\label{SECcsmink}
\begin{figure}[t]
  \centering
  \epsfig{file=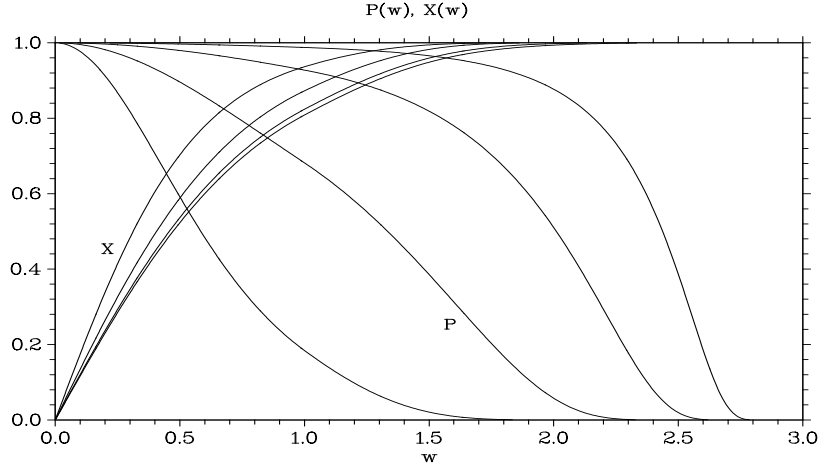, height=350pt, width=200pt, angle=-90}
  \caption{The cosmic string variables $P$ and $X$ are plotted for
           $\alpha =$ 10, 1, 0.1, 0.01 (from ``left to right''). The two
           families are labelled in the plot. As $\alpha$ increases,
           both, $P$ and $X$ become more concentrated towards the origin.
           Note that $w=3$ corresponds to $r\to \infty$ [cf. Eq.\,(\ref{w})].}
\label{csmink}
\end{figure}

In Eqs.\,(\ref{nuur})-(\ref{xur})
we set the metric variables to their Minkowskian values and all time
derivatives to zero to obtain the equations for the static cosmic string in
Minkowski spacetime (cf.\ \citeNP{Garfinkle1985})
\begin{align}
  r\frac{d}{dr} \left( r^{-1} \frac{dP}{dr} \right) &= \alpha X^2 P
        \label{mink_Prr}, \\[10pt]
  r\frac{d}{dr}\left( r\frac{dX}{dr} \right) &= X \left[ P^2 + 4r^2 (X^2-1)
        \right] . \label{mink_Xrr}
\end{align}
The boundary conditions are (see \citeNP{Garfinkle1985})
\begin{alignat}{4}
  P(0) &= 1, & \qquad \qquad & \lim_{r\rightarrow \infty} P(r) &= 0, 
  \nonumber \\
  X(0) &= 0, & &\lim_{r \rightarrow \infty} X(r) &= 1.
  \label{mink_bound}
\end{alignat}
In order to cover the whole spacetime with a finite coordinate range,
we divide the computational domain into two regions in the same way
as in section \ref{start_int}. In the inner region ($0\le r\le 1$) we
use the coordinate $r$, while in the outer region we introduce the compactified
radius $y$ defined by equation (\ref{y})
which covers the range $1 \ge y \ge 0$. This corresponds to the region
$1 \le r < \infty$ with infinity mapped to $y=0$. 
Again we combine $r$ and $y$ into the
single radial variable $w$ defined by (\ref{w}).
In terms of the coordinate $y$ Eqs.\,(\ref{mink_Prr}),
(\ref{mink_Xrr}) take the form
\begin{eqnarray}
  &&\frac{d}{dy} \left( y^5\frac{dP}{dy} \right) = 4\alpha \frac{X^2P}{y},
      \\[10pt]
  &&\frac{d}{dy} \left( y \frac{dX}{dy} \right)  =
  4X\left[\frac{P^2}{y}+ 4\frac{(X^2-1)}{y^5}\right].
\end{eqnarray}
The number of grid points in each region may differ, but
\begin{figure}[t]
\centering
%------------------------------------------------------------------------
\begin{picture}(0,0)%
\epsfig{file=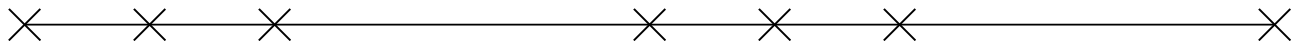}%
\end{picture}%
\setlength{\unitlength}{3947sp}%
\begingroup\makeatletter\ifx\SetFigFont\undefined%
\gdef\SetFigFont#1#2#3#4#5{%
  \reset@font\fontsize{#1}{#2pt}%
  \fontfamily{#3}\fontseries{#4}\fontshape{#5}%
  \selectfont}%
\fi\endgroup%
\begin{picture}(6312,2028)(976,-1606)
\put(3226,-436){\makebox(0,0)[lb]{\smash{\SetFigFont{11}{13.2}{\rmdefault}{\mddefault}{\updefault}...}}}
\put(2401,-436){\makebox(0,0)[lb]{\smash{\SetFigFont{11}{13.2}{\rmdefault}{\mddefault}{\updefault}3}}}
\put(1801,-436){\makebox(0,0)[lb]{\smash{\SetFigFont{11}{13.2}{\rmdefault}{\mddefault}{\updefault}2}}}
\put(976,-436){\makebox(0,0)[lb]{\smash{\SetFigFont{11}{13.2}{\rmdefault}{\mddefault}{\updefault}$k$=1}}}
\put(976,-61){\makebox(0,0)[lb]{\smash{\SetFigFont{11}{13.2}{\rmdefault}{\mddefault}{\updefault}$r$=0}}}
\put(3976,-1186){\makebox(0,0)[lb]{\smash{\SetFigFont{11}{13.2}{\rmdefault}{\mddefault}{\updefault}$K_1$+1}}}
\put(4051,-436){\makebox(0,0)[lb]{\smash{\SetFigFont{11}{13.2}{\rmdefault}{\mddefault}{\updefault}$K_1$}}}
\put(4051,-61){\makebox(0,0)[lb]{\smash{\SetFigFont{11}{13.2}{\rmdefault}{\mddefault}{\updefault}$r$=1}}}
\put(4051,-1561){\makebox(0,0)[lb]{\smash{\SetFigFont{11}{13.2}{\rmdefault}{\mddefault}{\updefault}$y$=1}}}
\put(4651,-1186){\makebox(0,0)[lb]{\smash{\SetFigFont{11}{13.2}{\rmdefault}{\mddefault}{\updefault}$K_1$+2}}}
\put(5251,-1186){\makebox(0,0)[lb]{\smash{\SetFigFont{11}{13.2}{\rmdefault}{\mddefault}{\updefault}$K_1$+3}}}
\put(5701,314){\makebox(0,0)[lb]{\smash{\SetFigFont{11}{13.2}{\rmdefault}{\mddefault}{\updefault}outer region}}}
\put(2101,314){\makebox(0,0)[lb]{\smash{\SetFigFont{11}{13.2}{\rmdefault}{\mddefault}{\updefault}inner region}}}
\put(6901,-1186){\makebox(0,0)[lb]{\smash{\SetFigFont{11}{13.2}{\rmdefault}{\mddefault}{\updefault}$K_1$+$K_2$}}}
\put(7051,-1561){\makebox(0,0)[lb]{\smash{\SetFigFont{11}{13.2}{\rmdefault}{\mddefault}{\updefault}$y$=0}}}
\put(6376,-1186){\makebox(0,0)[lb]{\smash{\SetFigFont{11}{13.2}{\rmdefault}{\mddefault}{\updefault}...}}}
\end{picture}

\vspace{0.5cm}
%------------------------------------------------------------------------
\caption{The combined grid of the inner and the outer region. Note that
         both grid points, $K_1$ and $K_1+1$, correspond to the position
         $r=1 \Leftrightarrow y=1$.
         These points form the interface of the code and facilitate
         transformation of the variables from the coordinate system using $r$
         into that using $y$.}
\label{grid}
\end{figure}
each half-grid is uniform. Thus we use
a total of $K:=K_1 + K_2$ grid points where the points labelled $K_1$ and
$K_1+1$ both correspond to the position $r=1=y$.
The points $K_1$, $K_1+1$ form the interface between the two regions
(see Fig.\,\ref{grid}). One point will contain
the variables in terms of $r$, the other in terms of $y$.
With the computational grid covering the whole spacetime, we now face a
two point boundary value problem. Due to the existence of unphysical solutions
diverging at $y=0$
% shooting methods turned out to be unsuitable for solving
%this problem. On the other hand numerical relaxation,
we have chosen to solve the equations with a numerical relaxation scheme
as described in section \ref{relaxation} which
allows us to directly control the behaviour of $P$ and $X$ at infinity.
The form of Eqs.\,(\ref{mink_Prr}),\ (\ref{mink_Xrr}) suggests that
in order to write them as a first order system we should introduce the
auxiliary variables $Q=r^{-1}P_{,r}$ and $R=rX_{,r}$. The equations may then be
written in the form
\begin{align}
  P_{,r} &= rQ, \\[10pt]
  X_{,r} &= \frac{R}{r}, \\[10pt]
  Q_{,r} &= \alpha \frac{PX^2}{r}, \\[10pt]
  R_{,r} &= X\left[\frac{P^2}{r} + 4r(X^2-1)\right].
\end{align}
The corresponding equations in the outer region are given by
\begin{align}
  P_{,y} &= -2\frac{Q}{y^5}, \\[10pt]
  X_{,y} &= -2\frac{R}{y}, \\[10pt]
  Q_{,y} &= -2 \alpha \frac{X^2P}{y}, \\[10pt]
  R_{,y} &= -2X \left( \frac{P^2}{y}+4\frac{X^2-1}{y^5} \right).
\end{align}
Standard second order centred finite differencing
according to Eqs.\,(\ref{relax_FDE1}), (\ref{relax_FDE2})
% Note, however, that in this case
% we obtain $4(N-2)$ algebraic equations 
% (instead of $4(N-1)$ as in section \ref{relaxation}) 
% due to the presence of the interface. 
results in $4(K-2)$ non-linear algebraic equations which are supplemented
by the 4 boundary conditions 
(\ref{mink_bound}) and 4 interface relations
\begin{align}
  P_{K_1+1} &= P_{K_1}, \label{int_P} \\[10pt]
  X_{K_1+1} &= X_{K_1}, \\[10pt]
  Q_{K_1+1} &= Q_{K_1}, \\[10pt]
  R_{K_1+1} &= R_{K_1} \label{int_R}.
\end{align}
We then start with piecewise linear initial guesses for $P$ and $X$ (and the
corresponding derivatives $Q$ and $R$) and solve the $4K$ algebraic equations
as described in section \ref{relaxation}. \\
In order to check the code for convergence, we vary the grid resolution
$K$ (using $K_1=K_2$ points in both regions) from
$150$ to $2400$, halving the grid spacing each time.
Since we do not have an analytic solution, the results are
compared against the high-resolution case ($K=2400$). For doing this we
calculate the $\ell_2$ norm according to Eq.\,(\ref{l2norm}).
In this case the function $\Psi$ in Eq.\,(\ref{l2norm}) 
stands for $P$, $X$, $Q$ or $R$ and the norm does not depend on time
because of the static nature of the problem.
For second order convergence we expect the $\ell_2$ norm to decrease by
a factor of 4 each time we increase the grid resolution by a factor of 2.
However, we do not compare our results against the exact solution but against
a high resolution result which itself has a finite truncation error,
so that
\begin{align}
  \ell_2[\Psi^K] &= \left(\sum_k\Psi^K_k - \Psi^{2400}_k\right)^{1/2}.
    \label{CAUCHYNORM}
\end{align}
Therefore we do not expect the factor to be exactly $4$. Using a
\begin{table}[t]
  \caption{Convergence test for the cosmic string in Minkowski space-time
           for $\alpha = 1$. The norm of the deviation
           $\ell_2[\Delta \Psi^K]$ is defined by Eq.\,(\ref{CAUCHYNORM}).
           As the grid resolution is increased, the deviation from the
           high resolution result decreases quadratically to a good
           approximation (see text for details).}
  \begin{center}
  \label{convMink}
    \begin{tabular}{l|cccc}
      \hline \hline
      & P & X & Q & R \\
      \hline\\[-7pt]
      $\ell_2(\Delta \Psi^{1200})$ & $5.77\cdot 10^{-7}$ & $2.84\cdot 10^{-7}$
                       & $5.86\cdot 10^{-7}$ & $8.89\cdot 10^{-7}$ \\
      \hline\\[-7pt]
      $\ell_2(\Delta\Psi^{150})/\ell_2(\Delta\Psi^{300})$
                        & 4.05 & 4.05 & 4.04 & 4.05 \\
      $\ell_2(\Delta\Psi^{300})/\ell_2(\Delta\Psi^{600})$
                        & 4.20 & 4.20 & 4.20 & 4.20 \\
      $\ell_2(\Delta\Psi^{600})/\ell_2(\Delta\Psi^{1200})$
                        & 5.00 & 5.00 & 5.00 & 5.00 \\
      \hline \hline
    \end{tabular}
  \end{center}
\end{table}
grid resolution $K$ the truncation error is given by
\begin{align}
  \Psi^K &= \Psi + \mathscr{O}\left(\frac{1}{K^2}\right),
\end{align}
where $\Psi$ is the exact and $\Psi^K$ the numerical solution. For simplicity
we will assume that the truncation error is either $-1/K^2$
or $+1/K^2$. If we use a reference solution obtained for $4K$ grid points
and compare solutions $\Psi^K$ and $\Psi^{2K}$ the ratio of the corresponding
$\ell_2$-norms becomes
\begin{align}
  \left( \frac{\sum(\Psi_i^K-\Psi_i^{4K})^2}
         {\sum(\Psi_i^{2K} - \Psi_i^{4K})^2}\right)^{1/2}
     &= \left( \frac{\sum(\pm\frac{1}{K^2} \pm \frac{1}{16K^2})^2}
             {\sum(\pm\frac{1}{4K^2} \pm \frac{1}{16K^2})^2} \right)^{1/2}
     = \left| \frac{\pm16 \pm 1}{\pm4\pm1} \right| .
\end{align}
Considering the extreme cases, we expect a convergence factor between
3 and $5\frac{2}{3}$. The truncation error of the high resolution 
result will have significantly less influence on the comparison of 
lower resolution results and the
factors should be closer to 4. Table \ref{convMink} shows our results for
the cosmic string in Minkowski space-time and clearly indicates
second order convergence.
In Fig.\,\ref{csmink} we show the string variables $P$ and $X$
for various values of $\alpha$ as a function of $w$. 
Due to the rescaling (\ref{rescale_t})-(\ref{rescale_X}) the equations
for the cosmic string in Minkowski spacetime (\ref{mink_Prr}) and 
(\ref{mink_Xrr}) do not explicitly contain the parameter $\eta$,
so the shape of the cosmic string fields expressed in terms of the rescaled
variables is independent of $\eta$.
Below we will see that this is no longer true 
in curved spacetime where $\eta$, representing the mass of the string,
determines the strength of its coupling to gravity.
Fig.\,\ref{csmink} does, however, reveal a significant variation of the
profiles of the scalar and vector field
with the coupling ratio $\alpha$. As the scalar-vector
coupling becomes more dominant with respect to the self coupling of the
scalar field (larger $\alpha$), both $P$ and $X$ become more concentrated
towards the origin.

%========================================================================
\subsubsection{The static cosmic string coupled to gravity}
\label{CSstat}
The equations governing a static cosmic string in curved spacetime
are obtained from the general equations (\ref{nuur})-(\ref{xur}) by
setting all time derivatives to zero. If we combine first and second
spatial derivatives in a single operator as in equations (\ref{mink_Prr}),
(\ref{mink_Xrr}), we can write these equations as
\begin{align}
  (r\nu_{,r})_{,r} =&\,\, -r\nu_{,r} \mu_{,r} + r\frac{\nu_{,r}^2
                       - \tau_{,r}^2}{\nu} + 8\pi\eta^2\left[ e^{-2\mu}
                       \nu^2 \frac{P_{,r}^2}{\alpha r}
                       - 2e^{2(\gamma+\mu)}r(X^2-1)^2\right], \label{nurr} 
                       \\[10pt]
  (r\tau_{,r})_{,r} =&\,\, 2r\frac{\tau_{,r} \nu_{,r}}{\nu}-r\tau_{,r}
                       \mu_{,r}\ , \label{taueq}
\end{align}
\begin{align}
  (r^2\mu_{,r})_{,r} =&\,\, -r^2\mu_{,r}^2 - 8\pi \eta^2 \left[ e^{2\gamma}
                        X^2 P^2 + 2e^{2(\gamma+\mu)} \nu^{-1}
                        r^2 (X^2-1)^2\right], \\[10pt]
    \gamma_{,r} =&\,\, \frac{r}{2(r\mu_{,r}+1)} \left[ \mu_{,rr}-\mu_{,r}^2
                   + \frac{1}{2\nu^2}\left(\tau_{,r}^2+ \nu_{,r}^2\right) 
                   + 8\pi \eta^2 \left( X_{,r}^2 + \frac{1}{\alpha}
                   e^{-2\mu} \nu \frac{P_{,r}^2}{r^2} \right) \right],
                   \label{stat_gammar} \\[10pt]
  \left( \frac{1}{r}P_{,r}\right)_{,r}
                 =&\,\, \frac{P_{,r}\mu_{,r}}{r} -\frac{P_{,r} \nu_{,r}}{r\nu}
                    +\alpha e^{2(\gamma+\mu)} \nu^{-1} \frac{PX^2}{r}, \\[10pt]
  (rX_{,r})_{,r} =&\,\, -rX_{,r} \mu_{,r}
                    +X\left[ e^{2\gamma} \frac{P^2}{r}
                    + 4e^{2(\gamma+\mu)} \nu^{-1}r (X^2-1)\right]. \label{xrr}
\end{align}
After completing the code, we realised that in the
case of vanishing rotation $\tau$ the field equations 
(\ref{nurr})-(\ref{xrr}) imply a simple relation between $\nu$,
$\mu$ and $\gamma$. An appropriate linear combination of these
equations and their spatial derivatives can be written as
\begin{align}
  (\gamma + \mu - \ln{\nu})_{,rr} + \left(\frac{1}{r} + \mu_{,r}\right)
  (\gamma + \mu - \ln{\nu})_{,r} 
  &= 0,
\end{align}
which after some manipulation becomes
\begin{align}
  (\gamma + \mu - \ln{\nu})_{,r} &= C \frac{e^{-\mu}}{r}.
\end{align}
Here $C$ is a constant that has to vanish in order to ensure
finite derivatives at the origin.
In the static case we
adjust the functions $a_1$ and $a_2$ in the boundary conditions
(\ref{bound_nutr}), (\ref{bound_mutr}) so that $\nu=1$ and $\mu=0$
at the origin and consequently
\begin{align}
  \gamma + \mu - \ln{\nu} &= 0,  \label{gmn}
\end{align}
for all values of $r$. Even though $\tau$ will be zero in
the analysis in this section,
we will numerically solve the original system of
equations (\ref{nurr})-(\ref{xrr}) and
use (\ref{gmn}) as a test for the code. \\
In order to numerically solve the equations of a cosmic string coupled to 
gravity, we rewrite them again as
a first order system. The differential operators appearing on the right hand
side suggest that we introduce the auxiliary
quantities $N=r\nu_{,r}$, $T=r\tau_{,r}$, $M=r^2\mu_{,r}$, $Q=r^{-1}P_{,r}$
and $R=rX_{,r}$. The system can then be written in the form
\begin{align}
  \nu_{,r} &= \frac{N}{r},  \label{nur} \\[10pt]
  \tau_{,r} &= \frac{T}{r}, \\[10pt]
  \mu_{,r} &= \frac{M}{r^2}, \\[10pt]
  P_{,r} &= rQ, \\[10pt]
  X_{,r} &= \frac{R}{r},  \label{xr} \\[10pt]
  N_{,r} &= \frac{N^2-T^2}{r\nu} - \frac{NM}{r^2}
         - 16\pi \eta^2 e^{2\gamma + 2\mu} r(X^2-1)^2
         + 8\pi \frac{\eta^2}{\alpha} e^{-2\mu} \nu^2 rQ^2, \\[10pt]
  T_{,r} &= 2\frac{TN}{\nu r} - \frac{TM}{r^2}, \\[10pt]
  M_{,r} &= -\frac{M^2}{r^2} - 8\pi\eta^2 e^{2\gamma} X^2P^2
         - 16\pi\eta^2e^{2\gamma+2\mu} \nu^{-1}r^2 (X^2-1)^2, \\[10pt]
  2(r+M)\gamma_{,r} &=  M_{,r} - 2\frac{M}{r} - \frac{M^2}{r^2}
           + \frac{T^2+N^2}{2\nu^2} + 8\pi \eta^2 R^2
           + 8\pi \frac{\eta^2}{\alpha} e^{-2\mu}\nu r^2Q^2, \\[10pt]
  Q_{,r} &= \frac{QM}{r^2} - \frac{QN}{r\nu} +\alpha
           e^{2\gamma+2\mu} \nu^{-1} \frac{PX^2}{r}, \\[10pt]
  R_{,r} &= -\frac{RM}{r^2} + 4e^{2\gamma+2\mu} \nu^{-1} r X(X^2-1)
           + e^{2\gamma} \frac{XP^2}{r}.
\end{align}
The corresponding equations in terms of the compactified radial coordinate $y$
are
\begin{align}
  \nu_{,y} &= \frac{N}{y}, \\[10pt]
  \tau_{,y} &= \frac{T}{y}, \label{CSTAT_TAUY} \\[10pt]
  \mu_{,y} &= yM, \\[10pt]
  P_{,y} &= \frac{Q}{y^5}, \\[10pt]
  X_{,y} &= \frac{R}{y}, \\[10pt]
  N_{,y} &= \frac{N^2-T^2}{y\nu} - yNM - 64\pi \eta^2 e^{2\gamma + 2\mu}
         \frac{(X^2-1)^2}{y^5} + 8\pi \frac{\eta^2}{\alpha} e^{-2\mu} \nu^2
         \frac{Q^2}{y^5}, \\[10pt]
  T_{,y} &= 2\frac{TN}{y\nu} - yTM, \\[10pt]
  M_{,y} &= -yM^2 - 32\pi \eta^2 e^{2\gamma} \frac{X^2 P^2}{y^3} - 64\pi \eta^2
         e^{2\gamma + 2\mu} \nu^{-1} \frac{(X^2-1)^2}{y^7}, \\[10pt]
  2(y^2M-2) \gamma_{,y} = & \,\,\, y^2M_{,y} 
         + 4yM - y^3M^2 + \frac{N^2+T^2}{2y\nu^2}
         +8\pi\eta^2 \frac{R^2}{y} + 8\pi \frac{\eta^2}{\alpha} e^{-2\mu}
         \nu \frac{Q^2}{y^5},
\end{align}
\begin{align}
  Q_{,y} &= yQM - \frac{QN}{y\nu} + 4\alpha e^{2\gamma+2\mu} \nu^{-1}
         \frac{PX^2}{y}, \\[10pt]
  R_{,y} &= -yRM + 4e^{2\gamma} \frac{XP^2}{y} + 16 e^{2\gamma+2\mu}\nu^{-1}
         \frac{X(X^2-1)}{y^5}.
\end{align}
From the numerical point of view, the problem of solving these equations
is virtually identical to that of a static string in Minkowski spacetime.
The only difference is the much higher degree of complexity
of the equations due to the appearance of
$\nu$, $\tau$, $\mu$ and $\gamma$
as extra variables. We will discuss the numerical implementation of the
boundary conditions at the origin and at infinity in the next
section when we consider the case of a dynamic cosmic string. The
boundary conditions are given by
equations (\ref{bound_dynin}), (\ref{bound_dynout}). In the static case
we replace the conditions for $N$, $T$ and $M$ in (\ref{bound_dynin}) by
\begin{align}
  \begin{split}
  \nu = 1, \\[5pt]
  \tau = 0, \\[5pt]
  \mu = 0,
  \end{split}
  \label{stat_bound}
\end{align}
but otherwise use the same boundary conditions.
The solution is then obtained using the
relaxation method described in the previous section.
As our initial guess for the metric
variables we use Minkowskian values, and for the string variables $X$
and $P$ we use the previously calculated values for a Minkowskian string with
the same string parameters. Due to the appearance of $\tau$ or its derivatives
in all terms of (\ref{taueq}) the Geroch potential will
\begin{table}[t]
  \caption{Convergence test for the static cosmic string in curved
           space-time for $\alpha = 1$. The norm of the deviation
           $\ell_2[\Delta \Psi^K]$ is defined by Eq.\,(\ref{CAUCHYNORM}).
           As the grid resolution is increased, the deviation from the
           high resolution result decreases quadratically to a good
           approximation (see text for details).}
  \begin{center}
  \label{convStat}
    \begin{tabular}{l|ccccc}
      \hline \hline
      & $\nu$ & $\mu$ & $\gamma$ & P $ X $ \\
      \hline\\[-7pt]
      $\ell_2(\Delta \Psi^{1200})$ & $1.28\cdot 10^{-7}$ & $2.51\cdot 10^{-6}$
                        & $2.39\cdot 10^{-6}$ & $5.95\cdot 10^{-7}$
                        & $4.16\cdot 10^{-7}$ \\
      \hline\\[-7pt]
      $\ell_2(\Delta\Psi^{150})/\ell_2(\Delta\Psi^{300})$
                        & 3.56 & 3.59 & 3.58 & 4.04 & 3.37 \\
      $\ell_2(\Delta\Psi^{300})/\ell_2(\Delta\Psi^{600})$
                        & 3.76 & 3.79 & 3.78 & 4.19 & 3.60 \\
      $\ell_2(\Delta\Psi^{600})/\ell_2(\Delta\Psi^{1200})$
                        & 4.58 & 4.61 & 4.60 & 4.98 & 4.44 \\
      \hline \hline
    \end{tabular}
  \end{center}
\end{table}
stay zero in the relaxation
process and our solution has no rotation.\\
We have checked the code for convergence in the way described in section
\ref{SECcsmink}. We have chosen the unphysically large value
$\eta = 0.2$ here in order to guarantee convergence even for strong
coupling between matter and geometry. $\alpha$ is set to 1
as in the Minkowski case. The results are given in Table \ref{convStat}.
For convenience we only display the results for the fundamental variables
$\nu$, $\mu$, $\gamma$, $P$ and $X$. Since we do not incorporate rotation,
the result for $\tau$ is, as expected, exactly 0 and we do
not include it in Table \ref{convStat}.
Again the code is shown to be second order convergent.
\begin{figure}[b]
  \centering
  \epsfig{file=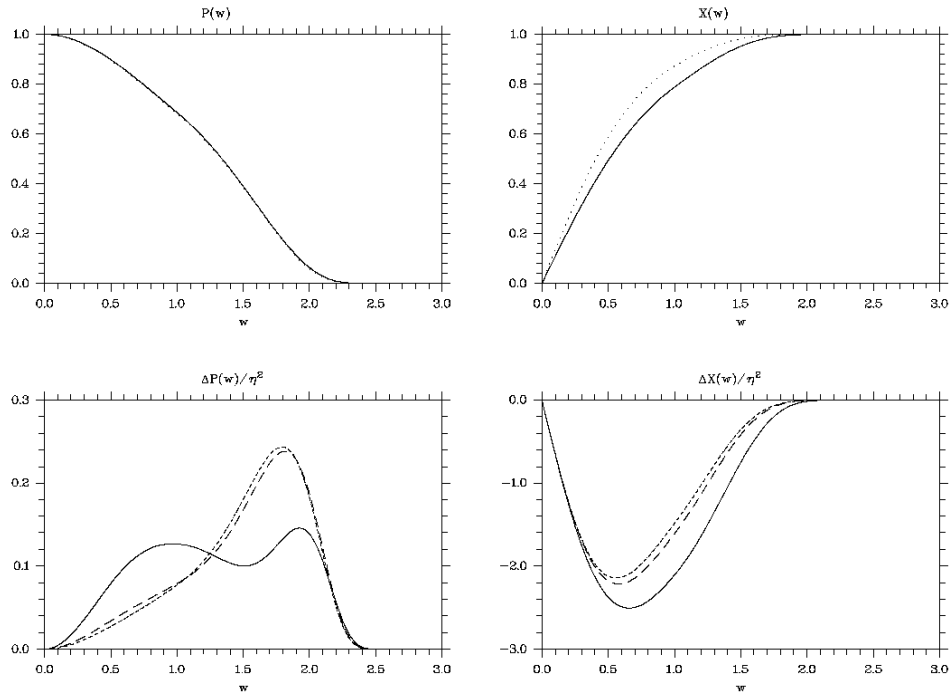, height=400pt, width=283pt, angle=-90}
  \caption{a) In the upper two panels we plot the string variables for
              $\eta =$ 0.001 (dotted) and 0.2 (solid) as a function of
              the radial variable $w$.
           b) In the lower panels we have plotted the deviation from the
              Minkowskian values rescaled by $\eta^2$ for $\eta = $
              0.001 (dotted), 0.01 (dashed), 0.1 (long dashed) and 0.2
              (solid). Note that the curves for 0.001 and 0.01 almost exactly
              coincide, which indicates the validity of the linear regime.
              For larger $\eta$, however, the deviation shows a more
              complicated behaviour.}
\label{stat_string}
\end{figure}
\begin{figure}[t]
  \centering
  \epsfig{file=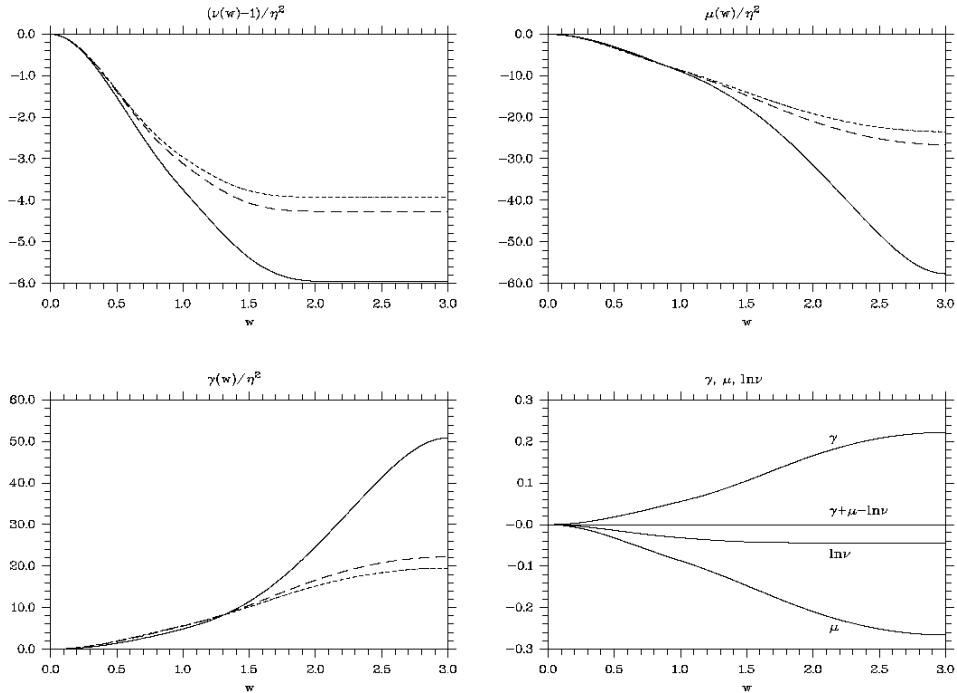, height=400pt, width=283pt, angle=-90}
  \caption{The deviation of the metric variables $\nu$, $\mu$ and
           $\gamma$ from Minkowskian values
           rescaled by $\eta^2$ is plotted as a function of $w$ 
           for $\eta=$ 0.001 (dotted), 0.01 (dashed), 0.1 (long dashed) and
           0.2 (solid). The dotted and the dashed curves almost exactly
           coincide indicating the linear regime. As in the case of the
           string variables we find a more complicated dependence for
           $\eta \ge 0.1$. In the lower right panel we plot $\gamma$
           $\mu$, $\ln{\nu}$ and their sum for $\eta=0.1$
           which vanishes in accordance with Eq.\,(\ref{gmn})
           to high accuracy.}
\label{stat_metric}
\end{figure}
In Fig.\,\ref{stat_string} and \ref{stat_metric} we plot the results obtained
for $N=2400$ grid points. In all these plots the relative coupling strength
is $\alpha = 1$, but qualitatively similar results are obtained for different
values of $\alpha$. We have already mentioned that the
% order of magnitude of the
effect of the string on the spacetime geometry is determined by
$\eta$. Therefore
we have compared the deviation of both the string variables and the metric
from the Minkowskian case for $\eta = $ 0.001, 0.01, 0.1 and 0.2.
In Fig.\,\ref{stat_string} we plot the string variables $P$ and $X$ for the
two extreme values and the deviation from the Minkowskian string
rescaled by $\eta^2$ for all four values. For small $\eta$ we
see that $\Delta P/\eta^2$ and $\Delta X/\eta^2$ is essentially independent of
$\eta$. In this case the deviation from Minkowskian values
can be treated as a small perturbation and
a linear dependence of $\Delta P$ and $\Delta X$ on $\eta^2$ is to be expected.
In the range $\eta =0.1 \ldots 0.2$ on the other hand, 
we clearly leave the linear regime
and the deviation depends on $\eta$ in a much more complicated way. 
These values, however, are 2 orders of magnitude larger than the value
$10^{-3}$ predicted
in current GUT theories (\citeNP{Vilenkin1994}). 
The deviation of the
metric variables $\nu$, $\mu$ and $\gamma$ is plotted in the first three
panels of Fig.\,\ref{stat_metric}. Again we see the
linear behaviour for small $\eta$ and the transition to the non-linear
regime at $\eta \approx 0.1$. In the fourth panel of Fig.\,\ref{stat_metric}
we check Eq.\,(\ref{gmn}) for $\eta = 0.1$. We clearly see that
$\gamma + \mu + \ln{\nu}$ is approximately zero. Indeed (\ref{gmn}) is
satisfied to within $\approx 10^{-8}$ as compared with the order of
magnitude of the individual terms $10^{-1}$. \\

%=======================================================================
\subsubsection{The dynamic cosmic string}
\label{dynstring}
In the dynamic case all variables $\nu$, $\tau$, $\mu$, $\gamma$, $P$ and $X$
are functions of $u,r$ and we have to solve the system
(\ref{nuur})-(\ref{xur}) of partial differential equations.
In order to
control the behaviour of the solution at infinity, we need a
generalisation for PDEs of the relaxation scheme applied to ordinary
differential equations. In view of the characteristic feature of the
relaxation scheme, namely the simultaneous calculation of new function values
at all grid points, this generalisation leads directly to implicit
evolution schemes as used for hyperbolic or parabolic PDEs.
Therefore, the dynamic code is based on the implicit,
second order in space and time Crank-Nicholson scheme described in 
section \ref{FDE_CN}. For this purpose we rewrite the
dynamic equations (\ref{nuur})-(\ref{xur}) as a first order system.
These equations involve radial
derivatives which may be written in terms of second order operators
exactly as in the static case (\ref{nurr})-(\ref{xrr}). This 
naturally leads to the auxiliary quantities introduced in
Eqs.\,(\ref{nur})-(\ref{xr}). In terms of these variables the equations for
the dynamic cosmic string become
\begin{align}
   \nu_{,r}  =& \frac{N}{r}, \label{dyn_nur} \\[10pt]
   \tau_{,r} =& \frac{T}{r}, \\[10pt]
   \mu_{,r}  =& \frac{M}{r^2}, \\[10pt]
   P_{,r}    =& rQ,          \\[10pt]
   X_{,r}    =& \frac{R}{r}, \\[10pt]
   \begin{split}
   2N_{,u}   =& N_{,r} + \frac{T^2-N^2}{r\nu} + \frac{NM}{r^2}
                + 2 \frac{\nu_{,u} N - \tau_{,u} T}{\nu} -\nu_{,u}
                - \frac{\nu_{,u} M}{r} - N\mu_{,u} \\[10pt]
             & + 8\pi\eta^2 \left[ 2e^{2(\gamma + \mu)} r(X^2-1)^2
                + \frac{1}{\alpha} e^{-2\mu} \nu^2(2 P_{,u}Q - rQ^2)
                \right],
   \end{split}  \\[10pt]
   2T_{,u}   =& T_{,r} - 2\frac{TN}{r \nu} + 2\frac{\tau_{,u} N+\nu_{,u}T}{\nu}
                + \frac{TM}{r^2} - \tau_{,u} - \frac{\tau_{,u} M}{r}
                -T\mu_{,u}, \\[10pt]
   2M_{,u}   =& M_{,r} + \frac{M^2}{r^2} - 2\mu_{,u} M - 2 r \mu_{,u}
                + 8\pi\eta^2 \left[ e^{2\gamma} X^2P^2 +2
                \frac{e^{2(\gamma+\mu)}}{\nu} r^2 (X^2-1)^2 \right], \\[10pt]
   2(r+M)\gamma_{,r} =& M_{,r} - 2\frac{M}{r} - \frac{M^2}{r^2}
                + \frac{T^2+N^2}{2\nu^2} + 8\pi\eta^2 \left[ R^2
                + \frac{1}{\alpha} e^{-2\mu}\nu r^2Q^2 \right],
                \label{1gammar} \\[10pt]
   2Q_{,u}   =& Q_{,r} - \frac{QM}{r^2} + Q \mu_{,u} - \frac{Q\nu_{,u}}{\nu}
                - \frac{P_{,u} N}{r^2\nu} + \frac{QN}{r\nu} + \frac{P_{,u}}{r^2}
                + \frac{P_{,u} M}{r^3} -\alpha \frac{e^{2(\gamma+\mu)}}{\nu}
                \frac{PX^2}{r}, \\[10pt]
   2R_{,u}   =& R_{,r} - X_{,u} - \frac{X_{,u} M}{r} + \frac{RM}{r^2}
                - R\mu_{,u} - 4 \frac{e^{2(\gamma+\mu)}}{\nu} r X(X^2-1)
                - e^{2\gamma} \frac{XP^2}{r} \label{Rur}.
\end{align}
The corresponding first order system in the outer region is given by
\begin{align}
  \nu_{,y}  =& -2 \frac{N}{y}, \label{nuy}  \\[10pt]
  \tau_{,y} =& -2 \frac{T}{y}, \\[10pt]
  \mu_{,y}  =& -2 yM,          \\[10pt]
  P_{,y}    =& -2 \frac{Q}{y^5}, \\[10pt]
  X_{,y}    =& -2 \frac{R}{y}, \\[10pt]
  \begin{split}
  2N_{,u}   =& -\frac{1}{2}y^3 \left( N_{,y} -2yNM -2\frac{T^2-N^2}{y\nu}
               \right) -y^2 \nu_{,u} M
                +2\frac{\nu_{,u} N - \tau_{,u} T}{\nu}
               \\[10pt]
            & - N\mu_{,u} - \nu_{,u} +8\pi\eta^2 \left[2e^{2(\gamma + \mu)}
               \frac{(X^2-1)^2}{y^2} +\frac{1}{\alpha} e^{-2\mu} \nu^2
               \left(  2P_{,u} Q - \frac{Q^2}{y^2} \right) \right],
  \end{split} \\[10pt]
  2T_{,u}   =& -\frac{1}{2}y^3 \hspace{-0.025cm}
               \left( T_{,y} -2y TM+4\frac{TN}{y\nu}\right) \hspace{-0.025cm}
               - T\mu_{,u} -y^2 \tau_{,u} M
               + 2\frac{\tau_{,u} N + \nu_{,u} T}{\nu} - \tau_{,u}, \\[10pt]
  \begin{split}
  2M_{,u}   =& -\frac{1}{2} y^3 (M_{,y} -2yM^2) - 2\frac{\mu_{,u}}{y^2}
               -2\mu_{,u} M \\[10pt]
             & + 8\pi\eta^2 \left[ e^{2\gamma} X^2P^2 
               + 2 \frac{e^{2(\gamma+\mu)}}{\nu} \frac{(X^2-1)^2}{y^4} \right],
  \end{split}  \\[10pt]
  2(y^2M+1) \gamma_{,y} =& \,\,\, y^2M_{,y} + 4yM +2y^3M^2
               - \frac{N^2+T^2}{y\nu^2}
               -16\pi\eta^2 \left[\frac{R^2}{y} + \frac{1}{\alpha} e^{-2\mu}
               \nu \frac{Q^2}{y^5} \right], \\[10pt]
  \begin{split}
  2Q_{,u}   =& -\frac{1}{2}y^3 \left( Q_{,y} +2yQM -2\frac{QN}{y\nu} \right)
               +y^4P_{,u} - y^4\frac{P_{,u} N}{\nu} - \frac{\nu_{,u} Q}{\nu}
               \\[10pt]
             & +y^6 P_{,u} M + Q\mu_{,u}
               -\alpha e^{2(\gamma + \mu)} \nu^{-1} y^2PX^2,
  \end{split} \\[10pt]
  \begin{split}
  2R_{,u}   =& -\frac{1}{2}y^3 (R_{,y} -2yRM) -X_{,u} -R\mu_{,u}
               - y^2X_{,u} M - e^{2\gamma} y^2 XP^2 \\[10pt]
             & - 4e^{2(\gamma+\mu)}\nu^{-1} \frac{X(X^2-1)}{y^2}.
               \label{Ruy}
  \end{split}
\end{align}
The derivation of these equations and a number of other calculations
in this work have been carried out with the algebraic computing package
GRTensorII (\citeNP{Musgrave1996}).
In order to solve these equations we must supplement them by
appropriate initial and boundary conditions. We have already mentioned the
boundary conditions on the axis (\ref{bound_nutr})-(\ref{bound_xtr}). 
In general we find that the dynamic code performs better
if one imposes boundary conditions on the radial derivatives
rather than the variables themselves. For the variables $\nu$, $\mu$, $\tau$,
$P$ and $X$ we therefore impose the required
boundary conditions on the initial data according to (\ref{mink_bound}) and
(\ref{stat_bound}). In the subsequent
evolution we impose the weaker condition that the radial derivatives 
$N$, $T$ and $R$ are finite on the axis. This ensures that the evolution
equations propagate the axial conditions given on the initial
data. For the variable $\mu$ we impose the condition that $M$ is zero
on the axis which is equivalent to the rather weak condition that
$r^2\mu_r$  vanishes there. The inverse power of $r$ in the
definition of $Q$ makes it unsuitable to specify the value of this
quantity at $r=0$ so in this case we work with the variable directly
and require that $P=1$ on the axis.
Finally the variable $\gamma$ is given by a purely radial equation and in
this case we specify the value on the axis where $\gamma$ vanishes
by virtue of Eq.\,(\ref{gammatr}). Therefore at $r=0$ we require
\begin{align}
  \begin{split}
  N &= 0, \\
  T &= 0, \\
  M &= 0, \\
  \gamma &= 0, \\
  P &= 1, \\
  R &= 0.
  \end{split}
  \label{bound_dynin}
\end{align}
For the boundary conditions at null infinity we know that regular
solutions of the cylindrical wave equation have radial derivatives that decay
faster than $1/r$ so that
we may take the variables $N$, $T$, and $R$, which satisfy a wave
type equation, to vanish at $y=0$. The asymptotics of $\mu$ are
slightly different due to the additional power of $r$ in the radial
derivative (similar to the spherically symmetric wave equation) but for a
regular solution $\mu_{,y}$ vanishes at null infinity.
The equation for $P$ does not satisfy a wave type equation due to the
inverse power of $r$ but has
asymptotic behaviour given by a modified Bessel function. The
physically relevant finite solution has exponential decay so in this
case one may impose the condition that $Q=0$ at $y=0$. Hence we
require  the solution to satisfy the following boundary conditions at $y=0$
\begin{align}
  \begin{split}
  N &= 0, \\
  T &= 0, \\
  \mu_{,y} &= 0, \\
  Q &= 0, \\
  R &= 0.
  \end{split}
  \label{bound_dynout}
\end{align}
These boundary conditions are sufficient to determine the solution of
the first order system (\ref{dyn_nur})-(\ref{Ruy}) while 
suppressing the unphysical
solutions which are singular on the axis or null infinity. Note that
$\gamma$ is determined by the constraint equation
(\ref{gammar}), which is a first order ODE, and thus only needs one
boundary condition. \\
We finally note that all variables are related at the interface
in the form $f_{K_1+1}=f_K$ analogous to Eqs.\,(\ref{int_P})-(\ref{int_R})
in the case of a static string in Minkowski spacetime.

%=======================================================================
\subsection{Testing the dynamic code}
\label{CStest}
%
%
%%
%\begin{figure}[t]
%  \centering
%  \epsfig{file=check_ww_delta.ps, height=400pt, width=150pt, angle=-90}
%  \caption{The deviation of the numerical $\nu$ and $\gamma$ from the
%           Weber-Wheeler solution as a function of $u$ and $w$ %(\ref{w})
%           obtained for 1920 grid points ($n_1=320$,
%           $n_2 = 1600$). The wave parameters are $a=2$, $b=0.5$.
%           Note that the error is amplified by $10^5$ and $10^7$ respectively.}
%\label{plot_ww_delta}
%\end{figure}
%
In this section we will describe four independent tests of the implicit code
for the dynamic cosmic string, namely
\begin{list}{\rm{(\arabic{count})}}{\usecounter{count}
             \labelwidth1cm \leftmargin1.5cm \labelsep0.4cm \rightmargin1cm
             \parsep0.5ex plus0.2ex minus0.1ex \itemsep0ex plus0.2ex}
\item reproducing the non-rotating vacuum solution of \lcite{Weber1957},
\item reproducing the rotating vacuum solution of \scite{Xanthopoulos1986},
\item using the results for the static cosmic string 
      as initial data and verifying that the system stays in its static
      configuration,
\item convergence analysis for the string hit by a Weber-Wheeler wave.
\end{list}
Two additional tests arise in a natural way from the field equations and
the numerical scheme. As described above there are two additional
field equations which are consequences of the other equations.
We have verified that these equations are satisfied to second order accuracy
($\sim \Delta r^2$). Furthermore the numerical scheme calculates the residuals
of the algebraic equations to be solved, which have thus been monitored
in test runs. They are satisfied to a much higher accuracy
(double precision
machine accuracy), so the total error is dominated by the truncation error
of the second order differencing scheme. Another independent test is the
comparison with the explicit CCM code which yields good
agreement as long as the latter remains stable.
The four main tests are now described in more detail.

%========================================================================
\subsubsection{The Weber-Wheeler wave}
\begin{figure}[t]
  \centering
  \epsfig{file=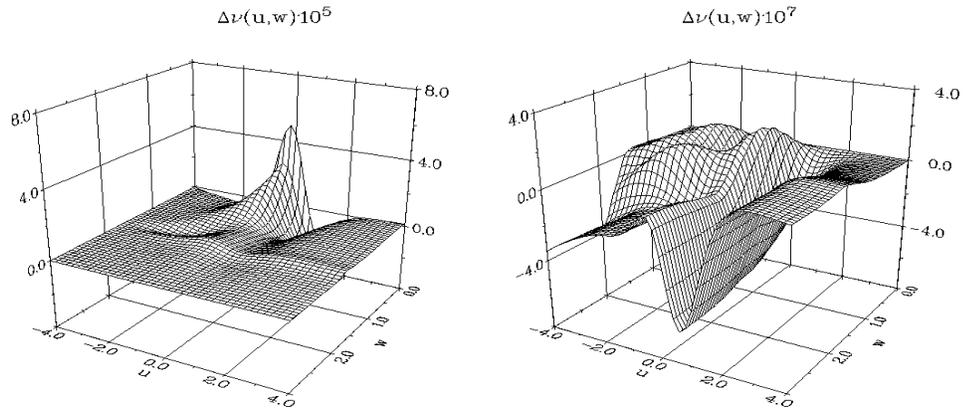, height=400pt, width=200pt, angle=-90}
  \caption{The deviation of the numerical $\nu$ and $\gamma$ from the
           Weber-Wheeler solution as a function of $u$ and $w$ %(\ref{w})
           obtained for 1920 grid points ($K_1=320$,
           $K_2 = 1600$). The wave parameters are $a=2$, $b=0.5$.
           Note that the error is amplified by $10^5$ and $10^7$ respectively.}
\label{plot_ww_delta}
\end{figure}
%
%%
%\begin{figure}[t]
%  \centering
%  \epsfig{file=check_as_delta1.ps, height=400pt, width=150pt, angle=-90}
%  \epsfig{file=check_as_delta2.ps, height=200pt, width=150pt, angle=-90}
%  \caption{The deviation of the numerical $\nu$, $\tau$, $\gamma$ from
%           Xanthopoulos' analytic solution as a function of $u$ and $w$
%           obtained for 1920 grid points
%           ($n_1=320$, $n_2 = 1600$).
%           Note that the error is amplified by $10^5$ and $10^6$
%           respectively.}
%\label{plot_as_delta}
%\end{figure}
%%
%%
%\begin{figure}[t]
%  \centering
%  \epsfig{file=es_evol_nm.ps, height=400pt, width=150pt, angle=-90}
%  \epsfig{file=es_evol_gp.ps, height=400pt, width=150pt, angle=-90}
%  \epsfig{file=es_evol_s.ps, height=200pt, width=150pt, angle=-90}
%  \caption{The deviation of the metric and matter variables from the initial
%           data in the case of evolving a static cosmic string with
%           $\alpha=1$, $\eta=0.2$. For our standard grid with
%           $1920$ points, the
%           configuration stays static to an accuracy of $\approx 10^{-5}$      
%           over a range of more than 30000 time steps.}
%  \label{evolstat}
%\end{figure}
%%
In the first test we evolve the analytic solution given by \lcite{Weber1957},
which describes a gravitational pulse of the ``+'' polarisation mode. We have
given the analytic expressions in section \ref{CCM_ww} in terms of $t$, $r$ 
[Eqs.\,(\ref{ww_X})-(\ref{ww_gamma})] and in terms of $u$, $y$
[Eqs.\,(\ref{ww_yX})-(\ref{ww_ygamma})]. 
The corresponding equations in characteristic
coordinates $u$, $r$ in the inner region 
are easily obtained from the coordinate transformation $t=u+r$. 
We prescribe $\nu$ as initial data according
to the analytic expressions obtained for $a=2$ and $b=0.5$ and set the other
free variables to zero, while $\gamma$ is calculated
via quadrature from the constraint equation (\ref{gammar}).
In Fig.\,\ref{plot_ww_delta} we show
the deviation of the numerical results from the analytic one for
$K=1920$ grid points (320 points in the inner, 1600
points in the outer region) and a Courant factor of 0.5 with respect to
the inner region. The convergence analysis (see below)
shows that this number of points provides sufficient resolution in the outer
region while still keeping computation times at a tolerable level.
All computations presented in the remainder of section \ref{cstr} have 
been obtained with these grid parameters, unless stated otherwise.
The code stays stable for much longer time intervals than shown in the figure,
but does not reveal any further interesting features as the analytic solution
approaches its Minkowskian values and the error goes to zero.

%======================================
\subsubsection{The rotating solution of Xanthopoulos}
\begin{figure}[t]
  \centering
  \epsfig{file=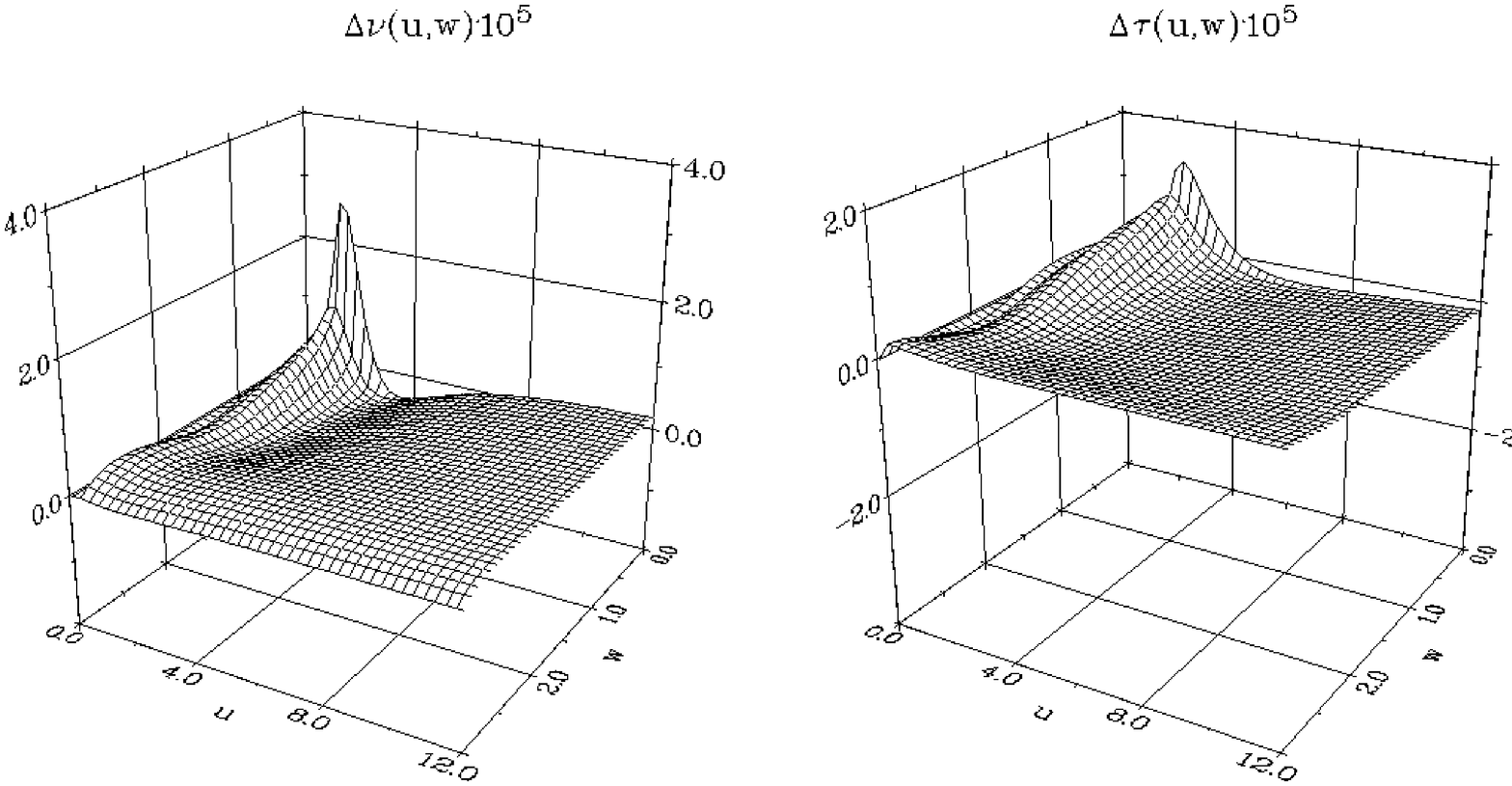, height=400pt, width=175pt, angle=-90}
  \epsfig{file=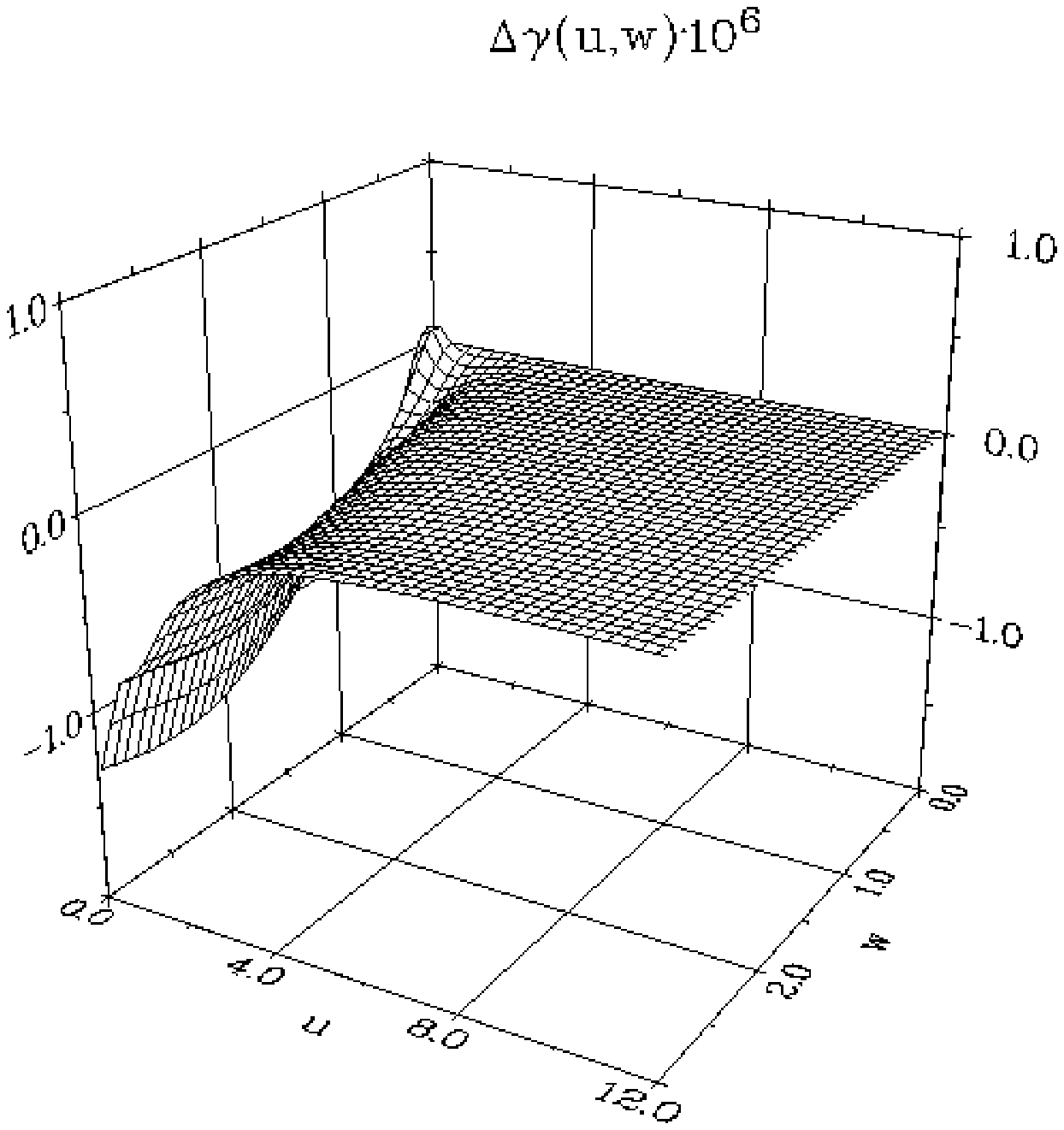, height=200pt, width=175pt, angle=-90}
  \caption{The deviation of the numerical $\nu$, $\tau$, $\gamma$ from
           Xanthopoulos' analytic solution as a function of $u$ and $w$
           obtained for 1920 grid points
           ($K_1=320$, $K_2 = 1600$).
           Note that the error is amplified by $10^5$ and $10^6$
           respectively.}
\label{plot_as_delta}
\end{figure}
\scite{Xanthopoulos1986} has derived an analytic vacuum solution for Einstein's
field equations in cylindrical symmetry containing both the ``+'' and
``$\times$'' polarisation mode. Its analytic form in terms of
our metric variables is given by Eqs.\,(\ref{as_Q})-(\ref{as_ygamma}) 
in section \ref{as}. Again
the transformation to coordinates $u$, $r$ in the inner region 
results straightforwardly from $t=u+r$. The solution has one free parameter
$a$ set to one in this calculation.
The error of our numerical results is displayed in Fig.\,\ref{plot_as_delta},
where we have used the same grid parameters as in the Weber-Wheeler case.
Again we have run
the code for longer times and found that the error approaches zero.
We conclude that the code reproduces both analytic vacuum solutions
with excellent accuracy comparable to that of the CCM code
and exhibits long term stability.

%========================================
\subsubsection{Evolution of the static cosmic string}
\begin{figure}[t]
  \centering
  \epsfig{file=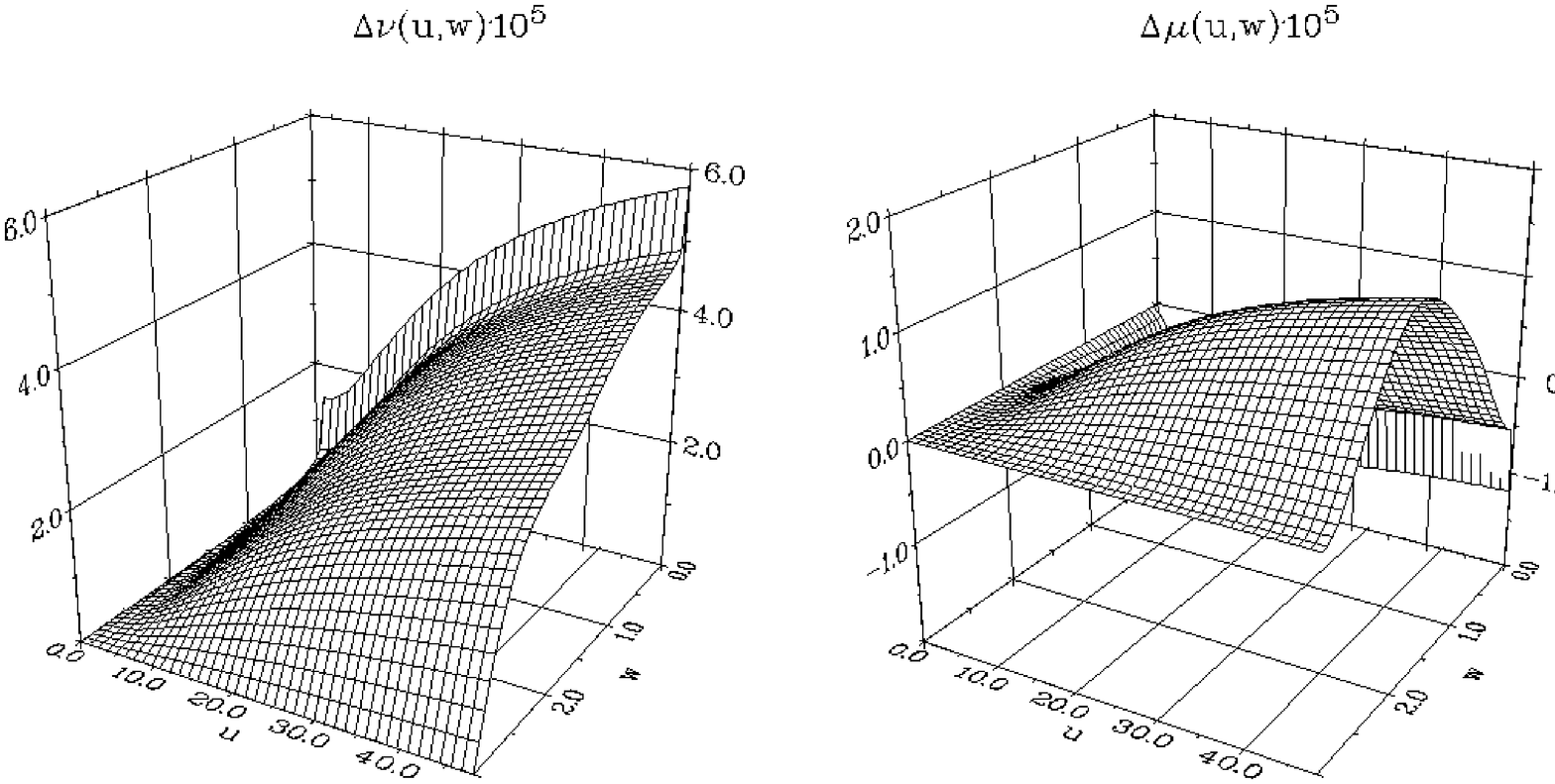, height=400pt, width=175pt, angle=-90}
  \epsfig{file=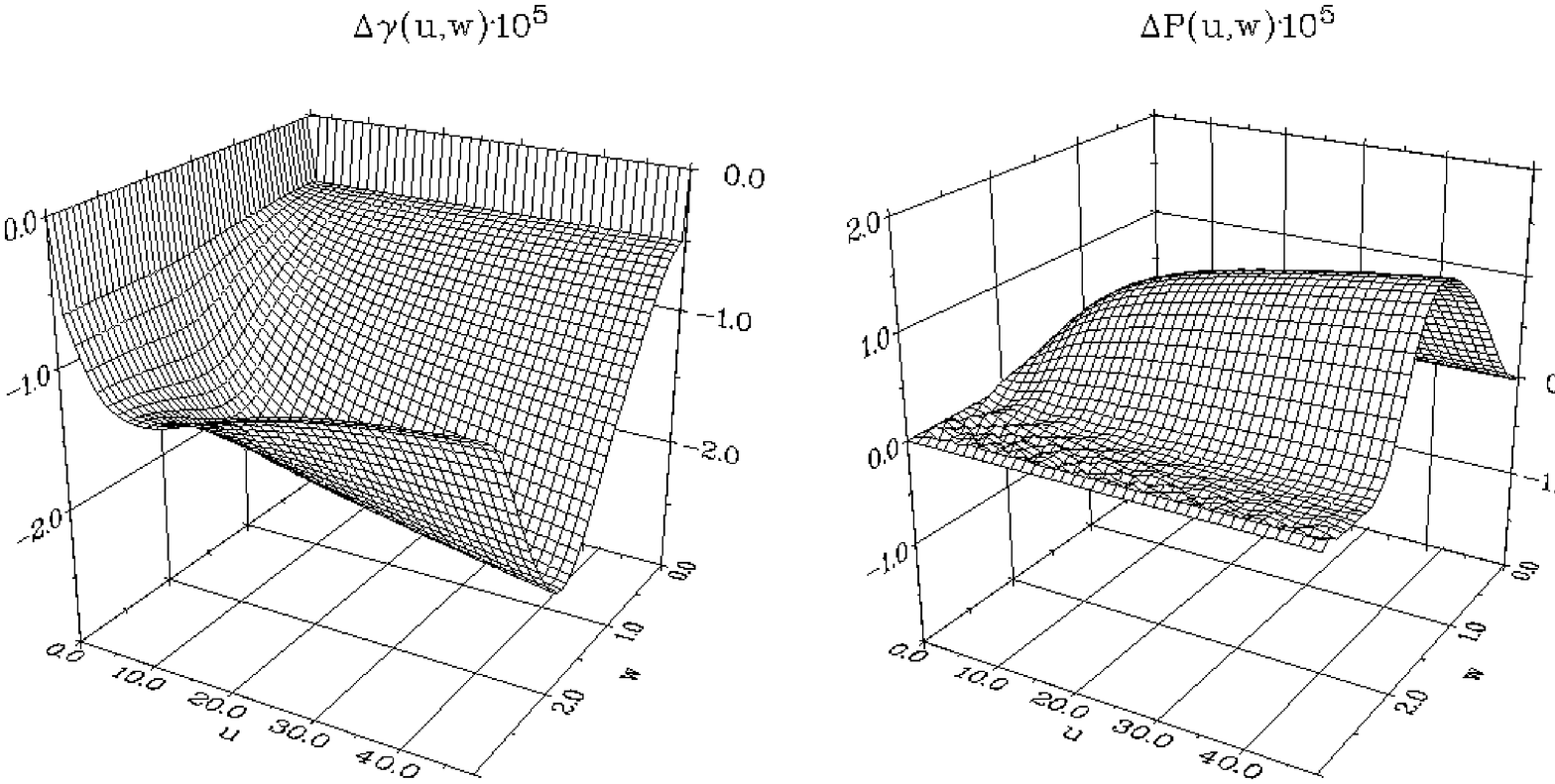, height=400pt, width=175pt, angle=-90}
  \epsfig{file=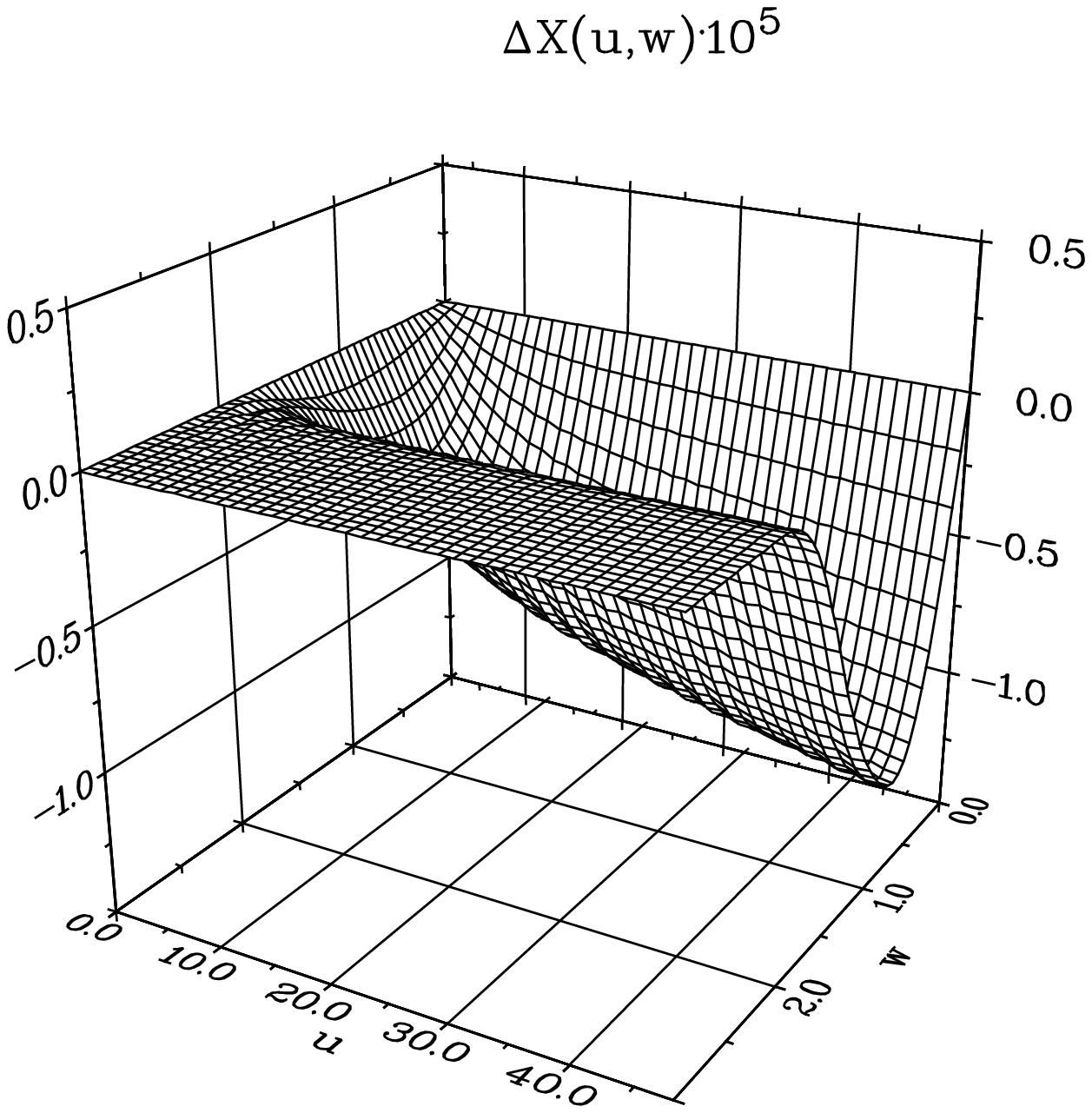, height=200pt, width=175pt, angle=-90}
  \caption{The deviation of the metric and matter variables from the initial
           data in the case of evolving a static cosmic string with
           $\alpha=1$, $\eta=0.2$. For our standard grid with
           $1920$ points, the
           configuration stays static to an accuracy of $\approx 10^{-5}$      
           over a range of more than 30000 time steps.}
  \label{evolstat}
\end{figure}
The tests described above only involve vacuum solutions, so the matter
part of the
code and the interaction between matter and geometry has not been tested.
An obvious test involving matter and geometry is to use the result for the
static cosmic string in curved spacetime as initial
data and evolve this scenario. All variables should, of course, remain
at their initial values. We have evolved the static string data for
our standard grid and the parameter set, $\alpha = 1$ and $\eta=0.2$,
which corresponds to a strong back-reaction of the string on the
metric. The results are shown in Fig.\,\ref{evolstat}. The system stays
in its static configuration with high accuracy over a time interval
which is significantly longer than the dynamically relevant
timescale of the vacuum solutions discussed above.

%=======================================
\subsubsection{Convergence analysis}
\begin{figure}[t]
  \centering
  \epsfig{file=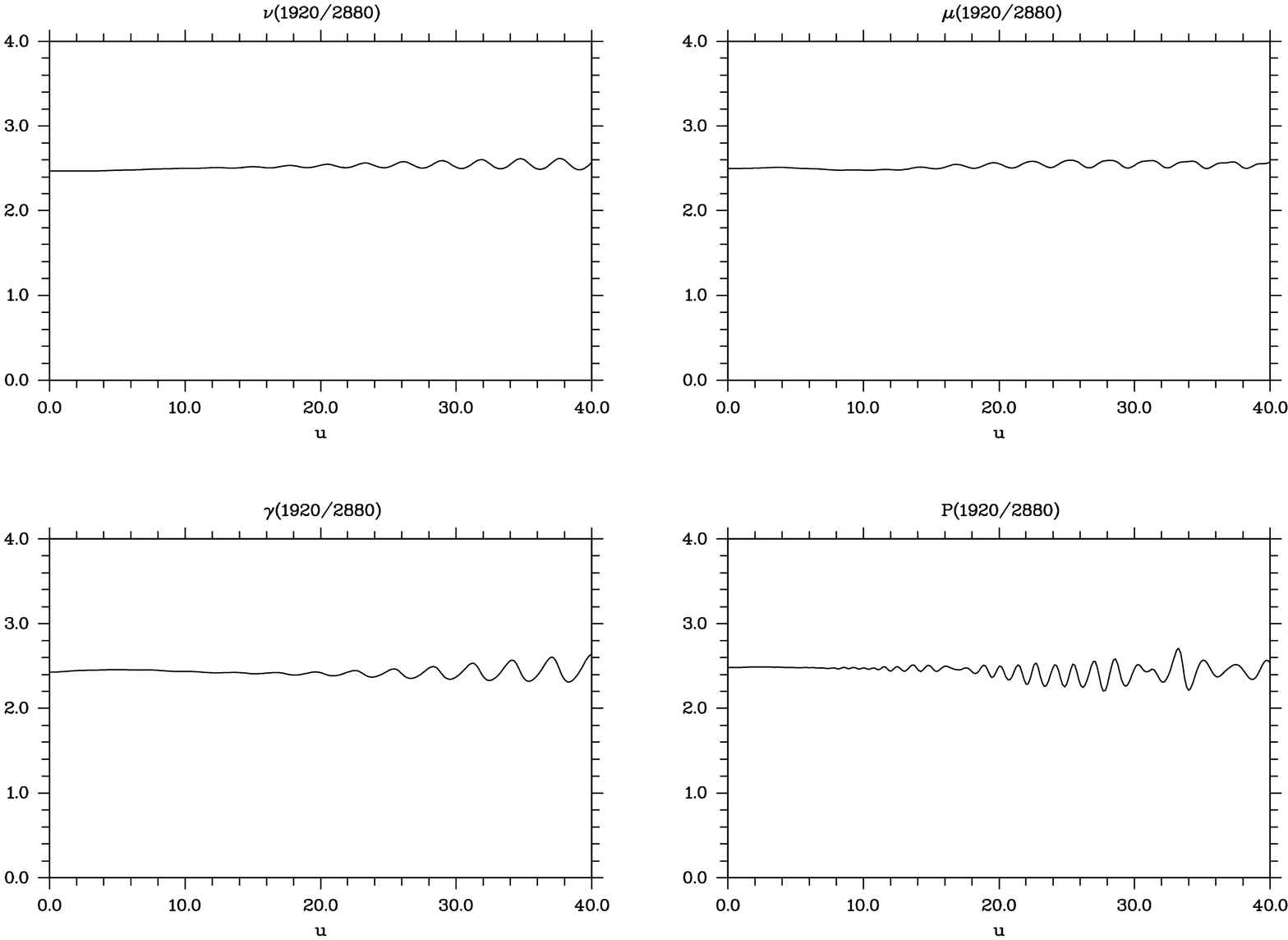, height=400pt, width=283pt, angle=-90}
  \epsfig{file=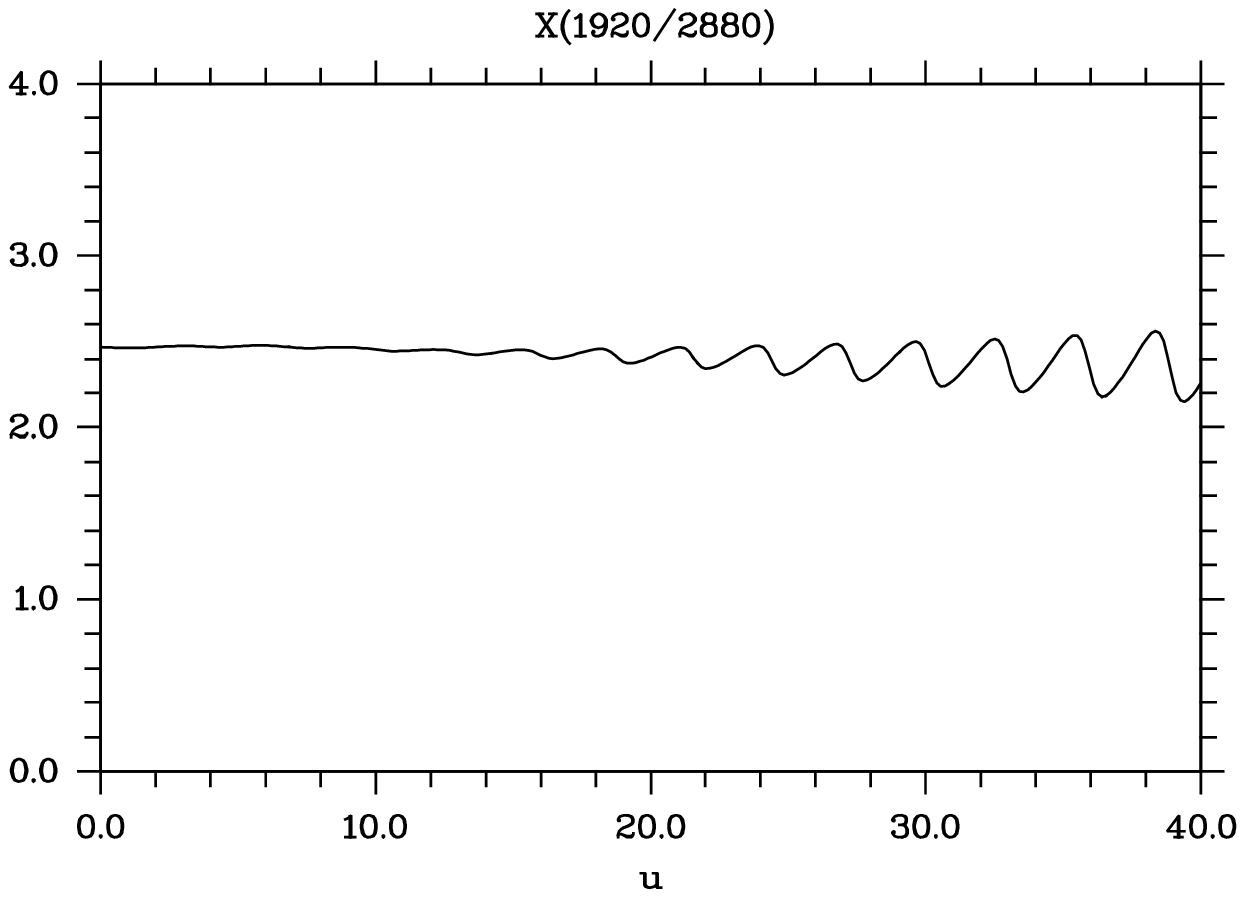, height=200pt, width=141pt, angle=-90}
  \caption{The convergence factor $\ell_2[\Psi^{1920}] / \ell_2[\Psi^{2880}]$
           is plotted as a function of $u$.
           We expect a convergence factor of 2.25 since the number of
           grid points is multiplied by 1.5.
           Even though our results show weak variability at later times,
           second order convergence is maintained
           throughout long runs (more than 30000 time steps with $K=1920$).}
\label{cs_conv}
\end{figure}
Our investigation of the interaction between the cosmic string and gravitational
waves will focus on the string being hit by a wave of the Weber-Wheeler type.
In order to check this scenario for convergence we have run
the code for the parameter set $\eta=0.2$, $\alpha=1$, $a=2$, $b=0.5$
for different grid resolutions, where $a$ and $b$
are again the width and amplitude
of the Weber-Wheeler wave. In our case it is of particular interest to
investigate the time dependence of the convergence to see what resolution we
need in order to obtain reliable results for long runs.
We calculate the convergence rate again according to equation (\ref{l2norm}). 
The high
resolution reference solution has been calculated for $K=4320$ grid points.
In Fig.\,\ref{cs_conv} we show the convergence factor
$\ell_2[\Psi^{1920}]/\ell_2[\Psi^{2880}]$ as a function of $u$
for $\nu$, $\mu$, $\gamma$, $P$ and $X$. The initial data for $\tau$ is
identically zero for this scenario and stays zero during the evolution.
The number of grid points is increased by a factor of 1.5 here (instead of the
more commonly used 2) to reduce the computation time.
Only points common to all grids have been used in the
sum in equation (\ref{l2norm}). For second order convergence we would expect
a convergence factor of $1.5^2$. Although the results in
Fig.\,\ref{cs_conv}
show weak variations with $u$, second order convergence is clearly maintained
for long runs.
In each case the outer region contains 5 times as many grid points as
the inner region (e.g. $K_1=320$, $K_2=1600$ for the $K=1920$ case). The
reason for this is that in the dynamic evolutions
$X$ and especially $P$ exhibit significant spatial variations
out to large radii. Due to the compactification, the spatial resolution of
our grid decreases towards null infinity and in order
to resolve the spatial
variations of the string variables out to sufficiently large radii we
therefore have
to introduce a large number of grid points in the outer region.
No such problems occur in the inner region. If significantly
fewer grid points are used in the outer region for this analysis, the
convergence properties of the string variables can deteriorate
to roughly first order level.

%=========================================================================
\subsection{Time dependence of the string variables}
\label{CSevol}
\subsubsection{Static cosmic strings excited by gravitational waves}
\begin{figure}[t]
  \centering
  \epsfig{file=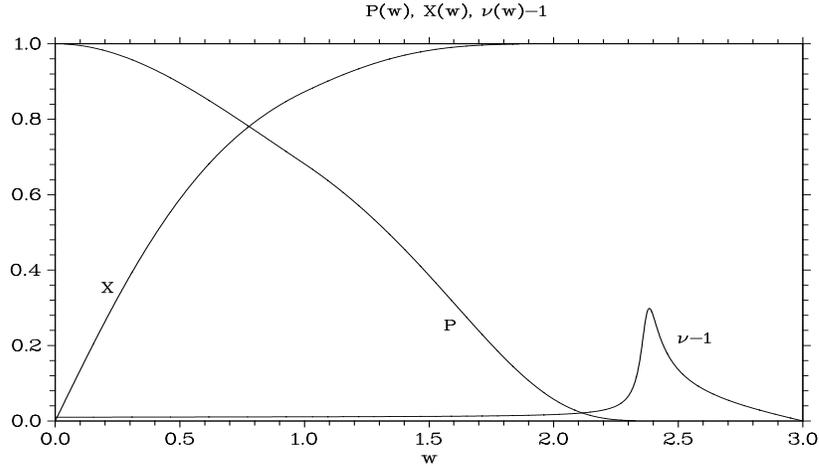, height=350pt, width=200pt, angle=-90}
  \caption{The initial data for $\nu$, $P$ and $X$ at $u_0=-20$
           for the standard parameters $\alpha=1$, $\eta=0.001$,
           $a=2$, $b=0.5$.
           The gravitational wave pulse is located in a region
           where the string fields $P$ and $X$ have almost fallen off
           to their asymptotic values.}
\label{ini_nps0001}
\end{figure}
\begin{figure}[t]
  \centering
  \epsfig{file=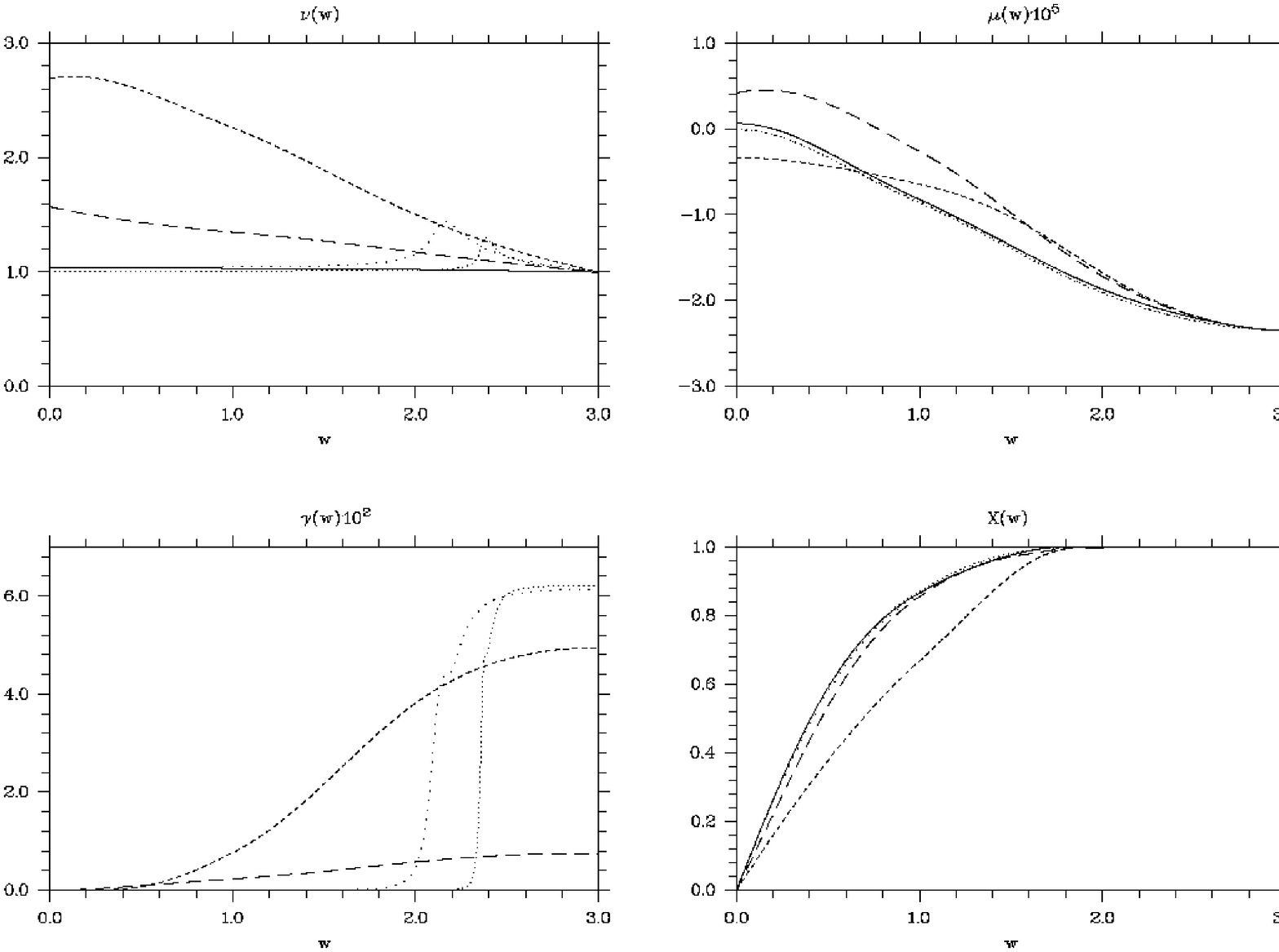, height=450pt, width=300pt, angle=-90}
  \epsfig{file=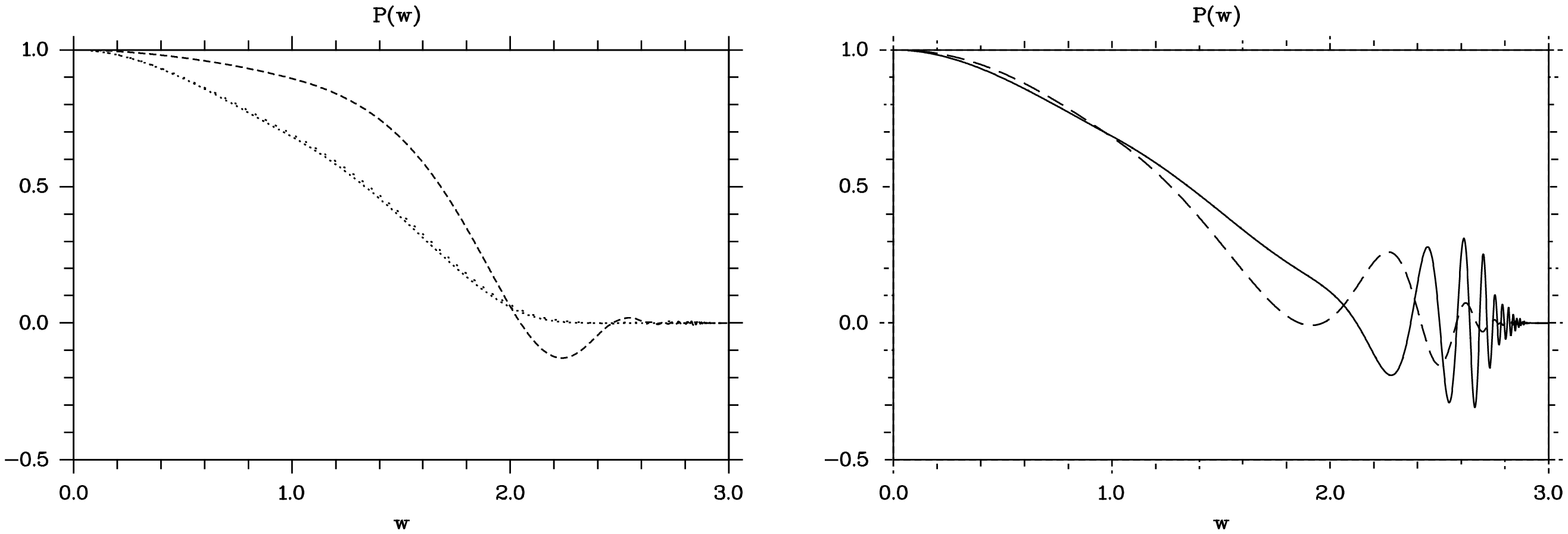, height=450pt, width=150pt, angle=-90}
  \caption{The metric and string variables are plotted as functions
           of $w$ at $u=-20$ (dotted), $u=-10$ (long dotted),
           $u=0$ (dashed), $u=2$ (long dashed) and $u=10$ (solid line).
           For clarity the graphs of $P$ are distributed over two panels.
           The wave pulse (in $\nu$) initially moves inwards. It excites
           the string, is reflected at the origin and travels outwards.
           After $u=10$ only $P$ differs significantly from the
           static configuration as the oscillations
           slowly decay and propagate towards larger
           radii (cf. Fig.\,\ref{ringing}).}
\label{mplot}
\end{figure}
\begin{figure}[t]
  \centering
  \epsfig{file=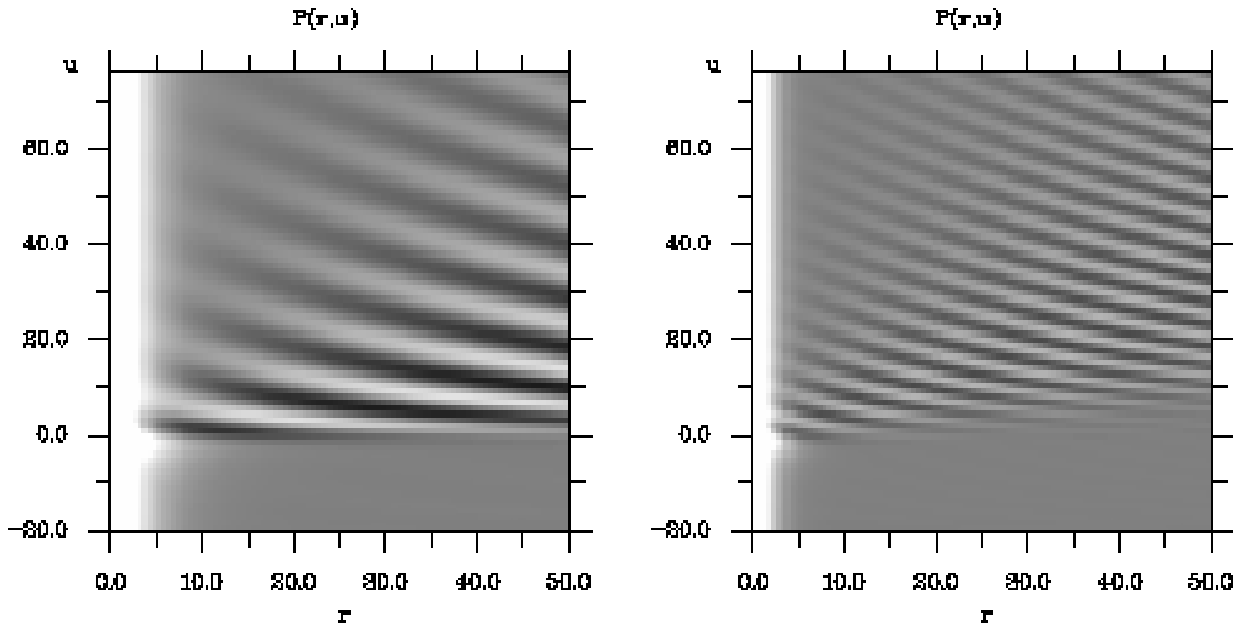, height=450pt, width=200pt, angle=-90}
  \caption{The cosmic string variable $P$ is shown as a function
           of radius and time for $\alpha = 0.2$ (left) and
           $\alpha = 1$ (right) (all other parameters have standard
           values). Note that we use the radial
           variable $r$ out to $r=50$ here. The ringing can clearly
           be seen and shows a lower frequency for smaller $\alpha$.}
\label{ringing}
\end{figure}
The scenario we are going to investigate in this section is an initially
static cosmic string hit by a gravitational wave of
Weber-Wheeler type. For this purpose we use the static results with
two modifications as initial data.
Firstly the static metric function $\nu_0$ is multiplied
by the exact Weber-Wheeler solution to simulate the gravitational wave
pulse. Thus we guarantee that initially the cosmic string is indeed
in an equilibrium configuration provided the wave pulse is located
sufficiently far away from the origin and its interaction with the string
is negligible.
%If the pulse is located sufficiently far away from the origin
%its interaction with the cosmic string is negligible and the initial
%string configuration is indeed static.
Ideally the numerical calculation would start with the incoming wave located
at past null infinity. In order to approximate this scenario, we
found it was sufficient to use the large negative initial time $u_0=-20$.
The second modification is to calculate $\gamma$ from the constraint
equation (\ref{gammar}) to preserve consistency with the Einstein field
equations.
In Fig.\,\ref{ini_nps0001}
the corresponding initial data for $\nu$, $P$ and $X$ are shown for
the parameter set $\eta=0.001$, $\alpha=1$, $a=2$ and $b=0.5$.
From now on we will refer to these values as ``standard parameters''
and only specify parameters if they take on non-standard values.
%Note that $\tau$ vanishes on the initial slice in this case and stays
%identically zero throughout the evolution. 
% The case of rotating gravitational
% waves hitting a cosmic string will be analysed in a future publication.
The time evolution of the ``standard configuration'' is shown in
Fig.\,\ref{mplot} where we plot $\nu$, $\mu$, $\gamma$, $P$ and $X$
as functions of $w$ at times $-20$, $-10$, $0$, $2$ and $10$. While the wave
pulse is still far away from the origin, its interaction with the cosmic
string is negligible (dotted lines).
When it reaches the core region it excites
the cosmic string and the scalar and vector field start oscillating (dashed
curves).
After being reflected at the origin, the wave pulse travels along the
outgoing characteristics and the metric variables
$\nu$, $\mu$ and $\gamma$ quickly settle down into their static configuration
which is close to Minkowskian values for $\eta = 10^{-3}$.
The vector and scalar field of the cosmic string, on the other hand,
continue ringing albeit with a different character. Whereas the
oscillations of the scalar field $X$ are dominant in the range $r\le 2$
and have significantly decayed at $u=10$ as shown in the figure,
the vector field oscillations propagate to large radii and fall off very
slowly (solid curves). This behaviour is also
illustrated in the right panel of Fig.\,\ref{ringing} which shows
a contour plot of $P$ as a function of $(u,r)$ out to $r=50$.
We shall see below, that the oscillations of
$P$ will also decay eventually and the
cosmic string will asymptotically settle back into its equilibrium
configuration.

%=====================================
\subsubsection{Frequency analysis}
\begin{figure}[t]
  \centering
  \epsfig{file=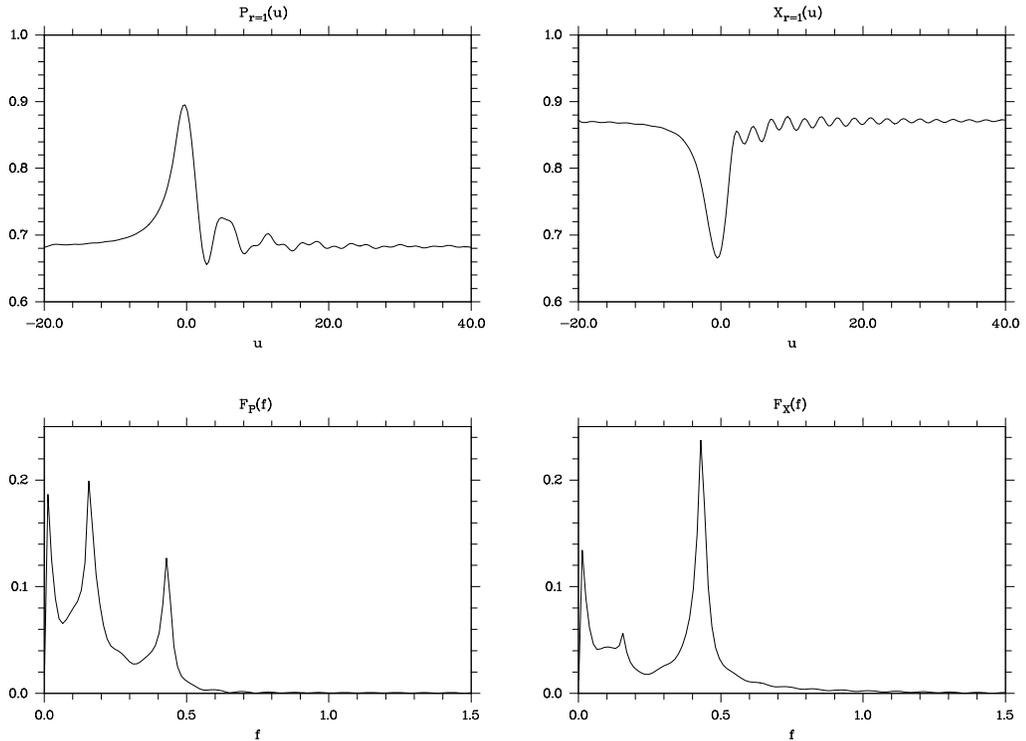, height=400pt, width=283pt, angle=-90}
  \caption{Upper panels: The variables $P$ and $X$ at $r=1$ are plotted
           as functions of $u$ for $\alpha=1$, $\eta=0.001$, $a=2$ and $b=0.5$.
           Lower panels: The corresponding power
           spectra show three peaks each. That near $f=0$ is merely due
           to a constant offset and the variation of the fields on
           long time scales and thus not associated with the oscillations.
           From the linear equations one can infer that the peaks
           at $f=0.45$ can be identified with the oscillation of the
           scalar field, the peaks at $f=0.16$ with those of the vector
           field. Note that due to our rescaling of the coordinates,
           $u$ is measured in units of $1/\sqrt{\lambda} \eta$.}
\label{fourier}
\end{figure}

\begin{figure}[t]
  \centering
  \epsfig{file=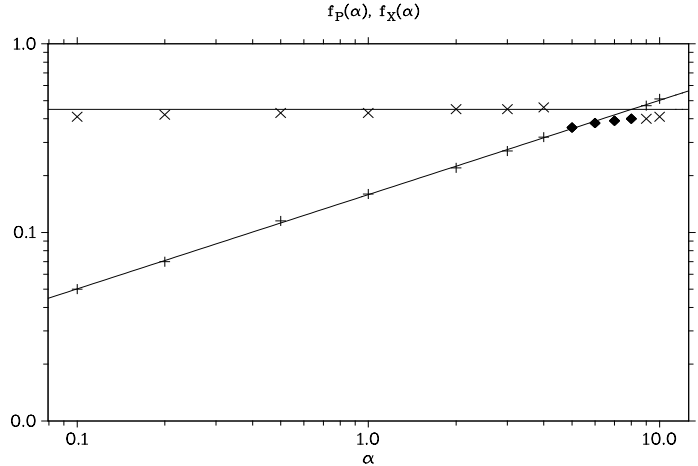, height=300pt, width=200pt, angle=-90}
  \caption{The frequencies obtained from the Fourier analysis of the
           oscillations of the scalar and vector fields
           are plotted as functions of $\alpha$.
           The curves show the frequencies of the scalar and vector field
           predicted by an analysis of the linearised equations.
           For $5 \le \alpha \le 8$ the predicted values for
           $P$ and $X$ are similar and we find only one frequency.
           These values are plotted as filled lozenges.}
\label{f_alpha}
\end{figure}

We will now quantitatively analyse the oscillations of the scalar and
vector field of the cosmic string. Since we are working in rescaled
coordinates, physical time and distance are
measured in units of $1/\sqrt{\lambda} \eta$ and
frequency in its inverse. To avoid complicated notation, however,
we will omit the units from now on unless the meaning is unclear.
In order to measure frequencies, we Fourier analyse the time evolution
of the scalar and vector field for fixed radius $r$. Fig.\,\ref{fourier}
shows $P$ and $X$ for standard parameters
as functions of time at $r=1$ together with the corresponding
power spectra. In each Fourier spectrum we can see three peaks. Those very
close to $f=0$ are merely due to the offset of the data and the gradual
change of the fields over long times. We therefore do not attribute these
peaks to the oscillations of the fields and will not consider them in the
ensuing discussion. We have calculated similar
power spectra for a large class of parameter sets and in most cases
found two peaks at non-zero frequencies. In order to interprete
the frequencies, it is convenient to plot them as functions of the
relative coupling strength $\alpha$. The result is shown in
Fig.\,\ref{f_alpha}. In this figure the solid lines show the frequency values
calculated for the scalar and vector field from the linearised equations
(see \citeNP{Sjodin2001})
\begin{align}
  f_X &= \frac{\sqrt{2}}{\pi}, \label{CSTR_LINFX} \\
  f_P &= \frac{\sqrt{\alpha}}{2\pi}. \label{CSTR_LINFP}
\end{align}
We can thus associate the
constant frequency $f=0.45$ with the scalar field $X$ and the
$\alpha$ dependent frequency with the vector field $P$.
We will refer to these frequencies as $f_X$ and $f_P$ from now on.
The $\alpha$-dependency of $f_P$ is also illustrated in Fig.\,\ref{ringing}
where we show contour plots of $P$ obtained for $\alpha=0.2$ and $\alpha=1$.
The oscillation frequency is significantly larger for $\alpha=1$. \\
In Fig.\,\ref{f_alpha} we can see that the
frequencies associated with the scalar
and vector field become similar near $\alpha=8$. For this value
it can be shown that the masses associated with the scalar and vector field
become equal (see for example \citeNP{Sjodin2001}). The frequencies are
difficult to resolve in these cases and we have only found one peak in the
Fourier spectra. The resulting values are shown as filled lozenges in
the figure. In this context it is worth mentioning that
the accuracy of the measurements of the
frequencies is limited by the resolution of the Fourier spectra
which again is limited by the time interval covered in the evolution.
In Fig.\,\ref{fourier} we can see however, that the oscillations of
both $P$ and $X$ gradually die out in later stages of the evolutions,
so that it becomes increasingly difficult to extract more information
about the frequencies by using larger integration times.
The evolutions used for this analysis provide an
accuracy $\Delta f \approx 0.01$ which corresponds approximately to one bin
in the frequency spectra. \\
It is interesting
to see that in the non-linear evolution the distinction between
the oscillations of the vector and the scalar field is not as clear
as in the linear case which is demonstrated by the presence of
two peaks in the Fourier spectra.
We attribute this feature to the interaction between the scalar and vector
component of the cosmic string. Concerning the radial dependence
of the spectra we have in general found that the
characteristic mode of $X$ resulted in stronger peaks at smaller radius,
that of $P$ was stronger at larger radii.
This variation of the relative strength of
the oscillations with radius
confirms the corresponding observation in Fig.\,\ref{mplot}.
In order to investigate the dependency of the oscillations on
$\eta$, $a$, $b$ and the radial position $r$, we have varied each
parameter over at least two orders of magnitude while keeping the other
parameters at standard values. We have found the following dependencies:
\begin{list}{\rm{(\arabic{count})}}{\usecounter{count}
             \labelwidth1cm \leftmargin1.5cm \labelsep0.4cm \rightmargin1cm
             \parsep0.5ex plus0.2ex minus0.1ex \itemsep0ex plus0.2ex}
\item The frequencies of both $X$ and $P$ do not show any variations
      with $\eta$ for $\eta < 0.1$. (Note that $\eta$ does, however,
      appear in the units). For larger values of $\eta$, the non-linear
      interaction between string and geometry becomes dominant and we did
      not detect a simple relation between frequency maxima and parameters.
\item The variation of the parameters $a$ and $b$,
      the width and amplitude of the Weber-Wheeler pulse, has
      no measurable effect on the frequencies of $X$ and $P$, but
      only determined the amplitude of the oscillations. A narrow, strong pulse
      leads to larger amplitudes.
\item For small $r$ the oscillations in $X$ are stronger, whereas
      those for $P$ dominate at large $r$. The frequency values, however,
      do not depend on the radius. For radii greater than 10
      the oscillations
      in $X$ had decayed so strongly that we could no longer measure its
      frequency.
\end{list}
We have also checked the empirical relation between the coupling constant
$\alpha$ and the frequencies $f_X$ and $f_P$. For this purpose we
have performed a linear regression analysis of the double logarithmic
data of Fig.\,\ref{f_alpha} excluding the cases where only one
frequency is observed. We obtain power law indices $\sigma_X = 0.00$
and $\sigma_P = 0.50$, so that
\begin{align}
  f_X & \sim {\rm const}, \label{fX}\\
  f_P & \sim \sqrt{\alpha}, \label{fP}
\end{align}
which agrees with Eqs.\,(\ref{CSTR_LINFX}), (\ref{CSTR_LINFP}).
If we transform this result back
into  physical units using $\alpha=e^2/\lambda$, we arrive at the following
relations for the physical variables
\begin{align}
  f_X & \sim \sqrt{\lambda} \eta, \\[10pt]
  f_P & \sim e \eta.
\end{align}
As shown in \lcite{Shellard1994} up to constant factors 
$\sqrt{\lambda} \eta$ and $e \eta$ are the masses
of the scalar and the vector field, $m_X$ and $m_P$, so that
$X$ and $P$ have characteristic frequencies
\begin{align}
  f_X & \sim m_X, \\[10pt]
  f_P & \sim m_P.
\end{align}
Since the frequencies for $X$ and $P$ seem only to depend upon the
respective masses we have attempted to confirm these results by
considering the oscillations of a cosmic string in two further scenarios.
Firstly since the frequencies do not depend upon the shape of the Weber-Wheeler
\begin{figure}[b]
  \centering
  \epsfig{file=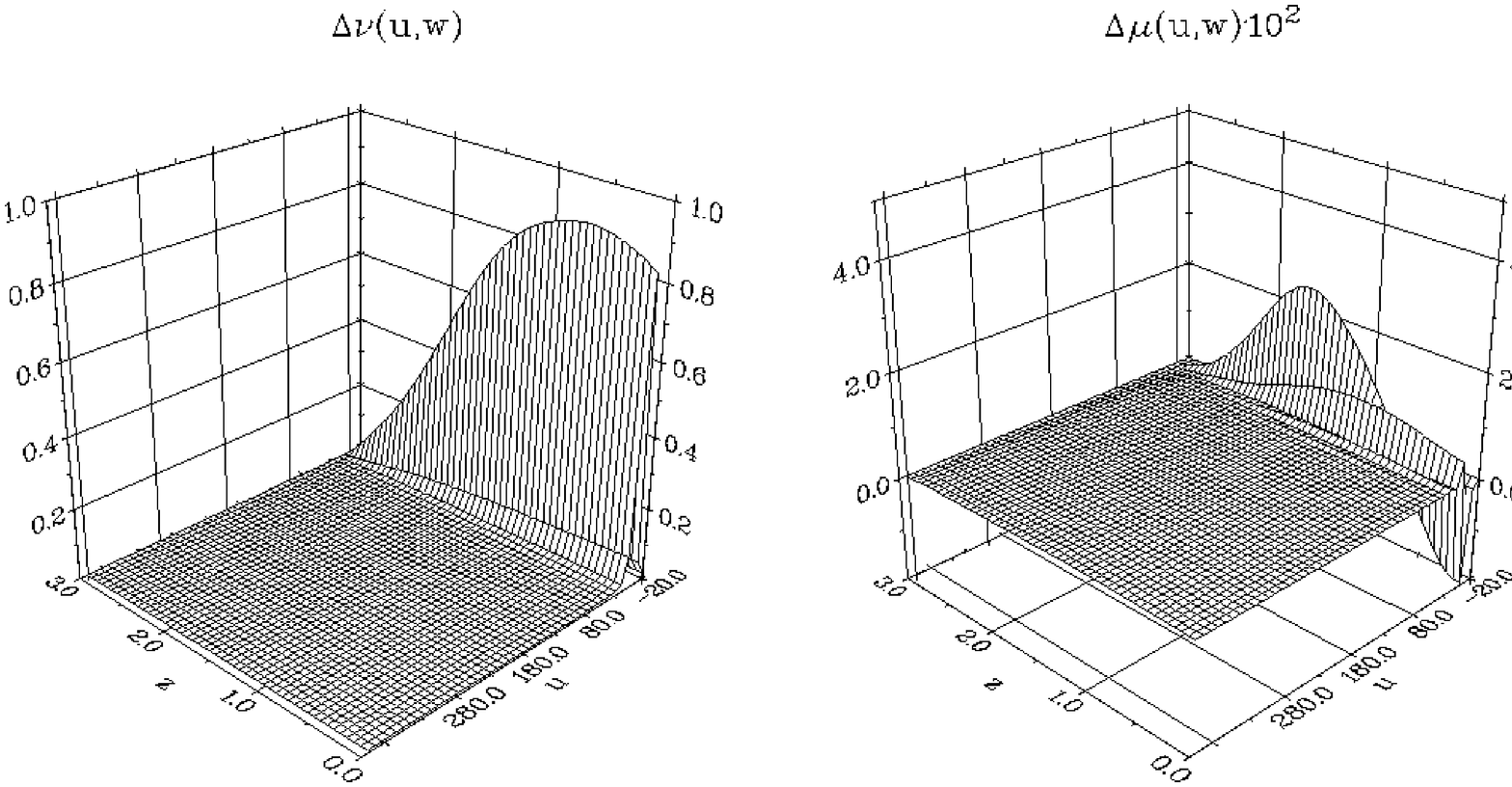, height=400pt, width=175pt, angle=-90}
  \epsfig{file=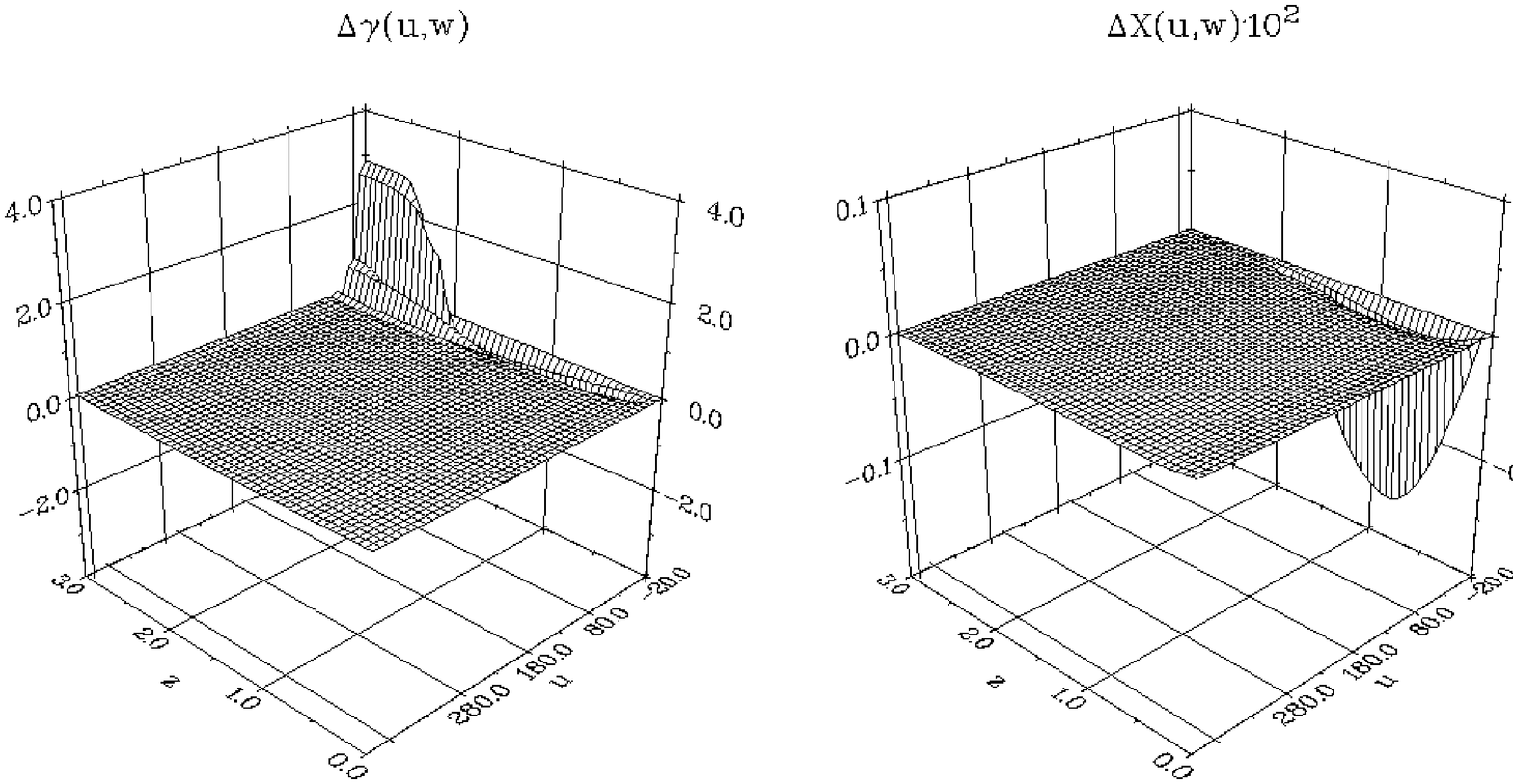, height=400pt, width=175pt, angle=-90}
  \epsfig{file=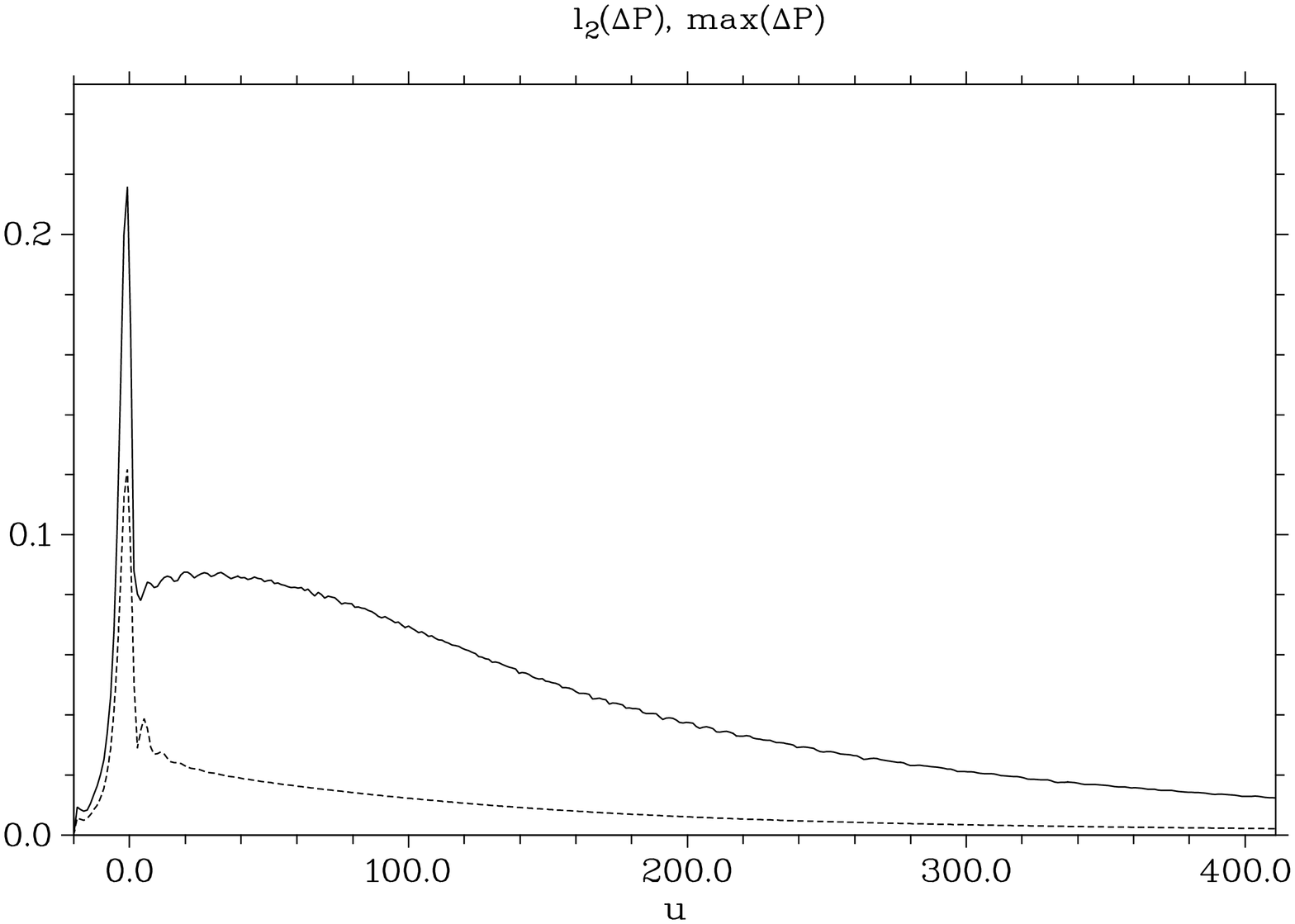, height=250pt, width=150pt, angle=-90}

  \vspace{0.3cm}
  \caption{The upper four plots show the difference between the evolved
           functions $\nu$, $\mu$, $\gamma$ and $X$ and their corresponding
           static results. For $P$ a similar 3-dimensional plot is not suitable
           since it fails to resolve the oscillations of the vector field.
           Therefore we plot the $\ell_2$-norm (dashed line)
           and the maximum (solid line) of $\Delta P$ as
           a function of time. $\nu$, $\mu$, $\gamma$ and $X$ quickly settle
           down in their equilibrium configuration to numerical accuracy.
           The decay of the oscillations of $P$ takes significantly more
           time but eventually $P$ also approaches its equilibrium state.}
\label{settle}
\end{figure}
pulse we take as initial data the static values for the metric
variables but excite the string by adding a Gaussian perturbation to
either the $X$ or $P$ static initial values. The evolution is then
computed using the fully coupled system. Secondly since the
frequencies do not seem to depend upon the strength of the coupling to
the gravitational field we have completely decoupled the gravitational
field and considered the evolution of a cosmic string in Minkowski
spacetime. The initial data is taken to be that for a static string in
Minkowski spacetime with a Gaussian perturbation to either the $X$ or
$P$ values. The evolution is then computed using the equations for a
dynamical string in a Minkowskian background [equations (\ref{pur}) and 
(\ref{xur}) with the metric variables set to Minkowskian values]. 
In both cases we obtain the same frequencies, to within an
amount $\Delta f = 0.01$, that we find in the original case of the
fully coupled system excited by a Weber-Wheeler pulse. Furthermore the
frequencies did not depend on the location or shape of the field perturbation
nor upon the choice of $X$ or $P$ as the perturbed field.

%========================================================================
\subsubsection{The long term behaviour of the dynamic string}
The time evolution shown in Figs.\,\ref{mplot} and \ref{ringing} indicates
that the oscillations of the cosmic string excited by gravitational waves
gradually decay and metric and string settle down into an equilibrium state.
In order to investigate the long term behaviour in detail
we have evolved the variables for a much longer time
($-20\le u \le 410$) than in the numerical evolutions discussed above.
The unphysically large value of $\eta = 0.1$ is chosen for this
calculation in order to guarantee a
strong interaction between spacetime geometry and the cosmic string.
In Fig.\,\ref{settle} we show the difference $\Delta f :=
f_{\rm evol} - f_{\rm stat}$
between the time-dependent $\nu$, $\mu$, $\gamma$ and $X$ and their
corresponding static results obtained for the same parameters.
For the vector field $P$ a similar 3-dimensional plot would require
an extreme resolution to properly display the oscillations of the vector
field (cf.\ Fig.\,\ref{mplot}). For this reason we proceed differently
and calculate the $\ell_2$-norm and the maximum of $\Delta P$ for each
slice $u={\rm const}$. Both functions
are plotted versus time in Fig.\,\ref{settle}.
The incoming wave pulse can clearly be seen as a strong deviation of $\nu$
from the static function. The pulse excites the cosmic
string and is reflected at the origin at $u=0$. The metric variables and
the scalar field $X$ then quickly reach their equilibrium values. The
oscillations in $P$ decay on a significantly longer time scale
which is also evident in Figs.\,\ref{mplot} and \ref{ringing}
and the $\ell_2$-norm of $\Delta P$ slowly approaches 0.
Significantly longer runs than shown here are prohibited by the required
computation time, but the results indicate that $P$ will also eventually
reach its equilibrium configuration.

%========================================================================
\subsection{Discussion}
In the previous two sections we have studied numerical problems
in cylindrical symmetry with particular emphasis on the use of
characteristic methods and the compactification of spacetime. \\
This work has completed the 1-dimensional stage of the
Southampton Cauchy-characteristic matching project by presenting
for the first time a long-term stable second order convergent code
for general cylindrically symmetric vacuum spacetimes with both the
$+$ and $\times$ polarisation mode. In order to obtain long-term
stability it was crucial to formulate the problem in a way that
simplifies the relations at the interface where information
is transferred between the interior Cauchy and the exterior
characteristic region. In this particular case we achieved the
simplification by applying the Geroch decomposition to both regions
which contrasts with the less successful previous attempts where
the Killing direction was factored out in the exterior region only.
In view of the numerical subtleties involved with the interpolation
techniques at the interface the importance of a suitable
choice of variables may not be too surprising. Nevertheless we stress
the significance of this result concerning Cauchy-characteristic
matching codes in higher dimensions. The structure of the null geodesics
will inevitably become much more complicated if the restriction of
cylindrical symmetry is dropped and the physical variables
are allowed to depend on the angular coordinates. Correspondingly the
transformation between variables at the interface will also be more
complicated. In view of our results it seems important to carefully
choose the variables describing the spacetime in both regions and
aim for ``simple'' transformation laws. \\
The inclusion of matter in the form of cosmic strings resulted in
qualitatively new numerical problems that finally were solved by the
use of specially adapted numerical methods. The incorporation of
null infinity proved to be important here for the specification
of outer boundary conditions on the matter variables. It was the existence
of unphysical exponentially diverging solutions of the equations
for a cosmic string that required a special numerical treatment.
We were able to suppress the unphysical diverging solutions by
solving the equations for a cosmic string with
a relaxation scheme in the static case and an
implicit evolution scheme in the dynamic case. We have thus been able
to develop the first fully non-linear simulation of a static and
a dynamic cosmic string
coupled to gravity which implements the exact boundary conditions
at both the origin and infinity. The resulting codes have been used
to study the interaction between a cosmic string and a gravitational wave
pulse. We have found that the gravitational wave pulse excites the
cosmic string which then starts oscillating with frequencies proportional
to the particle masses associated with the scalar and vector field.
The same frequencies have been observed if we excite the cosmic
string with a Gaussian perturbation to the scalar or vector field. \\
From a numerical point of view an interesting result of the numerical
solution of the equations for a dynamic cosmic string
concerns the transfer of information at the interface. We
have illustrated this in Fig.\,\ref{grid} where two grid points
$K_1$, $K_1+1$ have been used for the spatial position $r=1$. The grid
point $K_1$ contains the variables of the interior region at $r=1$, whereas
the variables of the exterior region are specified at the same position on grid
point $K_1+1$. The corresponding implementation of the interface
is remarkably simple as illustrated by
Eqs.\,(\ref{int_P})-(\ref{int_R}) which represent the interface for the
static cosmic string in Minkowski spacetime. The corresponding
equations in the dynamic case coupled to gravity are equally trivial.
Even if different variables are used in the interior and exterior
region, one is still able to transform the variables locally at
the grid points $K_1$ and $K_1+1$ and there is no need to use
interpolation techniques as in the case of the explicit numerical methods
used in section \ref{ccm}. We attribute the possibility of using this
simple implementation of the interface to the fact that all
function values are calculated simultaneously on the new time slice
in an implicit scheme,
so that there is no hierarchical order according to which the
function values have to be calculated. We have probed such a ``local
interface'' in an implicit Cauchy-characteristic matching code
for cylindrically symmetric, non-rotating vacuum spacetimes and
achieved a long term stable evolution with an accuracy comparable to the
explicit code described in section \ref{ccm}. Even though an interface
based on interpolation performs satisfactorily in cylindrical symmetry
this may no longer be the case in higher dimensional problems where
the interpolation techniques will be significantly more complicated.
On the other hand we can see no obvious reason why a
``local interface'' in
combination with an implicit numerical scheme should differ significantly
from that used in the 1-dimensional case.

\newpage
%=========================================================================
\section{Non-linear oscillations of spherically symmetric stars}
\label{pert}
%
%
%=========================================================================
\subsection{Introduction}
In this section we will turn our attention towards the study of compact
stars in the framework of general relativity. The discovery of stars
significantly more compact than the sun goes back to observations of the
binary star Sirius in the middle of the 19th century.
Sirius is the brightest star in
the sky as viewed from the earth. From an astrophysical point of view,
however, the faint companion of the bright main star, Sirius B has provoked
much more interest. The astronomer Bessel was the first to
infer the existence of an unseen companion of Sirius from a wobble in
the motion of the main star. It took another twenty years
before Alvin Clark managed to optically resolve Sirius B.
By the early twentieth century it became clear from the analysis of
its electromagnetic spectrum that Sirius B has a rather high surface
temperature of about 25,000$\,K$. In view of this result the extraordinarily
low luminosity of Sirius B lead to the conclusion that the star is very
small, about the size of the earth. This type of
high density star was consequently named a {\em white dwarf}.
It was understood at the time that white dwarfs mark the final stage in the
evolution of stars, but it remained a puzzle how such compact objects
were able to support themselves against gravitational contraction.
The answer was first provided by Eddington and
Fowler who suggested that the
star is supported by the degenerate electron pressure, a quantum
effect arising from the Pauli-exclusion principle. When Chandrasekhar
worked out the corresponding theory for a relativistic electron gas
he made the remarkable discovery that
the degenerate electron pressure will never be sufficient
to support white dwarfs above a mass of about $1.4\,M_{\odot}$.
In his words: {\em ``For a star of small mass the natural white dwarf
stage is an initial step towards complete extinction. A star of
large mass cannot pass into the white dwarf stage and one is left
speculating on other possibilities.''} It did not take long before
such speculations were directed towards the existence of neutron stars.\\
The first suggestion that stars made up of nucleons may exist
came from Landau in 1932 just two years after the discovery of the neutron.
Two years later Baade and Zwicky proposed the idea that neutron stars
may be the product of supernova explosions and thus mark the final stage in
the evolution of stars of large mass. The first theoretical models of
neutron stars were calculated in 1939 by Tolman, Oppenheimer and Volkoff.
It took another thirty years, however, before neutron stars were
actually discovered observationally. Furthermore this
discovery came in a completely unexpected way. In 1967 the then
Cambridge graduate student Jocelyne Bell and her supervisor Antony Hewish
were looking for scintillations of radio sources produced
by the interstellar medium. On the 28th of November 1967 they discovered
a source with an exceptionally regular pattern of radio pulses
which at the time even gave
rise to the speculation of an extra-terrestrial, intelligent origin.
These speculations were quickly abandoned, however, when
three more ``pulsars'' were found within the next few weeks. The extremely
short duration of the pulses and the high pulse frequencies
lead to the conclusion that these sources must be significantly smaller than
white dwarfs. An explanation for this phenomenon was finally found
when a pulsar was
detected at the centre of the crab nebula. From historical records it is
known that the crab nebula marks a supernova explosion that was observed
in the year 1054. Pulsars are therefore identified with neutron stars,
the remnants of supernova explosions. In the same way that the degeneracy
pressure of the electrons
supports white dwarfs against gravitational collapse,
the internal pressure in neutron stars
arises from the degenerate nucleons. A great deal
of work has gone into the observational and theoretical study of these
compact objects. From these studies it is known that neutron stars
have masses of about $1.4$ solar masses and radii of about
$10\,{\rm km}$. Neutron stars are believed to have a solid crust in which the
density increases from about $10^4\,{\rm g/cm}^3$ to a few times
$10^{11}\,{\rm g/cm}^3$. In this density range the matter is assumed to
consist of a degenerate electron gas and atomic nuclei that form a
crystal-like structure. At larger densities the atomic nuclei gradually
dissolve and at about $2\cdot10^{14}\,{\rm g/cm}^3$ the matter largely
consists of a highly incompressible neutron fluid with small amounts
of protons and electrons. An interesting property of this fluid
arises from the thermal temperature which is commonly believed to
be smaller than
$10^{8}\,{\rm K}$. Compared with the Fermi-temperature of the nucleons of about
$3\cdot 10^{11}\,{K}$ this implies that the matter behaves like
a fluid at zero temperature and becomes superfluid and, in the case of
the protons, super-conductive.
The structure of matter and the resulting equations
of state at higher densities become increasingly unclear and are
subject to ongoing research. Near the centre of a neutron star the density
is assumed to be of the order of $10^{15}\,{\rm g/cm}^3$ and the matter may
at least partly consist of hyperons, pions or quarks, so-called {\em strange
matter}. \\
The extreme compactness of neutron stars makes them particularly interesting
from a relativistic point of view. We have already mentioned the significance
of neutron stars in the context of the search for gravitational waves.
In this respect the importance of neutron star oscillations
arises from the
discovery of secularly unstable oscillation modes that increase
in amplitude due to the spin down of the neutron star
while energy is radiated away in the form of gravitational waves.
%This class of oscillations, called {\em r-modes},
%(\ref{Andersson1993})
%gains their energy from the rotational energy
%of the neutron star.
If the attempts to measure gravitational waves
are indeed successful, a whole new window for astrophysical observations
may be opened and facilitate a unique opportunity to directly observe the
interior of astrophysical objects such as neutron stars.
In this work, however, we will not directly study
neutron star oscillations in the context of gravitational radiation.
Instead we use the simpler case of spherically symmetric dynamic
neutron stars in order to probe a new numerical approach
which enables us to
numerically evolve non-linear oscillations of arbitrary amplitude with
high accuracy. While these evolutions will not lead to the generation
of gravitational waves because of the spherical symmetry, the numerical
results, the new techniques and the discussion of numerical
difficulties encountered
in the course of this work may still be relevant for numerical
simulations of more general types of neutron star oscillations. \\
The use of oscillations as a diagnostic tool to obtain information about
the interior structure of an object is an old idea and by no means restricted
to the realm of distant stars. For example the same technique has been
applied to the earth where the study of artificially induced oscillations
and, in particular, earthquakes has lead to invaluable insight into
the internal structure of our planet. In the same way
a great deal of knowledge has been obtained about the sun and more distant
stars by investigating their oscillations which reveal themselves in the
electromagnetic spectra of these objects.
Whereas Newtonian theory is perfectly adequate for studying ``normal''
stars, i.e. stars that gain their energy from continuous nuclear burning
of hydrogen and other light elements,
accurate modelling of compact objects like neutron stars
requires a general relativistic description. \\
%An accurate description of these objects can only be achieved in the
%framework of general relativity.
%Probably the most interesting aspect of studying oscillations of compact
%objects is the potential generation of gravitational waves sufficiently
%strong to be identified with gravitational detectors that are currently
%under construction or have recently gone online. \\
The first type of neutron star oscillations to be studied extensively
were linear radial oscillations
(see for example \shortciteNP{Chandrasekhar1964a},
\citeyearNP{Chandrasekhar1964b})
which today represent a well understood
problem that is described in the standard literature. The same is
not true, however, for nonlinear radial oscillations which lead to
qualitatively new problems. We have already mentioned that 
% It is a well known result of general relativity
spherically symmetric spacetimes do not admit radiative solutions.
Instead the generation of gravitational waves
requires a time varying quadrupole or higher multi-pole ($l \ge 2$) moment
of the neutron star inertia.
%Consequently radial oscillations will not give rise to gravitational
%waves and from that point of view their study is not immediately
%interesting.
From that point of view, the study of radial oscillations is
not immediately interesting.
There are, however, several other important aspects
associated with radial oscillations. In the work mentioned above,
Chandrasekhar first revealed the existence of relativistic instability.
In the framework of radial oscillations this instability manifests itself
in the instability of the fundamental radial oscillation mode. If the frequency
of this mode becomes imaginary, an exponential growth of
physical quantities results and the star collapses or evaporates.
A fully non-linear evolution code
based on spectral methods has been developed by \lcite{Gourgoulhon1991}
and has been used to study various aspects of the stability of neutron stars
and their collapse into black holes (\citeNP{Gourgoulhon1993},
\shortciteNP{Gourgoulhon1994}).
Radial oscillations have also been considered from the point of view
of astrophysical observations.
The discovery of quasi-periodic radio sub-pulses in the spectra of pulsars
and periodicities in X-ray sources has lead to the suggestion that
radial oscillations of neutron stars may give rise to these features
(\citeNP{Boriakoff1976}, \citeNP{vanHorn1980}), which in turn
has stimulated further
research in this direction (see for example \shortciteNP{Marti1988},
\citeNP{Vaeth1992}). Furthermore the influence of radial oscillations
on the electromagnetic spectrum of neutron stars and their dependence
on the structure of matter at super-nuclear densities may provide
valuable information about the equation of state in the
high density range (\citeNP{Glass1983}).
The study of radial oscillations
is frequently carried out in the linear regime,
where all physical quantities have a harmonic time dependence
$f=f(r)e^{i\omega t}$ and the radial profiles $f(r)$ are determined
by an eigenvalue problem. In this work we will present explicit time
evolutions of the physical variables in the fully non-linear case. These
evolutions will serve two purposes. First we will be able to study deviations
from the known linearized behaviour, such as mode coupling and shock formation.
Secondly the spherically symmetric case can be used to investigate
numerical difficulties that are also expected in the more complicated
time evolutions in two or three spatial dimensions. A detailed analysis in
the computationally less expensive 1-dimensional case may lead to the
development of new advantageous numerical techniques or other types
of solutions to these problems.
% An explicit evolution
%of the time dependent quantities might, however, reveal unexpected numerical
%problems that are also present
%in more general scenarios with lower degrees of symmetry. This is
%particularly relevant for fully non-linear evolutions and the numerical
%investigation of non-linear radial oscillations will thus serve, among
%other purposes, also for probing numerical techniques with a wider range of
%applicability.
The work of \lcite{Gourgoulhon1991} for example has shown
among other results that the use of momentum densities as fundamental
variables may lead to
computation errors in passing from the momentum densities to the
velocity fields which can be avoided if velocity variables are used
in the first place. \\
In our discussion we will start with a
static spherically symmetric star which is governed by the
Tolman-Oppenheimer-Volkoff equations (\citeNP{Tolman1939},
\citeNP{Oppenheimer1939}). In section \ref{STATIC} we
will investigate these equations and describe the numerical methods
we use to calculate the resulting neutron star models. In section
\ref{DYNAMIC} we will use the static results in order to obtain a
fully non-linear perturbative formulation of dynamic spherically
symmetric stars. As a subclass we will discuss the linearized limit of
these equations in section \ref{PERT_LIN} and numerically
calculate the corresponding eigenmode solutions. It is interesting to
see that the surface of the star turns out to be a problematic
area even in this comparatively simple case. After a more detailed
discussion of the general problems one faces at the surface
in an Eulerian formulation we describe the numerical implementation of the code.
Even though the code is shown to perform well
in the linear regime for a large variety of neutron star models
in section \ref{EULER_LIN}, the surface problem is shown to give rise
to spurious results in some special cases.
In order to circumvent these problems
we use a simplified neutron star model in section
\ref{MODECOUPLING} to test the code in the non-linear regime and
to investigate the non-linear coupling of eigenmodes.
We conclude this work with the development of a fully non-linear
perturbative Lagrangian code in section \ref{LAGR}. We demonstrate how the
difficulties at the surface are resolved in such a formulation and
extensively test this code in the linear and non-linear regime. We
use this code to address the question whether non-linear effects
are present near the surface of the neutron star models in the
case of low amplitude oscillations.

%=========================================================================
\subsection{Spherically symmetric static stars}
\label{STATIC}
In the fully non-linear perturbative approach to the study of
radial oscillations
we will decompose the time dependent physical quantities into static
background contributions and time dependent perturbations. The background
quantities will obey the corresponding static set of equations which will
then be used to remove terms of zero order from the fully non-linear
evolution equations in the time dependent case. In our studies we have 
two principal choices for the static background: vacuum flat space in
which case we recover the standard non-perturbative formulation of the
problem and a static self-gravitating perfect fluid in spherical
symmetry which is described by the Tolman-Oppenheimer-Volkoff equations.
It is the second case which enables us to obtain highly accurate numerical 
solutions for any given amplitude of the oscillations. 
We will therefore first discuss in detail the Tolman-Oppenheimer-Volkoff (TOV) 
equations as well as their numerical solution.\\

%=========================================================================
\subsubsection{The Tolman Oppenheimer Volkoff equations}
\label{TOV_EQ}
In the framework of the ``3+1'' formalism described in section
\ref{threep1}, we start by choosing coordinates $r$, $\theta$, $\phi$ on each 
spatial hypersurface $\Sigma$. $\theta$ and $\phi$ are standard angular 
coordinates and the radius $r$ is defined by the radial gauge condition,
so that the area of a surface $r={\rm const}$ is $4\pi r^2$.
The 3-dimensional line element is then given by
\begin{align}
  ds^2 &= \mu^2 dr^2 + r^2(d\theta^2 + \sin^2\theta d\phi^2),
\end{align}
where in spherical symmetry $\mu$ is a function of $r$ only. If we label the 
hypersurfaces $\Sigma$ by the coordinate $t$ we can apply the polar
slicing condition which combined with radial gauge can be shown to imply
a vanishing shift vector in spherical symmetry.
The 4-dimensional metric is then given by
\begin{align}
  ds^2 &= -\lambda^2 dt^2 + \mu^2 dr^2 + r^2(d\theta^2 + \sin^2\theta d\phi^2).
  \label{TOV_LINEELEMENT}
\end{align}
Here the lapse function $\lambda$ is also a function of $r$. Alternatively
this metric can be described by the variables $m$ and $\phi$ defined by
\begin{align}
  \mu^2 &= \left( 1-\frac{2m}{r} \right) ^{-1}, \label{TOV_MUOFM} \\[10pt]
  \lambda^2 &= e^{2\phi}.
\end{align}
In the Newtonian limit $\phi$ becomes the gravitational potential and $m$
the gravitating mass.\\
Our description of the matter is based on three simplifying assumptions, which
we will discuss in order. 

\vspace{0.5cm}
1) \parbox[t]{15cm}
  {
  We will describe the matter as a single component perfect fluid. 
  This means that the fluid is seen as isotropic by a comoving observer.
  %and it moves with a 4-velocity $\hbox{\vec u}^{\mu}$
  %which may vary from point to point.
  In particular no heat conduction, no shear stresses, anisotropic pressures
  or viscosity must be present. The deviation from the
  perfect fluid equilibrium due to anisotropic stresses resulting from the
  solid crust are found to be $<10^{-5}$ even for rotating stars
  (\citeNP{Friedman1992}). It is, however, not entirely clear to what
  extent the treatment of the neutron star matter as a {\em single}
  perfect fluid is too restrictive. It was suggested as early as 1959
  by \citeANP{Migdal1959} that nucleons might be present in the form of
  superfluids in the interior of neutron stars. In order to obtain
  more realistic descriptions of neutron stars it might therefore be
  necessary to describe the matter as a multicomponent fluid.
  These issues are subject to ongoing research (see for example
  \citeNP{Andersson2001}) and their investigation would exceed the scope
  of this work.
  We will therefore focus our discussion on single
  component perfect fluids in which case we can write the energy-momentum
  tensor in the form
  \begin{align}
    \hbox{\vec{T}}^{\mu \nu} &= (\rho + P) \hbox{{\vec u}}^{\mu} 
    \hbox{{\vec u}}^{\nu} + P \hbox{\vec{g}}^{\mu \nu}, \label{TOV_EMTENSOR}
  \end{align}
  }

\hspace{0.5cm}\parbox[t]{15cm}
  { 
  where $\rho$ is the energy density and $P$ the pressure measured by
  a comoving observer. In the static spherically symmetric case $\rho$
  and $P$ are functions of the radius $r$ and the 4-velocity 
  has a non-vanishing time component only. The normalisation condition
  $\hbox{{\vec u}}^{\mu}\hbox{{\vec u}}_{\mu}=-1$ then implies
  \begin{align}
    \hbox{{\vec u}}^{\mu} &= \left[ \lambda^{-1},0,0,0 \right].
    \label{TOV_4VELOCITY}
  \end{align}
  } \\

\vspace{0.5cm}
2) \parbox[t]{15cm}
  {
  The neutron star matter is assumed to be at zero temperature. This is
  justified by comparing the thermal temperature of the stellar interior,
  which is assumed to be smaller than $10^8\,{\rm K}$ in mature neutron stars,
  with the relevant temperature scale given by the Fermi
  temperature of the matter. Even though the thermal temperature is
  large compared with terrestrial standards, it is orders of magnitude
  below the Fermi temperature of matter at nuclear density ($\approx 3\cdot
  10^{11}$\,K), so that the thermal degrees of freedom are frozen out.
  As a consequence the single component perfect fluid is described
  by a 1-parameter equation of state which is
  commonly chosen to be of the form $P=P(\rho)$.
  } \\

\vspace{0.5cm}
3) \parbox[t]{15cm}
  {
  The equation of state (EOS) is assumed to be given by a polytropic law
  \begin{align}
    P &= K \rho^{\gamma}, \label{POLYTROPE}
  \end{align}
  where $K$ and $\gamma$ are constants. Instead of the polytropic exponent
  $\gamma$ sometimes the polytropic index $n$ is used which is defined by
  \begin{align}
    \gamma = 1+\frac{1}{n}. \label{POLYTROPICINDEX}
  \end{align}
  The suitability of such an EOS
  is certainly a debatable issue and the determination of realistic
  equations of state of matter at super-nuclear densities represents
  an entire branch of physical research. Conclusive answers have yet to be
  obtained, however, and by using polytropes with different indices $n$
  one is able to study the qualitative differences in the behaviour of
  neutron stars with equations of state of varying stiffness. Furthermore
  polytropes are given in analytic form
  so that no additional numerical error arises from their use.
  } \\

\vspace{0.5cm}
We have got all ingredients now to derive the equations governing the static
spherically symmetric neutron star model. Starting with the metric
(\ref{TOV_LINEELEMENT}) and the energy-momentum tensor given by
Eq.\,(\ref{TOV_EMTENSOR}) with the 4-velocity (\ref{TOV_4VELOCITY})
the Einstein field equations $\hbox{\vec{G}}_{\mu \nu} = 8\pi
\hbox{\vec{T}}_{\mu \nu}$ result in two independent equations
\begin{align}
  \frac{\lambda_{,r}}{\lambda} &= 
      \frac{\mu^2-1}{2r} + 4\pi r \mu^2 P, \label{TOV_LAMBDAR} \\[10pt]
  \frac{\mu_{,r}}{\mu} &= -\frac{\mu^2-1}{2r} + 4\pi r \mu^2 \rho. 
      \label{TOV_MUR}
\end{align}
All other field equations are consequences of these two equations, their
derivatives and the matter equation (\ref{TOV_PR}).
In terms of the alternative variable $m(r)$ defined by
Eq.\,(\ref{TOV_MUOFM}), the equation for $\mu$ can be rewritten as
\begin{align}
  m_{,r} &= 4\pi r^2 \rho. \label{TOV_MR}
\end{align}
From now on we will therefore refer to $m$ as the ``mass'' or
``mass function'' of the neutron star.
Conservation of energy and momentum
$\nabla_{\mu}\hbox{\vec{T}}^{\mu \nu}=0$ results
in a single equation describing the hydrostatic equilibrium
\begin{align}
  P_{,r} &= -\frac{\la_{,r}}{\la} (\rho + P). \label{TOV_PR}
\end{align}
The system of ODEs (\ref{TOV_LAMBDAR}), (\ref{TOV_MUR}), (\ref{TOV_PR}) was
first derived by \lcite{Tolman1939} and \lcite{Oppenheimer1939}
and is thus known as the Tolman-Oppenheimer-Volkoff or TOV equations.
Together with an
equation of state which we choose to be the polytropic law (\ref{POLYTROPE})
they describe a self-gravitating perfect fluid in spherical symmetry. \\
We finally need to specify appropriate boundary conditions for these
equations.
The condition for the radial component of the metric is $\mu=1$ at the origin
$r=0$ in order to avoid a conical singularity. This is also illustrated by
the requirement of a finite energy density $\rho$ at the centre which
according to Eq.\,(\ref{TOV_MR}) implies
that $m_{,r} = \mathscr{O}(r^2)$ near the centre. Consequently
$M= \mathscr{O}(r^3)$ and Eq.\,(\ref{TOV_MUOFM}) leads to $\mu=1$.
The lapse function $\lambda$ on the other hand appears in the equations
in the form $\lambda_{,r}/\lambda$ and is therefore only defined 
up to a constant factor. Normally this factor is
chosen so that $\lambda$ takes on the value $\sqrt{1-2m/r}$ at the stellar 
surface which matches the interior metric (\ref{TOV_LINEELEMENT}) to an
exterior Schwarzschild metric
\begin{align}
  ds^2 &= -\left( 1-\frac{2M}{r} \right) dt^2 + \left( 1-\frac{2M}{r}
          \right)^{-1} + r^2 d\theta^2 + r^2 \sin^2\theta d\phi^2,
\end{align}
where $M=m(R)$ and $R$ is the radius of the star.
Finally the surface of the star is defined by the vanishing of the
pressure $P$ which for the polytropic equation of state is 
equivalent to $\rho=0$. We note that for some equations of state
the fluid extends to infinity and the energy density will vanish nowhere.
%Even though these equations of state do not describe neutron stars 
%their study may still lead to valuable insights and 
%an alternative boundary condition has to be used, for example by prescribing
%the energy density at some finite radius.
In this work, however, we will restrict ourselves to equations of state which
lead to stars of
finite size. We therefore summarise the boundary conditions as
\begin{align}
  \mu &= 1 \label{TOV_MUBC}
\end{align}
at the origin $r=0$ and
\begin{align}
  \lambda &= \sqrt{1-\frac{2m}{r}} = \frac{1}{\mu}, \label{TOV_LAMBDABC} \\[10pt]
  \rho &= 0 \label{TOV_RHOBC}
\end{align}
at the surface $r=R$, i.e. three boundary conditions for the three
first order ODEs (\ref{TOV_LAMBDAR}), (\ref{TOV_MUR}), (\ref{TOV_PR}).
At first glance this seems to completely specify the physical scenario.
We have to note one subtlety however: the location of the stellar surface,
i.e. the extension of the numerical grid, is not determined at this
stage. For any given equation
of state we therefore expect a 1-parameter family of solutions parameterised
by the radius $R$. As we will see below we can alternatively parameterise the
family of solutions by the central density $\rho_{\rm c}$ of the star. Which
of these parameters we choose and therefore have to specify in addition to 
the boundary conditions (\ref{TOV_MUBC})-(\ref{TOV_RHOBC}) depends on the 
numerical approach we take towards solving the TOV-equations. There
are two main approaches to this problem.
%one strictly correct and one convenient.
% Before we turn our attention towards the numerical solution,
% however, we will briefly discuss the choice of physical units.

%=========================================================================
\subsubsection{The numerical treatment of the TOV-equations}
\label{TOV_NUM}
The problem we have to solve numerically is given by the TOV equations
(\ref{TOV_LAMBDAR}), (\ref{TOV_MUR}), (\ref{TOV_PR}), the boundary conditions
(\ref{TOV_MUBC})-(\ref{TOV_RHOBC}) and the prescription of the free parameter.
From a numerical point of view this is a two-point boundary value
problem and should be solved accordingly with shooting or relaxation methods.
This is the first of the two approaches we mentioned in the previous section.
Here we will discuss a relaxation algorithm.
In this case we set up a numerical grid, thus specifying the free parameter
in the form of the stellar radius, and finite difference the equations as
described in section \ref{relaxation}.
The three boundary conditions then provide
the remaining three algebraic equations and having specified an initial guess
the code relaxes to the solution of the TOV-equations. The main advantages
of this approach are:
\begin{list}{\rm{(\arabic{count})}}{\usecounter{count}
             \labelwidth1cm \leftmargin1.5cm \labelsep0.4cm \rightmargin1cm
             \parsep0.5ex plus0.2ex minus0.1ex \itemsep0ex plus0.2ex}
\item all boundary conditions are exactly satisfied,
%\item if a neutron star model with a given radius is required the prescription
%      of the radius is straightforward.
\item a neutron star model with a specified radius is obtained
      straightforwardly by appropriately setting up the numerical grid.
\end{list}
This code suffers from some drawbacks, however, which can be summarized
as follows:
\begin{list}{\rm{(\arabic{count})}}{\usecounter{count}
             \labelwidth1cm \leftmargin1.5cm \labelsep0.4cm \rightmargin1cm
             \parsep0.5ex plus0.2ex minus0.1ex \itemsep0ex plus0.2ex}
\item the specification of initial data is non-trivial and the convergence
      of the code depends on a ``good'' initial guess,
\item obtaining high accuracy via a higher ($>2^{\rm nd}$) order 
      finite differencing scheme results in more complicated coefficient 
      matrices and inversion routines,
\item it is not clear how to obtain a neutron star model with a specified
      central density,
\end{list}
It is quite remarkable that the second numerical approach has exactly the
opposite properties in that the advantages and drawbacks are reversed. In this
approach the outer boundary conditions are ignored initially
and instead one starts with three boundary conditions at the centre
\begin{align}
  \mu &= 1, \\[10pt]
  \lambda &= 1, \\[10pt]
  \rho &= \rho_{\rm c}.
\end{align}
The TOV-equations can then be integrated outwards straightforwardly until the
energy density becomes negative and the out-most grid point will define the 
surface of the star. Even though the energy density will not vanish exactly
at this point but take on a small positive value, the accuracy thus obtained
is good enough for most practical purposes. The remaining freedom to multiply
the lapse function $\lambda$ with an arbitrary constant
is used to enforce the
boundary condition (\ref{TOV_LAMBDABC}). Alternatively one can first integrate
Eqs.\,(\ref{TOV_MUR}), (\ref{TOV_PR}) for $\mu$ and $P$ 
which decouple from $\lambda$ and afterwards obtains $\lambda$ from
inward integration of Eq.\,(\ref{TOV_LAMBDAR}).\\
In a sense the two methods complement each other and for 
example we use the quadrature approach to obtain an initial guess for the 
relaxation scheme. Throughout this work we will use both numerical methods
and specify in each case how the TOV solutions were calculated. \\
Before we
investigate the solutions thus obtained, however, we have to discuss two
technical issues, the choice of physical units and a transformation to
a new radial coordinate which will provide higher resolution near
the surface of the star. Below we will see that sufficient resolution
in this region can be crucial for an accurate numerical
evolution in the time dependent case.

%=========================================================================
\subsubsection{Physical units}
\label{TOV_UNITS}
Throughout this work we have worked with natural
units, i.e. $c=1=G$. This choice can be written in the form
\begin{align}
  1\,{\rm s} &= 2.9979 \cdot 10^{10}\, {\rm cm},  \label{C1} \\[5pt]
  1\,{\rm g} &= 7.4237 \cdot 10^{-29}\, {\rm cm}. \label{G1}
\end{align}
In astrophysics energy density is commonly measured in g/cm$^3$ and
pressure in dyne/cm$^2$, where 1\,dyne=1\,erg/cm. However, we prefer to
measure all quantities in km or corresponding powers thereof. Using
Eqs.\,(\ref{C1}) and (\ref{G1}) we can calculate that
\begin{align}
  1\, {\rm km}^{-2} &= 1.3477 \cdot 10^{18}\,\frac{{\rm g}}{{\rm cm^3}},\\[10pt]
  1\, {\rm km}^{-2} &= 1.2106 \cdot 10^{30}\,\frac{{\rm dyne}}{{\rm cm^2}}
                   = 1.2106 \cdot 10^{30}\, \frac{{\rm g}}{{\rm cm\, s^2}}.
\end{align}
The metric variables $\mu$ and $\la$ are dimensionless and it is obvious then
from Eqs.\,(\ref{TOV_MUOFM}) and (\ref{TOV_MUR}) that radius $r$ and
mass $m$ are
measured in km.
For example a typical central density for neutron stars is
$10^{15}$ g/cm$^3$ which in our units becomes 0.000742 km$^{-2}$.
We can also compare our results for radius and mass with the solar values
\begin{align}
  M_{\odot} &= 1.4766\, {\rm km},  \\[10pt]
  R_{\odot} &= 6.960 \cdot 10^5\, {\rm km}.
\end{align}
In contrast to these values typical radii and masses of neutron stars
are given by
\begin{align}
  M_{\rm NS} &\approx 2\,{\rm km}, \\[10pt]
  R_{\rm NS} &\approx 10\,{\rm km}.
\end{align}
It is a well known result that relativistic correction terms to
a Newtonian description of stars generally appear in terms of the ratio
$M/R$, so that this quotient describes the importance of relativistic effects.
In view of this result and the quotient $M_{\odot}/R_{\odot}=2.1\cdot 10^{-6}$
it is immediately obvious why a Newtonian description of the sun
and other ``normal'' stars is
perfectly adequate. In contrast we find $M/R \approx 0.2$ for neutron stars,
so that relativistic effects will play an important role in their behaviour
and accurate models need to be developed in the framework of
general relativity.

%=========================================================================
\subsubsection{Transformation to a new radial coordinate}
\label{TOV_RYTRAFO}
\begin{figure}[t]
  \centering
  \epsfig{file=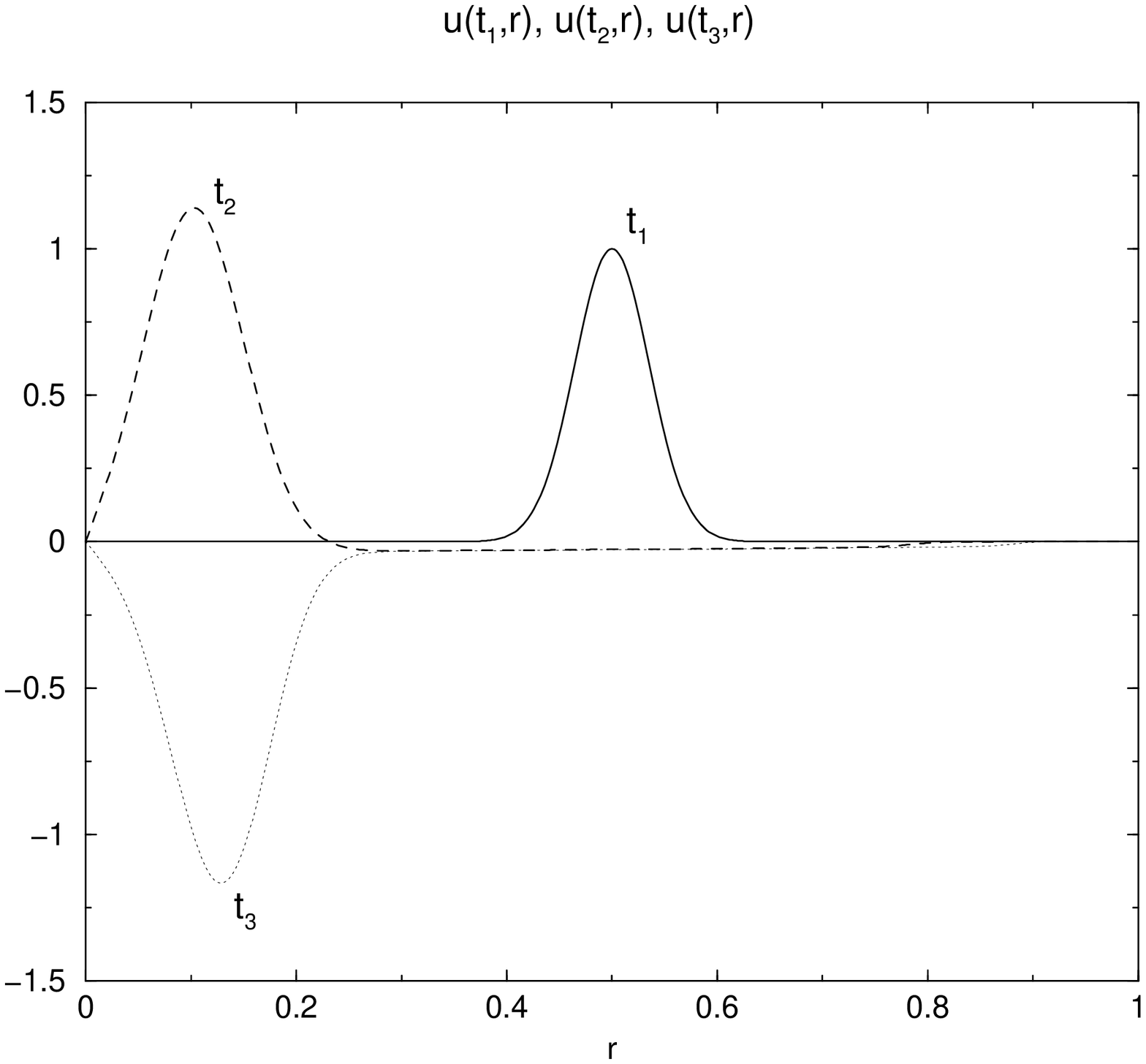, height=200pt, width=150pt,angle=-90}
  \epsfig{file=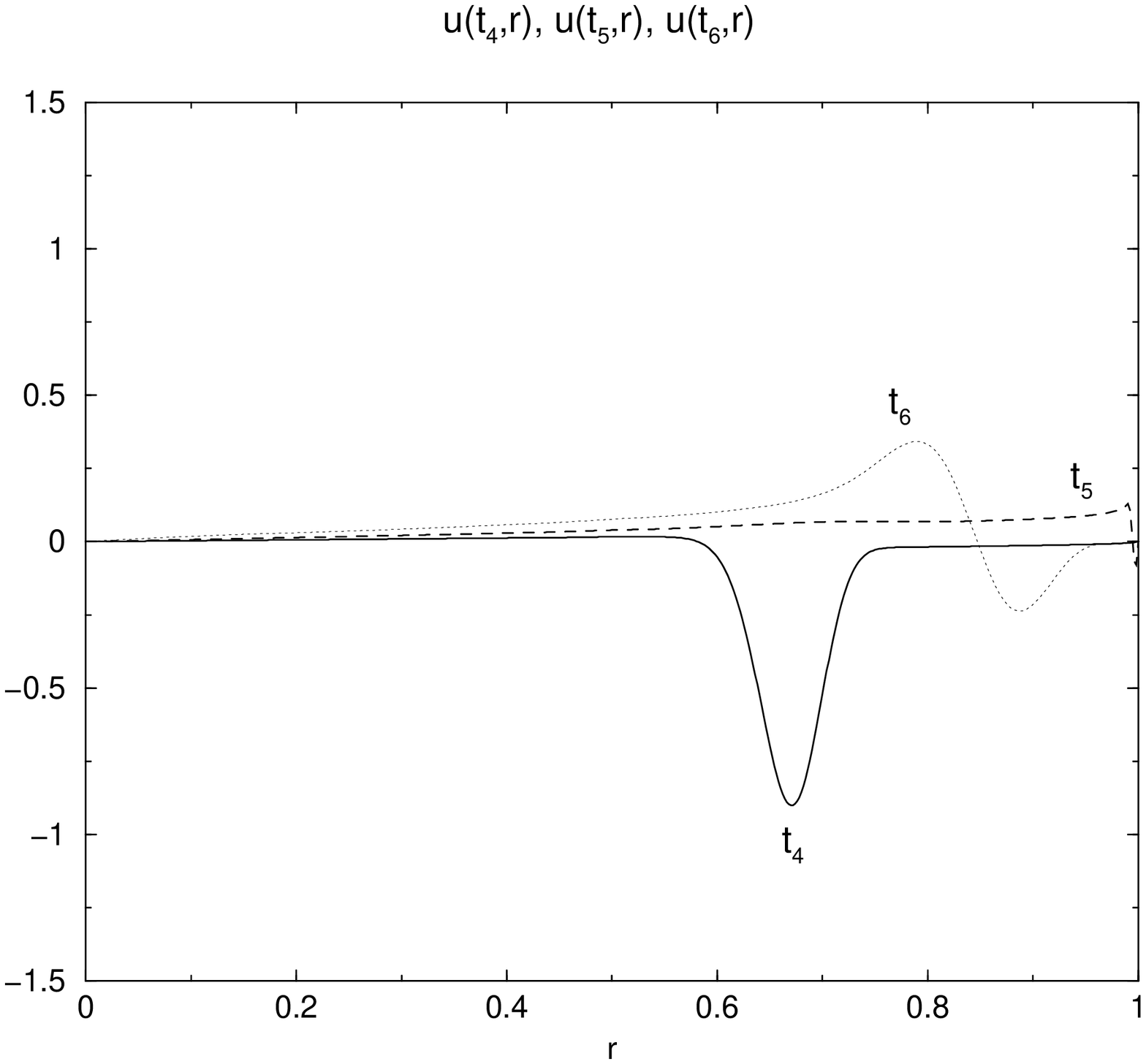, height=200pt, width=150pt,angle=-90}
  \epsfig{file=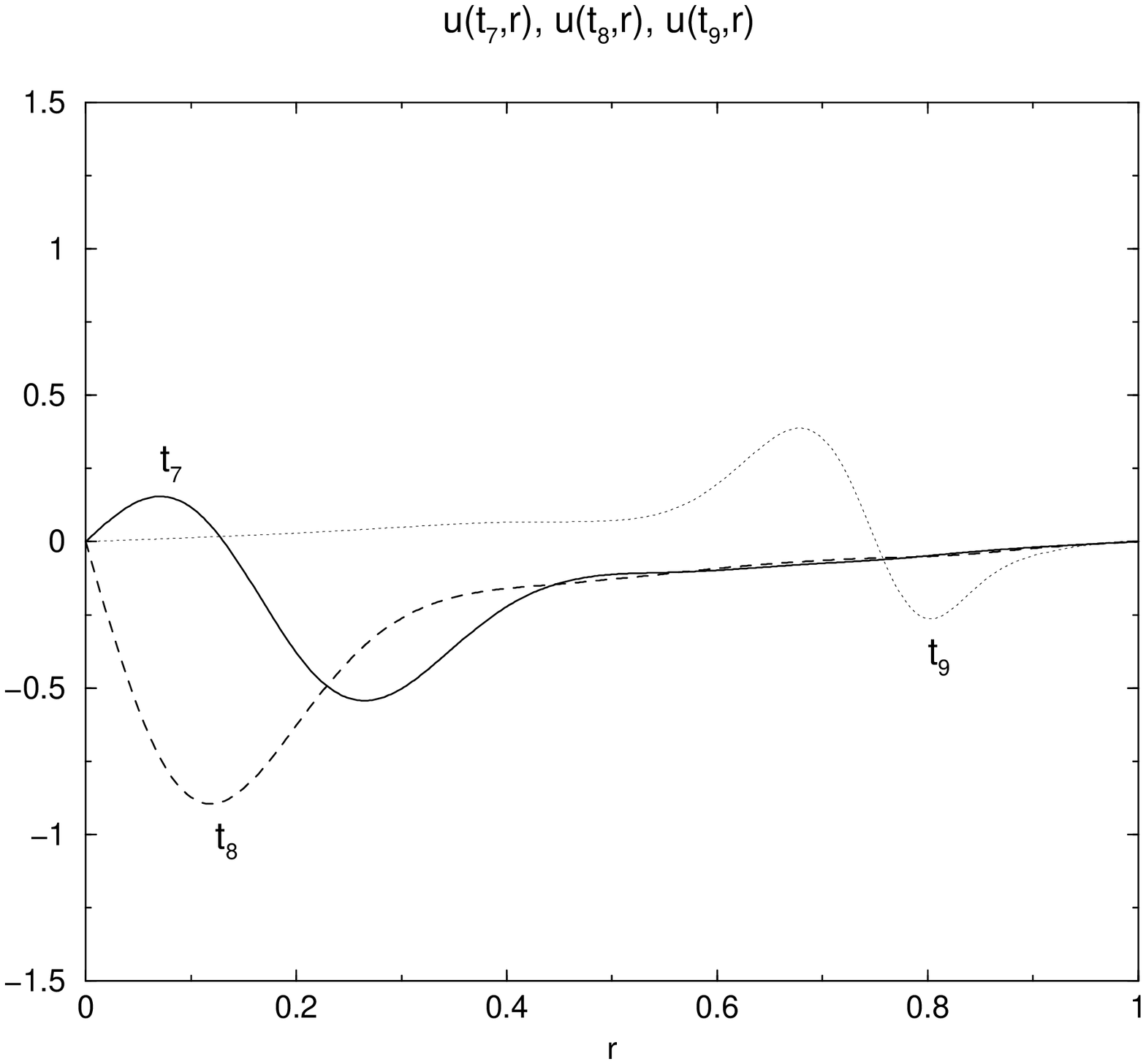, height=200pt, width=150pt,angle=-90}
  \epsfig{file=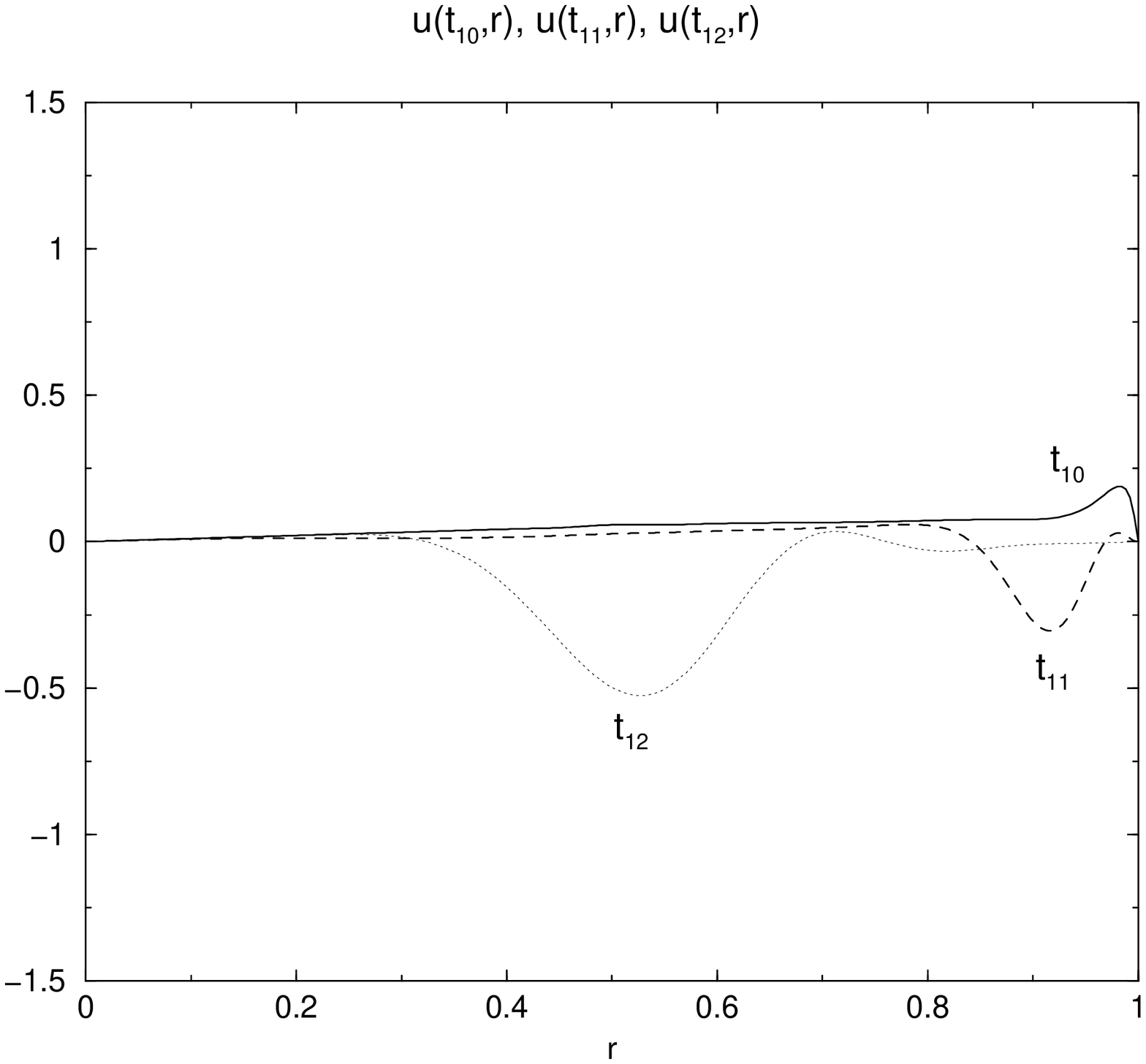, height=200pt, width=150pt,angle=-90}
  \caption{The numerical evolution of an initial
           Gaussian pulse according to the 
           wave equation in terms of the coordinate $r$ as obtained for
           the varying propagation speed given by Eq.\,(\ref{WAVE_C})
           which vanishes at $r=1$. The Snapshots are
           shown for the times $t_1,\ldots,t_{12}$.}
  \label{WAVES}
\end{figure}

\begin{figure}[t]
  \centering
  \epsfig{file=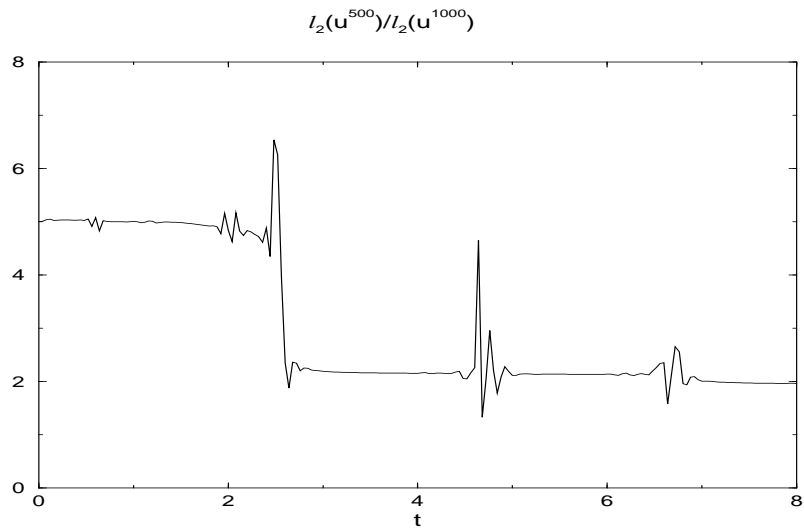, height=300pt, width=200pt, angle=-90}
  \caption{The convergence factor obtained for 500 and 1000
           grid points as a function of time. At $t\approx 2.5$ the
           convergence drops to first order.}
  \label{CONV_WAVES}
\end{figure}

We have already mentioned that the surface of the star is defined by the
vanishing of the pressure which in the case of a polytropic
equation of state is equivalent to a zero energy density.
A dependent quantity frequently introduced in the study of neutron stars
is the speed of sound defined by
\begin{align}
  C^2 &= \frac{\partial P}{\partial \rho}, \label{TOV_C2}
\end{align}
which in the polytropic case (\ref{POLYTROPE}) becomes
\begin{align}
  C^2 &= K \gamma \rho^{\gamma-1}.
\end{align}
Consequently the speed of sound will also vanish at the surface if
$\gamma>1$ as will always be the case
for a star of finite mass. In particular we will show below that the 
asymptotic behaviour of the speed of sound near the surface is given by
\begin{align}
  C \sim \sqrt{R-r}. \label{TOV_ASSYMPTOTICC}
\end{align}
Taking into account the vanishing of the propagation speed
of sound waves at $r=R$ we now consider the qualitative behaviour
of a localized pulse travelling towards
the surface. As a result of the decreasing sound speed $C$
the front of the pulse will
in general travel more slowly than its tail and we would expect the pulse
to narrow. In particular the numerical resolution near the surface might
be inadequate to accurately evolve the pulse in this region and it might be 
beneficial to work with a radial coordinate in terms of
which the propagation speed
is by and large independent of the position within the star.
In order to study the implications of a locally vanishing 
propagation speed we consider the simpler scenario of the
1-dimensional wave equation with variable propagation speed
\begin{align}
  u_{,tt} &= c(r)^2 u_{,rr}, \label{WAVEEQ}
\end{align}
on a physical domain $0\le r\le R$. 
Without loss of generality we will set $R=1$ for the rest
of this discussion. Eq.\,(\ref{TOV_ASSYMPTOTICC}) then suggests
to choose a propagation speed of the form
\begin{align}
  c(r) &= \sqrt{1-r}. \label{WAVE_C}
\end{align}
For the numerical implementation we introduce the
auxiliary variables $F=u_{,t}$ and $G=u_{,r}$ and rewrite Eq.\,(\ref{WAVEEQ})
as a system of two first order PDEs
\begin{align}
  F_{,t} &= c^2 G_{,r}, \label{WAVE1_FT} \\[10pt]
  G_{,t} &= F_{,r}, \label{WAVE1_GT}
\end{align}
and impose the boundary conditions $u=0$, $F=0$ at both
boundaries. The system (\ref{WAVE1_FT}), (\ref{WAVE1_GT})
is linear and can be written in vectorial form
\begin{align}
  \hbox{\vec{v}}_{,t} + A \hbox{\vec{v}}_{,r} &=0, \\[10pt]
  \hbox{\vec{v}} &= \begin{pmatrix} F \\ G \end{pmatrix}, \\[10pt]
  A &= \begin{pmatrix} 0 & -c^2 \\ -1 & 0 \end{pmatrix}.
\end{align}
The characteristics of the PDE are then given by
\begin{align}
  \frac{dr}{dt} &= \Lambda_i,
\end{align}
where $\Lambda_1=c$, $\Lambda_2=-c$ are the eigenvalues of the 
matrix $A$. At the outer boundary the slopes of the characteristics
collapse because of the vanishing of the wave speed $c$. \\
% It is
% this behaviour which gave rise to our suspicion in the first place.\\
This system has been evolved with the second order in space
and time McCormack finite differencing scheme described in section
\ref{McCormack} using a grid of 500 points.
In Fig.\,\ref{WAVES} we show the time evolution of $u$ obtained
for initial data in the form of a Gaussian pulse. Snapshots of $u$
are plotted at
$t_1 = 0.00$, $t_2=0.48$, $t_3=0.72$, $t_4=1.44$, $t_5=2.52$, $t_6=3.40$,
$t_7=4.44$, $t_8=4.60$, $t_9=5.60$, $t_{10}=6.56$, $t_{11}=7.20$
and $t_{12}=8.00$. 
%
%Nothing obviously wrong is seen in the plots although the broadening of 
%the pulse after reflection at the outer boundary may be unexpected.
In order to shed light on the quality of the numerical evolution we
analyse the convergence properties of the code.
For this purpose we have performed the same runs
using 1000 and 2000 grid points and calculated the time dependent convergence
factor according to the method described in section \ref{CCM_CONVERGENCE}.
Again we use a high resolution reference solution obtained for 2000 grid points
in place of the analytic solution. The results shown
in Fig.\,\ref{CONV_WAVES} demonstrate that
the convergence of the code drops to
first order at about $t=2.5$ which coincides with the snapshot at $t_5$
when the pulse is reflected at the outer boundary for the first time. This
result is confirmed by high resolution runs in which no
broadening of the pulse similar to that shown in Fig.\,\ref{WAVES} is observed
after reflections at either boundary. We
conclude that a naive numerical evolution can lead to spurious results
in regions with a vanishing propagation speed and that this problem is
due to an insufficient spatial resolution. \\
A solution to this problem is obtained by transforming to a new
spatial coordinate $y$ in terms of which the slopes of the characteristics
do not vary as drastically over the numerical domain and in particular do
not vanish at the boundary. A simple recipe is to define this new coordinate
by
\begin{align}
  y = \int_0^r{\frac{1}{c(\tilde{r})}d\tilde{r}}, \label{WAVE_RY}
\end{align}
which implies
\begin{align}
  \frac{\partial}{\partial r} &= \frac{1}{c} \frac{\partial}{\partial y},
       \\[10pt]
  dr &= c\,dy.
\end{align}
In the special case where the propagation speed is given by Eq.\,(\ref{WAVE_C})
the coordinates $r$ and $y$ are related by
\begin{align}
  y &= 2-2\sqrt{1-r}, \\[10pt]
  r &= y-\frac{y^2}{4},
\end{align}
so that the interval $r \in [0,1]$ is mapped to $y \in [0,2]$.
In terms of the new coordinate $y$ the system (\ref{WAVE1_FT}), 
(\ref{WAVE1_GT}) can be rewritten as
\begin{align}
  F_{,t} &= c G_{,y}, \label{WAVE2_FT} \\[10pt]
  G_{,t} &= \frac{1}{c} F_{,y}, \label{WAVE2_GT}
\end{align}
and the characteristic curves are given by
\begin{align}
  \frac{dy}{dt} &= \pm 1.
\end{align}
In order to compare the new scheme with the original approach, we evolve the
\begin{figure}[t]
  \centering
  \epsfig{file=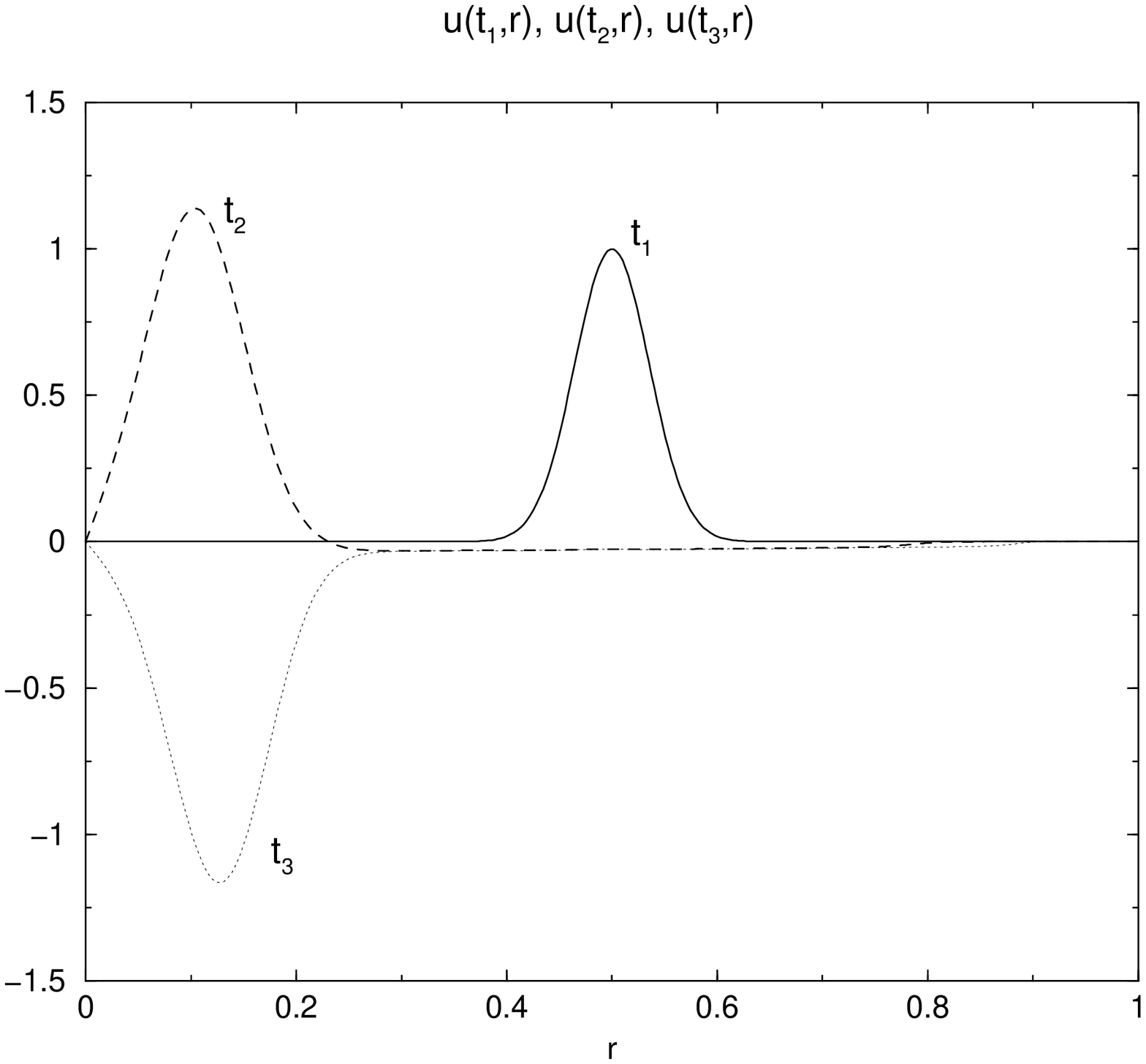, height=200pt, width=150pt,angle=-90}
  \epsfig{file=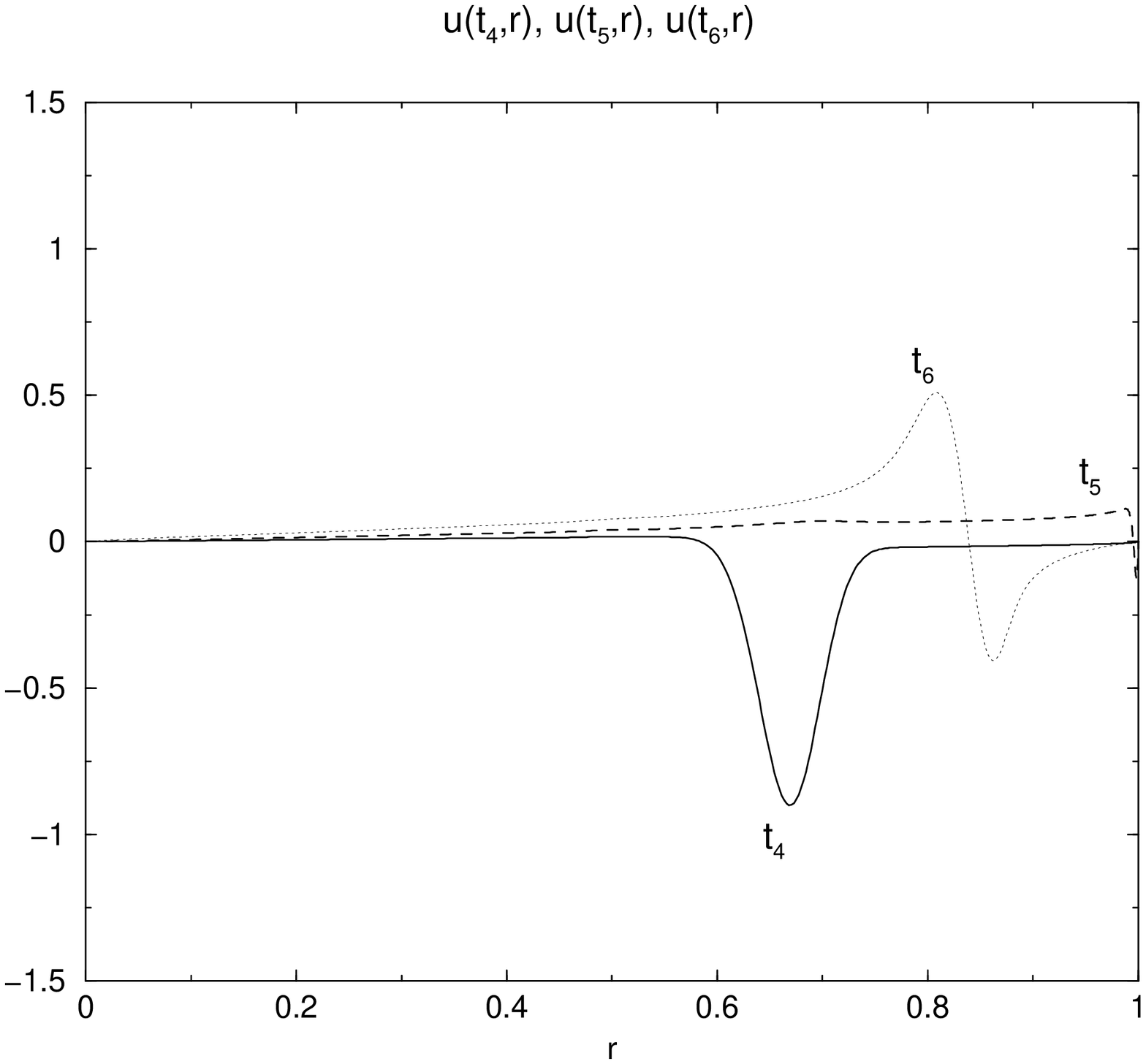, height=200pt, width=150pt,angle=-90}
  \epsfig{file=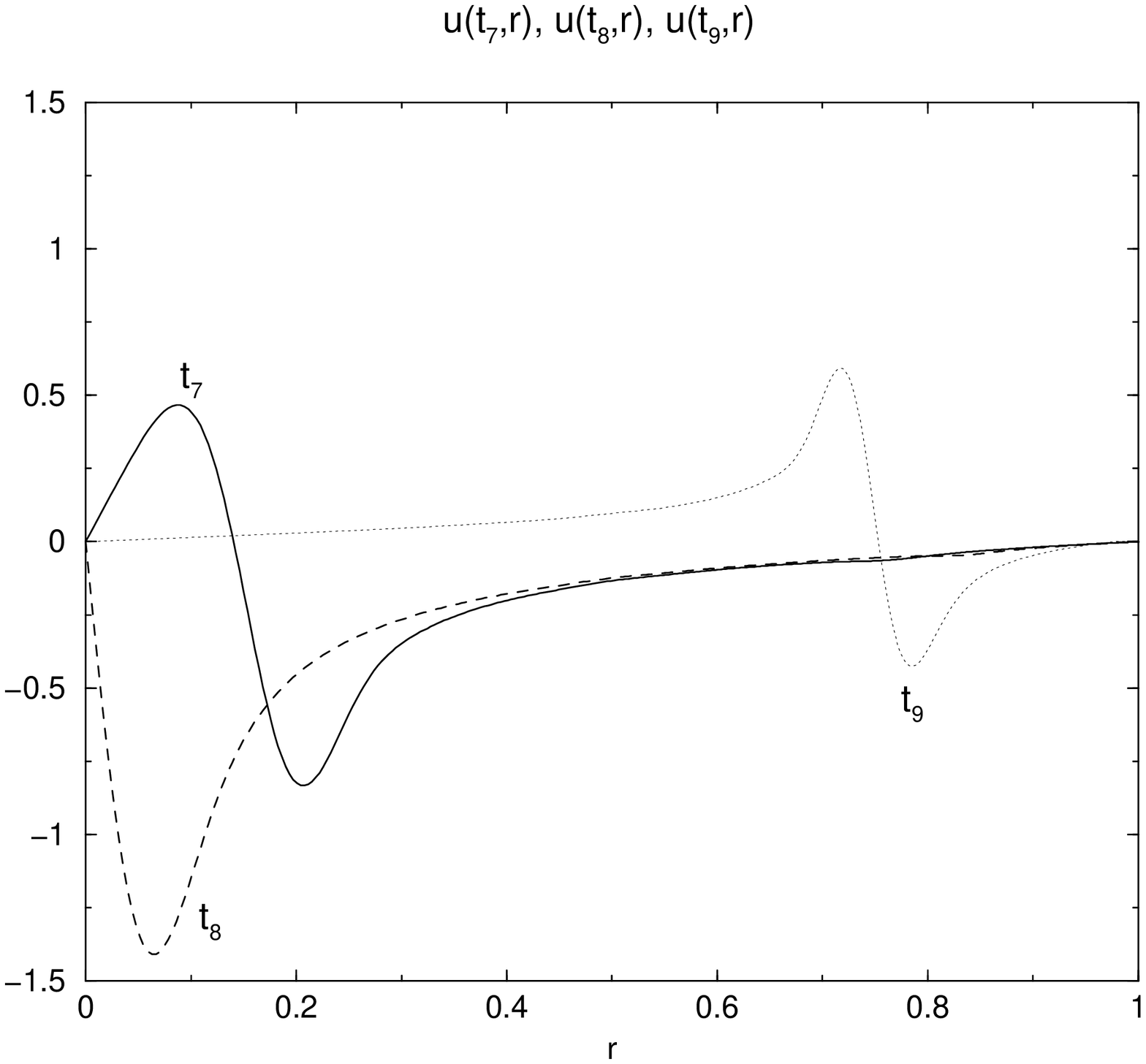, height=200pt, width=150pt,angle=-90}
  \epsfig{file=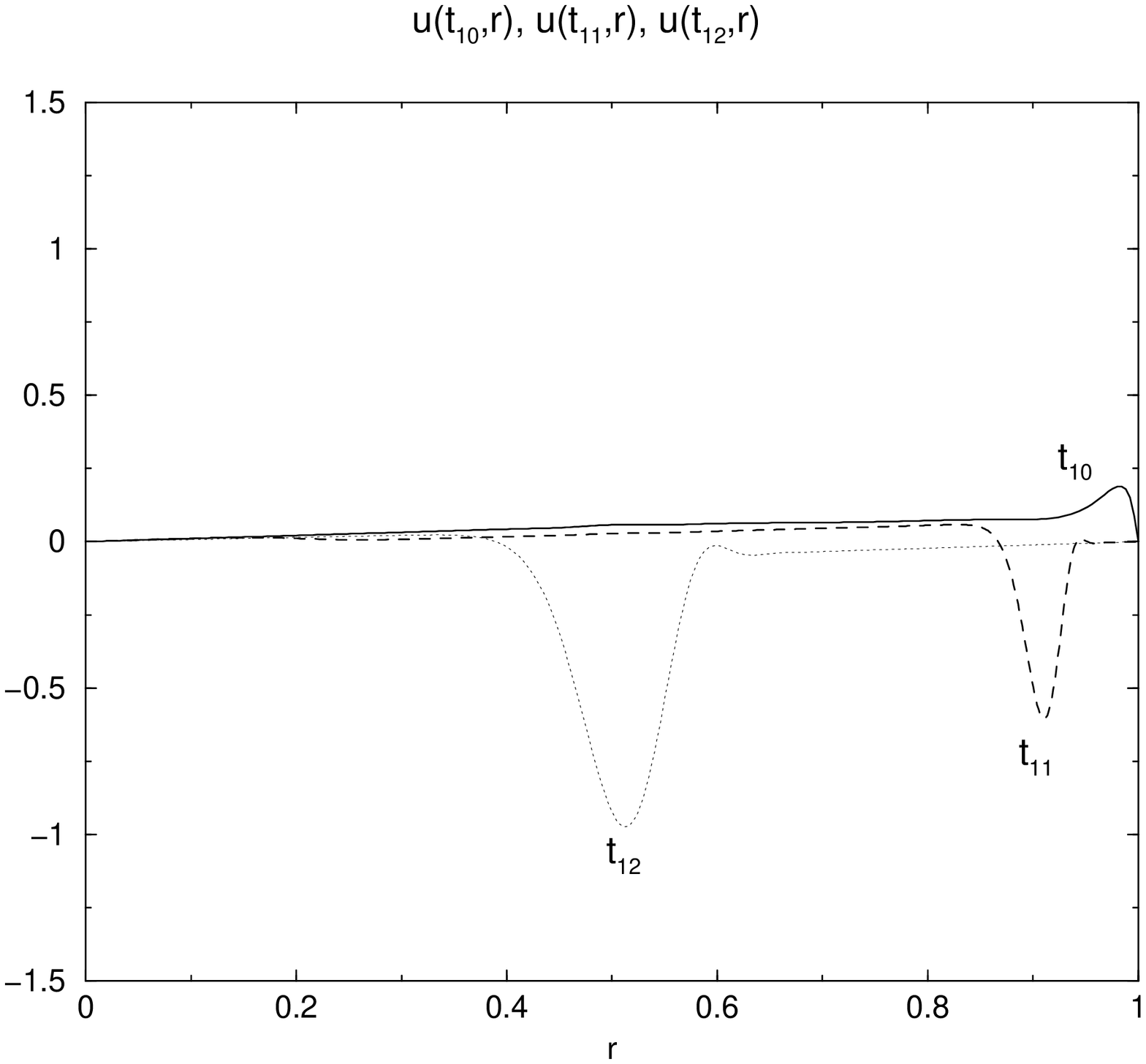, height=200pt, width=150pt,angle=-90}
  \caption{The same evolution as in Fig.\,\ref{WAVES}, but obtained with
           the new coordinate $y$ which results in a higher density
           of grid points near the outer boundary $r=1$.}
  \label{WAVEX}
\end{figure}
same initial data as above using the system (\ref{WAVE2_FT}), (\ref{WAVE2_GT})
on a $y$-grid again with 500 grid points and the same boundary conditions.
\begin{figure}[t]
  \centering
  \epsfig{file=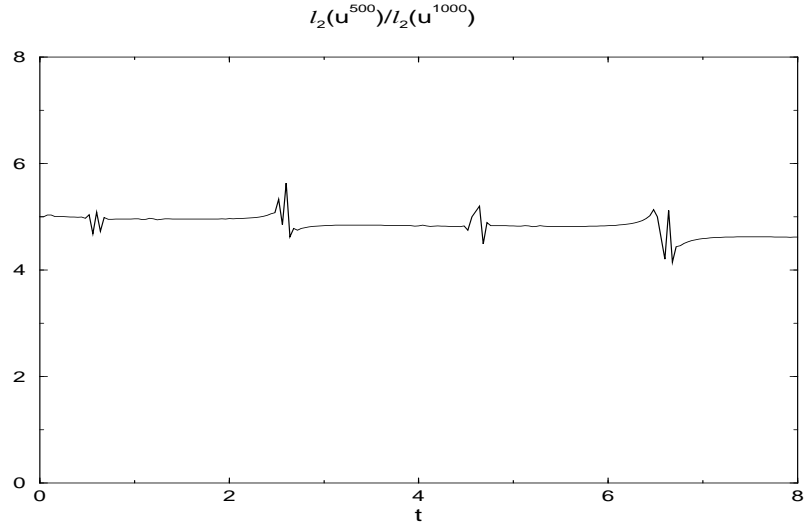, height=300pt, width=200pt, angle=-90}
  \caption{The time dependent convergence factor obtained for the
           numerical evolution of the wave equation on a $y$-grid with
           500 and 1000 grid points. Second order convergence is clearly 
           maintained throughout the evolution.}
  \label{CONV_WAVEX}
\end{figure}
The result is shown in Fig.\,\ref{WAVEX} where we plot 
the same snapshots as in Fig.\,\ref{WAVES}.
For comparison purposes the plots show $u$ as a function of the coordinate
$r$ but as a result of the computation on the $y$-grid,
the density of grid points is higher towards $r=1$ in Fig.\,\ref{WAVEX} whereas
the grid points are distributed homogeneously in Fig.\,\ref{WAVES}. In
contrast to the above evolution no broadening
of the pulse after reflection at the outer boundary is observed.
The time dependent convergence analysis shown 
in Fig.\,\ref{CONV_WAVEX}
demonstrates second order convergence throughout the run even though
small variations in the convergence factor are visible when the pulse is 
reflected at either boundary. We conclude that a transformation of the type
(\ref{WAVE_RY}) provides the necessary resolution in a region of
vanishing propagation speed and leads to satisfactory results at reasonable
grid resolutions. \\
We now have to apply this idea to the case of a static, spherically symmetric
neutron star. The role of the wave speed $c$ is now assumed by the speed
of sound $C$ defined in Eq.\,(\ref{TOV_C2}) and we introduce the new radial
coordinate
\begin{align}
  y &= \int_0^r{\frac{1}{C(\tilde{r})}d\tilde{r}}. \label{TOV_YOFR}
\end{align}
This transformation has also been successfully used by \lcite{Ruoff2000} in the
linearized time evolution of radial oscillations for more realistic
equations of state.
The asymptotic behaviour of the sound speed
in the Tolman-Oppenheimer-Volkoff case given by Eq.\,(\ref{TOV_ASSYMPTOTICC}) 
is identical to that of the wave speed
in the toy problem. Consequently the radial 
interval $r\in[0,R]$ of the
star will be mapped to a finite interval $y \in [0,Y]$.
% The numerical results
%shown below confirm this expectation.
In order to obtain a formulation which
includes both possible choices of the radial coordinate, we introduce
the variable $x$ in terms of which the TOV equations are written as
\begin{align}
  r_{,x} &= \left\{ \parbox{7cm}
            {
            $1\hspace{1cm}{\rm if}\,\,\,x=r$\\
            $C\hspace{0.9cm}{\rm if}\,\,\,x=y$,
            } \right. \label{TOV_RY} \\[10pt]
  \frac{\lambda_{,x}}{\lambda} &= 
      r_{,x} \left(\frac{\mu^2-1}{2r} 
      + 4\pi r \mu^2 P \right), \label{TOV_LAMBDAY}
      \\[10pt]
  \frac{\mu_{,x}}{\mu} &= r_{,x} \left( -\frac{\mu^2-1}{2r} + 4\pi r \mu^2 \rho
      \right), \label{TOV_MUY} \\[10pt]
  P_{,x} &= -\frac{\lambda_{,x}}{\lambda} (\rho + P) \label{TOV_PY}.
\end{align}
In the numerical code we are thus able to switch between the two alternative
modes of calculation by assigning the derivative $r_{,x}$ according to
either possibility of Eq.\,(\ref{TOV_RY}).
% No further modification is
%required.
In either case the boundary conditions are given by
Eqs.\,(\ref{TOV_MUBC})-(\ref{TOV_RHOBC}) supplemented with the requirement
that $r$ and $x$ vanish simultaneously at the origin
\begin{align}
  r &= 0 \,\,\,{\rm at}\,\,\,x=0.
\end{align}
%
%Unless specified otherwise, from now on we will use the rescaled
%radial coordinate and thus set $r_{,x}=C$.
One subtlety concerning the relaxation method of calculating
TOV solutions has to be mentioned. In this case we need to specify the
radius of the star. If we use the rescaled radial coordinate, however, the
surface value $x_{\rm s}$ is not a priori known. In practice we
therefore specify the free parameter in the form of the 
central density and solve the TOV equations via the quadrature
method first. This provides us with the outer boundary value of the
coordinate $x$ for the stellar model in question and we can solve the
TOV equations in a second step with the relaxation method.
% In the next section we will
%present numerical solutions for different polytropic models and also
%investigate the asymptotic behaviour of the solutions at both
%boundaries in detail.

%=========================================================================
\subsubsection{Asymptotic properties of the TOV equations}
\label{TOV_ASYMPTOTICS}
%
%
%For a given system of ordinary differential equations it is always worth
%studying the asymptotic behaviour of the solutions if the complexity
%of the equations allows such a study. In the case of the equations for
%a static cosmic string in section \ref{SECcsmink} for example the
%analysis of the
%asymptotic behaviour revealed the existence of exponentially diverging 
%unphysical solutions. This result enabled us to choose a suitable numerical
%technique which automatically selected the non-diverging physical solutions.
%In the case of the TOV equations no such exponentially diverging solutions
%do exist which makes the quadrature method of solving the equations such
%a successful tool, but we will see below that the asymptotic behaviour
The asymptotic behaviour of the solutions of the TOV equations
(\ref{TOV_RY})-(\ref{TOV_PY}) at the surface of the star has
serious implications for the
simulation of dynamic neutron stars with certain equations of state in a
strictly Eulerian framework. We will therefore discuss the asymptotic 
behaviour first and then compare the results with the numerically obtained 
solutions. Since the introduction of the rescaled radial coordinate resulted
from numerical requirements only, we use $r_{,x}=1$ i.e. the original system 
(\ref{TOV_LAMBDAR})-(\ref{TOV_PR}) for the asymptotic analysis.
We start with the behaviour at the origin, where we assume that
\begin{list}{\rm{(\arabic{count})}}{\usecounter{count}
             \labelwidth1cm \leftmargin1.5cm \labelsep0.4cm \rightmargin1cm
             \parsep0.5ex plus0.2ex minus0.1ex \itemsep0ex plus0.2ex}
\item the energy density and thus the pressure are finite and positive,
\item the lapse function $\lambda$ is finite and positive.
\end{list}
We have already seen that the central value of the energy density is a
free parameter and the pressure follows from the equation of state.
The central value of the lapse function, on the other hand,
is determined by matching $\lambda$ to an exterior 
Schwarzschild metric.
We also know from section \ref{TOV_EQ} that our assumptions imply
$\mu=1$ and $m=\mathscr{O}(r^3)$ at the origin. From Eq.\,(\ref{TOV_MUOFM})
we therefore conclude that $\mu=1+\mathscr{O}(r^2)$. Inserting this result into
Eq.\,(\ref{TOV_LAMBDAR}) and using the second assumption we find that
$\lambda_{,r}/\lambda \sim r$ and thus $\lambda =\lambda_{\rm c}
+\mathscr{O}(r^2)$. Using this result in Eq.\,(\ref{TOV_PR}) leads to
$P_{,r}\sim r$, i.e. $P=P_{\rm c} + \mathscr{O}(r^2)$ and the equation
of state then shows that the energy density has the same behaviour.
In summary the results near the origin are
\begin{align}
  \lambda(r) &= \lambda_{\rm c} + \mathscr{O}(r^2), \\[10pt]
  \mu(r)     &= 1 + \mathscr{O}(r^2), \\[10pt]
  \rho(r)    &= \rho_{\rm c} + \mathscr{O}(r^2), \\[10pt]
  P(r)       &= K \rho_{\rm c}^{\gamma} + \mathscr{O}(r^2).
\end{align}
%
%We also note that $\lambda$, $\mu$, $\rho$ and $P$ are components of rank
%2 tensors and consequently their series expansion should contain 
%only even powers of $r$ in spherical symmetry for regularity reasons. Our
%results are compatible with this requirement. \\
The corresponding analysis for the surface is more complicated and the results
will later prove to be of more significance. For this analysis it is convenient
to work with the radial variable
\begin{align}
  z &:= R-r.
\end{align}
We start with the following assumptions.
\begin{list}{\rm{(\arabic{count})}}{\usecounter{count}
             \labelwidth1cm \leftmargin1.5cm \labelsep0.4cm \rightmargin1cm
             \parsep0.5ex plus0.2ex minus0.1ex \itemsep0ex plus0.2ex}
\item The metric function $\mu$ is finite at the surface and also satisfies
      the inequality $\mu > 1$. This follows from Eq.\,(\ref{TOV_MUOFM})
      and the requirement that the mass satisfies
      the condition $0 < 2m(R) < R$. The first inequality follows
      from Eq.\,(\ref{TOV_MR}) for any non vacuum model and the
      second implies that
      the neutron star extends beyond its Schwarzschild radius.
\item The lapse $\lambda$ is finite and positive at the surface.
\item The energy density and the pressure vanish at the surface and their
      leading order terms are given by some positive powers of $z$.
\end{list}
We write these assumptions as
\begin{align}
  \mu &= \mu_{\rm s} + \mathscr{O}(z^{\epsilon_1}),
         \label{TOV_ASSYMPTOTICANSATZMU} \\[10pt]
  \lambda &= \lambda_{\rm s} + \mathscr{O}(z^{\epsilon_2}), \\[10pt]
  \rho &= \rho_{\rm s} z^{\alpha} + \mathscr{O}(z^{\alpha+\epsilon_3}),
          \\[10pt]
  P &= P_{\rm s} z^{\beta} + \mathscr{O}(z^{\beta+\epsilon_4}),
          \label{TOV_ASSYMPTOTICANSATZP}
\end{align}
where $\alpha$, $\beta$ and $\epsilon_1 , \ldots, \epsilon_4$ are positive
constants we have yet to determine and $\mu_{\rm s}$, $\lambda_{\rm s}$,
$\rho_{\rm s}$ and $P_{\rm s}$ are non vanishing constants subject to the
restrictions mentioned above. We first insert the expressions for
$\rho$ and $P$ into the equation of state (\ref{POLYTROPE}). Comparison of
the leading order terms then leads to
\begin{align}
  \beta &= \alpha \gamma, \label{TOV_ALPHABETA1} \\[10pt]
  \epsilon_3 &= \epsilon_4,
\end{align}
where $\gamma$ is the polytropic exponent. Similarly the leading order in
Eq.\,(\ref{TOV_MUR}) results in
\begin{align}
  \epsilon_1 = 1.
\end{align}
We then combine Eqs.\,(\ref{TOV_LAMBDAR}) and (\ref{TOV_PR}) to eliminate the
lapse function and insert
(\ref{TOV_ASSYMPTOTICANSATZMU})-(\ref{TOV_ASSYMPTOTICANSATZP}).
The result of comparing the two leading orders is
\begin{align}
  \alpha+1 &= \beta, \\[10pt]
  \epsilon_4 &= 1.
\end{align}
This provides a second condition for $\alpha$ and $\beta$ and 
with Eq.\,(\ref{TOV_ALPHABETA1}) we conclude that
\begin{align}
  \alpha &= \frac{1}{1-\gamma} = n, \\[10pt]
  \beta &= n+1,
\end{align}
where $n$ is the polytropic index defined in (\ref{POLYTROPICINDEX}).
Finally we use these results in Eq.\,(\ref{TOV_LAMBDAR}) for
the lapse function and obtain
\begin{align}
  \epsilon_2 &= 1.
\end{align}
We summarise the asymptotic behaviour at the surface:
\begin{align}
  \mu &= \mu_{\rm s} + \mathscr{O}(z), \label{TOV_ASSYMPTOTICMU}
         \\[10pt]
  \lambda &= \lambda_{\rm s} + \mathscr{O}(z), \\[10pt]
  \rho &= \rho_{\rm s} z^{n} + \mathscr{O}(z^{n+1}),
          \label{TOV_ASSYMPTOTICRHO} \\[10pt]
  P &= P_{\rm s} z^{n+1} + \mathscr{O}(z^{n+2}).
          \label{TOV_ASSYMPTOTICP}
\end{align}
As a consequence we will not be able to Taylor expand $\rho$ and $P$
about the surface $z=0$ unless a polytropic equation of state with
integer index $n$ is chosen. Indeed a more extensive analysis carried out with
the algebraic computing package GRTensor II
shows that higher
order terms containing the polytropic index $n$ also appear in the expansions
of $\lambda$ and $\mu$ so that these functions are subject to the same 
limitations regarding Taylor expansion. \\
The most important result of the asymptotic analysis concerns the
behaviour of the energy density $\rho$ near the surface given by
Eq.\,(\ref{TOV_ASSYMPTOTICRHO}). In particular we note that for a polytropic
index $n < 1$ or exponent $\gamma>2$ the gradient of $\rho$
with respect to the areal
radius $r$ will be infinite at the surface. The case $n=1$, i.e. $\gamma=2$
is the limiting case where $\rho$ has a finite gradient. This special case
also implies
that no fractional powers appear in the series expansions of $\lambda$,
$\mu$, $\rho$ and $P$. $\gamma=2$ is considered to provide a
qualitatively good description of the average stiffness of the equation
of state of neutron stars and thus a popular choice for the polytropic
exponent. For $n>1$ or $\gamma<2$ the energy density will have
a vanishing gradient at the surface. \\
It remains to check the asymptotic
behaviour in terms of the rescaled radial coordinate $y$. From the definition
of the speed of sound (\ref{TOV_C2}) and the results above we conclude that 
near the surface
\begin{align}
  C(z) &= \mathscr{O}(z^{1/2}), \label{TOV_ASSYMPTOTICC2}
\end{align}
which implies that
\begin{align}
  \frac{\partial \rho}{\partial y} &= C \frac{\partial \rho}{\partial r}
        = \mathscr{O}(z^{n-1/2}).
\end{align}
All other functions have vanishing gradients with respect to $y$ near the 
surface.
Consequently the rescaled coordinate allows us to calculate neutron star models
for polytropic exponents up to $\gamma=3$ without encountering infinite 
gradients and the corresponding numerical inaccuracies.
% Since models with such
%high polytropic exponents are not considered very realistic, we will restrict
%ourselves to the range $\Gamma<3$. \\

%=========================================================================
\subsubsection{Solutions of the TOV equations}
\begin{figure}[t]
  \centering
  \epsfig{file=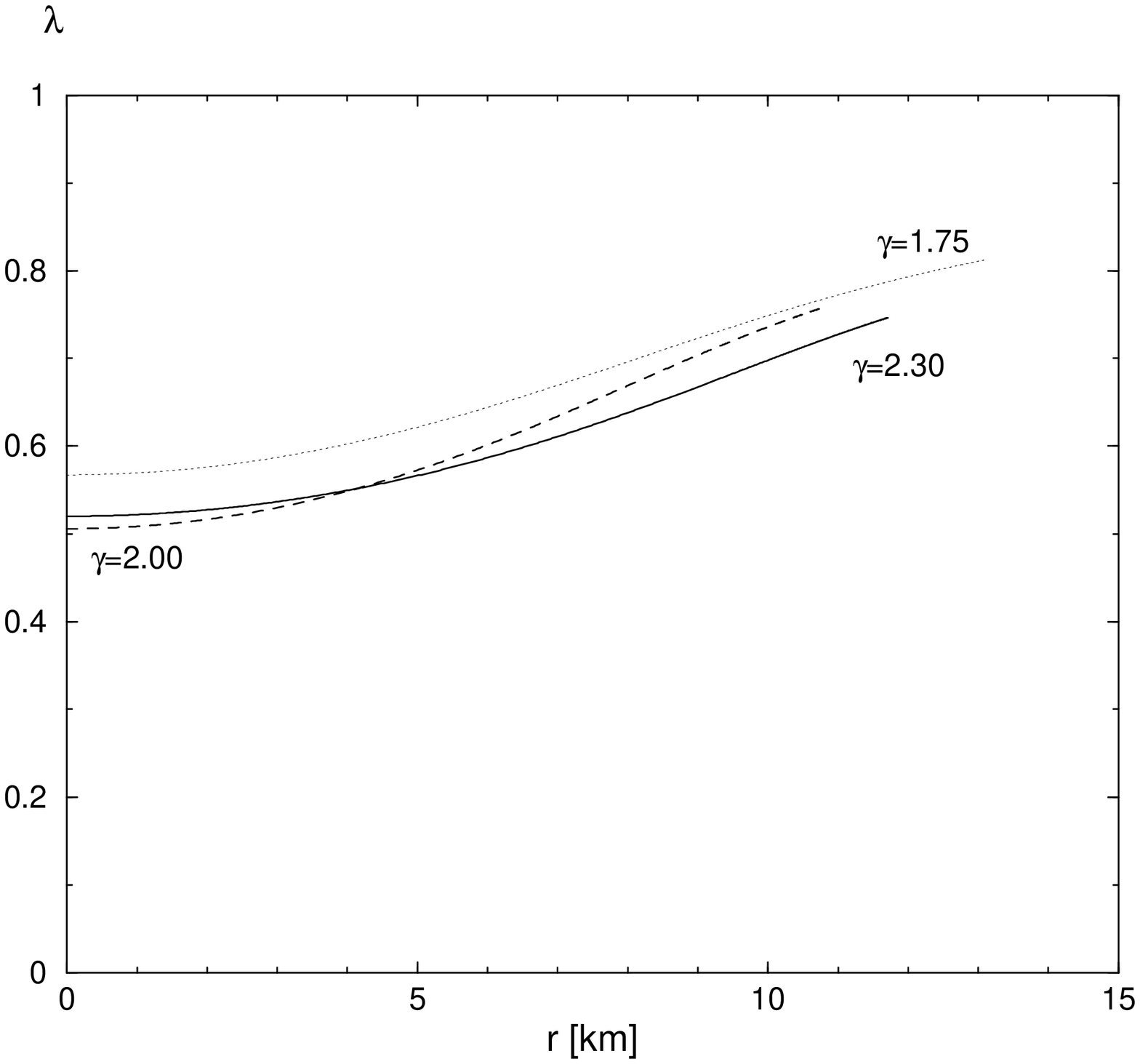, height=200pt, width=150pt, angle=-90}
  \vspace{0.5cm}
  \hspace{0.5cm}
  \epsfig{file=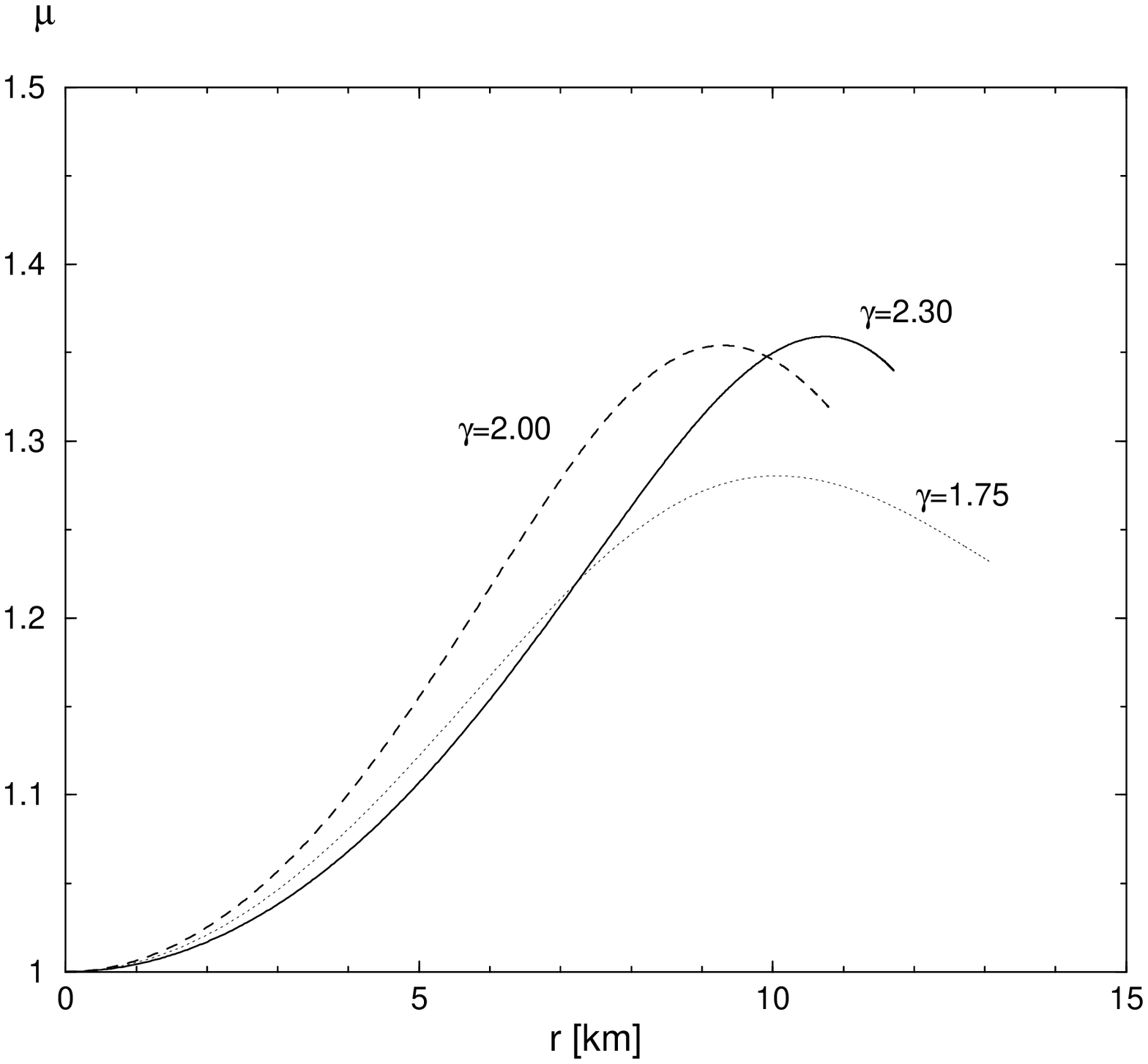, height=200pt, width=150pt, angle=-90}
  \vspace{0.5cm}
  \epsfig{file=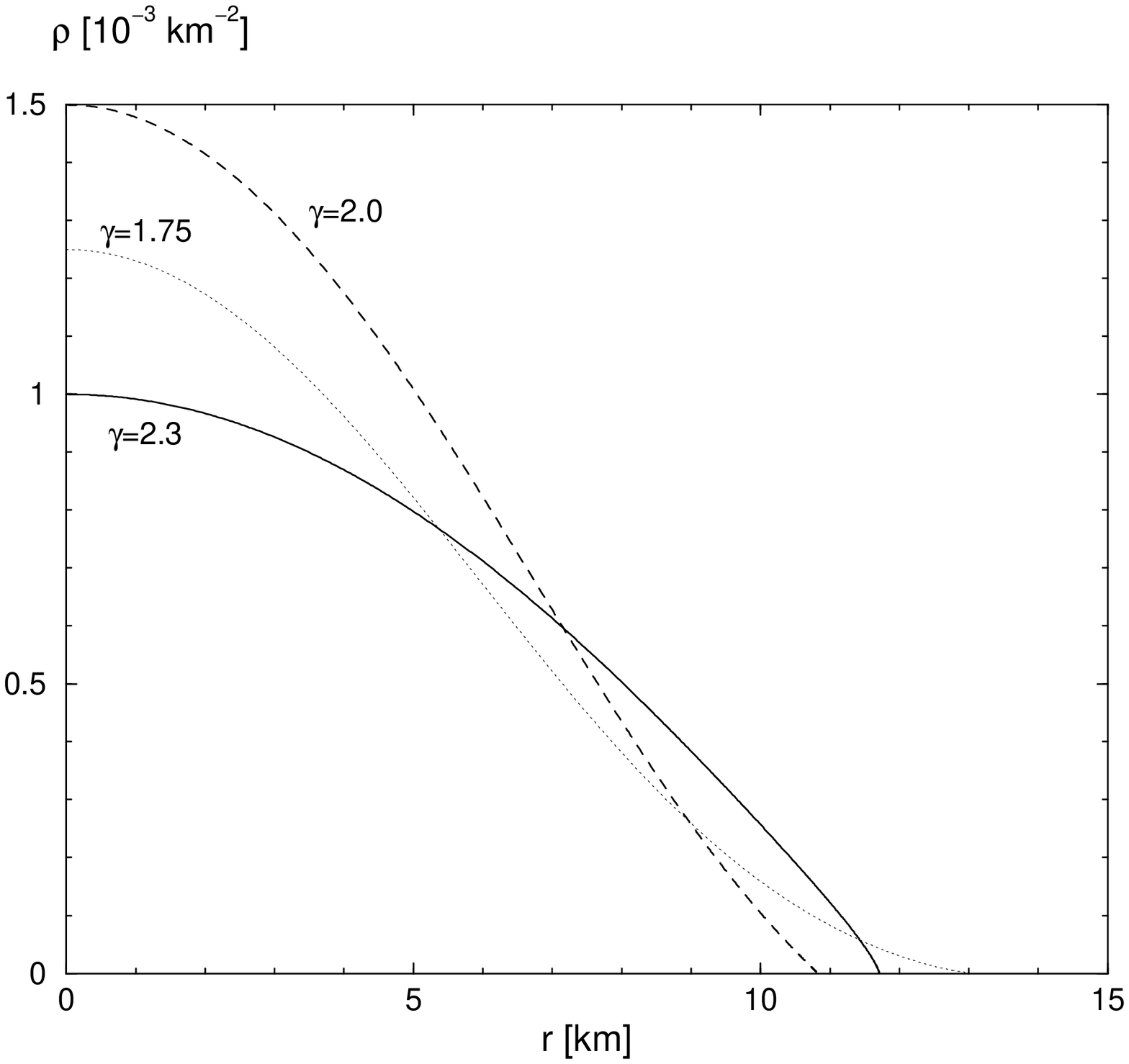, height=200pt, width=150pt, angle=-90}
  \hspace{0.5cm}
  \epsfig{file=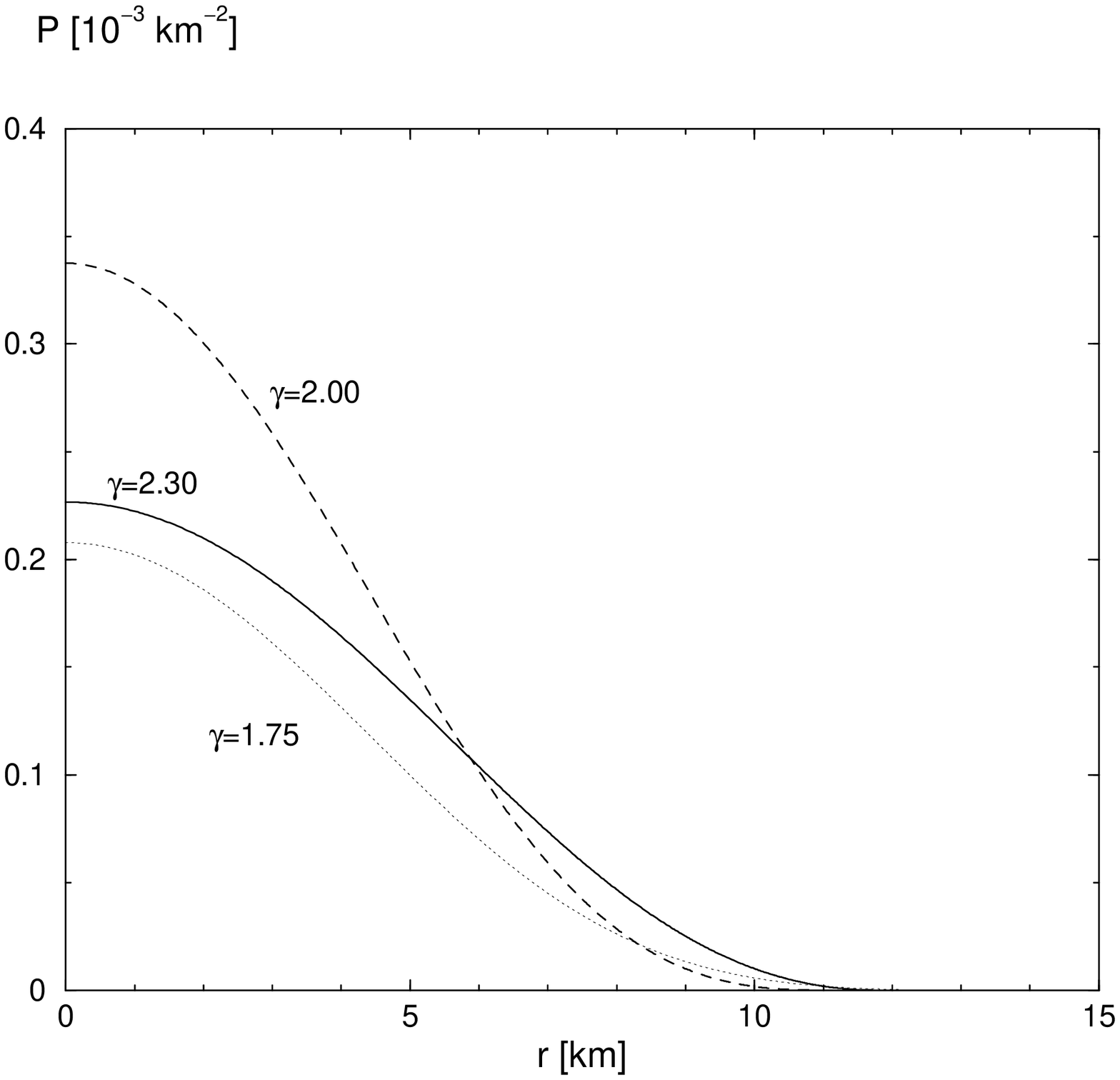, height=200pt, width=150pt, angle=-90}
  \epsfig{file=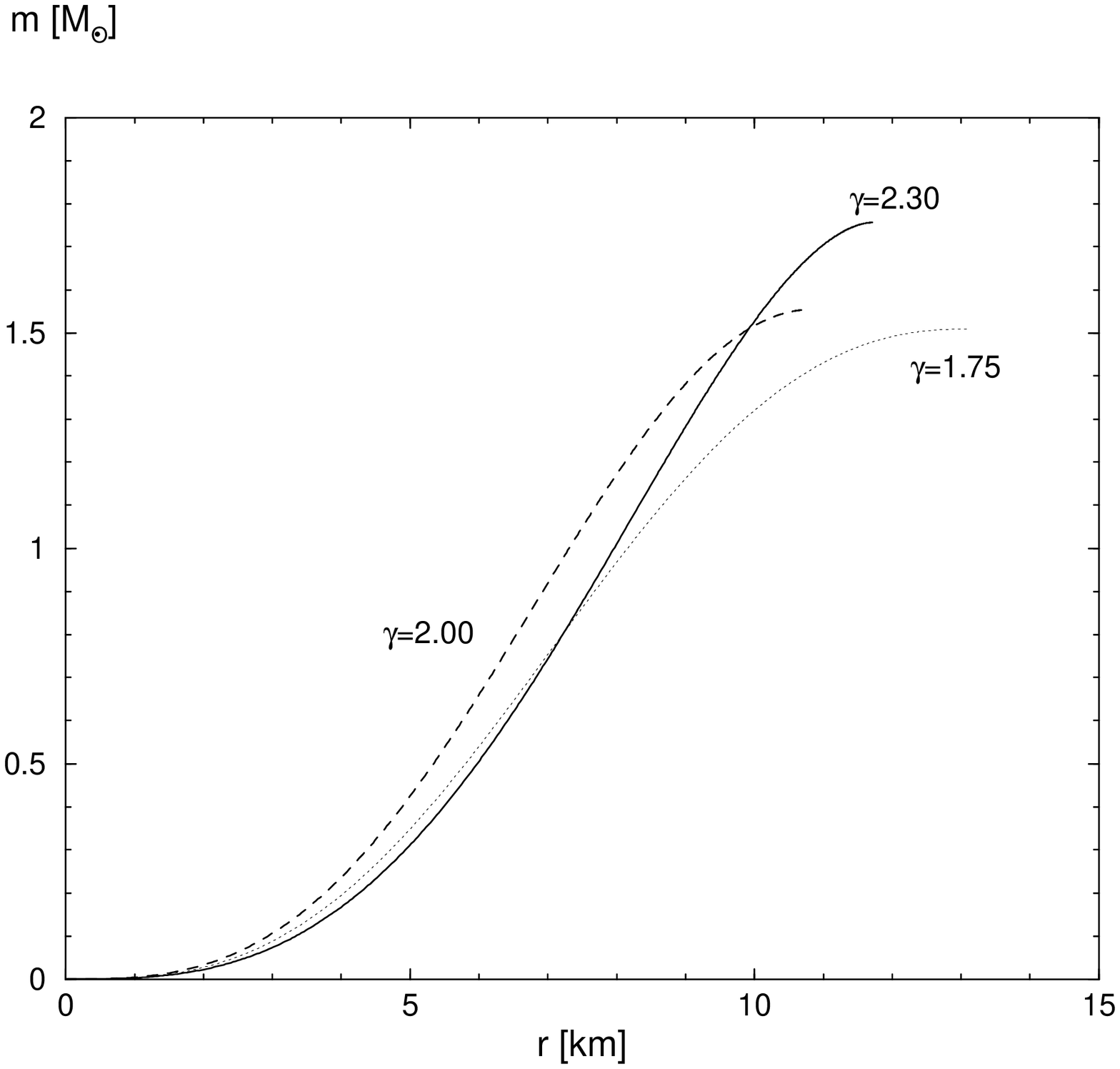, height=200pt, width=150pt, angle=-90}
  \hspace{0.5cm}
  \epsfig{file=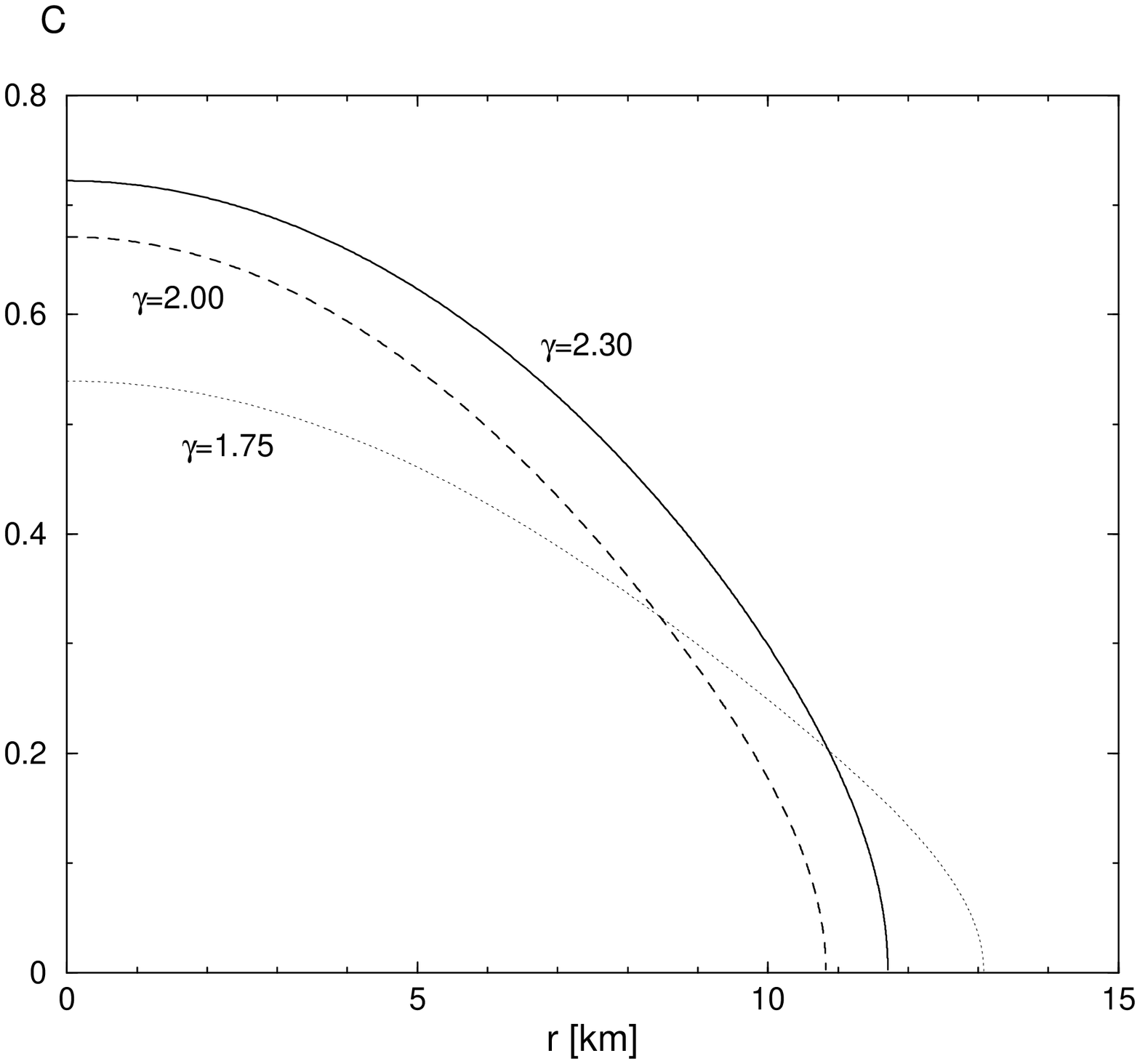, height=200pt, width=150pt, angle=-90}
  \caption{The metric functions $\lambda$, $\mu$, the energy density $\rho$,
           the pressure $P$, the mass $m$ and the speed of 
           sound $C$ are plotted
           as functions of radius for different polytropic
           indices $\gamma=1.75$ (model 1), $\gamma=2.00$ (model 3)
           and $\gamma=2.3$ (model 5).}
  \label{TOV_GAMMA}
\end{figure}

\begin{figure}[t]
  \centering
  \epsfig{file=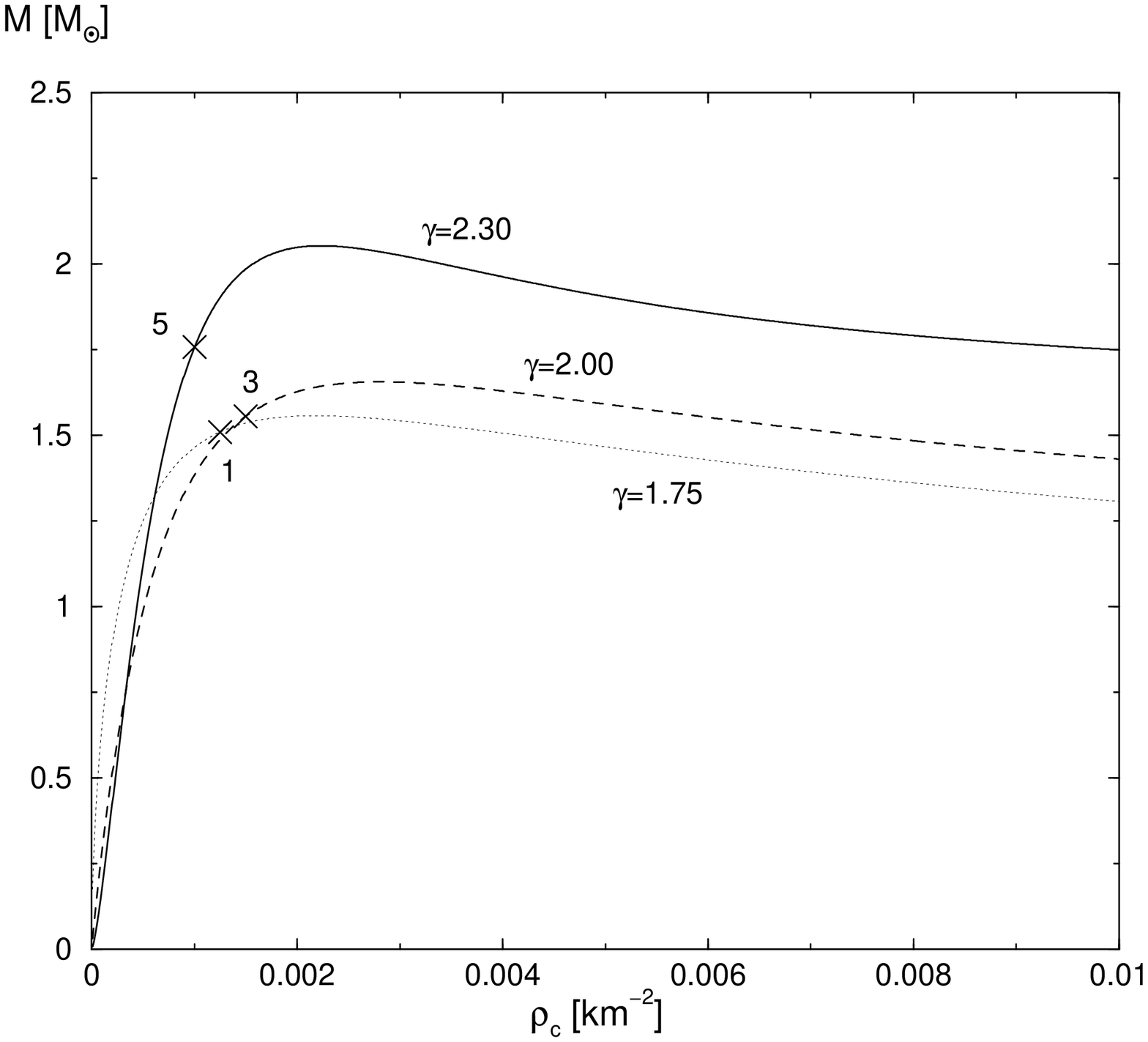, height=200pt, width=150pt, angle=-90}
  \epsfig{file=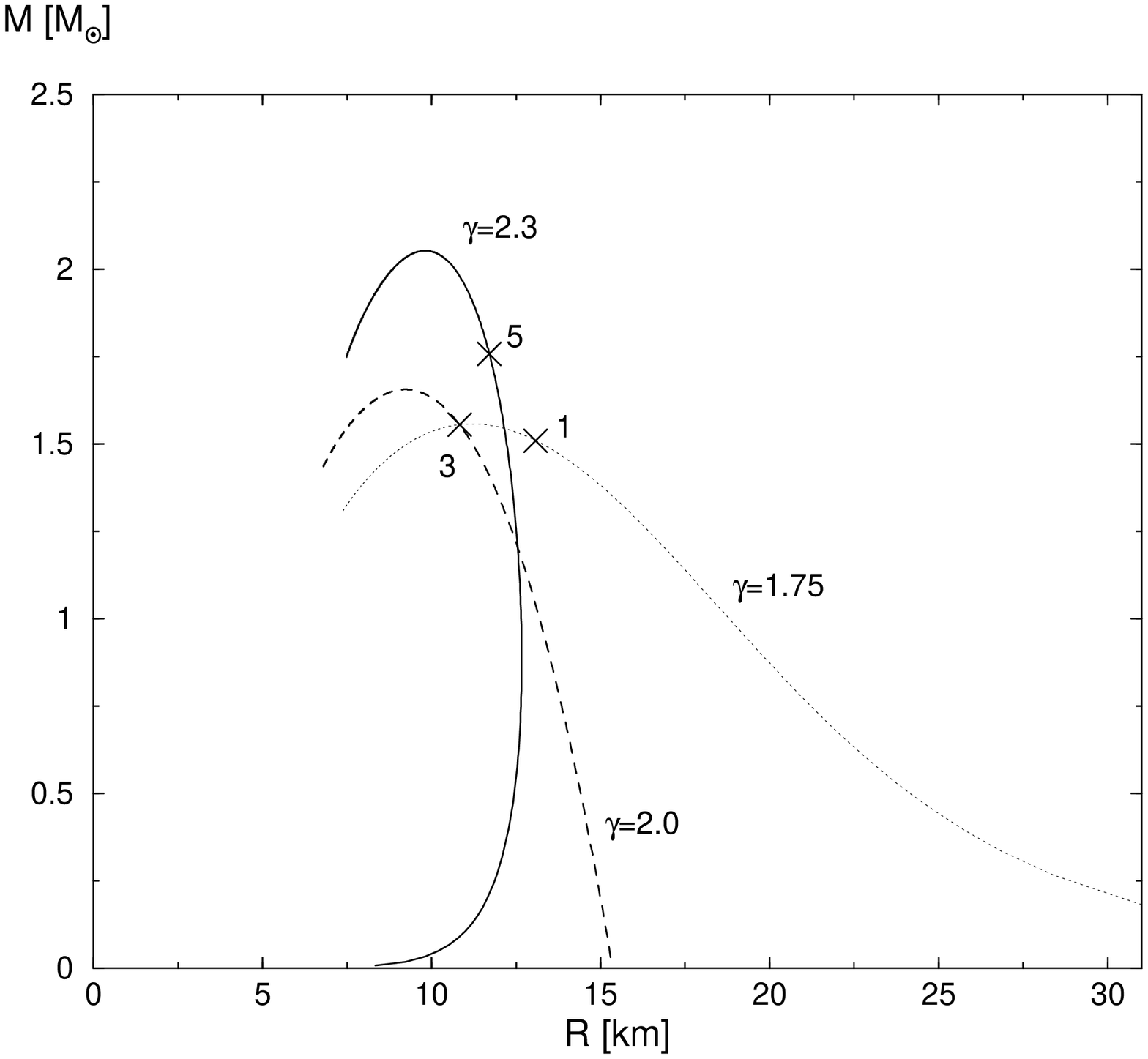, height=200pt, width=150pt, angle=-90}
  \epsfig{file=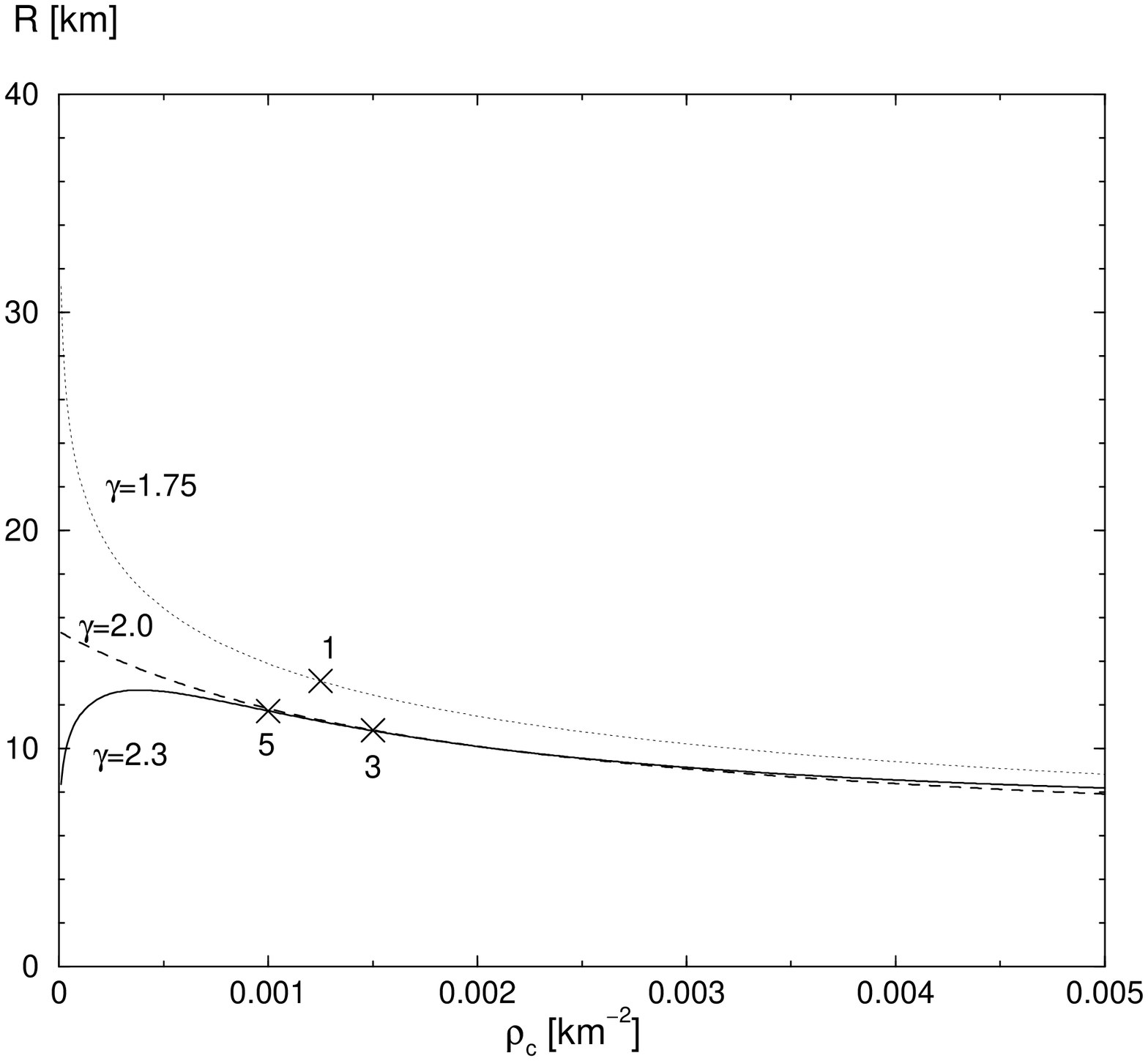, height=200pt, width=150pt, angle=-90}
  \caption{The 1-parameter families of static spherically symmetric neutron
           star models corresponding to models 1, 3 and 5
           are graphically illustrated by plotting the
           relations between the total mass, the central energy density
           and the radius of the star. The locations of neutron star
           models 1, 3 and 5 are indicated by crosses.}
  \label{TOV_FAMILIES}
\end{figure}

In view of the results of the asymptotic analysis we have numerically
solved the TOV-equations for neutron star models with different polytropic
exponents $\gamma<2$, $\gamma=2$ and $\gamma>2$.
The corresponding models are listed
in Table \ref{MODELS15} where we have included two further models
with $\gamma=2$ but different polytropic factor $K$, which we will
use to also study the variation of the solutions with $K$.
%
%\begin{table}[t]
%\begin{center}
%\caption{The parameters for five different neutron star models. We will
%         refer to these as models 1-5 in this work.}
%\begin{tabular}{c|ccc}
%  \hline
%  \hline
%  model & $\gamma$ & $K$ & $\rho_{\rm c}$ \\
%  \hline
%  1 & $1.75$ & $25\,\, {\rm km}^{1.5}$ &
%       $1.25\cdot 10^{-3}\, {\rm km}^{-2}$ \\
%  2 & $2.00$ & $100\, {\rm km}^2$ &
%       $1.5\cdot 10^{-3}\, {\rm km}^{-2}$ \\
%  3 & $2.00$ & $150\,\, {\rm km}^2$ &
%       $1.5\cdot 10^{-3}\, {\rm km}^{-2}$ \\
%  4 & $2.00$ & $200\, {\rm km}^2$ &
%       $1.5\cdot 10^{-3}\, {\rm km}^{-2}$ \\
%  5 & $2.30$ & $1800\, {\rm km}^{2.6}$ &
%       $1.0\cdot 10^{-3}\, {\rm km}^{-2}$ \\
%  \hline
%  \hline
%\end{tabular}
%\label{MODELS15}
%\end{center}
%\end{table}
%
%
%
\begin{table}[t]
\begin{center}
\caption{The parameters for five different neutron star models. We will
         refer to these as models 1-5 in this work.}
\begin{tabular}{c|ccccc}
  \hline
  \hline
  model & $\gamma$ & $K$ & $\rho_{\rm c}$ [km$^{-2}$] & $M[M_{\odot}]$
  & $R$ [km] \\
  \hline
  1 & $1.75$ & $25\,\, {\rm km}^{1.5}$ &
       $0.00125$ & 1.506 & 12.593  \\
  2 & $2.00$ & $100\, {\rm km}^2$ &
       $0.0015$ & 1.130 & 9.653 \\
  3 & $2.00$ & $150\,\, {\rm km}^2$ &
       $0.0015$ & 1.554 & 10.828 \\
  4 & $2.00$ & $200\, {\rm km}^2$ &
       $0.0015$ & 1.878 & 11.646 \\
  5 & $2.30$ & $1800\, {\rm km}^{2.6}$ &
       $0.0010$ & 1.756 & 11.710 \\
  \hline
  \hline
\end{tabular}
\label{MODELS15}
\end{center}
\end{table}
In the remainder of this work we will refer to these stellar models as
models 1-5.
The code we have used for the calculation is based on the quadrature method
described in section \ref{TOV_NUM} and uses a fourth order
Runge-Kutta scheme for the integration (see for
example \shortciteNP{Press1989}). We note, however,
that the results of the relaxation
method agree with those of the quadrature scheme with high precision and the
corresponding plots are indistinguishable from those we show in this section.
For the calculations in this section we use the rescaled coordinate $y$ and
set $r_{,x}=C$ in Eq.\,(\ref{TOV_RY}).
The code has been checked for convergence by calculating
models 1-5 for different grid resolutions starting with 250 grid points.
The resulting convergence factors $Q$ for the variables $\lambda$,
$\mu$ and $\rho$ obtained for doubling the grid resolution is shown in
Table \ref{CONV_TOV} for all 5 models. The high resolution 
reference solution has
been calculated for 2000 grid points in all cases. For the fourth order
Runge-Kutta scheme we would expect a convergence factor of 16. Even though
the results show some variation around this value they are compatible
with fourth order convergence.\\
The numerical results obtained for the 5 stellar models we will now discuss
\begin{table}[t]
  \caption{The convergence factors obtained for doubling the grid resolution
           in a fourth order Runge-Kutta scheme for solving the TOV-equations
           via quadrature. The high resolution reference solution has been
           calculated for 2000 grid points.}
\begin{center}
  \begin{tabular}{c|ccc}
    \hline \hline
    model & $Q_{\lambda}$ & $Q{\mu}$ & $Q_{\rho}$ \\
    \hline
    1 & 14.23 & 15.55 & 9.69 \\
    2 & 12.85 & 13.72 & 16.23 \\
    3 & 17.98 & 18.40 & 18.76 \\
    4 & 17.81 & 18.14 & 17.94 \\
    5 & 11.64 & 16.51 & 21.13 \\
    \hline \hline
  \end{tabular}
  \label{CONV_TOV}
\end{center}
\end{table}
have all been calculated by
using about 600 grid points.
In Fig.\,\ref{TOV_GAMMA} we plot the metric functions $\lambda$, $\mu$, 
the energy density $\rho$, the pressure $P$,
the mass $m$ and the sound speed $C$ as functions of the areal radius
$r$ for models 1, 3 and 5. We note that the different central densities of 
these
models have no impact on the qualitative behaviour of the solutions and have
only been chosen to obtain neutron star models of similar size.
The results demonstrate the dependence of the behaviour of the star near
its surface on the polytropic
exponent $\gamma$. According to the asymptotic analysis we expect the
gradient of the energy density to be zero for $\gamma=1.75$ in model 1, 
finite for the critical case $\gamma=2$ in model 3 and infinite for 
model 5 where $\gamma=2.3$. This result is compatible with the plots
of $\rho(r)$ in the middle left panel of Fig.\,\ref{TOV_GAMMA}.
The pressure gradient
on the other hand vanishes at the surfaces for any equation of state
with positive $n$ according to Eq.\,(\ref{TOV_ASSYMPTOTICP})
which agrees with the numerical results in the middle right panel. The
speed of sound shows the opposite behaviour and has an infinite gradient
independent of the polytropic index which is in
agreement with the asymptotic result given by
Eq.\,(\ref{TOV_ASSYMPTOTICC2}). With respect to the metric
we note that the radial component $\mu$ has a local maximum, while the 
lapse $\lambda$ is monotonically increasing in the stellar interior.
This behaviour becomes clear if we look at the corresponding equations
for $\lambda$ and $\mu$. We already know that $\mu_{,r}$ vanishes
at the centre. If we differentiate Eq.\,(\ref{TOV_MUR}) with respect to
$r$ only one term on the right hand side is non zero at the centre,
so that
\begin{align}
  \mu_{,rr}|_{r=0} &= 4\pi \rho_{\rm c},
\end{align}
and $\mu_{,r}$ will
\begin{figure}[t]
  \centering
  \epsfig{file=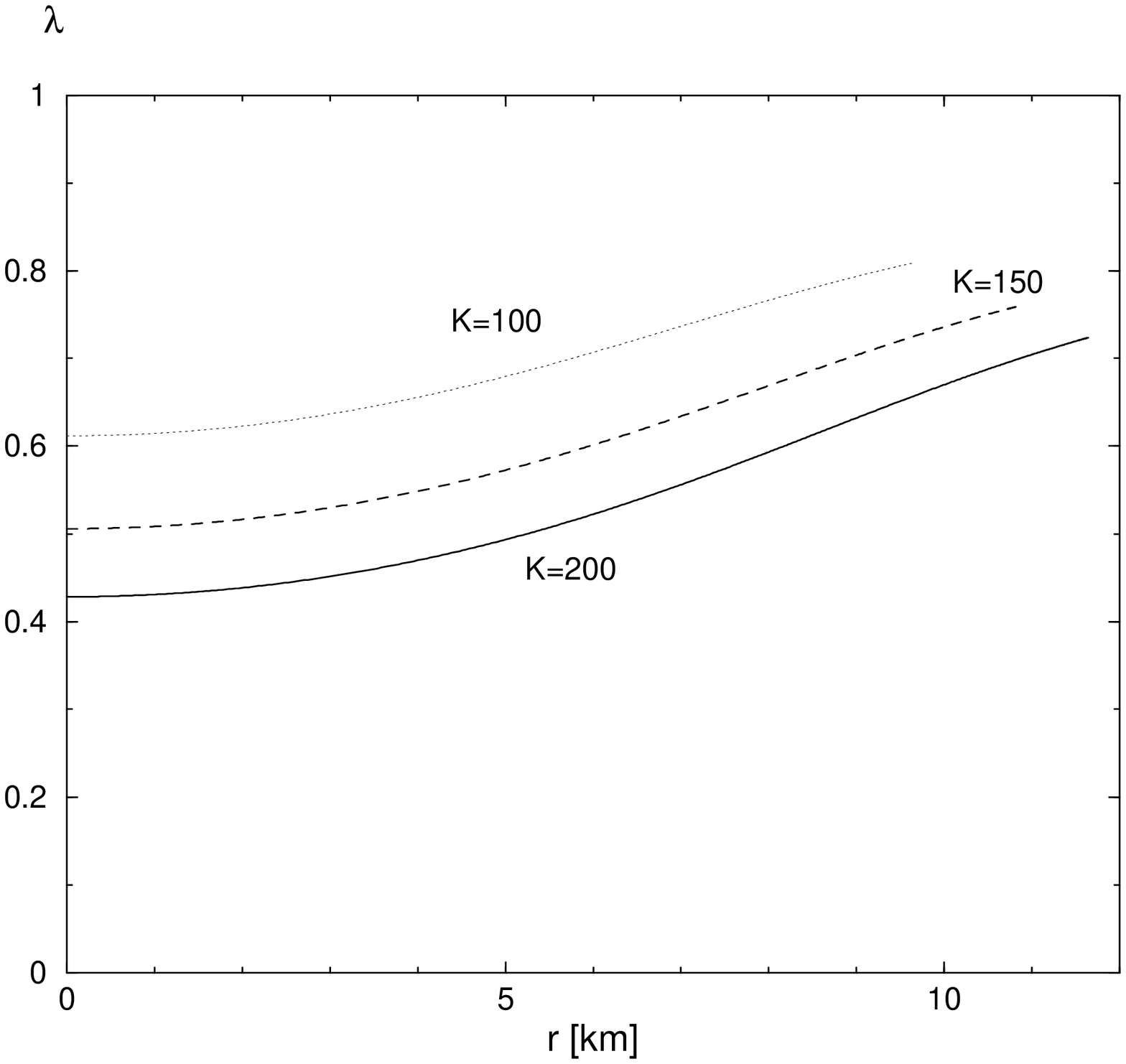, height=200pt, width=150pt, angle=-90}
  \vspace{0.5cm}
  \hspace{0.5cm}
  \epsfig{file=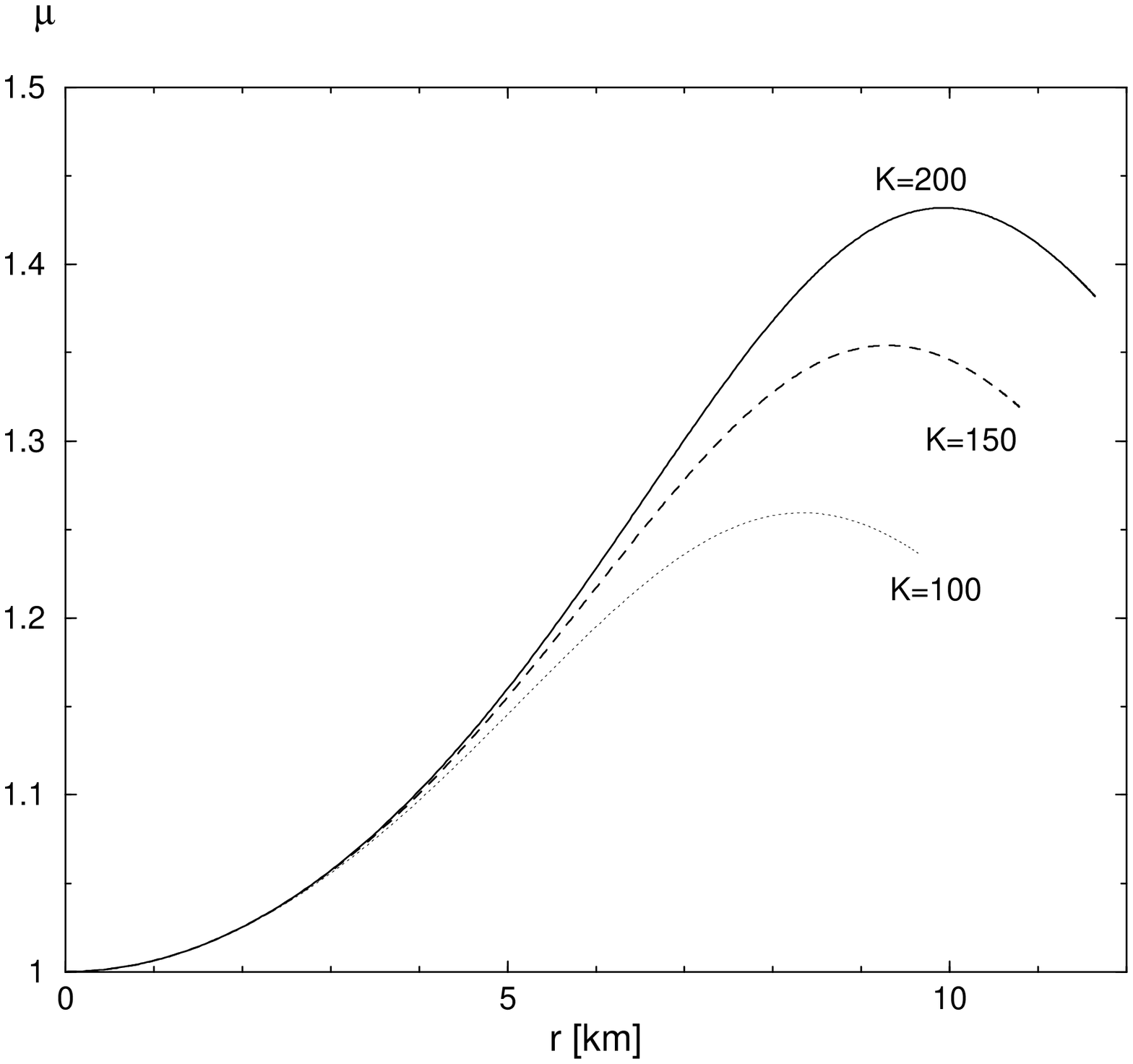, height=200pt, width=150pt, angle=-90}
  \vspace{0.5cm}
  \epsfig{file=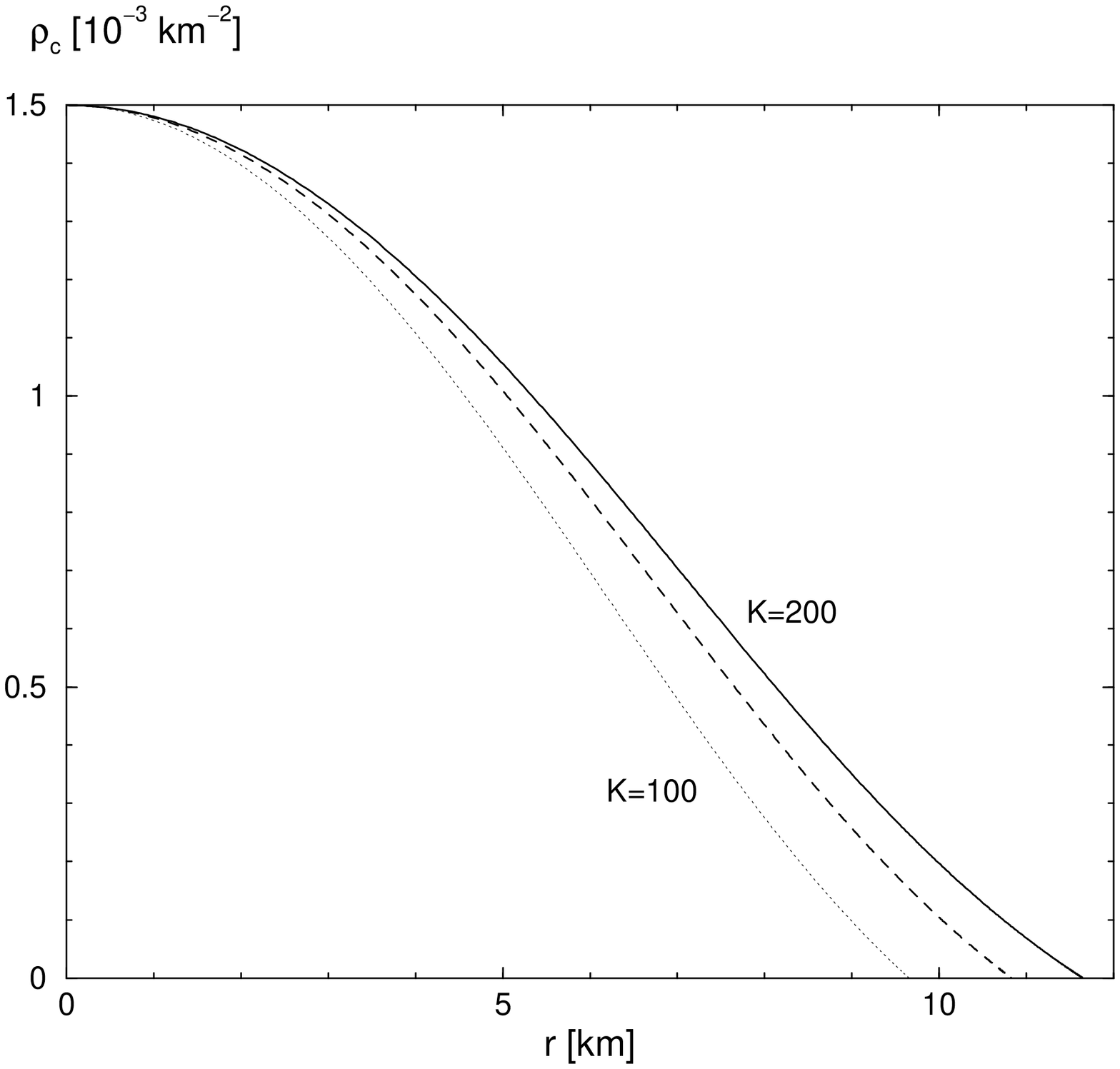, height=200pt, width=150pt, angle=-90}
  \hspace{0.5cm}
  \epsfig{file=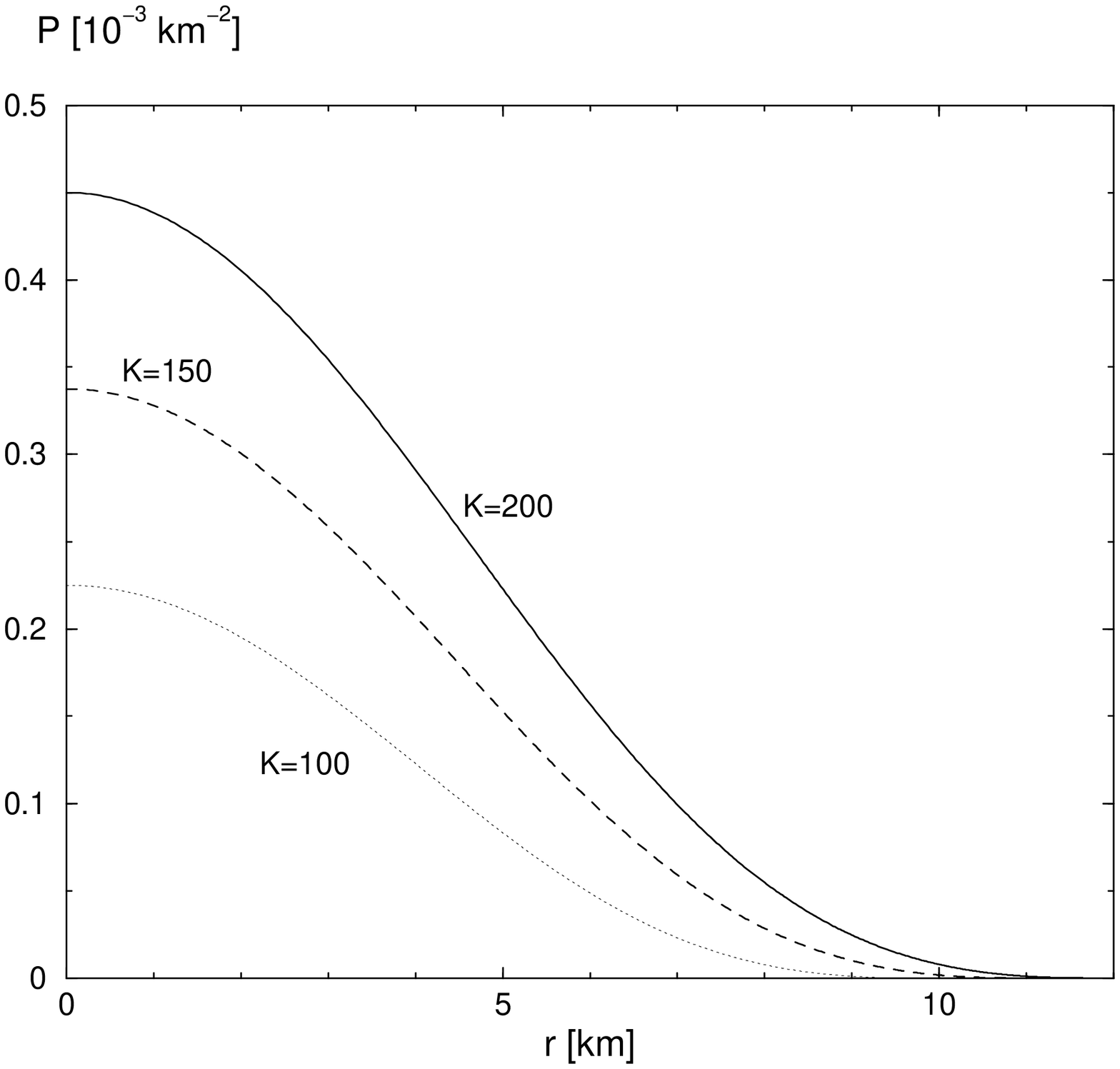, height=200pt, width=150pt, angle=-90}
  \epsfig{file=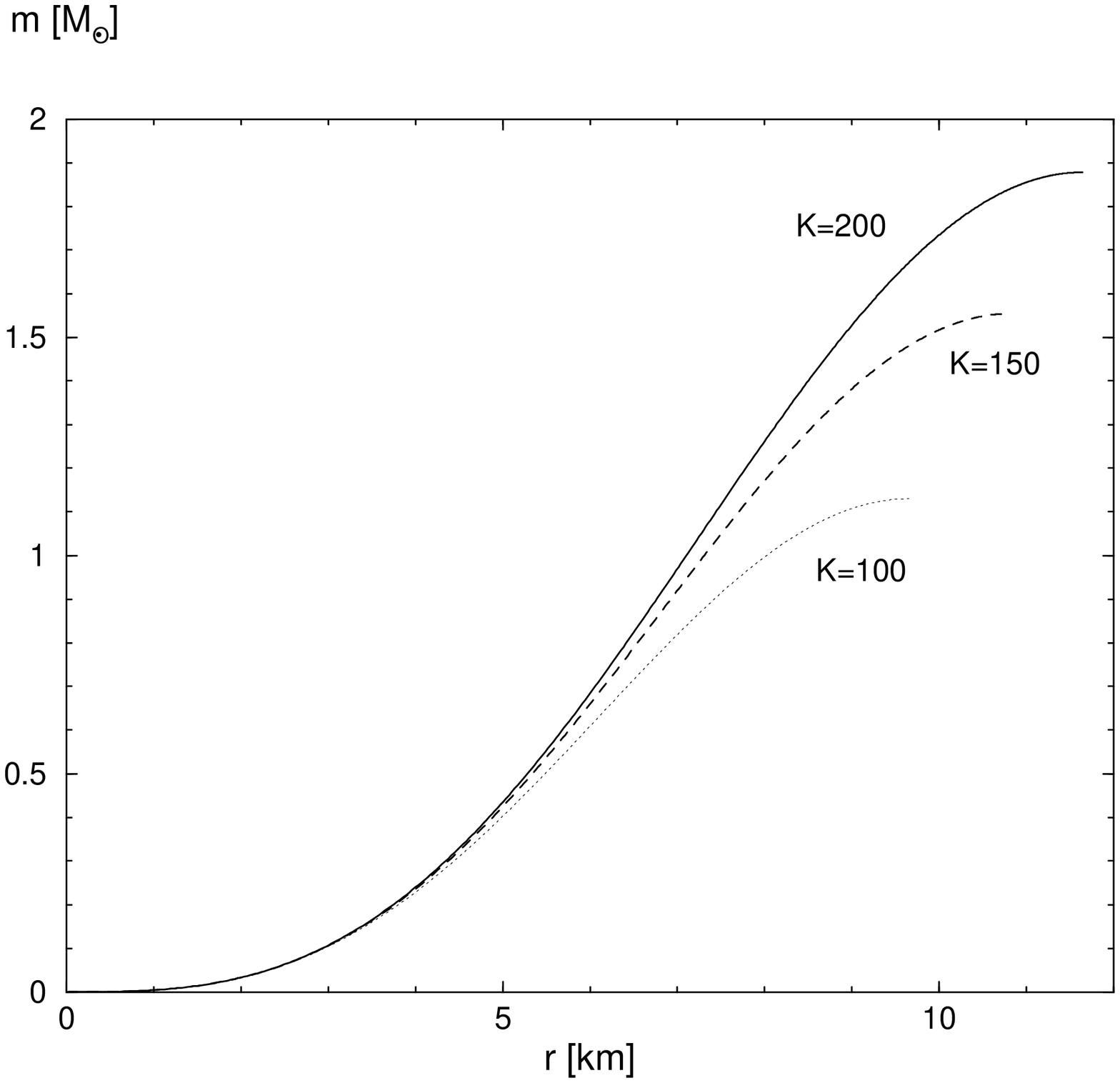, height=200pt, width=150pt, angle=-90}
  \hspace{0.5cm}
  \epsfig{file=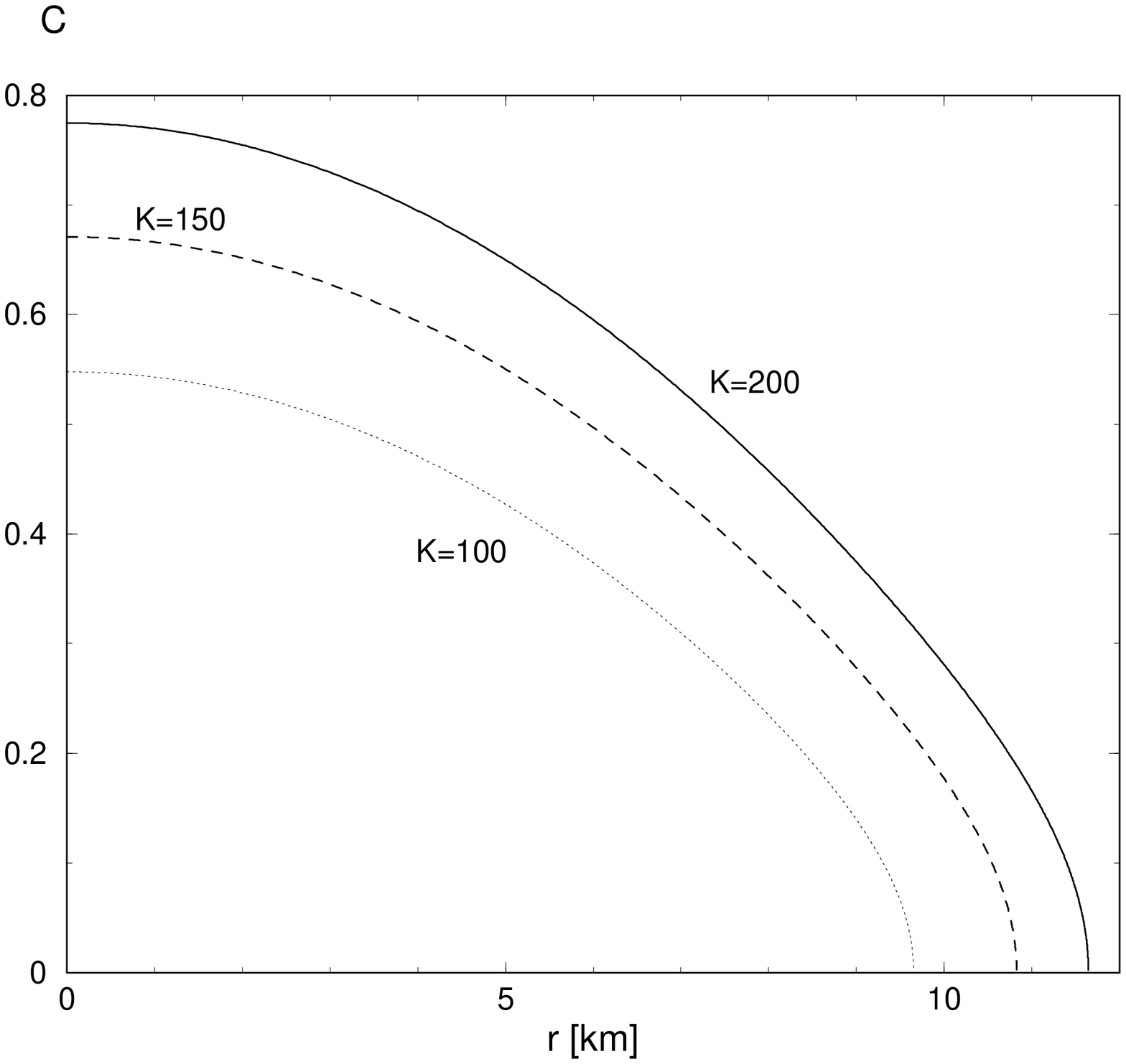, height=200pt, width=150pt, angle=-90}
  \caption{The metric functions $\lambda$, $\mu$, the energy density
           $\rho$, the pressure $P$, the mass $m$ and the speed of sound $C$
           are plotted as functions of $r$ for different polytropic
           factors $K=100\,\,{\rm km}^2$ (model 2), $K=150\,\,{\rm km}^2$
           (model 3) and $K=200\,\,{\rm km}^2$ (model 4).}
  \label{TOV_K}
\end{figure}
become positive as $r$ increases. At some point in the star,
however, the negative first term on the right hand side of Eq.\,(\ref{TOV_MUR})
will dominate the positive second term which goes to zero at the
surface and $\mu_{,r}$ will become negative. Since Eq.\,(\ref{TOV_MUR})
admits only one positive solution for $\mu$ if $\mu_{,r}=0$, $\mu$ will
monotonically decrease beyond this point.
We have already seen, however, that it cannot decrease to
$1$ or below inside the star since this conflicts with the nonzero 
mass $m$ in
Eq.\,(\ref{TOV_MUOFM}). Consequently $\mu>1$ inside the star and the
right hand side of Eq.\,(\ref{TOV_LAMBDAR}) will be positive
throughout the star which explains the monotonic behaviour of $\lambda$. \\
In order to study the dependence of the solutions on the polytropic
factor $K$ we compare the
numerical results for models 2, 3 and 4 in Fig.\,\ref{TOV_K}. In contrast to
the polytropic exponent, a variation of $K$ does not qualitatively change
the results. A larger factor $K$ leads to a larger mass
and radius of the neutron star model if all other parameters are kept fixed.
This behaviour has been observed for various polytropic models
and central densities
and can be attributed to the larger pressure that follows from a larger $K$
according to Eq.\,(\ref{POLYTROPE}). The star will thus be able to support more
mass against self gravitation and extend to larger radii. \\
We conclude the analysis of the TOV equations by studying the 1-parameter
families of solutions corresponding to the five stellar models. For this purpose
numerous solutions of the TOV equations with equations of state as given
in Table \ref{MODELS15} have been
calculated for various central densities. In Fig.\,\ref{TOV_FAMILIES} we 
plot the results in the form
of relations between central density $\rho_{\rm c}$, total radius $R$,
and total mass $M$ of the star. 
One obvious result is the maximum of the mass curves $M(R)$ and
$M(\rho_{\rm c})$ in the upper panels of the figure. It is a well known
result that these maxima separate the stable and unstable branches of the
neutron star families for a given equation of state
(see for example \citeNP{Shapiro1983}).
The stable branches consist of models with central densities below the
critical value
i.e. larger radii and the unstable branches correspond
to larger central densities
and smaller radii. In this context instability means that the frequency of 
the fundamental radial oscillation mode of the neutron star becomes imaginary
and thus its amplitude will grow exponentially in time and the
neutron star is unstable against arbitrarily small radial perturbations. The
eigenmode spectrum of radial oscillations will be discussed in the next 
section when we look at dynamic spherically symmetric stars. \\
Another interesting result is shown in the lower panel 
of Fig.\,\ref{TOV_FAMILIES} where
we plot the radius as a function of the central density. The polytropic
exponent $\gamma=2$ again appears as a critical value for which
the radius converges to a finite value as the central density goes
to zero. For smaller exponents the radius diverges in this 
limit whereas it goes to zero for exponents $\gamma>2$. We also discover
this behaviour in the upper right panel where the mass is plotted as a 
function of radius. For $\gamma<2$ a unique value of $M$ can be assigned to
any sufficiently large radius $R$.
In the critical case $\gamma=2$ equilibrium models
are only found for radii below a maximal value, but the relation $M(R)$ is 
still one to one near this maximum.
For $\gamma>2$ this is no longer the case and
for radii just below the maximal equilibrium radius we find two models
with different mass. No such qualitatively different behaviour has been found
when the polytropic factor $K$ is varied instead of $\gamma$.
It is interesting to compare the mass-radius relation for
$\gamma=2$ with the Newtonian case, where $\gamma=2$ is also a critical
value and leads to the relation $R\sim M^0={\rm const}$
(\citeNP{Shapiro1983}). The results in Fig.\,\ref{TOV_FAMILIES} indicate
that relativistic effects break this kind of degeneracy. \\
This completes our analysis of static spherically symmetric stars and in the
next section we turn our attention to the dynamic case. The equations and
results of this section will then be used to derive a fully
non-linear perturbative formulation of radial oscillations on a static
TOV background.

%=========================================================================
\subsection{Spherically symmetric dynamic stars in Eulerian coordinates}
\label{DYNAMIC}
%
%
%The study of radial oscillations of neutron stars is frequently carried
%out using the linearized form of the equations. We will see in this section,
%however, that a fully non-linear formulation of the problem leads to 
%a whole range of qualitatively new difficulties as well as
% interesting features.
%We have good reason to expect many of these difficulties and features to
%also occur in more complicated oscillation modes in 2 or 3-dimensional
%simulations. It is therefore well worth studying fully non-linear
%radial oscillations with particular emphasis on non-linear effects and the
%numerical problems resulting thereof.
In this section we will develop
an Eulerian formulation of a dynamic spherically symmetric neutron star.
For code testing purposes
it is interesting to also look at the corresponding scenario in the
Cowling approximation, i.e. with the metric frozen at its equilibrium values.
We will then use the results of the previous section to obtain a fully
non-linear perturbative formulation of the problem. In this new
approach to studying non-linear neutron star oscillations we eliminate
terms of zero order in the perturbations but keep all higher order terms
and thus obtain a formulation of the dynamic star which is equivalent to the
original non-perturbative set of equations.
% that
%allows us to simulate oscillations of arbitrary amplitude with high accuracy.
From the non-linear perturbative formulation it is easy to 
derive the linearized equations which we will use to investigate the eigenmode
spectrum of radial neutron star oscillations. After describing the numerical
methods used to evolve the dynamic neutron star in the non-linear case 
we have to discuss the ``surface problem'' which is intrinsic to any
Eulerian formulation of non-linear oscillations that involve a radial 
displacement of the stellar surface. The numerical methods we have
used to circumvent this problem will then be tested by comparing the
numerical results obtained in the linear regime with the analytic solution of
the linearized equations. By using vacuum flat space as the background,
we can emulate a non-perturbative ``standard'' approach to the numerical
evolution and compare the results with the perturbative scheme using
the TOV background. Even though the perturbative
scheme leads to highly accurate results for most stellar
models, we have not been able to find a perfectly satisfactory solution
to the surface problem.
We have therefore decided to follow a more cautious approach and use a
simplified neutron star model to investigate non-linear effects in the
evolution of radial oscillations. This model has also been used to
further test the performance of the code. The surface problem will be 
re-addressed with a Lagrangian approach in section \ref{LAGR}.

%=========================================================================
\subsubsection{The equations in the dynamic case}
\label{NONP_EQ}
We start the Eulerian formulation of the dynamic case with
the line element in radial gauge and polar slicing
\begin{align}
    ds^2 &= -\hat{\lambda}^2 dt^2 + \hat{\mu}^2 dr^2
          + r^2 (d\theta^2 + \sin^2 \theta d\phi^2),
    \label{NONP_LINEELEMENT}
\end{align}
where $\hat{\lambda}$ and $\hat{\mu}$ are now functions of $t$ and $r$ and
the ``hat'' has been introduced to distinguish them from their static 
counterparts. As in the static case we describe the matter as a perfect fluid
at zero temperature with a polytropic equation of state. As we have seen
in section \ref{TOV_EQ} this enables us to write the energy 
momentum tensor in the form
\begin{align}
  \hbox{\vec{T}}^{\mu \nu} &= (\hat{\rho} +\hat{P}) \hbox{\vec{u}}^{\mu}
  \hbox{\vec{u}}^{\nu} + \hat{P} \hbox{\vec{g}}^{\mu \nu},
  \label{NONP_EMTENSOR}
\end{align}
where again the ``hat'' on the functions $\hat{\rho}$, $\hat{P}$ means that
they are functions of $t$ and $r$. The time dependent pressure and
energy density are related by the polytropic law
\begin{align}
  \hat{P} &= K\hat{\rho}^{\gamma}, \label{DYNAMICPOLYTROPE}
\end{align}
where the polytropic parameters $\gamma$ and $K$ are the same as in the
static case. The time dependent speed of sound is defined in analogy to
Eq.\,(\ref{TOV_C2}) by
\begin{align}
  \hat{C}^2 &= \frac{\partial \hat{P}}{\partial \hat{\rho}}. \label{PERT_C2}
\end{align}
In contrast to the static case the
4-velocity will now have a non-vanishing radial component
\begin{align}
  \hbox{\vec{u}}^{\mu} &= (v,w,0,0),
\end{align}
where $v=v(r,t)$ and $w=w(r,t)$. We have
not denoted these quantities by a ``hat'' since we do not use
static counterparts in their case. The normalisation condition
$\hbox{\vec{u}}^{\mu}
\hbox{\vec{u}}_{\mu}=-1$ relates these functions by
\begin{align}
  \hat{\lambda}^2 v^2 &= 1+\hat{\mu}^2 w^2. \label{PERT_NORMU}
\end{align}
With the line element (\ref{NONP_LINEELEMENT}) and the energy momentum tensor
(\ref{NONP_EMTENSOR}) the Einstein field equations
$\hbox{\vec{G}}_{\mu \nu} = 8\pi \hbox{\vec{T}}_{\mu \nu}$ 
result in two independent constraint equations
\begin{align}
  \frac{\hat{\lambda}_{,r}}{\hat{\lambda}} &= \frac{\hat{\mu}^2-1}{2r} + 4\pi r
      \hat{\mu}^2 \left[ \hat{P} + (\hat{\rho} + \hat{P})
      \,\hat{\mu}^2w^2 \right], \label{NONP_LAMBDAR} \\[10pt]
  \frac{\hat{\mu}_{,r}}{\hat{\mu}} &= -\frac{\hat{\mu}^2-1}{2r} + 4\pi r
      \hat{\mu}^2 \left[ \hat{\rho} + (\hat{\rho} + \hat{P})
      \hat{\mu}^2w^2 \right]. \label{NONP_MUR}
\end{align}
It is a well known result that there are no gravitational degrees of freedom
in spherical symmetry and we therefore expect to be able to determine
the metric functions on each time slice without knowledge of their history.
This is compatible with the result that the field equations can be given
in the form of constraint equations only. The degrees of freedom of the 
scenario are thus entirely contained in the matter variables, whose evolution
is determined by the equations of hydrodynamics 
$\nabla_{\mu} \hbox{\vec{T}}^{\mu \nu}=0$. In our case we can write these
equations as a quasi linear system of PDEs
\begin{align}
  \hat{\rho}_{,t} + \tilde{\alpha}_{11} \hat{\rho}_{,r} 
                  + \tilde{\alpha}_{12} w_{,r} &= \tilde{b}_1,
  \label{NONP_RHOT}  \\[10pt]
  w_{,t}          + \tilde{\alpha}_{21} \hat{\rho}_{,r} 
                  + \tilde{\alpha}_{11} w_{,r} &= \tilde{b}_2,
  \label{NONP_WT}
\end{align}
where the coefficients are given by
\begin{align}
  D =&\,\, v\left( 1- \hat{C}^2\frac{\hat{\mu}^2 w^2}{1+ \hat{\mu}^2 w^2}
           \right), \label{NONP_D} \\[10pt]
  \tilde{\alpha}_{11} =&\,\, \frac{w(1-\hat{C}^2)}{D}, \\[10pt]
  \tilde{\alpha}_{12} =&\,\, \frac{\hat{\rho} + \hat{P}}{(1+\hat{\mu}^2 w^2) D},
                 \\[10pt]
  \tilde{\alpha}_{21} =&\,\, \frac{\hat{C}^2}{(\hat{\rho} + \hat{P})
               \hat{\mu}^2 D}, \label{NONP_ALPHA21} \\[10pt]
  \tilde{b}_1         =& \,\,-\frac{1}{D}(\hat{\rho} + \hat{P})
                 \left(\frac{w\hat{\mu}_{,r}/\hat{\mu}
                            +v\hat{\mu}_{,t}/\hat{\mu}}
                 {1+\hat{\mu}^2 w^2} + 2\frac{w}{r} \right), \\[10pt]
  \tilde{b}_2        =& \,\, -\frac{1}{D}\left[
                 w^2\left( \frac{\hat{\mu}_{,r}}{\hat{\mu}}
                 +\frac{\hat{\lambda}_{,r}}{\hat{\lambda}}
                 -\frac{2}{r}\hat{C}^2
                 +2\frac{v}{w} \frac{\hat{\mu}_{,t}}{\hat{\mu}}
                 \right)
                 +\frac{\hat{\lambda}_{,r}/\hat{\lambda}}{\hat{\mu}^2} 
                 \right]. \label{NONP_B2}
\end{align}
In practice we calculate the derivatives of the metric functions 
that appear in these coefficients from the constraint equations
(\ref{NONP_LAMBDAR}), (\ref{NONP_MUR}) and a third field equation
\begin{align}
  \frac{\hat{\mu}_{,t}}{\hat{\mu}} &= -4\pi r \hat{\mu}^2 \hat{\lambda}^2
      v w (\hat{\rho} + \hat{P}), \label{NONP_MUT}
\end{align}
which is an automatic consequence of the two constraints, their derivatives
and the matter equations.
We therefore calculate the coefficients $\tilde{\alpha}_{ij}$ and
$b_{i}$ without approximating any derivatives with finite difference
expressions. \\
We have already mentioned in the discussion of the static case that a
numerically superior performance is obtained if we transform to a new
radial coordinate $y$ defined by Eq.\,(\ref{TOV_YOFR}). We note however that
we need to calculate the corresponding static model first to obtain the
static sound speed $C$. In the perturbative approach which we will discuss
below that is done as a matter of course. There we will provide
a formulation of the perturbative equations that includes both choices
for the radial coordinate analogous to Eqs.\,(\ref{TOV_RY})-(\ref{TOV_PY}).
In the Cowling approximation the set of equations corresponding 
to (\ref{NONP_RHOT})-(\ref{NONP_B2}) describes
a dynamic, spherically symmetric perfect fluid in a fixed gravitational 
potential. We obtain these equations by the following modifications:
\begin{list}{\rm{(\arabic{count})}}{\usecounter{count}
             \labelwidth1cm \leftmargin1.5cm \labelsep0.4cm \rightmargin0cm
             \parsep0.5ex plus0.2ex minus0.1ex \itemsep0ex plus0.2ex}
\item the constraint equations for the dynamic metric functions
      (\ref{NONP_LAMBDAR}), (\ref{NONP_MUR}) are replaced by the
      corresponding TOV equations (\ref{TOV_LAMBDAR}), (\ref{TOV_MUR})
      which have to be solved only at the start of the evolution,
\item in the coefficients $\tilde{\alpha}_{11}$, $\tilde{\alpha}_{12}$,
      $\tilde{\alpha}_{21}$ and $\tilde{b}_1$ all occurrences
      of $\hat{\mu}$, $\hat{\lambda}$,
      $\hat{\lambda}_{,r}$/$\hat{\lambda}$ and $\hat{\mu}_{,r}/\hat{\mu}$
      are replaced with their static
      analogues $\lambda$, $\mu$, $\lambda_{,r}/\lambda$ and $\mu_{,r}/\mu$
      respectively and $\hat{\mu}_{,t}/\hat{\mu}$ is set to zero,
\item the coefficient function $\tilde{b}_2$ is replaced with the 
      slightly modified version
      \begin{align}
        \bar{b}_2         =& \,\, -\frac{1}{D}\left\{
                       w^2\left[ \left( \frac{\mu_{,r}}{\mu}
                       +\frac{\lambda_{,r}}{\lambda} \right)
                       \left(1-\hat{C}^2 \right) -\frac{2}{r}\hat{C}^2 \right]
                       +\frac{\lambda_{,r}/\lambda}{\mu^2}
                       \right\}.
      \end{align}
\end{list}
These modifications are rather simple so that we incorporate both options, the
evolution with time dependent metric and the Cowling
approximation in one code. A user specified initial parameter determines
which version is to be run. Before we describe the numerical 
implementation, we need to rewrite the equations of this subsection
in a perturbative form.

%=========================================================================
\subsubsection{A fully non-linear perturbative formulation}
\label{PERT_EQ}
In this section we will decompose the time dependent quantities
$\hat{\lambda}$, $\hat{\mu}$ and $\hat{\rho}$ into static 
background contributions and time dependent perturbations. We will see
that the TOV equations are still present in the dynamic equations,
for example in the terms
$\alpha_{21} \hat{\rho}_{,r} - b_2$ in Eq.\,(\ref{NONP_WT}). It is the 
elimination of these zero order terms and the ensuing numerical
inaccuracies which provides the motivation for our perturbative formulation.
We start by decomposing the time dependent functions into a static
background plus a time dependent perturbation
\begin{align}
  \hat{\lambda}(t,r) &= \lambda(r) + \delta \lambda(t,r), \label{PERT_LAMBDA}
  \\[10pt]
  \hat{\mu}(t,r) &= \mu(r) + \delta \mu(t,r), \label{PERT_MU} \\[10pt]
  \hat{\rho}(t,r) &= \rho(r) + \delta \rho(t,r), \label{PERT_RHO} \\[10pt]
  \hat{P}(t,r) &= P(r) + \delta P(t,r). \label{PERT_P} 
\end{align}
The radial velocity component $w$ vanishes in the static limit and therefore
represents a perturbation in itself. The time dependent functions
$\hat{P}$, $\hat{C}$ and $v$ are dependent variables and
thus considered functions of the fundamental variables
$\hat{\lambda}$, $\hat{\mu}$, $\hat{\rho}$ and $w$ according to
Eqs.\,(\ref{DYNAMICPOLYTROPE}), (\ref{PERT_C2}) and (\ref{PERT_NORMU}).
We stress that the perturbations are finite and that no assumption with
regard to their size has been made. \\
We start rewriting the dynamic
equations with the constraint equation for $\hat{\lambda}$. If we
insert Eqs.\,(\ref{PERT_LAMBDA})-(\ref{PERT_RHO}) into (\ref{NONP_LAMBDAR})
and multiply with $\hat{\lambda}$ we obtain
\begin{align}
  \begin{split}
  \lambda_{,r} + \delta \lambda_{,r} =& \,\,\, \lambda
      \frac{\mu^2-1}{2r} + 4\pi r
      \lambda \mu^2 P + 4\pi r \lambda \mu^2
      \left[\delta P + (\rho + \delta \rho + \hat{P}) \hat{\mu}^2 w^2 \right]
      + \lambda \frac{2\mu \delta \mu + \delta \mu^2}{2r}
      \\[10pt]
  &   + \delta \lambda \frac{\hat{\mu}^2-1}{2r}
      +4\pi r \left[\lambda(2\mu \delta \mu + \delta \mu^2) + \delta \lambda
      \hat{\mu}^2 \right] \left[ \hat{P}+(\rho + \delta \rho + \hat{P})
      \hat{\mu}^2 w^2 \right]. \label{COMPLETE_LAMBDAR}
  \end{split}
\end{align}
The crucial terms are the first on the left and the first
two terms on the right hand side. We know that these terms
will cancel each other
identically according to Eq.\,(\ref{TOV_LAMBDAR}) if a solution of the
static equations is chosen as a background. Numerically, however,
this will not be the case because of truncation errors.
% This becomes immediately
%obvious if we assume $\lambda$, $\mu$ and $P$ given as an exact solution
%of the TOV equations.
This residual error will inevitably contaminate
the numerical evolution of the dynamic scenario. In other words the numerical
accuracy we will obtain is limited by the numerical accuracy of the static
background and not by that of the dynamic signal we are interested in.
The severeness of this effect will depend on the relative size of the 
perturbations with respect to the background. For very large perturbations
the numerical contamination will be less significant and for very small
perturbations we may satisfy ourselves with a linearized code. For
perturbations of intermediate strength, however, which are still 
smaller than the background but are large enough to give rise to non-linear 
effects, the numerical contamination will severely affect the
evolution and may give rise to spurious phenomena. \\
% The scenario many studies
%of neutron star oscillations will ultimately target is an
%oscillation mode which is unstable to the emission of gravitational waves.
%Such a mode, starting with a small amplitude, will increase 
%in strength and gradually enter the non-linear regime. It is therefore
%of particular interest to investigate the impact of non-linear effects
%in the weakly and moderately non-linear regime. Such a study will necessarily
%be a study in 2 or 3 spatial dimensions and is beyond the scope of this
%work. The development of a fully non-linear numerical technique which 
%provides high accuracy for perturbations of any given amplitude will,
%however, pave some of the way towards achieving this goal. A further benefit
%of the high accuracy of this method is the possibility of testing the code
%in the linear regime, the only regime where an exact solution is known
%with (for all practical purposes) arbitrary accuracy. \\
We return to Eq.\,(\ref{COMPLETE_LAMBDAR}) and continue the
perturbative formulation of the dynamic case. Since we know that the
zero order terms cancel each other, we can simply subtract them
from the equation.
% We thus keep higher order terms in the equations only
%and ensure that the numerical accuracy we obtain will always be the
%accuracy of the signal we are interested in.
The perturbative equation for $\hat{\lambda}$ then becomes
\begin{align}
  \begin{split}
  \frac{\delta \lambda_{,x}}{r_{,x}} 
      =& \,\,\, \lambda \frac{2\mu \delta \mu + \delta \mu^2}{2r}
      + \delta \lambda \frac{\hat{\mu}^2-1}{2r} + 4\pi r \lambda \mu^2
      \left[\delta P + (\rho + \delta \rho + \hat{P}) \hat{\mu}^2 w^2 \right]
      \\[10pt]
  & +4\pi r \left[\lambda(2\mu \delta \mu + \delta \mu^2) + \delta \lambda
      \hat{\mu}^2 \right] \left[ \hat{P}+(\rho + \delta \rho + \hat{P})
      \hat{\mu}^2 w^2 \right], \label{PERT_LAMBDAR}
  \end{split}
\end{align}
where we have also implemented the transformation to the generalised
radial coordinate $x$.
Proceeding in the same way we rewrite the constraint equation for $\hat{\mu}$
\begin{align}
  \begin{split}
  \frac{\delta \mu_{,x}}{r_{,x}} 
      =&\,\,\, -\mu \frac{2\mu \delta \mu + \delta \mu^2}{2r}
      -\delta \mu \frac{\hat{\mu}^2-1}{2r}
      + 4\pi r \mu^3 \left[ \delta \rho + (\rho + \delta \rho + \hat{P})
      \hat{\mu}^2 w^2 \right] \\[10pt]
  & +4\pi r (3 \mu^2 \delta \mu + 3 \mu \delta \mu^2 + \delta \mu^3)
      \left[\hat{\rho} + (\rho + \delta \rho + \hat{P})\hat{\mu}^2 w^2
      \right]. \label{PERT_MUR}
  \end{split}
\end{align}
The reformulation of the matter equations (\ref{NONP_RHOT}) and (\ref{NONP_WT})
is particularly simple due to their quasi linear nature. We obtain
\begin{align}
  \delta \rho_{,t} + \alpha_{11} \delta \rho_{,x} + \alpha_{12} w_{,x} &= b_1,
                      \label{PERT_DRHOT} \\[10pt]
  w_{,t}           + \alpha_{21} \delta \rho_{,x} + \alpha_{11} w_{,x} &= b_2,
                      \label{PERT_WT}
\end{align}
with the coefficient functions
\begin{align}
  D =&\,\, v\left( 1- \hat{C}^2\frac{\hat{\mu}^2 w^2}{1+ \hat{\mu}^2 w^2}
           \right), \label{PERT_D} \\[10pt]
  \alpha_{11} =&\,\, \frac{w(1-\hat{C}^2)}{r_{,x}D},
           \label{PERT_ALPHA11} \\[10pt]
  \alpha_{12} =&\,\, \frac{\rho + \delta \rho
           + \hat{P}}{(1+\hat{\mu}^2 w^2)r_{,x} D},
                 \label{PERT_ALPHA12} \\[10pt]
  \alpha_{21} =&\,\, \frac{\hat{C}^2}{(\rho +\delta \rho + \hat{P})
               \hat{\mu}^2r_{,x} D}, \label{PERT_ALPHA21} \\[10pt]
  b_1         =& \,\,-\frac{1}{D}\left[(\rho + \delta \rho + \hat{P})
                 \left(\frac{w\hat{\mu}_{,r}/\hat{\mu}
                            +v\hat{\mu}_{,t}/\hat{\mu}}
                 {1+\hat{\mu}^2 w^2} + 2\frac{w}{r} \right)
                 + \rho_{,r} w (1-C^2)\right], \\[10pt]
  \begin{split}
  b_2         =& \,\, -\frac{1}{D}\left\{
                 w^2\left( \frac{\hat{\mu}_{,r}}{\hat{\mu}}
                 +\frac{\hat{\lambda}_{,r}}{\hat{\lambda}}
                 -\frac{2}{r}\hat{C}^2
                 +2\frac{v}{w} \frac{\hat{\mu}_{,t}}{\hat{\mu}}
                 \right) +
                 \frac{1}{\hat{\mu}^2 (\rho + \delta \rho + \hat{P})}
                  \right. \\[10pt]
               &  \left. \left[ (\hat{C}^2 - C^2) \rho_r 
                 + \frac{\delta \lambda}{\hat{\lambda}}
                 C^2 \rho_r + \frac{\delta \lambda_r}{\hat{\lambda}} (\rho + P)
                 + \frac{\hat{\lambda}_r}{\hat{\lambda}}(\delta \rho 
                 + \delta P) \right] \right\}. \label{PERT_B2}
  \end{split}
\end{align}
Except for the coefficient $b_2$ where background terms have been eliminated
by using the TOV-equations
we note the similarity with the coefficients given
in Eqs.\,(\ref{NONP_D})-(\ref{NONP_B2}) in the non-perturbative formulation. \\
In order to derive the equations in the Cowling approximation we
have to proceed in analogy to the previous section.
\begin{list}{\rm{(\arabic{count})}}{\usecounter{count}
             \labelwidth1cm \leftmargin1.5cm \labelsep0.4cm \rightmargin0cm
             \parsep0.5ex plus0.2ex minus0.1ex \itemsep0ex plus0.2ex}
\item The metric perturbations $\delta \mu$ and $\delta \lambda$ are
      set to zero.
\item All occurrences of $\hat{\lambda}_{,r}/\hat{\lambda}$ and
      $\hat{\mu}_{,r}/\hat{\mu}$ are replaced with $\lambda_{,r}/\lambda$
      and $\mu_{,r}/\mu$ which are given by the TOV equations
      (\ref{TOV_LAMBDAR}), (\ref{TOV_MUR}).
\item $\hat{\mu}_{,t}/\hat{\mu}$ is set to zero.
\item The coefficient $b_2$ is replaced by 
      \begin{align}
      \begin{split}
      b_2         =& \,\, -\frac{1}{D}\left\{
                     w^2\left[ \left( \frac{\hat{\mu}_{,r}}{\hat{\mu}}
                     +\frac{\hat{\lambda}_{,r}}{\hat{\lambda}} \right)
                     (1-\hat{C}^2)-\frac{2}{r}\hat{C}^2
                     +2\frac{v}{w} \frac{\hat{\mu}_{,t}}{\hat{\mu}}
                     \right] +
                     \frac{1}{\hat{\mu}^2 (\rho + \delta \rho + \hat{P})}
                      \right. \\[10pt]
                   &  \left. \left[ (\hat{C}^2 - C^2) \rho_r 
                     + \frac{\delta \lambda}{\hat{\lambda}} C^2 \rho_r 
                     +\frac{\delta \lambda_r}{\hat{\lambda}} (\rho + P)
                     + \frac{\hat{\lambda}_r}{\hat{\lambda}}(\delta \rho 
                     + \delta P) \right] \right\}.
      \end{split}
      \end{align}
\end{list}
This completes our derivation of the equations for a dynamical spherically
symmetric neutron star. In later sections we will numerically investigate
the system of partial differential equations
(\ref{PERT_LAMBDAR})-(\ref{PERT_WT}) with the
coefficient functions (\ref{PERT_D})-(\ref{PERT_B2}) and
the corresponding system in the Cowling approximation. Before that, we will
turn our attention towards the linearized equations and the resulting
eigenmode spectrum. These results will not only be used as initial data,
but also provide one of the fundamental test beds for the code.

%=========================================================================
\subsubsection{The linearized equations and the eigenmode spectrum}
\label{PERT_LIN}
{\em (a) The equations} \\[5pt]
In this section we will discuss the linearized equations for a dynamic
spherically symmetric neutron star. For this purpose we will
explicitly assume that the background is given by a non-vacuum solution
of the TOV equations. If we further assume that all perturbations are small
compared with their background
values and the radial velocity $w$ is small compared with the speed of light,
i.e. $w \ll 1$, the higher order terms in
Eqs.\,(\ref{PERT_LAMBDAR})-(\ref{PERT_B2}) become negligible and can
be omitted from the equations. It is convenient to follow e.g.
\lcite{Misner1973} and introduce the variable $\xi$ which measures 
the displacement of the fluid elements. An observer who is comoving with the
fluid and is located at $r_0$ in the equilibrium case will find herself at
position $r_0+\xi(t,r_0)$ during the evolution. The displacement vector $\xi$
is therefore related to our variables by
\begin{align}
  \xi_{,t} &= \lambda w. \label{LIN_XI}
\end{align}
We note that the background value of the lapse function is used in this
equation because higher order terms have been neglected.
Another variable which facilitates a particularly simple formulation of the
resulting equations is the rescaled displacement $\zeta$ defined by
\begin{align}
  \zeta &= \frac{r^2}{\lambda} \xi \label{LIN_ZETA}.
\end{align}
If we insert this definition into the linearized form of
equation (\ref{PERT_WT}) and use the linearized versions of
Eqs.\,(\ref{PERT_LAMBDAR})-(\ref{PERT_DRHOT}) to eliminate the perturbations
$\delta \lambda$, $\delta \mu$ and $\delta \rho$ we obtain the second order
in time and space differential equation
\begin{align}
  W \zeta_{,tt} &= \frac{1}{r_{,x}}\left( \frac{\Pi}{r_{,x}}
      \zeta_{,x} \right)_{,x}
      + Q \zeta, \label{LIN_ZETATT}
\end{align}
where the auxiliary functions $W$, $\Pi$ and $Q$ are defined by
\begin{align}
  \Pi &= C^2(\rho + P) \frac{\mu \lambda^3}{r^2}, \label{LIN_PI} \\[10pt]
  W   &= (\rho +P) \frac{\mu^3 \lambda}{r^2}, \label{LIN_W} \\[10pt]
  Q   &= \frac{\mu \lambda^3}{r^2}(\rho + P) \left[ \left(\frac{\lambda_{,r}}
         {\lambda} \right)^2 + 4\frac{\lambda_{,r}}{r\lambda} -8\pi \mu^2P
         \right].
         \label{LIN_Q}
\end{align}
These equations describe the dynamics of a
spherically symmetric neutron star in the linearized limit.
If we insert the ansatz $\zeta(t,x)={\zeta(x)} f(t)$ into
Eq.\,(\ref{LIN_ZETATT}) we find that
the solution has harmonic time dependence
\begin{align}
  \zeta(t,x) &= {\zeta}(x) e^{i \omega t}.
\end{align}
and the spatial profile is determined by the
ordinary differential equation
\begin{align}
  \frac{1}{r_{,x}} \left( \frac{\Pi}{r_{,x}} \zeta_x\right)_x
     + (\omega^2 W + Q) \zeta=0. \label{LIN_ZETARR}
\end{align}
%
%where we have omitted the ``tilde'' from the spatial function $\zeta(r)$.
For the ensuing discussion it is convenient
to work with the areal radius
$r$ and therefore set $r_{,x}=1$. The ordinary differential equation
(\ref{LIN_ZETARR}) can then be written in the form
\begin{align}
  \mathcal{L} \zeta &= -\omega^2 \zeta,
\end{align}
where the differential operator $\mathcal{L}$ is defined by
\begin{align}
  \mathcal{L} &= \frac{1}{W}\left[ \frac{d}{dr} \left( \Pi \frac{d}{dr} \right)
                 - Q\right].
\end{align}
This type of ODE is called an {\em eigenvalue problem} and the
particular structure of the differential operator $\mathcal{L}$
classifies it as a {\em Sturm-Liouville problem} if the function $\zeta$
satisfies so-called homogeneous boundary conditions
(see for example \citeNP{Simmons1991}).
Due to the asymptotic behaviour of the background
solutions the functions $\Pi$, $W$ and $Q$ will either diverge or vanish
at the boundaries, however, and the problem we are facing is 
a {\em singular Sturm-Liouville problem}. An important subclass of this type
of problems is the {\em self-adjoint eigenvalue problem}
which is defined by the requirement that
\begin{align}
  \langle \mathcal{L}u,v \rangle &= \langle \mathcal{L}v,u \rangle,
  \label{SELFADJOINED}
\end{align}
for all solutions $u$, $v$. Here the inner product is defined by the
{\em weighting function} $W(r)$
\begin{align}
  \langle f,g \rangle &= \int_a^b{W(r)\, f(r)\, g(r)\,dr},
        \label{LIN_INNERPRODUCT}
\end{align}
where $a$ and $b$ are the boundaries, i.e. the centre and surface of the
star in our case. A short calculation shows that condition
(\ref{SELFADJOINED}) is ensured if the solutions satisfy the
self-adjoint boundary condition
\begin{align}
  \left[ \vphantom{\frac{1}{x}} \Pi (v u_{,r} - u v_{,r}) \right]_a^b &= 0.
      \label{BCSELFADJOINED}
\end{align}
Below we shall see that any solution $\zeta$ of the eigenvalue problem
(\ref{LIN_ZETARR}) will be $\mathscr{O}(r^3)$ at the origin and
be finite at the surface. In combination with the asymptotic behaviour of the
TOV solutions determined in section \ref{TOV_ASYMPTOTICS} we can see that
Eq.\,(\ref{BCSELFADJOINED}) is satisfied so that
the differential equation (\ref{LIN_ZETARR}) represents a self-adjoint
eigenvalue problem. For this type of equations one can show the
following properties (see for example \citeNP{Coddington1955})
\begin{list}{\rm{(\arabic{count})}}{\usecounter{count}
             \labelwidth1cm \leftmargin1.5cm \labelsep0.4cm \rightmargin0cm
             \parsep0.5ex plus0.2ex minus0.1ex \itemsep0ex plus0.2ex}
\item There exist an infinite number of solutions $\zeta_1(r)$, 
      $\zeta_2(r)$, $\zeta_3(r),\ldots$ which are called eigenfunctions
      and the corresponding eigenvalues are real and can be ordered
      \begin{align}
        (\omega^2)_1 < (\omega^2)_2 < (\omega^2)_3 <\ldots.
      \end{align} 
      We note that in our case the real eigenvalues are $\omega^2$ and
      the corresponding frequencies will be imaginary if $\omega^2<0$.
%      Below we will see that this will indeed be the case for the fundamental
%      eigenmodes of gravitationally unstable neutron stars.
\item After appropriate normalisation the eigenfunctions form an orthonormal
      set, i.e.
      \begin{align}
        \langle \zeta_i, \zeta_j \rangle &= \delta_{i,j}.
        \label{LIN_ORTHONORMALITY}
      \end{align}
\item The eigenfunctions $\zeta_i$ form a complete set, i.e. any function
      $f(r)$ which satisfies the self-adjoint boundary conditions
      (\ref{SELFADJOINED}) can be expanded in a series of eigenmodes
      \begin{align}
        f(r) &= \sum_i{A_i \zeta_i(r)},
      \end{align}
      where the eigenmode coefficients of the function $f$ are given by
      \begin{align}
        A_i &= \langle f, \zeta_i \rangle.
      \end{align}
\end{list}
Before we investigate Eq.\,(\ref{LIN_ZETARR}) numerically, we
consider the asymptotic behaviour of the solutions. At the
origin the displacement vectors $\xi$ and $\zeta$ have to vanish because
of the spherical symmetry. If we therefore assume $\zeta(r) \sim r^{\alpha}$
near the origin where $\alpha>0$, insert this ansatz into 
Eq.\,(\ref{LIN_ZETARR}) and use the asymptotic behaviour of the TOV solution,
we obtain the leading order
\begin{align}
  \zeta(r) \sim \mathscr{O}(r^3). \label{LIN_ASSYMPTOTICZETA}
\end{align}
At the surface we only require $\xi$ and $\zeta$ to be finite but allow
for non-zero displacements
\begin{align}
  \zeta(z) \sim \mathscr{O}(z^0).
\end{align}
It is of particular interest to consider the impact of these results
on the asymptotic behaviour of the energy density perturbation
$\delta \rho$ which is related to the displacement by the linearized
version of Eq.\,(\ref{PERT_DRHOT})
\begin{align}
  \delta \rho &= -\frac{\lambda}{r^2} \left[(\rho + P) \zeta_{,r} 
      + \rho_{,r} \zeta \right]. \label{LIN_DRHOOFZETA}
\end{align}
At the centre the $r^3$ behaviour of the displacement $\zeta$ results in
\begin{align}
  \delta \rho \sim \mathscr{O}(r^0),
\end{align}
so that the condition we imposed on $\zeta$ also guarantees a finite
energy density perturbation at the origin. At the surface, however, the 
leading term on the right hand side of Eq.\,(\ref{LIN_DRHOOFZETA}) is the
term involving the derivative of the background energy density. This
term is responsible for the asymptotic behaviour of $\delta \rho$ at
the surface
\begin{align}
  \delta \rho &\sim \mathscr{O}(z^{n-1}). \label{LIN_ASSYMPTOTICDRHO}
\end{align}
Consequently the energy density perturbation is zero at the surface
for $n > 1$, finite
for $n=1$ and it diverges for $n<1$ i.e. $\gamma>2$. Even worse we
also obtain the result
\begin{align}
  \frac{\delta \rho}{\rho} \sim \mathscr{O}(z^{-1})\label{DRHOOVERRHO}
\end{align}
independent of the polytropic index.
The energy density perturbation will therefore necessarily be larger than
the background $\rho$ in a finite interval around the surface.
This is in obvious conflict with the initial assumption
$\delta \rho \ll \rho$ we used in the linearisation process and raises
doubts about the validity of the results. Below we will see, however,
that the linearized equations can be derived without any implicit
contradiction from the fully non-linear Lagrangian formulation of
the problem. This is already illustrated by a closer investigation of
Eq.\,(\ref{LIN_DRHOOFZETA}) which can be rewritten as
\begin{align}
  \delta \rho &= \Delta \rho - \xi \rho_{,r}. \label{LIN_DRHOEULERLAGR}
\end{align}
Here $\Delta \rho$ is the Lagrangian energy density perturbation measured
by an observer moving with the fluid and is given by
\begin{align}
  \Delta \rho &= -\frac{\lambda}{r^2}(\rho + P) \zeta_{,r}.
  \label{LIN_LAGRDRHO}
\end{align}
[cf. Eq.\,(\ref{LLIN_DRHO})].
The asymptotic behaviour of $\Delta \rho$ is perfectly regular
$\Delta \rho \sim x^{n}$ and the difficulties purely originate from the term
$\xi \rho_{,r}$ on the right hand side of Eq.\,(\ref{LIN_DRHOEULERLAGR}).
This correction term which facilitates the transformation
between the Eulerian and 
Lagrangian perturbations is based on a Taylor expansion of $\rho$ which, as 
we have already seen above, is not generally permissible. For polytropic
indices $n<1$ the derivative of $\rho$ does indeed diverge
and Eq.\,(\ref{LIN_DRHOEULERLAGR}) is not a valid relation between the Eulerian
and Lagrangian quantities. This is the first indication that a Lagrangian
formulation is a somewhat more natural way of describing radial oscillations
of neutron stars. From this point of view it is a remarkable fact
that the linearisation of the Eulerian case leads to the ``correct''
equations in spite of the internal inconsistency of the derivation. 
Finally it is worth pointing out that the irregular behaviour of
$\delta \rho$ is not merely down to a poor choice of dependent variables.
It is certainly possible to formulate the problem in Eulerian coordinates
in terms of regular variables such as $\zeta$
or $\xi$. We have seen, however, that such a regular formulation of the
problem still leads to the unphysical result of a diverging
total energy density $\rho+ \delta \rho$
if the equations of state has an asymptotic power law behaviour $P\sim
\rho^n$ with $n<1$. In view of these difficulties one may ask the question
why we have decided to use an Eulerian rather than a Lagrangian
formulation in the first place. Our main motivation for studying
Eulerian schemes is to probe a method in spherical symmetry
which enables one to accurately
model a wide range of different types of non-linear neutron star oscillations.
Below we shall see that the Lagrangian approach is a very powerful tool
for the study of dynamic stars in spherical symmetry. However, it is a
generic problem of Lagrangian methods that it is not clear how to generalise
them to two or three spatial dimensions, where the paths of fluid elements
may intersect and give rise to caustics. The vast majority of neutron star
oscillations on the other hand will only be present if one drops
the assumption of spherical symmetry, so that their numerical simulation
requires the use of two or three spatial dimensions. In default
of higher dimensional generalisations of Lagrangian techniques these
simulations are generally performed in an Eulerian framework. \\
We will now turn our attention towards the numerical solution of the
linearized equations. From the asymptotic behaviour, we expect, however, that
the results we obtain for $n<1$ will diverge at the surface
and thus not represent a physical solution.
From a numerical point of view it turns out to be beneficial to reformulate
Eq.\,(\ref{LIN_ZETARR}) in terms of the displacement vector $\xi$. This is due
to the asymptotic behaviour of $\zeta$ at the origin given
by Eq.\,(\ref{LIN_ASSYMPTOTICZETA}).
Below we will use the numerically calculated eigenmodes as initial data for
the fully non-linear evolutions and for that purpose the solution for
$\zeta$ would have to be converted into data for $w$ or in the Lagrangian
code discussed in section \ref{LAGR} for $\xi$. The corresponding division by
$r^2$ combined with the second order accuracy of the numerical eigenmode 
solutions results in poor accuracy of these initial data near the origin.
We therefore rewrite Eq.\,(\ref{LIN_ZETARR}) in terms of $\xi$
and introduce the auxiliary variable $A$
to write the result as a first order system
\begin{align}
  &\Pi \xi_{,x} - A = 0, \label{LIN_XIY} \\[10pt]
  & A_{,x} + (r_{,x})^2\frac{\lambda^2}{r^4} \left( \frac{r^2}{\lambda r_{,x}} 
      \right)_{,r} A + (r_{,x})^2\left\{ \frac{\lambda}{r^2}
      \left[ \Pi \left(\frac{r^2}{\lambda}\right)_{,r} 
      \right]_{,r} + \omega^2 W + Q \right\} \xi =0.
      \label{LIN_AY}
\end{align}
We note that the occurrence of $r$-derivatives in equation (\ref{LIN_XIY}) 
is purely a convenient notation. In practice all these
derivatives are eliminated
via the TOV equations. If we use the rescaled radial coordinate, we have
$r_{,x}=C$ and the $r$-derivative of $r_{,x}$ can
be calculated from the relation
\begin{align}
  C^2_{,r} &= (\gamma-1) \frac{P_{,r}}{\rho},
\end{align}
which is a consequence of the equation of state and the definition of 
the sound speed. The only derivatives
in Eqs.\,(\ref{LIN_XIY}), (\ref{LIN_AY}) that have to be represented
by finite differencing are the $x$-derivatives of $\xi$ and
$A$. \\
In the Cowling approximation all these results remain unchanged except for
the function $Q$ which has to be replaced by
\begin{align}
  \tilde{Q} &= \lambda^2 (\rho + P) \left[ \left( \frac{\lambda_{,r} 
         C^2 \mu}{r^2} \right)_r - \mu \left( \frac{\lambda_{,r}}{r^2}
         \right)_r + \lambda \mu
         \left( \frac{C^2 \mu_r}{r^2 \mu} \right)_r \right].
\end{align}
and the relation between displacement and energy density perturbation which
becomes
\begin{align}
  \delta \rho &= -(\rho + P) \frac{\lambda}{r^2} \left[
      \left(\frac{\lambda_{,r}}{\lambda} + \frac{\mu_{,r}}{\mu} \right)
      \zeta +\zeta_{,r} \right] - \frac{\lambda}{r^2} \rho_{,r} \zeta.
\end{align}
It is an interesting fact that in both cases the results are simpler
due to the cancellation of terms if gravity is included. \\

{\em (b) The numerical implementation} \\[5pt]
We have numerically calculated solutions of the eigenvalue problem
(\ref{LIN_XIY}), (\ref{LIN_AY}) using a relaxation method. % as described
%in section\,\ref{relaxation}.
For this purpose
we introduce an additional differential equation for the eigenvalues
\begin{align}
  (\omega^2)_{,x} &= 0, \label{LIN_OMEGAY}
\end{align}
which states that the eigenmode frequency is constant throughout the star.
The value of $\omega$ is not known at this stage but will result from the
relaxation algorithm.
In order to solve the system (\ref{LIN_XIY}), (\ref{LIN_AY}),
(\ref{LIN_OMEGAY}) we need to supply three boundary conditions. At the centre
we require that
\begin{align}
  \xi(0) &= 0, \\[10pt]
  A(0) &= \mathrm{const} \ne 0.
\end{align}
The vanishing of the displacement $\xi$ at the origin is a necessary
condition in spherical symmetry. The value of $A$ at the origin is allowed
to take on any non-zero value because an eigenfunction is only defined
up to a constant factor. At the outer boundary we have the condition
\begin{align}
  A &= 0, \label{LIN_BCOUTA}
\end{align}
which follows from the definition of $A$ and the vanishing of the energy
density at the surface of the star. 
An initial guess for $\omega$ enables us to calculate
the initial functions $\xi$ and $A$ by integrating Eqs.\,(\ref{LIN_XIY}),
(\ref{LIN_AY}) outwards. The solution including the eigenvalue $\omega^2$
is then obtained by relaxation as described in section\,\ref{relaxation}. \\

{\em (c) Testing the code} \\[5pt]
For sufficiently low eigenmodes both alternative choices of the radial
coordinate lead to good agreement between the predicted frequencies
up to the fourth significant digit. As we will see below high order
\begin{table}[t]
  \caption{The convergence factors obtained for doubling the grid resolution
           in the relaxation code for calculating the eigenmodes of
           the neutron star models 1 - 5. Grid resolutions
           of 500, 1000 and 2000 points have been used.}
\begin{center}
  \begin{tabular}{c|cc}
    \hline \hline
    model & fundamental mode & $10^{\rm th}$ eigenmode \\
    \hline
    1 & 4.75  & 5.05  \\
    2 & 4.76  & 4.85   \\
    3 & 4.80  & 3.97   \\
    4 & 4.75  & 4.82   \\
    5 & 4.75  & 4.82   \\
    \hline \hline
  \end{tabular}
  \label{CONV_EIG}
\end{center}
\end{table}
eigenmode profiles show rapid oscillations near the surface of the star
\begin{table}[t]
  \caption{Radius, mass and frequencies of the lowest three eigenmodes
           for three randomly chosen models of Kokkotas and Ruoff have
           been recalculated with our codes and agree well with their values.}
\begin{center}
  \begin{tabular}{l|ccccccccc}
    \hline \hline
    & $\gamma$ & $K$ & $\rho_{\rm c}$ & $R$ & $M$ 
    & $\nu_1$ & $\nu_2$ & $\nu_3$ \\
    & & & $[10^{15}\,\,{\rm g/cm}^3]$ 
    & $[{\rm km}]$ & $[M_{\odot}]$ 
    & [kHz] & [kHz] & [kHz]  \\
    \hline \\[-10pt]
    Kokkotas \& Ruoff & 2.00 & $100\,\,{\rm km}^2$ & 5.000 & 7.787
    & 1.348 & 1.129 & 7.475 & 11.365 \\
    this work & 2.00 & $100\,\,{\rm km}^2$ & 5.000 & 7.788
    & 1.348 & 1.128 & 7.470 & 11.355 \\
    \hline \\[-10pt]
    Kokkotas \& Ruoff & 2.25 & $700\,\,{\rm km}^{2.5}$ & 4.000 & 8.199
    & 1.600 & 1.455 & 7.610 & 11.573 \\
    this work & 2.25 & $700\,\,{\rm km}^{2.5}$ & 4.000 & 8.200
    & 1.600 & 1.443 & 7.594 & 11.544 \\
    \hline \\[-10pt]
    Kokkotas \& Ruoff & 3.00 & $2\cdot 10^5\,\,{\rm km}^4$ & 2.200 & 9.419
    & 1.988 & 2.716 & 8.305 & 12.516 \\
    this work & 3.00 & $2\cdot 10^5\,\,{\rm km}^4$ & 2.200 & 9.419
    & 1.988 & 2.637 & 8.215 & 12.389 \\
    \hline \hline
  \end{tabular}
  \label{COMPEIGS}
\end{center}
\end{table}
which
% depending on the number of grid points
may not be well resolved if we
work with the areal radius $r$. The frequencies deviate more significantly
in these cases. In the rest of this section we will therefore work with
the rescaled coordinate and set $r_{,x}=C$.
The resulting code has been checked in four independent ways. First
we have computed the eigenfunctions of the fundamental and the tenth
mode for the neutron star models listed in Table \ref{MODELS15}
and checked for convergence using 500, 1000 and 2000 grid points. The results
shown in Table \ref{CONV_EIG} clearly demonstrate second order convergence
as expected for the second order finite differencing scheme applied
in the relaxation algorithm. \\
Next we have randomly chosen three of the stellar models
listed in \lcite{Kokkotas2001} and recalculated radius, mass of the neutron
stars as well as the frequencies of the lowest three eigenmodes. The
results are compared in Table \ref{COMPEIGS} and show good agreement. \\
For the third test we recall the 1-parameter families of neutron stars
shown in Fig.\,\ref{TOV_FAMILIES}. We have already
mentioned that the maxima in the mass vs. central density plots separate the
stable and unstable branches of neutron star models and that the frequency
of the fundamental eigenmode becomes zero at the critical point
and imaginary on the unstable branch. We have therefore
determined the critical central densities for the five neutron star
models of Table \ref{MODELS15}
and calculated the frequency of the fundamental modes
just below and above the critical densities. The numerical results 
are shown in Table \ref{CRITFREQS} and confirm this picture. The frequencies
of the fundamental mode are very small but real for central densities
\begin{table}[t]
  \caption{The critical central densities corresponding to
           the neutron star models 1-5
           are given to four significant digits together with the
           frequency of the fundamental mode just below and above the
           critical point. Above the critical density the frequencies
           become imaginary as expected.}
\begin{center}
  \begin{tabular}{l|ccccccccc}
    \hline \hline
    model & $\rho_{\rm c,crit}$ 
    & $\nu (\rho_{\rm c,crit}-10^{-6})\,{\rm km^{-2}}$
    & $\nu (\rho_{\rm c,crit}+10^{-6})\,{\rm km^{-2}}$ \\
    & [km$^{-2}$] & [kHz] & [kHz] \\
    \hline \\[-10pt]
    1 & 0.002179 & 0.0294 & 0.0477$\,i$ \\
    2 & 0.004205 & 0.0578 & 0.0429$\,i$ \\
    3 & 0.002804 & 0.0629 & 0.0350$\,i$ \\
    4 & 0.002103 & 0.0409 & 0.0592$\,i$ \\
    5 & 0.002233 & 0.0591 & 0.0627$\,i$ \\
    \hline \hline
  \end{tabular}
  \label{CRITFREQS}
\end{center}
\end{table}
just below the critical value and become imaginary for larger densities.\\
A further test for the eigenmode frequencies arises from a relation
between the period of the fundamental mode $T_1$ of a neutron star model
and the deviation of the radius $R$ from the critical radius $R_{\rm c}$
that has been suggested by \scite{Harrison1965} [see their Eq.\,(155)]
\begin{align}
  (R-R_{\rm c})\cdot T_1^{\,2} &= \mathrm{const}. \label{LIN_HARRISON}
\end{align}
In Table \ref{LIN_RCT0} we show the results obtained for neutron star
models identical to model 1 and 3 with central densities as indicated.
\begin{table}[t]
  \caption{Equation\,(\ref{LIN_HARRISON}) is checked for neutron star models
           1 and 3 for various central densities.}
\begin{center}
  \begin{tabular}{ccc|ccc}
    \hline \hline
    \multicolumn{3}{c|}{model 1} & \multicolumn{3}{c}{model 3} \\
    $\rho_{\rm c}$ & $\omega_1$ & $(R-R_{\rm crit})/\omega_1^2$ 
    & $\rho_{\rm c}$ & $\omega_1$ & $(R-R_{\rm crit})/\omega_1^2$ \\
    $[{\rm km}^{-2}]$ & [km$^{-1}$] & [km$^3$] & [km$^{-2}$] & [km$^{-1}$]
    & [km$^3$] \\
    \hline
    0.0021785 & \hspace{0.3cm} 0.000187 & 187.96 &
    0.0028035 & \hspace{0.3cm} 0.000172 & 97.55 \\
    0.0021780 & \hspace{0.3cm} 0.000616 & 192.72 &
    0.0028030 & \hspace{0.3cm} 0.000773 & 85.21 \\
    0.0021775 & \hspace{0.3cm} 0.000850 & 192.34 &
    0.0028025 & \hspace{0.3cm} 0.001080 & 84.89 \\
    0.0021750 & \hspace{0.3cm} 0.001563 & 192.59 &
    0.0028020 & \hspace{0.3cm} 0.001317 & 84.77 \\
    0.0021700 & \hspace{0.3cm} 0.002424 & 193.00 &
    0.0028000 & \hspace{0.3cm} 0.002002 & 84.87 \\
    0.0020000 & \hspace{0.3cm} 0.010939 & 207.93 &
    0.0027000 & \hspace{0.3cm} 0.010838 & 86.97 \\
    0.0015000 & \hspace{0.3cm} 0.020261 & 236.69 &
    0.0020000 & \hspace{0.3cm} 0.029622 & 106.64 \\
    0.0011775 & \hspace{0.3cm} 0.023606 & 235.61 &
    0.0015000 & \hspace{0.3cm} 0.036427 & 128.95 \\
    \hline \hline
  \end{tabular}
  \label{LIN_RCT0}
\end{center}
\end{table}
Even though a deviation from Eq.\,(\ref{LIN_HARRISON}) up to $20\,\%$
is observed for both models,
this is rather small if one considers the variation of the frequency
$\omega_1$ over several orders of magnitude. \\

\begin{figure}[t]
  \centering
  \epsfig{file=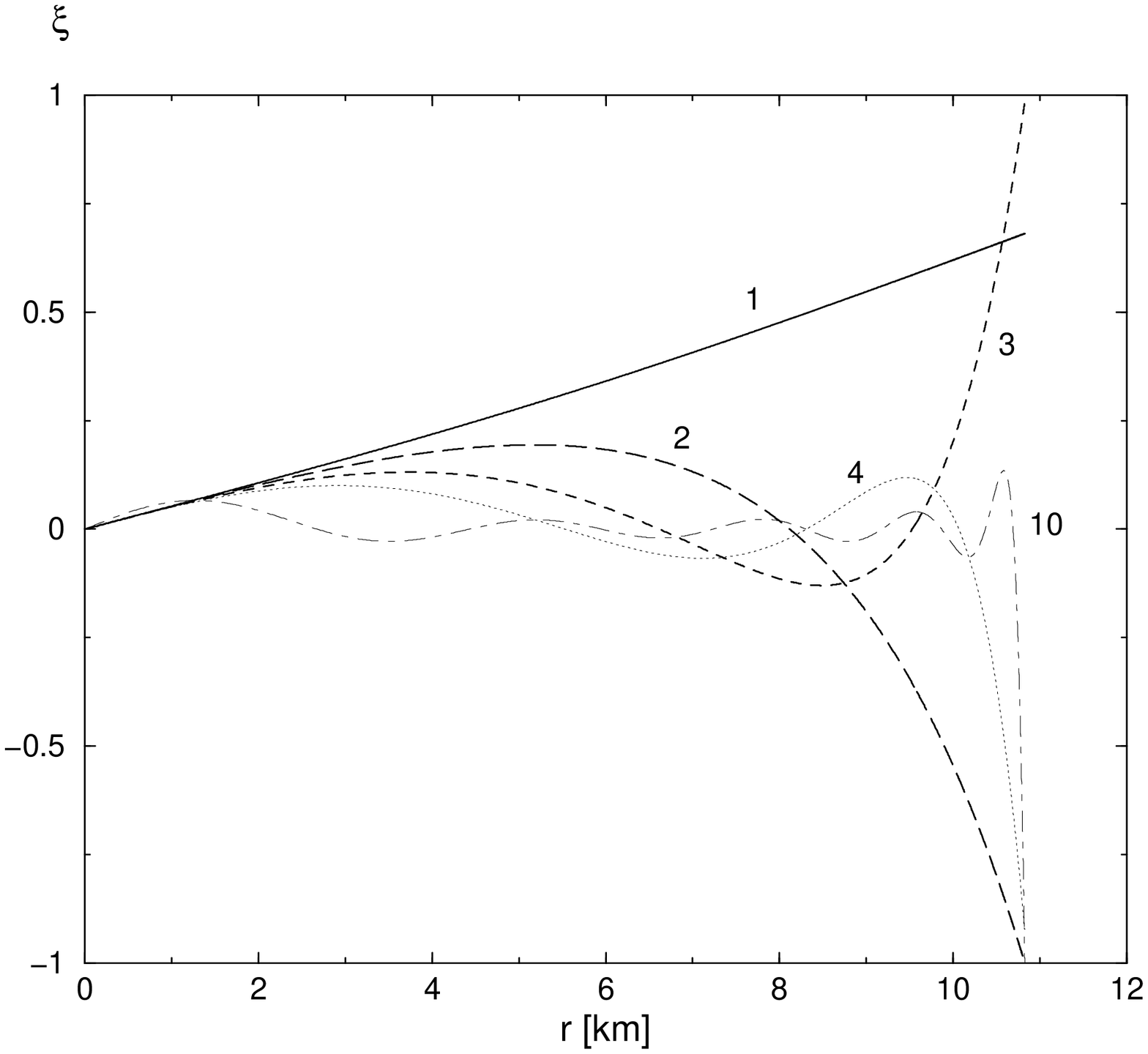, height=200pt, width=150pt, angle=-90}
  \epsfig{file=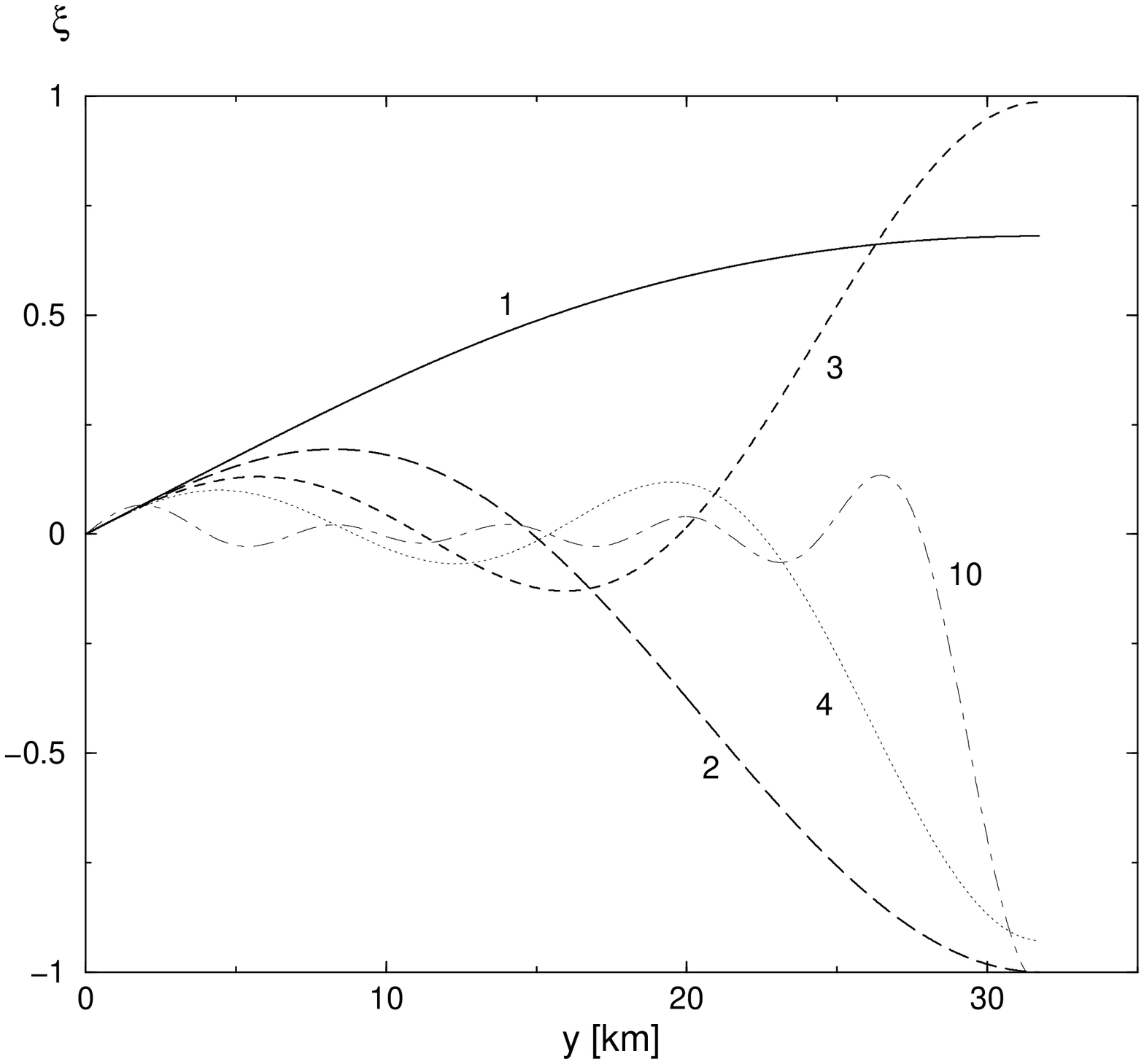, height=200pt, width=150pt, angle=-90}
  \epsfig{file=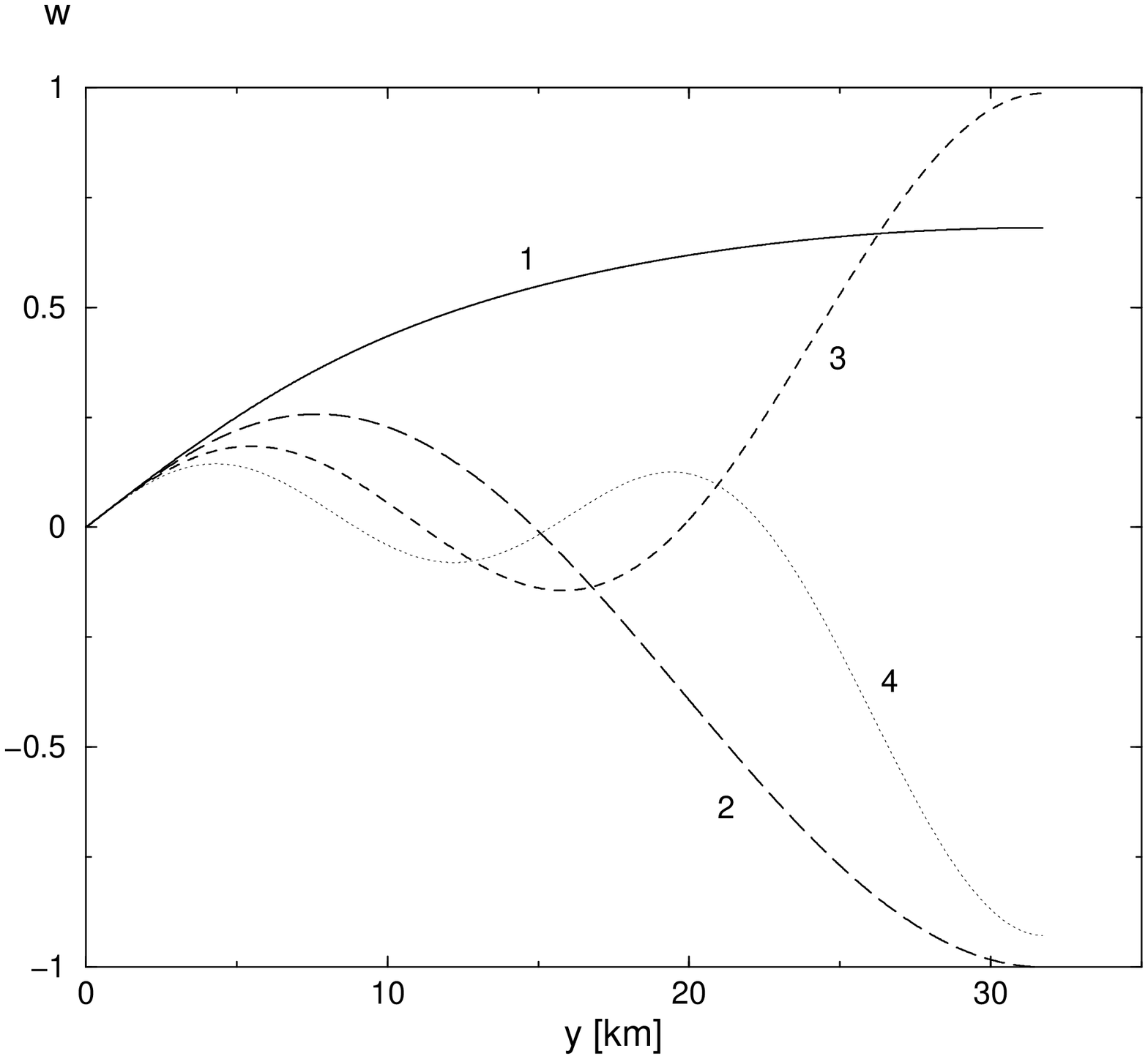, height=200pt, width=150pt, angle=-90}
  \epsfig{file=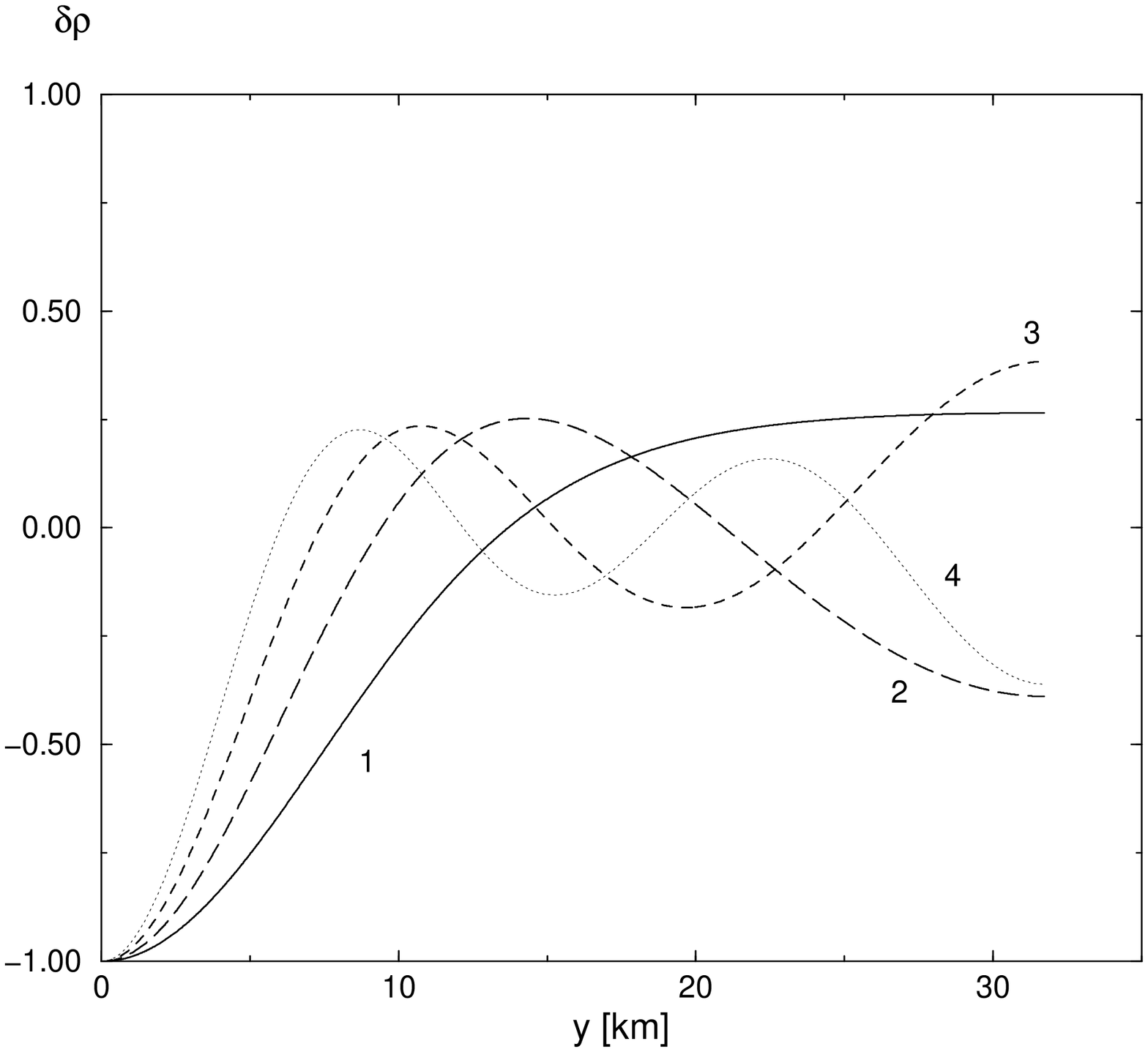, height=200pt, width=150pt, angle=-90}
  \caption{The displacement $\xi$ as a function of the areal radius $r$
           and the rescaled radius $y$ as well as the velocity $w$ and
           the energy density $\delta \rho$ as a function of $y$ are
           shown for the first four eigenmodes of model 3. 
           For $\xi$ we have also
           plotted mode 10 to illustrate the concentration of oscillations
           towards larger $r$.}
  \label{MODES1_4}
\end{figure}

{\em (d) The eigenmode solutions} \\[5pt]
We will now turn our attention to the eigenmode profiles of the
physical variables.
We have already noted that the eigenvalue problem has an enumerable infinite
set of solutions which can be ordered with respect to their eigenvalues.
This order is also reflected in the spatial profiles of the corresponding 
eigenfunctions. We have numerically calculated the first four 
eigenmodes in terms of the displacement vector $\xi$ for model 3
with polytropic exponent $\gamma=2$. The velocity $w$, 
the rescaled displacement $\zeta$ and the energy density perturbation
$\delta \rho$ then follow from Eqs.\,(\ref{LIN_XI}) where we use harmonic time 
dependence, (\ref{LIN_ZETA}) and (\ref{LIN_DRHOOFZETA}).
The results are shown in Fig.\,\ref{MODES1_4}, 
where we have also included the solution
for $\xi$ corresponding to the tenth eigenmode.
Since the eigenmode solutions are determined
up to a constant factor only, we have rescaled them to about unit amplitude.
For all variables we see that the number of nodes is given by the order
of the mode and the number of local maxima or minima is given by
the order minus one.
This behaviour remains valid for higher modes and is characteristic of the
eigenmode solutions.
In order to illustrate the significance of the transformation to the
rescaled radius $y$ we have plotted $\xi$ as a function of $r$ as well.
In the upper panels of Fig.\,\ref{MODES1_4} we can see that the
oscillations in the spatial profile of the eigenmodes become more
concentrated towards larger radii $r$ the higher the order of the mode.
In terms of the rescaled radius $y$, however, the oscillations are evenly
distributed over the entire interval. This behaviour is reminiscent of the
narrowing of the wave pulse we observed in section \ref{TOV_RYTRAFO} 
and illustrates why
a superior numerical performance is obtained when using the coordinate $y$,
especially when higher order modes are present in the evolution. \\
\begin{figure}[ht]
  \centering
  \epsfig{file=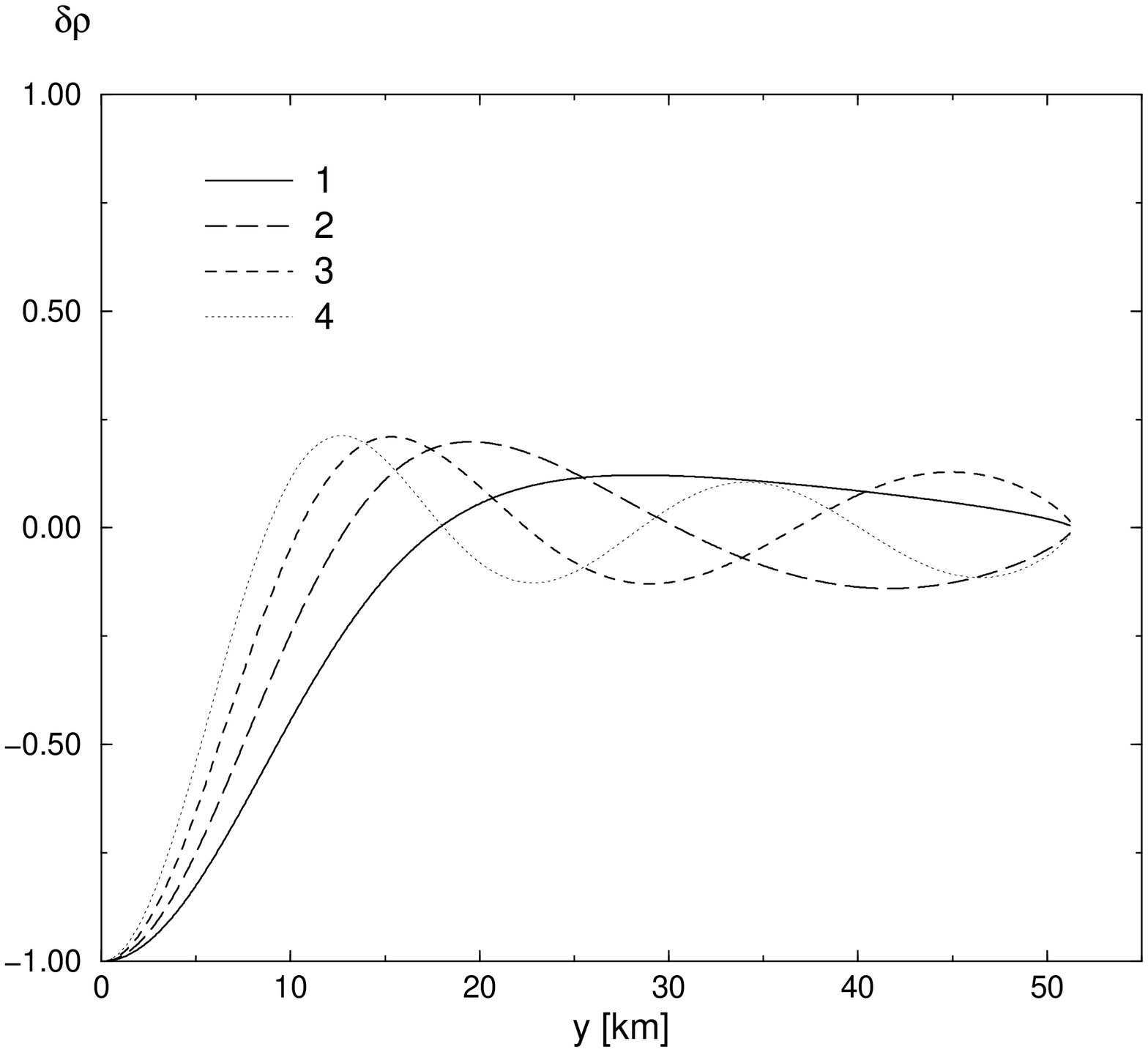, height=200pt, width=150pt, angle=-90}
  \epsfig{file=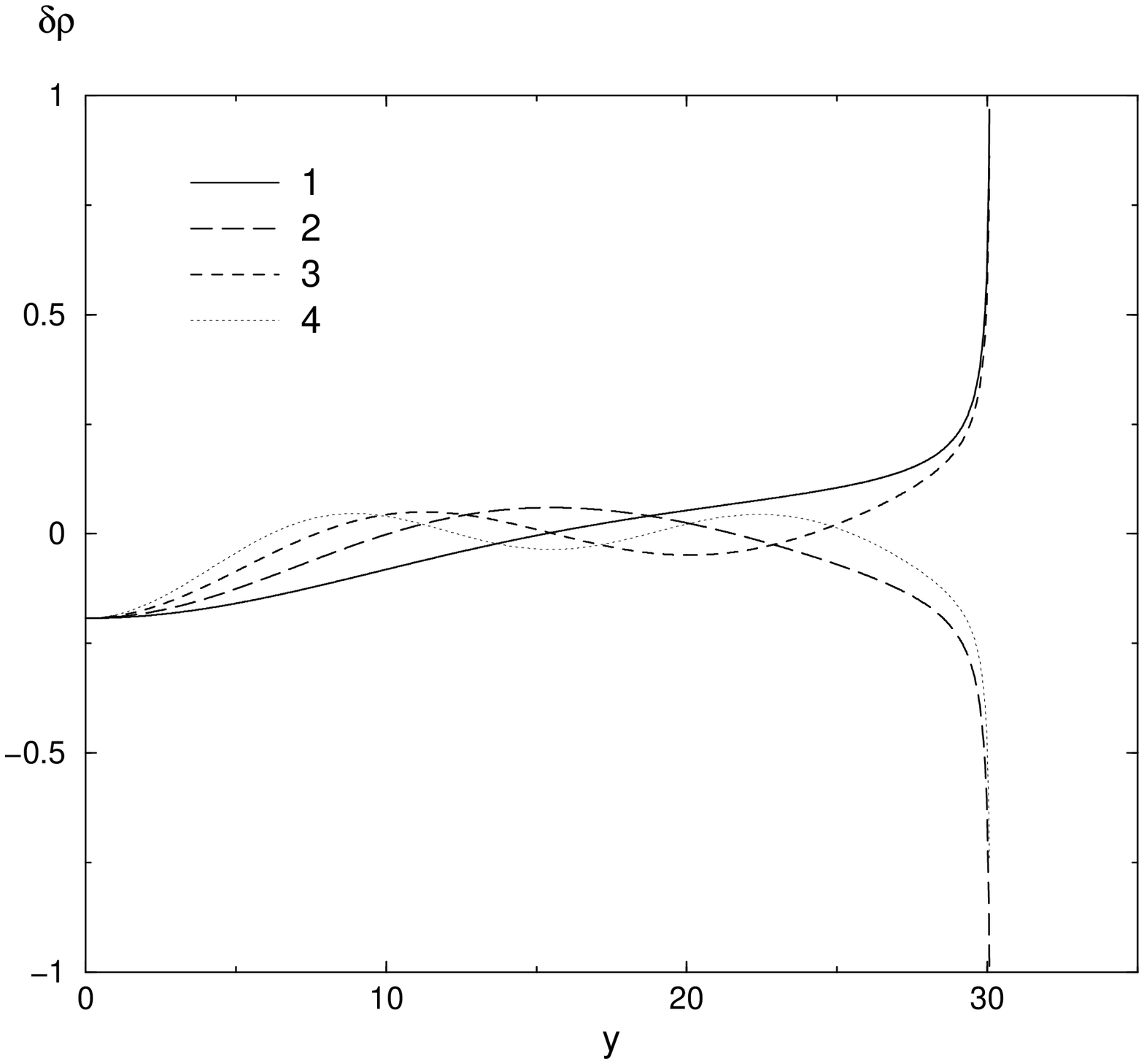, height=200pt, width=150pt, angle=-90}
  \caption{The energy density perturbation $\delta \rho$ obtained for the
           first four eigenmodes of stellar model 1 (left panel) and 5 
           (right panel) is plotted as a function of $y$.}
  \label{EIG_DRHO}
\end{figure}
The corresponding eigenmodes obtained for the other stellar models look 
qualitatively similar in all variables except for the energy density 
perturbation $\delta \rho$. We have already noted that the asymptotic 
behaviour of $\delta \rho$ depends on the polytropic exponent $\gamma$.
This is confirmed by the numerical solutions shown in Fig.\,\ref{EIG_DRHO} 
where we plot the profiles of the energy density perturbation obtained
for the stellar models 1 and 5 with polytropic exponents $\gamma=1.75$
and $2.3$ respectively. For model 1 the energy density perturbation
goes to zero at the surface, although with a non-zero gradient. In
comparison the gradient of the background density of the same model vanishes
in Fig.\,\ref{TOV_GAMMA} and the 
quotient $\delta \rho/\rho$ can indeed be shown to diverge
in agreement with Eq.\,(\ref{DRHOOVERRHO}).
For the larger polytropic index 2.3 the 
perturbation $\delta \rho$ itself diverges at the surface as expected
from Eq.\,(\ref{LIN_ASSYMPTOTICDRHO}). \\
The corresponding results obtained in the Cowling approximation are
very similar to those shown above. The only notable difference is the
frequency of the fundamental mode which does not decrease towards
zero as the central density approaches the critical value but instead
remains real and positive. This result is to be expected since a fluid
will not become gravitationally unstable if the gravitational field is
kept fixed. \\
The eigenmode solutions obtained in this section will be used extensively
as initial data in the non-linear evolutions. We have seen, however, 
that the stellar surface represents a problematic area even in the linearized
case. The difficulties are more pronounced in the non-linear case
and need to be investigated in more detail before we can study the
fully non-linear numerical evolutions.

%=========================================================================
\subsubsection{The surface problem}
\label{SURFACE}
When we formulated the description of non-linear radial oscillations
of neutron stars in section \ref{NONP_EQ} we consciously omitted the issue
\begin{figure}[t]
  \centering
  \epsfig{file=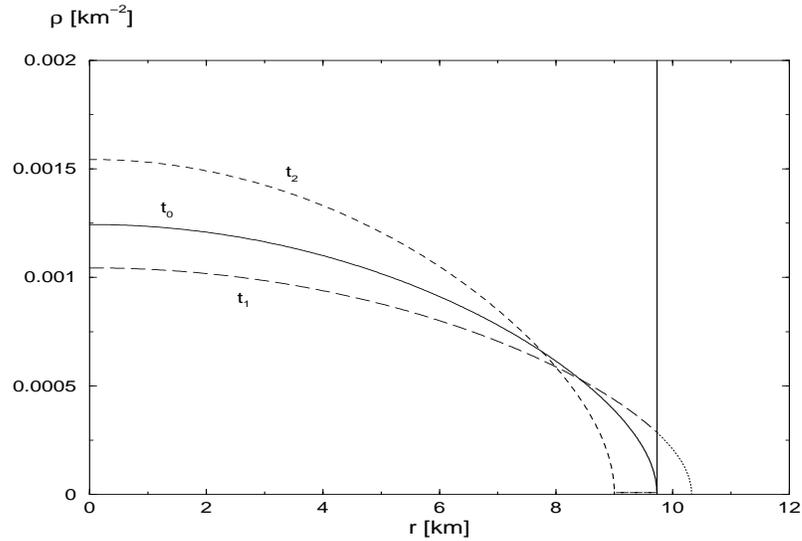, height=300pt, width=200pt, angle=-90}
  \caption{The energy density profile of an oscillating neutron
           star is schematically plotted at three different
           stages of one oscillation.
           Initially the stellar radius is at its equilibrium value,
           at the later time $t_1$ the star has expanded and at $t_2$
           it has shrunk below its initial radius. The vertical
           line indicates the extension of the numerical grid.}
  \label{NONL_SURFACE}
\end{figure}
of boundary conditions. The difficulties involved in specifying outer boundary
conditions in an Eulerian code are so complex that we dedicate a whole
subsection to this topic.
We have already mentioned that the surface is defined by the condition
$\hat{P}=0$ which is equivalent to $\hat{\rho}=0$ for a polytropic
equation of state. With respect to the fixed numerical grid,
however, the surface of the star is moving and we cannot apply this condition
at the outer grid boundary. This is a further indication that one may
have to emulate a Lagrangian treatment of the surface in order to
accurately model neutron star oscillations involving radial displacements
of the surface.
The situation is graphically illustrated in
Fig.\,\ref{NONL_SURFACE} where the total energy density profile is
schematically plotted as a function of radius. At time
$t_0$ an equilibrium star (solid curve)
is perturbed with a velocity field that causes the star to
expand. The initial
configuration also determines the extension of the numerical grid indicated
by the vertical line. At a later time $t_1$ the star has expanded
(long dashed curve). The outer part of the star has therefore moved out of
the numerical grid (dotted part of the curve) and the corresponding
information would be lost in a non-linear numerical evolution.
At time $t_2$ the star 
has shrunk and is completely contained inside the numerical grid.
Outside of the star the energy density will be zero. In general,
therefore, the energy density profile or its derivatives will have a
discontinuity at the stellar surface. Worse from a numerical point of view
is the region between the stellar surface and the outer
grid boundary.
Even though the energy density will be zero at these points theoretically, 
numerically this will not exactly be the case. At some of these points the
total energy density will have small negative values due to numerical
noise, unless the
values are manipulated in some form. A negative energy density, however,
means that the pressure can no longer be calculated from the equation of 
state which normally terminates the evolution. There are several 
possibilities for dealing with these difficulties. We will discuss
four methods and implement two of them in the course of this work. \\
\begin{list}{\rm{\arabic{count}.)}}{\usecounter{count}
             \labelwidth1cm \leftmargin1.0cm \labelsep0.4cm \rightmargin0cm
             \parsep0.5ex plus0.2ex minus0.1ex \itemsep0ex plus0.2ex}
\item
The first method consists in embedding the star in an atmosphere
of low density. In this method the numerical grid extends
well beyond the size of the neutron star and no information is lost at
any stage of the evolution. The boundary conditions are then applied
to the atmosphere whereas the star will always be confined to the interior
numerical grid and the surface of the star is entirely described by
the interior numerical evolution, for example by shock capturing methods.
It is a non-trivial question, however, to what extent the atmosphere
and the numerical treatment of the surface discontinuities
will affect the evolution of the neutron star.
% especially at reasonably
%small amplitudes.
For this reason it seems plausible to use an atmosphere
of low density. A low density, however, will in general be accompanied by a
small speed of sound and we have already seen in the discussion of the 
wave equation in section \ref{TOV_RYTRAFO} that such regions require
a careful numerical treatment. An insufficient resolution
may result in spurious phenomena. In terms of a rescaled radius such as
the coordinate $y$ defined in Eq.\,(\ref{TOV_YOFR}) we have been able to obtain
a sufficient resolution, but a large number of grid points would be required
to simulate an atmosphere of significant spatial extension. \\
An interesting variation of this method consists in viewing the surface
of the star as an interface to an exterior vacuum region and explicitly
tracking the movement of the interface. Sophisticated techniques such as
{\em level set methods} and {\em fast marching methods}
have been developed for these
purposes (see for example \citeNP{Sethian1999}) and may provide an answer to
the surface problem in Eulerian formulations. One may even go a step further
and recall the strikingly similar concept of Cauchy-characteristic
matching and, thus, consider a combination of these ideas. It is, however,
well beyond the scope of this work to investigate these methods in more
detail and we will therefore focus on simpler techniques.
\item
The second method is a modified version of the atmosphere approach discussed
above. Instead of using an external atmosphere, we
modify the equation of state of the neutron star at low densities
and thus view the outer layers of the neutron star itself as an atmosphere.
For that purpose we use an equation of state given by
\begin{align}
  P &= K\,\rho^{\gamma}\hspace{4.17cm} \mathrm{if }\,\,
      \rho > \rho_{\rm t}, \\[10pt]
  P &= a_1 \rho + a_2\rho^2 + a_3 \rho^3 + a_4 \rho^4 \hspace{0.8cm} 
      \mathrm{if }\,\, \rho \le \rho_{\rm t},
\end{align}
where $a_2$, $a_3$ and $a_4$ are coefficients determined by the continuity
of $P$ and its first two derivatives with respect to $\rho$. The
coefficient $a_1$ and the transition density $\rho_{\rm t}$ are free
parameters that are specified by the user. A consequence of this definition
is that $P\sim \rho$ at low densities and the behaviour will
be similar to that of a $\gamma=1$ polytrope in this region, i.e.
extend beyond the surface of the original purely polytropic model. The
low density part of the neutron star can thus be viewed as an atmosphere
smoothly attached to a polytropic neutron star truncated at $\rho_{\rm t}$.
Whenever the energy density falls below a threshold value $\rho_{\rm min}$
during the evolution, it is set to this threshold value. The
parameter $\rho_{\rm min}$ also needs to be specified by the user.
This requirement avoids the occurrence of negative total energy densities,
but introduces ad hoc discontinuities in the $\delta \rho$ profile. We take
care of these discontinuities by introducing artificial viscosity of the
%``standard'' modification of the 
modified von Neumann-Richtmyer form (see for example \citeNP{Fox1962})
\begin{align}
  q &= \left\{ \parbox{7cm}
                 {
                 $\displaystyle{b \Delta y^2 \hR w_{,y}^2}
                 \hspace{1.5cm} {\rm if}\,\,\, w_{,y} < 0 \\[10pt]
                 \displaystyle{0 \hspace{2.9cm} {\rm if} \,\,\, w_{,y} \ge 0},$
                 } \right. \label{qrv2}
\end{align}
where $b$ is the viscosity parameter. In many
cases $b=2$ leads to satisfactory results.
This viscosity term is added to the pressure perturbation $\delta P$
wherever it occurs in the equations. With careful choices of the
free parameters $a_1$, $\rho_{\rm t}$, $\rho_{\rm min}$ and $b$ we have
obtained long term stable evolutions of localised wave pulses. The particular
values we have to choose for a stable evolution, 
especially the density values $\rho_{\rm t}$
and $\rho_{\rm min}$, do however depend sensitively on the initial data.
Furthermore the manipulation of the energy density perturbation $\delta \rho$
in cases of a negative total energy density leads to a contamination
of the evolution of eigenmodes in the low density range. The resulting
disturbances then travel into the stellar interior within a few
oscillation periods. In view of these difficulties we have decided to
use a different treatment of the stellar surface. \\
\item
A fully satisfactory solution to the surface problem in one spatial
dimension can be obtained with
a Lagrangian formulation either of the surface or the whole star. 
In the first case this can be implemented by rescaling to a new
radial coordinate
\begin{align}
  s&:= \frac{r}{R(t)},
\end{align}
where $R$ is the time dependent total radius of the star. This transformation
leads to a few extra terms in the equations in the radial gauge,
but is more complicated
to implement in terms of the rescaled coordinate $y$. For this reason
and because of the wider range of applications we have chosen instead
to reformulate the non-linear radial oscillations entirely within
a Lagrangian framework. Combined with the singularity avoiding properties
of the polar slicing condition the resulting code can not only be used for 
the simulation of radial oscillations but also allows high resolution
studies of spherically symmetric gravitational collapse. This code and the
corresponding testing will be discussed in detail in section \ref{LAGR}. \\
Even though Lagrangian methods represent a formidable
tool for 1-dimensional problems,
we have already mentioned that there is no
straightforward generalisation to two or three spatial
dimensions, where the paths of fluid elements may intersect and give
rise to caustics. \\
\item
The method we will be using in the remainder of this section can be considered
the inverse of the atmospheric treatments discussed above. Instead
of adding matter in the form of an atmosphere the outer layers of the
star are removed. In this context it is worth remembering that
the solution of the TOV equations via quadrature does not go all the way
out to $\rho=0$ and a fully non-linear perturbative code working with
such a background intrinsically describes a truncated neutron star.
The percentage of mass that we will remove from the star will be very
small in most cases ($\ll 1 \%$). We will see below that the
resulting code behaves well in the linearized limit in most cases.
% The truncation
% density below which the outer layers are removed from the background star,
%has to be chosen larger for higher amplitude perturbations, however, which
%leads to spurious numerical effects in the evolution of marginally
%stable neutron stars, i.e. stars with a central density just below
%the critical value. Consequently we will follow a more conservative approach
%and study non-linear effects in a simplified neutron star model. Such
%a model, albeit less realistic, is numerically under control in the sense
%that no modifications of the model are required throughout the
%amplitude range of interest and thus spurious numerical effects are
%avoided. This model will enable us to study non-linear effects
%in radial oscillations of a neutron star like dynamic system. \\
\end{list}

%=========================================================================
\subsubsection{The numerical implementation in Eulerian coordinates}
\label{PERT_NUMERICS}
In section \ref{PERT_EQ} we have derived the equations for a fully non-linear
perturbative formulation of a dynamic spherically
symmetric star in terms of the generalised coordinate $x$. In the
remainder of the Eulerian discussion we will restrict ourselves to the
rescaled version and set $r_{,x} = C$ and $x=y$.
In order to numerically solve these equations, we
also have to specify appropriate boundary conditions. 
We start with the origin and recall that the displacement $\xi$ of a 
fluid element at the centre of a spherically symmetric star
vanishes. As a consequence the radial velocity will also vanish at
the origin.
As far as the energy energy density is concerned, we note that
$\hat{\rho}$ is a component of a rank 2 tensor
and therefore the spatial derivative $\hat{\rho}_{,y}$ will vanish
in spherical symmetry. The same is true for the background density
$\rho$ and therefore we obtain the inner boundary condition
$\delta \rho_{,y}=0$. Finally we require the vanishing of
$\delta \mu$ to avoid a conical singularity. \\
At the outer boundary we match
the lapse function to an exterior Schwarzschild metric as in the static
case which results in the condition $\hat{\lambda}\cdot \hat{\mu}=1$. As
far as the matter variables are concerned, the situation is a bit
more complicated. For the velocity we use the regularity condition $w_{,y}=0$.
In view of the definition of the radial coordinate $y$ this is equivalent
to demanding that the velocity
has a finite gradient with respect to $r$ at the surface. 
This condition is satisfied by the eigenmode solutions obtained
in section \ref{PERT_LIN}.
In Fig.\,\ref{MODES1_4} we can see that the gradient $w_{,y}$ vanishes
for all three polytropic exponents $\gamma=1.75$, $2.00$
and $2.3$.
In contrast to the velocity gradient the derivative of the energy density
perturbation $\delta \rho_{,y}$ will in general not vanish at the surface.
If we consider the stellar models listed in Table \ref{MODELS15}
it can be shown that $\delta \rho_{,y}$ will only vanish in the
case $\gamma=2$ which is also illustrated in Figs.\,\ref{MODES1_4}
and \ref{EIG_DRHO}. In summary the boundary conditions are
\begin{align}
  \delta \rho_{,y} &= 0, \label{PERT_BCINDRHO} \\[10pt]
  w &= 0, \label{PERT_BCINW} \\[10pt]
  \delta \mu &= 0. \label{PERT_BCINDMU}
\end{align}
at the origin and
\begin{align}
  w_{,y} &= 0, % \hspace{0.5cm} \mathrm{or} \hspace{0.5cm} w =0,
  \label{PERT_BCOUTW} \\[10pt]
  \hat{\lambda} \cdot \hat{\mu} &= 1
\end{align}
at the surface. \\
In this context it is worth mentioning a subtlety concerning
second order finite differencing schemes used for evolution
equations such as (\ref{PERT_DRHOT}), (\ref{PERT_WT}).
In general this system of equations 
has one ingoing and one outgoing characteristic at each boundary
and physical information has to be specified in the form of one
condition for either $w$ or $\delta \rho$ at either boundary. The centred
finite differencing scheme (or variation thereof) used in second order 
techniques, however, cannot be applied at the grid boundaries and the
variables must be evolved in an alternative way. The physical
boundary conditions do not necessarily provide enough information
for this. In our case, for example, we have two variables $\delta \rho$, $w$ 
that need to be updated at
two grid points respectively which requires four conditions, but only 
two conditions are required to provide information for the
characteristics entering the numerical grid.
The remaining boundary values not determined by these two conditions
have to be obtained in alternative ways,
for example by extrapolation or the use of one sided derivatives in the
evolution equations. We have obtained optimal performance in the evolution
of $\delta \rho$ and $w$ by using conditions (\ref{PERT_BCINDRHO})
and (\ref{PERT_BCINW}) at the centre and (\ref{PERT_BCOUTW}) at the
surface. The outer boundary value of $\delta \rho$ is then obtained
by extrapolation on each new time slice.
It is worth pointing out that this problem is not apparent in the implicit
finite difference methods applied to the cosmic string in section \ref{cstr}
or the Lagrangian code in section \ref{LAGR}. \\
Before we schematically outline the computational steps involved in
the time evolution we need to discuss one final numerical issue,
the CFL stability condition. We have mentioned in section \ref{stability}
that the stability criterion of Courant, Friedrichs and Lewy requires
the physical domain of dependence to be included in the numerical
domain of dependence. A standard method to ensure that this criterion
is met in a hydrodynamical evolution is based on calculating the
slopes of the characteristics at each point on the numerical grid.
In our case we consider the system of evolution equations
(\ref{PERT_DRHOT}), (\ref{PERT_WT}). The quasi-linear nature of this
system enables us to calculate the characteristics from
\begin{align}
  \frac{dy_i}{dt} &= \Lambda_i,
\end{align}
where $\Lambda_i$ are the eigenvalues of the coefficient matrix
and are defined by the equation
\begin{align}
  \left[ \begin{pmatrix} \alpha_{11} & \alpha_{12} \\
                  \alpha_{21} & \alpha_{11} \\
  \end{pmatrix} -\Lambda \cdot \hbox{\oneone{1}} \right]
  \begin{pmatrix} \delta \rho \\ w
  \end{pmatrix} &= 0.
\end{align}
The solution for the coefficient functions 
(\ref{PERT_ALPHA11})-(\ref{PERT_ALPHA21}) is given by
\begin{align}
  \Lambda &= \frac{1}{r_{,x}D} \left( w(1-\hat{C}^2) \pm \frac{\hat{C}}
             {\hat{\mu}\hat{\lambda}v} \right).
\end{align}
If the characteristics are straight lines, the Courant-Friedrichs-Lewy
condition is satisfied if the time step $dt$ obeys the inequality
\begin{align}
  dt &\le \frac{dy}{\max{|\Lambda_i|}}. \label{PERT_DT}
\end{align}
We therefore calculate the eigenvalue fields $\Lambda_1$, $\Lambda_2$ on each
time slice and determine the value of $\max{|\Lambda_i|}$.
Even though the characteristics will in general not be straight lines,
the deviation is small on time scales of $dt$ and we allow for
this effect by multiplying the resulting time step by a factor of 0.9. With
that choice and about 500 grid points
we have obtained stable evolutions over several 100000
time steps which corresponds to more than $0.1$ s of proper time
as measured by an observer at infinity.\\
We have got all ingredients now to summarise the individual steps involved in
the fully non-linear numerical evolution.
\begin{list}{\rm{(\arabic{count})}}{\usecounter{count}
             \labelwidth1cm \leftmargin1.5cm \labelsep0.4cm \rightmargin1cm
             \parsep0.5ex plus0.2ex minus0.1ex \itemsep0ex plus0.2ex}
\item A static background model is calculated according to the
      TOV equations (\ref{TOV_RY})-(\ref{TOV_PY}), where
      we set $r_{,x}=C$. For this
      purpose the polytropic exponent $\gamma$, the polytropic factor
      $K$, the central density $\rho_{\rm c}$ and the surface density
      $\rho_{\rm s}$ need to be specified by the user. A non-zero
      surface density will result in a truncated neutron star model.
      The results are given in the form of data
      files containing the background variables $\lambda$, $\mu$,
      $\rho$ and $r$ as functions of $y$.
\item If initial data is required in the form of eigenmode profiles,
      the eigenmode solutions can be calculated according to the
      method described in section \ref{PERT_LIN}. The order of the eigenmode is
      determined by the initial guess for the frequency which needs
      to be specified. The amplitude of the eigenmode is a free parameter
      in the evolution code.
\item There are several alternative choices for the initial data. Among
      these are localised perturbations of Gaussian shape and
      linear combinations of different eigenmodes.
\item With the initial velocity $w$ and energy  
      density $\delta \rho$ specified, the metric 
      perturbations follow from the constraint
      equations (\ref{PERT_LAMBDAR}) and (\ref{PERT_MUR}).
      These equations are numerically integrated with a fourth
      order Runge-Kutta scheme.
\item The initial data is evolved according to the second order in
      space and time McCormack scheme described in section \ref{McCormack}.
      One evolution cycle
      consists of the following steps. \\[4pt]
      a) Calculation of the Courant factor, \\
      b) predictor step for $\delta \rho$ and $w$,\\
      c) application of the inner boundary conditions for
         $\delta \rho$ and $w$,\\
      d) \parbox[t]{12.7cm}{integration of the constraint equations 
         to obtain preliminary values for $\delta \lambda$ and $\delta \mu$
         on the new time slice,}\\[10pt]
      e) corrector step for $\delta \rho$ and $w$,\\
      f) application of boundary conditions for $\delta \rho$ and $w$,\\
      g) \parbox[t]{12.7cm}{integration of the constraint equation on the 
         new slice to obtain the final values of $\delta \lambda$ 
         and $\delta \mu$.}
\end{list}

%=========================================================================
\subsubsection{The performance of the code in the linear regime}
\label{EULER_LIN}
\begin{figure}[t]
  \centering
  \epsfig{file=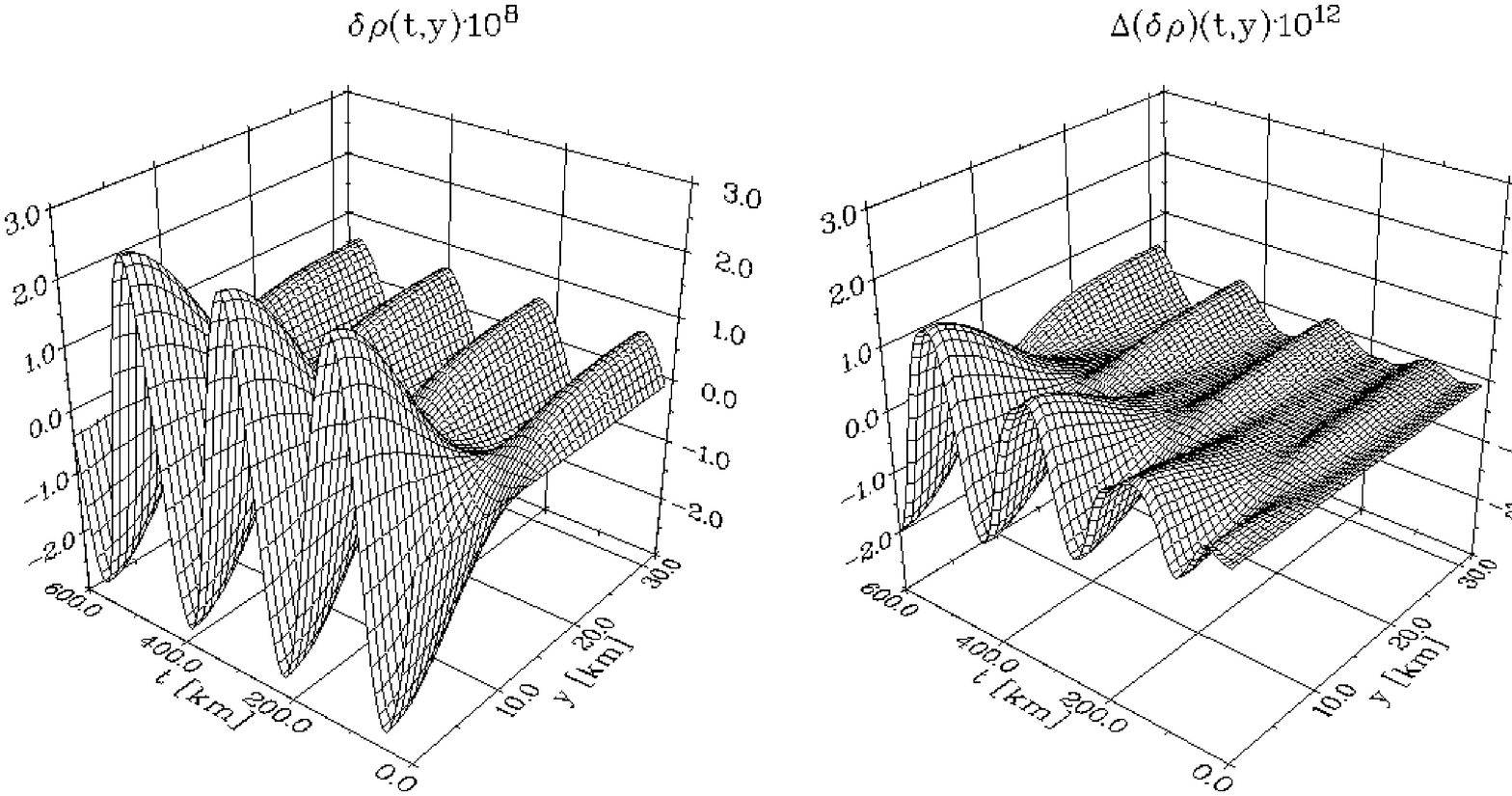, height=400pt, width=175pt, angle=-90}
  \epsfig{file=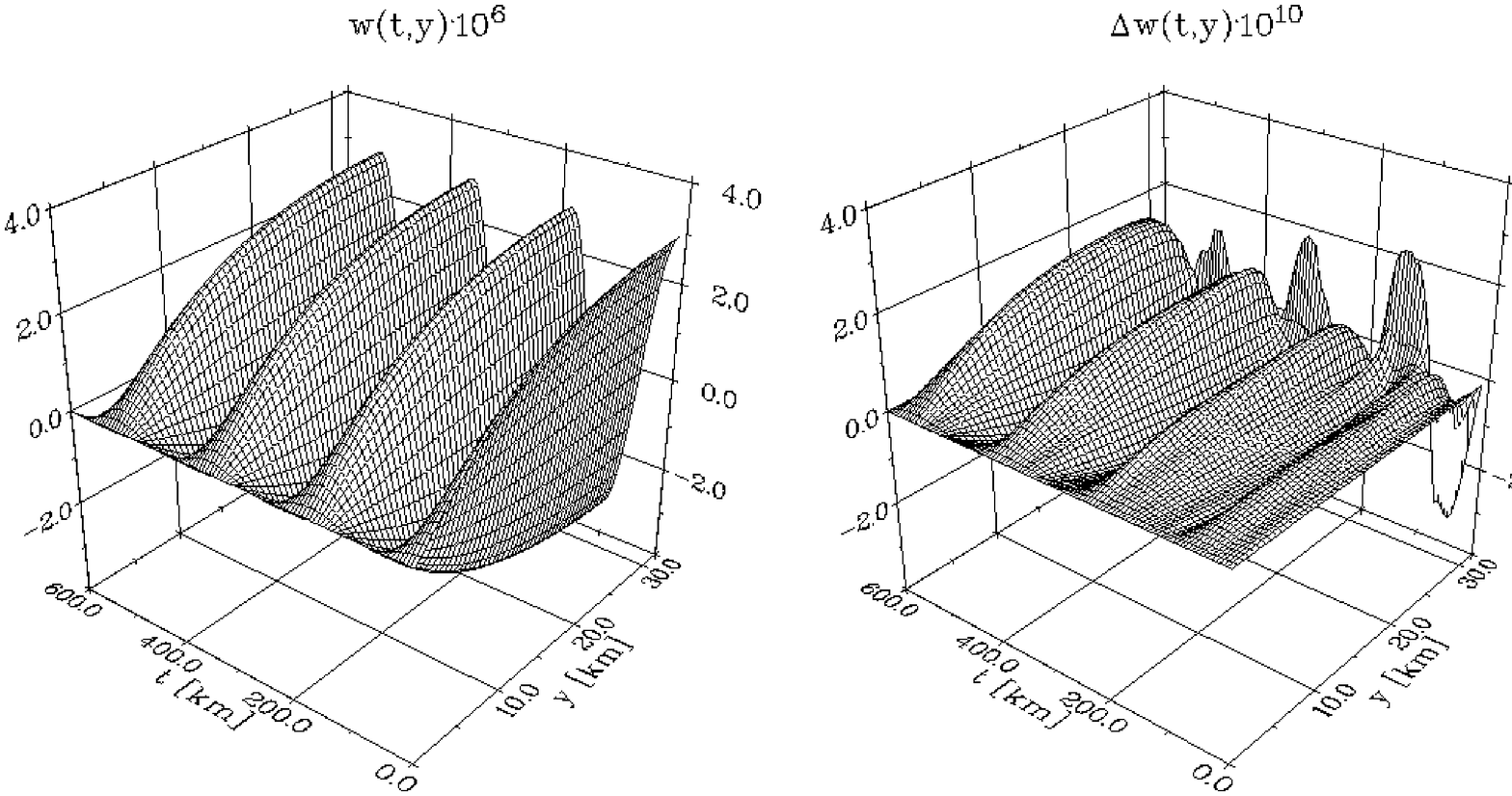, height=400pt, width=175pt, angle=-90}
  \caption{The left panels show the time evolution of $\delta \rho$ and $w$
           obtained for neutron star model 3 with $\gamma=2.00$.
           The initial perturbation
           is given in the form of the fundamental mode in the velocity field
           $w$. The right panels show the deviation from the exact
           solution of the linearized equations.}
  \label{PERT_LINEVOL}
\end{figure}

\begin{figure}[t]
  \centering
  \epsfig{file=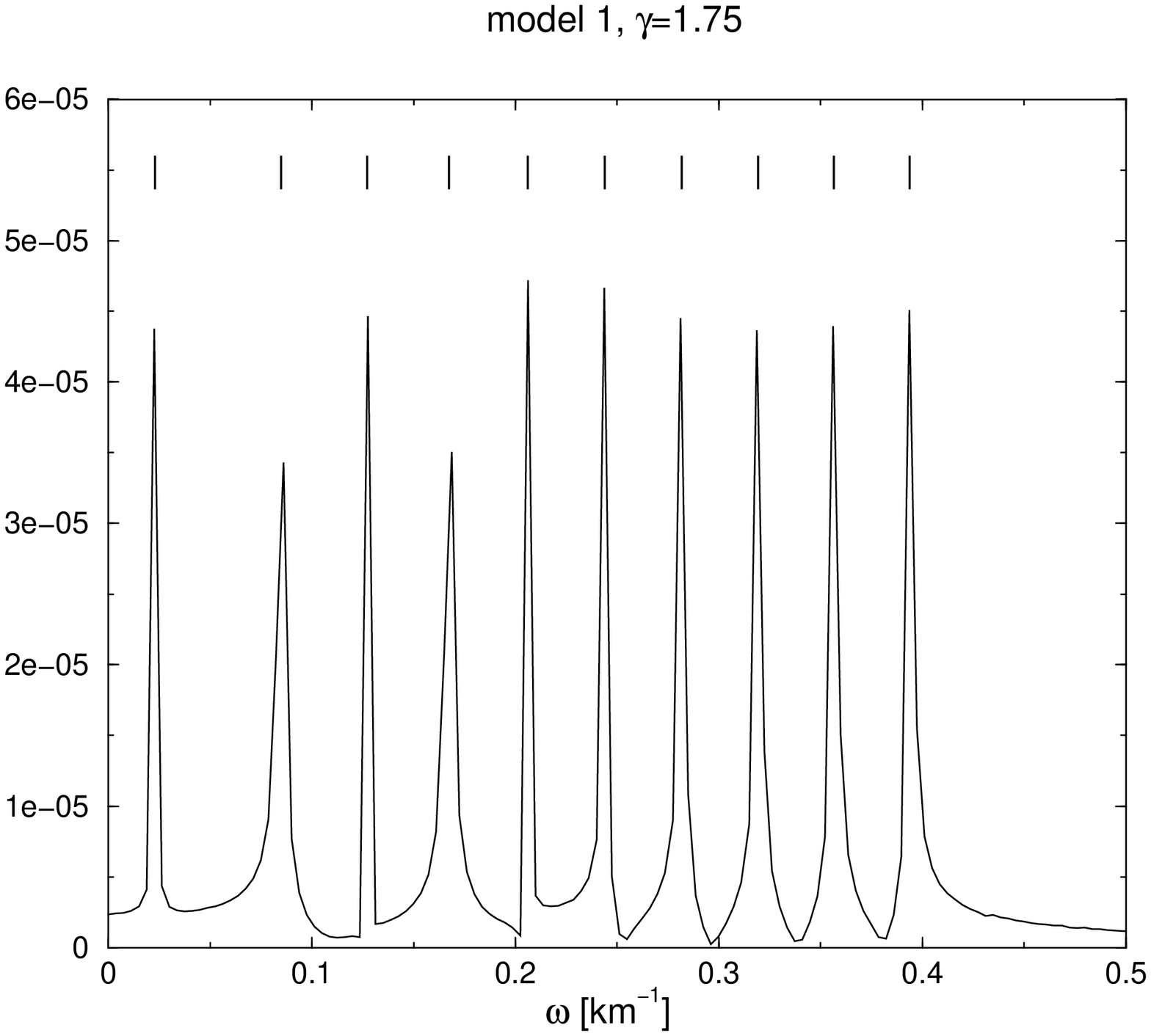, height=200pt, width=150pt, angle=-90}
  \epsfig{file=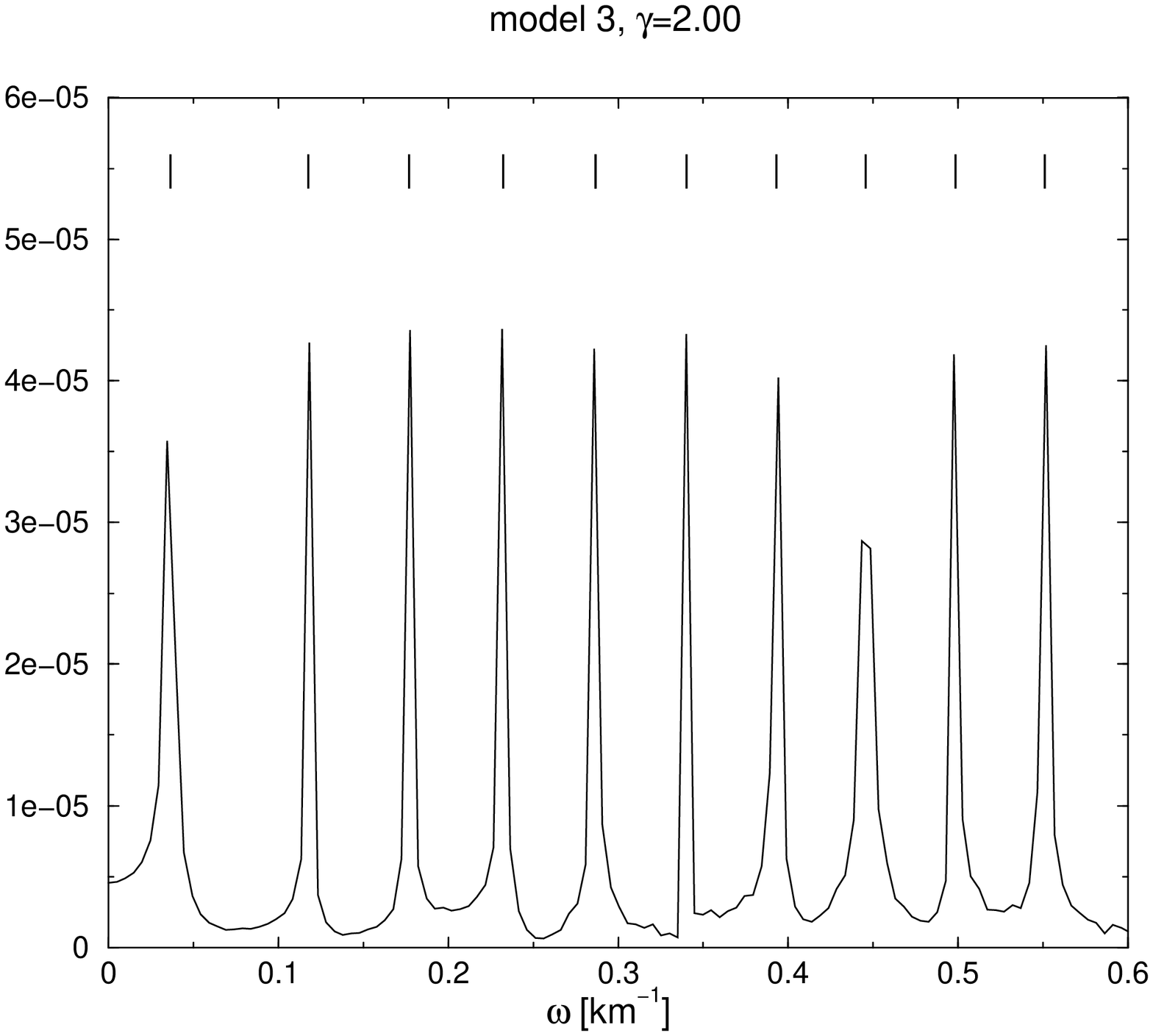, height=200pt, width=150pt, angle=-90}
  \caption{The power spectra of the time evolution of the central
           density for neutron star models 1 and 3. The vertical bars
           indicate the frequencies predicted by the linear analysis.}
  \label{PERT_LINFOUR}
\end{figure}

\begin{figure}[t]
  \centering
  \epsfig{file=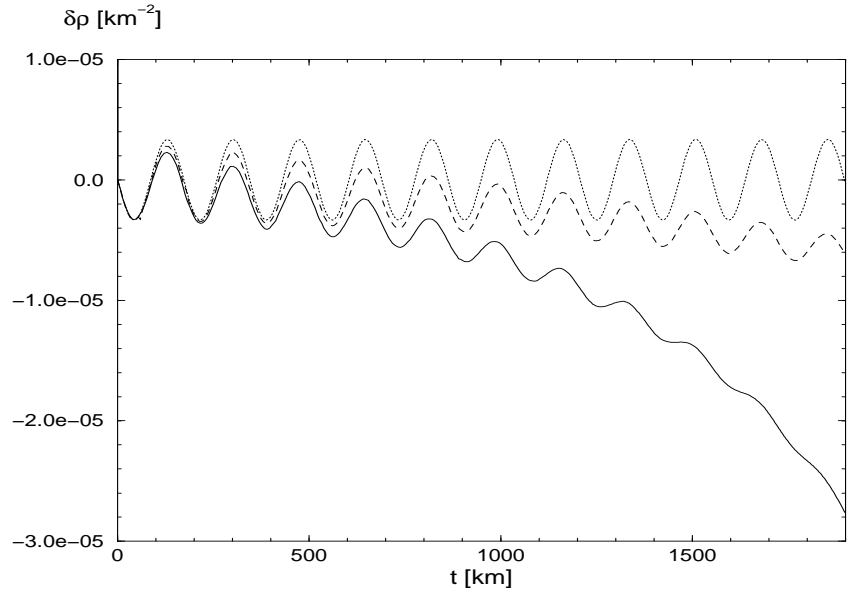, height=320pt, width=220pt,
          angle=-90}
  \caption{The central density perturbation corresponding to the
           fundamental oscillation mode of model 3 as obtained
           with a perturbative method for 600 grid points
           (dotted curve) and a non-perturbative method for 600
           (solid) 
           and 1200 grid points (dashed curve). See text for details.}
  \label{PERT_NONP}
\end{figure}

We will now investigate the performance of the code in the linear regime,
where we know the exact solution with high accuracy.
If initial data is provided in the form of an eigenmode profile $w_i(y)$
and zero $\delta \rho$, we know that
the time dependent solution in the linear regime is given by
\begin{align}
  \delta \rho(t,y) &= -\delta \rho_i(y) \sin{\omega_i t}, \label{LIN_DRHOOFT}
  \\[10pt]
  w(t,y) &= w_i(y) \cos{\omega_i t}. \label{LIN_WOFT}
\end{align}
For finite amplitudes this solution is not exact, but for 
sufficiently small amplitudes
the deviation of the exact solution from (\ref{LIN_DRHOOFT}),
(\ref{LIN_WOFT}) is negligible
compared with the truncation error of the numerical scheme.
We have therefore calculated the fundamental mode for stellar model
3 of Table \ref{MODELS15} using 1600 grid points and a truncation density
$\rho_{\rm s}=1.0\cdot 10^{-7}\,\,{\rm km}^{-2}$. This
density corresponds to the removal of about $3\cdot 10^{-8}$ of the
neutron star mass which is one order of magnitude smaller than the accuracy
of the numerically calculated total mass.
The amplitude of the eigenmode corresponds to an oscillation
of the stellar radius of about $10\,\,{\rm cm}$, i.e. a relative
displacement of about $10^{-5}$. In Fig.\,\ref{PERT_LINEVOL}
we show the time evolution
of $\delta \rho$ and $w$ together with the deviation from
the analytic solution (\ref{LIN_DRHOOFT}), (\ref{LIN_WOFT}).
The numerical evolution reproduces the expected harmonic time dependence
with high accuracy. Because of its low frequency the fundamental mode
is particularly suitable for this graphical illustration. The code
reproduces the sinusoidal evolution of higher modes with comparable
accuracy, but the large number of oscillations is not well resolved in plots
similar to Fig.\,\ref{PERT_LINEVOL}.
For the same reason we have shown the earlier stages
of the evolution up to $t=600\,\,{\rm km}$ only in the figure. The whole run
lasts more than ten times longer and shows no significant loss
of accuracy. It is worth
mentioning that the accuracies obtained here are limited not only by
the evolution code but also by the results for the static background,
the eigenmode profiles and, most importantly, the eigenmode
frequencies used in the calculation of the analytic solution.
The same long term stability and high accuracy has been observed in
similar evolutions for a variety of different neutron star models with 
polytropic indices $\gamma\le 2$. Below we will see, however, that the code 
does not perform equally satisfactorily if we use a larger truncation density
in combination with a marginally stable
neutron star model with a central density just below the
critical value. \\
For neutron star models sufficiently far away from the stability limit,
we can also check the performance of the code in the linear regime by
comparing the frequency spectrum of the time evolution with the
values predicted by the eigenmode calculations of section \ref{PERT_LIN}.
For this purpose initial velocity fields have been calculated 
for models 1 and 3 by
adding the first ten eigenmode profiles whereas the initial density
perturbation is set to zero. The combined amplitude of the perturbations is
similar to that used above for determining the
deviation from harmonic time dependence. 
In Fig.\,\ref{PERT_LINFOUR} we show the Fourier spectra for the 
corresponding time evolutions of the
central density perturbation $\delta \rho(t,0)$. The frequencies
predicted by the eigenmode analysis are indicated by vertical bars
and show good agreement with the peaks in the Fourier spectra. \\
Next we compare the performance of the perturbative
approach with that of a ``standard'' non-perturbative method.
We have already mentioned that we can simulate a non-perturbative approach
by using vacuum flat space for the background variables. In this case we
only use the TOV-model to determine the numerical grid as well as the areal
radius $r$ and the sound speed $C$ as functions of $y$.
The background variables, however, are
specified as $\lambda=1$, $\mu=1$ and $\rho=0$. If we insert these
values into the perturbative equations
(\ref{PERT_LAMBDAR})-(\ref{PERT_B2}) they will become identical to the
non-perturbative system (\ref{NONP_LAMBDAR})-(\ref{NONP_B2})
(after transformation to the radial coordinate $y$) with
$\hat{\lambda}$, $\hat{\mu}$, $\hat{\rho}$ and $\hat{P}$ replaced by
$1+\delta \lambda$, $1+\delta \mu$, $\delta \rho$ and $\delta P$. The
occurrence of the constant 1 in the metric variables has no implications
on the numerical performance. We have thus evolved
initial data in the form of a fundamental eigenmode profile in the
velocity field $w$ for neutron star model 3.
First we have used the TOV-background and a resolution of 600
grid points. We have then repeated the evolution with a flat space
background using 600 and 1200 grid points in order to check the dependence
of the non-perturbative results on the spatial resolution.
It is important
to note that the same code and the same evolution algorithm has been
used in both cases. 
%We have only altered the outer boundary
%conditions in the non-perturbative case where
%we use the first order evolution of the predictor
%step. With this choice we have obtained the best performance of 
%our non-perturbative code.
%If we use the original boundary conditions we obtain
%virtually the same result but on a shorter time scale.
The amplitude of the perturbation corresponds
to an oscillation of the surface of several metres. For this
amplitude we still expect the evolution to be dominated by the
harmonic time dependence, although the results of section \ref{MODECOUPLING}
below indicate the presence of weak non-linear effects.
The numerical results are shown in Fig.\,\ref{PERT_NONP}, where the
central density perturbation is plotted as a function of time.
We clearly see that the perturbative evolution results in the expected
sinusoidal time dependence. In the non-perturbative case the
central density shows similar oscillations but simultaneously
the mean value decreases significantly.
In longer runs this decrease is revealed
to be exponential and thus indicates a starting evaporation of the star.
Neutron star model 3, however, is located on the
stable branch as we can clearly see in Fig.\,\ref{TOV_FAMILIES}
and no collapse or evaporation is expected.
Indeed the higher resolution run indicates that the non-perturbative scheme
converges to the harmonic solution. In order
to understand this behaviour of the non-perturbative scheme we recall
the presence of background terms in the evolution equation
for $w_{,t}$. If we consider the coefficients $\tilde{b}_2$ and 
$\tilde{\alpha}_{21}$ given by Eqs.\,(\ref{NONP_ALPHA21}), (\ref{NONP_B2}),
we can see that the evolution equation (\ref{NONP_WT}) contains the
background in the form
\begin{align}
  e(y) &:= \frac{1}{\mu^2 D} \left[ \frac{\lambda_{,r}}{\lambda} 
  + \frac{C^2 \rho_{,r}}{(\rho + P)} \right]. \label{NONP_ERROR}
\end{align}
We know that this term vanishes by virtue of the TOV equation
\begin{figure}[t]
  \centering
  \epsfig{file=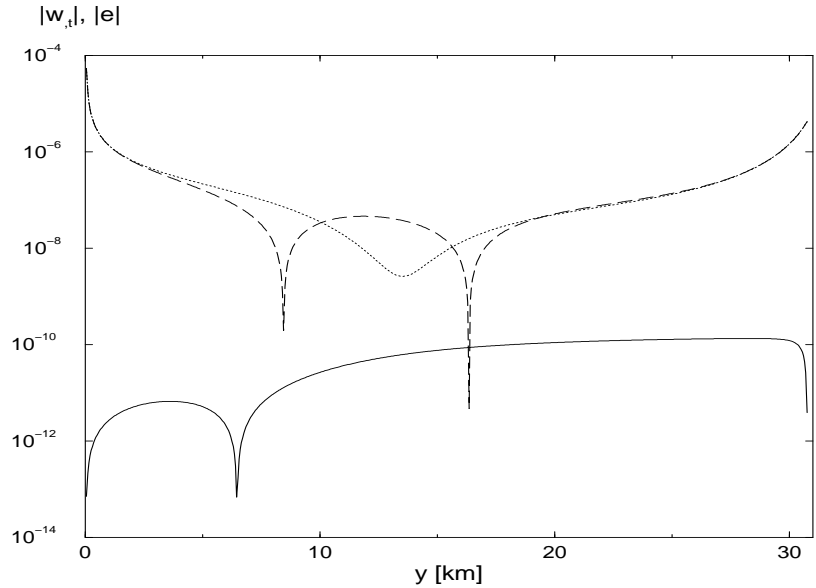, height=300pt, width=220pt,
          angle=-90}
  \caption{The source terms of the evolution equation for $w$
           (\ref{PERT_WT}) at the
           first computational step are shown for the perturbative (solid curve)
           and the non-perturbative scheme (dashed curve). The dotted
           curve shows the numerical error of the background terms
           and demonstrates the significance of the spurious source terms.}
  \label{SOURCEW}
\end{figure}
(\ref{TOV_PR}) and it has been removed from the equations in the perturbative
formulation. In the non-perturbative case, however, it will manifest itself
in the form of a residual numerical error.
This error is shown in Fig.\,\ref{SOURCEW} for the first
step in the evolution with
600 grid points together with the entire source terms of $w_{,t}$ as
given by Eq.\,(\ref{NONP_WT}) in the non-perturbative and
Eq.\,(\ref{PERT_WT}) in the perturbative case. Because of the cosine time
dependence of the velocity the source terms should nearly vanish at $t=0$.
It can be seen, however, that
the source terms are dominated by the residual numerical
error in the non-perturbative scheme which is particularly large
at the centre and the surface. On the time scale of one oscillation period,
about $150\,\,{\rm km}$, the spurious acceleration of up to
$10^{-4}\,\,{\rm km}^{-1}$ will have a significant impact on the 
oscillation of several metres of the star. A closer investigation
of the velocity field reveals that the integral effect of the residual
error is an increase in the velocity field near the surface. We
attribute the gradual evaporation of the star to this disturbance
in the velocity field which gradually radiates matter off the numerical 
grid. \\
Considering that the same code has been used for the comparison just described,
it is necessary to check the perturbative scheme for similar
spurious effects. After all the main advantage of the perturbative
scheme lies in higher accuracy which may postpone the onset
of a spurious collapse or evaporation but not necessarily avoid it.
We have already mentioned, however, that no significant deviation
from the harmonic time dependence has been observed in the case of model 3
and initial data in the form of eigenmodes over very long times.
In order to avoid even longer integration times and the associated
computational costs, we have
chosen an alternative way of testing the code for this behaviour. We use a
stellar model identical to model 3 but with a central density of
$\rho_{\rm c}=0.002802\,\,{\rm km}^{-2}$ which is just below the
critical value given in Table \ref{CRITFREQS}.
The initial data consist of the
fundamental velocity mode with an amplitude corresponding to a
surface displacement of about $10\,\,{\rm cm}$
and we use a numerical grid with 600
grid points. In the first calculation we have imposed a truncation density
of $\delta \rho_{\rm s}=5\cdot 10^{-6}\,\,{\rm km}^{-2}$ and in a second run
the intrinsic value of the TOV 
\begin{figure}[t]
  \centering
  \epsfig{file=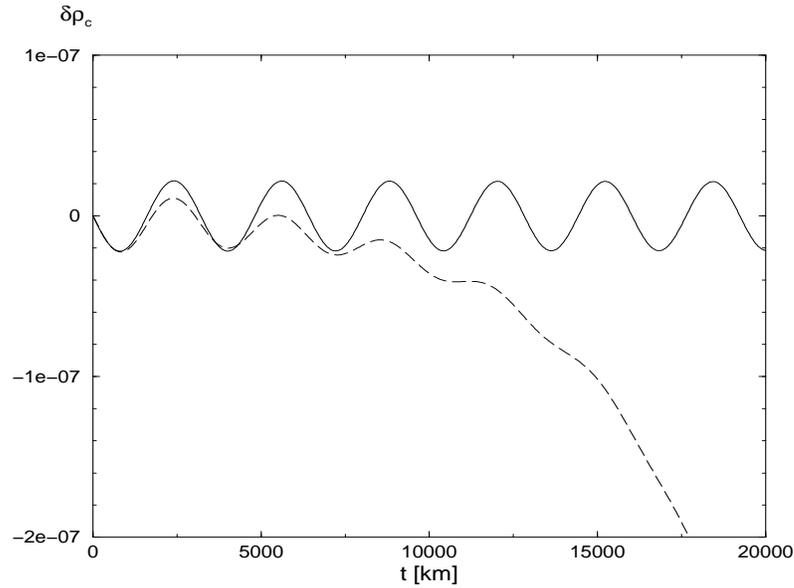, height=300pt, width=220pt,
          angle=-90}
  \caption{The central density resulting from the evolution of the
           fundamental eigenmode of 
           a neutron star corresponding to model 3 with a central density just
           below the critical value is plotted for a truncation
           density of $5\cdot 10^{-6}\,\,{\rm km}^{-2}$ (dashed curve)
           and $2.5\cdot 10^{-8}\,\,{\rm km}^{-2}$ (solid curve).
           }
  \label{PERT_MARGSTABLE}
\end{figure}
code $\delta \rho_{\rm s}=2.5\cdot 10^{-8}\,\,{\rm km}^{-2}$ is used.
In Fig.\,\ref{PERT_MARGSTABLE}
we show the resulting central density perturbation as a function of time.
For the small truncation density we obtain the expected sinusoidal
time dependence whereas the larger value significantly affects the evolution,
even though only a fraction of $10^{-5}$ of the stellar mass has been neglected
in this case. This result demonstrates the limitations of the code in its 
current form. For larger truncation densities it does not necessarily
guarantee mass conservation which we attribute to the boundary condition
(\ref{PERT_BCOUTW}) which is strictly valid only if the numerical grid
extends to $\rho=0$. For sufficiently small truncation densities
the resulting numerical error is
negligible and has no significant effect on the evolution. For larger
truncation densities, however, it can result in spurious phenomena similar
to those observed in the non-perturbative case. This is particularly
problematic
since the investigation of non-linear effects will require perturbations
of larger amplitudes and consequently larger truncation densities are
necessary in order to avoid total negative energy densities. From here
on we will therefore proceed in two different ways. In the remainder
of section \ref{DYNAMIC} we will investigate a simplified
neutron star model for
which the code ensures mass conservation for arbitrary amplitudes
and negative energy densities are still avoided. This model will
necessarily provide a less realistic description of a neutron star, but
the general structure of the eigenmodes remains the same and it
is not unrealistic to expect that non-linear effects such as mode coupling
will be qualitatively similar in
more realistic models. Considering the sensitivity of the numerical
evolutions to the treatment of the surface, it is, however, desirable to
develop a formulation of the dynamic neutron star which unambiguously 
provides a correct treatment of the surface. This will be done in
section \ref{LAGR} where we develop a fully non-linear perturbative Lagrangian
code.

%=========================================================================
\subsubsection{A simplified neutron star model}
\label{NEWMODEL}
\begin{figure}[t]
  \centering
  \epsfig{file=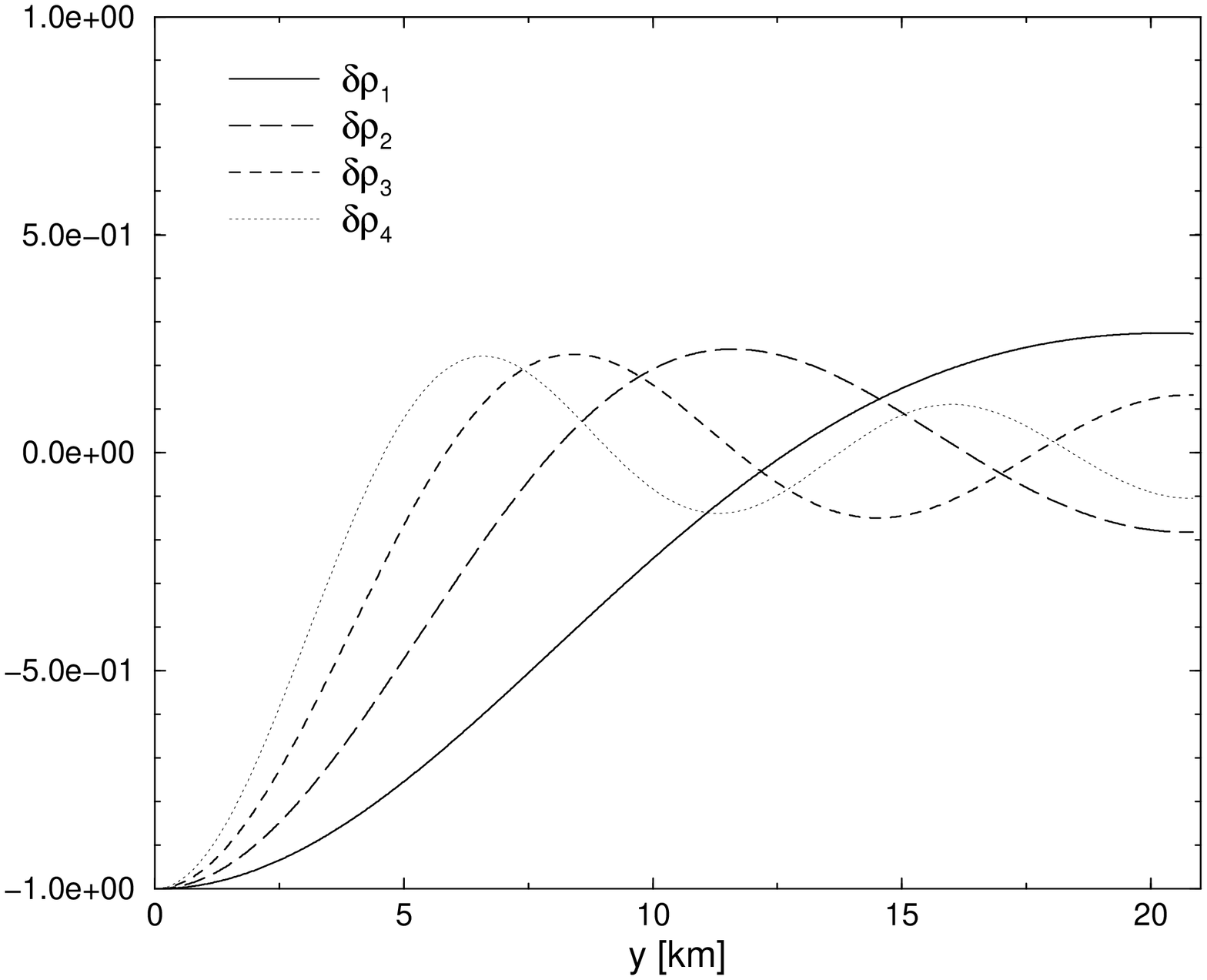, height=200pt, width=150pt, angle=-90}
  \hspace{0.5cm}
  \epsfig{file=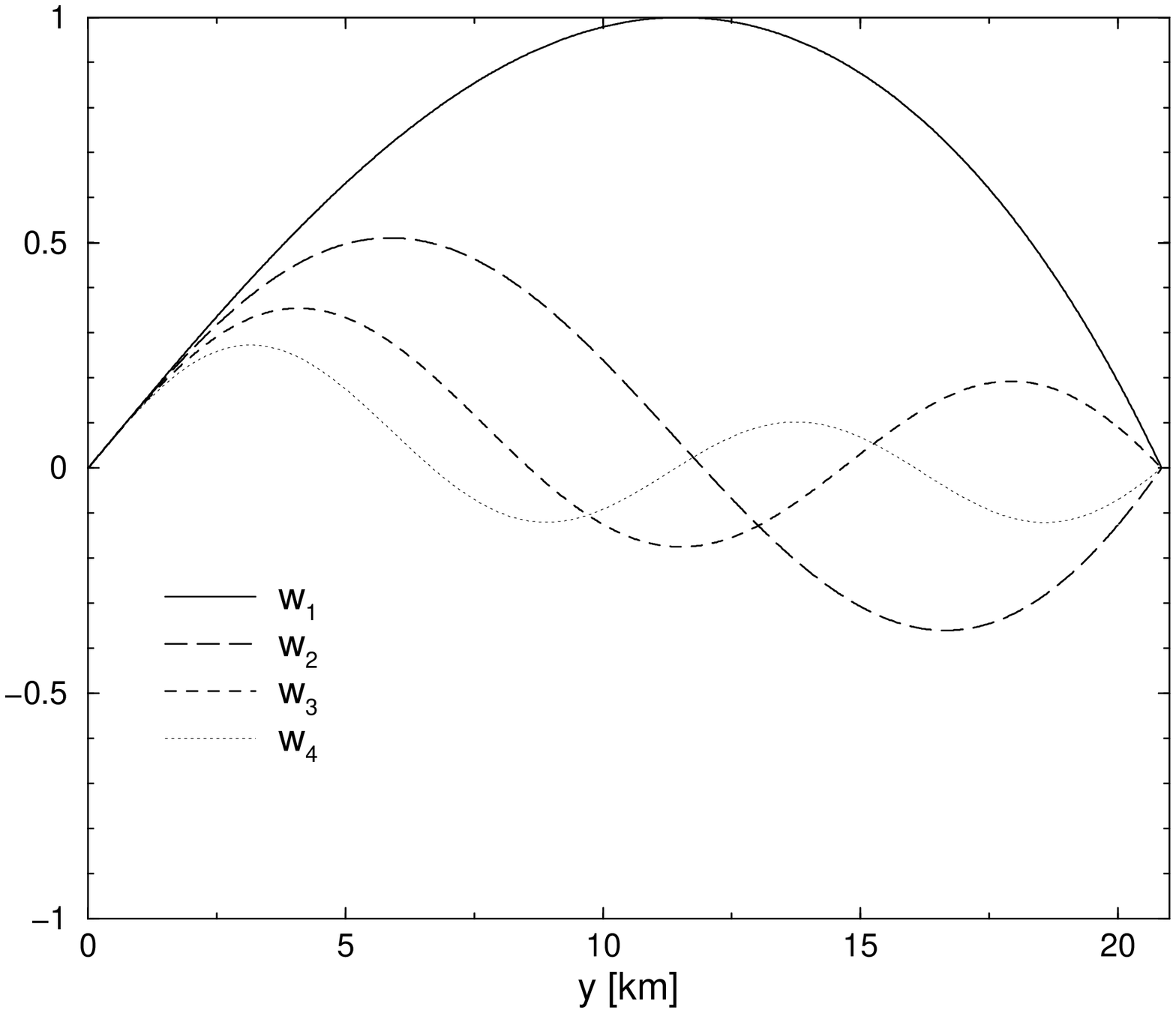, height=200pt, width=150pt, angle=-90}
  \caption{The profiles of the lowest four eigenmodes in $\delta \rho$
           and $w$ for the simplified neutron star model.}
  \label{MOD_MODES}
\end{figure}

In the previous section we have seen that a sufficiently large
truncation density in combination with the boundary condition
(\ref{PERT_BCOUTW}) may result in a continuous loss or gain of mass.
In order to avoid total negative energy densities, however, we have to use
sufficiently large truncation densities when we study
non-linear effects in the time evolution of large amplitude
perturbations. We have therefore decided to ensure mass conservation
by using the alternative boundary condition
\begin{align}
  w &= 0 \label{PERT_BCOUTWMOD}
\end{align}
at the surface instead of Eq.\,(\ref{PERT_BCOUTW}). This means that
the surface of the star remains at a fixed position in space and
only fluid elements in the interior of the star are displaced during the
evolution. It is the fixed location of the surface
which avoids the main problems
we have encountered with the Eulerian formulation so far.
The model we use for the following analysis
has the same equation of state as model 3 of Table \ref{MODELS15}
and a central density $\rho_{\rm c}=1.224\cdot 10^{-3}
\,\,{\rm km}^{-2}$ which implies a radius $R=11.34\,\,{\rm km}$
and a total mass $M=2.18\,\,{\rm km}$.
The truncation density is fixed at $\rho_{\rm s}=2.0\cdot 10^{-4}
\,\,{\rm km}^{-2}$ which means that the simplified model contains
$90\,\%$ of the mass of the original star and extends to $84\,\%$ of the
original radius. Apart from changing the
truncation density in the calculation of the TOV-background
and implementing the new boundary condition in the evolution code
only one further modification in the numerical setup described in
section \ref{PERT_NUMERICS} is required. The outer
boundary condition (\ref{LIN_BCOUTA})
in the calculation of the eigenmodes is replaced by
\begin{align}
  \xi(R) &= 0.
\end{align}
The resulting eigenmodes can be ordered in the same way as described in 
section \ref{PERT_LIN} and the evolution of eigenmodes 
in the linear regime again results
in harmonic time dependence as in the original case with
the frequencies predicted by the eigenmode calculation. The first
four eigenmode profiles of $\delta \rho$ and $w$ for the model mentioned 
above are shown in Fig.\,\ref{MOD_MODES}.
The plots show that the number of local maxima and minima of the
\begin{figure}[b]
  \centering
  \epsfig{file=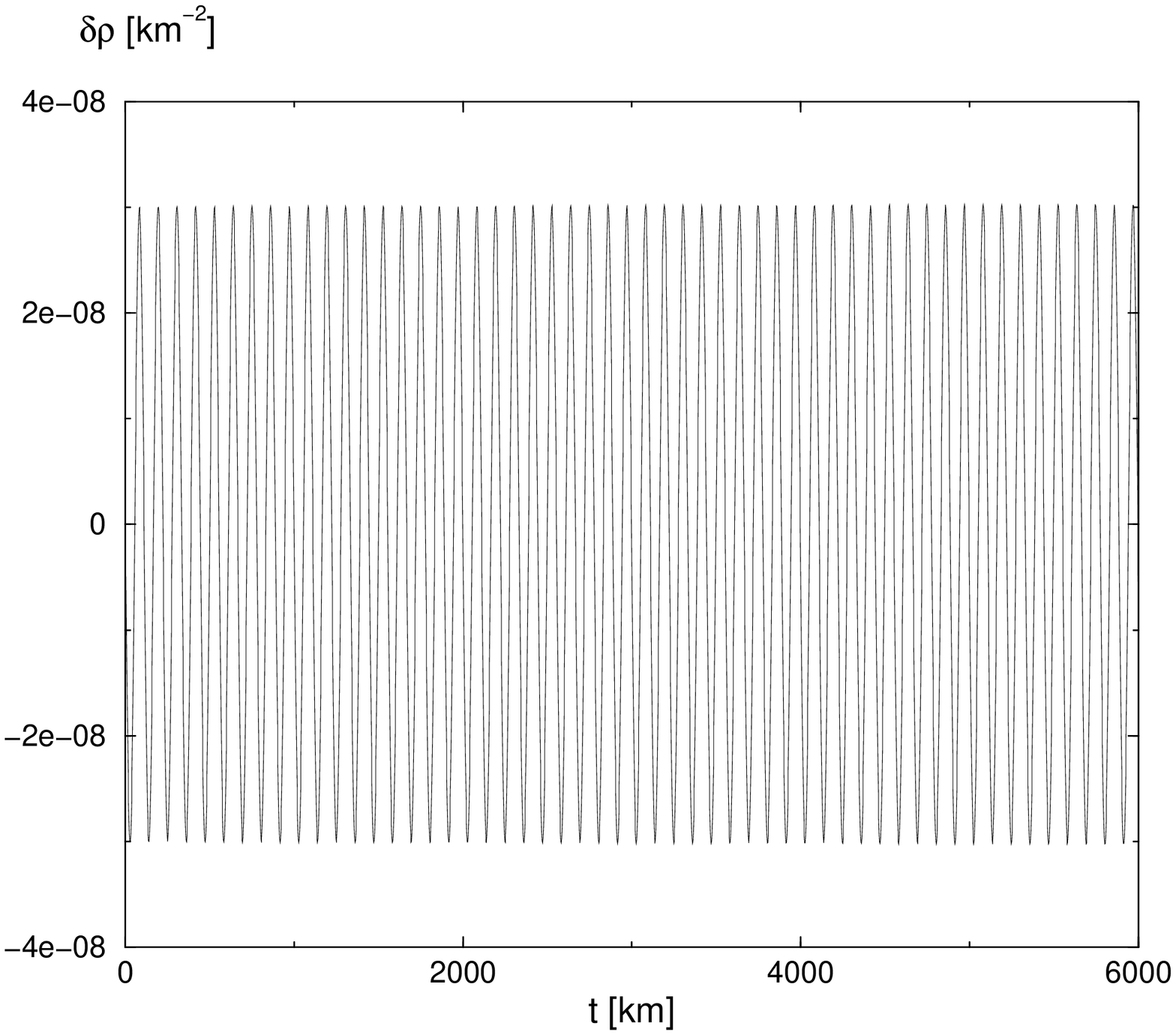, height=220pt, width=150pt, angle=-90}
  \hspace{0.5cm}
  \epsfig{file=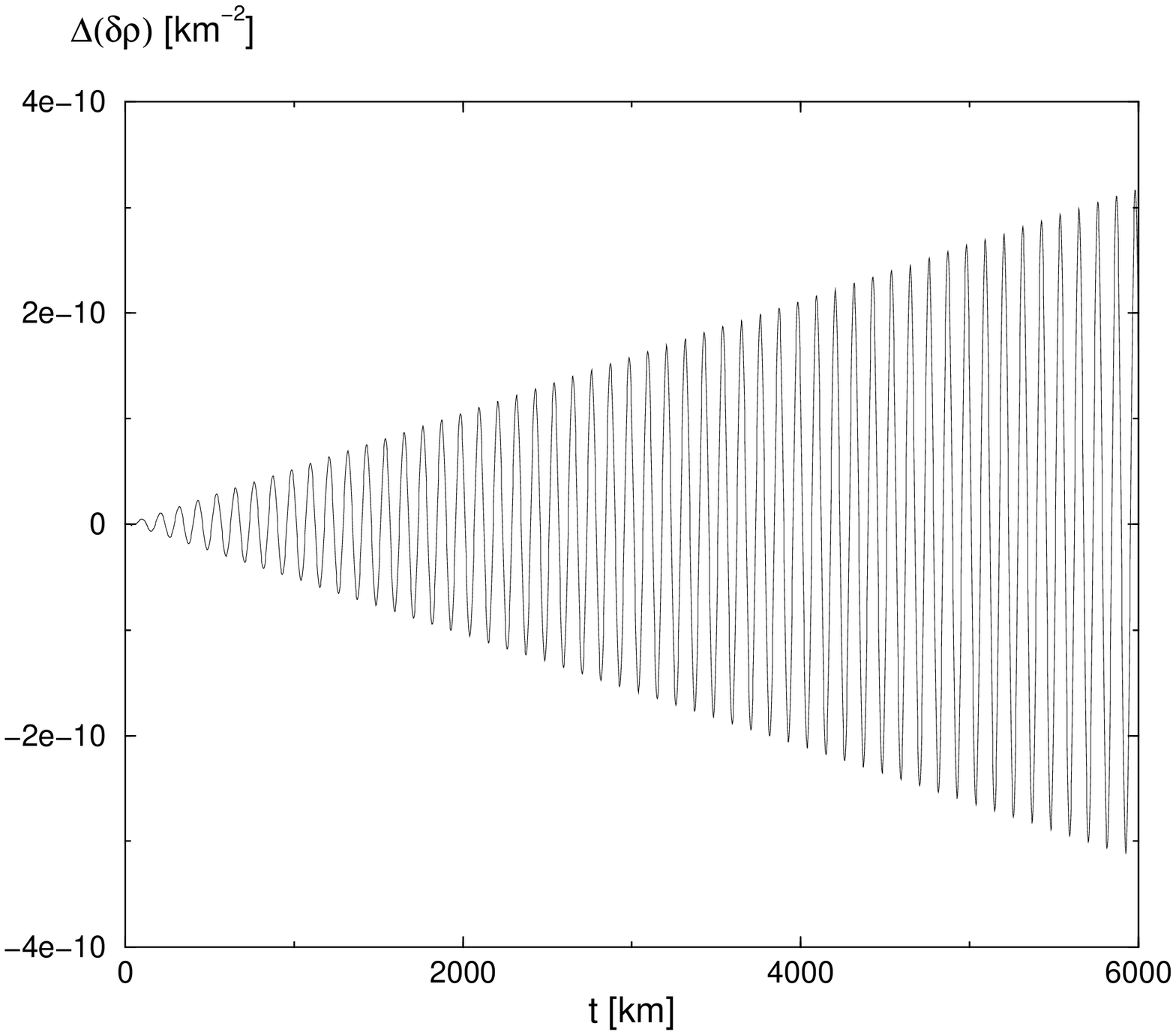, height=200pt, width=150pt, angle=-90}
  \caption{The evolution of the central density for initial data in the form
           a fundamental eigenmode in the velocity field for model 3
           with a central density $\rho_{\rm c}=0.002802\,\,{\rm km}^{-2}$.}
  \label{MOD_DRHOC}
\end{figure}
profiles still corresponds to the order of the mode.

%=======================================================================
\subsubsection{Testing the code with the new model}
The only modification of the code that needed to be implemented for the
new model is the outer boundary condition (\ref{PERT_BCOUTWMOD}).
The performance of the code in the linear regime is thus well established
by the results of section \ref{EULER_LIN} and we merely have to demonstrate
that no spurious results are obtained for larger truncation densities.
This is the only case where we will depart from the model parameters listed
in the previous section and use a central density
$\rho_{\rm c}=0.002802\,\,{\rm km}^{-2}$ instead. We thus recover the
parameters of the model which lead to a spurious evaporation of the star in
Fig.\,\ref{PERT_MARGSTABLE}.
For this model we have again evolved initial data in the form of the
fundamental mode of the velocity with an amplitude of $10\,{\rm cm}$
using 600 grid points. In Fig.\,\ref{MOD_DRHOC} we show the resulting
central density $\delta \rho_{\rm c}$
together with the deviation from the harmonic
solution of the linearized case. For presentation purposes
we only show the evolution up to $t=6000\,\,{\rm km}$. The harmonic
time dependence is reproduced with reasonable accuracy as the deviation
increases linearly up to about $1\,\%$. In general we have found the
eigenmode frequency the quantity most vulnerable to numerical error as
can be seen for example by varying the resolution. Because of this
observation and the oscillatory character of the deviation in
the figure
we attribute the error mainly to the limited accuracy of the frequency
rather than the numerical error of the time evolution itself.
The increasing phase shift between the numerical and the analytic solution
arising from the limited accuracy of the frequency
\begin{figure}[t]
  \centering
  \epsfig{file=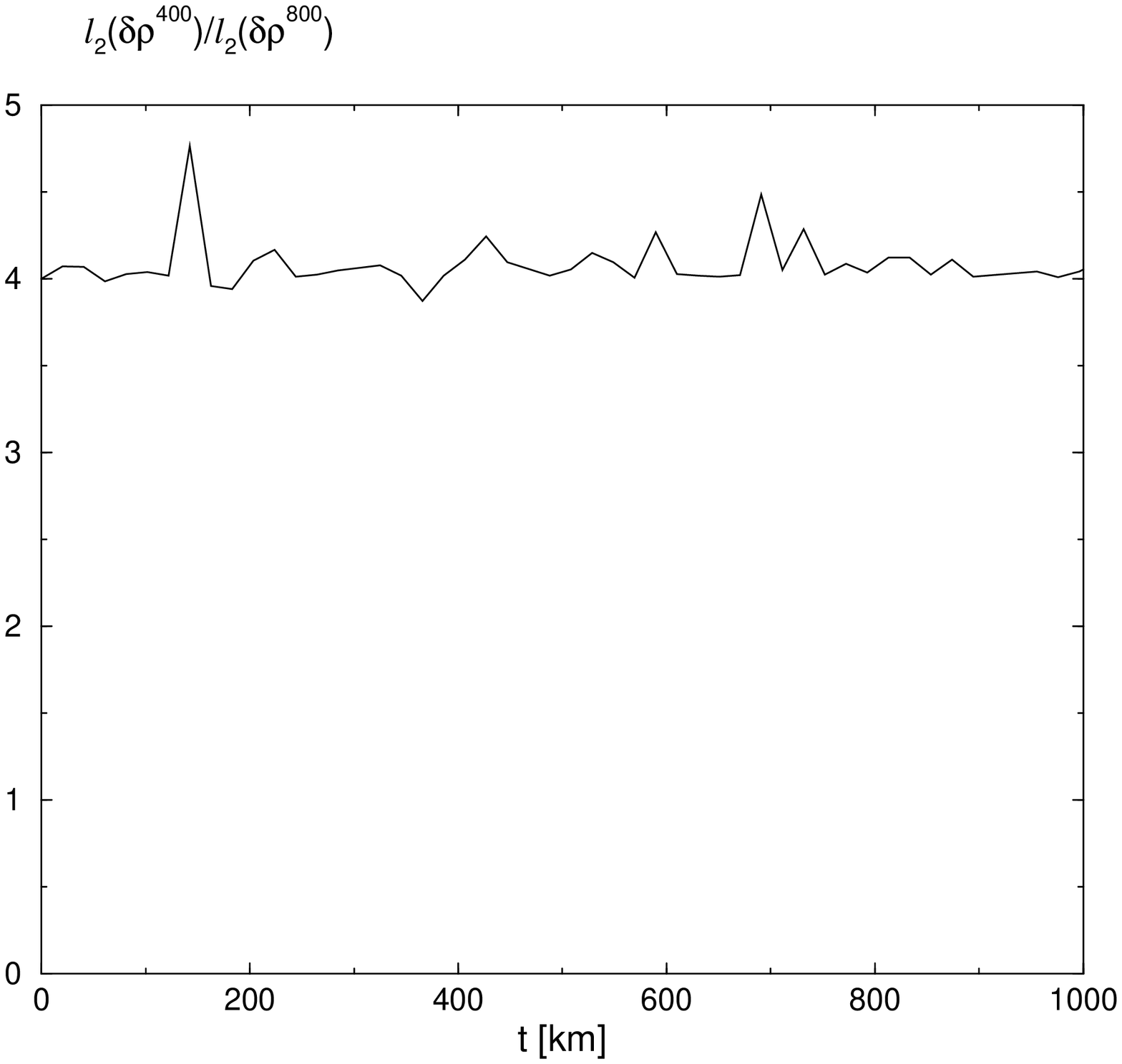, height=200pt, width=175pt, angle=-90}
  \epsfig{file=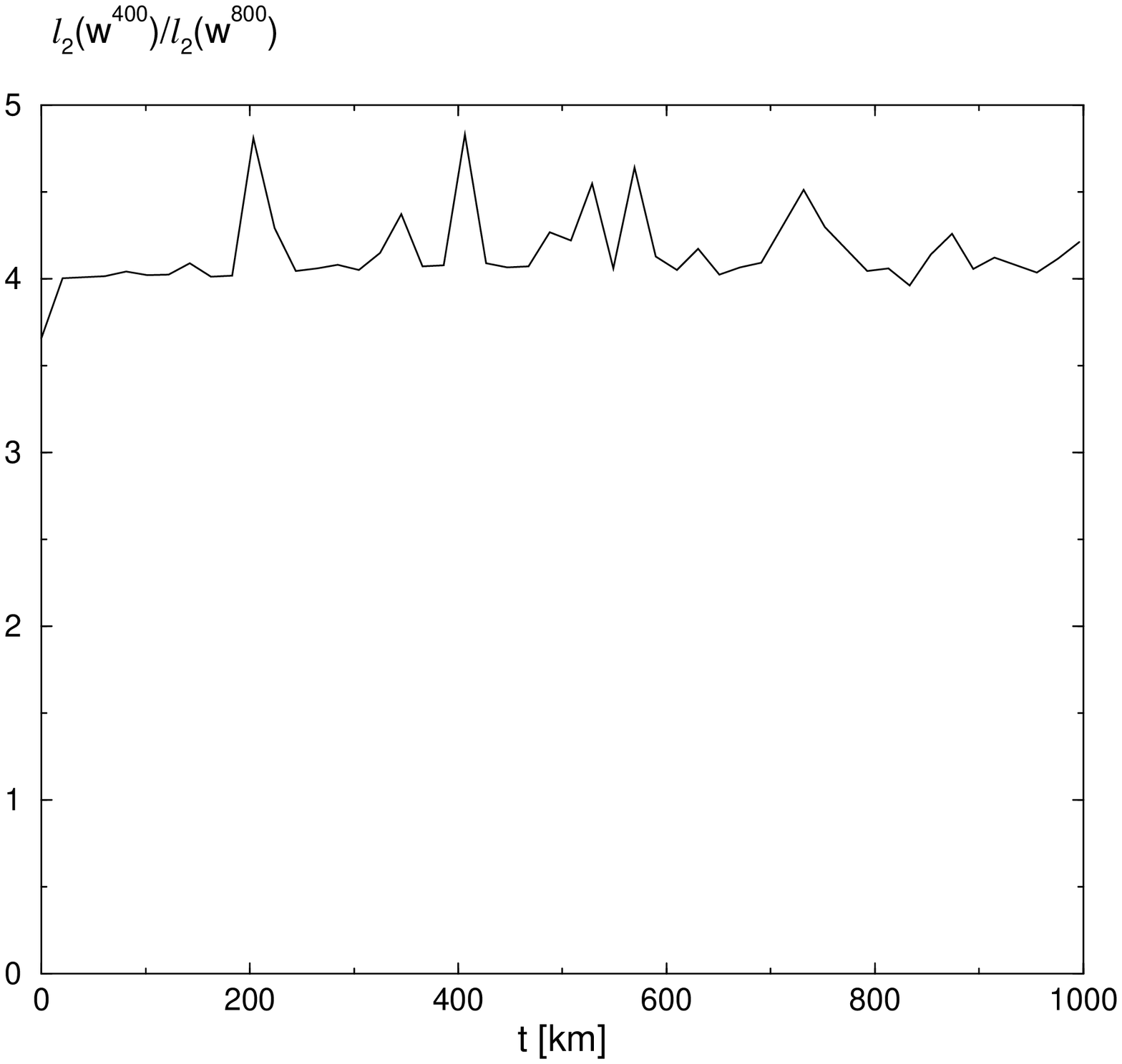, height=200pt, width=175pt, angle=-90}
  \epsfig{file=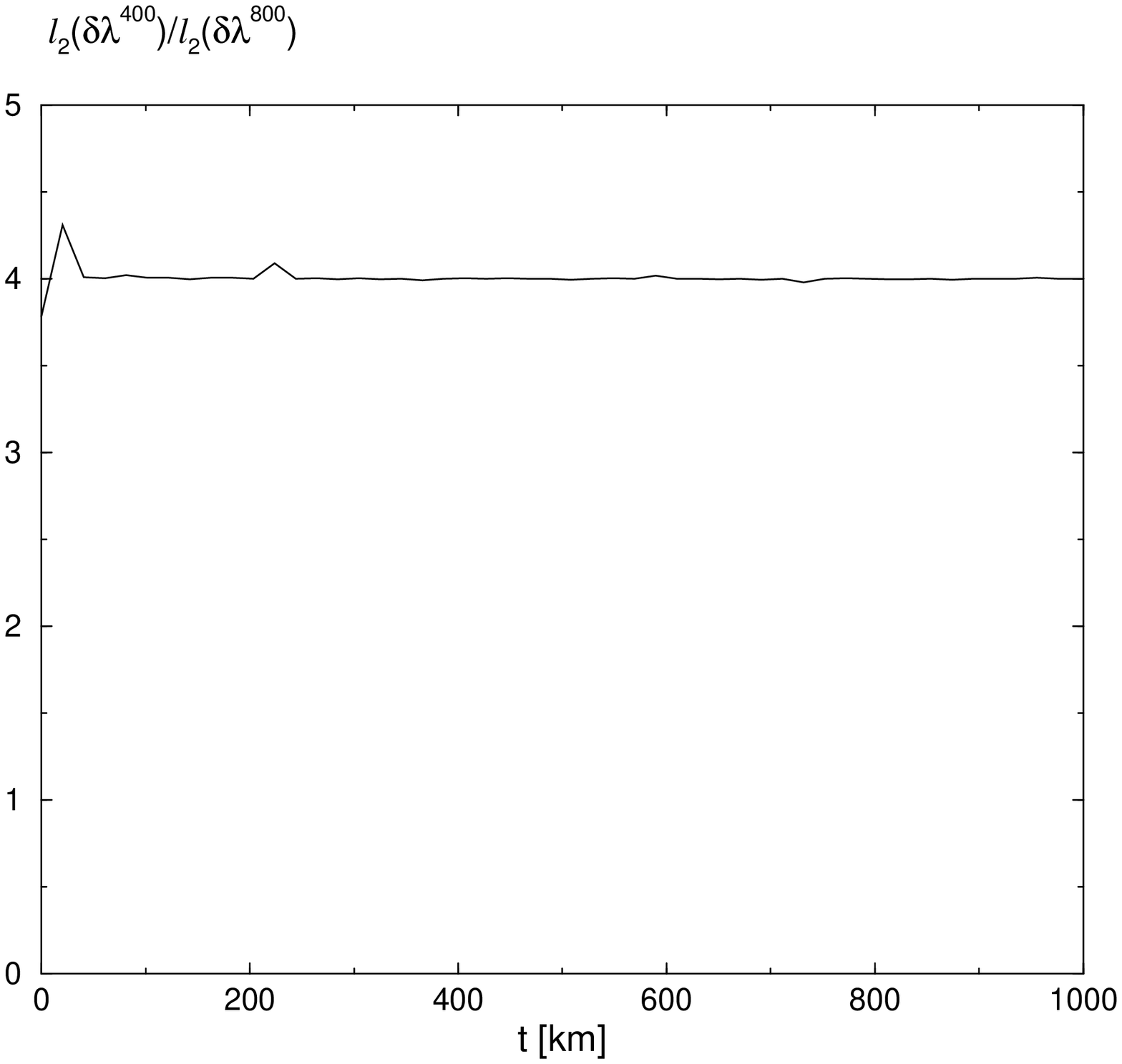, height=200pt, width=175pt, angle=-90}
  \epsfig{file=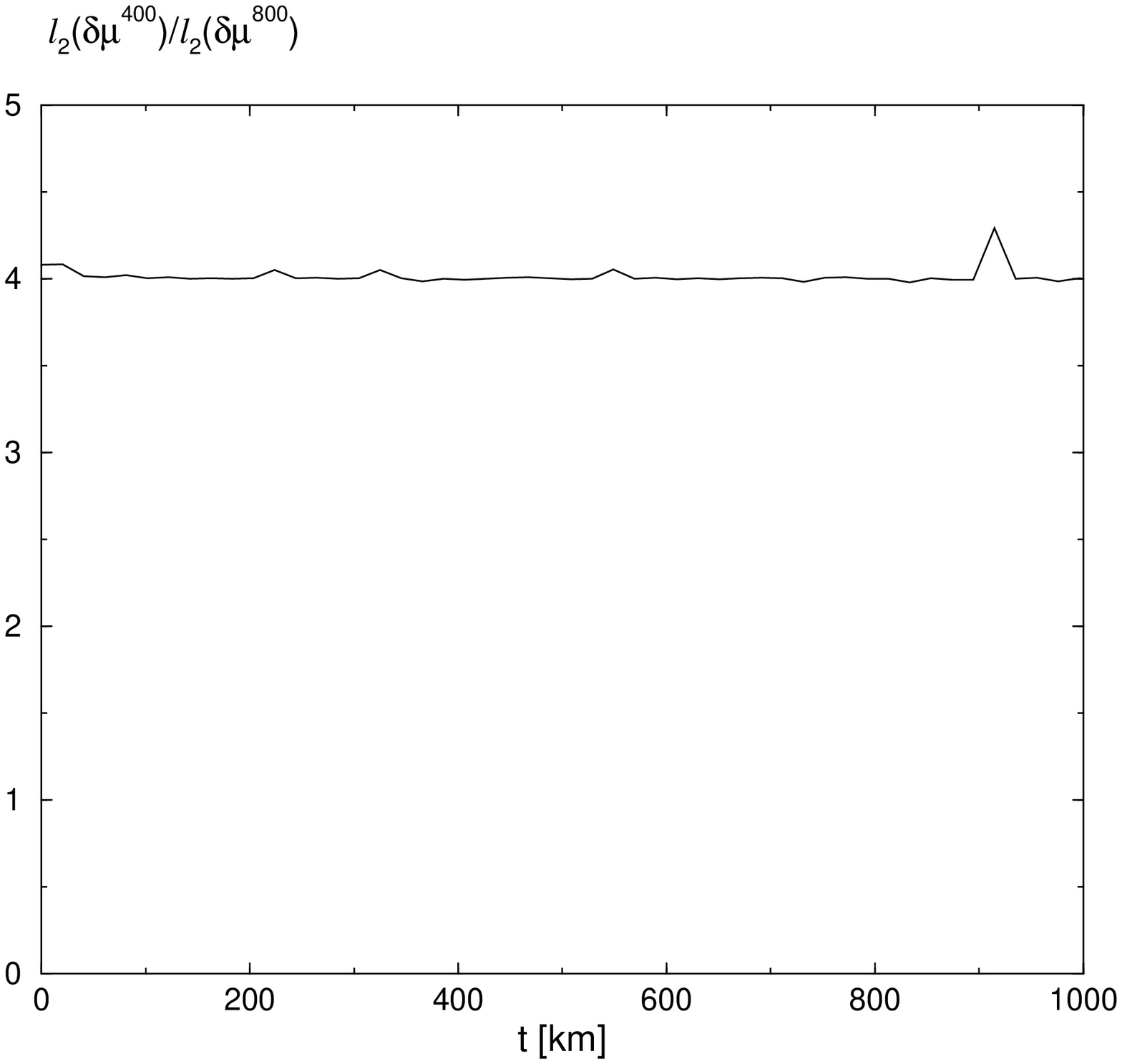, height=200pt, width=175pt, angle=-90}
  \caption{The convergence factor obtained for evolving the second
           eigenmode with an amplitude of $70\,\,{\rm m}$ is shown
           for the variables $\delta \rho$, $w$, $\delta \lambda$
           and $\delta \mu$.}
  \label{MOD_CONV}
\end{figure}
will result in a linear increase
of the deviation as observed in Fig.\,\ref{MOD_DRHOC}.
In spite of the small deviation this calculation
is in sharp contrast with that shown in Fig.\,\ref{PERT_MARGSTABLE}, where
a much smaller truncation density resulted in an exponential decay of the
central density. We conclude that using a large truncation density
in combination with the boundary condition (\ref{PERT_BCOUTWMOD})
the code performs well in the linearized regime. \\
We now return to the model parameters of the previous section and use
$\rho_{\rm c}=0.001224\,\,{\rm km}^{-2}$.
In order to test the code for convergence in the non-linear regime
we have evolved the second eigenmode with an amplitude
corresponding to a maximal displacement of fluid elements of $70\,\,{\rm m}$.
% For this amplitude strong non-linear effects are present as we will
%see below.
The calculation has been carried out with 400, 800 and
1600 grid points and the resulting convergence factors
are shown in Fig.\,\ref{MOD_CONV}. In spite of variations
around the expected value 4, the results for all variables are compatible
with second order convergence. \\
For the next test we will use the code in the Cowling approximation, since
the static metric provides a straightforward recipe to calculate
conserved quantities. We have seen in section \ref{PERT_EQ} that 
only minor modifications are required
to switch between the Cowling approximation and a dynamic metric.
The conservation properties with a fixed metric will therefore represent
a good test for the matter evolution in the general case.
The first step in the derivation of a conserved quantity is to find a time-like
Killing field. The existence of such a vector
field follows from the static nature of the metric in the Cowling
approximation. The Killing vector can be found by looking at the
Killing equation
\begin{align}
  \nabla_{\mu} \hbox{\vec X}_{\nu} + \nabla_{\nu} \hbox{\vec X}_{\mu} &= 0.
\end{align}
The resulting 10 differential equations can be solved rather easily
and define the solution up to a constant
factor. We choose this factor so that the Killing field can be written as
\begin{align}
  \hbox{\vec X}_{\mu} &= \left[ \lambda^2,0,0,0 \right].
\end{align}
The conserved quantity then follows from contraction of the Killing field
with the energy momentum tensor
\begin{align}
  \hbox{\vec J}^{\mu} &= \hbox{\vec{T}}^{\mu \nu} \hbox{\vec X}_{\nu}.
\end{align}
By virtue of conservation of energy momentum this vector satisfies the
condition
\begin{align}
  \nabla_{\mu} \hbox{\vec J}^{\mu} &= 0.
\end{align}
With the metric (\ref{TOV_LINEELEMENT}) and the energy momentum tensor
\begin{figure}[t]
  \centering
  \epsfig{file=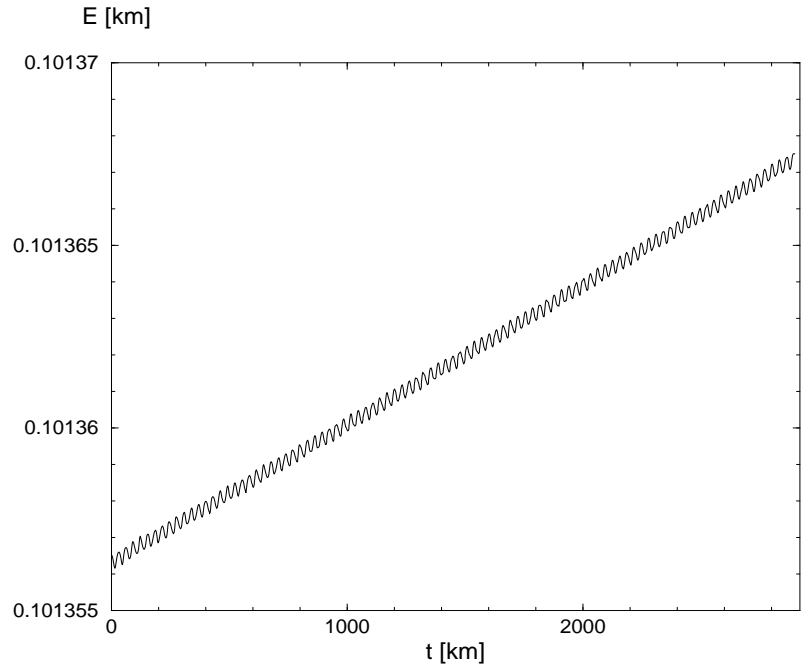, height=300pt, width=250pt, angle=-90}
  \caption{The numerical evolution of the function $E$ obtained in the
           Cowling approximation. The quantity is conserved with an accuracy
           better than $10^{-4}$.}
  \label{CONSERVED}
\end{figure}
(\ref{NONP_EMTENSOR}) this equation can be written in conservative form
\begin{align}
  \partial_t \left( \lambda \mu r^2 \hbox{\vec J}^t \right) + \partial_r \left(
      \lambda \mu r^2 \hbox{\vec J}^r \right) &= 0,
\end{align}
where the $t$ and $r$ components of $\hbox{\vec{J}}$ are given by
\begin{align}
  \hbox{\vec J}^t &= (1+\mu^2 w^2) \hat{\rho} + \mu^2 w^2 \hat{P}, \\[10pt]
  \hbox{\vec J}^r &= \frac{w}{v} (1+\mu^2 w^2) (\hat{\rho} +\hat{P}).
\end{align}
If we consider a general conservation law in one dimension
\begin{align}
  u_{,t} + F(u)_{,r} &= 0,
\end{align}
we obtain after integration over $t$ and $r$
\begin{align}
  \int_0^R{[u(T,r)-u(0,r)]}dr + \int_0^T{[F(t,R)-F(t,0)]}dr &= 0.
\end{align}
In our case the flux function is given by $F = \lambda \mu r^2 \hbox{\vec J}^r$
and vanishes at $r=0$ and $r=R$ because the velocity $w$ vanishes
at both boundaries. Consequently
\begin{align}
  E &= \int_0^R{\lambda \mu r^2 \hbox{\vec J}^t}dr
\end{align}
is a conserved quantity. \\
In order to test the conservation properties of the code
we have evolved the same initial data as in the convergence
analysis with the metric fixed at the background values.
In Fig.\,\ref{CONSERVED}
we show $E$ as a function of time as calculated with 800 grid points.
The quantity
is conserved with a relative accuracy better than
$10^{-4}$. Even higher accuracy is 
obtained for smaller amplitudes of the initial data. 
We have thus demonstrated that the code performs well in the linear
as well as the non-linear regime. The applicability of the
code to a wide range of amplitudes will be crucial when we study
non-linear effects in the evolution of eigenmodes in the next subsection.

%=========================================================================
\subsubsection{Non-linear mode coupling}
\label{MODECOUPLING}
{\em (a) Measuring the eigenmode coefficients} \\[5pt]
We will now use the simplified neutron star model described in
section\,\ref{NEWMODEL} to study the coupling of eigenmodes
in non-linear evolutions of
radial oscillations.
% It is necessary therefore to find a way of measuring
%the presence of the individual eigenmodes in the evolution. For this
%purpose we recall the {\em Sturm-Liouville problem} (\ref{LIN_ZETARR})
%which determines the eigenmode solutions in terms of the displacement
%vector $\zeta$.
In order to measure the presence of the individual eigenmodes
in the evolution we recall the {\em Sturm-Liouville problem}
(\ref{LIN_ZETARR}) which determines the eigenmode solutions
in terms of the rescaled displacement vector $\zeta$.
In section \ref{PERT_LIN} we have seen that the solutions
$\zeta_i$ form a complete orthonormal system with respect to
the inner product defined in Eq.\,(\ref{LIN_INNERPRODUCT}). This
property enables us to quantify the contributions of the different
eigenmodes in the evolution at any given time.
We need to calculate the displacement $\zeta(t,r)$ of a
fully non-linear evolution from the fundamental variables $\delta \rho$
and $w$. For this purpose we eliminate $\xi$ from Eqs.\,(\ref{LIN_XI})
and (\ref{LIN_ZETA}) and obtain
\begin{align}
  \zeta_{,t} &= r^2 w. \label{MC_ZETAT}
\end{align}
The initial values of $\zeta$ follow from the initial data
which we provide in the
form of an eigenmode in the velocity field $w$ and
zero energy density perturbation $\delta \rho$. We can see from
Eq.\,(\ref{LIN_DRHOOFZETA}) that the initial displacement $\zeta$
vanishes as a consequence.
At any time $t$ we can then expand the non-linear
displacement $\zeta(t,r)$ in terms of the eigenmodes
\begin{align}
  \zeta(t,r) &= \sum_i{A_i(t) \zeta_i(r)}, \label{ZETAEXPANSION}
\end{align}
where the time dependent coefficients are given by the inner product
\begin{align}
  A_i(t) &= \langle \zeta(t,r),\zeta_i(r) \rangle .
\end{align}
In practice we prefer to calculate the eigenmode coefficients from
the time derivative of this equation
\begin{align}
  \frac{\partial A_i}{\partial t} &= \langle \zeta_{,t}, \zeta_i \rangle,
\end{align}
where we have dropped the $t$ and $r$ dependence for convenience.
If we substitute Eq.\,(\ref{MC_ZETAT}) for $\zeta_{,t}$ we obtain the
final result
\begin{align}
  \frac{\partial A_i}{\partial t} &= \langle r^2 w, \zeta_i \rangle.
\end{align}
We can thus calculate the time derivative of the coefficients
and use the initial values to obtain the coefficients at
any given time $t$. In our case all coefficients are zero initially
because of the vanishing of $\zeta$. The integral appearing
in the definition of the inner product is calculated with the
fourth order Simpson method (see for example \shortciteNP{Press1989}). \\
It is also interesting to consider the relative coefficients defined by
\begin{align}
  R_i(t) &= \frac{\langle \zeta , \zeta_i \rangle}
                 {\langle \zeta, \zeta \rangle},
\end{align}
whenever $\zeta$ is a non-zero function. If we multiply this equation
\begin{figure}[t]
  \centering
  \epsfig{file=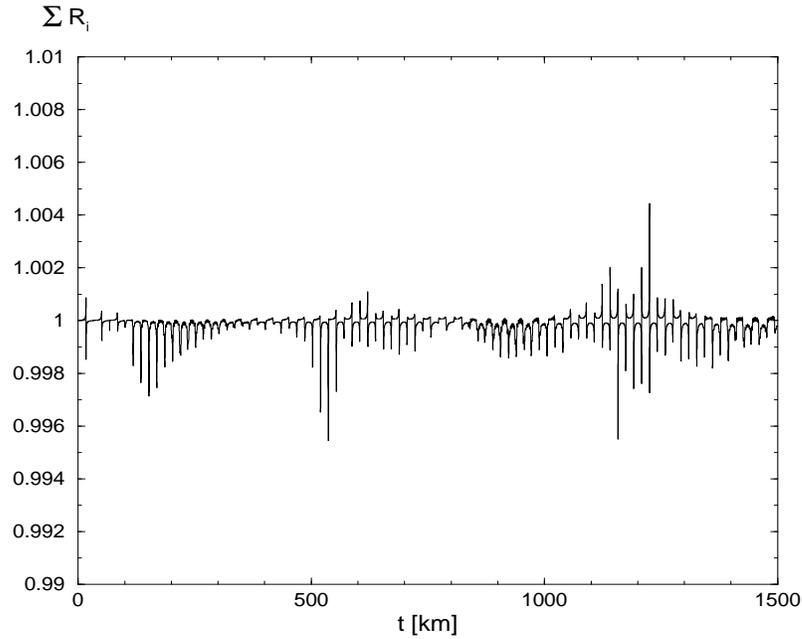, height=300pt, width=250pt, angle=-90}
  \caption{The sum of the first ten $R_i$ has been calculated
           for evolving the second eigenmode with an amplitude of
           $70\,\,{\rm m}$.}
  \label{MOD_SUMR}
\end{figure}
by $A_i$ and sum over $i$, we can use Eq.\,(\ref{ZETAEXPANSION}) to
obtain the relation
\begin{align}
  \sum_i{R_i} &= 1, \label{MC_RSUM}
\end{align}
which can be used to check the completeness of the numerically
calculated eigenmodes. For this purpose we have
evolved the second eigenmode with a large amplitude corresponding to
a maximum displacement of
$70\,\,{\rm m}$ and calculated the sum of the first ten weighted coefficients
$R_i$ using 600 grid points. The result is shown in Fig.\,\ref{MOD_SUMR}
and demonstrates
that Eq.\,(\ref{MC_RSUM}) is satisfied to within less than one per cent.
This does not only confirm the completeness of the system of eigenmodes,
but also indicates that the energy essentially remains within the lowest
ten eigenmodes.
In order to check the orthonormality we have calculated the inner products
of the eigenmodes. The results for
the lowest five eigenmodes are shown in Table \ref{MOD_ORTHONORMALITY}
and demonstrate that
\begin{table}[t]
  \caption{The inner product $\langle \zeta_i, \zeta_j \rangle$ between
           the five lowest eigenmodes.}
\begin{center}
  \begin{tabular}{c|ccccc}
    \hline \hline
    & $\zeta_1$ & $\zeta_2$ & $\zeta_3$ & $\zeta_4$ & $\zeta_5$\\
    \hline
    $\zeta_1$ & 1.0 & $-2.1\cdot 10^{-6}$ & $-6.3\cdot 10^{-6}$ 
              & $-1.2\cdot 10^{-6}$ & $-2.6\cdot 10^{-6}$ \\
    $\zeta_2$ & $-2.1\cdot 10^{-6}$ & 1.0 & $-8.0\cdot 10^{-6}$
              & $-1.5\cdot 10^{-5}$ & $-6.3\cdot 10^{-6}$ \\
    $\zeta_3$ & $-6.3\cdot 10^{-6}$ & $-8.0\cdot 10^{-6}$ & 1.0 
              & $-1.8\cdot 10^{-5}$ & $-2.7\cdot 10^{-5}$ \\
    $\zeta_4$ & $-1.2\cdot 10^{-6}$ & $-1.5\cdot 10^{-5}$ 
              & $-1.8\cdot 10^{-5}$ & 1.0 & $-3.2\cdot 10^{-5}$ \\
    $\zeta_5$ & $-2.6\cdot 10^{-6}$ & $-6.3\cdot 10^{-6}$
              & $-2.7\cdot 10^{-5}$ & $-3.2\cdot 10^{-5}$ & 1.0 \\
    \hline \hline
  \end{tabular}
  \label{MOD_ORTHONORMALITY}
\end{center}
\end{table}
the orthonormality condition (\ref{LIN_ORTHONORMALITY}) is satisfied with high
accuracy. \\

\begin{figure}[t]
  \centering
  \epsfig{file=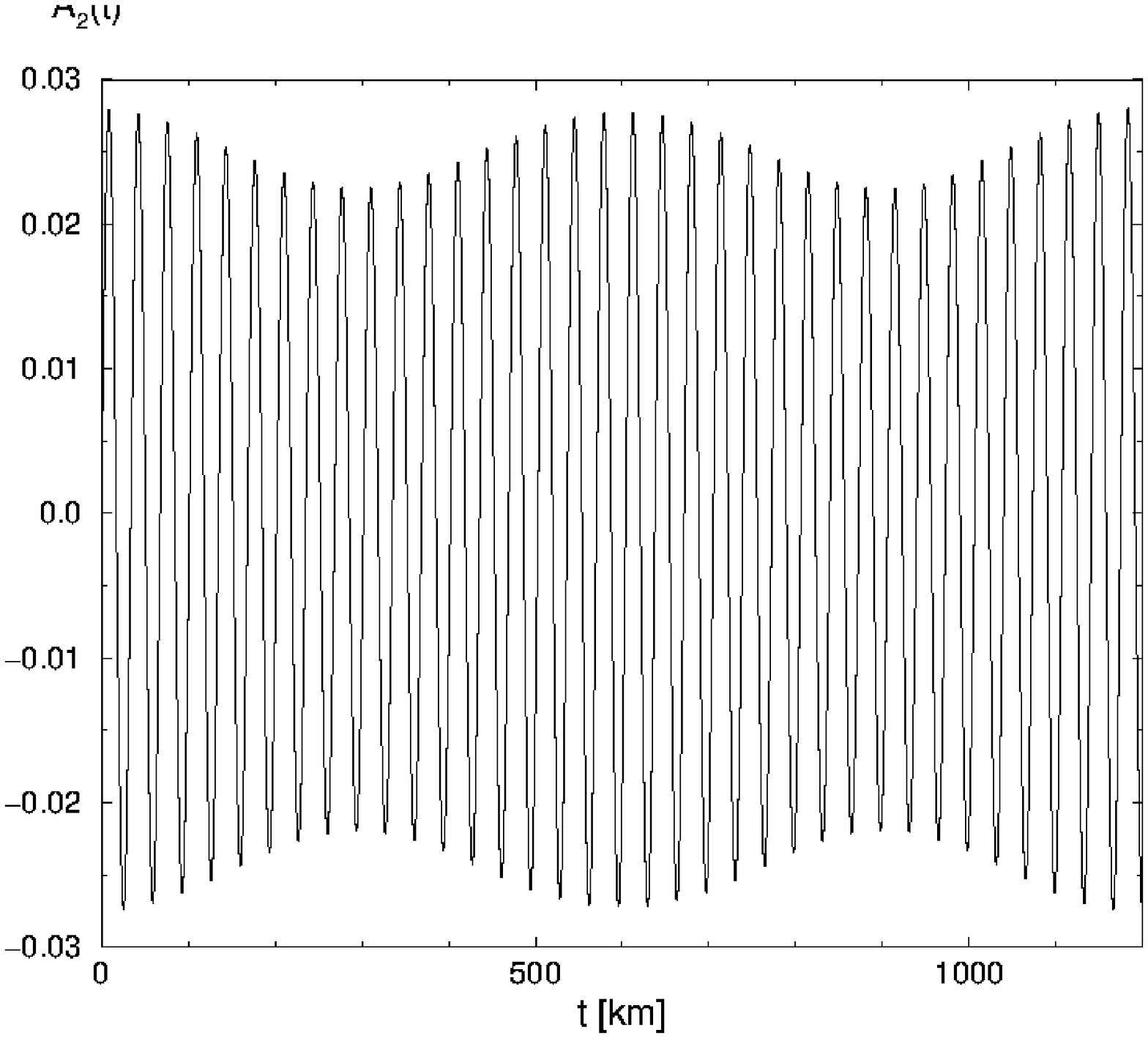, height=200pt, width=150pt, angle=-90}
  \epsfig{file=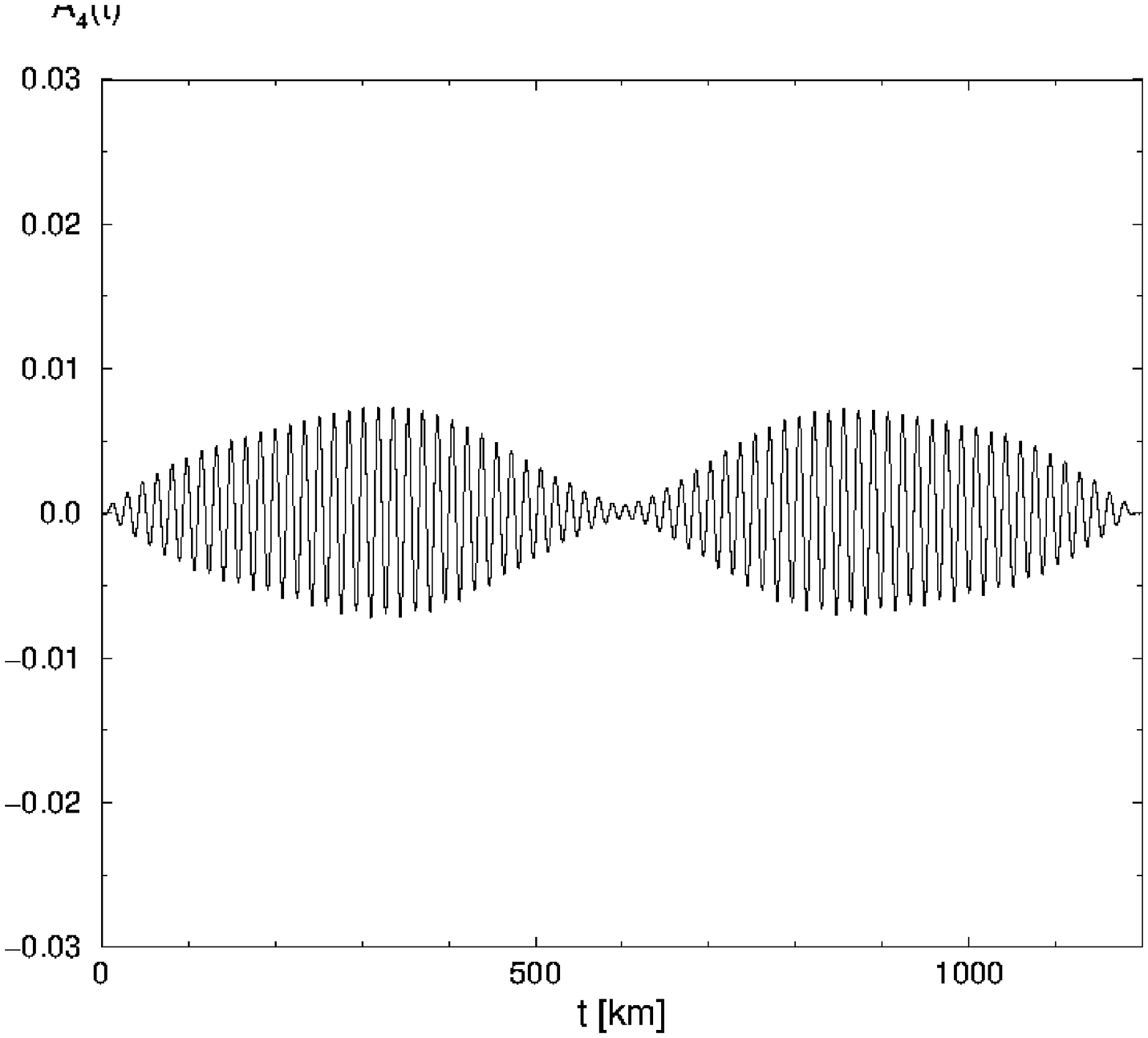, height=200pt, width=150pt, angle=-90}
  \caption{The coefficients $A_2(t)$ and $A_4(t)$ obtained for initial
           data in the form of the second eigenmode with an amplitude
           of $70\,{\rm m}$.}
  \label{MOD_EVOLA2A4}
\end{figure}

{\em (b) Non-linear coupling between eigenmodes} \\[5pt]
In order to study the coupling of modes due to non-linear effects we have
provided initial data in the form of one velocity eigenmode. The order
of the eigenmode $j$ and the amplitude of the initial data $K_j$ are
free parameters that determine the physical setup. We will specify the
amplitude of the initial perturbation by the maximum value of the
eigenmode profile of the displacement vector $\xi$ corresponding to the
initial velocity perturbation. This is a measure
for the maximum displacement a fluid element of the interior of the
star will undergo. During the evolution
we calculate the eigenmode coefficients
$A_i(t)$ with $1\le i \le 10\,\,{\rm or}\,\,15$
according to the method described above. Due to the oscillatory character
of the modes, the coefficients will also oscillate during the evolution.
This is shown in Fig.\,\ref{MOD_EVOLA2A4}
where we plot the coefficients $A_2(t)$ and
$A_4(t)$ for evolving the second eigenmode. A large amplitude
corresponding to a maximum displacement of $70\,\,{\rm m}$ has been used
for this calculation and we can clearly see the transfer of energy
between the second and the fourth mode. It is interesting to see
that the energy transferred to the fourth mode does not remain there
but instead is periodically passed back and forth between the two modes.
We observe
a qualitatively similar behaviour for the other eigenmodes, although these are
excited less efficiently.
If we want to investigate this coupling between eigenmodes more systematically,
we need to quantify the degree to which a particular mode has been excited
in an evolution. For this purpose we will use
the maximum value of the corresponding coefficient 
obtained during that evolution. We will refer to these maxima by
$A_i$ as opposed to $A_i(t)$ used for the time dependent coefficients.
We have thus evolved the eigenmodes $i=1$, 2 and 3, referred to
\begin{figure}[t]
  \centering
  \epsfig{file=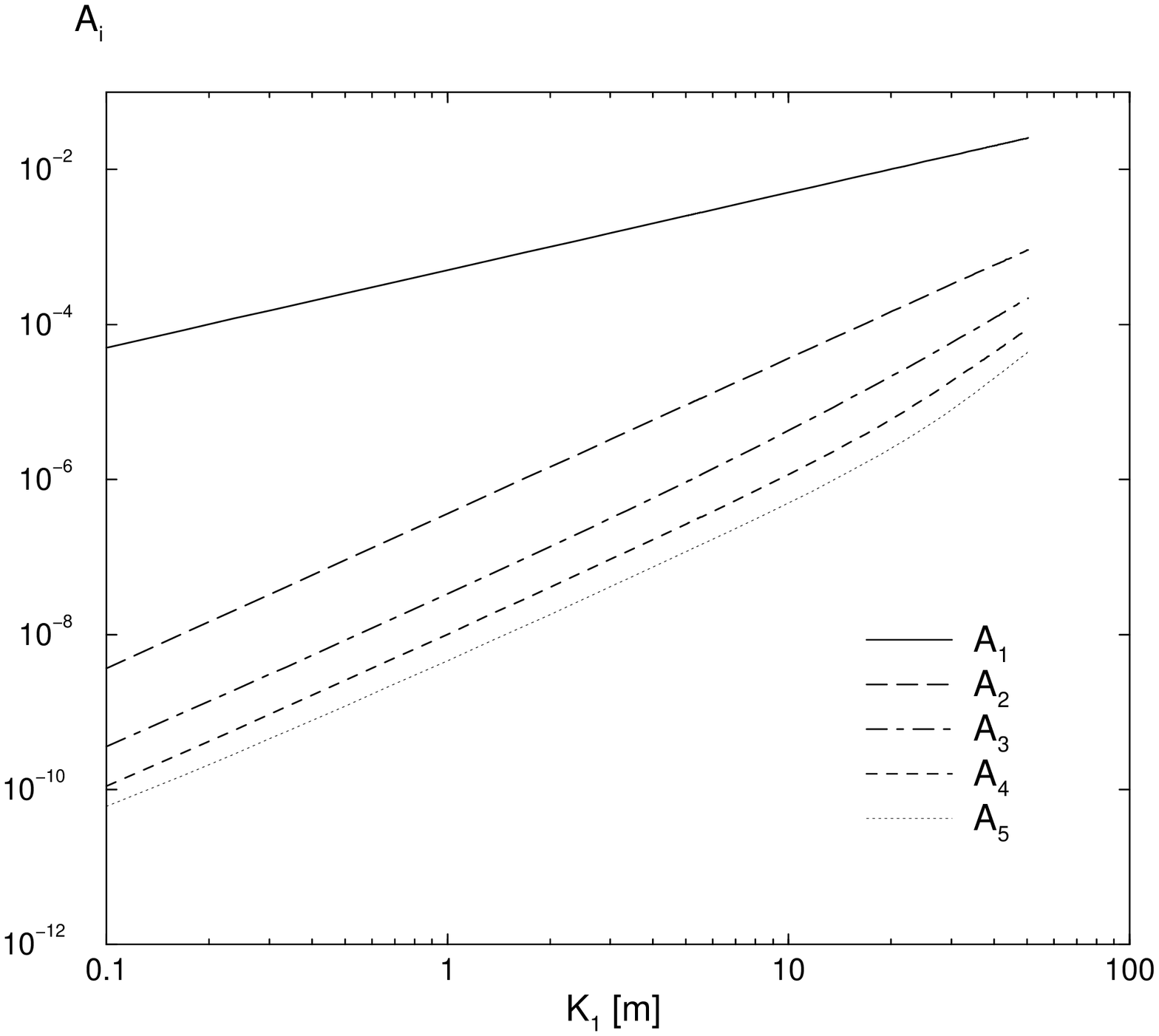, height=375pt, width=250pt, angle=-90}
  \epsfig{file=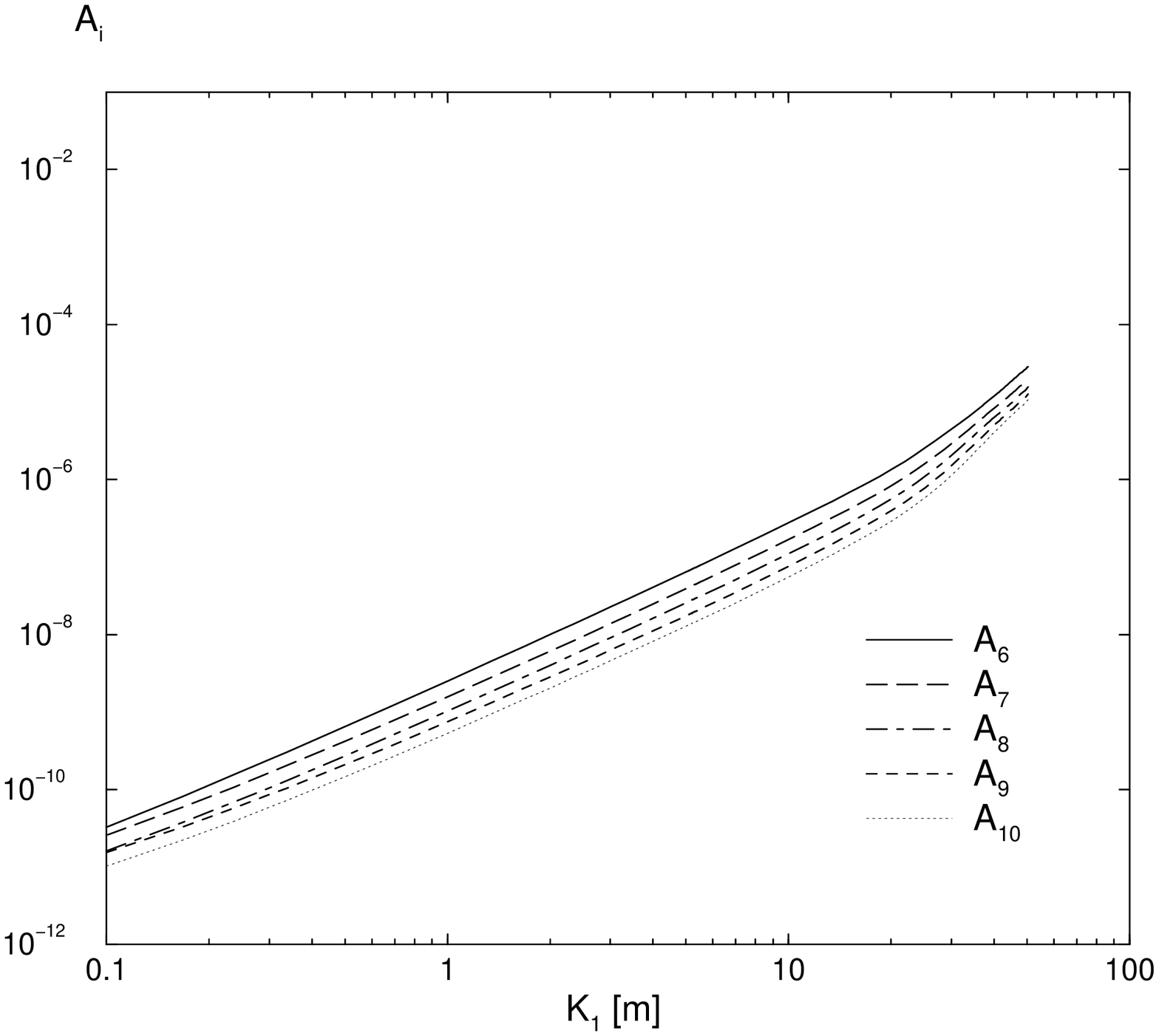, height=375pt, width=250pt, angle=-90}
  \caption{The eigenmode coefficients for the first ten eigenmodes are
           shown as a function of the amplitude $K_1$
           for initial data in the form of the fundamental velocity
           mode.}
  \label{MOD_COUPLING1A}
\end{figure}
as case 1, 2 and 3 from now on, with amplitudes
ranging between $1\,\,{\rm cm}$ and $100\,\,{\rm m}$.
At some stage in the range between about $50\,{\rm m}$
and $100\,{\rm m}$ we observed the onset of shock formation.
The accuracy of the eigenmode coefficients resulting from these
evolutions is not clear. In this discussion we have therefore only
used amplitudes for which no discontinuities are observed.
\begin{figure}[h]
  \centering
  \epsfig{file=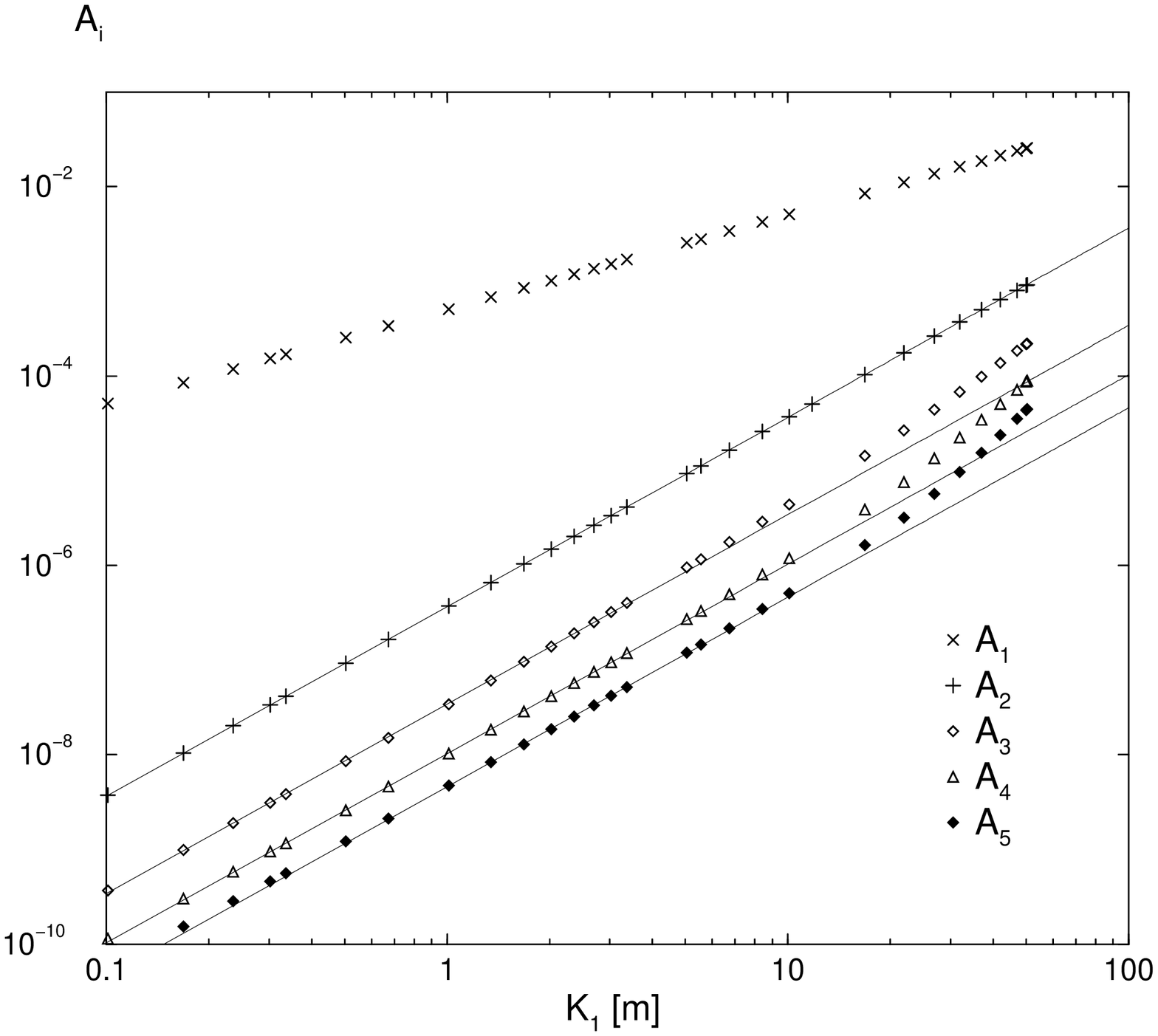, height=375pt, width=250pt, angle=-90}
  \epsfig{file=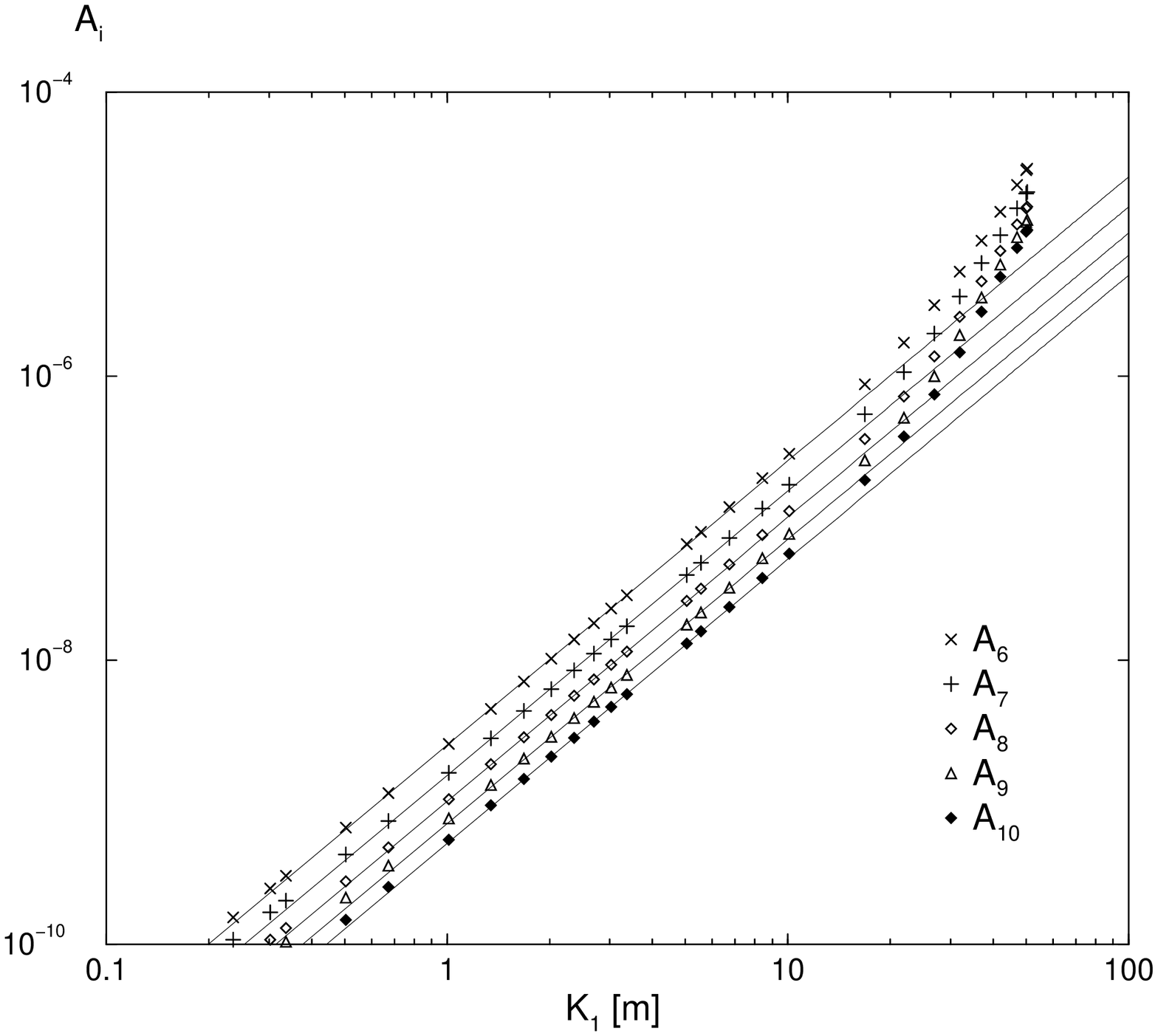, height=375pt, width=250pt, angle=-90}
  \caption{The excitation of eigenmodes has
           been fitted with quadratic power
           laws in the range between
           $K_1=1\,\,{\rm m}$ and $10\,\,{\rm m}$.}
  \label{MOD_COUPLING1B}
\end{figure}
For the numerical runs we have used 3200 grid points and an integration time
of $1500\,\,{\rm km}$. Test runs over significantly longer times
did not lead to significantly different results for the $A_i$
which is compatible with the periodic exchange of energy shown in
Fig.\,\ref{MOD_EVOLA2A4}.
The high grid resolution on the other hand enables us to measure small
eigenmode coefficients with good accuracy. \\

\begin{figure}[t]
  \centering
  \epsfig{file=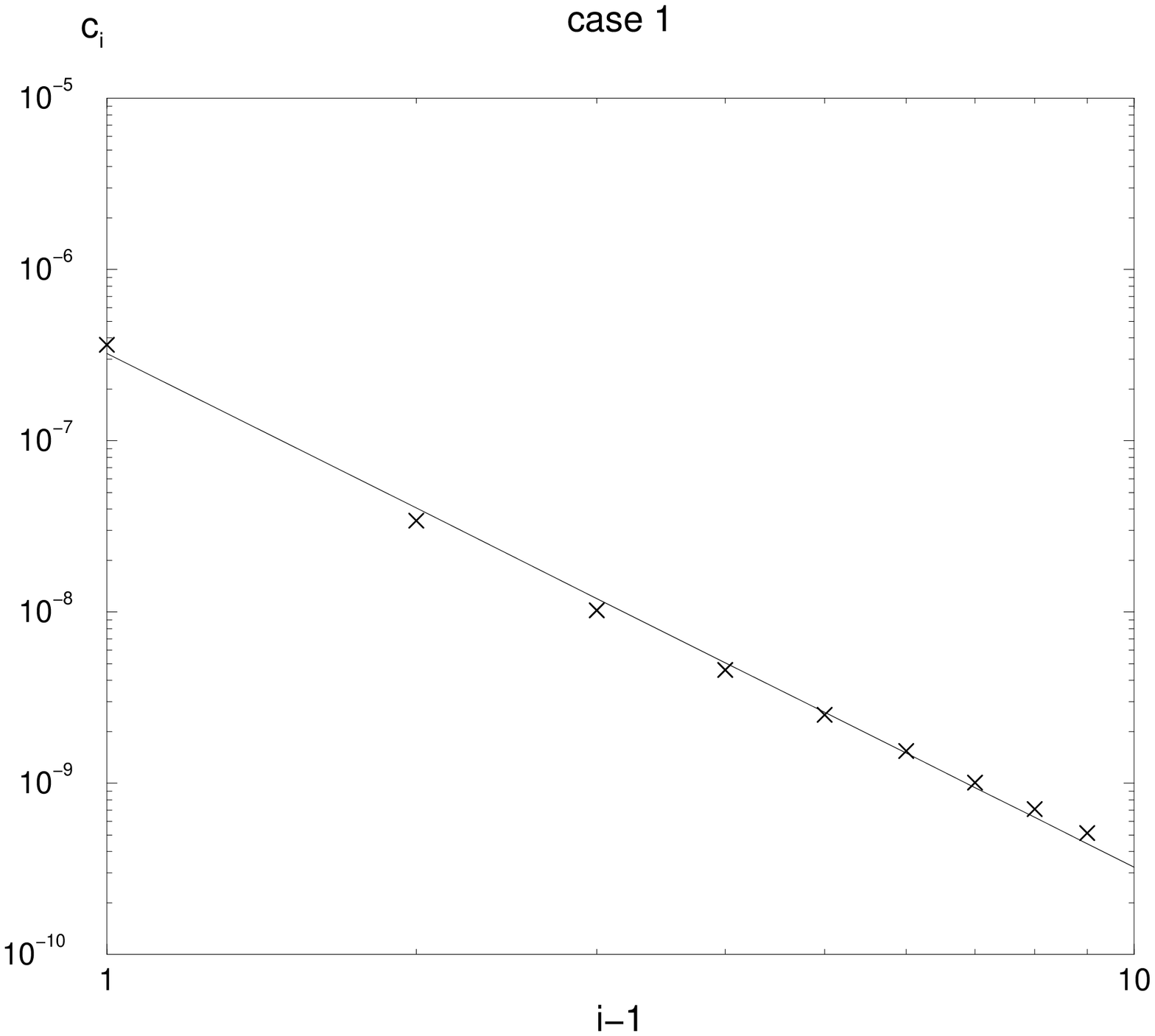, height=215pt, width=175pt, angle=-90}
  \epsfig{file=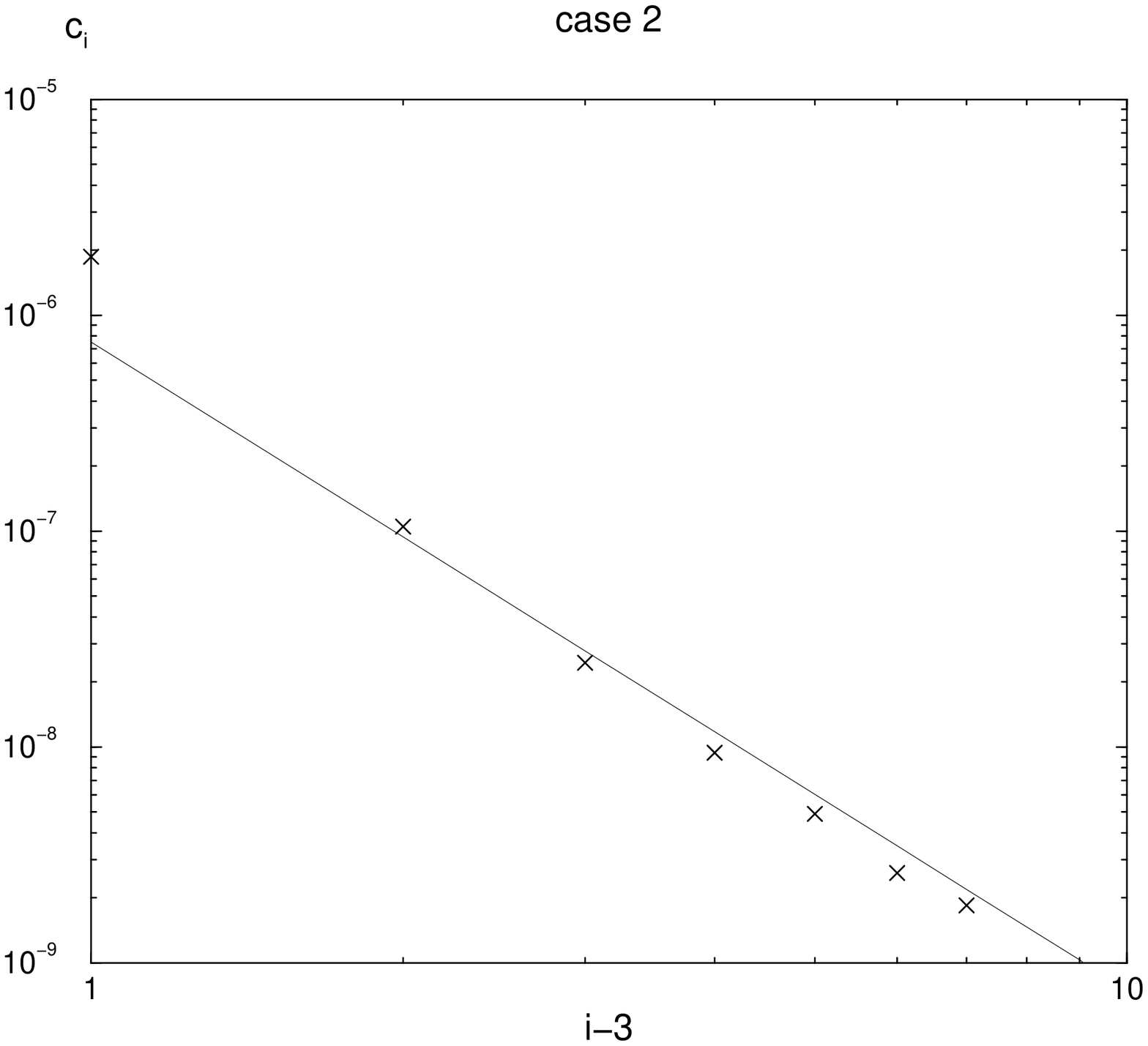, height=215pt, width=175pt, angle=-90}
  \epsfig{file=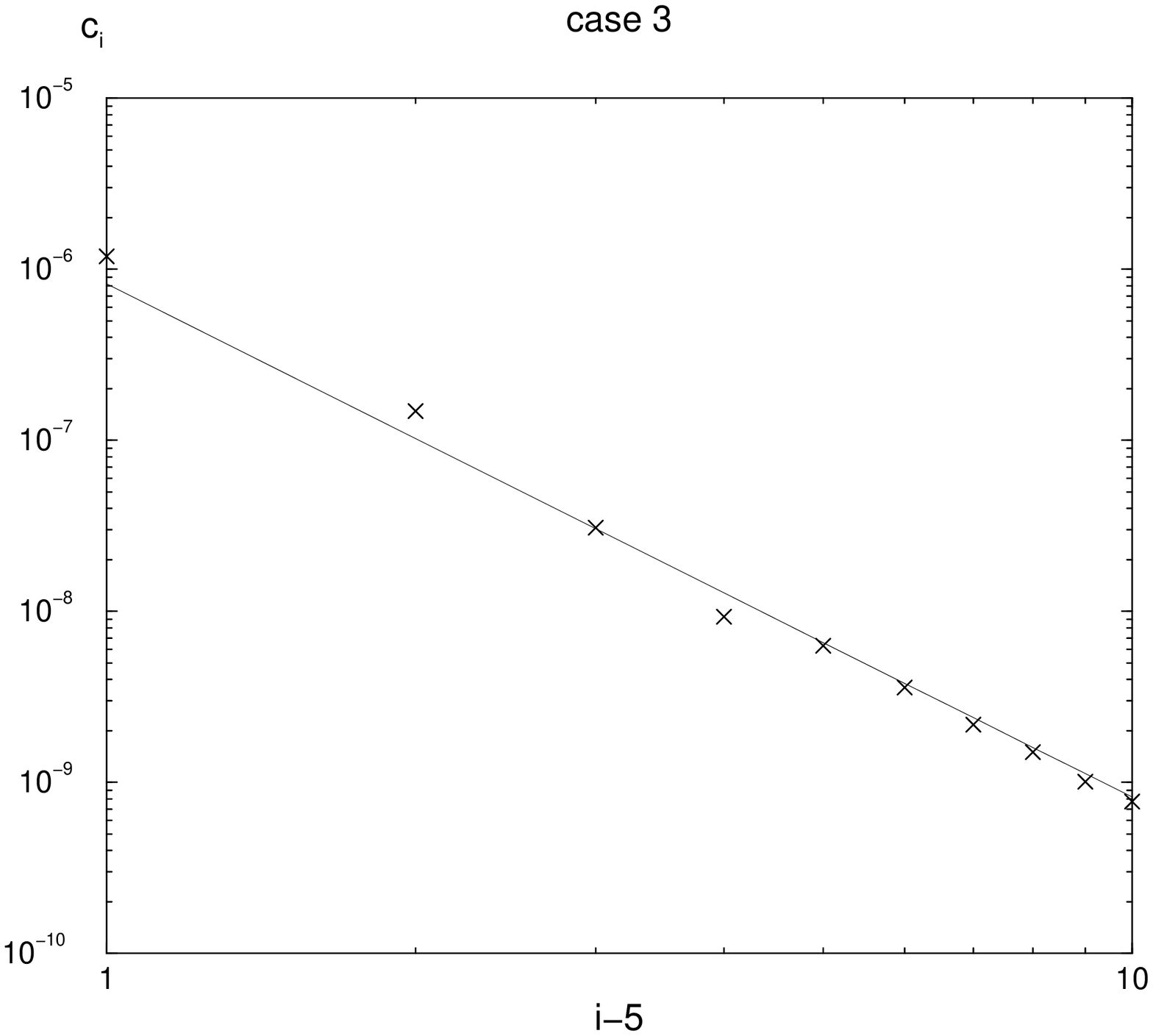, height=215pt, width=175pt, angle=-90}
  \caption{The coupling coefficients $c_i$ defined in Eq.\,\ref{MOD_SQUARE01}
           are plotted as a function of the mode number $i-1$ for case 1
           in the upper left panel. In the upper right and lower panel
           we plot the corresponding
           coefficients for case 2 and 3
           as a function of the mode number $i-3$ and $i-5$ respectively.
           In all cases the coefficients can be approximated with
           inverse cubic power laws as indicated by the solid lines.}
  \label{MOD_CI}
\end{figure}

{\sl Case 1:} \\[5pt]
We start our analysis with case 1, where the fundamental mode is excited
initially. In Fig.\,\ref{MOD_COUPLING1A} we plot the coefficients
$A_i$ as a function of the initial amplitude $K_1$ for the first ten
\begin{figure}[h]
  \centering
  \epsfig{file=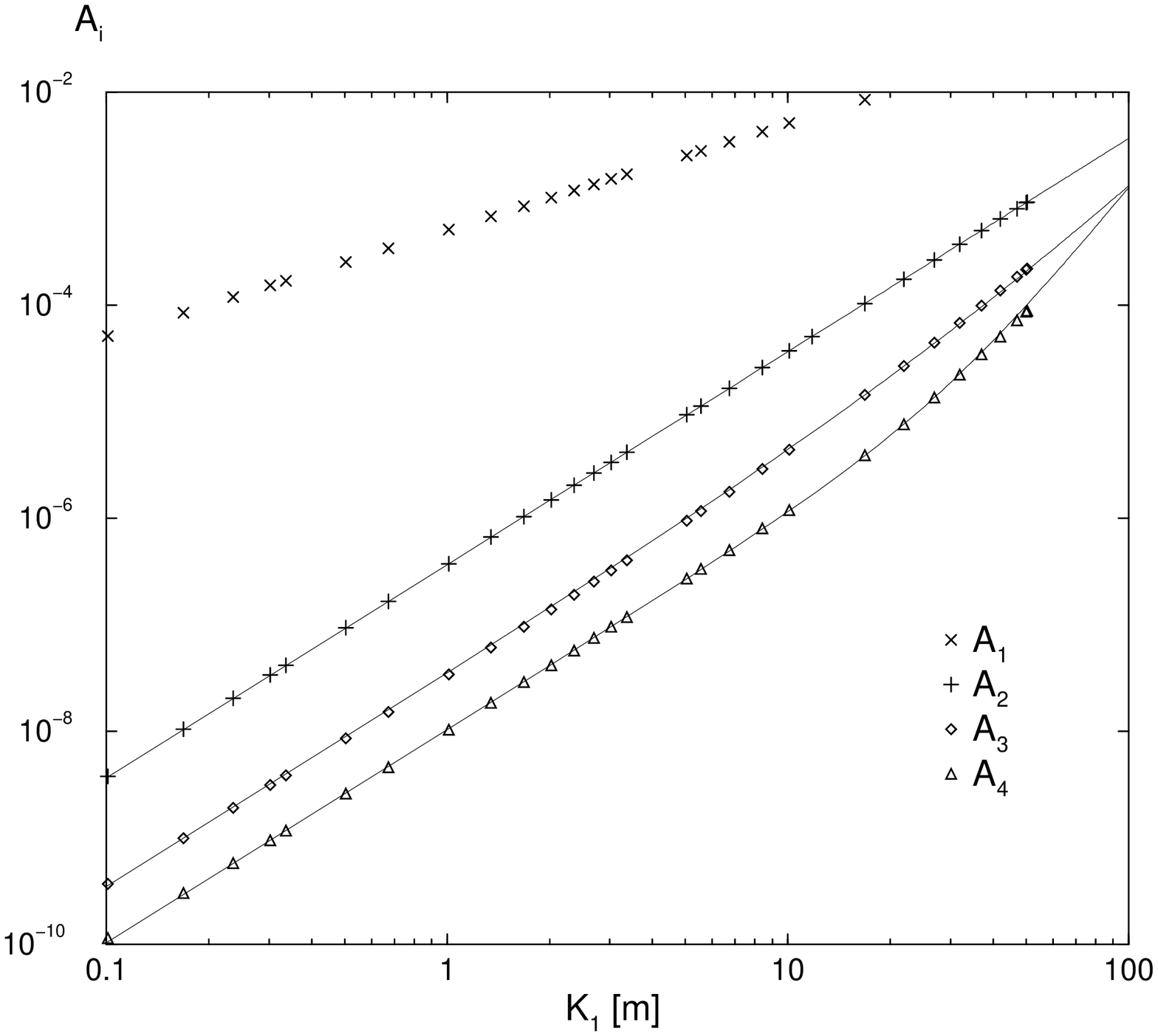, height=375pt, width=250pt, angle=-90}
  \caption{The eigenmode coefficients $A_2$, $A_3$
           and $A_4$ are fitted with linear combinations of power laws
           according to Eqs.\,(\ref{FIT_01A2})-(\ref{FIT_01A4}).}
  \label{MOD_COUPLING1C}
\end{figure}
eigenmodes.
We find that the coefficient $A_1$ increases linearly with the amplitude
$K_1$ as expected. A closer investigation of the higher eigenmode
coefficients, however, reveals the presence of two distinct regimes.
\begin{list}{\rm{(\arabic{count})}}{\usecounter{count}
             \labelwidth1cm \leftmargin1.5cm \labelsep0.4cm \rightmargin1cm
             \parsep0.5ex plus0.2ex minus0.1ex \itemsep0ex plus0.2ex}
\item In a weakly non-linear regime
      for amplitudes up to about $10\,\,{\rm m}$ all coefficients
      $A_2,\ldots A_{10}$ increase quadratically with the amplitude $K_1$.
      Deviations from this quadratic power law at very small amplitudes
      are due to the limited numerical accuracy in calculating
      the coefficients.
\item At larger amplitudes all eigenmode coefficients except for $A_2$
      show a transition to power laws with larger exponent which marks
      a moderately non-linear regime.
\end{list}
We have illustrated this behaviour in
Fig.\,\ref{MOD_COUPLING1B}
where the eigenmode coefficients have been approximated with
quadratic power laws
\begin{align}
  A_i &= c_i\cdot K_1^{\,\,2}. \label{MOD_SQUARE01}
\end{align}
The coupling coefficients $c_i$ which represent the coupling strength in the
weakly non-linear regime have been obtained from least square fits of
quadratic power laws to the eigenmode coefficients in amplitude ranges
between $0.1\,\,{\rm m}$ and $10\,\,{\rm m}$. It is interesting to investigate
the dependence of the coupling coefficients
on the order of the eigenmodes. This is shown in the upper left panel
of Fig.\,\ref{MOD_CI}, where we plot $c_i$ over the order $i-1$.
The solid line in this figure shows a power law fit for these coupling
coefficients given by
\begin{align}
  c_i &= 3.2\cdot 10^{-7}\cdot (i-1)^{-3}. \label{CI}
\end{align}
This result is compatible with the expectation that $c_i \rightarrow 0$
as $i\rightarrow \infty$. Otherwise an infinite number of modes would each be
excited with a finite amount of energy.
In the moderately non-linear regime the eigenmode coefficients
$A_3,\ldots,A_{10}$ show a higher order growth with the amplitude
$K_1$. For the most efficiently excited modes 2, 3 and 4 we have been able
to approximate the eigenmode coefficients with the following combinations
of power laws
\begin{align}
  A_2 &= 3.6\cdot10^{-7}\cdot K_1^{\,2}, \label{FIT_01A2} \\[10pt]
  A_3 &= 3.4\cdot 10^{-8}\cdot K_1^2+9.7\cdot 10^{-10}\cdot K_1^3,
     \label{FIT_01A3} \\[10pt]
  A_4 &= 1.0\cdot 10^{-8}\cdot K_1^2+1.2\cdot 10^{-11}\cdot K_1^4.
     \label{FIT_01A4}
\end{align}
Here the higher order power laws have been obtained from fitting
the eigenmode coefficients after subtracting the quadratic contributions.
The resulting fits are shown in Fig.\,\ref{MOD_COUPLING1C}.
The higher order contributions for the higher eigenmodes is rather
weak so that it is difficult to obtain accurate measurements of the
\begin{figure}[t]
  \centering
  \epsfig{file=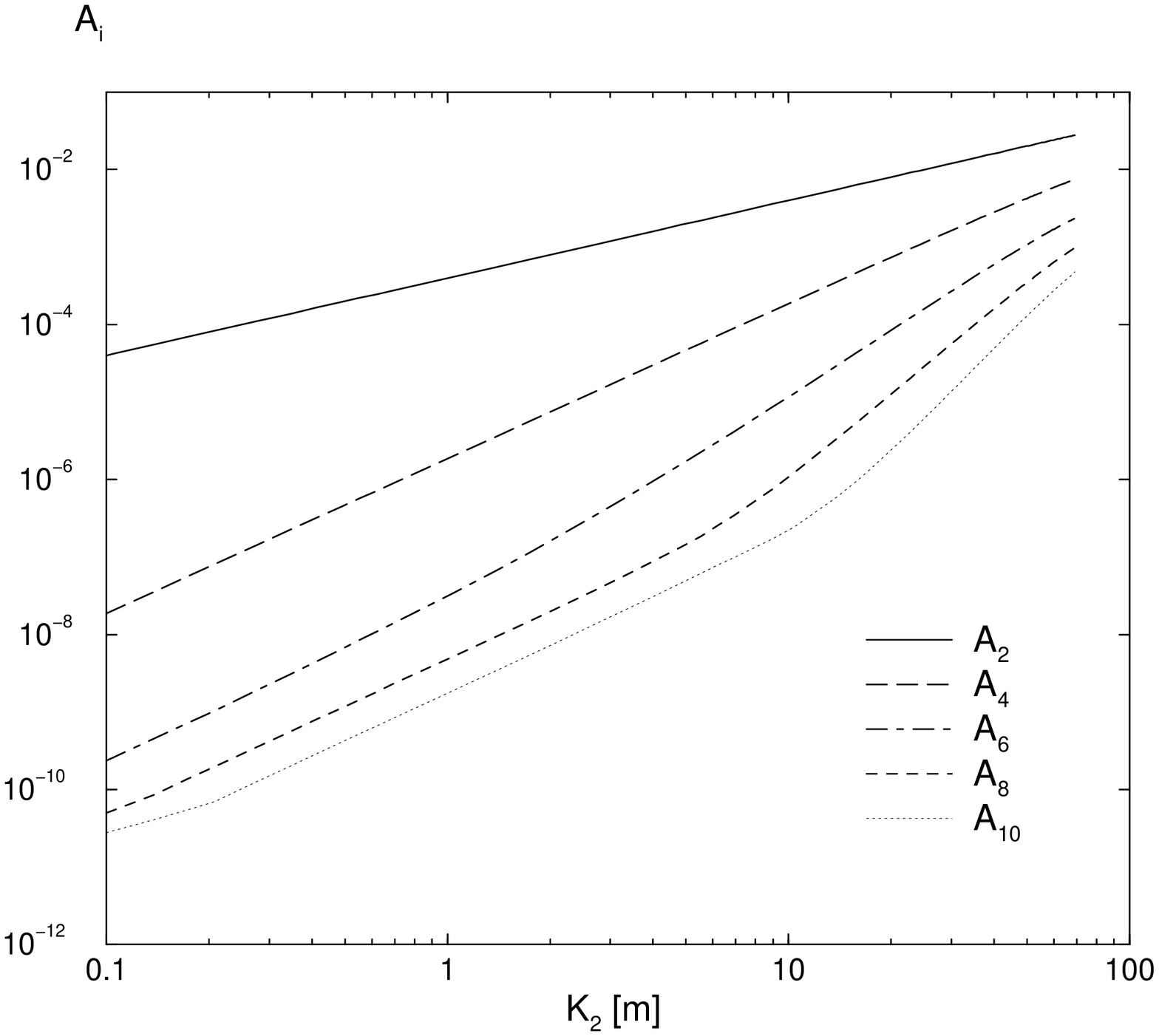, height=375pt, width=250pt, angle=-90}
  \epsfig{file=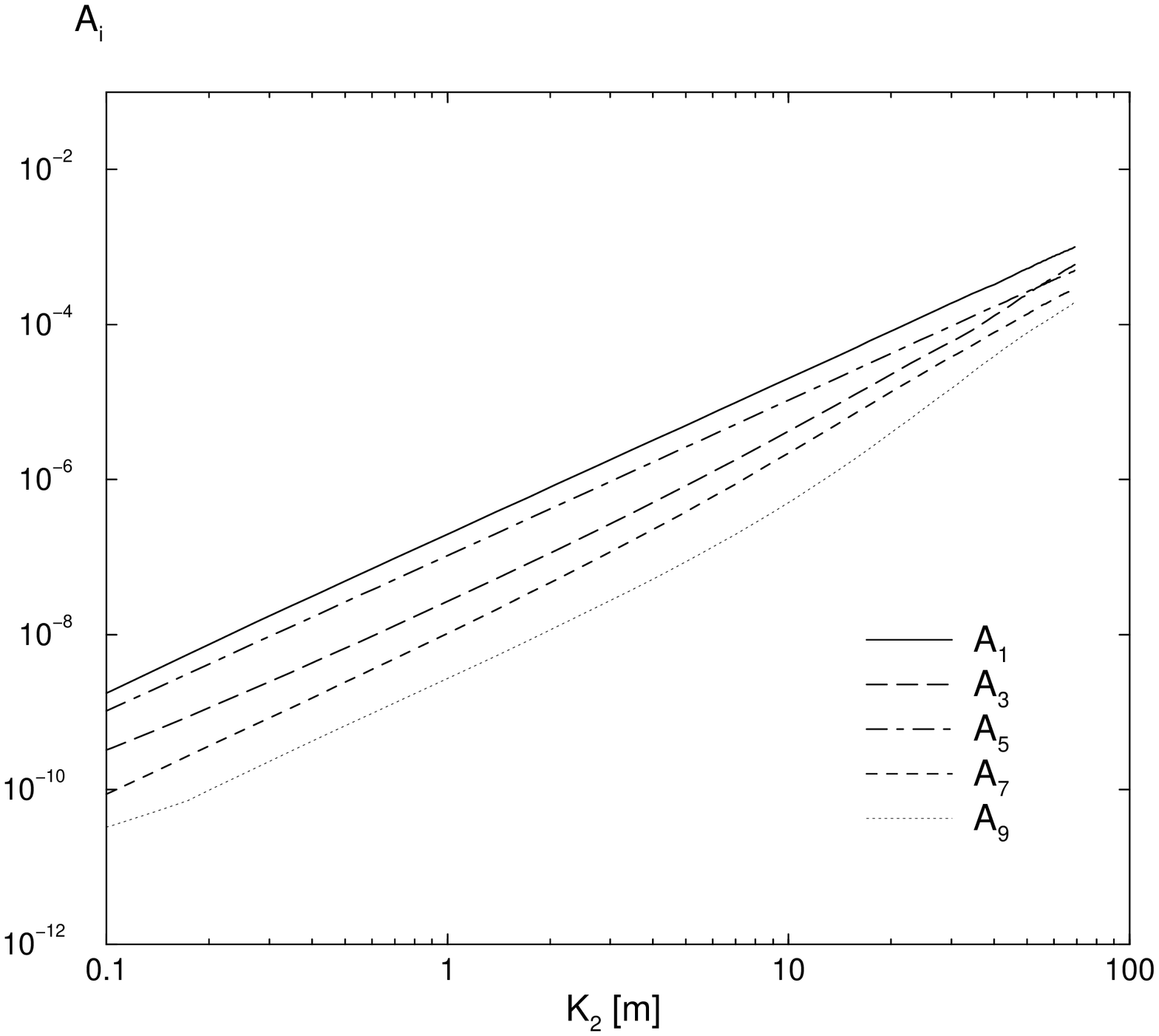, height=375pt, width=250pt, angle=-90}
  \caption{The eigenmode coefficients for the first ten eigenmodes are
           shown for initial data in the form of the second velocity mode.}
  \label{MOD_COUPLING2A}
\end{figure}
corresponding power law exponents. It is thus not clear whether the regular
pattern suggested by Eqs.\,(\ref{FIT_01A2})-(\ref{FIT_01A4}) remains valid
for higher modes. The steepening of the curves in the
\begin{figure}[t]
  \centering
  \epsfig{file=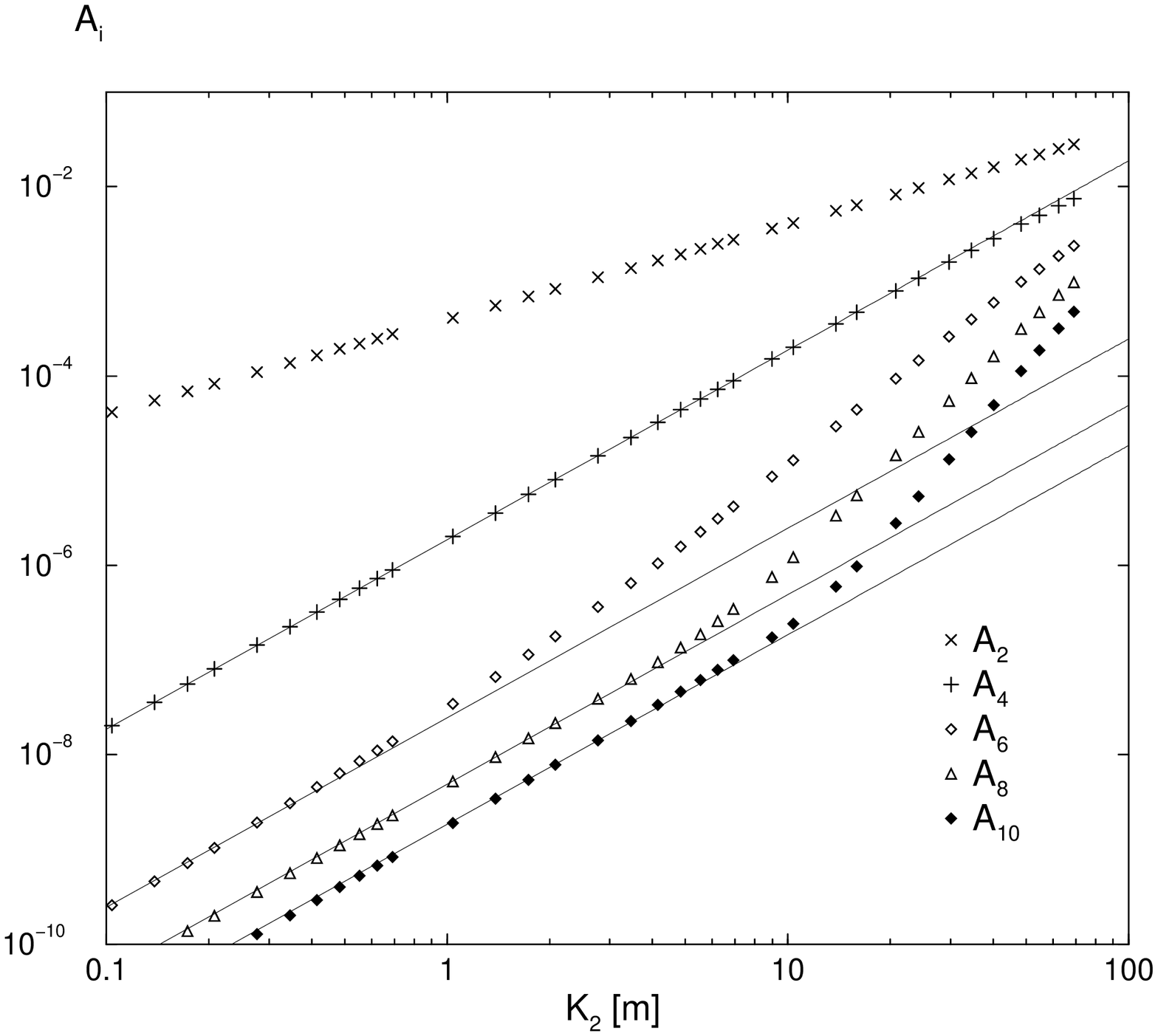, height=375pt, width=250pt, angle=-90}
  \epsfig{file=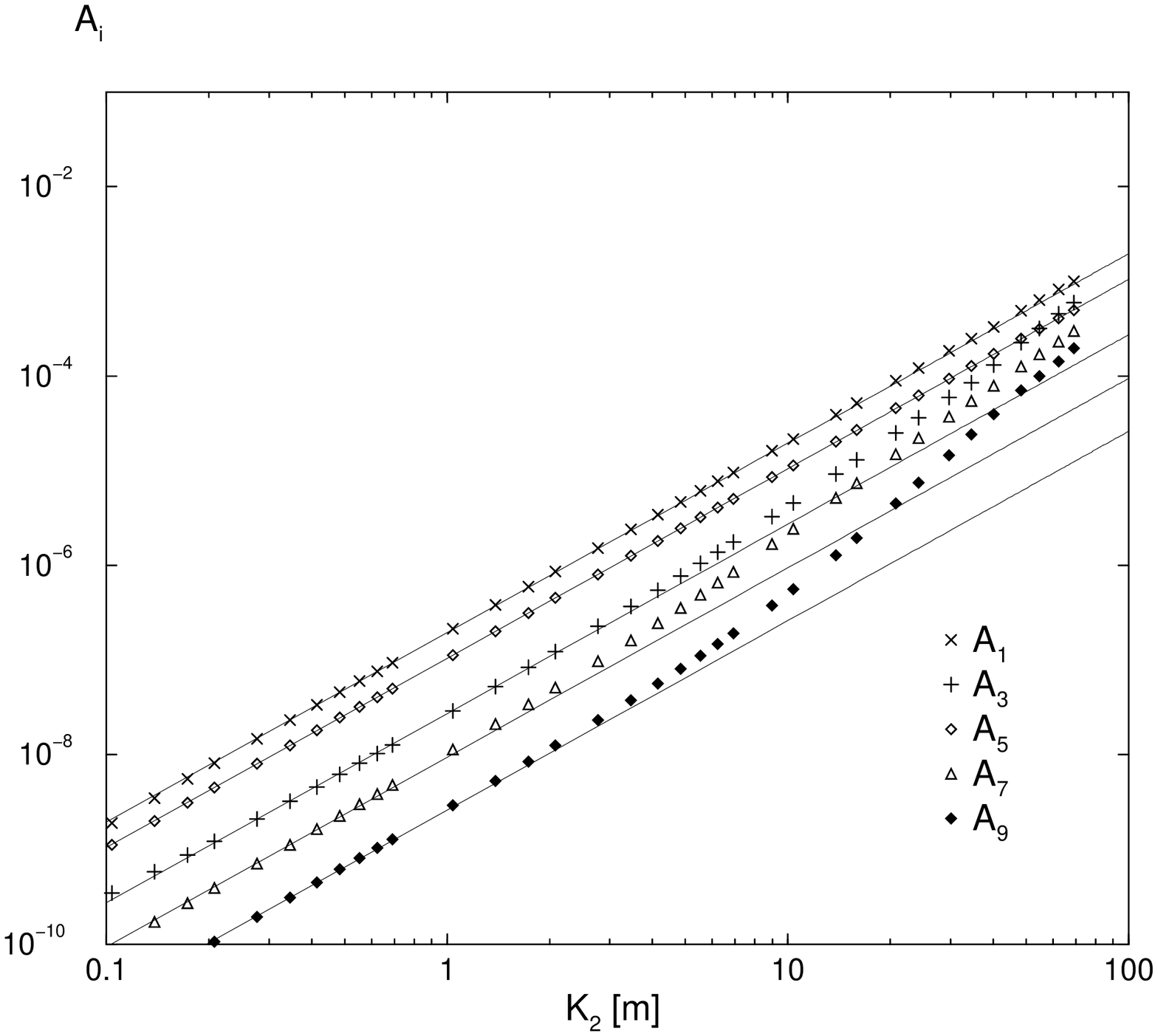, height=375pt, width=250pt, angle=-90}
  \caption{The excitation of eigenmodes in case 2 has been fitted with
           quadratic power laws in the range between $K_1=0.1\,\,{\rm m}$
           and $10\,\,{\rm m}$.}
  \label{MOD_COUPLING2B}
\end{figure}
moderately non-linear regime, however, can be clearly seen in
Fig.\,\ref{MOD_COUPLING1B}. \\

{\sl Case 2:} \\[5pt]
We will now address the question to what extent these results remain valid
if we initially excite higher modes. For this purpose we have repeated the
\begin{figure}[t]
  \centering
  \epsfig{file=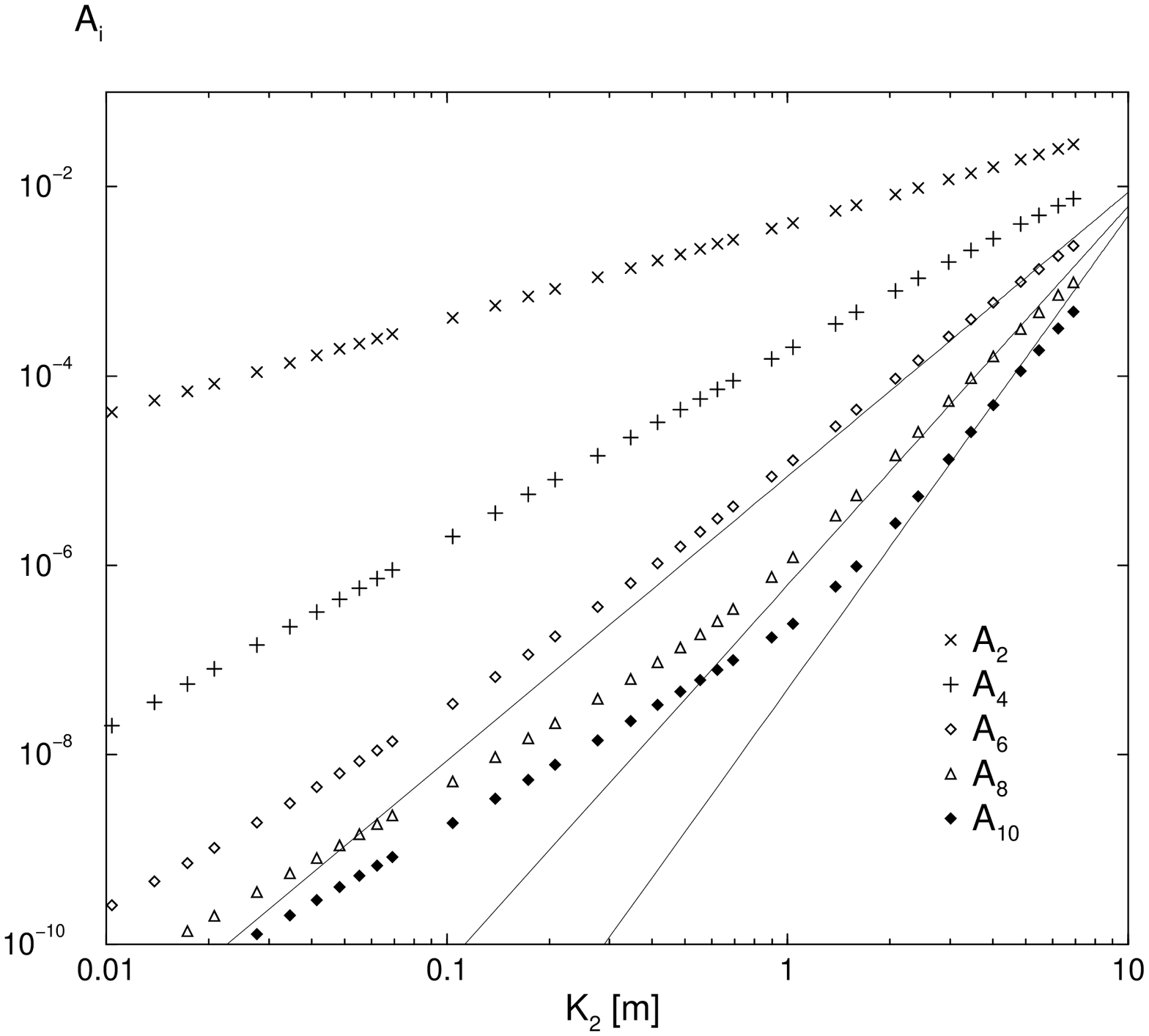, height=375pt, width=250pt, angle=-90}
  \epsfig{file=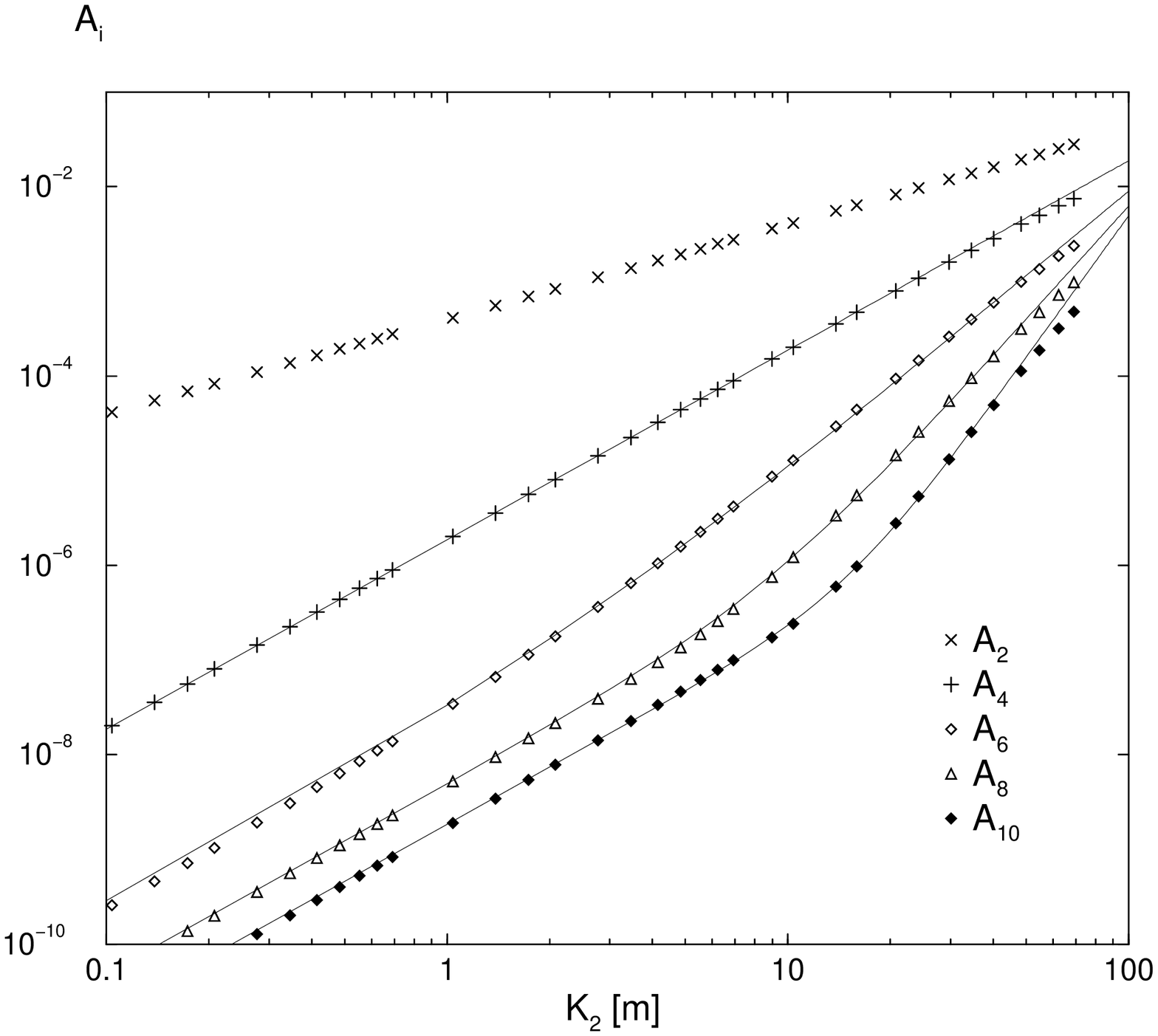, height=375pt, width=250pt, angle=-90}
  \caption{In the upper panel we show the higher order power law contributions
           of Eqs.\,(\ref{FIT02_04})-(\ref{FIT02_10}) which fit the
           even eigenmode coefficients rather well in the moderately
           non-linear regime. The lower panel shows the
           resulting fits obtained from the sum of the quadratic and the higher
           order power laws according to the same equations.}
  \label{MOD_COUPLING2C}
\end{figure}
numerical analysis by providing initial data in the form of the second
velocity mode. The resulting eigenmode coefficients are shown
as a function of the amplitude $K_2$ in Fig.\,\ref{MOD_COUPLING2A}.
The presence of the two distinct regimes is again clearly demonstrated by the
figures and a closer investigation confirms the quadratic growth of the
eigenmode coefficients in the weakly non-linear regime. This is demonstrated
in Fig.\,\ref{MOD_COUPLING2B} where the corresponding quadratic power law
fits are shown for the eigenmodes. We also observe a similar dependence
of the quadratic coupling coefficients $c_i$ on the mode number. In
case 1 we observed a power law relation given by Eq.\,(\ref{CI}) between the
coefficients $c_i$ and the mode number $i-1$. In case 2 we can also
approximate the coefficients $c_i$ reasonably well with
an inverse cubic power law if we use the number $i-3$ instead which is
demonstrated in the right panel of Fig.\,\ref{MOD_CI}. The lower order
modes 1 and 3 do not fit into this pattern and we shall
%In Fig.\,\ref{MOD_COUPLING2B} we furthermore observe the
%super-quadratic dependence of the eigenmode
%coefficients $A_i$ on the amplitude $K_2$ in the moderately non-linear
%region. \\
readdress their behaviour in the quadratic regime below when we discuss
case 3. \\
Apart from these similarities there are some interesting
differences between case 1 and case 2:
\begin{list}{\rm{(\arabic{count})}}{\usecounter{count}
             \labelwidth1cm \leftmargin1.0cm \labelsep0.4cm \rightmargin0cm
             \parsep0.5ex plus0.2ex minus0.1ex \itemsep0ex plus0.2ex}
\item The transition from the weakly to the moderately non-linear
      regime occurs at smaller amplitudes than in case 1. This
      is particularly pronounced in the case of
      mode 6 (see Fig.\,\ref{MOD_COUPLING2B}).
\item The regular pattern observed in case 1 in the moderately non-linear
      regime for the strongly excited modes 2, 3
      and 4, which is expressed in Eqs.\,(\ref{FIT_01A2})-(\ref{FIT_01A4}),
      is now being observed for the eigenmodes of even order
      $2n$. We obtain excellent fits for the data if we model the even
      eigenmode coefficients with the following linear combinations of
      power laws.
      \begin{align}
        A_4&=1.9\cdot10^{-6}\cdot K_1^2, \label{FIT02_04} \\[10pt]
        A_6&=2.5\cdot10^{-8}\cdot K_1^2+8.7\cdot10^{-9}\cdot K_1^3,
      \end{align}
      \begin{align}
        A_8&=4.9\cdot10^{-9}\cdot K_1^2+6.2\cdot10^{-11}\cdot K_1^4, \\[10pt]
        A_{10}&=1.8\cdot10^{-9}\cdot K_1^2+4.9\cdot10^{-13}\cdot K_1^5.
            \label{FIT02_10}
      \end{align}
      In Fig.\,\ref{MOD_COUPLING2C} we show the curves resulting from the
      higher order power laws as well as those corresponding to the
      linear combinations.
      For the odd modes the higher order contributions are rather
      small so that we cannot accurately measure the corresponding power
      law indices. The steepening of the curves and thus the onset
      of the moderately non-linear regime, however, is clearly visible.
\item Whereas the quadratic coupling coefficients $c_i$ shown in the right
      panel of Fig.\,\ref{MOD_CI} show a continuous decrease with the
      order of the mode starting with mode 4,
      a clear preference of the second mode to
      couple to modes
      of even order $2n$ is observed in the moderately non-linear regime.
      This is indicated by the rather efficient coupling to mode 4
      and the significantly steeper increase of the
      eigenmode coefficients $A_6$, $A_8$ and $A_{10}$ for
      larger amplitudes $K_2$ in Fig.\,\ref{MOD_COUPLING2B}.
\item A small flattening of the even eigenmode coefficients at
      large amplitudes in Fig.\,\ref{MOD_COUPLING2C} may indicate the onset
      of saturation effects. A possible mechanism for saturation is
      the formation of discontinuities. As we have already mentioned
      we have chosen an amplitude range in which no shock formation is
      observed. At the high end of our amplitude range, it may be
      possible, however, that
      similar dissipative effects due to the strong non-linearity
      start having an effect on the coupling of eigenmodes.
\end{list}
{\sl Case 3:} \\[5pt]
Next we consider case 3 where we perturb the star with the third velocity
mode. The fundamental observations we have made in the previous two cases
\begin{table}[t]
  \begin{center}
  \caption{The quadratic coupling coefficients $c_i$ for the lower modes
           in case 3.}
    \vspace{0.5cm}
  \begin{tabular}{c|c}
    \hline \hline
    $i$ & $c_i$ \\
    \hline
     1 & $2.0\cdot10^{-7}$ \\
     2 & $1.2\cdot10^{-7}$ \\
     4 & $6.7\cdot10^{-8}$ \\
     5 & $3.0\cdot10^{-8}$ \\
        \hline \hline
  \end{tabular}
  \label{MOD_CI03}
  \end{center}
\end{table}
are confirmed by the results in this case. In the weakly
non-linear regime all eigenmode coefficients (except for $A_3$) grow
quadratically with the amplitude $K_3$. The corresponding quadratic
coupling coefficients can once more be approximated with a power law
with exponent $-3$. We find, however, that the relevant mode number is
now $i-5$. This behaviour is graphically illustrated in the lower panel
of Fig.\,\ref{MOD_CI} where the coupling coefficients are shown together
with the power law approximation.
The results of this figure suggest the following
regular pattern: For initial data in the form of eigenmode $j$ the quadratic
coupling coefficients starting with mode $2j$
are well approximated by an inverse cubic power law of a
relative mode number $i+1-2j$ which is $1$ for mode $2j$, $2$ for
mode $2j+1$ and so on. \\
We still have to analyse the quadratic coupling
coefficients of the modes below $2j$. In case 1 and 2 we did not have
enough data to derive any results for these modes. For case 3 we have listed
the corresponding coefficients $c_i$ in Table \ref{MOD_CI03}.
The coefficients $c_i$ are approximately reduced by a factor of 2 each time the
mode number is increased which may indicate an exponential decrease of the
quadratic coupling coefficients for the low order modes. 
\begin{figure}[t]
  \centering
  \epsfig{file=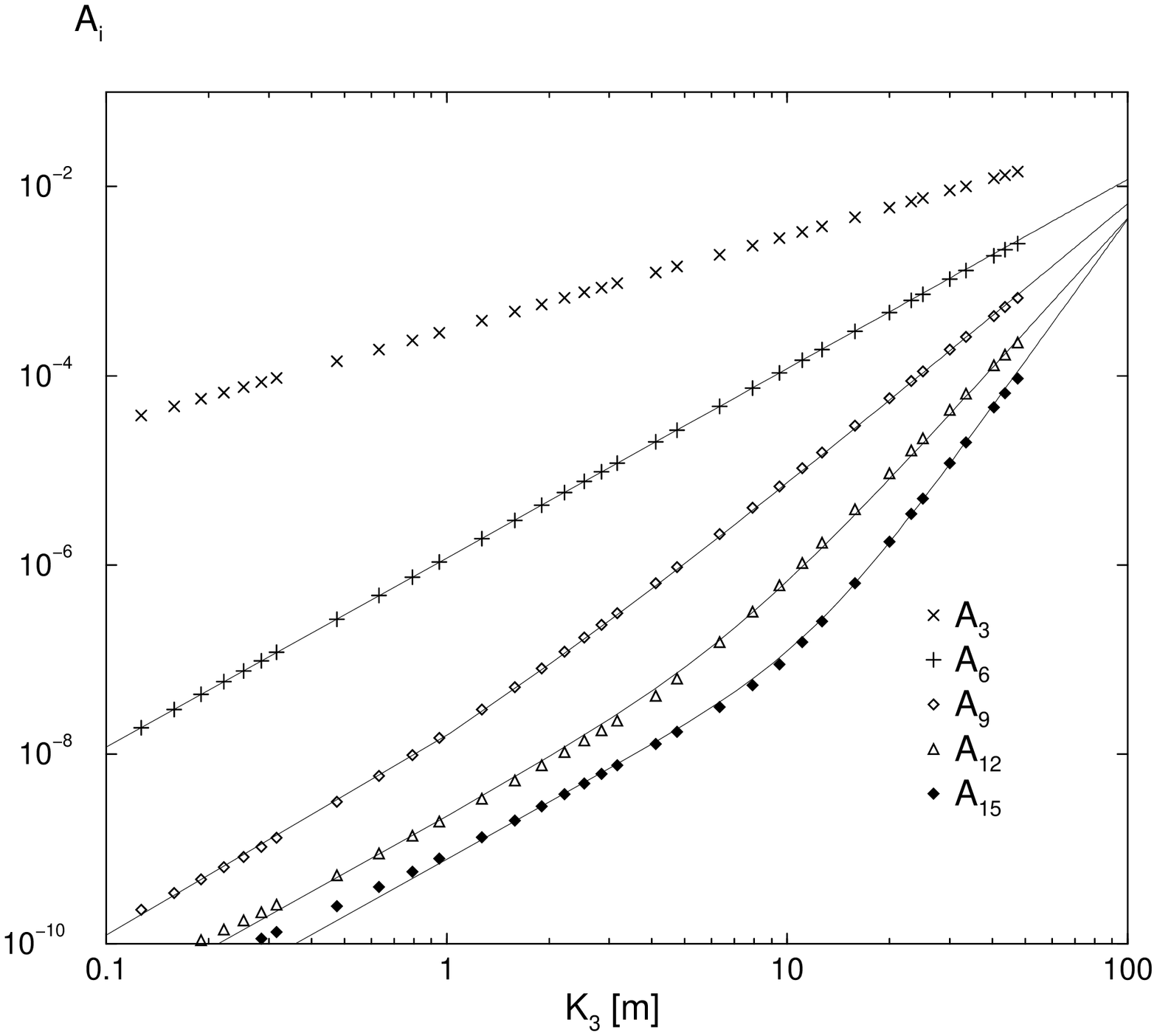, height=375pt, width=250pt, angle=-90}
  \caption{The eigenmode coefficient $A_3$, $A_6$,$\ldots$,$A_{15}$ are
           shown for case 3 together with the resulting fits
           according to Eqs.\,(\ref{FIT03_06})-(\ref{FIT03_15}).}
  \label{MOD_COUPLING3A}
\end{figure}
This is only a vague conclusion from a small data set, however, and needs to
be confirmed by studies of higher eigenmodes. \\
In the moderately non-linear regime we have seen for case 2
a preferred coupling to modes with an even order
$2n$. In analogy we find that the
third eigenmode couples more efficiently to modes of order $3n$ for larger
amplitudes. Again
we can approximate the results with good accuracy with combinations of
two power laws analogous to Eqs.\,(\ref{FIT02_04})-(\ref{FIT02_10})
\begin{align}
  A_6 &= 1.2\cdot 10^{-6}\cdot K_3^2, \label{FIT03_06} \\[10pt]
  A_9 &= 0.9\cdot 10^{-8}\cdot K_3^2+6.5\cdot 10^{-9}\cdot K_3^3, \\[10pt]
  A_{12} &= 2.2\cdot 10^{-9}\cdot K_3^2+4.6\cdot 10^{-11}\cdot K_3^4, \\[10pt]
  A_{15} &= 7.8\cdot 10^{-10}\cdot K_3^2+4.5\cdot 10^{-13}\cdot K_3^5.
       \label{FIT03_15}
\end{align}
We recognise the same pattern of increasing integer power law indices in the
higher order terms that we have already found in case 1 and 2.
These results are graphically illustrated in
Fig.\,\ref{MOD_COUPLING3A}. Again the higher order contributions in the
other eigenmodes is clearly present but too weak to facilitate an accurate
measurement of the exponents. \\
We conclude the study of non-linear mode coupling with a summary of the
key results.
\begin{list}{\rm{(\arabic{count})}}{\usecounter{count}
             \labelwidth1cm \leftmargin1.0cm \labelsep0.4cm \rightmargin0cm
             \parsep0.5ex plus0.2ex minus0.1ex \itemsep0ex plus0.2ex}
\item We clearly observe two distinct regimes in the non-linear
      coupling of eigenmodes. In the weekly non-linear regime,
      normally up to amplitudes of several metres, all eigenmode
      coefficients grow
      quadratically with the amplitude $K_j$. In the moderately
      non-linear regime we observe a steeper increase of
      the coefficients $A_i$.
      % which in many cases can be reasonably well modelled
      % with higher order power laws.
\item In the quadratic regime the coupling coefficients $c_i$ generally
      decrease with increasing order of the eigenmodes. If the initial
      perturbation is given in the form of mode $j$, we can model the
      behaviour of the quadratic coupling coefficients with an inverse
      cubic power law of the mode number
      starting with mode $2j$. The coupling to lower
      modes does not obey the same pattern, but we also observe a decrease
      of the $c_i$ with increasing mode number for these modes.
      This decrease may have exponential character.
\item In the moderately non-linear regime an initially present mode $j$
      shows a preference to couple to modes of order $n\cdot j$ where
      $n\ge 2$ is an integer number. In these
      cases we can accurately model the dependence of the eigenmode
      coefficients on
      the amplitude $K_j$ with the sum of a quadratic and a higher order power
      law with exponent $n$: $A_i=c_i\cdot K_j^{\,2} + d_i \cdot K_j^{\,n}$
      for $i=n\cdot j$.
\item In some cases we observe a flattening of the eigenmode coefficients
      at amplitudes of about $50\,{\rm m}$ which may indicate the onset
      of saturation.
\end{list}
%

%=========================================================================
\subsubsection{Discussion of the non-linear mode-coupling}
In the previous section we have studied the coupling of eigenmodes due
to non-linear effects by evolving a single eigenmode with varying
amplitude. Concerning the transfer of energy to other modes
we have found two distinct regimes, a weakly non-linear regime where the
excitation of modes grows quadratically with the initial amplitude and
a moderately non-linear regime, where this increase can be reasonably
well described by power laws of higher order. \\
% Given an initial mode
%$j$ we have also seen that the coupling to modes $n\cdot j$ is particularly
%efficient in the moderately non-linear regime. \\
In the analytic study of non-linear mode coupling one normally views the
eigenmode coefficients as harmonic oscillators and the non-linear interaction
between eigenmodes is represented in the form of driving terms which are
quadratic or of higher order in the amplitudes (see for example
\citeNP{VanHoolst1996})
\begin{align}
  \frac{d^2 A_i}{dt^2} + \omega_i^2 A_i &= c_i^{jk} A_jA_k + d_i^{jkl}
      A_j A_k A_l + \ldots,
      \label{FORCEDOSCILLATOR}
\end{align}
where the $c_i^{jk}$, $d_i^{jkl},\ldots$ are the quadratic, cubic
and higher order coupling coefficients and summation over $j,k,l$
is assumed. In our analysis the initial data
consists in one isolated eigenmode $j$, so that the right hand side
can be approximated by $c_i A_j^2 + d_i A_j^3 + \ldots$
In analytic studies this series expansion is normally truncated at second
or third order. In view of our results
the omission of higher order terms
seems to be justified in the weakly non-linear regime, where
our fully non-linear simulations confirm that quadratic terms in the
initial amplitude dominate the coupling between eigenmodes.
This is no longer true,
however, in the moderately non-linear regime, where higher order terms
are more important. In particular the regular pattern suggested
for example by Eqs.\,(\ref{FIT02_04})-(\ref{FIT02_10}) indicates that
the excitation of higher order modes is dominated by increasingly
higher order powers of the initial amplitude. It is not clear how 
this behaviour can be modelled in the framework of a finite
series expansion of the type (\ref{FORCEDOSCILLATOR}). It rather seems
that the use of fully non-linear methods such as the numerical
technique described in this work is necessary in order to obtain
a comprehensive description of the coupling between eigenmodes
in the moderately non-linear regime. In terms of the maximum displacement
of fluid elements in the star this corresponds to initial amplitudes
as low as a couple of metres. \\
We have also observed that given an initial mode
$j$ the coupling to modes $n\cdot j$ is particularly
efficient in the moderately non-linear regime. We interprete this as a
resonance effect, which we illustrate in the simple case of a forced oscillator
\begin{align}
  \frac{d^2 A_i}{dt^2} + \omega_i^2 A_i = F\sin{\Omega t},
\end{align}
where $\Omega$ is the frequency and $F$ the amplitude of the external force.
The particular
integral of this ordinary differential equation is
\begin{align}
  A_i(t) &= \frac{F}{\omega_i^2 - \Omega^2} \sin{\Omega t},
\end{align}
which implies resonance if $\omega_i=\Omega$.
If we assume that resonance occurs for any integer multiple
of the frequency $\Omega$ in the general non-linear case,
we can schematically write the eigenmode coefficients in the form
\begin{align}
  A_i(t) &= \sum_n \frac{F_n}{\omega_i^2 - (n\Omega)^2}, \label{RESONANCE}
\end{align}
where the $F_n$ may depend on the frequencies.
The analytic study of non-linear mode coupling up to
cubic order leads to eigenmode coefficients which resemble this pattern
[see for example Eqs.\,(18), (19) of \citeNP{VanHoolst1996}].
In our case the
external force is provided by the non-linear coupling to the initial mode $j$,
so that $\Omega=\omega_j$. We therefore obtain
resonance in Eq.\,(\ref{RESONANCE}) if $\omega_i = n \omega_j$. As can
be seen for example in Fig.\,\ref{PERT_LINFOUR}, the eigenfrequencies
of radial neutron star oscillations
are fairly equally spaced in the frequency domain with the exception of the
fundamental mode and we can reasonably well approximate
$\omega_i\approx (i \omega_j)/j$ for $i,j\ge 2$. The condition for resonance
then becomes
\begin{align}
  i &= n\cdot j,
\end{align}
which is exactly the relation we have observed in section \ref{MODECOUPLING}. \\
From the relativistic point of view
the non-linear coupling of eigenmodes in the weak and moderately
non-linear regime is of particular interest in the discussion of
unstable modes of rotating neutron stars. The underlying principle of
these unstable oscillation modes is the increase in amplitude of the
oscillation due to the emission of gravitational waves. The
increased amplitude in turn gives rise to stronger gravitational radiation
and so on. The
conservation of energy is ensured in this case by the spin-down of the
neutron star and the resulting decrease of rotational energy which sets
a natural upper limit on this run-away effect.
The physical mechanism which facilitates this remarkable instability
is known as the
CFS-instability (\citeNP{Chandrasekhar1970}, \citeNP{Friedman1978}).
In order for a neutron star oscillation mode to be subject to the
CFS-instability two conditions must be satisfied: (1) the mode must
be retrograde with respect to the star but prograde with respect to a
distant inertial observer and (2) the energy loss in the rotating frame
due to dissipative effects must be smaller than the amount of energy
gained from the gravitationally driven instability.
The particular importance of
the so-called $r$-modes in this respect arises from the fact that the
dominating $l=m=2$ $r$-mode satisfies the first
CFS-condition for arbitrarily small
values of the angular frequency of the neutron star
(\citeNP{Andersson1998}). One of the most important questions
raised in connection
with the $r$-modes concerns the efficiency with which energy is dissipated for
example due to viscosity or non-linear effects. \\
Considering the gradual increase in the oscillation amplitude, it is
important to understand how the instability of the mode is affected
in the weakly non-linear regime. To our knowledge the
numerical studies presented
in this work provide the first fully non-linear time evolutions
of neutron star oscillations with high accuracy for amplitudes going
all the way down to the weakly non-linear regime. Our results may
therefore pave some of the way towards understanding non-linear effects
in a wider class of neutron star oscillations. 
In particular we have managed to quantify the transfer of energy
from low into higher eigenmodes. The picture that emerges from these
evolutions is that only a rather small fraction of energy is shifted away
from the low eigenmodes. In particular the results shown in
Fig.\,\ref{MOD_EVOLA2A4} indicate that the energy shifted towards higher
eigenmodes does not accumulate in time but is rather transferred back and
forth between the initially present and the higher mode.
Correspondingly we do not observe an efficient cascade of
energy into higher modes.
It is not clear,
however, to what extent this picture will change if the energy residing
in the higher order modes is gradually dissipated.
In the context of $r$-modes it is expected that
the energy in higher order modes is dissipated on a much
shorter timescale than that of the dominating $l=m=2$ mode.
The numerical techniques and the code developed in this work may
facilitate a corresponding study in the framework of radial oscillations
by introducing an artificial damping of higher order modes
and an external force which drives the fundamental mode.
One may then look for steady state situations arising from
this model, where the amount of energy transfered to higher modes and thus
dissipated equals that gained from the external driving mechanism.\\
From a numerical point of view we emphasise the new perturbative
approach which enabled us to obtain highly accurate fully non-linear evolutions
over a large range of amplitudes. This technique can be applied for
any physical problem where there exists a non-trivial static limit.
The dynamic evolution can always be considered a finite perturbation
of the static case and a corresponding perturbative formulation will
provide a numerical accuracy that is determined by the amplitude of the
perturbation rather than the static background. We expect this method to
be particularly effective in higher dimensional evolutions where
the grid resolution is rather limited by computational costs and
the ensuing residual error arising from background terms in a non-perturbative
formulation will be more significant.

\newpage
%=========================================================================
\subsection{Radial oscillations in a Lagrangian formulation}
\label{LAGR}

In the previous section we have seen that an Eulerian description of
radial oscillations encounters difficulties at the stellar surface for several
reasons. For certain equations of state the eigenmode profiles
predicted by the linearized theory result in a diverging energy density
perturbation. A purely numerical problem arises from the movement of the 
stellar surface with respect to the numerical grid. Highly sophisticated 
techniques may be required to adequately describe the surface of a neutron star
in Eulerian coordinates
and it is not clear to what extent these will lead to a fully satisfactory
performance in the linear regime where the exact solution is
known to high accuracy and facilitates a quantitative test for the code.
It is interesting to see that these problems
vanish immediately once the problem is described in a formalism where
the coordinates follow the movement of the fluid elements. Even though
it is not obvious how to generalise a Lagrangian approach to
scenarios in two or three spatial dimensions, it still seems to
be the natural choice for the 1-dimensional case. Lagrangian codes
have often been based on the formulation of \lcite{May1966}
and \citeyear{May1967}
who following \lcite{Misner1964} use a vanishing shift vector and define
the radial coordinate in terms of the interior rest mass. 
In order to facilitate a simple comparison with the Eulerian code
discussed in section \ref{DYNAMIC}, however, it will be convenient
for us to use as similar a gauge choice to the Eulerian case as possible.
For this purpose we will follow \scite{Schinder1988} and use a Lagrangian
gauge in combination with the polar slicing condition which is also
implemented in the Eulerian code (cf. section \ref{NONP_EQ}).
As a particularly useful consequence the singularity avoiding
properties of this condition in combination with the Lagrangian gauge make
this code highly suitable for studying spherically symmetric
gravitational collapse. We will
not exhaustively study this type of scenarios in this work, but will
use the analytic solution by \lcite{Oppenheimer1939b} which
describes the collapse of a homogeneous dust sphere
for testing the code. \\

%=========================================================================
\subsubsection{The equations in the Lagrangian formulation}
The derivation of the Lagrangian equations for a dynamic spherically symmetric
neutron star was largely inspired by the work of \scite{Schinder1988}.
We will, however, slightly deviate from their approach
and work with a different set of variables and equations. \\
We start by considering the line element of a spherically
symmetric space time in polar slicing and Lagrangian gauge.
As a result of the polar slicing condition, we are able to choose the same
time coordinate $t$ as in the Eulerian case. The radial coordinate $x$ will
label the fluid elements and generally differ from the
areal radius $r$ which is intrinsically not comoving with the matter.
Finally we choose standard angular coordinates
$\theta$ and $\phi$ as above. Below we will see that the polar slicing 
condition implies a non-vanishing shift vector so that the line
element
% in polar slicing and Lagrangian gauge
becomes
\begin{align}
  ds^2 &= \left(-\hat{\lambda}^2+\frac{\hat{\Gamma}\beta^2}{\hat{r}_{,x}^2}
          \right) dt^2
          +2\beta dt\, dx + \frac{\hat{r}_{,x}^2}{\hat{\Gamma}} dx^2 
          + \hat{r}^2 (d\theta^2 +\sin^2{\theta} d\phi^2).
          \label{LAGR_LINEELEMENT}
\end{align}
It turns out to be convenient for our discussion if we introduce the variables
\begin{align}
  w &= \frac{\hat{r}_{,t}}{\hat{\lambda}},    \label{LAGR_U} \\[10pt]
  \Omega^2 &= \hat{\Gamma} - w^2,  \label{LAGR_OMEGA2} \\[10pt]
  \hat{m} &= \frac{\hat{r}}{2}(1-\hat{\Gamma}), \label{LAGR_M}
\end{align}
where the velocity is identical to that used in the Eulerian case.
As before we use the ``hat'' to distinguish
between the time dependent variables and their counterparts in the static 
case. We note that we need to distinguish between the time dependent
areal radius $\hat{r}$ and the static value $r$, since the areal
radius corresponding to the position of a fluid element
is a dependent variable and will generally vary with time. In
the Eulerian case the areal radius was a coordinate and therefore 
intrinsically independent of time. 
If we compare the Lagrangian line element 
(\ref{LAGR_LINEELEMENT}) with the Eulerian one given by
Eq.\,(\ref{NONP_LINEELEMENT}) we therefore have to use the time dependent
$\hat{r}$ in the latter line element instead of $r$. The coordinate 
transformation relating the two line elements is described by
\begin{align}
  \hat{r} &= \hat{r}(t,x).
\end{align}
The transformation law for the metric components corresponding to the
transformation from coordinates $\hbox{\vec $x$}^{\mu}=(t,x,\theta, \phi)$ to
$\hbox{\vec $x$}^{'\mu}=(t,r,\theta,\phi)$ is given by
\begin{align}
  \hbox{\vec g}'_{\mu\nu} &= \frac{\partial \hbox{\vec $x$}^{\rho}}
             {\partial \hbox{\vec $x$}^{'\mu}} 
             \frac{\partial \hbox{\vec $x$}^{\sigma}}
             {\partial \hbox{\vec $x$}^{'\nu}} \hbox{\vec g}_{\rho \sigma},
\end{align}
and leads to the two non-trivial equations
\begin{align}
  \beta &= \hat{r}_{,x} \hat{r}_{,t} \hat{\mu}^2, \\[10pt]
  \hat{\Gamma}&= \frac{1}{\hat{\mu}^2}. \label{LAGR_GAMMAOFMU}
\end{align}
As a consequence the shift vector $\beta$ is related to the 
components of the Lagrangian metric by
\begin{align}
  \beta = \hat{\lambda} \frac{w\,\hat{r}_{,x}}{\hat{\Gamma}}. \label{LAGR_BETA}
\end{align}
In terms of the extrinsic curvature defined in Eq.\,(\ref{BACKGR_EXTRCURV})
this relation can be
written as $\hbox{\vec{K}}^{\theta}_{\,\,\,\theta}=
\hbox{\vec{K}}^{\phi}_{\,\,\,\phi}=0$ and we
have recovered the polar slicing condition. The non-vanishing shift vector
(\ref{LAGR_BETA}) is the price we have to pay for keeping the
polar slicing condition in the Lagrangian gauge. \\
As far as the matter is concerned, we use again a single component
perfect fluid
and thus the energy momentum tensor given by Eq.\,(\ref{NONP_EMTENSOR}).
Since the fluid elements do not move with respect to the radial coordinate
$x$, the 4-velocity has zero spatial components and is determined by the
normalisation $\hbox{\vec u}^{\mu} \hbox{\vec u}_{\mu}=-1$
\begin{align}
  \hbox{\vec u}^{\mu} &= \left( \frac{\sqrt{\hat{\Gamma}}}{\hat{\lambda}
         \Omega}, 0,0,0 \right).
\end{align}
The resulting field equations
$\hbox{\vec{G}}_{\mu \nu} = 8\pi \hbox{\vec{T}}_{\mu \nu}$ can
be written as
\begin{align}
  \frac{\hat{\lambda}_{,x}}{\hat{\lambda}} &= \frac{\hat{r}_{,x}}
    {\hat{r} \hat{\Gamma}}
    \left( \frac{\hat{m}}{\hat{r}} + 4\pi \hat{r}^2 \frac{w^2 \hat{\rho}
    + \hat{\Gamma} \hat{P}} {\Omega^2} \right), \label{LAGR_LAMBDAX} \\[10pt]
  \hat{m}_{,x} &= 4\pi \hat{r}^2 \hat{r}_{,x} \frac{\hat{\Gamma} \hat{\rho}
         +w^2 \hat{P}}{\Omega^2} \label{LAGR_MX}, \\[10pt]
  \hat{m}_{,t} &= -4\pi \hat{r}^2 \hat{\lambda} w \hat{P} \label{LAGR_MT}.
\end{align}
Similarly the conservation of energy and
momentum $\nabla_{\mu} \hbox{\vec{T}}^{\mu}_{\,\,\,\,{\nu}}=0$
leads to two evolution equations for the matter variables
\begin{align}
  &\hat{\rho}_{,t} +(\hat{\rho}+\hat{P}) \frac{\Omega}{\hat{r}^2\hat{r}_{,x}}
    \left( \frac{\hat{r}^2 \hat{r}_{,x}}{\Omega} \right)_{,t}=0,
    \label{LAGR_RHOT} \\[10pt]
  &\Omega^4 \frac{\hat{P}_{,x}}{\hat{r}_{,x}} + \hat{P}_{,t} 
    \frac{w}{\hat{\lambda}} \Omega^2
    + (\hat{\rho} + \hat{P})\left[ \hat{\gamma}\frac{w_{,t}}{\hat{\lambda}}
    +(\hat{\gamma}-2w^2) \left( \frac{\hat{m}}{\hat{r}^2} + 4\pi \hat{r}\hat{P}
    \right) \right] = 0, \label{LAGR_UT}
\end{align}
and the system is closed by the polytropic equation
of state (\ref{DYNAMICPOLYTROPE}).
It is worth pointing out that the appearance of the time derivative in the 
field equation (\ref{LAGR_MT})
does not contradict the absence of gravitational degrees of 
freedom in spherical symmetry. This equation can be shown to be a consequence
of the constraints (\ref{LAGR_LAMBDAX}), (\ref{LAGR_MX}) and
the matter equations (\ref{LAGR_RHOT}), (\ref{LAGR_UT}).
In this sense the degrees of freedom still reside in
the matter variables and the metric is determined at each time irrespective
of its history. In practice, however,
we will use the rather simple equation (\ref{LAGR_MT}) to evolve the variable
$\hat{m}$ instead of evolving $\hat{\rho}$ via the matter equation
(\ref{LAGR_RHOT}). \\
If we consider the static limit of the system of equations 
(\ref{LAGR_LAMBDAX})-(\ref{LAGR_UT}) we expect to recover the 
Tolman-Oppenheimer-Volkoff equations (\ref{TOV_RY})-(\ref{TOV_PY}).
That this is indeed the case can be seen if we set all time derivatives
including the velocity $w$ to zero and assume that $x$ is identical to the
coordinate $x$ we used in the static case.
The second condition can always be satisfied since the fluid elements are
not moving and can be labelled by the areal radius of their position or
the rescaled coordinate $y$ defined in Eq.\,(\ref{TOV_YOFR}). 
Then Eq.\,(\ref{LAGR_MX}) directly reduces to Eq.\,(\ref{TOV_MR}) or the
transformed version thereof expressed in terms of the coordinate $y$.
From Eq.\,(\ref{LAGR_GAMMAOFMU}) we conclude that $\Gamma=1/\mu^2$ and the
constraint (\ref{LAGR_LAMBDAX}) becomes identical to (\ref{TOV_LAMBDAY}).
Finally the matter equation (\ref{LAGR_UT}) reduces to Eq.\,(\ref{TOV_PY}) and
the evolution equations (\ref{LAGR_MT}) 
and (\ref{LAGR_RHOT}) vanish identically.

%=========================================================================
\subsubsection{The linearized evolution equations}
We have seen that the static limit of the evolution equations
(\ref{LAGR_LAMBDAX})-(\ref{LAGR_RHOT}) is given by the TOV equations.
We can therefore linearise the dynamic equations around this background and
compare the results with the Eulerian case described in section \ref{PERT_LIN}.
In order to distinguish between Eulerian and Lagrangian perturbations
we will use a capital $\Delta$ in the Lagrangian case.
The only exception is the radial displacement which is identical
in both formulations so that we keep the variable name $\xi$. \\
We start the linearisation with the definition of the radial velocity
$w$ (\ref{LAGR_U}). In terms of the radial displacement this equation
becomes
\begin{align}
  w &= \frac{\xi_{,t}}{\lambda}. \label{LLIN_U}
\end{align}
We note that the background value of the lapse $\lambda$ appears
in the denominator instead of the time dependent $\hat{\lambda}$.
In the same way we will neglect higher order terms in the other equations.
If we substitute this expression for $w$ in the evolution
equation (\ref{LAGR_MT})
for $m$ and integrate over time, we obtain
\begin{align}
  \Delta m &= -4\pi r^2 P \xi. \label{LLIN_DM}
\end{align}
The constant of integration vanishes because a zero displacement $\xi$
of the fluid elements implies $\Delta m=0$.
We can use this expression for $\Delta m$ in the definition (\ref{LAGR_M})
to obtain the result for the auxiliary variable $\hat{\Gamma}$
\begin{align}
  \Delta \Gamma &= 8\pi r P \xi + \frac{\xi}{r} (1-\Gamma). \label{LLIN_DGAMMA}
\end{align}
The energy density perturbation then follows from substituting 
Eqs.\,(\ref{LLIN_U})-(\ref{LLIN_DGAMMA}) in the evolution equation
(\ref{LAGR_RHOT}) 
and integrating over time. With the constant of integration vanishing as
before the result is
\begin{align}
  \Delta \rho &= \frac{(\rho +P)}{r_{,x}}
        \left( \frac{\xi r^2}{\lambda} \right)_{,x}.
  \label{LLIN_DRHO}
\end{align}
From the definition of the speed of sound we can calculate the pressure
perturbation
\begin{align}
  \Delta P &= C^2 \Delta \rho. \label{LLIN_DP}
\end{align}
If we substitute the results (\ref{LLIN_U})-(\ref{LLIN_DP}) in the
evolution equation (\ref{LAGR_UT}) we get exactly the second order
differential equation (\ref{LIN_ZETATT}) of the Eulerian case with
the coefficient functions (\ref{LIN_PI})-(\ref{LIN_Q}). No substitution
for $\Delta \lambda$ is necessary here, because
all terms containing $\Delta \lambda$ drop out by virtue of the
TOV background equations. Writing the displacement as a product
$\xi(x) f(t)$ we obtain again harmonic time dependence
and finally arrive at the ordinary differential
equation (\ref{LIN_ZETARR}) so that we can use the whole machinery
developed in section \ref{PERT_LIN} to calculate the eigenmodes.
It is interesting,
however, to contrast Eq.\,(\ref{LLIN_DRHO}) for the Lagrangian $\Delta \rho$
with the Eulerian analogue Eq.\,(\ref{LIN_DRHOOFZETA}). We have seen in section
\ref{PERT_LIN} that the extra term in the Eulerian relation leads to the
problematic asymptotic behaviour of $\delta \rho$ at the surface. No
such problem occurs in the Lagrangian case which thus provides a 
self-consistent way of deriving the linearized equations.

%=========================================================================
\subsubsection{The equations for the numerical implementation}
The Lagrangian evolution of a dynamic neutron star in spherical
symmetry is described by the system of equations (\ref{LAGR_U}),
(\ref{LAGR_LAMBDAX})-(\ref{LAGR_MT}), (\ref{LAGR_UT}), where the auxiliary 
variables $\hat{\Gamma}$ and $\Omega$ are defined by Eqs.\,(\ref{LAGR_OMEGA2})
and (\ref{LAGR_M}). This choice of variables and equations, however, did
not lead to an entirely satisfactory performance of the code. This
became most obvious in the simulation of the Oppenheimer-Snyder dust
collapse where the energy density
showed an increasing deviation from the analytic solution near the centre
of the star. When the dust sphere had collapsed close to its
Schwarzschild radius,
the deviation was larger than $10\,\%$. In order to understand this
inaccuracy, we consider Eq.\,(\ref{LAGR_MX}) which relates
the energy density to the mass. If we solve this equation for
$\hat{\rho}$ we see that the mass appears in the form
$\hat{m}_{,x}/\hat{r}^2$, which will be of the order $\mathscr{O}(1)$
near the origin. The second order accuracy of the finite differencing
scheme we have used, however, implies that the variable
$\hat{m}$ is known with a local error
$\mathscr{O}(\Delta x^3)$ only and consequently
the numerical derivative $\hat{m}_{,x}$ has an error
$\mathscr{O}(\Delta x^2)$. Near the origin
the radius $\hat{r}$ is of the same order of magnitude as $\Delta x$ and
the error of $\hat{m}_{,x}/\hat{r}^2$ and thus the energy density $\hat{\rho}$
is large. This problem is a consequence of the $\hat{r}^3$ behaviour of the
mass $\hat{m}$ near the origin combined with the strong variation
of the variables in the dust collapse and persists in a perturbative
formulation. In the numerical evolution we therefore use the
variable
\begin{align}
  \hat{N} &= \frac{\hat{m}}{\hat{r}^2}, \label{NLAGR_NOFM}
\end{align}
instead of the mass $\hat{m}$.
The Lagrangian equations
(\ref{LAGR_U}), (\ref{LAGR_LAMBDAX})-(\ref{LAGR_MT}), (\ref{LAGR_UT})
then become
\begin{align}
  & \hat{\Gamma} \Omega^2 \hat{\lambda}_{,x} - \hat{r}_{,x} \hat{\lambda} \left[
    \Omega^2 \hat{N} + 4\pi \hat{r} (w^2 \hat{\rho} + \hat{\Gamma} \hat{P})
    \right] = 0, \label{NLAGR_LAMBDAX} \\[10pt]
  & \hat{r}\Omega^2 \hat{N}_{,x} + 2\Omega^2 \hat{r}_{,x} \hat{N}
    -4\pi \hat{r} \hat{r}_{,x}
    (\hat{\Gamma} \hat{\rho} + w^2 \hat{P}) = 0, \label{NLAGR_NX} \\[10pt]
  & \hat{r}\hat{N}_{,t} + 2\hat{\lambda} w \left( \hat{N} + 2\pi \hat{r}
    \hat{P} \right) =0, \\[10pt]
  & \hat{r}_{,t} - \hat{\lambda} w = 0, \\[10pt]
  & \hat{\lambda} \Omega^4 \hat{P}_{,x} + \hat{r}_{,x} w \Omega^2 \hat{P}_{,t}
    + \hat{r}_{,x} (\hat{\rho}+\hat{P}) \left[ \hat{\Gamma} w_{,t}
    + \hat{\lambda} (\hat{\Gamma}-2w^2)(\hat{N} + 4\pi \hat{r} \hat{P}) \right],
    \label{NLAGR_UT}
\end{align}
where $\hat{\Gamma}$ is now defined by
\begin{align}
  \hat{\Gamma} &= 1-2\hat{N}\hat{r}.
\end{align}
In the static limit these equations reduce to the TOV equations
\begin{align}
  & \Gamma \lambda_{,x} - r_{,x}\lambda (N + 4\pi r P ), \\[10pt]
  & r N_{,x} + r_{,x} \left( 2 N - 4\pi r \rho \right) = 0, \\[10pt]
  & \Gamma P_{,x} + r_{,x} (\rho+P) (N+4\pi r P)=0, \\[10pt]
  & \Gamma = 1-2Nr.\end{align}
In order to derive a fully non-linear perturbative formulation, we decompose
the time dependent quantities into static background contributions
and time dependent perturbations
\begin{align}
  \hat{r}(t,x) &= r(x) + \xi(t,x), \\[10pt]
  \hat{\lambda}(t,x) &= \lambda(x) + \Delta \lambda(t,x), \\[10pt]
  \hat{N}(t,x) &= N(x) + \Delta N (t,x), \\[10pt]
  \hat{\Gamma}(t,x) &= \Gamma(x) + \Delta \Gamma(t,x), \\[10pt]
  \hat{\rho}(t,x) &= \rho(x) + \Delta \rho(t,x).
\end{align}
With these definitions the fully non-linear perturbative version
of Eqs.\,(\ref{NLAGR_LAMBDAX})-(\ref{NLAGR_UT}) becomes
\begin{align}
  \begin{split}
  & \hat{\Gamma}^2 \Delta \lambda_{,x} + \Delta \Gamma (2\Gamma
    + \Delta \Gamma) \lambda_{,x} - (\xi_{,x} \lambda \Gamma
    + \hat{r}_{,x} \Delta \lambda \Gamma + \hat{r}_{,x} 
    \hat{\lambda}\Delta \Gamma ) (N+4\pi r P)\\[10pt]
  & + w^2 \left[ -\hat{\Gamma} \hat{\lambda}_{,x} + \hat{r}_{,x}\hat{\lambda}
    (\hat{N}-4\pi \hat{r} \hat{\rho}) \right] - \hat{r}_{,x} \hat{\lambda}
    \hat{\Gamma} \left[ \Delta N + 4\pi (\xi P + \hat{r} \Delta P)
    \right] = 0,
  \end{split} \label{NLAGR_DLAMBDAX} \\[20pt]
  \begin{split}
  & -w^2(\hat{r}\hat{N}_{,x} +2\hat{r}_{,x} \hat{N} + 4\pi \hat{r}
    \hat{r}_{,x} \hat{P}) + \Delta \Gamma (\hat{r}\hat{N}_{,x}
    + 2\hat{r}_{,x} \hat{N} -4\pi \hat{r}\hat{r}_{,x} \hat{\rho} )
    \\[10pt]
  & + \Gamma \left[ \xi N_{,x} + \hat{r} \Delta N_{,x} + 2\xi_{,x} N
    + 2\hat{r}_{,x} \Delta N - 4\pi( \xi \rho r_{,x} + \hat{r}\xi_{,x}
    \rho + \hat{r} \hat{r}_{,x} \Delta \rho ) \right] = 0,
  \end{split} \label{NLAGR_DNX} \\[20pt]
  & \hat{r}\hat{N}_{,t} + 2\hat{\lambda} w (\hat{N} + 2\pi \hat{r} \hat{P}) = 0,
    \label{NLAGR_DNT} \\[20pt]
  & \xi_{,t} - \hat{\lambda} w = 0, \label{NLAGR_XIT} \\[20pt]
  \begin{split}
  & \hat{\lambda}(-2\hat{\Gamma} w^2 + w^4) \hat{P}_{,x} + \hat{r}_{,x}
    w \Omega^2 \hat{P}_{,t} + (\hat{\rho}+\hat{P}) \hat{r}_{,x} \left[
    \hat{\Gamma} w_{,t} -2\hat{\lambda}w^2(\hat{N} + 4\pi \hat{r} \hat{P}) 
    \right] \\[10pt]
  & + (\Delta \lambda \Gamma + \hat{\lambda} \Delta \Gamma)
    \left[ \hat{\Gamma} \hat{P}_{,x} + (\hat{\rho} + \hat{P}) \hat{r}_{,x}
    (\hat{N} + 4\pi \hat{r} \hat{P}) \right] + \lambda \Gamma \left\{
    \Delta \Gamma P_{,x} + \hat{\Gamma} \Delta P_{,x} \right. \\[10pt]
  & \left. + \left[ (\Delta \rho
    + \Delta P)r_{,x} + (\hat{\rho} + \hat{P}) \xi_{,x} \right] (N+4\pi r P)
    + (\hat{\rho} + \hat{P}) \hat{r}_{,x} (\Delta N + 4\pi \xi P
    + 4\pi \hat{r}\Delta P) \right\} = 0.
  \end{split}  \label{NLAGR_DPT}
\end{align}
This is the final system of equations used in the numerical implementation.

%=========================================================================
\subsubsection{Initial data and boundary conditions}
In order to numerically evolve the system of partial differential 
equations (\ref{NLAGR_DLAMBDAX})-(\ref{NLAGR_DPT}) we have to specify
initial data and boundary conditions. We will start with the initial data. \\
In the Eulerian case we have determined the physical setup by providing
initial data for the matter variables $\hat{\rho}$ and $w$. This
gave us energy density and velocity at each radial position $\hat{r}$.
In order to provide the same information in the Lagrangian case
it is not sufficient to give initial data in the form of
$\hat{\rho}(x)$ and $w(x)$ because the meaning of the spatial
coordinate $x$ is not determined at this stage. Indeed it can easily be
seen that the system of equations
(\ref{NLAGR_LAMBDAX})-(\ref{NLAGR_UT}) is invariant under
any transformation $x\rightarrow \bar{x}(x)$ which corresponds to a relabelling
of the fluid elements. Consequently we also need to establish a relation
between the Lagrangian coordinate $x$ and the areal radius $\hat{r}$ on
the initial slice. The initial data for $\hat{r}(x)$ serve this purpose.
Alternatively this additional requirement becomes obvious if we
consider the structure of the system (\ref{NLAGR_LAMBDAX})-(\ref{NLAGR_UT}).
These equations contain the time derivatives of $\hat{r}$, $w$,
$\hat{N}$ and $\hat{P}$. In addition to the lapse function $\hat{\lambda}$
only one of these quantities is determined by the constraint equations
(\ref{NLAGR_LAMBDAX}), (\ref{NLAGR_NX}). The remaining three variables follow
from the time evolution and thus require the specification of initial
data. In the perturbative formulation the background functions
$r(x)$, $\rho(x)$, $N(x)$ and $\lambda(x)$ follow from the
solution of the TOV equations and we prescribe initial data for the
perturbations $\xi$, $w$ and $\Delta \rho$. The values of $\Delta N$
and $\Delta \lambda$ are then calculated from the constraint equations
(\ref{NLAGR_DLAMBDAX}) and (\ref{NLAGR_DNX}). For this purpose we use
an implicit second order scheme based on the finite differencing
given for these equations in appendix \ref{LAGR_FDE}. \\
The specification of boundary conditions, in particular at the
stellar surface, turned out to be the most problematic part in the
Eulerian formulation of the dynamic star. In contrast the boundary
conditions are well defined in the Lagrangian case. At the centre
we demand
\begin{align}
  \xi &= 0, \label{NLAGR_BCINXI} \\[10pt]
  w &= 0, \\[10pt]
  \Delta N &= 0.
\end{align}
The first two conditions guarantee that the centre of the star does not
move which immediately follows from the spherical symmetry and the third
condition
avoids the appearance of a conical singularity. At the surface we require
\begin{align}
  \Delta \rho &= 0, \label{NLAGR_BCOUTDRHO} \\[10pt]
  \hat{\lambda}^2 &= 1-2\hat{N}\hat{r}, \label{NLAGR_BCOUTLAMBDA}
\end{align}
which follows from the definition of the surface and the matching
to an exterior Schwarzschild metric.
% For the numerical evolution
%we will use an implicit second order in space and time finite
%differencing scheme.
If $K$ is the number of grid points used, the
finite differencing of the evolution equations
(\ref{NLAGR_DLAMBDAX})-(\ref{NLAGR_DPT}) results in $5K-5$ algebraic
relations between the $5K$ function values. The boundary
conditions (\ref{NLAGR_BCINXI})-(\ref{NLAGR_BCOUTLAMBDA}) provide
the remaining 5 relations to determine the evolution and no additional
treatment of boundary values is required.

%=========================================================================
\subsubsection{The finite differencing of the equations}
\begin{figure}[t]
  \centering
%%%%%%%%%%%%%%%%%%%%%%%%%%%%%%%%%%%%%%%%%%%%%%%%%%%%%%%%%%%%%%%%%%%%%%%%%
\begin{picture}(0,0)%
\epsfig{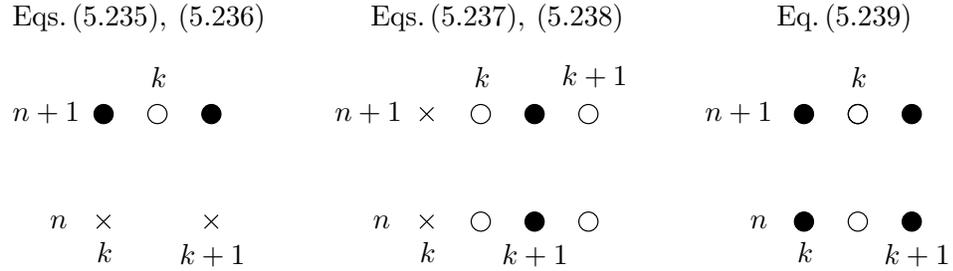}%
\end{picture}%
\setlength{\unitlength}{3947sp}%
\begingroup\makeatletter\ifx\SetFigFont\undefined%
\gdef\SetFigFont#1#2#3#4#5{%
  \reset@font\fontsize{#1}{#2pt}%
  \fontfamily{#3}\fontseries{#4}\fontshape{#5}%
  \selectfont}%
\fi\endgroup%
\begin{picture}(5717,1620)(301,-811)
\put(1170,299){\makebox(0,0)[lb]{\smash{\SetFigFont{11}{13.2}{\rmdefault}{\mddefault}{\itdefault}$k$}}}
\put(2861,-800){\makebox(0,0)[lb]{\smash{\SetFigFont{11}{13.2}{\rmdefault}{\mddefault}{\itdefault}$k$}}}
\put(5230,-800){\makebox(0,0)[lb]{\smash{\SetFigFont{11}{13.2}{\rmdefault}{\mddefault}{\itdefault}$k$}}}
\put(2566,-589){\makebox(0,0)[lb]{\smash{\SetFigFont{11}{13.2}{\rmdefault}{\mddefault}{\itdefault}$n$}}}
\put(4934,-589){\makebox(0,0)[lb]{\smash{\SetFigFont{11}{13.2}{\rmdefault}{\mddefault}{\itdefault}$n$}}}
\put(831,-800){\makebox(0,0)[lb]{\smash{\SetFigFont{11}{13.2}{\rmdefault}{\mddefault}{\itdefault}$k$}}}
\put(536,-589){\makebox(0,0)[lb]{\smash{\SetFigFont{11}{13.2}{\rmdefault}{\mddefault}{\itdefault}$n$}}}
\put(3200,299){\makebox(0,0)[lb]{\smash{\SetFigFont{11}{13.2}{\rmdefault}{\mddefault}{\itdefault}$k$}}}
\put(5568,299){\makebox(0,0)[lb]{\smash{\SetFigFont{11}{13.2}{\rmdefault}{\mddefault}{\itdefault}$k$}}}
\put(3376,-811){\makebox(0,0)[lb]{\smash{\SetFigFont{11}{13.2}{\rmdefault}{\mddefault}{\itdefault}$k+1$}}}
\put(5776,-811){\makebox(0,0)[lb]{\smash{\SetFigFont{11}{13.2}{\rmdefault}{\mddefault}{\itdefault}$k+1$}}}
\put(1351,-811){\makebox(0,0)[lb]{\smash{\SetFigFont{11}{13.2}{\rmdefault}{\mddefault}{\itdefault}$k+1$}}}
\put(3751,314){\makebox(0,0)[lb]{\smash{\SetFigFont{11}{13.2}{\rmdefault}{\mddefault}{\itdefault}$k+1$}}}
\put(301, 89){\makebox(0,0)[lb]{\smash{\SetFigFont{11}{13.2}{\rmdefault}{\mddefault}{\itdefault}$n+1$}}}
\put(2326, 89){\makebox(0,0)[lb]{\smash{\SetFigFont{11}{13.2}{\rmdefault}{\mddefault}{\itdefault}$n+1$}}}
\put(4651, 89){\makebox(0,0)[lb]{\smash{\SetFigFont{11}{13.2}{\rmdefault}{\mddefault}{\itdefault}$n+1$}}}
\put(301,689){\makebox(0,0)[lb]{\smash{\SetFigFont{11}{13.2}{\rmdefault}{\mddefault}{\updefault}Eqs.\,(\ref{NLAGR_DLAMBDAX}), (\ref{NLAGR_DNX})}}}
\put(2551,689){\makebox(0,0)[lb]{\smash{\SetFigFont{11}{13.2}{\rmdefault}{\mddefault}{\updefault}Eqs.\,(\ref{NLAGR_DNT}), (\ref{NLAGR_XIT})}}}
\put(5101,689){\makebox(0,0)[lb]{\smash{\SetFigFont{11}{13.2}{\rmdefault}{\mddefault}{\updefault}Eq.\,(\ref{NLAGR_DPT})}}}
\end{picture}
%%%%%%%%%%%%%%%%%%%%%%%%%%%%%%%%%%%%%%%%%%%%%%%%%%%%%%%%%%%%%%%%%%%%%%%%%
  \caption{The stencils used for the finite differencing of
           Eqs.\,(\ref{NLAGR_DLAMBDAX})-(\ref{NLAGR_DPT}).}
  \label{LAGR_STENCILS}
\end{figure}

We numerically solve the system of partial differential equations
(\ref{NLAGR_DLAMBDAX})-(\ref{NLAGR_DPT}) by using an implicit
second order in space and time finite differencing scheme. The
particular choice of stencils has been guided by the presence of
derivatives in the individual differential equations. This is illustrated
in Fig.\,\ref{LAGR_STENCILS} where the grid points $k$ and $k+1$
are shown for the time levels
$n$ and $n+1$. The filled circles indicate grid points that have been used
for the finite differencing, the crosses those points which have not 
been used. The constraint equations (\ref{NLAGR_DLAMBDAX}) and
(\ref{NLAGR_DNX}) contain spatial derivatives only. It is therefore
suitable to use two neighbouring grid points on the new time slice $n+1$.
In contrast Eqs.\,(\ref{NLAGR_DNT}) and (\ref{NLAGR_XIT}) contain time 
derivatives only and we use two grid points at spatial position $k+1$
on neighbouring time slices for the finite differencing. Both
kinds of derivatives are present in Eq.\,(\ref{NLAGR_DPT}) and we need to use
all four grid points as a consequence. Fig.\,{\ref{LAGR_STENCILS}
also illustrates an
extra option that has been included in the finite differencing.
In the case of the Oppenheimer-Snyder dust collapse it turns out
to be necessary to interpret the values of the energy density
$\rho^n_k$, $\Delta \rho^n_k$ as cell averages and correspondingly
use a staggered grid for these variables. This is indicated by the
empty circles in Fig.\,\ref{LAGR_STENCILS}.
In the finite differencing equations we will
therefore introduce a parameter $\sigma$ which allows us to switch between
a staggered and the ``normal'' grid for $\rho$ and $\Delta \rho$.
The staggering, however, is only needed for the dust collapse and will not
be used when we simulate neutron stars. \\
The resulting finite difference equations are listed in appendix
\ref{LAGR_FDE}
together with the additional relations we use to calculate auxiliary 
functions and derivatives of the background variables. The parameter
$\sigma$ will be zero in all cases except for the simulation
of the Oppenheimer-Snyder dust collapse, where we will use the staggered grid
for the energy density and set $\sigma=1$. Before we turn our attention
towards solving this system of algebraic equations, we need to comment on some
of its properties.
\begin{list}{\rm{(\arabic{count})}}{\usecounter{count}
             \labelwidth1cm \leftmargin1.5cm \labelsep0.4cm \rightmargin1cm
             \parsep0.5ex plus0.2ex minus0.1ex \itemsep0ex plus0.2ex}
\item If we use the staggered grid to calculate the energy density,
      the outer boundary condition (\ref{FDE_BCOUTDRHO}) is only a
      formal condition because $\Delta \rho_K$ decouples from the 
      remaining $5K-1$ variables. In the analysis of the dust collapse
      we will therefore use the interior values
      $\Delta \rho_k$ for $k=1,\ldots ,K-1$ only.
\item We also note that the finite difference expression (\ref{FDE_EQ5DRHOX})
      for $\Delta \rho_{,x}$ is only a first order accurate approximation
      if the staggered grid is used for the energy density. This does 
      not affect the accuracy of the numerical scheme, however, since 
      this derivative appears in the form of the pressure gradient
      $\Delta P_{,x}$ only in Eq.\,(\ref{NLAGR_DPT}). The only scenario where
      we use the staggering is the dust collapse, where
      the pressure and thus its gradient vanish identically.
\item Finally we note that the finite differencing scheme used here
      slightly differs from that used for the evolution of
      cosmic strings in section \ref{dynstring}. The scheme used here
      was partly inspired by the work of \scite{Schinder1988}
      and partly resulted
      from attempts to eliminate numerical noise that we encountered
      during the development of the code. It turned out, however, that
      this noise originated from the numerical inaccuracy associated with
      the $\hat{r}^3$ behaviour of the variable $\hat{m}$ we discussed above.
      We have no reason therefore to question the applicability
      of the Crank-Nicholson scheme described in section \ref{FDE_CN}.
\end{list}
In order to solve the system of $5K$ non-linear algebraic relations we
use the Newton-Raphson method described in section \ref{relaxation}.
The initial
guess is given by the values on the previous time slice and convergence 
is typically achieved after three iterations.

%=========================================================================
\subsubsection{Testing the code}
In order to check the performance of the code we subject it to
three independent tests. As in the Eulerian case, we will compare
the numerical results with the
approximative analytic solution obtained from the linearized equations of a
dynamic spherically symmetric neutron star. Secondly we will test the 
convergence properties of the code in the non-linear regime. Finally
we calculate the deviation of the numerical results from the analytic
solution by \lcite{Oppenheimer1939b} which describes the collapse of
a homogeneous dust sphere. \\
We start by testing the performance of the code in the linear regime.
In the Eulerian analysis we have seen that the eigenmodes for
stellar models with polytropic indices $\gamma > 2$
lead to a diverging energy density perturbation at 
the surface and thus could not be used for a time evolution. We have seen,
however, that this divergence results from a coordinate singularity
at the stellar surface and the Lagrangian energy density perturbation is
well behaved for any polytropic index. It is tempting therefore to
use a stellar model with a large polytropic index to test the performance
of the Lagrangian code in the linear regime. We choose a model with
polytropic exponent $\gamma=3.0$, polytropic factor $K=2\cdot10^5\,\,
{\rm km}^{-2}$ and central density $\rho_{\rm c}= 2.2\cdot
10^{15}\,\,{\rm g/cm}^3$. This is the third model of Table \ref{COMPEIGS} 
where we compared our results of the eigenmode frequencies with
those of \lcite{Kokkotas2001}. \\
In general we have achieved better performance with
the Lagrangian code if the outer boundary condition $\rho=0$ is satisfied
\begin{figure}[t]
  \centering
  \epsfig{file=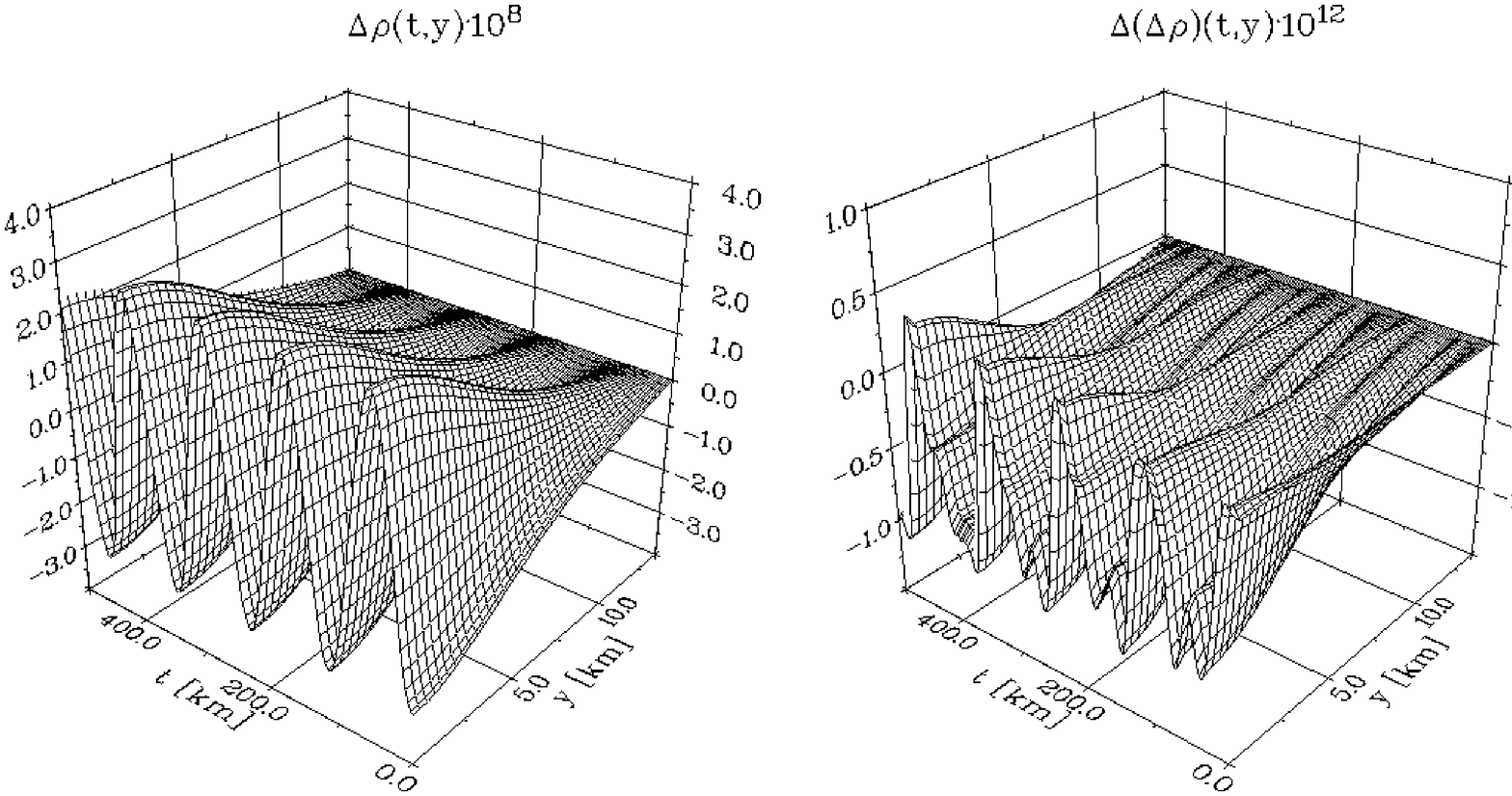, height=400pt, width=175pt, angle=-90}
  \epsfig{file=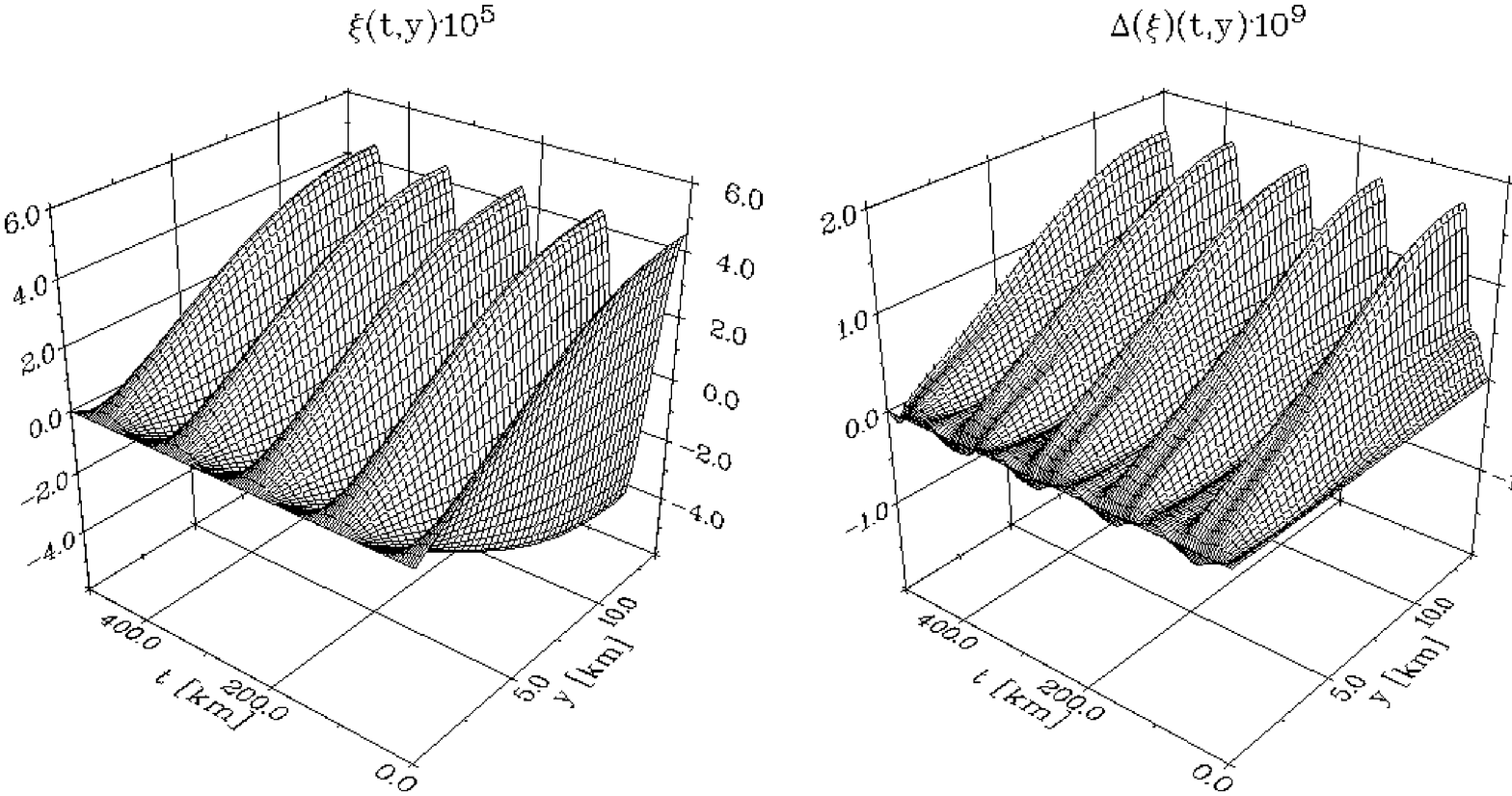, height=400pt, width=175pt, angle=-90}
  \epsfig{file=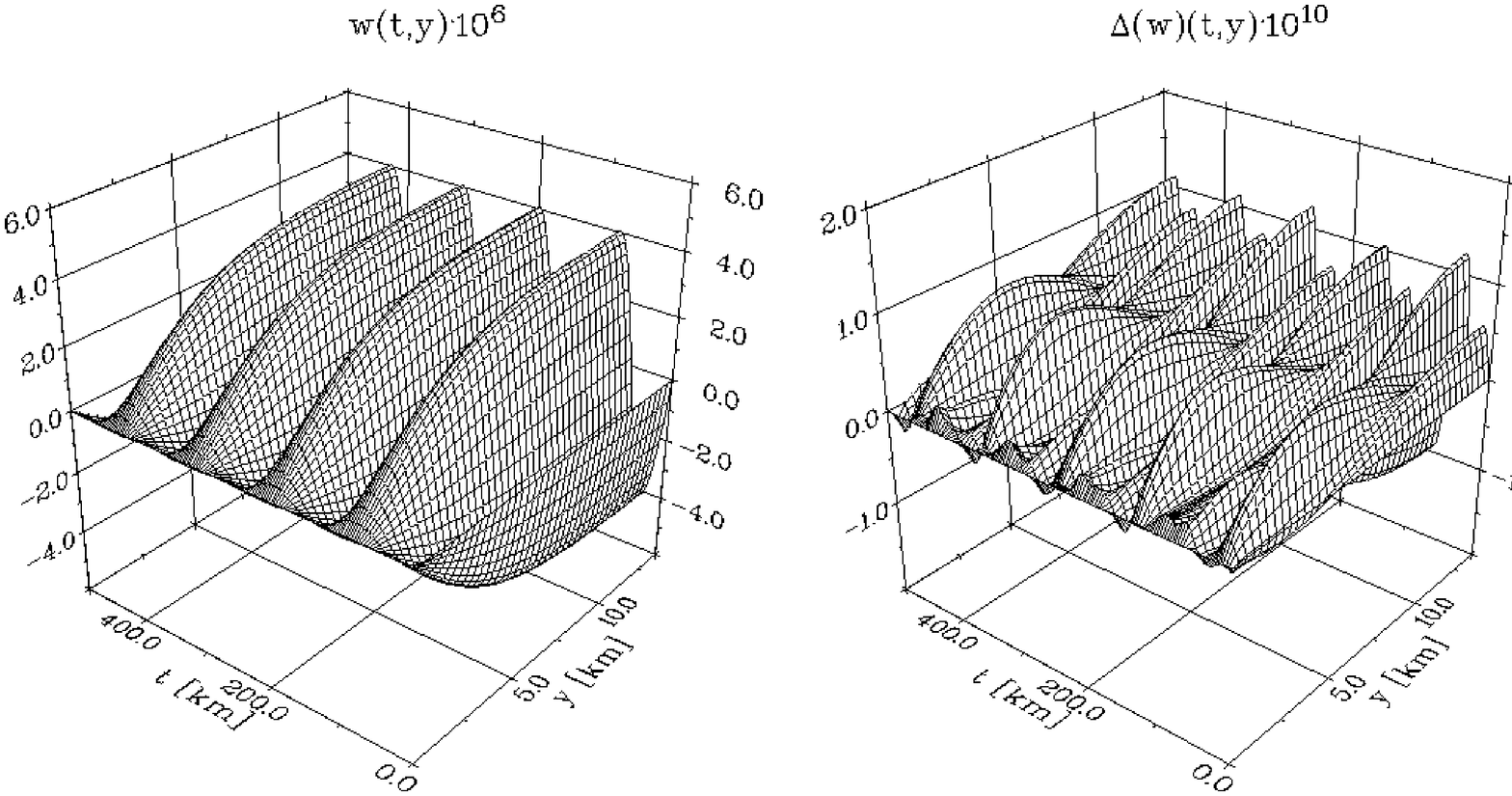, height=400pt, width=175pt, angle=-90}
  \caption{The left panels show the time evolution of $\Delta \rho$,
           $\xi$ and $w$ obtained for 1600 grid points. The initial
           perturbation is given as the fundamental mode in the
           displacement vector $\xi$. The right panels show the
           deviation from the exact solution of the linearized equations.}
  \label{LAGR_LINTEST}
\end{figure}
exactly. In the remainder of the Lagrangian discussion we will therefore
use the relaxation method described in section \ref{TOV_NUM}
to calculate the TOV background. Unless specified otherwise we will
use the rescaled coordinate $y$ for this calculation and the time evolution
and thus set $r_{,x}=C$.\\
The next step
consists in calculating the eigenmode profiles for the variables
$\xi$, $w$ and $\Delta \rho$. These results enable us to specify initial 
data and calculate the analytic solutions. In this case the initial
perturbation of the star consists in a displacement $\xi$ of the fluid
elements corresponding to the fundamental mode with a surface amplitude
of about $5\,\,{\rm cm}$. The initial velocity is set to zero and
the energy density corresponding to this eigenmode follows from
Eq.\,(\ref{LLIN_DRHO}). The remaining initial variables are calculated
from the constraint equations (\ref{NLAGR_DLAMBDAX}), (\ref{NLAGR_DNX}).
The resulting
data on the initial slice are then evolved in time according to the method
described in the previous section. The analytic solution for the
fundamental variables $\xi$, $w$, $\Delta \rho$ is given by
\begin{align}
  \xi(t,x) &= \xi_1(x) \cos{\omega t}, \\[10pt]
  w(t,x) &= -w_1(x) \sin{\omega t}, \\[10pt]
  \Delta \rho &= \Delta \rho_1(x) \cos{\omega t},
\end{align}
where $\omega$ is the frequency derived from the eigenmode calculation.
In Fig.\,\ref{LAGR_LINTEST} we show the numerical
results obtained for 1600 grid points together with their
deviation from the harmonic solutions.
These results show that the code reproduces the analytic solution
\begin{figure}[t]
  \centering
  \epsfig{file=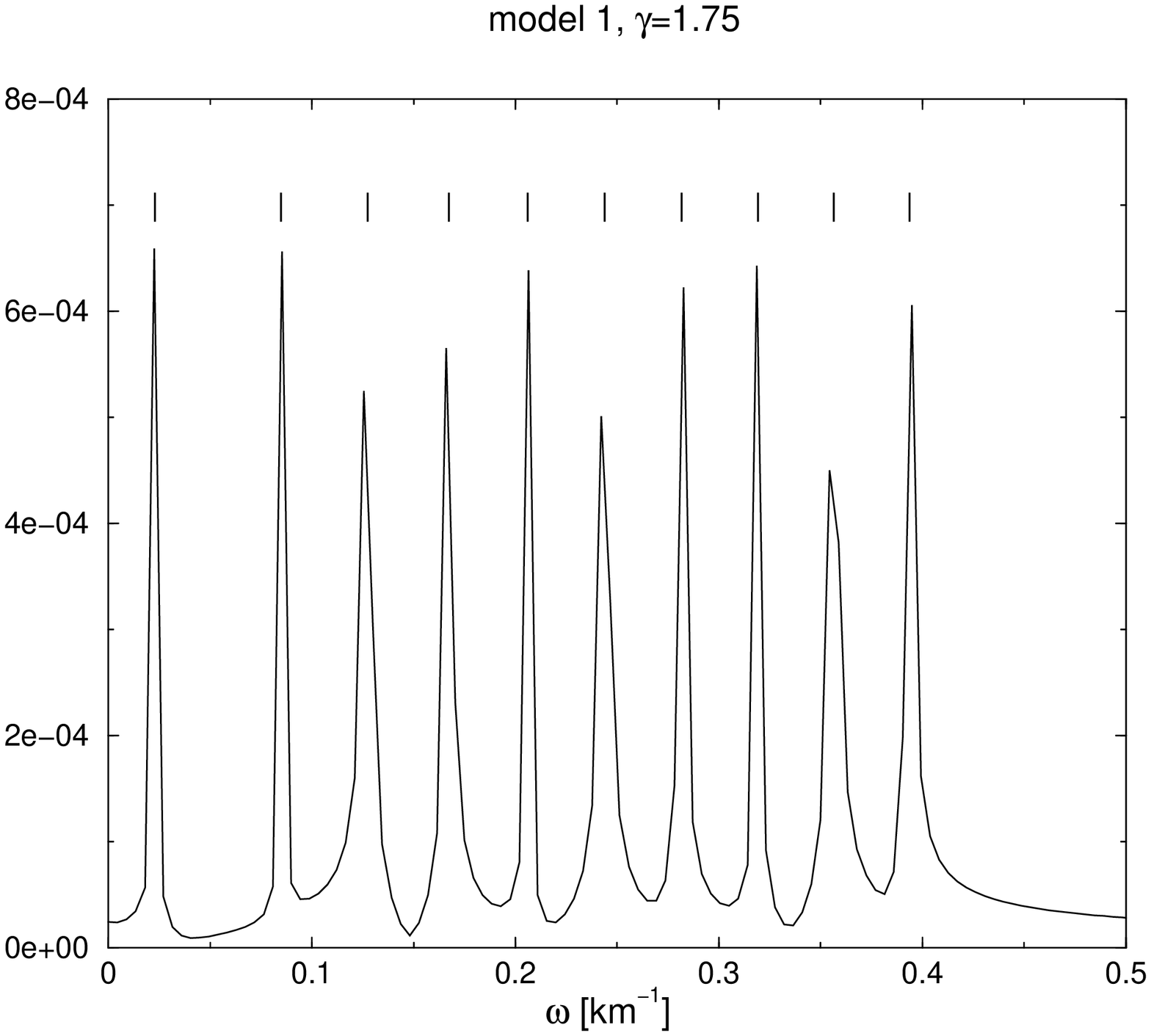, height=200pt, width=150pt, angle=-90}
  \epsfig{file=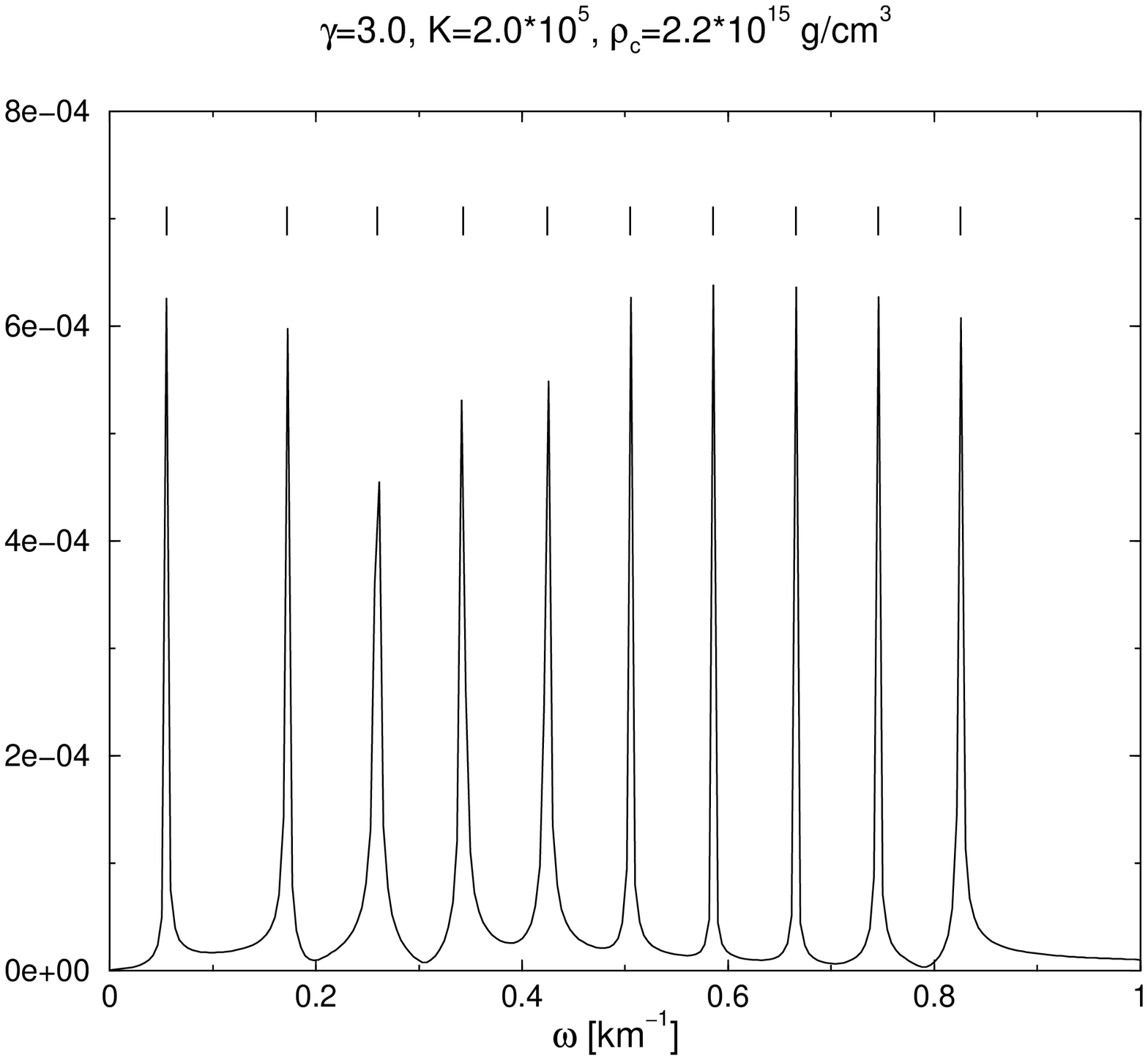, height=200pt, width=150pt, angle=-90}
  \caption{Frequency spectra obtained for stellar models with polytropic
           indices $\gamma = 1.75$ (left) and 3.0 (right). The initial
           data consists of a displacement $\xi$ given by the sum of the
           first 10 eigenmodes. The vertical bars indicate the frequencies
           predicted by the eigenmode calculations.}
  \label{LAGR_FOUR}
\end{figure}
with a relative accuracy of about $10^{-4}$. For presentation purposes
the time evolution is shown up to $t=500\,\,{\rm km}$ only. No
significant loss of accuracy has been observed for longer evolutions. \\
We have also compared the frequency spectrum resulting from time evolutions
with the corresponding predictions by the eigenmode calculation. For
this purpose we have used the same stellar model as in the previous test
as well as model 1 of Table \ref{MODELS15} which has a polytropic index 
$\gamma = 1.75$. In both cases the initial
perturbation is given by the sum of the first ten eigenmodes in
the displacement $\xi$. The combined amplitude is about $10\,\,{\rm cm}$
in both cases, so that the deviation from the linear approximation should
again be very small. In Fig.\,\ref{LAGR_FOUR} we show
the Fourier spectra of the central energy density perturbation
$\Delta \rho(t,0)$ obtained
for time evolutions over $1500\,\,{\rm km}$ using 600 grid points.
The vertical bars indicate the frequencies predicted for the first 10
eigenmodes and coincide well with the peaks in the power spectra. \\
In order to test the performance of the code in the non-linear regime
\begin{figure}[t]
  \centering
  \epsfig{file=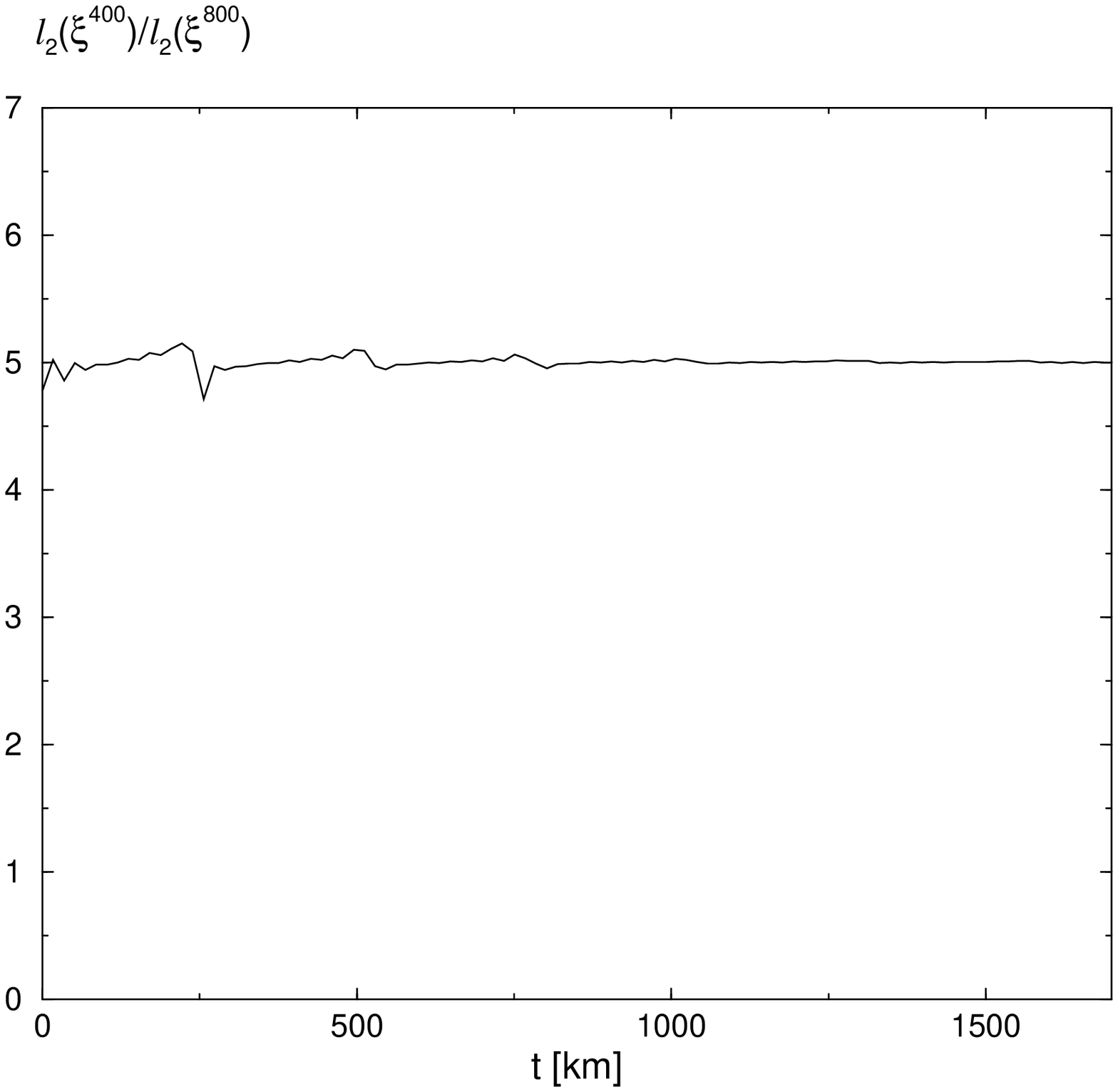, height=180pt, width=130pt, angle=-90}
  \hspace{1cm}
  \epsfig{file=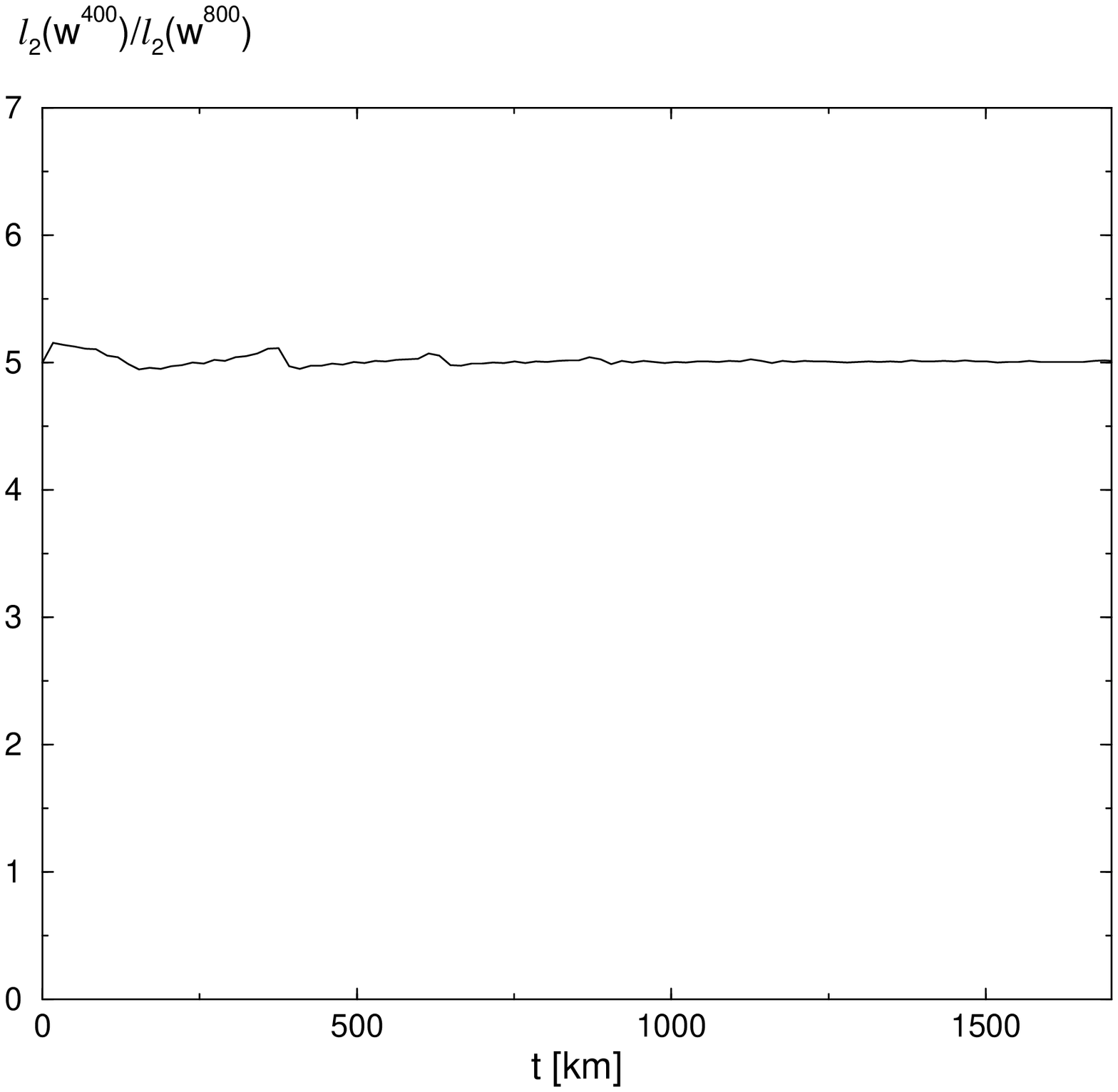, height=180pt, width=130pt, angle=-90}
  \epsfig{file=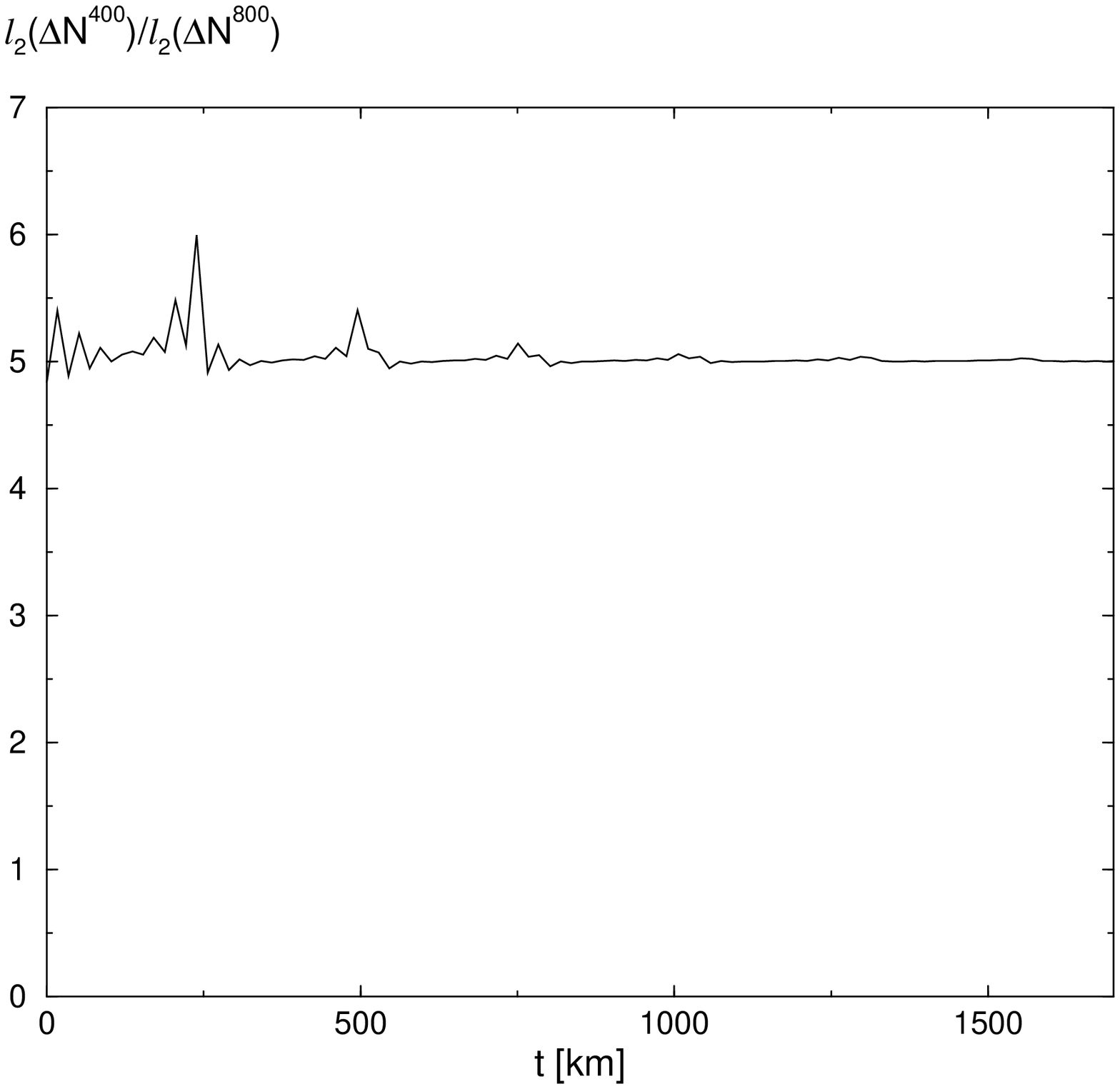, height=180pt, width=130pt, angle=-90}
  \hspace{1cm}
  \epsfig{file=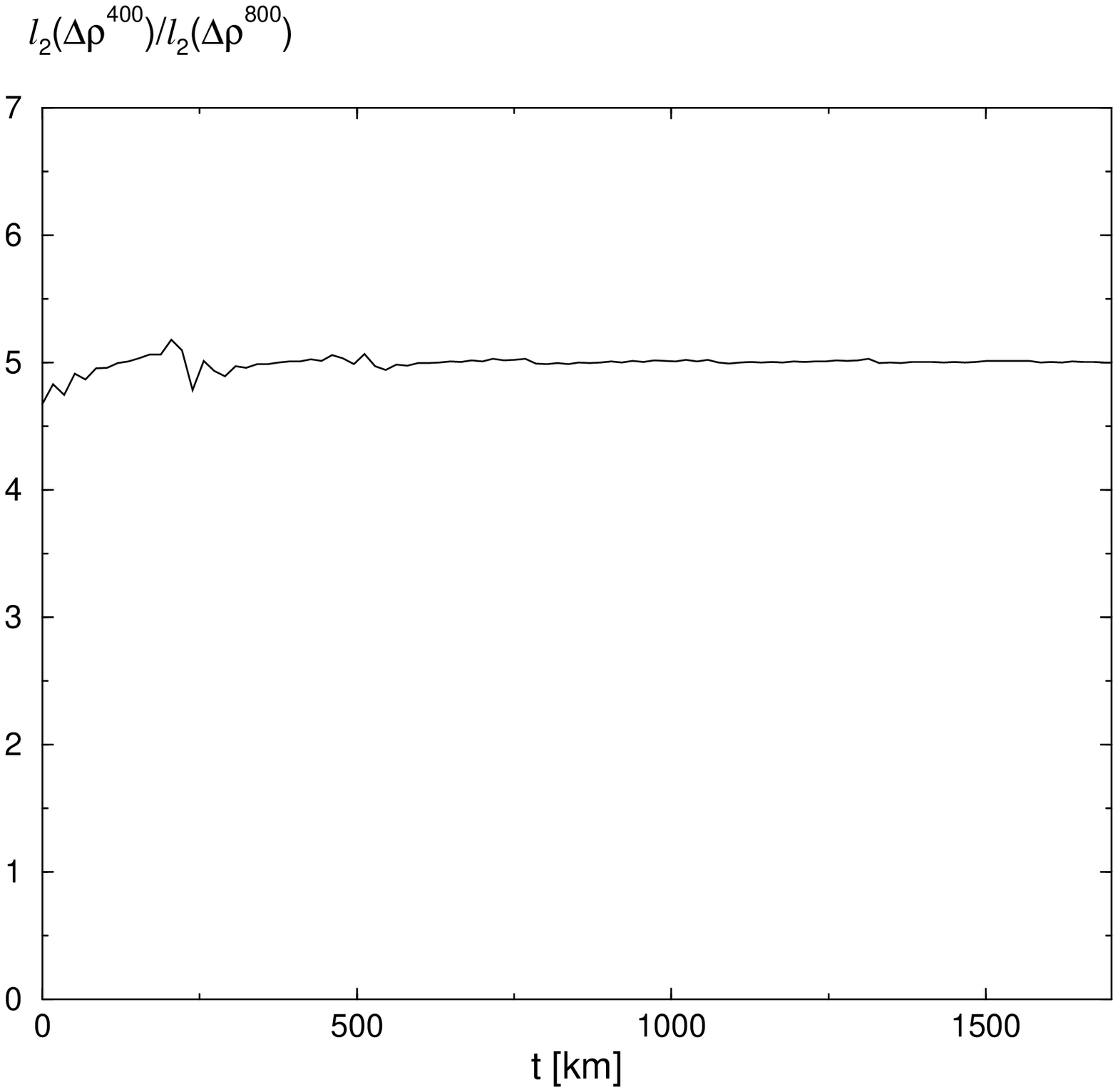, height=180pt, width=130pt, angle=-90}
  \epsfig{file=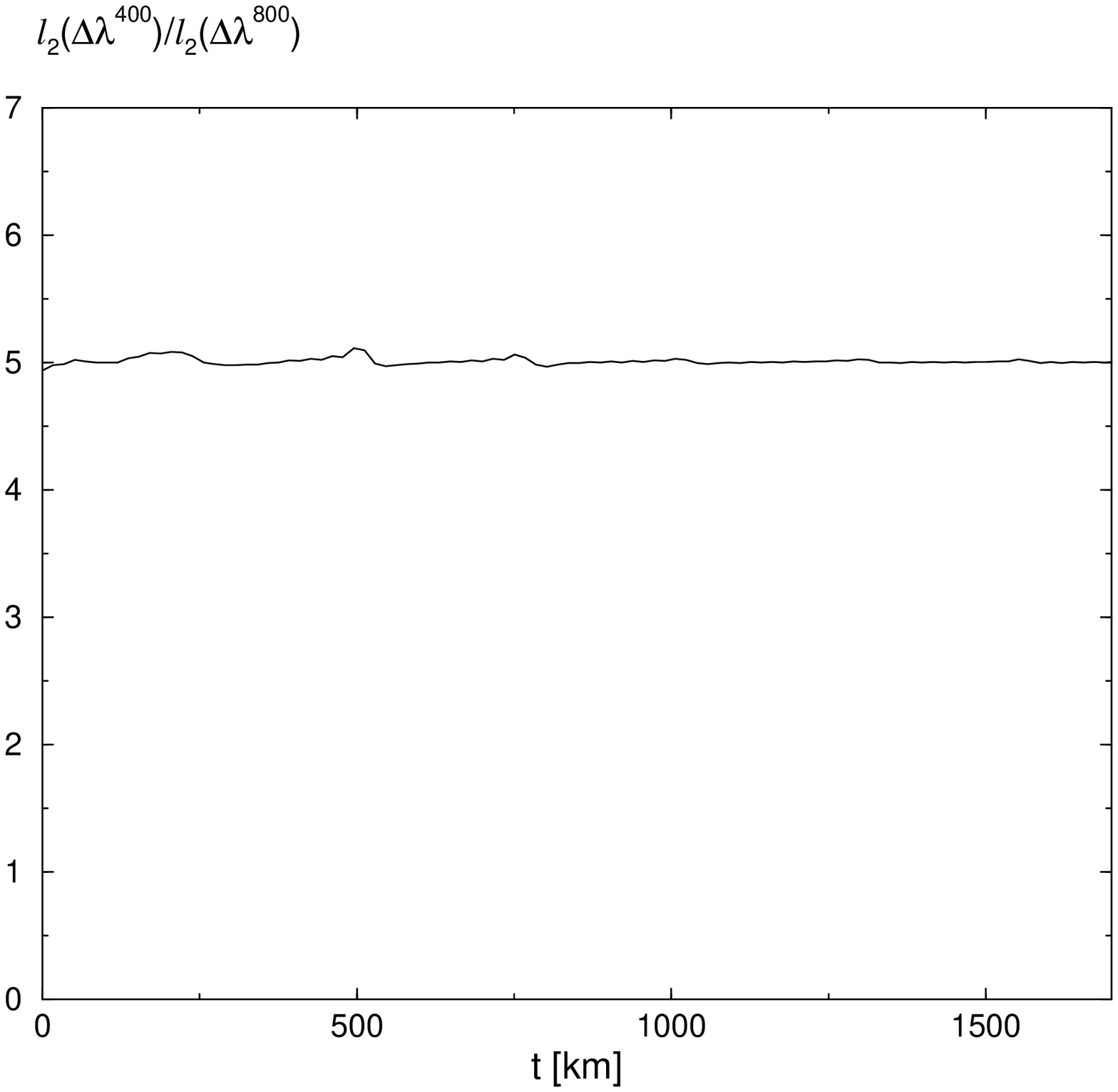, height=180pt, width=130pt, angle=-90}
  \caption{The convergence factor for $\xi$, $w$, $\Delta N$, $\Delta \rho$
           and $\Delta \lambda$ obtained for 400 and 800 grid points.
           The reference solution has been calculated for 1600
           grid points.}
  \label{LAGR_CONV}
\end{figure}
we have performed a convergence
analysis for an initial displacement with the profile of the second
eigenmode and an amplitude of about $50\,\,{\rm m}$ for the stellar model
with $\gamma=3$ and $K=2.0\cdot 10^5\,\,{\rm km}^{-2}$. In this amplitude range
non-linear effects are present, but shock formation is not yet expected
for initial data with sufficiently weak spatial variation. We have
evolved these initial data using 400, 800 and 1600 grid points and
have calculated the time dependent convergence factor according to the
method described in section \ref{CCM_CONVERGENCE}. Since the exact
solution is not known, we use the reference solution for 1600 grid points
in its place. The result obtained for the
variables $\xi$, $w$, $\Delta N$, $\Delta \rho$ and $\Delta \lambda$
is shown in Fig.\,\ref{LAGR_CONV} and
demonstrates second order convergence throughout the evolution. \\
Finally we have tested the code with the analytic solution
by \lcite{Oppenheimer1939b}
which describes the collapse of a homogeneous spherically symmetric
dust cloud. \scite{Petrich1986} have expressed this analytic solution in
polar slicing combined with radial or isotropic gauge. Even though
we are using a Lagrangian gauge condition here, we can use their
results for a comparison with our numerical simulation. \\
%In their calculation Petrich et al. consider the dust collapse from the
%point of view of the closed Friedmann metric
%%
%\begin{align}
%  ds^2 &= -d\tau^2 + a(\tau)^2 (d\chi^2 + \sin^2{\chi} d\Omega^2).
%\end{align}
%%
%Furthermore they use the conformal time $\eta$ 
%which is related to $\tau$ via
%%
%\begin{align}
%  \tau &= (\sin{\chi_{\rm s}})^{-3} (\eta - \sin{\eta}).
%\end{align}
%%
%Here $\chi_{\rm s}$ is the value of the coordinate $\chi$ at the surface
%of the dust cloud. The time parameter $\eta$ varies between $-\pi$ and $0$
%as the dust sphere collapses from initial radius to $r=0$. On given
%time slice $t=\mathrm{const}$, where $t$ is the time coordinate defined
%by polar slicing $\eta$ is given as a function of $\chi$ by
%
%
In their calculation of the analytic solution \shortciteANP{Petrich1986} use
a Lagrangian coordinate $\chi$ and a time parameter $\eta$ which
varies from $-\pi$ to $0$ as the dust sphere collapses from initial radius to
$\hat{r}=0$. On a given time slice $t=\mathrm{const}$, where $t$ is the time
coordinate defined by the polar slicing condition, $\eta$ is given as a 
function of $\chi$ by
\begin{align}
  \cos{\frac{\eta}{2}} &= \cos{\frac{\eta_{\rm s}}{2}}
      \sqrt{\frac{\cos{\chi_{\rm s}}}{\cos{\chi}}},
      \label{OS_ETAOFCHI}
\end{align}
where $\eta_{\rm s}$ and $\chi_{\rm s}$ are the values of 
$\eta$ and $\chi$ at the surface of the dust cloud. If we label
the initial slice by $\eta_{\rm s}=-\pi$, this equation implies that
$\eta=-\pi$ everywhere on the initial slice. 
At any given time $t$ the areal radius is then shown to be related to the
coordinate $\chi$ by
\begin{align}
  \hat{r} &= 2M \frac{\sin{\chi}}{\sin^3{\chi_{\rm s}}} \left( 1- \cos^2
       \frac{\eta_{\rm s}}{2} \cdot \frac{\cos{\chi_{\rm s}}}{\cos{\chi}}
       \right), \label{OS_ROFCHI}
\end{align}
where $M$ is the Schwarzschild mass of the dust cloud.
If we consider the special case of this equation at the surface and on
the initial slice we can calculate $\chi_{\rm s}$ from
\begin{align}
  \sin^2{\chi_{\rm s}} &= \frac{2M}{R},
\end{align}
where $R$ is the initial radius of the dust sphere. For reasons that will
be given below we will identify the radial coordinate $x$ with the
areal radius of the initial location of the fluid elements. We
can therefore set $\eta_{\rm s}=-\pi$ and $\hat{r}=x$ in Eq.\,(\ref{OS_ROFCHI})
and use the result to calculate $\chi(x)$ on the initial slice.
Since
both coordinates are comoving with the fluid elements, this relation
between $\chi$ and $x$
remains valid at any time $t$. In order to calculate $\eta(x)$ at a given time
$t$ we still need to find the value $\eta_{\rm s}$. This is done by
inverting the relation
\begin{align}
  t &= M \frac{\cos{\chi_{\rm s}}}{\sin^3{\chi_{\rm s}}} \left\{
         (\eta_{\rm s} - \sin{\eta_{\rm s}}) + 2 \sin^2{\chi_{\rm s}}
         \left[ \eta_{\rm s} - 2\tan{\chi_{\rm s}} \artanh{\left(
         \tan{\chi_{\rm s}} \cot{\frac{\eta_{\rm s}}{2}}\right)} \right]
         \right\},
\end{align}
for which we use a Newton-Raphson method. Once $\eta_{\rm s}$ has been 
calculated, we can use Eq.\,(\ref{OS_ETAOFCHI}) to calculate $\eta(x)$
on that time slice. The physical variables $\hat{r}$, $\hat{\rho}$,
$\hat{\Gamma}$ and $\hat{\lambda}$ then follow from Eq.\,(\ref{OS_ROFCHI})
and further relations by \shortciteANP{Petrich1986} which we write in the form
\begin{align}
  \hat{\rho} &= 6 \frac{a_0}{a^3} \frac{1}{8\pi M^2}, \\[10pt]
  \hat{\Gamma} &= \frac{\cos^3{\chi}-\cos^2{\frac{\eta_{\rm c}}{2}}}
       {\cos{\chi}-\cos^2{\frac{\eta_{\rm c}}{2}}}, \\[10pt]
  \hat{\lambda} &= -\frac{\hat{\lambda}_{\rm c}}{\sin{\frac{\eta_{\rm c}}{2}}}
       \frac{\cos{\chi}-\cos^2{\frac{\eta_{\rm c}}{2}}}
       {\sqrt{\cos^3{\chi}-\cos^2{\frac{\eta_{\rm c}}{2}}}},
\end{align}
where $\lambda_{\rm c}$ is the central value of the lapse function
and $a_0$ and $a$ are given by
\begin{align}
  a_0 &= \frac{1}{\sin^3{\chi_{\rm s}}}, \\[10pt]
  a &= a_0 (1-\cos{\eta}).
\end{align}
In practice we specify the initial energy density and radius 
of the dust sphere and set the velocity to zero.
The functions $\hat{N}$ and $\hat{\lambda}$ are
then calculated from the constraint equations and the total mass
of the sphere follows from the definition (\ref{NLAGR_NOFM}). \\
From the numerical point of view the dust collapse is a special
\begin{figure}[t]
  \centering
  \epsfig{file=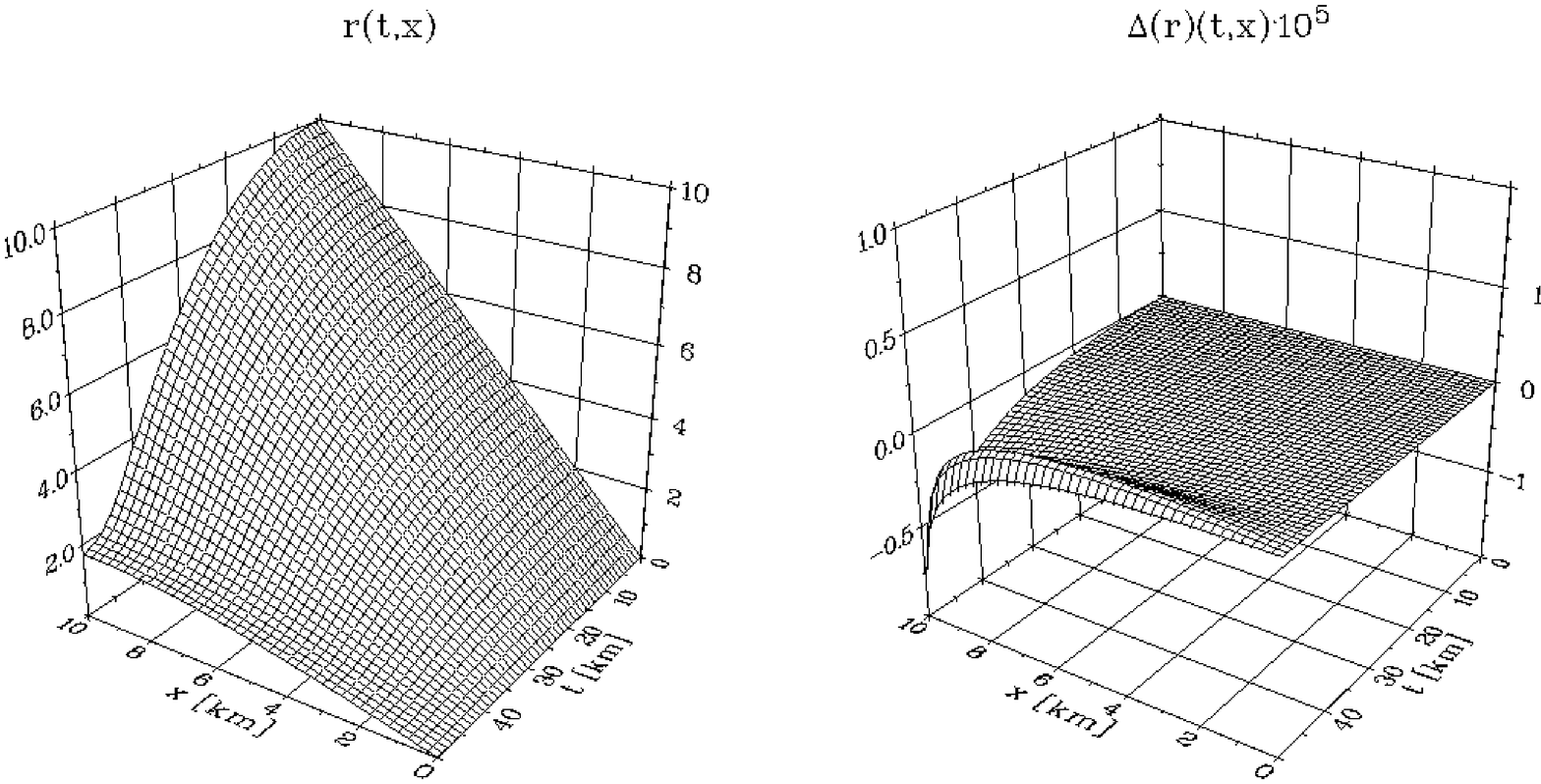, height=400pt, width=175pt, angle=-90}
  \epsfig{file=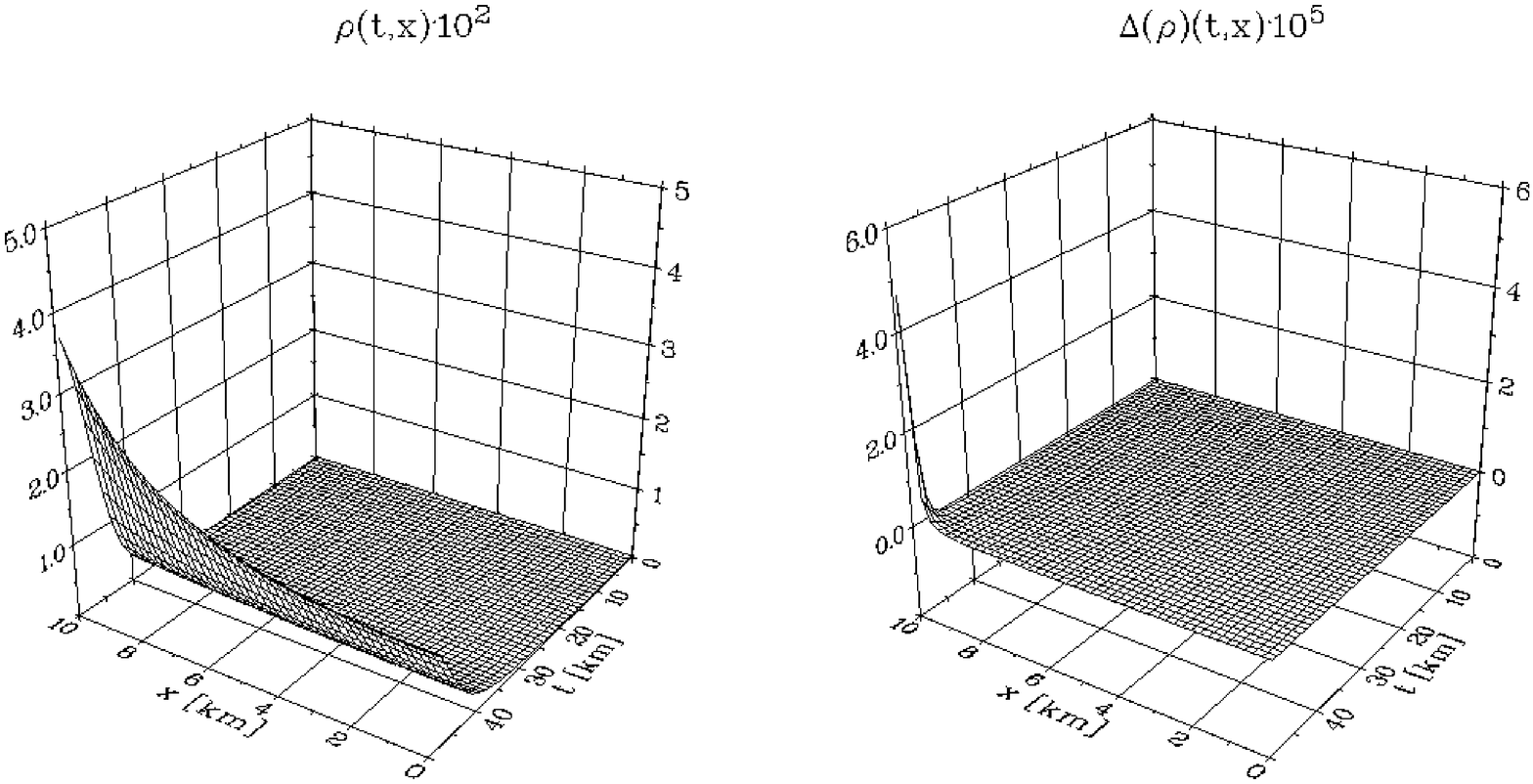, height=400pt, width=175pt, angle=-90}
  \caption{The numerical simulation of the Oppenheimer-Snyder
           dust collapse for a dust sphere of $10\,\,{\rm km}$ radius
           and initial density $2\cdot 10^{-4}\,\,{\rm km}^{-2}$.
           The left panels show the numerical results for the
           radius $\hat{r}$ and the energy density $\hat{\rho}$, the right
           panels the deviation from the analytic solution.}
  \label{OS_RRHO}
\end{figure}
\begin{figure}[t]
  \centering
  \epsfig{file=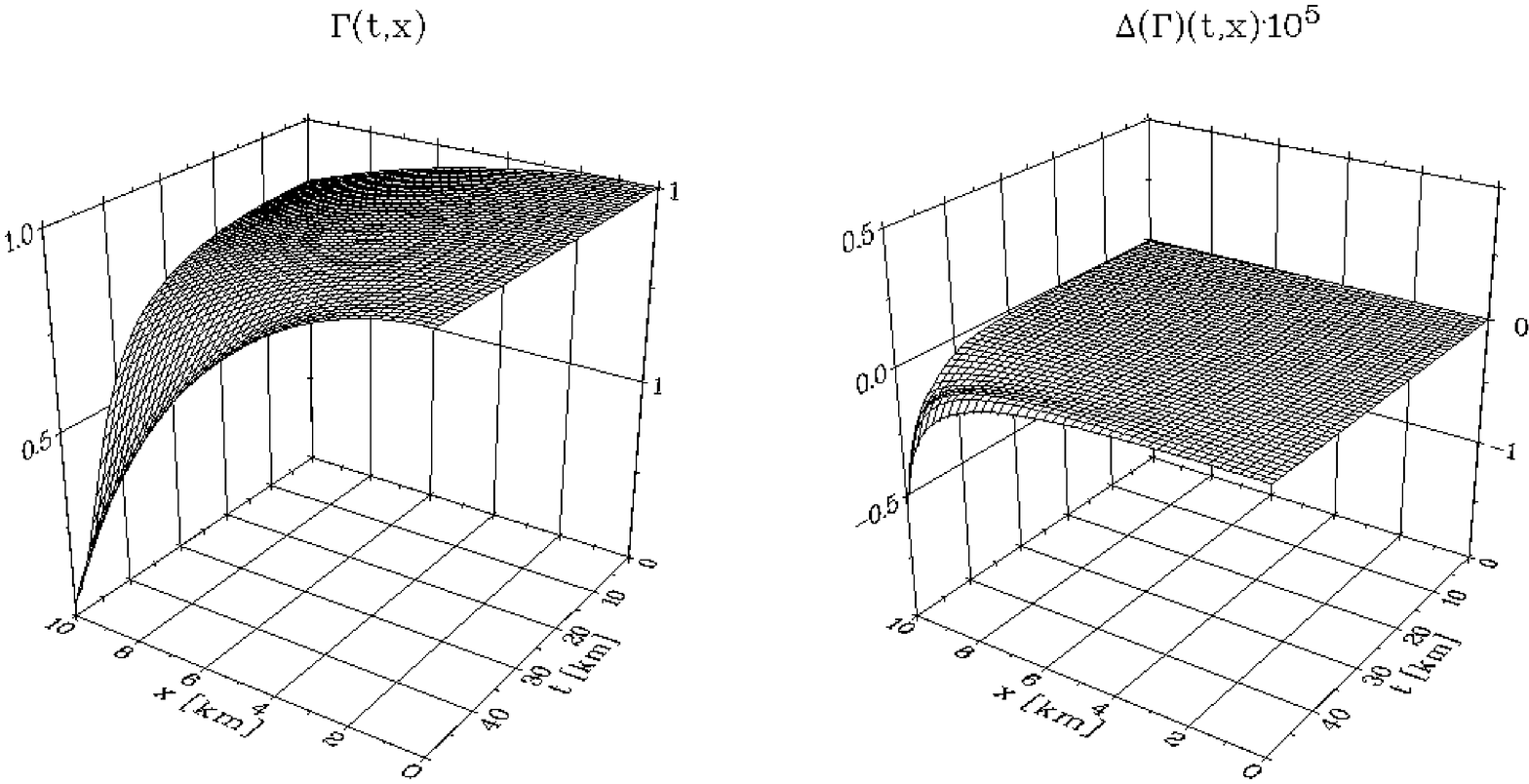, height=400pt, width=175pt, angle=-90}
  \epsfig{file=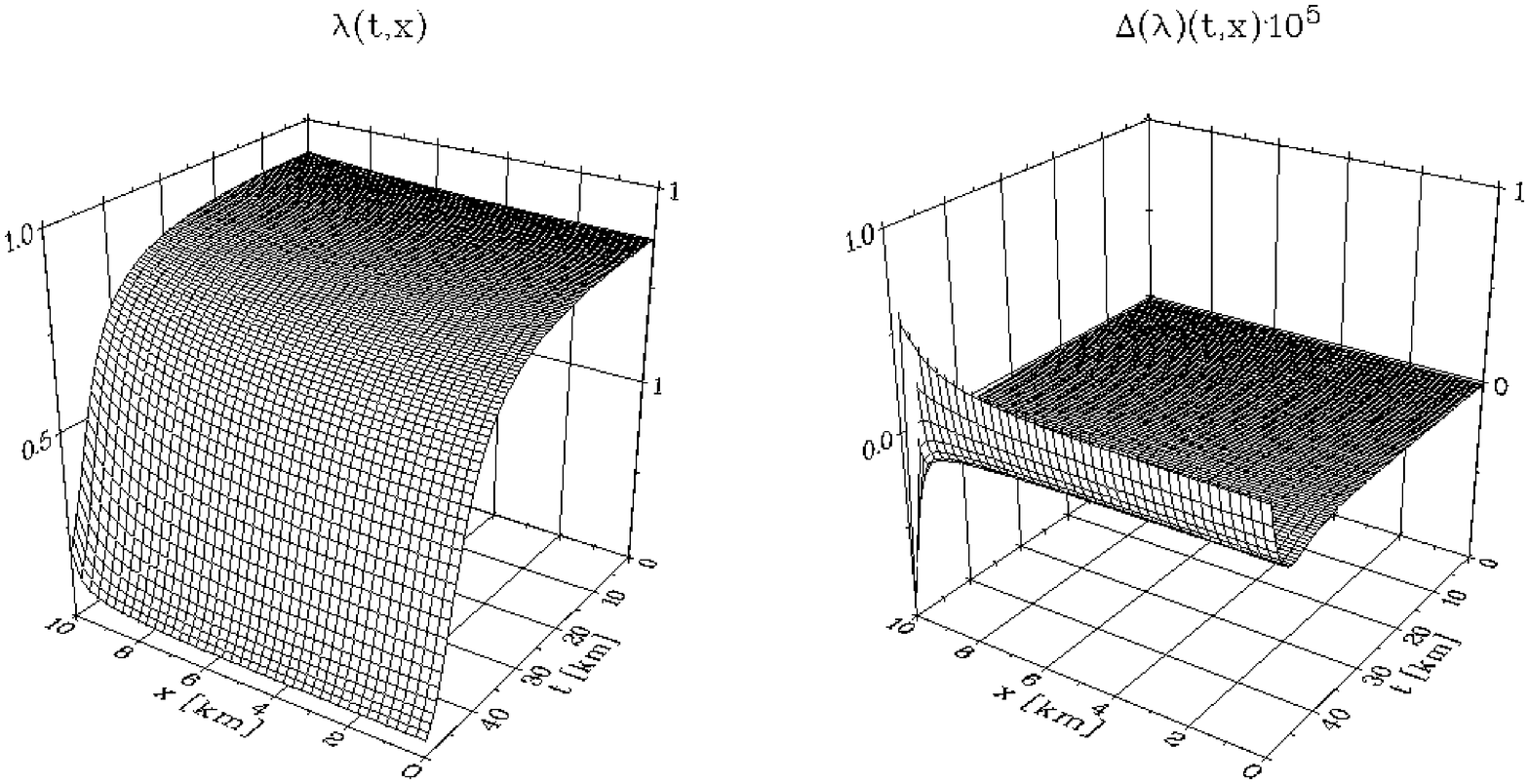, height=400pt, width=175pt, angle=-90}
  \caption{Same as Fig.\,\ref{OS_RRHO} for the metric variables
           $\hat{\Gamma}$ and $\hat{\lambda}$.}
  \label{OS_GAMLAM}
\end{figure}
case in several aspects which restricts our choice of the
available options of the code.
\begin{list}{\rm{(\arabic{count})}}{\usecounter{count}
             \labelwidth1cm \leftmargin1.5cm \labelsep0.4cm \rightmargin1cm
             \parsep0.5ex plus0.2ex minus0.1ex \itemsep0ex plus0.2ex}
\item By definition the pressure vanishes in the dust sphere. As
      a result there is no static configuration analogous to the
      static neutron star governed by the TOV equations. We therefore
      need to use vacuum flat space as the background and run the
      code in the non-perturbative mode.
\item The vanishing of the pressure also implies that the speed of sound
      is zero throughout the dust sphere so that it cannot be used
      to rescale the radial coordinate according to Eq.\,(\ref{TOV_YOFR}).
      The radial coordinate $x$ is therefore
      defined by the areal radius of the initial positions of the
      fluid elements and we use the condition $r_{,x}=1$ in the code.
\item The surface of a neutron star with a polytropic equation of
      state is defined by the vanishing of the energy density $\hat{\rho}$
      which provided the outer boundary condition in the numerical evolution.
      For the dust sphere this relation is not valid any more and the
      energy density is finite at the outer boundary. The exact value,
      however, is not known, so that we cannot use it to derive an
      alternative boundary condition. The boundary condition $\hat{P}=0$
      is trivially satisfied in the case of a dust sphere and does not
      provide any extra information either. If we consider the
      structure of equations (\ref{NLAGR_LAMBDAX})-(\ref{NLAGR_UT}),
      however, we can see that all spatial derivatives of the energy density
      appear in the form of pressure gradients. These terms are
      identically zero in this case and disappear from the equations.
      We can therefore use the staggered grid for the energy density
      and thus eliminate the need of a boundary condition for $\hat{\rho}$.
      For this purpose we set the parameter $\sigma$ to 1 in the evolution
      of the dust sphere.
\end{list}
In Figs.\,\ref{OS_RRHO} and \ref{OS_GAMLAM}
we show the results obtained for a dust sphere with initial 
density $\hat{\rho}_0 = 2\cdot10^{-4}\,\,{\rm km}^{-2}$ and radius
$R_0=10\,\,{\rm km}$ which corresponds to a total mass of 
$M=0.838\,\,{\rm km}$.
A grid resolution of 800 points has been used for this calculation.
The results demonstrate the good accuracy with which the code
reproduces the analytic solution. Near the surface of the dust sphere,
however, the numerical error increases significantly as the
sphere approaches its Schwarzschild radius. We attribute this behaviour
to the steep gradient of the lapse function near the surface that
arises in the late stages of the evolution. \\
This simulation also illustrates the singularity avoiding properties of the
polar slicing condition. As the dust sphere collapses towards its Schwarzschild
radius, the lapse function decreases towards zero and the evolution is
practically frozen. This effect, the so called {\em collapse of the lapse},
is responsible for the apparent slow down of the collapse of the
radial function $r$ that can be seen in the upper left panel of
Fig.\,\ref{OS_RRHO}. It is this property that makes polar
slicing a popular choice for
the numerical analysis of 1-dimensional gravitational collapse.

%=======================================================================
\subsection{Do shocks form at the surface for low amplitude oscillations?}
We will now address a question that implicitly arose in the discussion
of the linearized equations in the Eulerian formulation. We have seen
in Eq.\,(\ref{DRHOOVERRHO})
that the linearized equations predict a diverging ratio $\delta \rho /\rho$
at the surface. For polytropic indices $\gamma>2$ we know that the
divergence of $\delta \rho$ is a result of the Taylor expansion used
to relate the Eulerian energy density perturbation to the Lagrangian one
in Eq.\,(\ref{LIN_DRHOOFZETA}) and thus a non-physical result. For
polytropic exponents $\gamma\le 2$, however, Eq.\,(\ref{LIN_DRHOOFZETA})
represents a valid relation to first order in the perturbations,
so that the Eulerian density perturbation
will indeed be large compared with the background value near the surface.
This behaviour raises the question whether non-linear effects will
affect the evolution near the surface and give rise to
the formation of discontinuities. From a different point of view one may
consider the speed of sound which vanishes at the surface for a
polytropic exponent $\gamma > 1$ and the particle speed $w$ which
is finite because of the movement of the stellar surface. Consequently the
velocity of the fluid elements will exceed the speed of sound and
one may again ask whether this leads to shock formation.
We will investigate this by using the exact treatment of the
surface provided by the Lagrangian code. \\
For this purpose we consider the neutron star model 3 of Table \ref{MODELS15}
and provide initial data in the form of a displacement $\xi$ corresponding
to a single eigenmode. For reasonably low order eigenmodes and amplitudes up to
several metres we have not observed any significant deviation from
the expected harmonic time dependence. For eigenmodes of very high order,
however, this picture changes. We illustrate this in the case of an
initial displacement of the fluid elements corresponding to
a high order eigenmode (about 50) and an amplitude of about $1\,{\rm m}$
at the surface. The high resolution of 3200 grid points has been used for
this calculation to adequately resolve the high order mode. We stress that
this evolution is only possible because of the high resolution near the
surface provided by the rescaled variable $y$.
\begin{figure}[t]
  \centering
  \epsfig{file=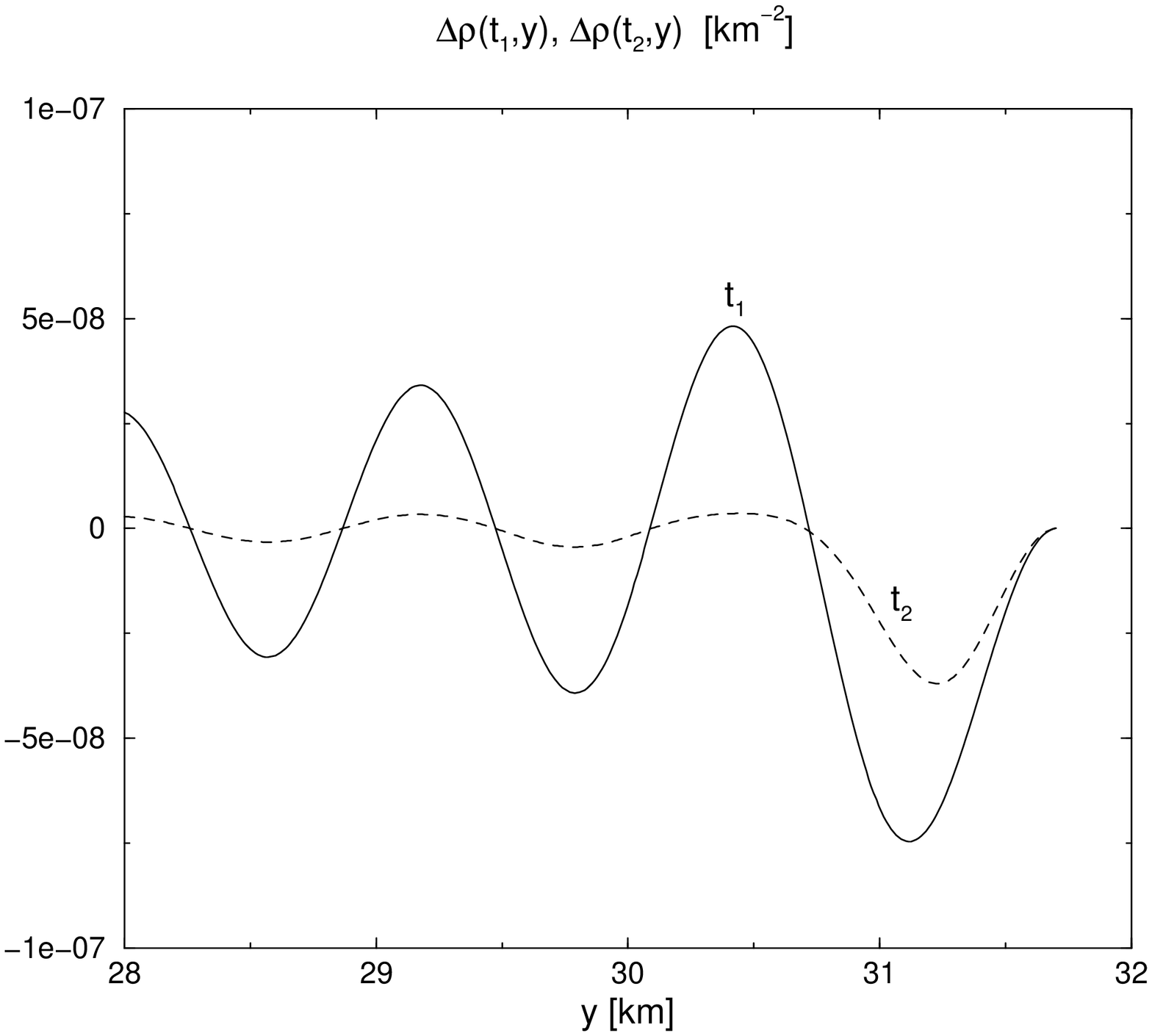, height=200pt, width=150pt, angle=-90}
  \epsfig{file=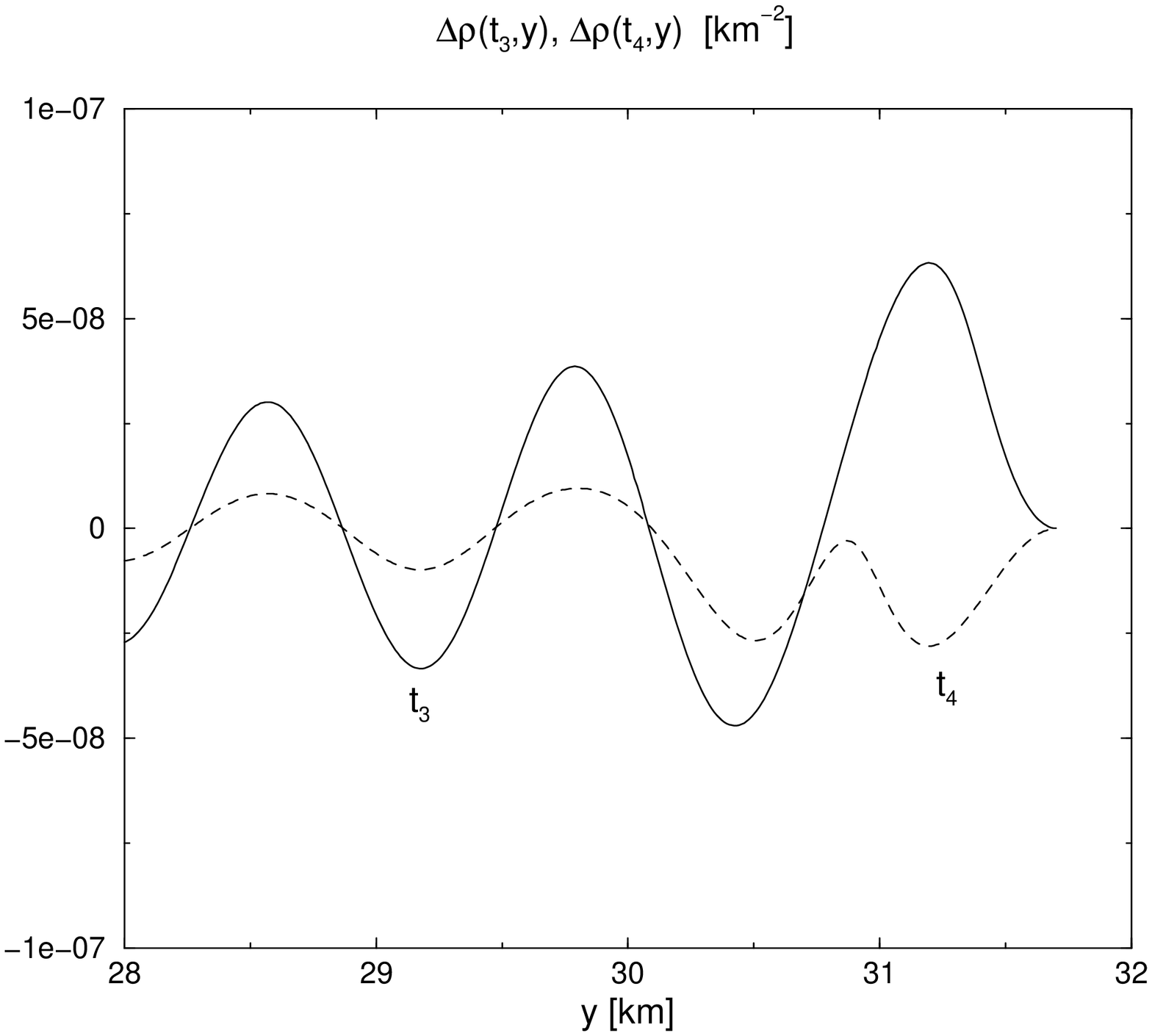, height=200pt, width=150pt, angle=-90}
  \epsfig{file=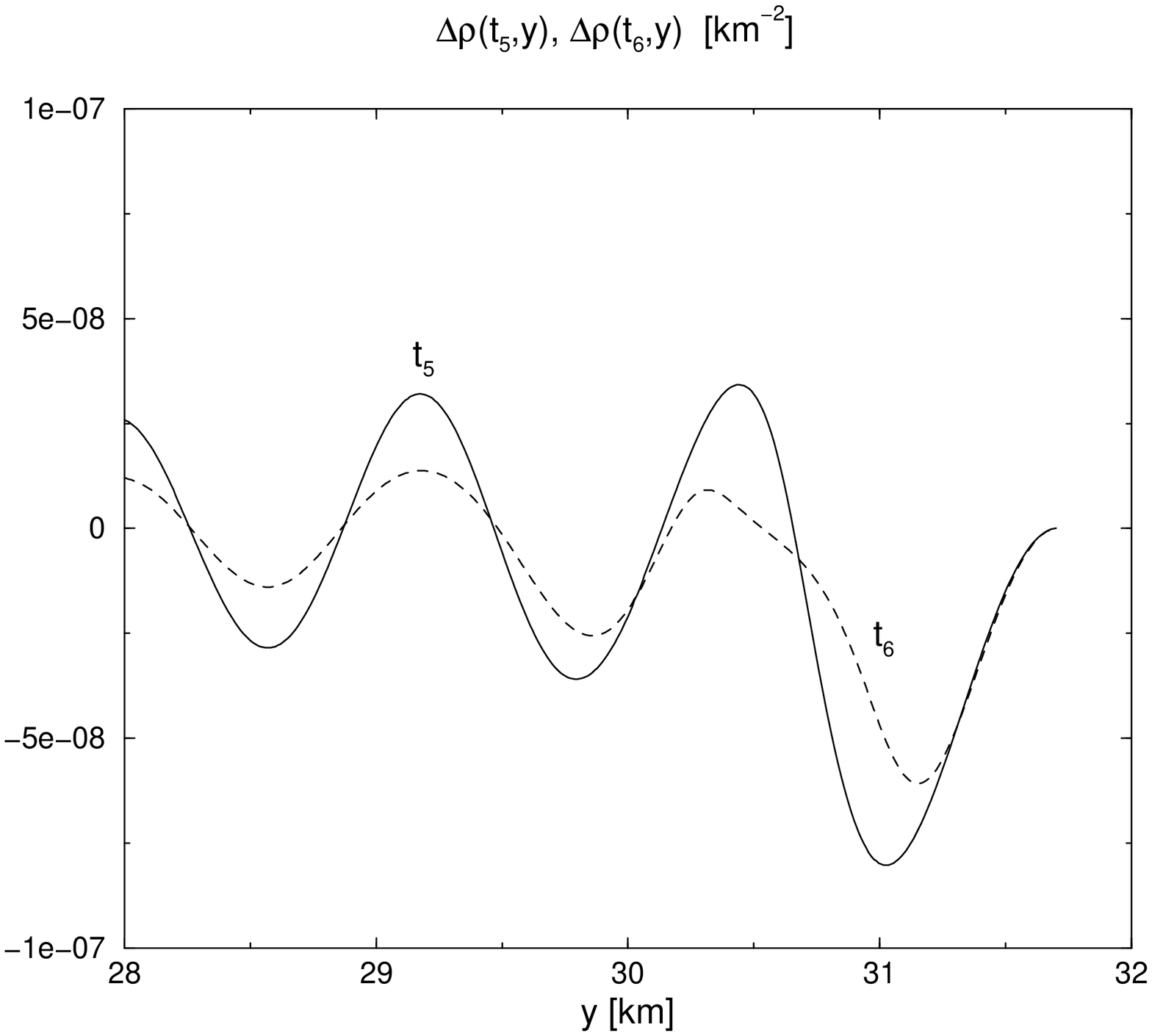, height=200pt, width=150pt, angle=-90}
  \epsfig{file=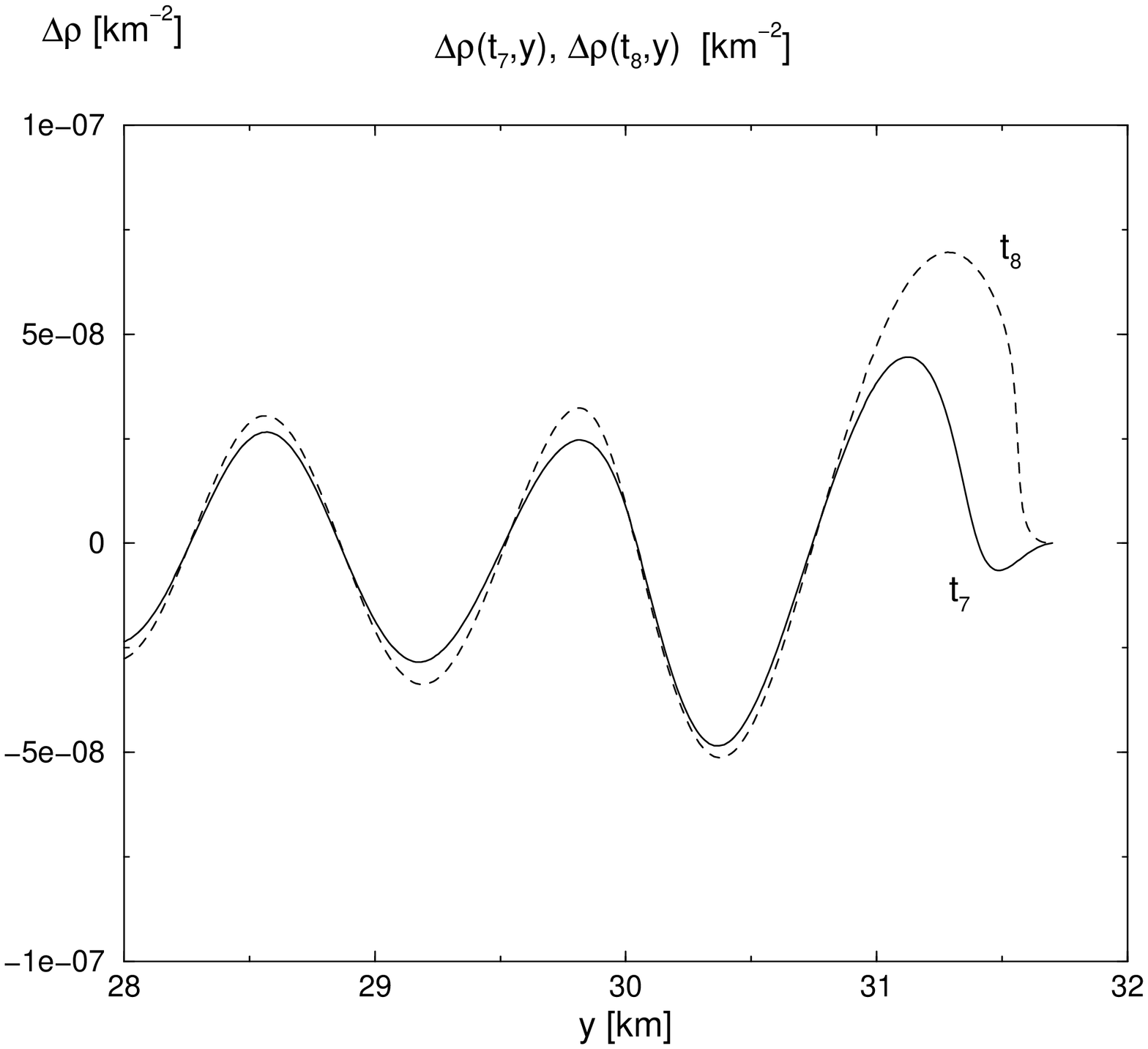, height=200pt, width=150pt, angle=-90}
  \caption{The numerical evolution of the energy density
           perturbation $\Delta \rho$ as a function of $y$
           obtained for an initial displacement corresponding to about the
           50th eigenmode with amplitude $1\,{\rm m}$.
           Snapshots are shown
           at times $t_1\ldots,t_8$.}
  \label{LAGR_SHOCKS}
\end{figure}
In Fig.\,\ref{LAGR_SHOCKS} we show snapshots of the time evolution of
the energy density perturbation at times $t_1=0.0$, $t_2=0.5$,
$t_3=1.0$, $t_4=1.5$, $t_5=2.0$, $t_6=2.5$, $t_7=3.0$ and
$t_8=3.1\,{\rm km}$. We note that only the small radial
range $28\,{\rm km} \le y \le 31.7\,{\rm km}$ is shown in the figure.
In terms of the areal radius this corresponds to a range of about
$120\,{\rm m}$ below the surface. We can see that for this small
amplitude a steep gradient forms near the surface
after about $t=3.1\,{\rm km}$ which
corresponds to less than two oscillation periods of the eigenmode.
This indicates the formation of a discontinuity. At later times than shown
here the code fails to converge which we attribute to the numerical
noise caused by the shock formation and the extreme sensitivity of the
code near the surface of the star.
In order to demonstrate that this result is not merely due to
numerical inaccuracies, we have evolved the same initial data with
the smaller amplitude of $1\,{\rm cm}$. In Fig.\,\ref{LAGR_NOSHOCKS}
we show the same snap shots for this evolution
as in Fig.\,\ref{LAGR_SHOCKS}. In this case we
obtain harmonic time dependence as expected in the linear limit.
\begin{figure}[t]
  \centering
  \epsfig{file=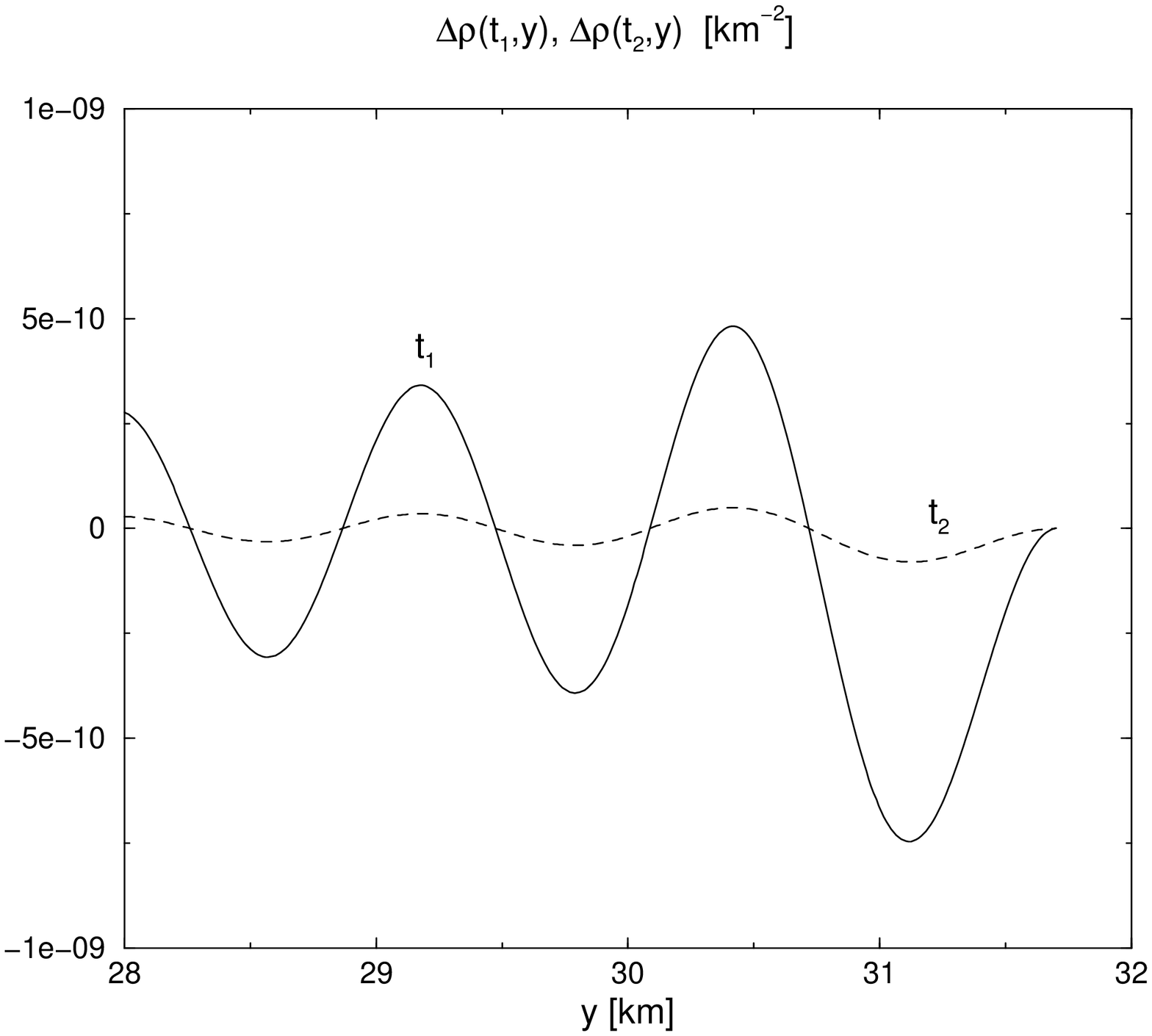, height=200pt, width=150pt, angle=-90}
  \epsfig{file=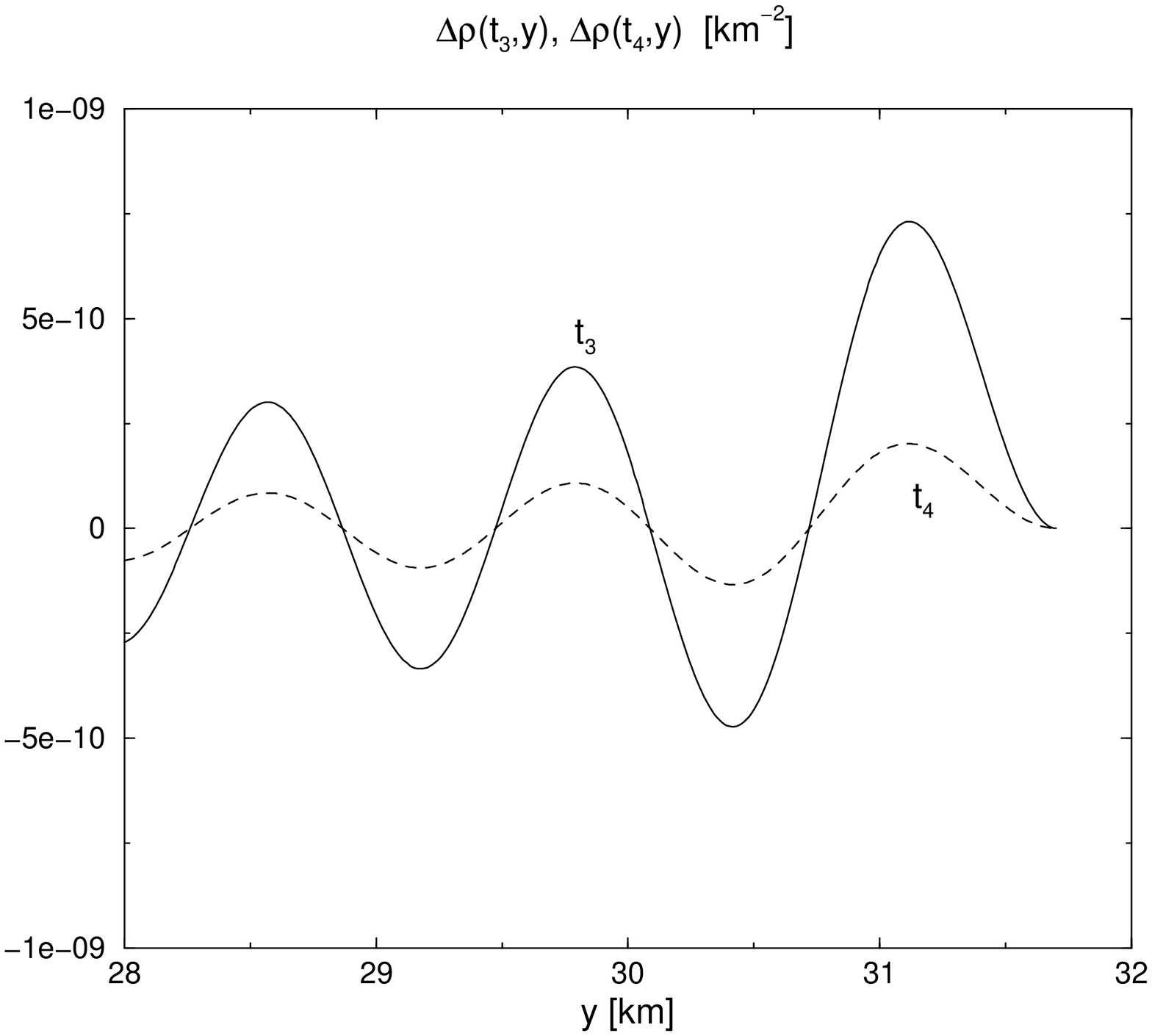, height=200pt, width=150pt, angle=-90}
  \epsfig{file=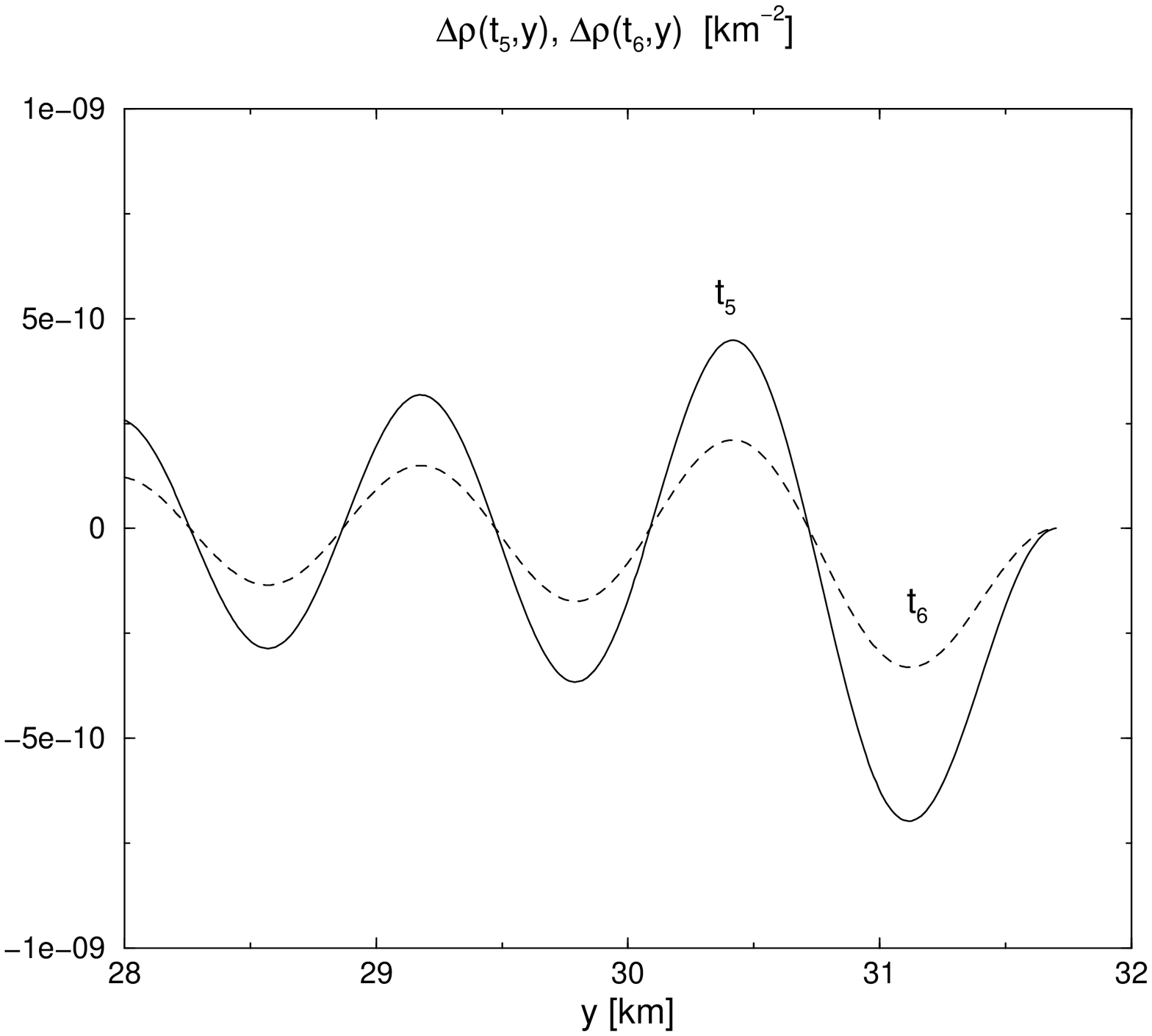, height=200pt, width=150pt, angle=-90}
  \epsfig{file=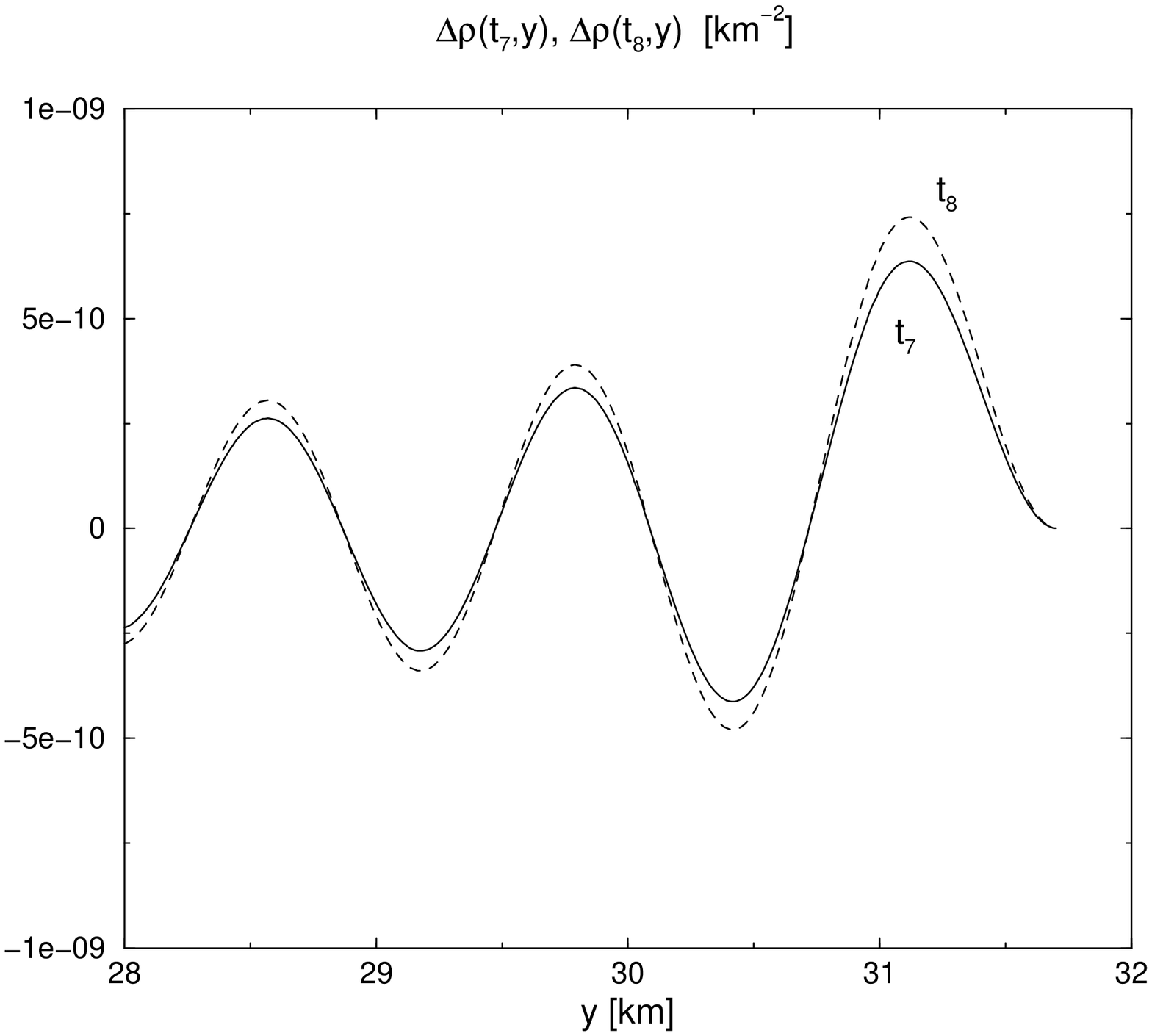, height=200pt, width=150pt, angle=-90}
  \caption{The same as Fig.\,\ref{LAGR_SHOCKS} but for an amplitude of
           $1\,{\rm cm}$.}
  \label{LAGR_NOSHOCKS}
\end{figure}
By using eigenmodes with even higher order we have observed shock formation
at the surface for smaller amplitudes.
% In the case of initial
%displacements in the form of lower order modes
%we can go up to amplitudes significantly larger than those used here
%without any indication of discontinuities.
In view of the results for low order modes where no significant non-linear
effects are observed for similar amplitudes,
we conclude that the magnitude
of non-linear effects is not only determined by the size of the perturbations
relative to the background variables, but also by the length scale on which
the perturbations vary significantly. We finally note that the surface of
a neutron star is too complicated to be accurately described by
the polytropic equation of state used for these evolutions. It is not clear
whether discontinuities will form in the same way for more realistic
descriptions of neutron stars. Nevertheless our results demonstrate that
the surface requires a careful numerical treatment.

\newpage
%=========================================================================
\section{Conclusions}
In this work we have presented the application of different numerical
techniques to solve Einstein's field equations. We have laid the foundation
for our discussion by describing in detail the ADM ``3+1'' and the
characteristic Bondi-Sachs formulation of the field equations together
with various aspects of numerical analysis. \\
In the framework of ``3+1'' formulations of the Einstein field equations the
restriction to a finite grid in numerical computations results in
difficulties concerning the specification of outer boundary conditions and
the interpretation of gravitational waves. These problems are naturally
resolved in a characteristic formulation, but the foliation of spacetime based
on the characteristic surfaces may break down in regions of strong curvature
due to the formation of caustics.
The combination of the two schemes in
the framework of Cauchy-characteristic matching enables one to make use of
the advantages of both methods while avoiding the respective drawbacks.
In this work
we have completed the cylindrically symmetric stage of the Southampton
Cauchy-characteristic matching project by providing a new long term stable
CCM code including
both gravitational degrees of freedom. A Geroch decomposition of
the 4-dimensional spacetime allows us to reformulate the problem in terms
of the norm of the axial Killing vector $\nu$ and the Geroch potential $\tau$
on an asymptotically flat 3-dimensional
quotient spacetime. These geometrical fields describe the gravitational
degrees of freedom in simple terms and appear to be a natural choice of
variables for the description of a cylindrically symmetric vacuum spacetime.
The conformal compactification of the resulting
3-dimensional spacetime allows us to impose exact boundary conditions at
null infinity. In contrast to the previous work we have also
applied the Geroch decomposition to the interior Cauchy region and thus
been able to use the same variables throughout the numerical grid. This
leads to a substantial simplification of the interface and the
evolution equations and facilitates a
long term stable evolution with both gravitational degrees of freedom
present. The effectiveness of the code has been demonstrated by reproducing
the analytic Weber-Wheeler solution and the vacuum spacetime
with two degrees of freedom due to Xanthopoulos. The code has been shown
to be second order convergent over the dynamically relevant time intervals. 
Our results demonstrate the importance of a ``good'' choice of variables
in order to obtain a stable, accurate code even in the 1-dimensional
case. For higher dimensional problems the structure of the null-geodesics
will be much more complicated because of the angular dependence.
As a consequence the transformation between the Cauchy and the
characteristic variables at the interface will also be more complicated
and thus more vulnerable to instabilities. In view of our results
it seems preferable to search for natural variables, such as
the Geroch variables in the cylindrically symmetric case, to
describe the two regions rather than follow the ``brute force''
calculations which arise for example from a direct application of the
ADM-formulation in the Cauchy region. \\

Next we have derived a characteristic formulation of
the equations governing a dynamic
cosmic string in cylindrical symmetry. A feature of
the cosmic string equations is that they admit exponentially diverging
unphysical solutions. By using the Geroch decomposition it is again possible
to reformulate the problem in terms of fields which describe the
string on an asymptotically flat $2+1$-dimensional spacetime and
the two auxiliary fields $\nu$ and $\tau$. As well as avoiding the need
to introduce artificial outgoing radiation boundary conditions
the inclusion of null infinity as part of the numerical grid
has the advantage that we can enforce outer boundary
conditions for the string variables which rule out
the unphysical solutions. As special cases of the dynamic equations
we also obtain the equations
for a static cosmic string in curved or Minkowski spacetime. These
sets of equations have been solved by using a relaxation scheme
in the static cases and an implicit method for the dynamic scenario. \\
A convergence analysis for all codes demonstrates clear second order
convergence. The dynamic code has also
been shown to reproduce the results of the two exact vacuum solutions
by Weber \& Wheeler and Xanthopoulos.
Finally the dynamic code
reproduces the results for the static cosmic string in that initial
data corresponding to a static solution do not change significantly when
evolved in time. For both the exact vacuum
solutions and the static initial data the code shows excellent long
term stability. \\
After demonstrating the reliability of the code we have used it to analyse
the interaction between an initially static cosmic string and a
Weber-Wheeler type pulse of gravitational radiation. We have found that the
gravitational wave excites the string and causes the string variables
$X$ and $P$ to oscillate. In terms of unphysical rescaled
variables we find that the frequencies of the oscillations are essentially
independent of the strength of the coupling between string and
gravity described by $\eta$ and of the
width and amplitude of the Weber-Wheeler pulse. We have also found that the
frequency of $X$ is independent of the relative coupling constant $\alpha$
while that of $P$ is
proportional to $\sqrt \alpha$. When this result is translated back
into the physical units we find that the frequency of the scalar field
is proportional to the mass of the scalar field and the
frequency of the vector field is proportional to the mass of
the vector field as predicted by the linearized theory.
This result is confirmed by investigating two
further scenarios. Firstly we consider the evolution of static initial
data for the string coupled to the gravitational field, but with a
Gaussian perturbation to one of the string variables, and secondly we
consider the same scenario but in a Minkowskian background with the
gravitational field decoupled. In both cases we obtain the same
relationship between the frequencies and the mass. \\
An interesting numerical result arising from the use of an implicit
numerical scheme concerns the structure of the interface between the
interior and the compactified outer region. In contrast to the rather
complicated interpolation techniques that were necessary to transform
between the Cauchy and characteristic variables in the explicit
vacuum CCM-code, we have been able to ``localise''
the interface in the implicit scheme by using two grid points for
the spatial position $r=1$, one containing the variables of the inner
region, one containing those used in the outer region. The interface
then merely consists in relating these variables and their
derivatives by using their definitions and applying the chain-rule.
In our case the resulting relations were trivial and could easily
be incorporated into the main evolution algorithm. We attribute this
substantial simplification to the simultaneous calculation in implicit
schemes of all function values on the new time-slice. In explicit schemes,
on the other hand the calculation of the new function values is normally
subject to a certain hierarchical order. \\

In the final part of this work we have presented a new numerical
approach which enables us to evolve radial oscillations
of neutron stars over a large amplitude range with high accuracy. In radial
gauge and polar slicing the dynamic star is described by two constraint
equations for the metric and a quasi-linear system of two evolution equations
for the matter variables. The crucial step in our approach is to decompose
the dynamic variables into static background contributions which are determined
by the Tolman-Oppenheimer-Volkoff equations and time dependent perturbations.
We have used this decomposition to rewrite the system
of equations in a perturbative
form. We do, however, keep all terms of higher order in the perturbations
and thus obtain a formulation equivalent to the original set of equations.
The motivation for our approach is given by the fact that background terms
(terms of zero order) are in general present in the dynamic equations.
These terms cancel each other analytically by virtue of the background
equations. Numerically, however, this is generally satisfied up to a residual
numerical error only which will constitute a spurious source term in the
evolution of the perturbations and contaminate the numerical results. In
order to avoid this effect, we use the background equations to remove all
zero order terms from the perturbative equations. We thus ensure that
the numerical accuracy is
determined by the perturbations instead of the static background. \\
We have compared the resulting perturbative code with a ``standard''
non-perturbative method by evolving the fundamental eigenmode of a dynamically
stable neutron star using an amplitude of several metres.  Whereas the
perturbative scheme reproduces the expected harmonic oscillations with high
accuracy, the non-perturbative scheme leads to an exponential decay of the
central energy density perturbation after a few oscillations which we
attribute to the numerical contamination caused by the background terms. \\
Even though the perturbative code performs well in the linear regime
for a wide variety of neutron star models, we have observed a
spurious exponential growth of the physical variables in the evolution
of marginally stable neutron star models if we truncate the neutron star
at a sufficiently large density and thus omit the outer low density
layers from the numerical evolution. The need to truncate the neutron
star at finite densities arises from the occurrence of negative energy
densities near the surface of the star due to numerical inaccuracies.
In a purely Eulerian formulation the outer grid boundary does
not coincide with the surface of the star in a non-linear evolution.
When the star shrinks inside the numerical grid negative energy densities
will occur because the numerical evolution is not able to accurately
model the vacuum region between the stellar surface and the outer
grid boundary. It is interesting to see that the surface represents
a problematic area even in the comparatively simple linearized case.
For equations of state with an asymptotic behaviour $P \sim \rho^{\gamma}$
and $\gamma > 2$ the Eulerian energy density perturbation
diverges at the surface of the star. We have shown how this problem arises
from the transformation between Lagrangian and Eulerian perturbations and
is not present in a Lagrangian formulation. \\
In order to alleviate the surface problem in the Eulerian case
in a simple manner we have used a fixed
boundary condition by setting the radial velocity $w=0$ at the outer grid
boundary. Furthermore we have truncated the outer layers of the neutron star,
so that the resulting model contains $90\,\%$ of the original mass.
We have thus demonstrated second order convergence
of the code in the non-linear
regime and checked the conservation properties of the code in the Cowling
approximation. We have finally used the simplified neutron star model to study
the coupling between eigenmodes due to non-linear effects. For this
purpose we have provided initial data in the form of an isolated eigenmode and
quantified the excitation of other modes in terms of the inner product,
defined by the self-adjoined eigenvalue problem of the linearized case,
between the non-linear data and the eigenmode solutions.
The high accuracy of the perturbative scheme enables us to
vary the amplitude of the initial data over a wide range
corresponding to a maximum displacement of fluid elements
between several cm and about $50\,{\rm m}$. For significantly
larger amplitudes we observe the formation of steep gradients which makes the
accurate measurement of the eigenmode coefficients problematic. \\
In our study we have provided initial data in the form of either of the
first three eigenmodes in the velocity field
while the energy density
perturbation has been set to zero. We have then measured
the maximum coefficients for
the first 10 or 15 eigenmodes. Our results clearly show the existence of
two distinct regimes. In the weakly non-linear regime with amplitudes
up to several metres all eigenmode coefficients increase quadratically
with the amplitude. If the order of the initially excited mode is $j$,
we have also found that the coupling coefficients in the weakly non-linear
regime decrease with increasing order of the eigenmodes starting with
mode $2j$. This decrease can be approximated well with an inverse cubic
power law. In the moderately non-linear regime we have observed a different
behaviour of the modes. An initially present mode $j$ has been found to couple
more efficiently to the eigenmodes $n\cdot j$, where $n=2$, 3, 4 and so on.
For these modes we can model the resulting eigenmode coefficients with a sum of
a quadratic power law and a power law of index $n$ with good accuracy.
The remaining eigenmode coefficients also show a steeper increase with amplitude than in the weakly non-linear regime, but the power law
contribution with exponent larger than two is generally too small to
facilitate an accurate measurement. \\
Finally we have developed a fully non-linear Lagrangian code for the evolution
of spherically symmetric dynamic neutron stars. We have demonstrated how
the numerical difficulties encountered in the Eulerian case are
resolved in the Lagrangian formulation. The code has been shown to accurately
reproduce the analytic solution of the linearized equations for low
amplitudes and the analytic solution of Oppenheimer and Snyder describing
the collapse of a spherically symmetric homogeneous dust sphere.
We have furthermore demonstrated second order convergence
of the code. The code has been used to investigate non-linear effects
near the stellar surface arising in low amplitude oscillations. Whereas
we do not observe a significant deviation from the linear regime for
low order eigenmodes and amplitudes of several metres, high order
eigenmodes of the order of 50 with amplitudes of $1\,{\rm m}$ lead
to the formation of steep gradients near the surface due to non-linear effects.
We conclude that the magnitude of non-linear effects is not only
determined by the relative size of the perturbations with respect to
the background but also on the length scale on which the perturbations
vary significantly. The high resolution at the surface required for these
evolutions has been obtained by the use of a rescaled radial
coordinate which naturally takes into account the vanishing of the
speed of sound at the surface and facilitates a formulation of the
equations in terms of which the slopes of the characteristics are
by and large independent of the position in the star.

\newpage
\appendix
%=================================================================
\section{The finite differencing of the Lagrangian equations}
\label{LAGR_FDE}
We use an implicit second order in space and time finite differencing scheme
for the numerical evolution of the fully non-linear perturbative
Lagrangian equations (\ref{NLAGR_DLAMBDAX})-(\ref{NLAGR_DPT}).
The parameter $\sigma$ enables us to use the energy density
$\rho$, $\Delta \rho$ on the ``normal'' grid ($\sigma = 0$) or
the staggered grid ($\sigma = 1$). The staggering, however, affects
the energy density only. It is therefore suitable to describe the finite
differencing for a general function $f$, $\Delta f$ and the energy 
density $\rho$, $\Delta \rho$.
The function $f$ always represents the background variables $r$, $N$
and $\lambda$. Similarly $\Delta f$ stands for the perturbations
$\xi$, $w$, $\Delta N$ and $\Delta \lambda$. \\
In that notation Eqs.\,(\ref{NLAGR_DLAMBDAX}) and (\ref{NLAGR_DNX}) 
are converted into finite differences by using
\begin{align}
  f &= \frac{1}{2} \left( f_k + f_{k+1} \right), \\[10pt]
  \rho &= \frac{1}{2} \left[ (1+\sigma) \rho_k + (1-\sigma) \rho_{k+1} \right],
          \\[10pt]
  \Delta f &= \frac{1}{2} \left( \Delta f_k^{n+1} + \Delta
          f_{k+1}^{n+1} \right), \\[10pt]
  \Delta \rho &= \frac{1}{2} \left[ (1+\sigma) \Delta \rho_k^{n+1}
          +(1-\sigma) \Delta \rho_{k+1}^{n+1} \right], \\[10pt]
  \Delta f_{,x} &= \frac{1}{\Delta x} \left( \Delta f_{k+1}^{n+1}
          - \Delta f_k^{n+1} \right). 
\end{align}
In order to calculate the derivatives of the background variables we
use the TOV equations to express them in terms of undifferentiated
variables
\begin{align}
  N_{,x} &= -2\frac{N}{r} + 4\pi \rho, \\[10pt]
  \lambda_{,x} &= \frac{\lambda}{\Gamma} \left( N + 4\pi r P\right).
\end{align}
The remaining auxiliary variables follow from the definitions
\begin{align}
  r_x &= \left\{ \parbox{6cm}
                   {
                   $1 \hspace{1.2cm} {\rm if}\,\,\, x=r$ \\
                   $C \hspace{1.1cm} {\rm if}\,\,\, x=y$,
                   } \right. \label{FDE_RX}  \\[10pt]
  \Gamma &= 1-2Nr, \\[10pt]
  \Delta \Gamma &= -2(\hat{N} \xi + r\Delta N), \\[10pt]
  P &= K \rho^{\gamma}, \\[10pt]
  \hat{P} &= K \hat{\rho}^{\gamma}, \label{FDE_PPH} \\[10pt]
  \Delta P &= \hat{P} - P, \\[10pt]
  C^2 &= \frac{\partial P}{\partial \rho} = \gamma K \rho^{\gamma-1}.
         \label{FDE_C2}
\end{align}
The finite difference expressions used for Eqs.\,(\ref{NLAGR_DNT})
and (\ref{NLAGR_XIT}) are given by
\begin{align}
  f &= f_{k+1}, \\[10pt]
  \rho &= \frac{1}{2} \left( \sigma \rho_k + (2-\sigma) \rho_{k+1}
          \right), \\[10pt]
  \Delta f &= \frac{1}{2} \left( \Delta f^{n+1}_{k+1}
         + \Delta f^n_{k+1} \right), \\[10pt]
  \Delta \rho &= \frac{1}{4} \left[ \sigma (\Delta \rho^{n+1}_k + \Delta
         \rho^n_k) + (2-\sigma)(\Delta \rho^{n+1}_{k+1} + \Delta
         \rho^n_{k+1} ) \right], \\[10pt]
  \Delta f_{,t} &= \frac{1}{\Delta t} \left( \Delta f^{n+1}_{k+1}
         - \Delta f^n_{k+1} \right),
\end{align}
where the total pressure $\hat{P}$ is defined by Eq.\,(\ref{FDE_PPH}).
Finally we finite difference Eq.\,(\ref{NLAGR_DPT}) according to
\begin{align}
  f &= \frac{1}{2} \left( f_k + f_{k+1} \right), \\[10pt]
  \rho &= \frac{1}{2} \left[ (1+\sigma) \rho_k + (1-\sigma) \rho_{k+1}
       \right], \\[10pt]
  \Delta f &= \frac{1}{4} \left( \Delta f^{n+1}_k + \Delta f^n_k 
       + \Delta f^{n+1}_{k+1} + \Delta f^n_{k+1} \right), \\[10pt]
  \Delta f_{,x} &= \frac{1}{2\Delta x} \left( \Delta f^{n+1}_{k+1}
       - \Delta f^{n+1}_k + \Delta f^n_{k+1} - \Delta f^n_k \right),
       \\[10pt]
  \Delta f_{,t} &= \frac{1}{2\Delta t} \left( \Delta f^{n+1}_k
       - \Delta f^n_k + \Delta f^{n+1}_{k+1} - \Delta f^n_{k+1}
       \right), \\[10pt]
  \Delta \rho_{,x} &= \frac{1}{2\Delta x} \left( \Delta \rho^{n+1}_{k+1}
       - \Delta \rho^{n+1}_k + \Delta \rho^n_{k+1} - \Delta \rho^n_k
       \right), \label{FDE_EQ5DRHOX} \\[10pt]
  \Delta \rho_{,t} &= \frac{1}{2\Delta t} \left[ (1+\sigma) (\Delta
       \rho^{n+1}_k - \Delta \rho^n_k) + (1-\sigma) (\Delta \rho^{n+1}_{k+1}
       - \Delta \rho^n_{k+1} \right).
\end{align}
The auxiliary variables are again defined 
by Eqs.\,(\ref{FDE_RX})-(\ref{FDE_C2}). We also use the relations
\begin{align}
  \hat{C}^2 &= \gamma K \hat{\rho}^{\gamma - 1}, \\[10pt]
  \rho_{,x} &= \frac{P_{,x}}{C^2}, \\[10pt]
  \Delta P_{,x} &= (\hat{C}^2-C^2) \rho_{,x} + \hat{C}^2 \Delta \rho_{,x},
         \\[10pt]
  \Delta P_{,t} &= \hat{C}^2 \Delta \rho_{,t}, \\[10pt]
  \frac{\partial \hat{C}^2}{\partial \hat{\rho}} &= (\gamma - 1) 
         \frac{\hat{C}^2}{\hat{\rho}}.
\end{align}
The last relation is needed for the Newton-Raphson method
we use to solve the resulting system of non-linear algebraic equations
(cf. section \ref{relaxation}). \\
These finite difference equations result in $5K-5$ algebraic relations,
where $K$ is the total number of grid points. In order to determine the
$5K$ variables $\xi_k$, $w_k$, $\Delta N_k$, $\Delta \rho_k$
and $\Delta \lambda_k$ we also need the 5 boundary conditions
(\ref{NLAGR_BCINXI})-(\ref{NLAGR_BCOUTLAMBDA}) which we now write
as
\begin{align}
  \xi_1 &= 0, \\[10pt]
  w_1 &= 0, \\[10pt]
  \Delta N_1 &= 0, \\[10pt]
  \Delta \rho_K &= 0, \label{FDE_BCOUTDRHO} \\[10pt]
  \hat{\lambda}_K - \sqrt{1-2\hat{N}_K \hat{r}_K} &=0.
  \label{FDE_BCOUTLAMBDA}
\end{align}

\newpage
%Test
%
%%
%\begin{figure}[t]
%  \centering
%  \epsfig{file=EIGMODES/m3_eigxi.eps, height=200pt, width=150pt, angle=-90}
%  \epsfig{file=EIGMODES/m3_eigw.eps, height=200pt, width=150pt, angle=-90}
%  \epsfig{file=EIGMODES/m3_eigrdrho.eps, height=200pt, width=150pt, angle=-90}
%  \epsfig{file=EIGMODES/m3_eigdrho.eps, height=200pt, width=150pt, angle=-90}
%  \caption{The displacement $\xi$ as a function of the areal radius $r$
%           and the rescaled radius $y$ as well as the velocity $w$ and
%           the energy density $\delta \rho$ as a function of $y$ are
%           shown for the first four eigenmodes of model 3. 
%           For $\xi$ we have also
%           plotted mode 10 to illustrate the concentration of oscillations
%          towards larger $r$.}
%\end{figure}
%%

%=========================================================================
%\newpage

\bibliographystyle{chicagoa}
\addcontentsline{toc}{section}{References}
\bibliography{refthesis.bib}

\end{document}